\documentclass[twoside]{article}

\pdfoutput=1

\usepackage{times}
\usepackage{amsmath,amssymb,mathrsfs,mathtools,xspace}
\usepackage{graphicx}
\usepackage{feynmp}
\usepackage{url} 
\usepackage{imakeidx}
\usepackage{hyperref} 
\usepackage{longtable}
\usepackage{soul}
\usepackage{enumitem}
\usepackage{booktabs}
\usepackage{subcaption}
\usepackage{tikz}
\usepackage{xcolor}
\usepackage{stackrel}
\usepackage{multirow}
\usepackage{centernot}
\usepackage{cancel}
\usepackage[utf8]{inputenc}

\makeatletter
\@ifundefined{pdfoutput}{}{\DeclareGraphicsRule{*}{mps}{*}{}}
\makeatother

\usepackage[bottom]{footmisc}

\graphicspath{{figs/}}

\numberwithin{equation}{section}

\hypersetup{
        colorlinks=true,                        
        linkcolor={red!50!black},       
        citecolor={blue!50!black},      
        urlcolor={blue!80!black}        
}

\usetikzlibrary{arrows, shapes.geometric, arrows.meta, shapes, decorations.pathreplacing, fit, patterns, patterns.meta}
\definecolor{Gcolor}{HTML}{3b528b}
\definecolor{Dcolor}{HTML}{e41a1c}
\definecolor{Rcolor}{HTML}{E99595}
\definecolor{G2color}{HTML}{C5E0B4}
\definecolor{Bcolor}{HTML}{9DC3E6}
\definecolor{Ycolor}{HTML}{FFE699}
\definecolor{Gcolor_light}{HTML}{F1F8ED}

\tikzstyle{embed} = [rectangle, rounded corners=0.3ex, minimum width=1.5cm, minimum height=1cm, text centered, align=center, inner sep=0, fill=Ycolor, font=\large, draw]
\tikzstyle{transformer} = [rectangle, rounded corners, minimum width=6cm, minimum height=2.4cm, font=\large, fill=Gcolor_light, draw]
\tikzstyle{attention} = [rectangle, rounded corners=0.3ex, minimum width=5.5cm, minimum height=1.2cm, align=center, fill=G2color, draw, font=\large]
\tikzstyle{small_cinn} = [double arrow, double arrow head extend=0cm, double arrow tip angle=130, inner sep=0, align=center, minimum width=1.1cm, minimum height=0.5cm, fill=Bcolor, draw]
\tikzstyle{generator} = [rectangle, rounded corners, minimum width=3cm, minimum height=1cm,text centered, draw=Gcolor]
\tikzstyle{discriminator} = [rectangle, rounded corners, minimum width=3cm, minimum height=1cm,text centered, draw=Dcolor]
\tikzstyle{io} = [circle, trapezium left angle=70, trapezium right angle=110, minimum width=1cm, minimum height=1cm, text centered, draw=black]

\tikzstyle{process} = [rectangle, minimum width=1cm, minimum height=1cm, text centered, draw=black]
\tikzstyle{decision} = [rectangle, minimum width=1cm, minimum height=1cm, text centered, draw=black]

\tikzstyle{arrow} = [thick,->,>=stealth]
\tikzstyle{blok} = [rectangle, minimum width=1.0cm, minimum height=0.75cm, thin, draw=black]
\tikzstyle{expr} = [circle, minimum width=1.8cm, minimum height=1.8cm, text centered, align=center, inner sep=0, draw,font=\LARGE]
\tikzstyle{txt_huge} = [align=center, font=\Huge, scale=2]
\tikzstyle{txt} = [align=center, font=\LARGE]
\tikzstyle{cinn} = [double arrow, double arrow head extend=0cm, double arrow tip angle=130, shape border rotate=90, inner sep=0, align=center, minimum width=2.1cm, minimum height=2.3cm, fill=Bcolor, draw,font=\LARGE]
\tikzstyle{cinn_black} = [cinn, minimum height=2.5cm, fill=black]
\tikzstyle{arrow} = [thick,-{Latex[scale=1.0]}, line width=0.2mm, color=black]
\tikzstyle{line} = [thick, line width=0.2mm, color=black]
\tikzstyle{loss} = [rectangle, align=center,  minimum width=1.8cm, minimum height=1.5cm,fill=Rcolor,font=\LARGE, rounded corners]
\tikzstyle{xt} = [rectangle, align=center,  minimum width=4cm, minimum height=1.5cm,fill=G2color,font=\Large, rounded corners]
\tikzstyle{xts} = [rectangle, align=center,  minimum width=1cm, minimum height=1.5cm,fill=G2color,font=\Large, rounded corners]

\setitemize{itemsep=0.5pt,topsep=0.5pt,parsep=0pt,partopsep=0pt,leftmargin=*}
\setenumerate{itemsep=0pt,topsep=2pt,parsep=0pt,partopsep=0pt,labelindent=3pt,leftmargin=*}
\newcommand{\dz}{\phantom{0}}
\newcommand{\result}[2]{{#1}~$\pm$ {#2}}
\newcommand{\lrgen}{\text{LR}_\text{gen}}
\newcommand{\relu}{\text{ReLU}}
\newcommand{\kl}{D_\text{KL}}

\newcommand{\psim}{p_\text{sim}}

\newcommand{\pdj}[1]{p_{\text{data},#1}}
\newcommand{\pmdj}[1]{p_{\text{model},#1}}
\newcommand{\pl}{p_\text{latent}}
\newcommand{\pd}{p_\text{data}}
\newcommand{\pcs}{p_{\text{cross section}}}
\newcommand{\pmd}{p_\text{model}}

\newcommand{\punf}{p_\text{unfold}}
\newcommand{\xp}{x_\text{parton}}

\newcommand{\xr}{x_\text{reco}}

\newcommand{\loss}{\mathcal{L}}
\newcommand{\argmin}{\text{argmin}}
\newcommand{\argmax}{\text{argmax}}
\DeclareMathOperator{\softmax}{Softmax}
\DeclareMathOperator{\sigmoid}{Sigmoid}
\newcommand{\Langle}{\big\langle}
\newcommand{\Rangle}{\big\rangle}
\newcommand{\XLangle}{\Big\langle}
\newcommand{\XRangle}{\Big\rangle}
\newcommand{\XXLangle}{\Bigg\langle}
\newcommand{\XXRangle}{\Bigg\rangle}
\newcommand{\XXXLangle}{\Biggl\langle}
\newcommand{\XXXRangle}{\Biggr\rangle}

\newcommand{\vp}{\phi}

\newcommand{\vpjt}{\mbox{${\vp^\dag i\,\raisebox{2mm}{\boldmath ${}^\leftrightarrow$}\hspace{-4mm} D_\mu^{\,a}\,\vp}$}}

\newcommand{\obs}{\mathcal{O}}
\newcommand{\ord}{\mathcal{O}}

\newcommand{\normal}{\mathcal{N}}

\newcommand\one{\leavevmode\hbox{\small1\normalsize\kern-.33em1}}
\newcommand{\p}{\partial}

\newcommand{\met}{\slashchar{E}_T}

\newcommand{\lag}{\mathscr{L}}

\newcommand{\ope}[1]{\ensuremath{\mathcal{O}_{#1}}}

\newcommand{\mat}{\mathcal{M}}
\newcommand{\qqquad}{\qquad \qquad}
\newcommand{\qqqquad}{\qquad \qquad \qquad}

\newcommand{\really}{\stackrel{!}{=}}

\newcommand{\gev}{\text{GeV}}
\newcommand{\tev}{\text{TeV}}

\newcommand{\br}{\text{BR}}
\newcommand{\sign}{\text{sign}}
\newcommand{\ifb}{\text{fb}^{-1}}

\def\slashchar#1{\setbox0=\hbox{$#1$}           
   \dimen0=\wd0                                 
   \setbox1=\hbox{/} \dimen1=\wd1               
   \ifdim\dimen0>\dimen1                        
      \rlap{\hbox to \dimen0{\hfil/\hfil}}      
      #1                                        
   \else                                        
      \rlap{\hbox to \dimen1{\hfil$#1$\hfil}}   
      /                                         
   \fi}

\def\ie{{\sl i.e.} \,}

\usepackage{makeidx}
\makeindex[intoc]
\begin{document}

\title{Modern Machine Learning for LHC Physicists}

\author{
  Tilman Plehn$^{a,b}$\footnote{plehn\@@uni-heidelberg.de},
  Anja Butter$^{a,c}$,
  Barry Dillon$^{a,d}$,\\[1mm]
  Theo Heimel$^{a,e}$,
  Claudius Krause$^{a,f}$, and 
  Ramon Winterhalder$^{a,g}$
  \\[6mm]
  $^a$ Institute for Theoretical Physics, Heidelberg University, Germany
  \\[2mm]
  $^b$ Interdisciplinary Center for Scientific Computing (IWR), Heidelberg University, Germany
  \\
  $^c$ LPNHE, Sorbonne Universit\'e, Universit\'e Paris Cit\'e, CNRS/IN2P3, Paris, France
  \\
  $^d$ Intelligent Systems Research Centre, Ulster University, Northern Ireland
  \\
  $^e$ CP3, Universit\'e catholique de Louvain, Louvain-la-Neuve, Belgium
  \\
  $^f$ HEPHY, Austrian Academy of Sciences, Vienna, Austria
  \\
  $^g$ TIFLab, Universit\'a degli Studi di Milano \& INFN Sezione di Milano, Italy
}

\maketitle
\thispagestyle{empty}

\begin{abstract}
 Depending on the point of view, modern machine learning is either
 providing an unprecedented boost to the numerical methods of particle
 physics, or it is transforming the way we do science with vast amounts of
 complex data.  In any case, it is crucial for young researchers to
 stay on top of this development and apply cutting-edge methods and
 tools to all LHC physics tasks.  These lecture notes lead students
 with basic knowledge of particle physics and significant enthusiasm
 for machine learning to relevant applications.  They start with an
 LHC-specific motivation and a non-standard introduction to neural
 networks and then cover classification, unsupervised classification,
 generative networks, data representations, and inverse
 problems. Three themes defining much of the discussion are
 statistically defined loss functions, uncertainties, and accuracy. To
 understand the applications, the notes include some aspects of
 theoretical LHC physics. All examples are chosen from particle
 physics publications of the last few years, and many of them come with 
 corresponding
   \href{https://github.com/heidelberg-hepml/ml-tutorials}{tutorials}.
 \footnote{Given that these
   notes are by definition always outdated, they will be
   \href{https://www.thphys.uni-heidelberg.de/~plehn/pics/modern_ml.pdf}{updated
     frequently}}
\end{abstract}

\clearpage
\thispagestyle{empty}

\tableofcontents 

\thispagestyle{empty}

\clearpage
\thispagestyle{empty}
\raggedright
\section*{Welcome}

These notes are based on lectures in the 2022 Summer term and the 2023
Winter term, both held at Heidelberg University. The lectures were held on the
black board, corresponding to the formula-heavy style, but
supplemented with
\href{https://github.com/heidelberg-hepml/ml-tutorials}{tutorials}. The
notes start with a very brief motivation why LHC physicists are
naturally driven towards modern machine learning. Many people are
pointing out that the LHC Run~3 and especially the HL-LHC are going to
be a new experiment rather than a continuation of the earlier LHC
runs. One reason for this is the vastly increased amount of data and
the opportunities for analysis and inference, inspired and triggered
by data science as a new common language of particle experiment and
theory.

The introduction to neural networks is meant for future particle
physicists, who know basic numerical methods like fits or Monte Carlo
simulations. We tell our ML-story through a series of original
publications. We start with supervised classification, which is how
everyone in particle physics is coming into contact with modern neural
networks. We then move on to non-supervised classification, since no
ML-application at the LHC is properly supervised, and the big goal of
the LHC is to find physics that we do not yet know about. A major part
of the lecture notes is then devoted to generative networks, because
one of the defining aspects of LHC physics is the combination of
first-principle simulations and high-statistics datasets. Finally, we
present some ideas how LHC inference can benefit from ML-approaches to
inverse problems. The last chapter on symbolic regression is meant to
remind us that numerical methods are driving much of physics research,
but that the language of physics remains formulas, not computer code.

As the majority of authors are German, we would like to add two
apologies before we start. First, many of the papers presented in the
different sections come from the Heidelberg group. This does not mean
that we consider them more important than other papers, but for those
we know what we are talking about and the corresponding tutorials are
based on the codes used in our papers. Second, we apologize that these
lecture notes do not provide a comprehensive list of references beyond
the papers presented in the individual chapters. Aside from copyright
issues, the idea of these references that it should be easy so switch
from a lecture-note mode to a paper-reading mode. For a comprehensive
list of references we recommend the Living Review of Machine Learning
for Particle Physics~\cite{Feickert:2021ajf}.

Obviously, these lecture notes are outdated before a given version appears on
the arXiv. Our plan is to update them regularly, which will also allow
us to remove typos, correct wrong arguments and formulas, and improve
discussions. Especially young readers who go through these notes from
the front to the back, please mark your questions and criticism in a
file and send it to us. We will be grateful to learn where we need to
improve these notes.

Talking about --- we are already extremely grateful to the people who
triggered the machine learning activities in our Heidelberg group:
Kyle Cranmer, Gregor Kasieczka, Ullrich K\"othe, and Manuel
Hau{\ss}mann. We are also extremely grateful to our machine learning
collaborators, including David Shih, Ben Nachman, Daniel Whiteson,
Michael Kr\"amer, Jesse Thaler, Stefano Forte, Martin Erdmann, Sven
Krippendorf, Jernej Kamenik, Peter Loch, Aishik Ghosh, Eilam Gross,
Tobias Golling, Michael Spannowsky, and many others.  The same holds
for all current and former group members in Heidelberg, by now just
too many to name.

Most importantly, we want to thank the great ML4Jets community, because
without those meetings machine learning at the LHC would be as
uninspiring as so many other fields, and nothing the unique science
endeavor it now is.

We hope that you will enjoy reading these notes, that you will
discover interesting aspects, and that you can turn your new ideas
into cool papers!
\bigskip
\bigskip

Tilman, Anja, Barry, Theo, Claudius, and Ramon

\clearpage
\begin{fmffile}{feynman}
\setcounter{page}{1}
\section{Basics}
\label{sec:basics}

\subsection{Particle physics}
\label{sec:basics_particle}

Four key ingredients define modern particle physics in general and
modern LHC physics in particular,
\begin{itemize}
\item fundamental physics questions;
\item huge datasets;
\item full uncertainty control;
\item precision simulations from first principles.
\end{itemize}
For the last aspect it is important that we distinguish between some
kind modelling and a proper first-principle simulation. Only the
latter leads to scientific insights.

What has changed after the first two LHC runs is that we are less and
less interested in testing pre-defined models for physics beyond the
Standard Model (BSM).  The last discovery that followed this kind of
analysis strategy was the Higgs boson in 2012. We are also not that
interested in measuring parameters of the Standard Model Lagrangian,
with very few notable exceptions linked to our fundamental physics
questions. What we care about is the particle content and the
fundamental symmetry structure\index{symmetries}, all encoded in the Lagrangian that
describes \underline{LHC data in its entirety}.

The multi-purpose experiments ATLAS and CMS, as well as the more
dedicated LHCb experiment are trying to get to this fundamental
physics goal.  During the LHC Runs~3 and~4, or HL-LHC, they will
record as many interesting scattering events as possible and
understand them in terms of quantum field theory predictions at
maximum precision. With the expected dataset, concepts and tools from
data science have the potential to transform LHC research.  Given that
the future LHC runs will collect 25 times the amount of Run~1 and
Run~2 data over the next 10-15 years, such a new approach is not only
an attractive option, it is the only way to properly analyze these new
datasets. With this perspective, the starting point of this lecture is
to understand LHC physics as \underline{field-specific data science},
unifying theory and experiment.

Before we see how modern machine learning can help us with many
aspects of LHC physics, we briefly review the main questions behind an
LHC analysis from an ML-perspective.

\begin{figure}[t]
    \centering
    \includegraphics[width=0.55\textwidth]{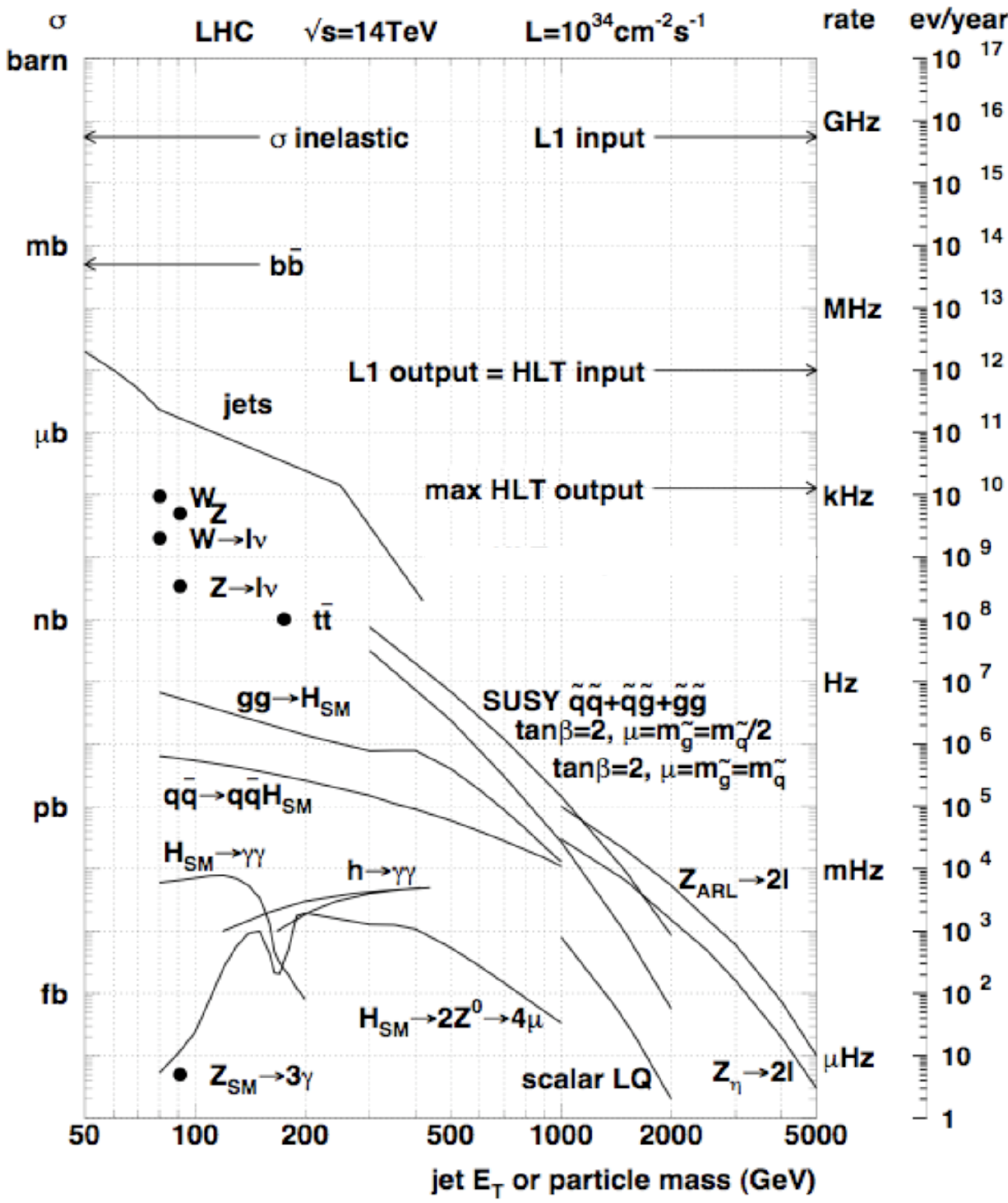}
    \caption{Production rates for the LHC. Figure from
      Ref.~\cite{ATLAS:2009zsq}.}
    \label{fig:lhc_rates}
\end{figure}

\subsubsection{Data recording}
\label{sec:basics_particle_data}

The first LHC challenge is the sheer amount of data produced by ATLAS
and CMS. The two proton beams cross each other every 25~ns or at
40~MHz, and a typical event consists of millions of detector channels,
requiring 1.6~MB of memory. This data output of an LHC experiments
corresponds to 1~PB per second. \underline{Triggering} is another word
for accepting that we cannot write all this information on tape and
analyze is later, and we are also not interested in doing that. Most
of the proton-proton interactions do not include any interesting
fundamental physics information, so in the past we have just selected
certain classes of events to write to tape. For model-driven searches
this strategy was appropriate, but for our modern approach to LHC
physics it is not. Instead, we should view triggering as some kind of
\underline{data compression} of the incoming LHC data, including
compressing events, compressing event samples, or selecting events.

In Fig.~\ref{fig:lhc_rates} we see that between the inelastic
proton-proton scattering cross section, or rate, of around 600~mb and
hard jet production at a rate around $1~\mu$b we can afford almost a
factor $10^{-6}$ in data reduction, loss-less when it comes to the
fundamental physics questions we care about.  This is why a first,
level-one (L1) trigger can reduce the rate from an input rate of 40~MHz to
an output rate around 100~kHz, without losing interesting physics,
provided we make the right trigger decisions. To illustrate this
challenge, the time a particle or a signal takes to cross a detector
at the speed of light is
\begin{align}
  \frac{10~\text{m}}{3 \cdot 10^8~\text{m/s}}
  \approx 3 \cdot 10^{-8}~\text{s}
  = 30~\text{ns} 
\end{align}
or around one bunch-crossing time.  As a starting point, the
L1-trigger uses very simple and mostly local information from the
calorimeters and the muon system, because at this level it is already
hard to combine different regions of the detector in the L1 trigger
decision. From a physics point of view this is a problem because
events with two forward jets are very different depending on the
question if they come with central jets as well. If yes, the event is
likely QCD multi-jet production and not that interesting, if no, the
event might be electroweak vector boson fusion and very relevant for
many Higgs and electroweak gauge boson analyses.

After the L1 hardware trigger there is a second, software-based
high-level (HL) or L2 trigger. It takes the L1 trigger output at
100~kHz and reduces the rate to 3~kHz, running on a farm with more
than 10.000 CPU cores already now. After that, for instance ATLAS runs
an additional software-based L3 trigger, reducing the data rate from
3~kHz to 200~Hz. For 1.6~MB per event, this means that the experiment
records 320~MB per second for the actual physics analyses. Following
Fig.~\ref{fig:lhc_rates} a rate of 200~Hz starts to cut into Standard
Model jet production, but covers the interesting SM processes as well
as typical high-rate BSM signals (not that we believe that they
describe nature).

This trigger chain defines the standard data acquisition by the LHC
experiments and its main challenges related to the amount of
information which needs to be analyzed. The big problem with
triggering by selection is that it really is data compression with
large losses, based on theoretical inspiration on interesting or
irrelevant physics. Or in other words, if theorists are right,
triggering by selection is lossless, but the track record of theorists
in guessing BSM physics at the LHC is not a success story.

Even for Run~2 there were ways to circumvent the usual triggers and
analyze data either by randomly choosing data to be written on tape
(prescale trigger) or by performing analyses at the trigger level and
without access to the full detector information (data scouting). One aspect that 
is,
for instance, not covered by standard ATLAS and LHC triggers are
low-mass di-jet resonances and BSM physics appearing inside jets.  The
prize we pay for this kind of event-level compression is that again it
is not lossless, for instance when we need additional information to
improve a trigger-level analysis later. Still, for instance LHCb is
running a big program of encoding even analysis steps on programmable
chips, FPGAs, to compress their data flow.

Before we apply concepts from modern data science to triggering we
should once again think about what we really want to
achieve. Reminding ourselves of the idea behind triggering, we can
translate the trigger task into deep-learning language as one of two
objectives,
\begin{itemize}
\item fast identification of events corresponding to a known,
  interesting class;
\item fast identification of events which are different from our
  Standard Model expectations;
\item compress data such that it can be used best.
\end{itemize}
The first of these datasets can be used to measure, for example, Higgs
properties, while the second dataset is where we search for physics
beyond the Standard Model. As mentioned above, we can use compression
strategies based on event selection, sample-wise compression, and
event-level compression to deal with the increasingly large datasets
of the coming LHC runs.  Conceptually, it is more interesting to think
about the \underline{anomaly-detection}\index{anomaly detection} logic behind the second
approach. While it is, essentially, a literal translation of the
fundamental goal of the LHC to explore the limitations of the Standard
Model and find BSM physics, is hardly explored in the classic
approaches. We will see that modern data science provides us with
concepts and tools to also implement this new trigger strategy.

\subsubsection{Jet and event reconstruction}
\label{sec:basics_particle_reco}

After recording an event, ATLAS and CMS translate the detector output
into information on the particles which leave the interaction points,
hadrons of all kind, muons, and electrons. Neutrinos can be
reconstructed as missing transverse momentum, because in the azimuthal
plane we know the momenta of both incoming partons.

To further complicate things, every bunch crossing at the HL-LHC consists
of 150-200 overlapping proton-proton interactions. If we assume that
one of them might correspond to an interesting production channel,
like a pair of top quarks, a Higgs boson accompanied by a hard jet or
gauge boson, or a dark matter particle, the remaining 149-199
interactions are refereed to as pileup and need to be removed. For
this purpose we rely on tracking information for charged particles,
which allow us to extrapolate particles back to the primary
interaction point of the proton-proton collision we are interested
in. Because the additional interaction of protons and of partons
inside a proton, as well as soft jet radiation, do not have
distinctive patterns, we can also get rid of them by subtracting
unwanted noise for instance in the calorimeter. Denoising is a
standard methodology used in data science for image analyses.

After reconstructing the relevant particles, the hadrons are clustered
into jets, corresponding to hard quarks or gluons leaving the
interaction points. These jets are traditionally defined by
\underline{recursive algorithms}, which cluster constituents into a
jet using a pre-defined order and compute the 4-momentum of the jet
which we use as a proxy for the 4-momentum of the quark of gluon
produced in the hard interaction. The geometric separation of two LHC
objects is defined in terms of the azimuthal angle $\phi_{ij} \in
[0,\pi]$ and the difference in rapidity, $\Delta \eta_{ij} = |\eta_i -
\eta_j|$. In the most optimistic scenario the LHC rapidity coverage is
$|\eta_i| \lesssim 4.5$, for a decent jet reconstruction or $b$-jet
identification is around $|\eta_i| \lesssim 2.5$.  This 2-dimensional
plane is what we would see if we unfolded the detector as viewed from
the interaction point.  We define the geometric separation in this
\underline{$\eta-\phi$ plane} as
\begin{align}
 R_{ij} = \sqrt{ (\Delta \phi_{ij})^2 + (\Delta \eta_{ij})^2 } \; .
\label{eq:def_r}
\end{align}
Jets can also be formed by hadronically decaying tau leptons or
$b$-quarks, or even by strongly boosted, hadronically decaying $W$,
$Z$, and Higgs bosons or top quarks. The top quark is the only quark
that decays before it hadronizes. In all of these cases we need to
construct the energy and momentum of the initial particle and its
particle properties from the jet constituents, including the
possibility that BSM physics might appear inside jets.  Identifying
the partonic nature of a jet is called \underline{jet tagging}\index{jet tagging}.  The
main information on jets comes from the hadronic and electromagnetic
calorimeters, with limited resolution in the $\eta-\phi$ plane. The
tracker adds more information at a much better angular resolution, but
only for the charged particles in the jet. The combination of
calorimeter and tracking information is often referred to as particle
flow.

When relating jets of constituents we need to keep in mind a
fundamental property of QFT: radiating a soft photon or gluon from a
hard electron or quark can, in the limit $E_{\gamma,g} \to 0$, have no
impact on any kinematic observable. Similarly, it cannot make a
difference if we replace a single parton by a pair of partons arising
from a collinear splitting. In both, the soft and collinear limits,
the corresponding splitting probabilities described by QCD are
formally divergent, and we have to resum these splittings to define a
hard parton beyond leading order in perturbation theory. Because any
detector has a finite resolution, and the calorimeter resolution is
not even that good, these divergences are not a big problem for many
standard LHC analyses, but when we define high-level kinematic
observables to compare to QFT predictions, these observables should
ideally be infrared and collinearly safe. An example for an unsafe
observable is the number of particle-flow objects inside a jet.

Finally, details on jets only help us understand the underlying hard
scattering through their correlations with other particles forming an
event. This means we need to combine the subjet physics information
inside a jet with correlations describing all particles in an
event. This combination allows us, for instance to reconstructs a
Higgs decaying to a pair of bottom quarks or a top decaying
hadronically, $t \to W^+ b \to jjb$, or leptonically $t \to W^+ b \to
\ell^+ \nu b$. However, fundamentally interesting information requires
us to understand complete \underline{events} like for instance
\begin{align}
  pp
  \to t \bar{t} H + \text{jets}
  \to (jj b) \; (\ell^- \bar{\nu} \bar{b}) \; (b\bar{b}) + \text{jets} \; .
\label{eq:tth}
\end{align}
Again applying a deep-learning perspective, the reconstruction of LHC
jets and events includes tasks like 
\begin{itemize}
\item fast identification of particles through their detector
  signatures;
\item data denoising to extract features of the relevant scattering;
\item jet tagging and reconstruction using calorimeter and tracker;
\item combination of low-level high-resolution with high-level
  low-resolution observables.
\end{itemize}
Event reconstruction and kinematic analyses have been using
multivariate analysis methods for a very long time, with great
success for example in $b$-tagging. Jet tagging is also the field of
LHC physics where we are making the most rapid and transformative
progress using modern machine learning, specifically classification
networks. The switch to event-level tagging, on the other hand, is an
unsolved problem.

\subsubsection{Simulations}
\label{sec:basics_particle_sim}

\begin{figure}[t]
    \centering
    \includegraphics[width=0.99\textwidth]{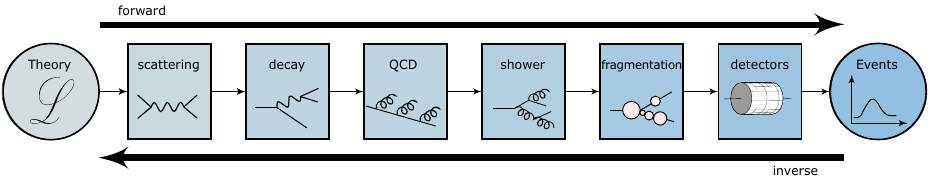}
    \caption{Illustrated forward and inverse simulation chain for LHC
      physics.}
    \label{fig:simchain}
\end{figure}

Simulations are the way we provide theory predictions for the LHC
experiments. A \underline{Lagrangian} encodes the structures of the
underlying quantum field theory. For the Standard Model this includes
a $SU(3)\times SU(2) \times U(1)$ gauge group with the known
fundamental particles, including one Higgs scalar. This theory
describes all collider experiments until now, but leaves open
cosmological questions like dark matter or the baryon asymmetry of the
Universe, which at some point need to be included in our
Lagrangian. It also ignores any kind of quantum gravity or
cosmological constant.  Because we use our LHC simulations not only
for background processes, but also for potential signals, the input to
an LHC simulation is the Lagrangian. This means we can simulate LHC
events in any virtual world, provided we can describe it with a
Lagrangian. From this Lagrangian we then extract known and new
particles with their masses and all couplings.  The universal
simulation tools used by ATLAS and CMS are Pythia as the standard
tool, Sherpa with its excellent QCD description, MadGraph with its
unique flexibility for BSM searches, and Herwig with its excellent
hadronization description.

The basic elements of the LHC simulation chain are illustrated in
Fig.~\ref{fig:simchain}, and many more details can be found in
Ref.~\cite{Plehn:2009nd}. Once we define our underlying theory
Lagrangian, meaning the Standard Model without or with hypothetical
new particles and interactions, we can compute the \underline{hard
  scattering} amplitude. Following our $t\bar{t}H$ example in
Eq.\eqref{eq:tth} the hard scattering can be defined in terms of top
and Higgs, but if we want to include angular correlations in the top
\underline{decays}, the hard process will include the decays and
include four $b$-quarks, a lepton, a neutrino, and two light-flavor
quarks. If decaying particles contribute to such a phase space
signature, we need to regularize the divergent on-shell propagators
through a resummation which leads to a Breit-Wigner propagator and
introduces a physical particle width to remove the on-shell
divergence. Breit-Wigner propagators are one example for localized and
strongly peaked feature in the matrix element as a function of phase
space. In our $t\bar{t}H$ example process two top resonances, one
Higgs resonance, and two $W$-resonances lead to numerical challenges
in the phase space description, sampling, and integration. Finally,
the transition amplitudes are computed in perturbative QCD. To leading
order or at tree level, these amplitudes can be generated extremely
fast by our standard generators. Nobody calculates those cross
sections by hand anymore, and the techniques used by the automatic
generators have little to do with the methods we tend to teach in our
QFT courses. At the one-loop or 2-loop level things can still be
automized, but the calculation of amplitudes including virtual QCD
corrections and additional jet radiation can be time-consuming.

Moving on in Fig.~\ref{fig:simchain}, strongly interacting partons can
radiate gluons and split off quarks into a \underline{collinear phase
  space}. For instance looking at incoming gluons, they are described
by parton densities beyond leading order in QCD only once we include
this collinear radiation into the definition of the incoming
gluons. The same happens for strongly interacting particles in the
final state, referred to as fragmentation. For our reference process
in Eq.\eqref{eq:tth} we are lucky in that heavy top quarks do not
radiate many gluons. The universal initial and final state radiation,
described by QCD beyond strict leading order gives rise to the
additional jets indicated in Eq.\eqref{eq:tth}. If we go beyond
leading order in $\alpha_s$, we need to include hard jet radiation,
which appears at the same order in perturbation theory as the virtual
corrections. From a QFT-perspective virtual and real corrections lead
to infrared-divergent predictions individually and therefore cannot be
treated separately. More precise simulations in perturbation theory
also lead to higher-multiplicity final states. From a machine learning
perspective this means that LHC simulations cannot be defined in terms
of a fixed phase-space dimensionality. In addition, it illustrates how
for LHC simulation we can often exchange \underline{complexity and
  precision}, which in return means that faster simulation tools are
almost automatically more precise.

Hard initial-state and final-state radiation is only one effect of
collinear and soft QCD splittings\index{QCD splittings}. The splitting of quarks and gluons
into each other can be described fairly well by the 2-body splitting
kernels, and these kernels describe the leading physics aspects of
parton densities. In addition, successive QCD splittings define the
so-called \underline{parton shower}, which means they describes how a
single parton with an energy of 100~GeV or more turns into a spray of
partons with individual energies down to 1~GeV. There are several
approaches to describing the parton shower, which share the simple set
of QCD splitting kernels, but differ in the way the collinear
radiation is spread out to fill the full phase space and in which order
partons split. Improving the precision of parton showers to match the
experimental requirements is one of the big challenges in theoretical
LHC physics.

Next, the transition from quarks and gluons to mesons and baryons, and
the successive hadron decays are treated by \underline{hadronization
  of fragmentation} tools. From a QCD perspective those models are the
weak spot of LHC simulations, because we are only slowly moving from
ad-hoc models to first-principle QCD predictions. A precise
theoretical description of many hadronization processes and hadron
decays is challenging, so many features of hadron decays are extracted
from data. Here we typically rely on the kinematic features of
Breit-Wigner resonances combined with continuum spectra and form
factors computed in low-energy QCD. The LHC simulation chain up to
this point is developed and maintained by theorists.

The finally step in Fig.~\ref{fig:simchain} is where the particles
produced in LHC collisions enter the \underline{detectors} and are
analyzed by combining many aspects and channels of ATLAS, CMS, or
LHCb.  From a physics perspective detectors are described by the
interaction of relativistic particles with the more or less massive
different detector components. This interaction leads to
electromagnetic and hadronic showers, which we need to describe
properly if we want to simulate events based on a hypothetical
Lagrangian. Because we do not expect to learn fundamental physics from
the detector effects, and because detector effects depend on many
details of the detector materials and structures, these simulations
are in the hands of the experimental collaborations. The standard full
simulations are based on the detailed construction plans of the
detectors and use the complex and quite slow Geant4 tool for the full
simulations. Fast simulations are based on these full simulations,
ignore the input of the geometric detector information, and just
reproduce the observed signals and measurements for a given particle
entering the detector. Historically, they have relied on Gaussian
smearing, but modern fast simulations are much more complex and
precise.

If we are looking for deep-learning applications, first-principle
simulations include challenges like
\begin{itemize}
\item optimal phase space coverage and mapping of amplitude features;
\item fast and precise surrogate models for expensive loop amplitudes;
\item variable-dimensional and high-dimensional phase spaces;
\item improved data- and theory-driven hadron physics, like
  heavy-flavor fragmentation;
\end{itemize}
Once we can simulate LHC events all the way to the detector output,
based on an assumed fundamental Lagrangian, and with high and
controlled precision, we can use these simulated events to extract
fundamental physics from LHC data. While not all LHC predictions can
be included in this forward simulation, the multi-purpose event
generators\index{event generators} and the corresponding detector simulations are the work
horses behind every single LHC analysis. They define LHC physics as
much as the fact that the LHC collides two protons (or more), and
there are infinitely many ways they can benefit from modern machine
learning~\cite{Butter:2022rso}.

\subsubsection{Inference}
\label{sec:basics_particle_inf}

LHC analyses are almost exclusively based on frequentist or
\underline{likelihood methods}, and we are currently observing a slow
transition from classic Tevatron-inherited analysis strategies to
modern LHC analysis ideas. In an ideal LHC world, we would just
compare observed events with simulated events based on a given theory
hypothesis. From the \underline{Neyman-Pearson lemma}\index{Neyman-Pearson lemma} we know that the
likelihood ratio is the optimal way to compare two hypotheses and
decide if a background-only model or a combined signal plus background
model describes the observed LHC data better. This means we can assign
any LHC dataset a confidence level for the agreement between
observations and first-principle predictions and either discover or
rule out BSM models with new particles and interactions. The
theory-related assumption behind such analyses is that we can provide
precise, flexible, and fast event generation for SM-backgrounds and
for all signals we are interested in.

If we compare observed and predicted datasets, a key question is how
we can set up the analysis such that it provides the best possible
measurement. For two hypotheses we know that 
an observable encoding the likelihood ratio combines all available information
into a sufficient statistics. The question is how we can first define
and then experimentally reconstruct this \underline{optimal
  observable}\index{optimal observable}. Going back to our $t\bar{t}H$ example, we can for
instance try to measure the top-Higgs Yukawa coupling. This coupling
affects the signal rate simply as $\sigma_\text{tot}(t\bar{t}H)
\propto y_t^2$ and does not change the kinematics of the signal
process, so we can start with a simple $\bar{t}H$ rate
measurement. Things get more interesting when we search for a
modification of the top-Higgs couplings through a higher-dimensional
operator, which changes the Lorentz structure of the coupling and
introduces a momentum dependence. In that case the effect of a shifted
coupling depends on the phase space position. Because the
signal-to-background ratio also changes as a function of phase space,
we need to find out which phase space regions work best to extract
such an operator-induced coupling shift. The answer will depend on the
luminosity, and we need an optimal observable framework to optimize
such a full phase-space analysis. Finally, we can try to test
fundamental symmetries\index{symmetries} in an LHC process, for instance the CP-symmetry
of the top-Higgs Yukawa coupling. For this coupling, CP-violation
would appear as a phase in the modified Yukawa coupling and affect the
interference between different Feynman diagrams over phase
space. Again, we can define an optimal observable for CP-violation,
usually an angular correlation.

From a simulation point of view, illustrated in
Fig.~\ref{fig:simchain}, the question of how to measure an optimal
observable points to a structural problem. While we can assume that
such an optimal observable is naturally defined at the parton or hard
scattering level, the measurement has to work with events measured by
the detector. The challenge becomes how to best link these different
levels of the event generation and simulation to define the
measurement of a Lagrangian parameter?

If we want to interpret measurements as model-independently as
possible and in the long term, we need to report experimental
measurements without detector effects. We can assume that the detector
effects are independent on the fundamental nature of an event, so we
translate a sample of detector-level events into a corresponding
sample of events before detector effects and therefore entirely
described by fundamental physics. From a formal perspective, we want
to use a forward detector simulation to define a \underline{detector
  unfolding}\index{unfolding} as an incompletely defined inverse problem.

Going back to Fig.~\ref{fig:simchain} the possibility of detector
unfolding leads to the next question, namely why we do not also unfold
other layers of the simulation chain based on the assumption that they
will not be affected by the kind of physics beyond the Standard Model
we aim for. It is safe to assume that testing fragmentation models
will not lead to a discovery of new particles, and the same can be
argued (or not) for the parton shower. This is why it is standard to
unfold or invert the shower and fragmentation steps of the forward
simulations through recursive jet algorithms\index{jet algorithm}.

Next, it is reasonable to assume that BSM features like heavy new
particles of momentum-dependent effective operators affect heavy
particle production in specific phase space regions much more than the
well-measured decays of gauge bosons or top quarks. For the
$t\bar{t}H$ signal we would then not be interested in the top and
Higgs decays, because they are affected with limited momentum
transfer, and we can unfold these decays. This method is applied very
successfully in the top groups of ATLAS and CMS, while other working
groups are less technically advanced.

Finally, we can remind ourselves that what we really want to compute
for any LHC process is a likelihood ratio. Modulo prefactors, the
likelihood for a given process is just the transition amplitude for
the hard process. This means that for example for two model or
model-parameter hypotheses based on the same hard process, the
likelihood ratio can be extracted easily by inverting the entire LHC
simulation chain and extracting the parton-level matrix elements
squared for a given observed events. This analysis strategy is called
the matrix element method and has in the past been applied to
especially challenging signals with small rates.

Altogether, \underline{simulation-based inference} methods immediately bear the
question how we want to compare simulation and data most precisely.
If our simulation chain only works in the forward direction, we have
no choice but to compare predictions and data at the event
level. However, if our simulation chain can be inverted, we generate
much more freedom. A very practical consideration might then also be
that we are able to provide precision-QCD predictions for certain
kinematic observables at the parton level, but not as part of a fast
multi-purpose Monte Carlo. In this situation we can choose the
\underline{point of optimal analysis} along the simulation chain shown
in Fig.~\ref{fig:simchain}.

In a data-oriented language some of the open questions related to
modern LHC inference concern
\begin{itemize}
\item precision simulations of backgrounds and signals in optimized
  phase space regions;
\item fast and precise event generation\index{event generators} for flexible signal
  hypotheses;
\item definition of optimal analysis strategies and optimal
  observables;
\item unfolding or inverted forward simulation using a consistent
  statistical procedure;
\item single-event likelihoods to be used in the matrix element
  method;
\end{itemize}
Many of these questions have a long history, starting from LEP and the
Tevatron, but for the vast datasets of the LHC experiment we finally
need to find conceptual solutions for a systematic inversion of our
established and successful simulation chain.

\subsubsection{Uncertainties}
\label{sec:basics_particle_unc}

Uncertainties\index{uncertainties} are extremely serious business in
particle physics, because if we ever declare a major discovery we need
to be sure that this statement lasts\footnote{This rule has
  traditionally not applied to the CDF experiment at the
  Tevatron.}. We define, most generally, two kinds of uncertainties,
\underline{statistical uncertainties}\index{statistical uncertainty} and \underline{systematic
  uncertainties}\index{systematic uncertainty}.  The first kind is defined by the fact that they
vanish for large statistics and are described by Poisson and
eventually Gaussian distributions in any rate measurement. The second
kind do not vanish with large statistics, because they come from some
kind of reference measurement or \underline{calibration}, because they
describe detector effects, or they arise from theory predictions which
do not offer a statistical interpretation. Some systematic
uncertainties are Gaussian\index{systematic uncertainty}, for example when they describe a
high-statistics measurement in a control region. Others just give a
range and no preference for a single central value, for instance in
the case of theory predictions based on perturbative QCD. Again, the
distribution of the corresponding nuisance parameter reflects the
nature of the systematic or theory uncertainties\index{theory uncertainty}.

The machine learning community also distinguishes between two kinds of
uncertainties, \underline{aleatoric uncertainties} related to the
(training) data and \underline{epistemic uncertainties} related to the
model we use to describe the data. This separation is similar to
statistical and systematic uncertainties: first, epistemic
uncertainties vanish when we train a network on more, assumed perfect,
data --- in physics we would call them statistical uncertainties;
aleatoric uncertainties arise from imperfect data, such that more of
the same data will not help --- in physics this defines systematic
uncertainties. One of the issues is that we typically work towards the
the limit of a perfectly trained network which reproduce all features
of the data. Deviations from this limit are reproduced by the
epistemic uncertainty, but the same limit requires increasingly large
networks which we need to train on correspondingly large datasets.

Technically, we include uncertainties in a likelihood analysis using
hundreds or thousands of nuisance parameters. Any statistical model
for a dataset $x$ then depends on nuisance parameters $\nu$ and
parameters of interest $g$, to define a likelihood
$p(x|\nu,g)$. Instead of using Bayes' theorem\index{Bayes' theorem} to extract a
probability distribution $p(g|x,\nu)$, we use profile likelihood
techniques to constrain $g$. These techniques do not foresee
priors, unless we can describe them reliably as nuisance parameters
reflecting measurements or theory predictions with a well-defined
likelihood distribution. Whenever nuisance parameters come from a
measurement, we expect their distribution to allow for a frequentist
interpretations.

If we start with the assumption that the definition of an observable
does not induce an uncertainty on the measurement, any observable will
first be analyzed using simulations. Here we can ensure its resilience
to detector effects, or, if needed, its appropriate QFT definition. In
the current LHC strategy, any numerically defined observable, like a
kinematic reconstruction, a boosted decision tree, or a neural network
will actually be defined on simulated data.  In the next step, this
observable needs to be calibrated by combining data and simulations. A
standard, nightmare task in ATLAS and CMS is the precise calibration
of QCD jets. As a function of the many parameters of the jet algorithm
this calibration will for instance determine the jet energy scale from
controlled reference data like on-shell $Z$-production. Calibration
can remove systematic biases, and it always comes with a range of
uncertainties in the reference data, which include statistical
limitations as well as theory uncertainties\index{theory uncertainty} from the way the reference
data is described by Monte Carlo simulations.  We will see that for
ML-based observables this second step can and should be considered
part of the training, in which case the training has to account for
uncertainties.

In the inference, we need to consider all kinds of uncertainties,
together with the best choice of observables, to provide the optimal
measurement. Because theory predictions are based on perturbative QFT,
they always come with an uncertainty which we can sometimes quantify
well, and which we sometimes know very little about. What we do know
is that this theory uncertainty cannot be defined by a frequentist
argument. Similarly, some systematic uncertainties are hard to
estimate, for example when they correspond to detector effects which
are not Gaussian or when we simply do not know the source of a bias
for example in a calibration. Quantifying, controlling, and reducing
all uncertainties is the challenge of any LHC analysis.

Finally, the uncertainty in defining an observable might not be part
of the uncertainty treatment at the analysis level, but it will affect
its optimality. This means that especially for numerically defined
observable we need to control underlying uncertainty like the
statistics of the training data, systematic affects related to the
training, or theoretical aspects related to the definition of an
observable. Historically, these uncertainties mattered less, but with
the rapidly growing complexity of LHC data and analyses, they should
be controlled.

This means that again in the language closer to machine learning we
have to work on
\begin{itemize}
\item controlled definitions and resilience of observables;
\item calibration leading to nuisance parameters for the uncertainty;
\item control of the features learned by a neural network;
\item uncertainties on all network outputs from classification to generation;
\item balance optimal observables with uncertainties.
\end{itemize}
These requirements are, arguably, the biggest challenge in applying
ML-methods to particle physics and the reason for conservative
reservations especially in the ATLAS experiment. Given that we have no
alternative to thinking of LHC physics as a field-specific application
of data science, we have to work on the treatment of uncertainties in
modern ML-techniques.

\subsection{Machine learning basics}
\label{sec:basics_deep}

After setting the physics stage, we will briefly review some
fundamental concepts which we need to then talk about machine learning
at the LHC. We will not follow the usual line of many introductions to
neural networks, but choose a particle physics path. This means we
will introduce many concepts and technical terms using multivariate
analyses and decision trees, then review numerical fits, and with that
basis introduce neural networks with likelihood-based training. We
will end this section with a state-of-the-art application of learning
transition amplitudes over phase space.

\subsubsection{Multivariate classification}
\label{sec:basics_deep_multi}

\begin{figure}[b!]
  \centering
  \includegraphics[width=0.32\textwidth]{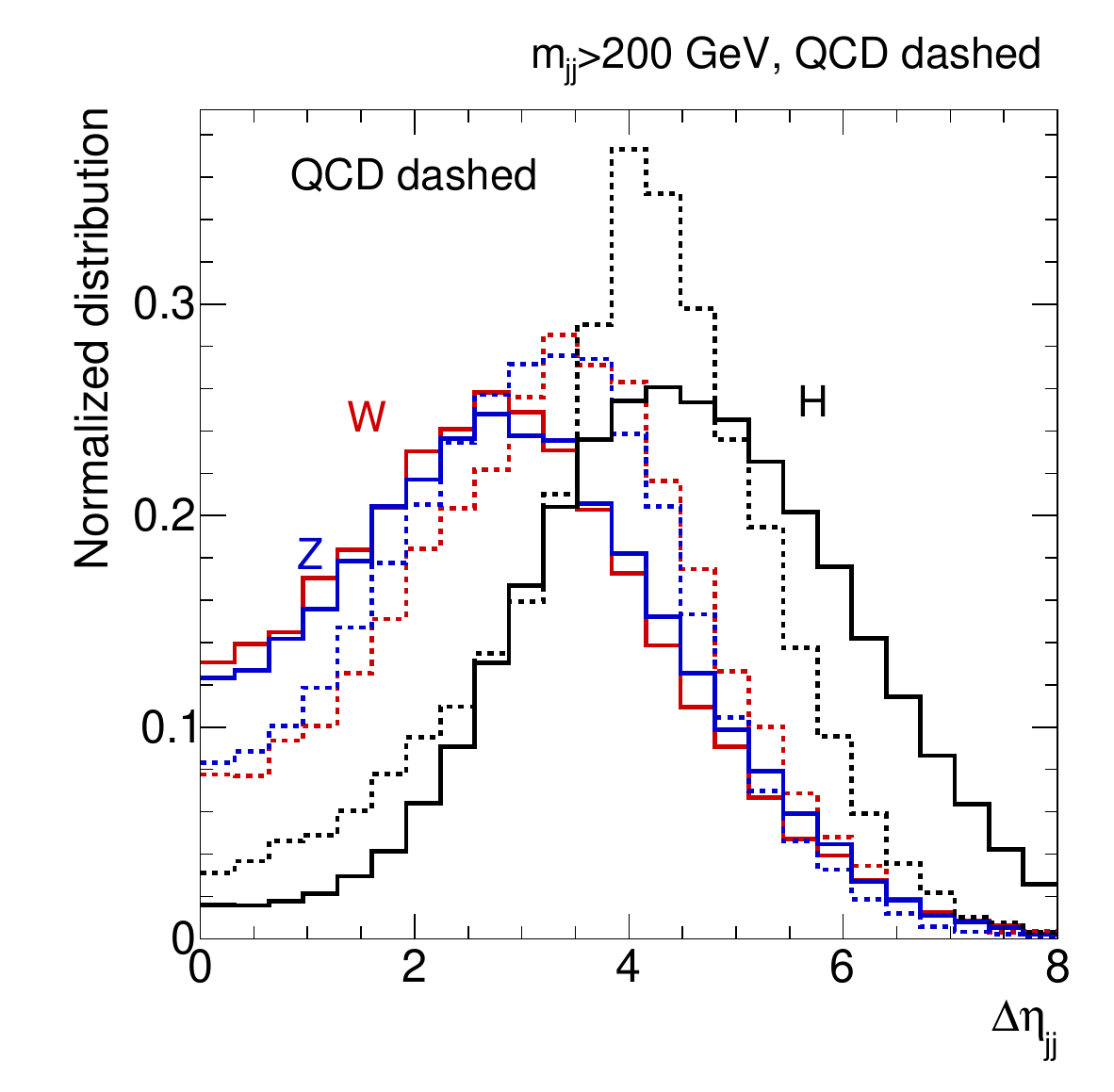}
  \includegraphics[width=0.32\textwidth]{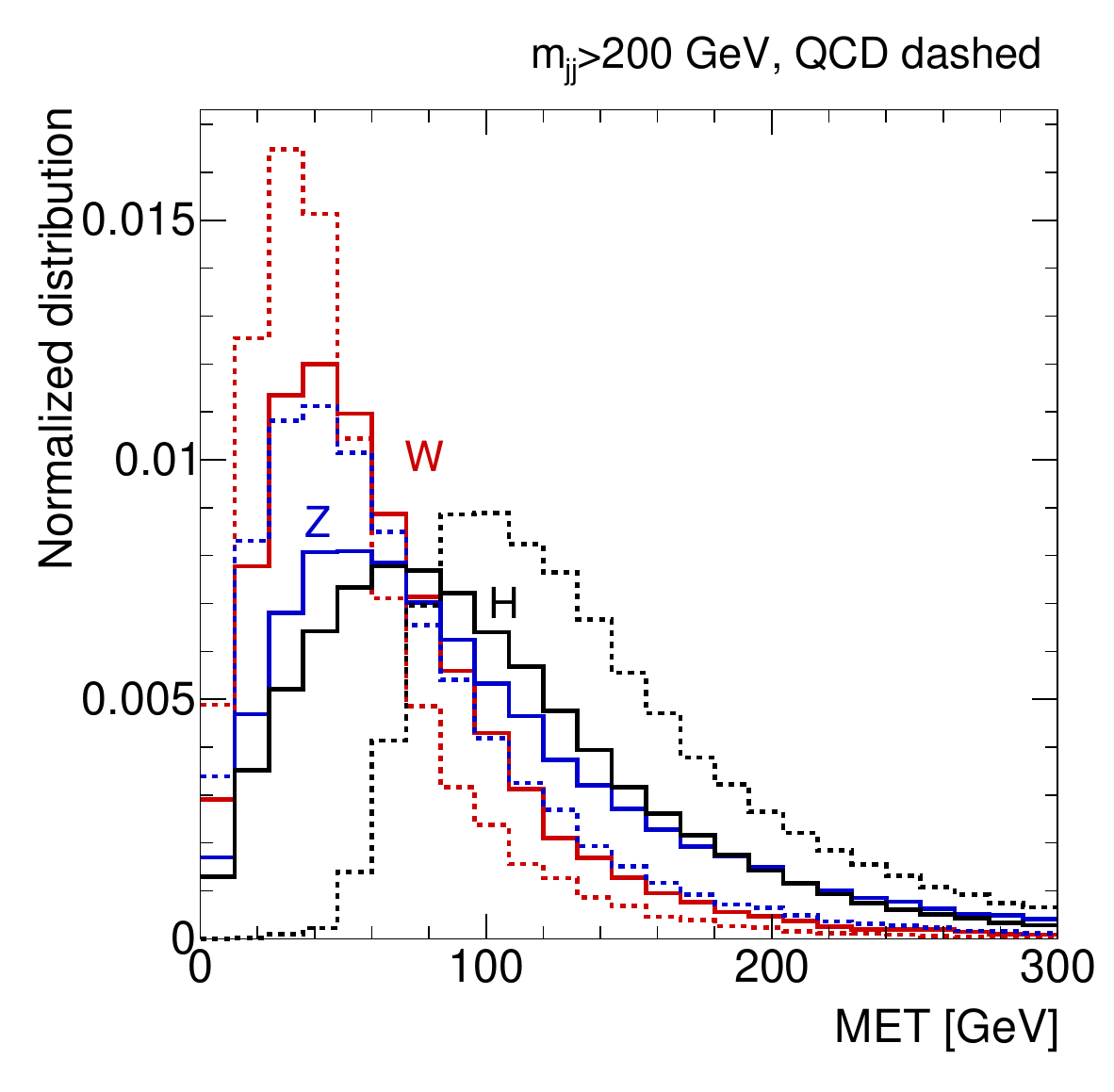}
  \includegraphics[width=0.32\textwidth]{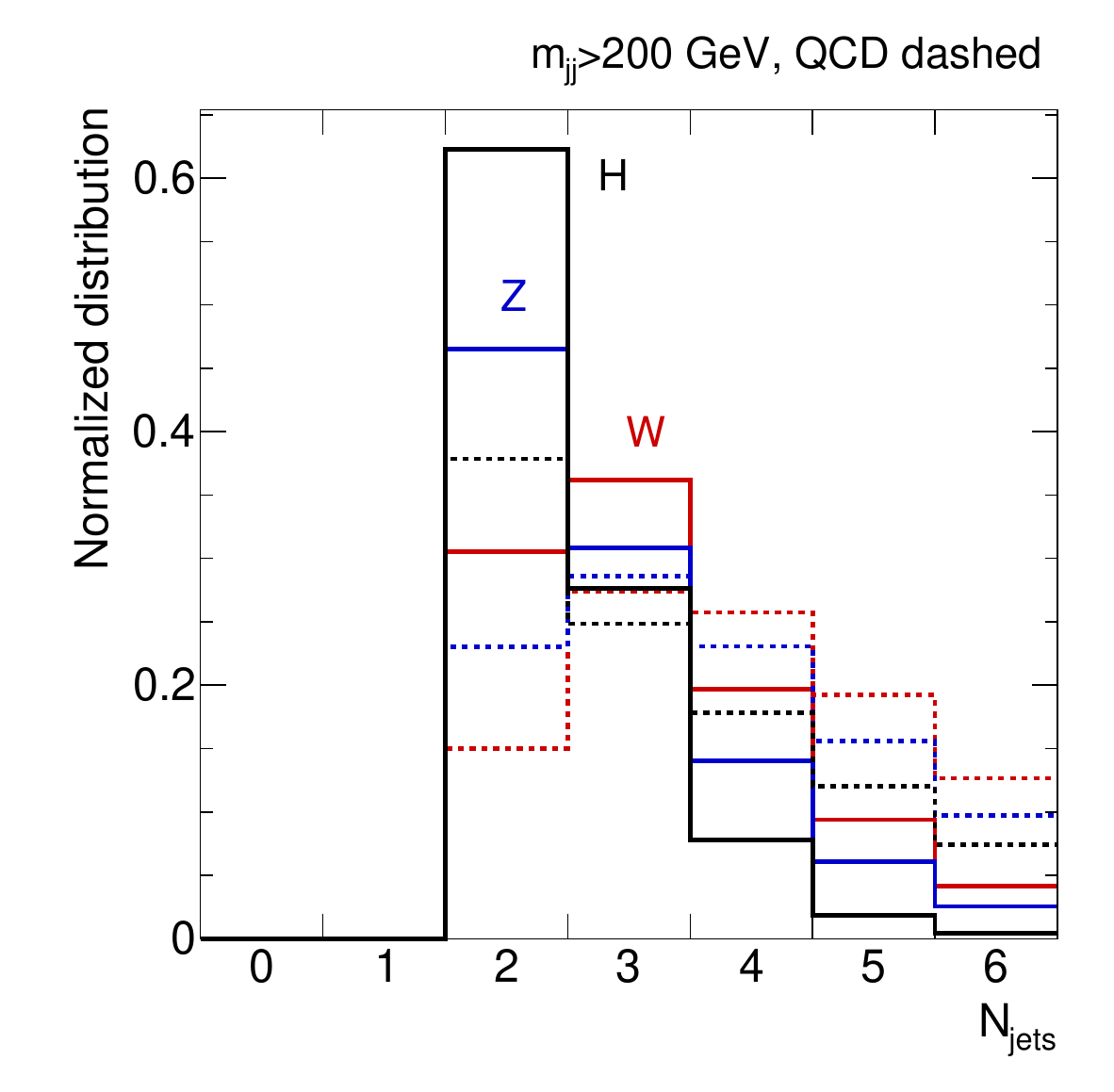} \\
  \includegraphics[width=0.32\textwidth]{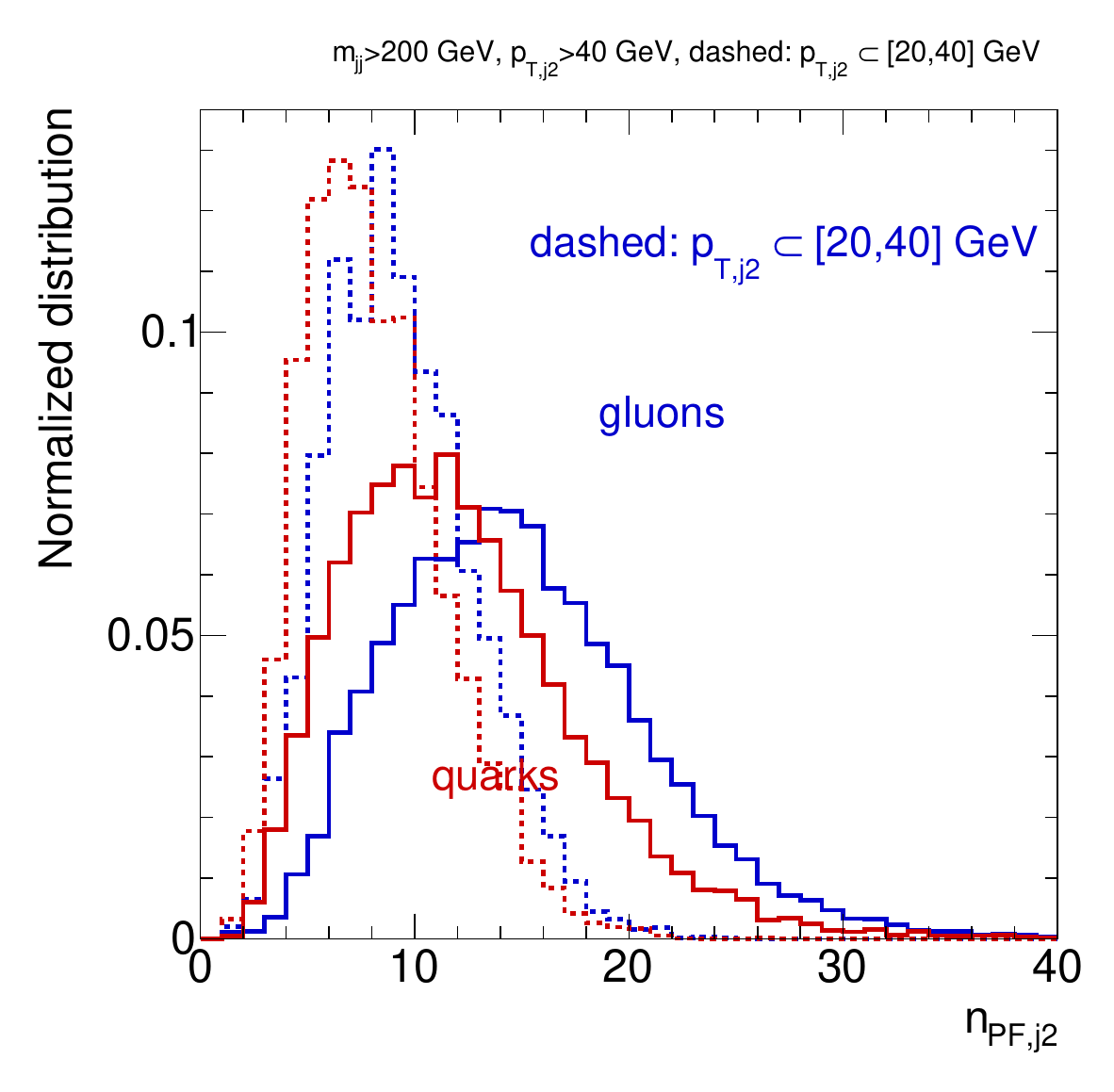}
  \includegraphics[width=0.32\textwidth]{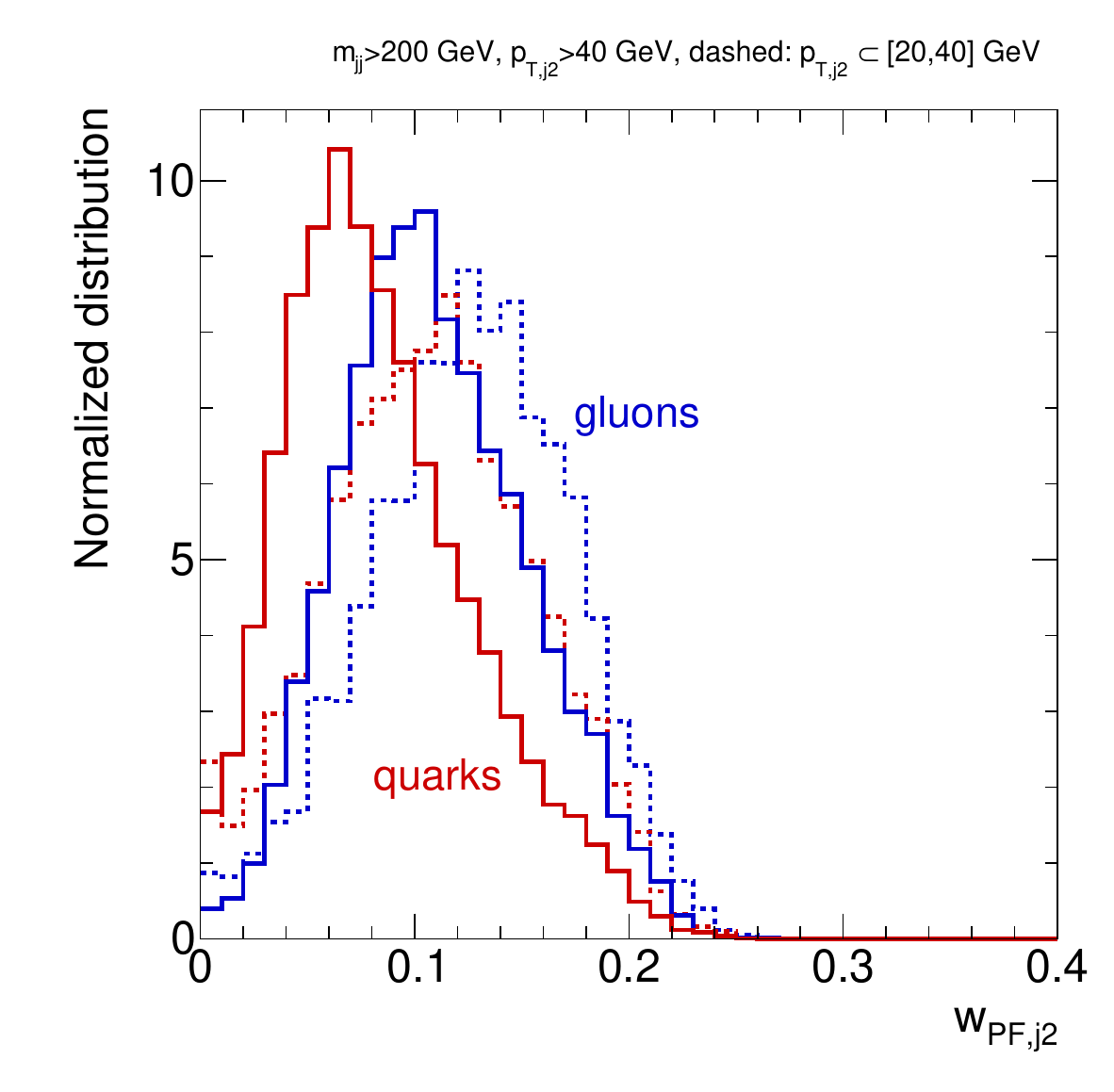}
  \includegraphics[width=0.32\textwidth]{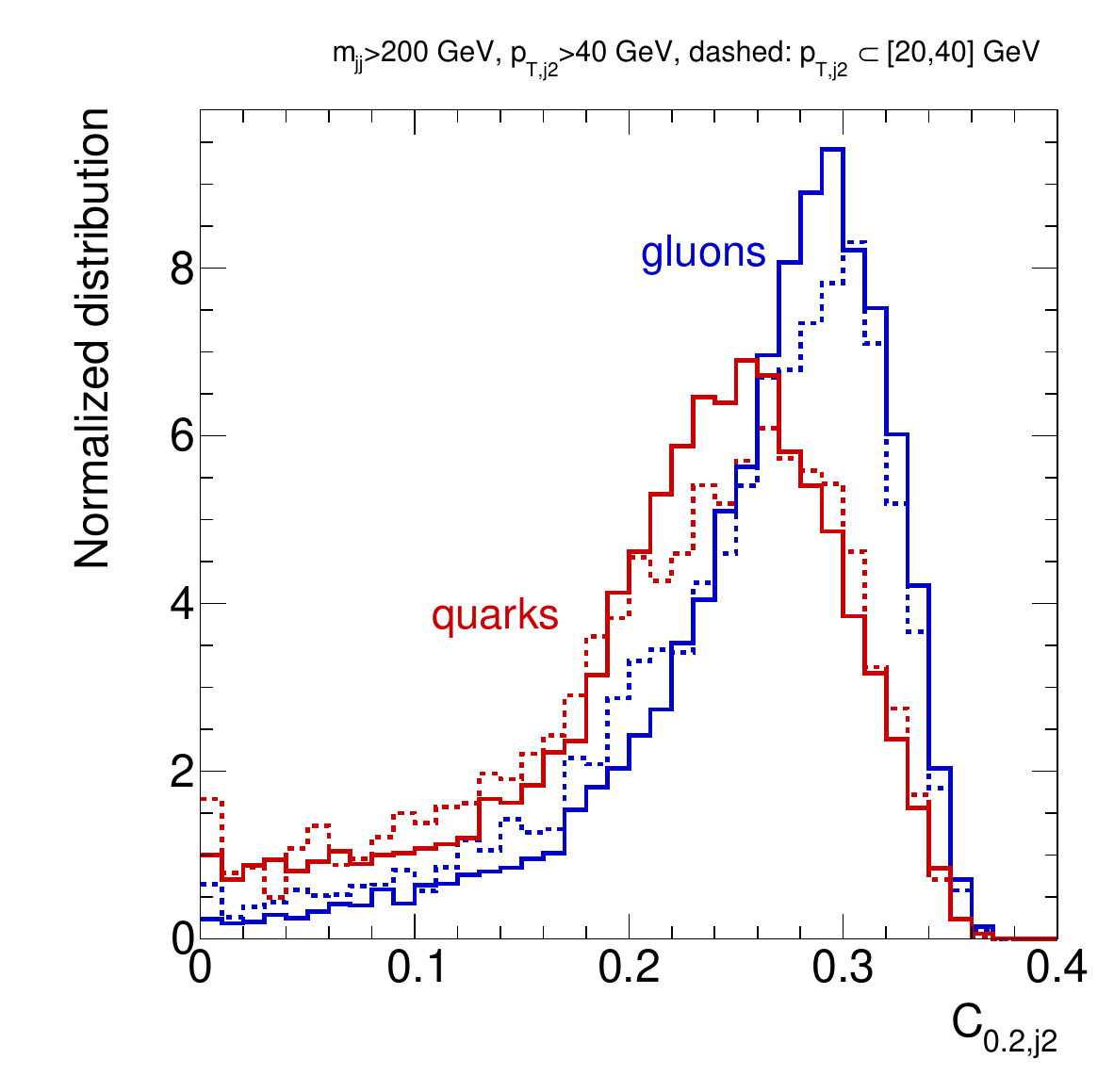}
\caption{Sample distributions for the WBF kinematics (upper) and quark
  vs gluon discrimination (lower). Figures from
  Ref.~\cite{Biekotter:2017gyu}.}
\label{fig:wbf_dist}
\end{figure}

Because LHC detector have a very large number of components, and
because the relevant analysis questions involve many aspects of a
recorded proton-proton collision, experimental analyses rely on a
large number of individual measurements. We can illustrate this
multi-observable strategy with a standard \underline{classification
  task}, the identification of invisible Higgs decays as a dark matter
signature. The most promising production process to look for this
Higgs decay is weak boson fusion, so our signature consists of two
forward tagging jets and missing transverse momentum from the decaying
Higgs,
\begin{align}
  qq' \to (H \to \text{inv}) \; qq'
  \qquad \text{or} \qquad
  pp \to (H \to \text{inv}) \; jj + \text{jets} \; .
\end{align}
Following Fig.~\ref{fig:simchain} we start with a $2\to 3$ hard
scattering, where the Higgs decay products cannot be observed, two
forward tagging jets with high energy and transverse momenta around
$p_T \sim 30~...~150$~GeV, and the key feature that the additional
jets in the signal process are not central because of fundamental QCD
considerations~\cite{Plehn:2009nd}. The main background is $Z$+jets
production, where the $Z$ decays to two neutrinos. A typical set of
basic kinematic cuts at the event level is
\begin{alignat}{7}
p_{T,j_{1,2}} &> 40 \, \gev  &\qqqquad
|\eta_{j_{1,2}}| &< 4.5 &\qqqquad 
\met &> 140 \, \gev \notag \\
\eta_{j_1} \; \eta_{j_2} &< 0 &
|\Delta \eta_{jj} | &> 3.5 &
m_{jj} &> 600 \, \gev \notag \\
p_{T,j_3} &> 20 \, \gev  &\qqqquad
|\eta_{j_3}| &< 4.5 &\qqqquad 
& \text{(if 3rd jet there)} \; .
\label{eq:acc_cuts}
\end{alignat}
On the sub-jet level we can exploit the fact that the electroweak signal only
includes quark jets, whereas the background often features one gluon
in the final state. Kinematic subjet variables which allow us to
distinguish quarks from gluons based on \underline{particle-flow (PF)
  objects} are
\begin{alignat}{9}
n_\text{PF} &= \sum_{i_\text{PF}} 1 &\qqqquad
w_\text{PF} &= \frac{\sum_{i_\text{PF}} \; p_{T,i} \; \Delta R_{i, \text{jet}}}
                   {\sum_{i_\text{PF}} \; p_{T,i}} \notag \\
p_{T}D &= \frac{ \sqrt{ \sum_{i_\text{PF}} \; p_{T,i}^2 } }
          { \sum_{i_\text{PF}} p_{T,i} }  &\qqqquad 
C &= \frac{\sum_{i_\text{PF},j_\text{PF}} \; p_{T,i} p_{T,j} \; \left(\Delta R_{ij} \right)^{0.2}}
           {\left( \sum_{i_\text{PF}} p_{T,i} \right)^2} \; .
\label{eq:qg_obs}
\end{alignat}
Altogether, there is a sizeable number of kinematic observables which
help us extract the signal. They are shown in Fig.~\ref{fig:wbf_dist},
and the main message is that none of them has enough impact to
separate signal and background on its own. Instead, we need to look at
\underline{correlations} between these observables, including
correlations between event-level and subjet observables. Examples for
such correlations would be the rapidity of the third jet, $\eta_{j3}$
as a function of the separation of the tagging jets, $\Delta
\eta_{jj}$, and combined with the quark or gluon nature of the tagging
jets.  More formally, what we need is a method to classify events
based on a combination of many observables.

As a side remark we can ask why we should limit ourselves to the set
of \underline{theory-defined observables} in Eq.\eqref{eq:acc_cuts}
and Eq.\eqref{eq:qg_obs}. When looking at classification with neural
networks in Sec.~\ref{sec:class} we will see that modern machine
learning allows us to instead use preprocessed calorimeter
output, but for now we will stick to the standard, high-level
observables in Eqs.\eqref{eq:acc_cuts} and~\eqref{eq:qg_obs}.

To target correlated observables, \underline{multivariate
  classification} is an established problem in particle physics. The
classical solution for it are or used to be a decision tree. Imagine
we want to classify an event $x_i$ into the signal or background
category based on a set of observables $\obs_j$. As a basis for this
decision we study a sample of training events $\{ x_i \}$, histogram
them for each observable $\obs_j$, and find the values
$\obs_{j,\text{split}}$ which give the most successful split between
signal and background for each distribution. To define such an optimal
split we start with the signal and background probabilities or
so-called impurities as a function of the signal event count $s$ and
the background event count $b$,
\begin{align}
  p_S = \frac{s}{s+b} \equiv p
  \qqquad \text{and} \qqquad 
  p_B = \frac{b}{s+b} \equiv 1-p  \; .
\label{eq:prob_sb}
\end{align}
These probabilities will allow us to define optimal cuts. If we look
at a histogrammed normalized observable $\obs \in [0~...~1]$ we can
compute $p$ and $1-p$ from number of expected signal and background
events following Eq.\eqref{eq:prob_sb}.  For instance, we can look at
a signal which prefers large $\obs$-values two background
distributions and first compute the signal and background
probabilities $p(\obs)$. We can then decide that a single event is
signal-like when its (signal) probability is $p(\obs) > 1/2$,
otherwise the event is background,
\begin{align}
  s(\obs) &= A \obs
  &\qqquad
  b(\obs) &= A ( 1 - \obs )
  &\quad \Rightarrow \quad
  p(\obs) &= \frac{s(\obs)}{s(\obs)+b(\obs)} = \obs \notag \\
  s(\obs) &= A \obs
  &\qqquad
  b(\obs) &= B
  &\quad \Rightarrow \quad
  p(\obs) &= \frac{\obs}{\obs + \dfrac{B}{A}} \; .
\end{align}
We then declare a single event signal-like when its (signal)
probability is $p(\obs) > 1/2$, otherwise the event is background-like.
This defines the splitting point
\begin{align}
  \obs_{j,\text{split}} 
  = \obs_j \Big|_{p=1/2}
  =
  \begin{cases}
     \dfrac{1}{2} \\[4mm]
     \dfrac{B}{A} 
  \end{cases}    \; .
\label{eq:opt_split1}
\end{align}
To organize our multivariate analysis we then need to evaluate the
performance of each optimized cut $\obs_{j,\text{split}}$, for example
to apply the most efficient cut first or at the top of a decision
tree.

Next, we formalize this condition in terms of statistics --- we want
to construct the $\obs_{j,\text{split}}$ such that they maximize the
the gain in information. This gain can be calculated in terms of the
\underline{information entropy}. We remind ourselves that for a system with
$2^N$ states with equal probabilities $p_j$ the
most efficient way to determine its state is  to step-wise split the possible
states into groups of equal size and then ask in which of the two
halves the system lives. This way we avoid any element of luck, and we
need
\begin{align}
    N = - \log_2 p_j = \frac{\ln p_j}{\ln 2}
\end{align}
answers. For a system with different probabilities
$p_j$ the branches of the decision tree correspond to different
probabilities, there will be an element of luck. The number of
necessary questions is only defined as an expectation value,
\begin{align}
    \langle - \log_2 p_j \rangle 
    = - \frac{1}{\ln 2} \sum_j p_j \ln p_j
    &\equiv \frac{H}{\ln 2} \notag \\
    H = - \sum_j p_j \ln p_j
    \label{eq:entropy}
\end{align}
The information entropy $H > 0$ measures the (logarithmic) uncertainty
in units of $\ln 2$, so-called bits. Uncertainty is the amount of
expected information needed to correctly identify the state of the
system.  For two outcomes with $p_1 = p$ and $ p_2 = 1- p$ we can
compute the maximum information entropy, 
\begin{align}
    H &= - \left[ p \ln p + (1-p) \ln (1-p) \right] \notag \\
    \frac{d H}{d p} & = - \left[ \ln p + \frac{p}{p} - \ln (1-p) - \frac{1-p}{1-p} \right]
    \notag \\
    &= - \ln p + \ln (1-p) = 0
    \qquad \Leftrightarrow \qquad p = 1 - p = \frac{1}{2} \; .
    \label{eq:def_info_entropy}
\end{align}
The information entropy vanishes for $p=0$ and $p=1$ and is symmetric
around its maximum at $p=1/2$. In general, the
entropy is maximal for equal probabilities,
  \begin{align}
    H \le - \sum_j p \ln p = - \ln p \; .
    \label{eq:min_entropy}
  \end{align}
If we split our system in two not independent systems, we can compute
the entropies in terms of the individual and joint probability
$p_{i,j}$,
\begin{align}
  H_1 =  - \sum_i  p_i \ln p_i
  \qqqquad
  H_2 =  - \sum_j p_j \ln p_j 
  \qqqquad 
  H_{12}
  =  - \sum_{i,j} p_{i,j} \; \ln p_{i,j} \; .
\end{align}
Because equally distributed systems maximize the entropy, this
combined entropy should be smaller than the sum of the independent
individual entropies. The difference is the \underline{mutual information}
$I_{12}$
\begin{align}
    I_{12}
    &= H_1 + H_2 - H_{12} \notag \\
    &=  - \sum_i p_i \left( \sum_j p_{j|i} \right) \ln p_i  
    - \sum_j \left( \sum_i p_{i|j} \right) p_j \ln p_j
    + \sum_{i,j} p_{i,j} \; \ln p_{i,j} \notag \\
    &=  - \sum_{i,j} \left[
      p_i p_{j|i} \ln p_i  
    + p_j p_{i|j} \ln p_j
    - p_{i,j} \; \ln p_{i,j} \right] \notag \\
    &=  - \sum_{i,j} p_{i,j} \ln \frac{p_i p_j}{p_{i,j}} \; .
    \label{eq:def_mutual}
\end{align}

In terms of the information entropy, we can write our construction of
split values in Eq.\eqref{eq:opt_split1} as
\begin{align}
  \obs_{j,\text{split}} 
  = \argmax_\text{splits} H[p(\obs_j)] \; .
\label{eq:opt_split2}
\end{align}
To build a \underline{decision tree} out of our observables we first
compute the best splitting for each observable individually and then
choose the observable with the most successful split. More precisely,
we want to maximize the difference of the information entropy before
the split and the sum of the information entropies after the split,
the so-called information gain. We choose the first observable of our
decision tree though
\begin{align}
  \max_j \Big[ H_\text{before split}[p(\obs_j)]
    - H_{\text{after split},1}[p(\obs_j)]
    - H_{\text{after split},2}[p(\obs_j)]
    \Big] \; .
\end{align}

\begin{figure}[t]
  \centering
  \includegraphics[width=0.35\textwidth]{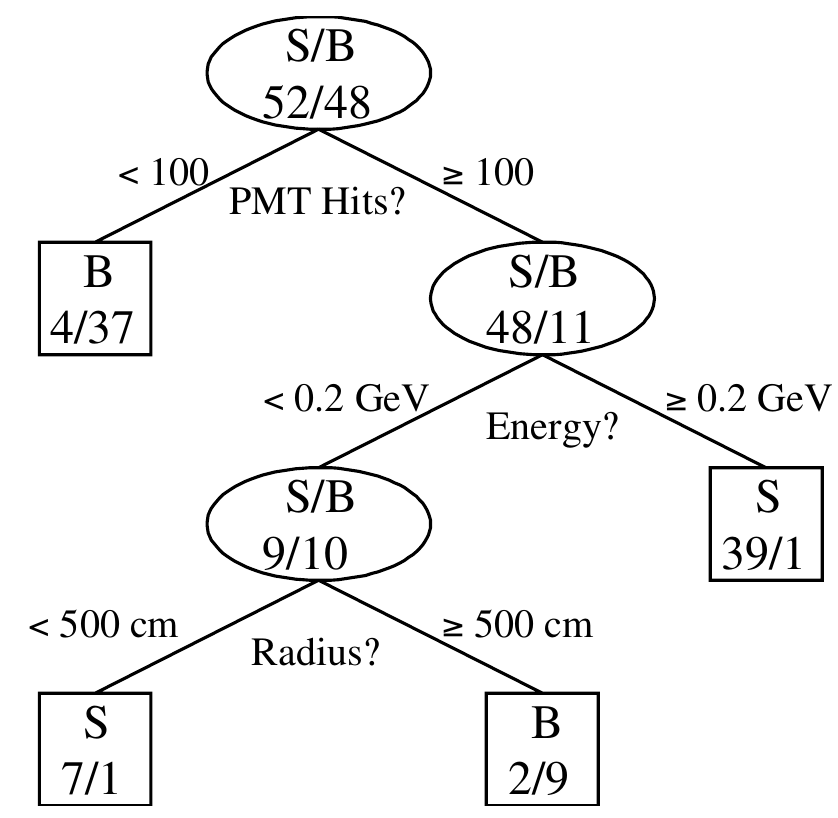}
  \hspace*{0.1\textwidth}
  \includegraphics[width=0.45\textwidth]{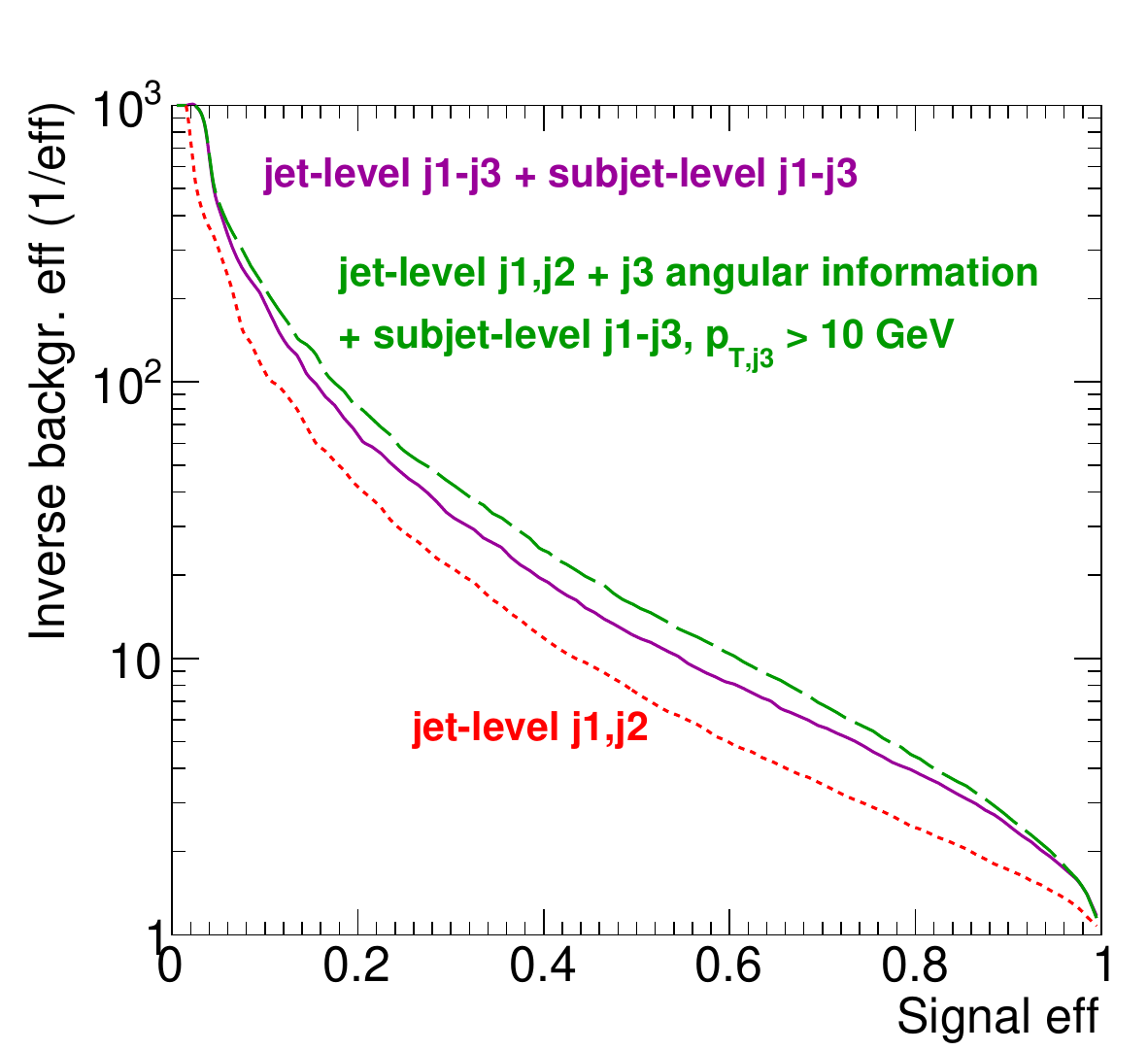}
\caption{Left: illustration of a decision tree from an early
  application in particle physics, selecting electron neutrinos to
  prove neutrino oscillations. Figure from
  Ref.~\cite{Roe:2004na}. Right: signal efficiency vs background
  rejection for WBF Higgs production and invisible Higgs decays, based
  on jet-level and additional subjet-level information shown in
  Tab.~\ref{tab:bdt}. Figure from Ref.~\cite{Biekotter:2017gyu}.}
\label{fig:bdt}
\end{figure}

A historic illustration for a decision tree used in particle physics
is shown in the left panel of Fig.~\ref{fig:bdt}. It comes from the
first high-visibility application of (boosted) decision
trees\index{boosted decision tree} in particle physics, to identify
electron-neutrinos from a beam of muon-neutrinos using the MiniBooNE
Cerenkov detector. Each observable defines a so-called node, and the
two branches below each node are defined as `yes' vs `no' or as
`signal-like' vs `background-like'.  The first node is defined by the
observable with the highest information gain among all the optimal
splits.  The two branches are based on this optimal split value, found
by maximizing the cross entropy. Every outgoing branch defines the
next node again through the maximum information gain, and its outgoing
branches again reflect the optimal split, etc.  Finally, the algorithm
needs a condition when we stop splitting a branch by defining a node
and instead define a so-called leaf, for instance calling all events
`signal' after a certain number of splittings selecting it as
`signal-like'. Such conditions could be that all collected training
events are either signal or background, that a branch has too few
events to continue, or simply by enforcing a maximum number of
branches.

No matter how we define the stopping criterion for constructing a
decision tree, there will always be signal events in background leaves
and vice versa. We can only guarantee that a tree is completely right
for the \underline{training sample}, if each leaf includes one
training event. This perfect discrimination obviously does not carry
over to an independent \underline{test sample}, which means our
decision tree is \underline{overtrained}. In general, overtraining
means that the performance for instance of a classifier on the
training data is so finely tuned that it follows the statistical
fluctuations of the training data and does not generalize to the same
performance on independent sample of test data.

If we want to further enhance the performance of the decision tree we
can focus on the events which are wrongly classified after we define
the leaves. For instance, we can add an event weight $w >1$ to every
mis-identified event (we care about) and carry this weight through the
calculation of the splitting condition. This is the simple idea behind
a \underline{boosted decision tree (BDT)}. Usually, the weights are chosen
such that the sum of all events is one. If we construct several
independent decision trees, we can also combine their output for the
final classifier. It is not obvious that this procedure will improve
the tree for a finite number of leaves, and it is not obvious that
such a reweighting will converge to a unique or event improved
boosted decision tree, but in practice this method has been shown to
be extremely powerful.

\def\arraystretch{1.2}
\begin{table}[b!]
\centering
  \begin{small} \begin{tabular}{lll}
\toprule
Set && Variables \\
\midrule
jet-level $j_1$, $j_2$ &\quad &    $p_{T,j_1} \quad p_{T,j_2} \quad \Delta\eta_{jj}
        \quad \Delta\phi_{jj} \quad m_{jj} \quad \met
        \quad \Delta\phi_{j_1,\met} \quad \Delta\phi_{j_2,\met}$ \\
subjet-level $j_1$, $j_2$  &\quad &  $n_{\text{PF},j_1} \quad n_{\text{PF},j_2} \quad 
                p_T D_{j_1} \quad p_T D_{j_2} \quad C_{j_1} \quad C_{j_2} $  \\[1mm] \midrule
$j_3$ angular information &\quad &   $ \Delta\eta_{j_1,j_3}
        \quad \Delta\eta_{j_2,j_3} \quad \Delta\phi_{j_1,j_3} \quad \Delta\phi_{j_2,j_3}$ \\
jet-level $j_1$-$j_3$ &\quad &   jet-level $j_1$, $j_2$ \quad $+$  \quad 
        $j_3$ angular information \quad $+$ \quad 
        $p_{T,j_3}$ \\
subjet-level $j_1$-$j_3$&\quad &   subjet-level $j_1$, $j_2$ \quad $+$  \quad  
$n_{\text{PF},j_3} \quad C_{j_3} \quad p_T D_{j_3}$ \\
\bottomrule
\end{tabular} \end{small}
\caption{Sets of variables used in a BDT study for a WBF Higgs search
  with invisible Higgs decays.  The subscript $jj$ refers to the two
  tagging jets, not all events have a third jet. Example from
  Ref.~\cite{Biekotter:2017gyu}.}
\label{tab:bdt}
\end{table}

Finally, we need to measure the performance of a BDT classification
using some kind of success metric. Obviously, a large signal
efficiency alone is not sufficient, because the signal-to-background
ratio $S/B$ or the Gaussian significance $S/\sqrt{B}$ depend on the
signal and background rates. For a simple classification task we can
compute four numbers
\begin{enumerate}
\item \underline{signal efficiency}, or true positive rate
  $\epsilon_S \equiv s^\text{(S-tagged)}/s^\text{(truth)}$;
\item background efficiency, or true negative rate
  $b^\text{(B-tagged)}/b^\text{(truth)}$;
\item \underline{background mis-identification rate}, or false positive rate
  $\epsilon_B \equiv b^\text{(S-tagged)}/b^\text{(truth)}$;
\item signal mis-identification rate, or false negative rate
  $s^\text{(B-tagged)}/s^\text{(truth)}$.
\end{enumerate}
If we tag all events we know the normalization conditions
$s^\text{(truth)} = s^\text{(S-tagged)}+s^\text{(B-tagged)}$ and
correspondingly for $b^\text{(truth)}$.  The signal efficiency is also
called recall or sensitivity in other fields of research. The
background mis-identification rate can be re-phrased as background
rejection $1 - \epsilon_B$, also referred to as specificity.

Once we have tagged a signal sample we can ask how many of those
tagged events are actually signal, defining the 
\begin{align}
\text{purity} = \text{precision} 
= \frac{s^\text{(S-tagged)}}{s^\text{(S-tagged)}+b^\text{(S-tagged)}} \; .
\end{align}
Finally,
we can ask how many of our decisions are correct and compute the 
\begin{align}
\text{accuracy} = \frac{s^\text{(S-tagged)}+b^\text{(B-tagged)}}{s^\text{(truth)}+b^\text{(truth)}} \; , 
\end{align}
reflecting the fraction of correct decisions.

In particle physics we usually measure the success of a classifier in
the plane $\epsilon_S$ vs $\epsilon_B$, where for the latter we either
write $1-\epsilon_B$ of $1/\epsilon_B$. In the right panel of
Fig.~\ref{fig:bdt} we show such a plane for our invisible Higgs decay
example. It is called \underline{receiver--operator characteristics}
(ROC) curve. The different sets of observables are shown in
Tab.~\ref{tab:bdt}. The lowest ROC curve corresponds to a BDT analysis
of the kinematic observables of the two tagging jets and the missing
transverse energy vector. For a signal efficiency $\epsilon_S = 40\%$
it gives a background rejection around $1/\epsilon_B = 1/10$. If we
expect a relatively large signal rate and at the same time need to
reject the background more efficiently, we can choose a different
\underline{working point} of the classifier, for instance $\epsilon_S
= 20\%$ and $1/\epsilon_B = 1/35$.

If a classifier gives us, for example, a continuous measure of
signal-ness of an event being signal we can choose different working
points by defining a cut on the classifier output. The problem with
such any such cut is that we lose information from all those signal
events which almost made it to the signal-tagged sample. If we can
construct our classifier such that its output is a probability, we can
also weight all events by their signal vs background probability score
and keep all events in our analysis.

Going back to LHC physics, in Fig.~\ref{fig:bdt} we see that adding
information on the potential third jet and subjet observables for all
three jets with the additional requirement of $p_{T,j} > 10$~GeV
improves the background rejection for $\epsilon_S = 40\%$ to almost
$1/\epsilon_B = 1/20$. Adding information from softer jets only has a
small effect. Such ROC curves are the standard tool for benchmarking
LHC analyses, reducing them to single performance values like the
integral under the ROC curve (AUC) or the background rejection for a
given signal efficiency is usually an oversimplification.

To summarize, data analysis without multivariate classification is
hard to imagine for particle physicists. Independent of the question
if we want to call boosted decision trees machine learning or not, we
have shown how they can be constructed or trained for multivariate
classification tasks, and we have taken the opportunity to define many
technical terms we will need later in these notes.  The great
advantage of decision trees, in addition to the great performance of
BDTs, is that we can follow their definition of signal and background
regions fairly easily. We can always look at graphics like the one in
the left panel of Fig.~\ref{fig:bdt}, at least before boosting, and
standard tools like TMVA provide a list of the most powerful
observables. The biggest disadvantage of decision trees is that by
construction they do not account for correlations properly, they only
break up the observable space into many rectangular leaves.

\subsubsection{Fits and interpolations}
\label{sec:basics_fit}

From a practical perspective, we start with the assumption or
observation that neural network are nothing but \underline{numerically
  defined functions}. As alluded to in the last section, some kind of
minimization algorithm on a \underline{loss function} will allow us to define
determine its underlying parameters $\theta$. The simplest case,
regression networks are scalar or vector fields defined on some space,
approximated by $f_\theta(x)$. Assuming that we have indirect or
implicit access to the truth $f(x)$ in form of a training dataset
$(x,f)_j$, we want to construct the approximation
\begin{align}
 \boxed{ f_\theta(x) \approx f(x) } \; .
\label{eq:def_fit}
\end{align}
Usually, approximating a set of functional values for a given set of
points can be done two ways. First, in a \textsl{fit} we start with a
functional form in terms of a small set of parameters which we also
refer to as $\theta$. To determine these network parameters, we
maximize the probability for the fit output $f(x_j)$ to correspond to
the training points $f_j$, with uncertainties
$\sigma_j$\index{uncertainties}. Formally, we can write this as an
estimation of the best-suited fit parameters
\begin{align}
  \theta = \argmax \; p(\theta|x) \; .
\label{eq:argmax_prob}
\end{align}
The problem with this posterior in model space is that we typically
canot evaluate it, \ie it is intractable. However, we can use Bayes'
theorem to re-write it in terms of the corresponding likelihood and
then assume that the prior on $\theta$ does not affect our
maximization,
\begin{align}
  \theta
  &= \argmax \; \frac{p(x|\theta) p(\theta)}{p(x)} \notag \\
  &= \argmax \; \left[ p(x|\theta) p(\theta) \right]  
  \approx \argmax \; p(x|\theta) \; .
\label{eq:argmax_like}
\end{align}

To maximize the likelihood we need to assume a functional form. An
obvious choice for statistical problems, backed up by the central
limit theorem, is to maximize the \underline{Gaussian likelihood}
\begin{align}
  p(x|\theta) &= \prod_j \frac{1}{\sqrt{2 \pi} \sigma_j}
  \exp \left( - \frac{|f_j - f_\theta(x_j)|^2}{2 \sigma_j^2} \right) \notag \\
  \Rightarrow \qquad 
  \log p(x|\theta) &= - \sum_j \frac{|f_j - f_\theta(x_j)|^2}{2\sigma_j^2} +
  \text{const}(\theta) \; .
\label{eq:def_gaussian}
\end{align}
In the Gaussian limit this log-likelihood is often referred to as
$\chi^2$. In this form the individual Gaussians have mean $f_j$, a
variance $\sigma_j^2$, a standard deviation $\sigma_j$, and a width of
the bell curve of $2 \sigma_j$.  For the definition of the best
parameters $\theta$ we can again ignore the $\theta$-independent
normalization.  The loss function for our fit, \ie the function we
minimize to determine the fit's model parameters is
\begin{align}
  \boxed{
    \loss_\text{fit} = \sum_j \loss_j = \sum_j \frac{|f_j - f_\theta(x_j)|^2}{2\sigma_j^2}
    } \; .
  \label{eq:likelihood_loss}
\end{align}
The fit function is not optimized to go though all or even some of the
training data points, $f_\theta(x_j) \ne f_j$. Instead, the
log-likelihood loss is a compromise to agree with all training data
points within their uncertainties. We can plot the values $\loss_j$
for the training data and should find a Gaussian distribution of mean
zero and standard deviation one, $\normal(\mu = 0,\sigma = 1)$.

An interesting question arises in cases where we do not know an
uncertainty $\sigma_j$ for each training point, or where such an
uncertainty does not make any sense, because we know all training data
points to the same precision $\sigma_j = \sigma$. In that case we can
still define a fit function, but the loss function becomes a simple
\underline{mean squared error},
\begin{align}
  \loss_\text{fit} = \frac{1}{2\sigma^2} \sum_j |f_j - f_\theta(x_j)|^2
    \equiv \frac{1}{2\sigma^2} \text{MSE} \; .
\label{eq:mse_loss}
\end{align}
Again, the prefactor is $\theta$-independent and does not contribute
to the definition of the best fit.  This simplification means that our
MSE fit puts much more weight on deviations for large functional
values $f_j$. This is often not what we want, so alternatively we
could also define the loss in terms of relative uncertainties. In
practical applications of machine learning we will instead apply
preprocessings of the data,
\begin{align}
  f_j \to \log f_j
  \qquad \text{or} \qquad
  f_j \to f_j - \langle f_j \rangle
  \qquad \text{or} \qquad
  f_i \to \frac{f_i}{\langle f_i \rangle} \cdots 
\label{eq:preprocs}
\end{align}
In cases where we expect something like a Gaussian distribution a
standard scaling would preprocess the data to a mean of zero and a
standard deviation of one.  In an ideal world, such preprocessings
should not affect our results, but in reality they almost always
do. The only way to avoid preprocessings is to add information like
the scale of expected and allowed deviations for the likelihood loss
in Eq.\eqref{eq:likelihood_loss}.

The second way of approximating a set of functional values is
\underline{interpolation}, which ensures $f_\theta(x_j) = f_j$ and is
the method of choice for datasets without noise. Between these
training data points we choose a linear or polynomial form, the latter
defining a so-called spline approximation. It provides an
interpolation which is differentiable a certain number of times by
matching not only the functional values $f_\theta(x_j) = f_j$, but
also the $n$th derivatives $f_\theta^{(n)}(x \uparrow x_j) =
f_\theta^{(n)}(x \downarrow x_j)$. In the machine learning language we
can say that the order of our spline defines an \underline{implicit
  bias} for our interpolation, because it defines a resolution in
$x$-space where the interpolation works. A linear interpolation is not
expected to do well for widely spaced training points and rapidly
changing functional values, while a spline-interpolation should
require fewer training points because the spline itself can express a
non-trivial functional form.

The main difference between a fit and an interpolation is their
respective behavior on unknown dataset.  For both, a fit and an
interpolation we expect our fit model $f_\theta(x)$ to describe the
training data.  To measure the quality of a fit beyond the training
data we can compute the loss function $\loss$ or the point-wise
contributions to the loss $\loss_j$ on an independent \underline{test
  dataset}.  If a fit does not generalize from a training to a test
dataset, it is usually because it has learned not only the smooth
underlying function, but also the statistical fluctuation of the
training data. While a test dataset of the same size will have
statistical fluctuations of the same size, they will not be in the
same place, which means the loss function evaluated on the training
data will be much smaller than the loss function evaluated on the test
data. This failure mode is called over-fitting or, more generally,
\underline{overtraining}. For interpolation this overtraining is a
feature, because we want to reproduce the training data perfectly. The
generalization property is left to choice of the interpolation
function. As a side remark, manipulating the training dataset while
performing one fit after the other is an efficient way to search for
outliers in a dataset, or a new particle in an otherwise smooth
distribution of invariant masses at the LHC.

More systematically, we can define a set of errors which we make when
targeting a problem by constructing a fit function through minimizing a
loss on a training dataset. First, an \underline{approximation error}
is introduced when we define a fit function, which limits the
expressiveness of the network in describing the true function we want
to learn. Second, an estimation or \underline{generalization error}
appears when we approximate the true training objective by a
combination of loss function and training dataset.  In practice,
these errors are related.  A justified limit to the expressiveness of
a fit function, or implicit bias, defines useful fits for a given
task. In many physics applications we want our fit to be smooth at a
given resolution.  When defining a good fit, increasing the class of
functions the fit represents leads to a smaller approximation error,
but increases the estimation error. This is called the
\underline{bias-variance trade off}, and we can control it by limiting
or regularizing\index{regularization} the expressiveness of the fit function and by ensuring
that the loss of an independent test dataset does not increase while
training on the training dataset.  Finally, any numerical optimization
comes with a \underline{training error}, representing the fact that a
fitted function might just live, for instance, in a sufficiently good
local minimum of the loss landscape. This error is a numerical
problem, which we can solve though more efficient loss minimization.
While we can introduce these errors for fits, they will become more
relevant for neural networks.

\subsubsection{Neural networks}
\label{sec:basics_deep_nn}

One way to think about a neural network is as a numerically defined
fit function, often with a huge number of model parameters $\theta$,
and written just like the fit of Eq.\eqref{eq:def_fit},
\begin{align}
  f_\theta(x) \approx f(x) \; .
\label{eq:define_nn}
\end{align}
As mentioned before, we minimize a loss function numerically to
determine the neural network parameters $\theta$. This procedure us
called network training and requires a training dataset $(x, f)_j$
representing the target function $f(x)$. To control and avoid
overtraining, we can compare the values of the loss function between
the training dataset and an independent test dataset.

We will skip the usual inspiration from biological neurons and instead
ask our first question, which is how to describe an unknown function
in terms of a large number of model parameters $\theta$ without making
more assumptions than some kind of smoothness on the relevant
scales. For a simple regression task We can write the mapping as
\begin{align}
  x \to f_\theta(x)
  \qquad \text{with} \quad x \in \mathbb{R}^D
  \quad \text{and} \quad f_\theta \in \mathbb{R} \; .
\label{eq:nn_mapping}
\end{align}
The key is to think of this problem in terms of building blocks which
we can put together such that simple functions require a small number
of modules or building blocks, and model parameters, and complex
functions require larger and larger numbers of those building
blocks. We start by defining so-called \underline{layers}, which in a
fully connected or dense network transfer information from all $D$ entries of
the vector $x$ defining one layer to all vector entries of the layer to its
left,
\begin{align}
  x \to x^{(1)} \to x^{(2)} \cdots \to x^{(N)} \equiv f_\theta(x)
\label{eq:forward_pass}
\end{align}
Counting the input $x$ this means our network consist of $N$ layers,
including one input layer, one output layer, and $N-2$ hidden layers.
If a vector entry $x_j^{(n+1)}$ collects information from all $x^{(n)}_j$, we
can try to write each step of this chain as
\begin{align}
  x^{(n-1)} \to x^{(n)} := W^{(n)} x^{(n-1)} + b^{(n)} \; ,
\label{eq:affine}
\end{align}
where the $D \times D$ matrix $W$ is referred to as the network
weights and the $D$-dimensional vector $b$ as the bias.  In general,
neighboring layers do not need to have the same dimension, which means
$W$ does not have to be a diagonal matrix. In our simple regression
case we already know that over the layers we need to reduce the width
of the network from the input vector dimension $D$ to the output
scalar $x^{(N)} = f_\theta(x)$.

Splitting the vector $x^{(n)}$ into its $D$ entries defines the
\underline{nodes} which form our network layer
\begin{align}
  x_i^{(n)} = W_{ij}^{(n)} x_j^{(n-1)} + b_i^{(n)} \; .
\label{eq:single_node}
\end{align}
For a fully connected network a node takes $D$ components
$x_j^{(n-1)}$ and transforms them into a single output $x_i^{(n)}$.
For each node the $D+1$ network parameters are $D$ matrix entries
$W_{ij}$ and one bias $b_i$.  If we want to compute the loss function
for a given data point $(x_j, f_j)$, we follow the arrows in
Eq.\eqref{eq:forward_pass}, use each data point as the input layer, $x
= x_j$, go through the following layers one by one, compute the
network output $f_\theta(x_j)$, and compare it to $f_j$ though a loss
function.

The transformation shown in Eq.\eqref{eq:affine} is an
\underline{affine transformation}. Just like linear transformations,
affine transformations form a group. This is equivalent to saying that
combining affine layers still gives us an affine transformation, just
encoded in a slightly more complicated manner. This means our network
defined by Eq.\eqref{eq:affine} can only describe linear functions,
albeit in high-dimensional spaces.

To describe non-linear functions we need to introduce some kind of
non-linear structure in our neural network. The simplest
implementation of the required nonlinearity is to apply a so-called
\underline{activation function} to each node. Probably the simplest
1-dimensional choice is the so-called rectified linear unit
\begin{align}
  \relu (x_j) :=
  \max (0,x_j) = 
  \begin{cases} 0 & x_j \le 0 \\ x_j & x_j>0 \end{cases} \; ,
\label{eq:def_relu}
\end{align}   
giving us instead of Eq.\eqref{eq:single_node}
\begin{align}
  \boxed{
    x^{(n-1)} \to x^{(n)} := \relu \left[ W^{(n)} x^{(n-1)} + b^{(n)} \right]
    } \; ,
\label{eq:def_node}
\end{align}
Here we write the ReLU transformation of a vector as the vector of
ReLU-transformed elements.  This non-linear transformation is the same
for each node, so all our network parameters are still given by the
affine transformations. But now a sufficiently deep network can
describe general, non-linear functions, and combining layers adds
complexity, new parameters, and expressivity to our network function
$f_\theta(x)$. There are many alternatives to ReLU as the source of
non-linearity in the network setup, and depending on our problem they
might be helpful, for example by providing a finite gradient over the
$x$-range. However, throughout this lecture we always refer to a
standard activation function as ReLU.

This brings us to the second question, namely, how to determine a
correct or at least good set of network parameters $\theta$ to describe
a training dataset $(x,f)_j$. From our fit discussion we know that one
way to determine the network parameters is by minimizing a
loss function. For simplicity, we can think of the MSE loss defined in
Eq.\eqref{eq:mse_loss} and ignore the normalization
$1/(2\sigma^2)$. To minimize the loss we have to compute its
derivative with respect to the network parameters. If we ignore the bias
for now, for a given weight in the last network layer we need to
compute
\begin{align}
  \frac{d \loss}{d W^{(N)}_{1j}}
  &= \frac{\partial \left| f - \relu [W^{(N)}_{1k} x^{(N-1)}_k] \right|^2}{\partial W^{(N)}_{1j}} \notag \\
  &= \frac{\partial \left| f - \relu [W^{(N)}_{1k} x^{(N-1)}_k] \right|^2}{\partial \relu [W^{(N)}_{1k} x^{(N-1)}_k]} \;
  \frac{\partial \relu [W^{(N)}_{1k} x^{(N-1)}_k]}{\partial [W^{(N)}_{1k} x^{(N-1)}_k]} \;
  \frac{\partial [W^{(N)}_{1k} x^{(N-1)}_k]}{\partial W^{(N)}_{1j}} \notag \\
  &= - 2 \left| f - \relu [W^{(N)}_{1k} x^{(N-1)}_k] \right|
  \times 1 \times \delta_{jk} x^{(N-1)}_k  \notag \\
  &\equiv - 2 \sqrt{\loss} \; x^{(N-1)}_j \; ,
\label{eq:backprop1}
\end{align}
provided $W^{(N)}_{ij} x_j > 0$, otherwise the partial derivative
vanishes. This form implies that the derivative of the loss with
respect to the weights in the $N$th layer is a function of the loss
itself and of the previous layer $x^{(N-1)}$. If we ignore the ReLU
derivative in Eq.\eqref{eq:backprop1} and still limit ourselves to the
weight matrix in Eq.\eqref{eq:def_node} we can follow the chain of
layers and find
\begin{align}
  \frac{d \loss}{d W^{(n)}_{ij}}
  &= \frac{\partial \left| f - \relu [W^{(N)}_{1k} x^{(N-1)}_k] \right|^2}{\partial [W^{(N)}_{1k} x^{(N-1)}_k]} \;
  \frac{\partial [(W^{(N)} \cdots W^{(n+1)})_{1k} W^{(n)}_{k \ell} x^{(n-1)}_\ell]}{\partial W^{(n)}_{ij}} \notag \\
  &= - 2 \sqrt{\loss} \; \left( W^{(N)} \cdots W^{(n+1)} \right)_{1i} x_j^{(n-1)} \; .
\label{eq:backprop2}
\end{align}
This means we compute the derivative of the loss with respect to the
weights in the reverse direction as the network evaluation shown in
Eq.\eqref{eq:forward_pass}. We have shown this only for the network
weights, but it works for the biases the same way. This
\underline{back-propagation} is the crucial step in defining
appropriate network parameters by numerically minimizing a loss
function. The simple back-propagation might also give a hint to why
the chain-like network structure of Eq.\eqref{eq:forward_pass}
combined with the affine layers of Eq.\eqref{eq:affine} have turned
out so successful as a high-dimensional representation of arbitrary
functions.

The output of the back-propagation in the network training is the
expectation value of the derivative of the loss function with respect
to a network parameter. We could evaluate this expectation value over
the full training dataset. However, especially for large datasets, it
becomes impossible to compute this expectation value, so instead we
evaluate the same expectation value over a small, randomly chosen
subset of the training data. This method is called stochastic gradient
descent, and the subsets of the training data are called minibatches
or batches
\begin{align}
  \XXLangle \frac{\partial \loss}{\partial \theta_j}
  \XXRangle_\text{minibatch}
  \qquad \text{with} \quad
  \theta_j \in \{ b, W \} \; .
\label{eq:derivative_loss}
\end{align}
Even through the training data is split into batches and the network
training works on these batches, we still follow the progress of the
training and the numerical value of the loss as a function of
\underline{epochs}, defined as the number of batch trainings required
for the network to evaluate the full training sample.

After showing how to compute the loss function and its derivative with
respect to the network parameters, the final question is how we actually
do the minimization. For a given network parameter $\theta_j$, we
first need to scan over possible values widely, and then tune it
precisely to its optimal value. In other words, we first scan the
parameter landscape globally, identify the global minimum or at least
a local minimum close enough in loss value to the global minimum, and
then descend into this minimum. This is a standard task in physics,
including particle physics, and compared to many applications of
Markov Chains Monte Carlos the standard ML-minimization is not very
complicated.  We start with the naive iterative optimization in time
steps
\begin{align}
  \theta_j^{t+1} = \theta_j^t
  - \alpha \XXLangle \frac{\partial \loss^t}{\partial \theta_j} \XXRangle \; .
\label{eq:optimization_naive}
\end{align}
The minus sign means that our optimization walks against the direction
of the gradient, and $\alpha$ is the learning rate. From our
description above it is clear that the learning rate should not be
constant, but should follow a decreasing schedule. 

One of the problems with the high-dimensional loss optimization is
that far away from the minimum the gradients are small and not
reliable. Nevertheless, we know that we need large steps to scan the
global landscape. Once we approach a minimum, the gradients will
become larger, and we want to stay within the range of the minimum. An
efficient adaptive strategy is given by
\begin{align}
  \theta_j^{t+1} = \theta_j^t
  - \alpha \frac{\XLangle \dfrac{\partial \loss^t}{\partial \theta_j} \XRangle}
  {\epsilon +  \sqrt{ \XLangle \dfrac{\partial \loss^t}{\partial \theta_j} \XRangle^2 }} \; .
\label{eq:optimization_adam}
\end{align}
Away from the minimum, this form allows us to enhance the step size
even for small gradients by choosing a sizeable value $\alpha$.
However, whenever the gradient grows too fast, the step size remains
below the cutoff $\alpha/\epsilon$.  Finally, we can stabilize the
walk through the loss landscape by mixing the loss gradient at the
most recent step with gradients from the updates before,
\begin{align}
  \XXLangle \frac{\partial \loss^t}{\partial \theta_j} \XXRangle
  \; \to \; 
  \beta \XXLangle \frac{\partial \loss^t}{\partial \theta_j} \XXRangle
  + (1-\beta) \XXLangle \frac{\partial \loss^{t-1}}{\partial \theta_j} \XXRangle
\end{align}
This strategy is called momentum, and now the complicated form of the
denominator in Eq.\eqref{eq:optimization_adam} makes sense, and serves
as a smoothing of the denominator for rapidly varying gradients.  A
slightly more sophisticated version of this adaptive scan of the loss
landscape is encoded in the widely used Adam optimizer.

Note that for any definition of the step size we still need to
schedule the learning rate $\alpha$. A standard choice for such a
\underline{learning rate scheduling} is an exponential decay of
$\alpha$ with the batch or epoch number. An interesting alternative is
a one-cycle learning rate where we first increase $\alpha$ batch after
batch, with a dropping loss, until the loss rapidly explodes. This
point defines the size of the minimum structure in the loss
landscape. Now we can choose the step size at the minimum loss value to
define a suitable constant learning rate for our problem, potentially
leading to much faster training. Finally, we need to mention that the
minimization of the loss function for a neural network only ever uses
first derivatives, differently from the optimization of a fit
function. The simple reason is that for the large number of network
parameters $\theta$ the simple scaling of the computational effort
rules out computing second derivatives like we would normally do.

Going back to the three errors or uncertainties\index{uncertainties} introduced in the last section, they can
be translated directly to neural networks. The approximation uncertainty is
less obvious than for the choice of fit function, but also the
expressiveness of neural network is limited through the network
architecture and the set of hyperparameters. The training uncertainty
becomes more relevant because we now minimize the loss over an
extremely high-dimensional parameter space, where we cannot expect to
find the global minimum and will always have to settle for a
sufficiently good local minimum.  To define a compromise between the
approximation and generalization uncertainties we usually divide a ML-related
dataset into three parts. The main part is the training data, anything
between 60\% and 80\% of the data. The above-mentioned test data is
then 10\% to 20\% of the complete dataset, and we use it to check how
the network generalizes to unseen data, test for overtraining, or
measure the network performance. The validation data can be used to
re-train the network, optimize the architecture or the different
settings of the network. The crucial aspect is that the test data is
completely independent of the network training.

\subsubsection{Likelihood loss}
\label{sec:basics_deep_like}

There are (at least) two ways to derive loss functions for neural
networks: maximizing a probability or likelihood, as described in
Sec.~\ref{sec:basics_fit}, or variational inference, which we will
introduce in Sec.~\ref{sec:basics_deep_bayes}

For the likelihood maximization we start with the conditional
probability $p(\theta|x_\text{train})$, which we do not have access to
during the training. Instead, we can use Bayes' theorem\index{Bayes'
  theorem} to relate it to the corresponding likelihood
\begin{align}
  p(\theta|x_\text{train})
  = \frac{p(x_\text{train}|\theta) \; p(\theta)}{p(x_\text{train})} \; ,
\end{align}
which we can compute by forward simulation using the network.
Because the evidence $p(x_\text{train})$ does not depend on the
network parameters, we can define the loss
in analogy to Eq.\eqref{eq:argmax_like} as
\begin{align}
  \loss
  &= - \log  p(x_\text{train}|\theta) \; .
  \label{eq:loss_like}
\end{align}
As before we assume that the prior is independent of the training
data, and it allows us to implement requirements on the weights
improving the numerics, for example a Gaussian weight
regularization. This means that in most cases maximizing a probability
or a likelihood loss are equivalent.

The big difference between typical fits and network training is that
for the latter we do not have access to the uncertainty $\sigma$ of
the data. However, it is included in the Gaussian likelihood, so we
will try to learn it as together with $f_\theta(x)$ from the same data
and using the same network,
\begin{align}
  \sigma_\theta(x) \approx \sigma(x) \; .
\end{align}
Unlike $f(x)$, the reference function $\sigma(x)$ is not known
explicitly, but only implicitly through the data structure and our
assumed Gaussian likelihood from Eq.\eqref{eq:def_gaussian}
\begin{align}
  p(x|\theta) &= \prod_{x_\text{train}} \frac{1}{\sqrt{2 \pi} \sigma_\theta(x)}
  \exp \left( - \frac{|f - f_\theta(x)|^2}{2 \sigma_\theta(x)^2} \right) \; .
\label{eq:def_gaussian2}
\end{align}
Unlike in Eq.\eqref{eq:likelihood_loss} we now have to include the
normalization prefactor of the Gaussian, giving us the so-called
\underline{heteroskedastic loss},\index{heteroskedastic loss}
\begin{align}
\boxed{
  \loss_\text{heteroskedastic}
  =  \frac{|f(x) - f_\theta(x)|^2}{2\sigma_\theta(x)^2}
  + \log \sigma_\theta(x) + \cdots }
\label{eq:bnn_hetero}
\end{align}
%
In this loss the minimization of the Gaussian exponent will drive
$\sigma_\theta(x)$ towards large values, while its explicit appearance
from the normalization will prefer a small width. Learning both, we
can then use $f_\theta(x)$ and $\sigma_\theta(x)$, point by point, for
a downstream task.  Additional loss terms can include, for instance,
the prior turned weight regularization. As a side remark, adding an
uncertainty can stabilize the network training even if we are not
interested in the uncertainty, because it allows the network to ignore
data it cannot describe and focus on aspects it can improve, and
because it comes with a well-defined regularization.

From the way the uncertainty is learned it is clear that it will
perfectly capture statistical limitations and noise in the
data. However, it will also reflect a limited expressivity of the
model, which does not allow the training to minimize the numerator of
the exponent beyond a certain cutoff, which will then turn into the
learned value for $\sigma$. In cases where the uncertainty is not
symmetric and therefore poorly described by the Gaussian likelihood,
we can replace $p(x|\theta)$ in the loss by a Gaussian mixture.

The one shortcoming of the heteroskedastic loss is that it assumes a
perfect training in the loss landscape, which means it does not
capture effects from a broad minimum in $\theta$-space or a set of
similar local minima rather than one well-defined global minimum.

\subsubsection{Repulsive ensembles}
\label{sec:basic_deep_repuls}


An alternative, and we will we see orthogonal approach to learned
uncertainties, is based on ensembles of networks trained on the same
data. Usually, the networks forming an ensemble are trained
independently. However, if we assume that all these trainings lead to
the same global minimum of the loss function, all members of perfectly
trained networks will be equivalent. This suggests that we have to
train ensembles of networks together, if we hope for them to describe
a posterior distribution $p(\theta|x_\text{train})$ of the network output.

From Eq.\eqref{eq:optimization_naive} we know that neural
networks are trained using gradient descent, where an update
rule minimizes a loss function. From Eq.\eqref{eq:loss_like} we
remember that the network training should maximize the probability of
the network weights given a training dataset $x_\text{train}$,
\begin{align}
\theta^{t+1} = \theta^t + \alpha \nabla_{\theta^t} \log p(\theta^t|x_\text{train}) \; .
\label{eq:update_rule1}
\end{align}
We can extend the update rule to an ensemble of networks by
applying the update step to all networks in the ensemble.

For our case of a single global minimum of the loss landscape the
coverage by the ensemble can be improved by a repulsive interaction in
the update rule.  Such an interaction should take into account the
proximity of the ensemble member $\theta$ to all other members. We
introduce a kernel $k(\theta,\theta_j)$, typically chosen as Gaussian,
and sum the interactions with all other weight configurations
\begin{align}
\theta^{t+1} = \theta^t + \alpha 
\nabla_{\theta^t} \left[ \log p(\theta^t|x_\text{train})
- \frac{1}{n} \sum_{j=1}^n  k(\theta^t,\theta^t_j) \right] \; .
\label{eq:update_rule2}
\end{align}
The question is if this update rule, for a given kernel, leads to
ensemble members sampling the weight probability,
\begin{align}
    \theta \sim p(\theta|x_\text{train}) \; .
\end{align}

To answer this question we need to relate the update rule, or the
discretized $t$-dependence of a weight vector $\theta(t)$, to a
time-dependent probability density $\rho(\theta,t)$.  We use a
mathematical structure that we will encounter again when deriving
conditional flow matching networks in Sec.~\ref{sec:gen_diff_cfm}:
there are two equivalent ways to describe the time evolution of a
system, an ODE or a continuity equation,
\begin{align}
  \frac{d\theta}{dt} = v(\theta,t)
  \qquad \text{or} \qquad 
\frac{\partial \rho(\theta,t)}{\partial t} 
= - \nabla_\theta \left[ v(\theta,t) \rho(\theta,t) \right] \; .
\label{eq:ode}
\end{align}
For a given velocity field $v(\theta,t)$ the individual paths
$\theta(t)$ describe the evolving density $\rho(\theta,t)$ and the two
conditions are equivalent.  If we choose the velocity field as
\begin{align}
v(\theta,t) = - \nabla_\theta \log \frac{\rho(\theta,t)}{\pi(\theta)} \; ,
\end{align}
these two equivalent conditions read
\begin{align}
\frac{d\theta}{dt} &= - \nabla_\theta \log \frac{\rho(\theta,t)}{\pi(\theta)} 
\notag \\
\frac{\partial \rho(\theta,t)}{\partial t} 
&= \nabla_\theta \left[ \rho(\theta,t) \nabla_\theta \log \frac{\rho(\theta,t)}{\pi(\theta)}\right] \notag \\
&= - \nabla_\theta \left[ \rho(\theta,t) \nabla_\theta \log \pi(\theta) \right]
+ \nabla_\theta^2 \rho(\theta,t) \; .
\label{eq:fp}
\end{align}
This continuity equation is the Fokker-Planck equation, for which
$ \rho(\theta,t) = \pi(\theta)$ is a stationary probability
distribution,
\begin{align}
\nabla_\theta \left[ \pi(\theta) \nabla \log \pi(\theta) \right]
&= \nabla_\theta \left[ \pi(\theta) \frac{1}{\pi(\theta)} \nabla_\theta \pi(\theta) \right] \notag \\
&= \nabla_\theta^2 \log \pi(\theta) 
\qquad \Rightarrow \qquad 
\frac{\partial \rho(\theta,t)}{\partial t} \Bigg|_{\rho = \pi} = 0 \; .
\end{align}
This stationary solution of the Fokker-Planck equation is even unique,
so we know that the evolution given by the first line of
Eq.\eqref{eq:fp} is guaranteed to converge to $\rho(\theta,t) \to
\pi(\theta)$.

Next, we relate the ODE in Eq.\eqref{eq:fp} to the update rule for
repulsive ensembles, Eq.\eqref{eq:update_rule2}.  The discretized
version of the ODE is
\begin{align}
\frac{\theta^{t+1}-\theta^t}{\alpha}
= - \nabla_{\theta^t} \log \frac{\rho(\theta^t)}{\pi(\theta^t)} \; .
\end{align}
If we do not know the density $\rho(\theta^t)$ explicitly, we again 
approximate it as a superposition of kernels with the correct normalization,
\begin{align}
\rho(\theta^t) \approx \frac{1}{n} \sum_{j=1}^n k(\theta^t,\theta_j^t)
\qquad \text{with} \qquad 
\int d\theta^t \rho(\theta^t) 
= \frac{1}{n} \sum_{j=1}^n \int d\theta^t k(\theta^t,\theta_j^t) 
= \frac{1}{n} \sum_{j=1}^n 1 = 1 \; .
\label{eq:normalize}
\end{align}
We can insert this kernel approximation into the discretized ODE,
\begin{align}
\frac{\theta^{t+1}-\theta^t}{\alpha}
&= - \nabla_{\theta^t} \left[ \log \rho(\theta^t) - \log \pi(\theta^t) \right]
\notag \\
&= \nabla_{\theta^t} \log \pi(\theta^t)
- \nabla_{\theta^t} \log \left[ \frac{1}{n} \sum_j k(\theta^t,\theta_j^t) \right]
\notag \\
&= \nabla_{\theta^t} \log \pi(\theta^t)
- \nabla_{\theta^t} \log \sum_j k(\theta^t,\theta_j^t) 
\notag \\
&= \nabla_{\theta^t} \log \pi(\theta^t)
- \frac{\nabla_{\theta^t} \sum_j k(\theta^t,\theta_j^t)}{\sum_i k(\theta^t,\theta_i^t)}
\label{eq:fp_update}
\end{align}
This form can be compared to the update rule in Eq.\eqref{eq:update_rule2}.
To make them identical, we first identify
\begin{align}
\pi(\theta) \equiv p(\theta|x_\text{train}) \; .
\label{eq:posterior}
\end{align}
This means that our target weight density is the probability of the
weights given the training data. Second, we add the normalization
term of Eq.\eqref{eq:fp_update} to our original kernel in
Eq.\eqref{eq:update_rule2},
\begin{align}
\nabla_{\theta^t} \sum_j k(\theta^t,\theta_j^t)
\; \to \; 
\frac{\nabla_{\theta^t} \sum_j k(\theta^t,\theta_j^t)}{\sum_i k(\theta^t,\theta_i^t)} \; .
\label{eq:normalized_grad}
\end{align}
Note that we cannot remove the normalization by ansatz, because it
collides with Eq.\eqref{eq:normalize}.

So far, we consider ensembles with a repulsive force in weight
space. However, we are interested in the function the network encodes
and not the latent or weight representation.  For contemporary network architectures 
we know that many points in $\theta$-space provides us with equivalent 
learned functions. 
This is the reason why during training we do not have to find the global 
loss minimum and instead work with a good enough local minimum.  For instance, two
networks encoding the same function could be constructed by permuting
the weights of the hidden layers, unaffected by a repulsive force in
weight space.  This is why we prefer a repulsive force in the space of
network outputs $f_\theta(x)$.

Symbolically, we can then write the update rule from
Eq.\eqref{eq:update_rule2} with the normalization of
Eq.\eqref{eq:normalized_grad} as
\begin{align}
\frac{f^{t+1} - f^t}{\alpha} = 
\nabla_{f^t} \log p(f|x_\text{train})
- \frac{\sum_j \nabla_{f^t} k(f,f_j)}{\sum_j k(f,f_j)} \; .
\label{eq:update_rule_fspace}
\end{align}
A typical choice for the kernel in function space is still
a Gaussian in the multi-dimensional function space, evaluated over
a sample,
\begin{align}
k(f_{\theta}(x),f_{\theta_j}(x)) \propto
\exp \left( - \frac{|f_{\theta}(x) - f_{\theta_j}(x)|^2}{h} \right) \; .
\label{eq:gaussian_kernel}
\end{align}
The width $h$ should be chosen such that the width of the distribution
is not overestimated while still ensuring that it is sufficiently
smooth.

No matter how we define the update step, the network training is
always in weight space, so we have to translate the function-space
update rule into weight space using the appropriate Jacobian
\begin{align}
\frac{\theta^{t+1}-\theta^t}{\alpha} 
&= \frac{\partial f^t}{\partial \theta^t} 
\left[ \nabla_{f^t} \log p(f_{\theta^t}|x_\text{train})
- \frac{\sum_j \nabla_{f} k(f_{\theta^t},f_{\theta_j^t})}{\sum_j k(f_{\theta^t},f_{\theta_j^t})} \right] \notag \\
&= 
\nabla_{\theta^t} \log p(\theta^t|x_\text{train})
- 
\frac{\sum_j \nabla_{\theta^t} 
k(f_{\theta^t},f_{\theta_j^t})}{\sum_j k(f_{\theta^t},f_{\theta_j^t})} \; .
\end{align}
Furthermore, we cannot evaluate the repulsive kernel in function
space, so we have to evaluate the function for a finite batch of
points $x$,
\begin{align}
\frac{\theta^{t+1} - \theta^t}{\alpha} 
\approx \nabla_{\theta^t} \log p(\theta^t|x_\text{train})
- \frac{\sum_j \nabla_{\theta^t} k(f_{\theta^t}(x),f_{\theta_j^t}(x))}{\sum_j k(f_{\theta^t}(x),f_{\theta_j^t}(x))}  \; .
\label{eq:update_rule_fspace2}
\end{align}

Finally, we turn the update rule in Eq.\eqref{eq:update_rule_fspace2}
into a loss function for the repulsive ensemble training.  We want to
use a tractable likelihood loss, which we get from Bayes' theorem.
%
%
In the loss function we neglect the evidence $p(x_\text{data})$, but
have to include the prior $p(\theta)$, which we assume to be Gaussian,
\begin{align}
    \log p(\theta|x_\text{train}) 
    = \log p(x_\text{train}|\theta) \; 
    - \frac{\theta^2}{2 \sigma^2} + \text{const} \; .
\end{align}
Given a training dataset of size $N$, we evaluate the likelihood on
batches of size $B$, so Eq.\eqref{eq:update_rule_fspace2} becomes
%
%
%
%
\begin{align}
    \frac{\theta^{t+1} - \theta^t}{\alpha}
\approx \nabla_{\theta^t} \frac{N}{B} \sum_{b=1}^B \log p(x_b|\theta)
- \frac{\sum_j \nabla_{\theta^t} k(f_{\theta^t}(x),f_{\theta_j^t}(x))}{\sum_j k(f_{\theta^t}(x),f_{\theta_j^t}(x))} 
- \nabla_{\theta^t} \frac{\theta^2}{2 \sigma^2}
\; .
\end{align}
Here, $f_{\theta^t}(x)$ is to be understood as evaluating the function
for all samples $x_1,\dots,x_B$ in the batch.

To turn the update rule into a loss function, we flip the sign of term
in the gradient, divide it by $N$ to remove the scaling with the size
of the training dataset, and sum over all members of the
ensemble. Since the gradients of the loss function are computed with
respect to the parameters of all networks in the ensemble, we need to
ensure the correct gradients of the repulsive term using a
stop-gradient operation, denoted with an overline
$\overline{f_{\theta_j}(x)}$. The loss function for repulsive
ensembles then reads
\begin{align}
  \boxed{
  \loss = \sum_{i=1}^n \left[- \frac{1}{B} \sum_{b=1}^B \log p(x_b|\theta_i)
    + \frac{1}{N} \frac{\sum_{j=1}^n k(f_{\theta_i}(x), \overline{f_{\theta_j}(x)})}{\sum_{j=1}^n k(\overline{f_{\theta_i}(x)}, \overline{f_{\theta_j}(x)})} + \frac{\theta_i^2}{2N \sigma^2}  \right]
  } \; .
\label{eq:loss_re}
\end{align}
The prior has just become an L2-regularization with prefactor
$1/(2N\sigma^2)$.

Looking at the learned uncertainty from the heteroskedastic loss in
Eq.\eqref{eq:bnn_hetero} and the uncertainty in model space from the
repulsive ensembles, it is not clear how the two are related and what
kind of uncertainties they really cover. We will look at this aspect
in the following sections in more detail.

\subsubsection{Bayesian networks}
\label{sec:basics_deep_bayes}

Another way to include uncertainties on the network parameters
$\theta$ systematically are so-called Bayesian neural
networks\index{Bayesian network}. They are a naming disaster in that
there is nothing exclusively Bayesian about them~\cite{bnn_early3},
while in particle physics Bayesian has a clear negative connotation.
The difference between a deterministic and a Bayesian network is that
the latter allow for distributions of network parameters, which then
define distributions of the network output and provide central values
$f(x)$ as well as uncertainties on $f(x)$ by sampling over
$\theta$-space.

The Bayesian loss follows from a statistics argument.  Let us start
with a simple amplitude regression over phase space,
\begin{align}
  f_\theta(x) \approx f(x) \equiv A(x)
  \qquad \text{with} \quad x \in \mathbb{R}^D \; .
\end{align}
The training data consists of pairs $(x,A)_j$.  We define $p(A)
\equiv p(A|x)$ as the probability distribution for possible amplitudes
at a given phase space point $x$ and omit the argument $x$ from now
on. The mean value for the amplitude at the point $x$ is
\begin{align}
\langle \, A \, \rangle 
= \int dA \; A \; p(A)
\qquad \text{with} \qquad 
  p(A) 
  = \int d \theta \; p(A | \theta) \; p(\theta |x_\text{train}) \; .
\label{eq:folding}
\end{align}
Here, we can think of $p(A|\theta)$ as a single model describing an
amplitude through a set of network parameters, while
$p(\theta|x_\text{train})$ weights this model by its level of
agreement with the training data $x_\text{train}$.  We do not know the
closed form of $p(\theta | x_\text{train})$, because it is encoded in
the training data. Training the network means that we approximate it
as a distribution using \underline{variational approximation} for the
integrand in the sense of a distribution and test function
\begin{align}
p(A)
= \int d \theta \; p(A | \theta) \; p(\theta | x_\text{train})
\approx \int d \theta \; p(A | \theta) \; q(\theta|x) 
\equiv \int d \theta \; p(A | \theta) \; q(\theta) \; .
\label{eq:first_bayes}
\end{align}
As for $p(A)$ we omit the $x$-dependence of $q(\theta|x)$. This
approximation leads us directly to the BNN loss function.

There are many ways we can compare two distributions, defining a problem
called optimal transport. We will come back to alternative ways of
combining probability densities over high-dimension spaces in
Sec.~\ref{sec:gen}. For now we just introduce the
\underline{Kullback--Leibler divergence}\index{KL-divergence}, which compares two probability
distributions, evaluated on a dataset corresponding to the first
distribution,
\begin{align}
    \kl [p_a, p_b] = \sum p_a \ln \frac{p_a}{p_b}
    \ne
    \kl [p_b, p_a] 
    \; .
\end{align}
For continuous distributions it reads
\begin{align}
  \boxed{
  \kl [p_a,p_b]
= \XXLangle \log \frac{p_a}{p_b} \XXRangle_{p_a}
\equiv \int d x \; p_a(x) \; \log \frac{p_a(x)}{p_b(x)} } \; .
\label{eq:def_kl}
\end{align}
It vanishes if two distributions agree everywhere.  Because the
KL-divergence is not symmetric in its two arguments we can evaluate a
forward and a reverse KL-divergence,
\begin{align}
\kl [p_a,p_b]
= \XXLangle \log \frac{p_a}{p_b} \XXRangle_{p_a} 
\qquad \text{or} \qquad 
  \kl [p_b,p_a]
= \XXLangle \log \frac{p_b}{p_a} \XXRangle_{p_b} \; .
\label{eq:kl_twofold}
\end{align}
The difference between them is which of the two distributions we
choose to sample the logarithm from, and we will always choose the
version that suits the problem better.

There is a nice way to show that otherwise the KL-divergence is always
positive. The numerator of the logarithm gives the negative
information entropy for $p_a$, which means the contribution from the
numerator is negative and has a well-defined minimum given in
Eq.\eqref{eq:min_entropy}. The contribution of the denominator is the
information entropy for a distribution $p_b$, which does not match
$p_a$, which is called cross entropy of $p_a$ and $p_b$. If the two
distributions are different, their respective encoding is not optimal,
which means the cross entropy is larger than the information entropy,
so the KL-divergence is positive. Mathematically, we can use the
relation $\log p \le p-1$, where the equal sign is true for $p=1$ to
compute
\begin{align}
  - \kl [p_a,p_b]
  &= \int_{p_a>0} d x \; p_a(x) \; \log \frac{p_b(x)}{p_a(x)} \notag \\
  &\le \int_{p_a>0} d x \; p_a(x) \; \left( \frac{p_b(x)}{p_a(x)} -1 \right) 
  = \int_{p_a>0} d x \; p_b(x) - 1 \le 0 \; ,
\end{align}
using the fact that $p_a(x)$ and $p_b(x)$ are normalized probability
distributions.

We now define the variational approximation using the KL-divergence
introduced in Eq.\eqref{eq:def_kl},
\begin{align}
  \kl [q(\theta),p(\theta|x_\text{train})]
= \XXLangle \log \frac{q(\theta)}{p(\theta|x_\text{train})} \XXRangle_q  
= \int d\theta \; q(\theta) \; \log \frac{q(\theta)}{p(\theta|x_\text{train})} \; .
\label{eq:def_kl2}
\end{align}
Using Bayes' theorem\index{Bayes' theorem} we can write the KL-divergence as
\begin{align}
 \kl [q(\theta),p(\theta|x_\text{train})]
&= \int d\theta \; q(\theta) \; \log \frac{q(\theta) p(x_\text{train})}{p(\theta) p( x_\text{train}|\theta)}
\notag \\
&= \kl [q(\theta),p(\theta)] 
 - \int d\theta \; q(\theta) \; \log p(x_\text{train}|\theta)
 + \log p(x_\text{train}) \int d\theta \; q(\theta) \; .
\label{eq:kl_bayes2}
\end{align}
The prior $p(\theta)$ describes the network parameters before
training; since it does not really include prior physics or training
information we will still refer to it as a prior, but we think about
it as a hyperparameter which can be chosen to optimize performance and
stability.  From a practical perspective, a good prior will help the
network converge more efficiently, but any prior should give the
correct results, and we always need to test the effect of different
priors.

The evidence $p(x_\text{train})$ guarantees the correct
normalization of $p(\theta | x_\text{train} )$ and is usually
intractable.  If we implement the normalization condition for
$q(\theta)$ by construction, we find
\begin{align}
 \kl [q(\theta),p(\theta|x_\text{train})]
&= \kl [q(\theta),p(\theta)] 
 - \int d\theta \; q(\theta) \; \log p(x_\text{train}|\theta)
 + \log p(x_\text{train}) \; .
 \label{eq:kl_bayes3}
\end{align}
The log-evidence in the last term does not depend on $\theta$, which
means that it will not be adjusted during training and we can ignore
when constructing the loss. However, it ensures that $\kl
[q(\theta),p(\theta|x_\text{train})]$ can reach its minimum at zero. Alternatively,
we can solve the equation for the evidence and find
\begin{align}
 \log p(x_\text{train})
&=\kl [q(\theta),p(\theta|x_\text{train})]
 - \kl [q(\theta),p(\theta)] 
 + \int d\theta \; q(\theta) \; \log p(x_\text{train}|\theta) \notag \\
&> \int d\theta \; q(\theta) \; \log p(x_\text{train}|\theta)
 - \kl [q(\theta),p(\theta)] 
\label{eq:def_elbo}
\end{align}
This condition is called the \underline{evidence lower bound (ELBO)}\index{evidence lower bound}, and
the evidence reaches this lower bound exactly when our training
condition in Eq.\eqref{eq:def_kl} is minimal. Combining all of this,
we turn Eq.\eqref{eq:kl_bayes3} or, equivalently, the ELBO into the
loss function for a Bayesian network\index{Bayesian network},
\begin{align}
\boxed{
  \loss_\text{BNN} 
= - \int d\theta \; q(\theta) \; \log p(x_\text{train}|\theta) 
+ \kl [q(\theta),p(\theta)]
} \; .
\label{eq:loss_bayes1}
\end{align}
The first term of the BNN loss is a \underline{likelihood}\index{likelihood loss} sampled
according to $q(\theta)$, the second enforces a (Gaussian) prior.
This Gaussian prior acts on the distribution of network weights. Using
an ELBO loss means nothing but minimizing the KL-divergence between
the probability $p(\theta|x_\text{train})$ and its network
approximation $q(\theta)$ and neglecting all terms which do not depend
on $\theta$. It results in two terms, a likelihood and a
KL-divergence, which we will study in more detail next.

The Bayesian network output is constructed in a non-linear way with a
large number of layers, to we can assume that Gaussian weight
distributions do not limit us in terms of the uncertainty on the
network output.  The log-likelihood $\log p(x_\text{train}|\theta)$ implicitly
includes the sum over all training points.

Before we discuss how we evaluate the Bayesian network in the next
section, we want to understand more about the BNN setup and
loss. First, let us look at the deterministic limit of our Bayesian
network loss. This means we want to look at the loss function of the
BNN in the limit
\begin{align}
  q(\theta) = \delta(\theta - \theta_0) \; .
\label{eq:deterministic}
\end{align}
The easiest way to look at this limit is to first assume a Gaussian
form of the network parameter distributions, as given in
Eq.\eqref{eq:def_gaussian}
\begin{align}
  q_{\mu,\sigma}(\theta) = \frac{1}{\sqrt{2 \pi}\sigma_q} e^{- (\theta-\mu_q)^2/(2 \sigma_q^2)} \; ,
\label{eq:gaussian}
\end{align}
and correspondingly for $p(\theta)$. The
KL-divergence\index{KL-divergence} has a closed form,
\begin{align}
  \kl [q_{\mu,\sigma}(\theta),p_{\mu,\sigma}(\theta)]  
  &= \frac{\sigma_q^2 - \sigma_p^2 + (\mu_q - \mu_p)^2}{2\sigma_p^2}
  + \log \frac{\sigma_p}{\sigma_q} \; .
\label{eq:kl_gaussian}
\end{align}
We can now evaluate this KL-divergence in the limit of $\sigma_q \to
0$ and finite $\mu_q(\theta) \to \theta_0$ as the one remaining
$\theta$-dependent parameter,
\begin{align}
  \kl [q_{\mu,\sigma}(\theta),p_{\mu,\sigma}(\theta)]  
  \to \frac{(\theta_0 - \mu_p)^2}{2\sigma_p^2} + \text{const} \; .
\label{eq:kl_gaussian2}
\end{align}
We can write down the deterministic limit of Eq.\eqref{eq:loss_bayes1},
\begin{align}
  \loss_\text{BNN} \to - \log p(x_\text{train}|\theta_0) + \frac{(\theta_0 - \mu_p)^2}{2\sigma_p^2} \; .
\label{eq:bnn_gaussian3}
\end{align}
The first term is again the likelihood defining the correct network
parameters, the second ensures that the network parameters do not
become too large. Because it include the squares of the network
parameters, it is referred to as an
\underline{L2-regularization}\index{regularization}. Going back to
Eq.\eqref{eq:loss_bayes1}, an ELBO loss is a combination of a
likelihood loss and a regularization. For the Bayesian network the
prefactor of this regularization term is fixed.

We generalize this idea, choose $\mu_p=0$, and apply the
L2-regularization with a freely chosen pre-factor
instead of $\sigma_p$.  In that case the deterministic and
L2-regularized likelihood loss reads\index{likelihood loss}
\begin{align}
  \boxed{
    \loss_\text{L2} = - \log p(x_\text{train}|\theta_0) + \lambda \theta_0^2
    } \; ,
\label{eq:bnn_gaussian}
\end{align}
with a free hyperparameter $\lambda$.

Sampling the likelihood following a distribution of the network
parameters, as it happens in the first term of the Bayesian loss in
Eq.\eqref{eq:loss_bayes1}, is something we can also generalize to
deterministic networks. Let us start with a toy model where we sample
over network parameters by either including them in the loss
computation or not. When we include an event, the network weight is
set to $\theta_0$, otherwise $q(\theta) = 0$. Such a random sampling
between two discrete possible outcomes is described by a
\underline{Bernoulli distribution}.  If the two possible outcomes are
zero and one, we can write the distribution in terms of the
expectation value $\rho \in [0,1]$,
\begin{align}
  p_\text{Bernoulli}(x)
  = \begin{cases}
    \rho^x (1-\rho)^{1-x} \quad  & x=0,1 \\
    0 & \text{else} \; .
    \end{cases} 
  \label{eq:def_bernoulli}
\end{align}
When we include an event, the network weight is set to $\theta_0$,
otherwise it is set to zero, or $\theta = x \theta_0$. We can use the
Bernoulli probability as a test function for our integral over the
log-likelihood $\log p(x_\text{train}|\theta)$ and find
\begin{align}
  \loss_\text{BNN}
  &= - \int dx \; \left[ \rho^x (1-\rho)^{1-x} \right] \; \log p(x_\text{train}|x\theta_0) \notag \\
  &= - \rho \log p(x_\text{train}|\theta_0) + \text{const}.
  \label{eq:loss_dropoout}
\end{align}
Such an especially simple sampling of weights by removing nodes is called
\underline{dropout} and is commonly used to avoid overfitting of
networks. For deterministic networks $\rho$ is a free hyperparameter
of the network training, while for Bayesian networks this kind of
sampling is a key result from the construction of the loss
function.

\subsection{Regression}
\label{sec:basics_regr}

Following our brief introduction to deep networks we can directly look
at specific application. Using a neural network for regression means
that we learn a function $f_\theta(x)$ over some kind of phase space
$x$. We will look at three applications relevant to particle
physics. In Sec.~\ref{sec:basics_regr_amp} we use the Bayesian network
introduced in Sec.~\ref{sec:basics_deep_bayes} to learn transition
amplitudes over phase space with high precision, as suggested by our
notation in the BNN introduction. In Sec.~\ref{sec:basics_regr_nnpdf}
we will briefly introduce the most influential introduction of machine
learning to particle physics, namely the NNPDF regression of parton
density. It defines the target in accuracy, uncertainty, and control that LHC
applications of machine learning need to follow. Finally, in
Sec.~\ref{sec:basics_regr_int} we discuss a creative new method to
numerically integrate functions using surrogate integrands.

\subsubsection{Amplitudes with uncertainties}
\label{sec:basics_regr_amp}

\begin{figure}[b!]
    \centering
    \includegraphics[width=0.75\textwidth]{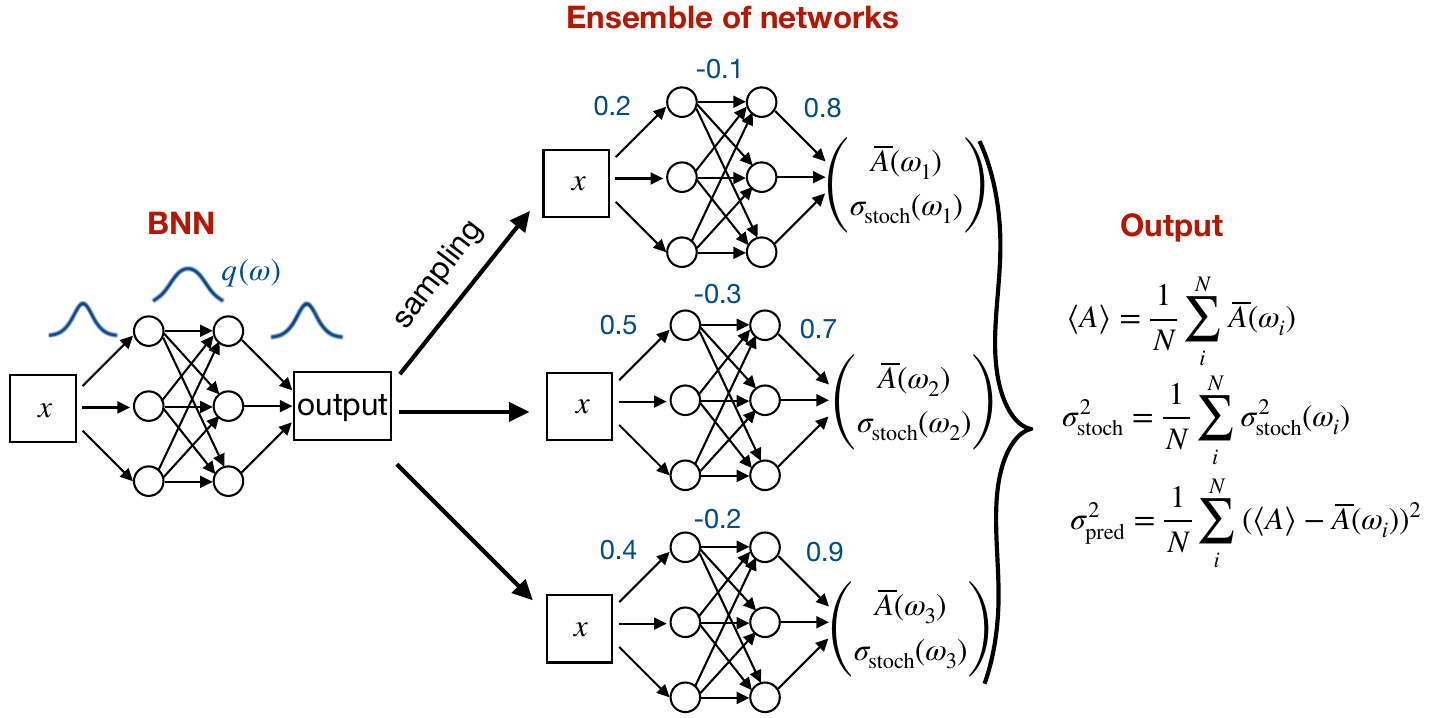}
    \caption{Illustration of a Bayesian network. Figure from
      Ref.~\cite{Kasieczka:2020vlh}.}
    \label{fig:bnn}
\end{figure}

After introducing BNNs using the notation of transition amplitude
learning, we still have to extract the mean and the
uncertainty\index{uncertainties} for the amplitude $A$ over phase
space. To evaluate the network we first exchange the two integrals in
Eq.\eqref{eq:folding} and use the variational approximation to write
the mean prediction $\overline{A}(\theta)$ for a given phase space
point as
\begin{align}
\langle A \rangle 
&= \int dA d \theta \; A \; p(A | \theta) \; p(\theta|x_\text{train}) \notag \\
&= \int dA d \theta \; A \; p(A | \theta) \; q(\theta) \notag \\
&\equiv \int d\theta \; q(\theta) \; \overline{A}(\theta)  
\qqquad \text{with $\theta$-dependent mean} \qquad
\overline{A}(\theta) 
= \int dA \; A \; p(A | \theta) \; .
\label{eq:expectations}
\end{align}
We can interpret this formula as a \underline{sampling over network
  parameters}, provided we assume uncorrelated variations of the
individual network parameters. Strictly speaking, this is an
assumption we make. We reproduce a standard, deterministic network in
the limit
\begin{align}
  q(\theta)\to \delta(\theta - \theta_0) \; .
\label{eq:bnn_delta_q}
\end{align}
In this case $p(A|\theta_0)$ returns the one correct value for the
amplitude.  For imperfect training the probability distribution
$p(A|\theta)$ describes a spectrum of amplitude labels for each phase
space point.  Corresponding to the definition of the
$\theta$-dependent mean $\overline{A}(\theta)$, the variance of $A$ is
\begin{align}
\sigma_\text{tot}^2
&= \int dA d \theta \; \left( A - \langle A \rangle \right)^2 \; p(A | \theta) \; q(\theta) \notag \\
&= \int dA d \theta \; \left( A^2 - 2 A \langle A \rangle + \langle A \rangle^2 \right) \; p(A | \theta) \; q(\theta) \notag \\
&= \int d\theta \; q(\theta) \left[ 
   \int dA \; A^2 \; p(A | \theta) 
 - 2 \langle A \rangle \; \int dA \; A \; p(A | \theta) 
 + \langle A \rangle^2 \; \int dA \; p(A | \theta) \right] 
\label{eq:def_sigmas1}
\end{align}
For the three integrals we can generalize the notation for the
$\theta$-dependent mean as in Eq.\eqref{eq:expectations} and write
\begin{align}
\sigma_\text{tot}^2
&= \int d\theta \; q(\theta) \left[ 
   \overline{A^2}(\theta) 
 - 2 \langle A \rangle \overline{A}(\theta) 
 + \langle A \rangle^2 \right] \notag \\
&= \int d\theta \; q(\theta) \left[ 
   \overline{A^2}(\theta) 
 - \overline{A}(\theta)^2 
 + \overline{A}(\theta)^2 
 - 2 \langle A \rangle \overline{A}(\theta)
 + \langle A \rangle^2 \right] \notag \\
&= \int d\theta \; q(\theta) \left[ 
   \overline{A^2}(\theta) - \overline{A}(\theta)^2
 + \left( \overline{A}(\theta) - \langle A \rangle \right)^2 \right] 
\equiv \sigma_\text{syst}^2 + \sigma_\text{stat}^2 \; .
\label{eq:def_sigmas2}
\end{align}
For this transformation we keep in mind that $\langle A \rangle$ is
already integrated over $\theta$ and $A$ and can be pulled out of the
integrals.

Equation~\eqref{eq:def_sigmas2} defines two contributions to the
variance or uncertainty.  First, $\sigma_\text{stat}$ is the
$\theta$-integrated expectation value
\begin{align}
 \sigma_\text{stat}^2
&= \int d\theta \; q(\theta) \;
   \Big[ \overline{A}(\theta) - \langle A \rangle \Big]^2 \; .
\label{eq:sig_pred}
\end{align}
Following the definition in Eq.\eqref{eq:expectations}, it vanishes in
the limit of perfectly trained network weights, $q(\theta)\to
\delta(\theta - \theta_0)$, where $\langle A \rangle =
\overline{A}(\theta_0)$.  That limit requires perfect training, which
means that $\sigma_\text{stat}$ decreases with more training data. In
physics this defines a \underline{statistical
  uncertainty}\index{statistical uncertainty}.

%

In contrast, $\sigma_\text{syst}$ is defined without sampling the
network parameters,
\begin{align}
\sigma_\text{syst}^2 
\equiv \langle \sigma_\text{syst}(\theta)^2 \rangle 
&= \int d\theta \; q(\theta) \; \sigma_\text{syst}(\theta)^2 \notag \\
&= \int d\theta \; q(\theta) \; \Big[
   \overline{A^2}(\theta) - \overline{A}(\theta)^2 \Big] \; ,
\label{eq:sig_stoch}
\end{align}
so it does not vanish for $q(\theta)\to \delta(\theta
- \theta_0)$.
However, from 
Eq.\eqref{eq:expectations} we see that it vanishes if the
amplitude is arbitrarily well known in terms of a sharp likelihood
$p(A|\theta) \to \delta(A - A_0)$. As the variance, for a Gaussian
likelihood it is just the $\sigma_\theta$ in the heteroskedastic loss
of Eq.\eqref{eq:bnn_hetero}.

While this systematic uncertainty might also be affected by too little
training data, it approaches a plateau for perfect training.  This
plateau value can reflect a stochastic training sample, limited
expressivity of the network, not-so-smart choices of hyperparameters
etc, in the sense of a \underline{systematic
  uncertainty}\index{systematic uncertainty}. In practice, the systematic
uncertainty can extractred as
\begin{align}
  \sigma_\text{syst}(x) =
  \lim_\text{large $x_\text{train}$} \sigma_\text{tot}(x) \; .
\end{align}  
For LHC applications, where networks are usually trained on
simulations, this limit often represents reality.

Because $\sigma_\text{syst}$ is $\theta$-dependent, we can read
Eq.\eqref{eq:expectations} and~\eqref{eq:sig_stoch} as sampling
$\overline{A}(\theta)$ and $\sigma_\text{syst}(\theta)^2$ over a
network parameter distribution $q(\theta)$ for each phase space point
$x$
\begin{align}
\boxed{
  \text{BNN}: x, \theta \to 
\begin{pmatrix}
\overline{A}(\theta)\\
\sigma_\text{syst}(\theta)
\end{pmatrix}  } \; .
\label{eq:bnn_output}
\end{align}
If we follow Eq.\eqref{eq:loss_bayes1} and assume $q(\theta)$ to be
Gaussian, we now have a network with twice as many
  parameters as a standard network to describe two outputs.  For a
given phase space point $x$ we can then compute the three global
network predictions $\langle A \rangle$, $\sigma_\text{syst}$, and
eventually $\sigma_\text{stat}$. Unlike the distribution of the
individual network weights $q(\theta)$, the amplitude output is not
Gaussian.

The evaluation of a BNN is illustrated in Fig.~\ref{fig:bnn}.  From
that graphics we see that a BNN works very much like an ensemble of
networks trained on the same data and producing a spread of
outputs. The first advantage over an ensemble is that the BNN is only
twice as expensive as a regular network, and less if we assume that
the likelihood loss leads to an especially efficient training. The
second advantage of the BNN is that it learns a function and its
uncertainty together, which will give us some insight into how
advanced networks learn such densities in
Sec.~\ref{sec:gen_inn_events}. The disadvantage of Bayesian networks over
ensembles is that the Bayesian network only cover local structures in
the loss landscape. Their advantage is that they cover this local structure
in a well-defined manner.

While we usually do not assume a Gaussian uncertainty on the Bayesian
network\index{Bayesian network} output, this might be a good approximation for
$\sigma_\text{syst}(\theta)$. Using this approximation we can write
the likelihood $p(x_\text{train}|\theta)$ in Eq.\eqref{eq:loss_bayes1} as a
Gaussian and use the closed form for the KL-divergence in
Eq.\eqref{eq:kl_gaussian}, so the BNN loss function turns into
\begin{align}
\loss_\text{BNN}
 &= \int d\theta \; q_{\mu,\sigma}(\theta) \; 
   \sum_\text{points $j$} \left[  \frac{\left| \overline{A}_j(\theta) - A_j^\text{truth} \right|^2}{2\sigma_{\text{syst},j}(\theta)^2} + \log \sigma_{\text{syst},j}(\theta)
   \right] \notag \\
  &+ \frac{\sigma_q^2 - \sigma_p^2 + (\mu_q - \mu_p)^2}{2 \sigma_p^2}
  + \log \frac{\sigma_p}{\sigma_q} \; .
\label{eq:loss_bayes2}
\end{align}
As always, the amplitudes and $\sigma_\text{syst}$ are functions of
phase space $x$. The loss is minimized with respect to the means and
standard deviations of the network weights describing $q(\theta)$.
This interesting aspect of this loss function is that, while the
training data just consists of amplitudes at phase space points and
does not include the uncertainty estimate, the network constructs a
point-wise uncertainty from the variation of the network
parameters. This means we can rely on a well-defined
\underline{likelihood loss}\index{likelihood loss}  rather than some kind of MSE loss even for
our regression network training, this is \underline{fudging genius}!

As shown in Eq.\eqref{eq:bnn_hetero} we might, for some applications,
want to use this advantage of the BNN loss in
Eq.\eqref{eq:loss_bayes2}, but without sampling the network parameters
$\theta$. In complete analogy to a fit we use a deterministic network
like in Eq.\eqref{eq:kl_gaussian}, described by $q(\theta) =
\delta(\theta - \theta_0)$, with $\mu_p= 0$. We then assume a Gaussian
likelihood loss including the proper normalization and learn the
uncertainty using\index{heteroskedastic loss}
\begin{align}
\loss_\text{heteroskedastic}
 = \sum_\text{points $j$} \left[  \frac{\left| \overline{A}_j(\theta_0) - A_j^\text{truth} \right|^2}{2\sigma_{\text{syst},j}(\theta_0)^2} + \log \sigma_{\text{syst},j}(\theta_0)
   \right] + \frac{\theta_0^2}{2\sigma_p^2} \; .
\label{eq:loss_hetero}
\end{align}
The interplay between the first two terms works in a way that the
first term can be minimized either by reproducing the data and
minimizing the numerator, or by maximizing the denominator. The second
term penalizes the second strategy, defining a correlated limit of
$\overline{A}$ and $\sigma_\text{syst}$ over phase space. Compared to
the full Bayesian network this simplified approach has the 
disadvantages that it only includes the systematic uncertainties.
Given what we have learned about repulsive ensembles and Bayesian networks,
this kind of network cannot be interpreted as something like 
efficient ensembling.

For a specific task, let us look at LHC amplitudes for the production
of two photons and a jet~\cite{Aylett-Bullock:2021hmo},
\begin{align}
  gg \to \gamma \gamma g (g)
\end{align}
The corresponding transition amplitude $A$ is a real function over
phase space, with the detector-inspired kinematic cuts
\begin{alignat}{9}
p_{T,j} &> 20~\gev
&\qqqquad
| \eta_j | &< 5
&\qqqquad 
R_{jj, j \gamma, \gamma \gamma} > 0.4 \notag \\
p_{T,\gamma} &> 40, 30~\gev
&\qqqquad
| \eta_\gamma | &< 2.37 \; .
\end{alignat}
The jet 4-momenta are identified with the gluon 4-momenta through a
jet algorithm\index{jet algorithm}. The standard computer program for this calculation is
called NJet, and the same amplitudes can also be computed with the
event generator\index{event generators} Sherpa at one loop. This amplitude has to be
calculated at one loop for every phase space point, so we want to
train a network once to reproduce its output much faster. Because
these amplitude calculations are a key ingredient to the LHC
simulation chain, the amplitude network needs to reproduce the correct
amplitude distributions including all relevant features and with a
reliable uncertainty estimate.

\begin{figure}[t]
    \includegraphics[width=0.45\textwidth,page=4]{dsi_vs_mlp}
    \hspace*{0.1\textwidth}
    \includegraphics[width=0.415\textwidth,page=1]{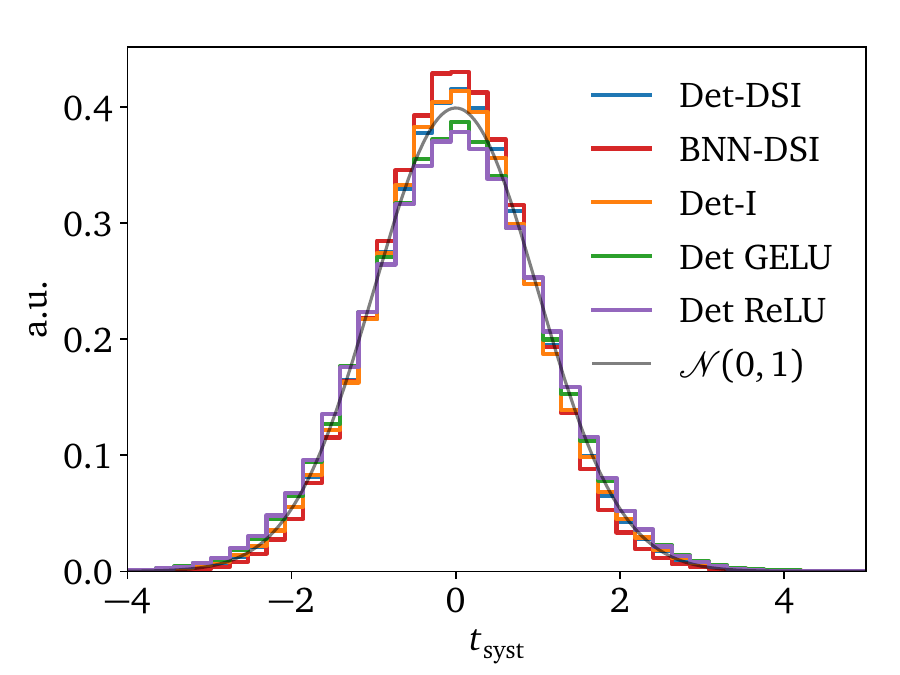}
    \caption{Performance of different neural networks learning the 
    $gg \to \gamma \gamma g$ amplitude. We show the relative 
    accuracy (left) and the systematic pull (right). 
    Figures modified from Ref.~\cite{Bahl:2024gyt}.}
    \label{fig:amplitudes}
\end{figure}

For one gluon in the final state, the most general phase space has $5
\times 4 = 20$ dimensions, but we can simplify this phase space by
requiring momentum conservation, on-shell particles, and Lorentz invariants. 
To estimate the performance of amplitude networks
we can use the fact that regression is a supervised task, which means
we know the true amplitudes. For the training or test
data we can compute the relative accuracy
\begin{align}
\Delta_j = \frac{\langle A \rangle_j - A_j}{A_j}
\label{eq:ampl_delta}
\end{align}
In the left panel of Fig.~\ref{fig:amplitudes} we show the
performance of different deterministic and BNN amplitude networks. 
The naive approach of learning amplitudes as a function of input 
and output 4-vectors limit the accuracy to the per-cent level. 
In Sec.~\ref{sec:gen_rl} we will discuss in detail how we can 
improve the power of neural networks by choosing better-suited 
data representations. For the amplitude we see that a standard 
network learning the amplitude as a function of Lorentz invariants
(Det-I) improves the accuracy. In Sec.~\ref{sec:gen_rl_eq} we will
introduce a general Lorentz-equivariant data representation (L-GATr),
which improves the amplitude accuracy to the $10^{-4}$ level. 
The best-performing DSI network combines invariants with a
deep sets representation which we will introduce in Sec.~\ref{sec:class_graph_sets}.
For this architecture the deterministic heteroskedastik network and
the BNN give relative accuracies around $10^{-5}$, with suppressed tails 
extending to 
the per-cent level, which should be sufficient for many LHC applications.

Once we are convinced that a network can learn a transition amplitude
precisely, the question is if the learned uncertainties, specifically the
dominant systematic uncertainties are calibrated correctly. 
For this purpose we look at the \underline{pull} distribution
\begin{align}
t_{\text{syst},j} = \frac{\langle A \rangle_j - A_j}{\sigma_{\text{syst},j}} \; .
\label{eq:ampl_pull}
\end{align}
If the learned $\sigma$ captured the absolute value of the deviation of
the learned amplitude from the truth exactly and for any phase space point, the
pull would be
\begin{align}
 \langle A \rangle_j = A_j \pm \sigma_j 
 \qquad \Leftrightarrow \qquad 
 t_j = \pm 1 \; .
\end{align}
Realistically, an uncertainty estimate will only capture the maximum
deviation, so the per-amplitude deviation from the truth 
will be smaller. For a stochastic source of uncertainties, the
learned values $\langle A \rangle_j$ will follow a Gaussian distribution
around the truth, where the width of this
Gaussian should be given by the learned uncertainty. This means the
pull will follow a unit Gaussian,
\begin{align}
 t_j \sim \normal(0,1) \; .
\end{align}
In the right panel of Fig.~\ref{fig:amplitudes} we show the 
systematic pull distributions for the different amplitude networks
and confirm that they follow a standard Gaussian for all accuracies.
This means that the heteroskedastic loss and the BNN learn calibrated
systematic uncertainties.

While the average relative accuracy of learned amplitudes, for LHC 
applications we also have to ensure that there are no tails of poorly 
learned amplitudes over phase space. This means we need to understand
the failure modes of the amplitude learning, and it turns out that 
a problem often occurs for the phase space points
with the largest amplitudes. The reason is that
there exist small phase space regions where the transition amplitude
increase rapidly, by several orders of magnitude. The network fails to
learn this behavior in spite of a $\log$-scaling following
Eq.\eqref{eq:preprocs}, because the training data in these regions is
sparse.  For an LHC simulation this kind of bias is a serious
limitation, because exactly these phase space regions need to be
controlled for a reliable rate prediction. This means we need to 
improved amplitude network training, by identifying and controlling
outliers in the $\Delta$-distribution of Eq.\eqref{eq:ampl_delta}.

The likelihood loss in Eq.\eqref{eq:loss_bayes2} with its
$\sigma_\text{syst}(\theta)$ has a great advantage, because we can
test the required Gaussian shape of the corresponding $\theta$-dependent pull
variable,
\begin{align}
\frac{\overline{A}_j(\theta) - A_j^\text{truth} }{\sigma_{\text{syst}, j}(\theta)} \; ,
\label{eq:bnn_pull}
\end{align}
as part of the training. If the source of the systematic uncertainty
is stochastic, the pull itself should form a standard, unit
Gaussian. For other systematic limitations, the Gaussian likelihood
might not be appropriate and the pull can have other shapes.

\begin{figure}[t]
    \centering
    \includegraphics[width=0.40\textwidth,page=2]{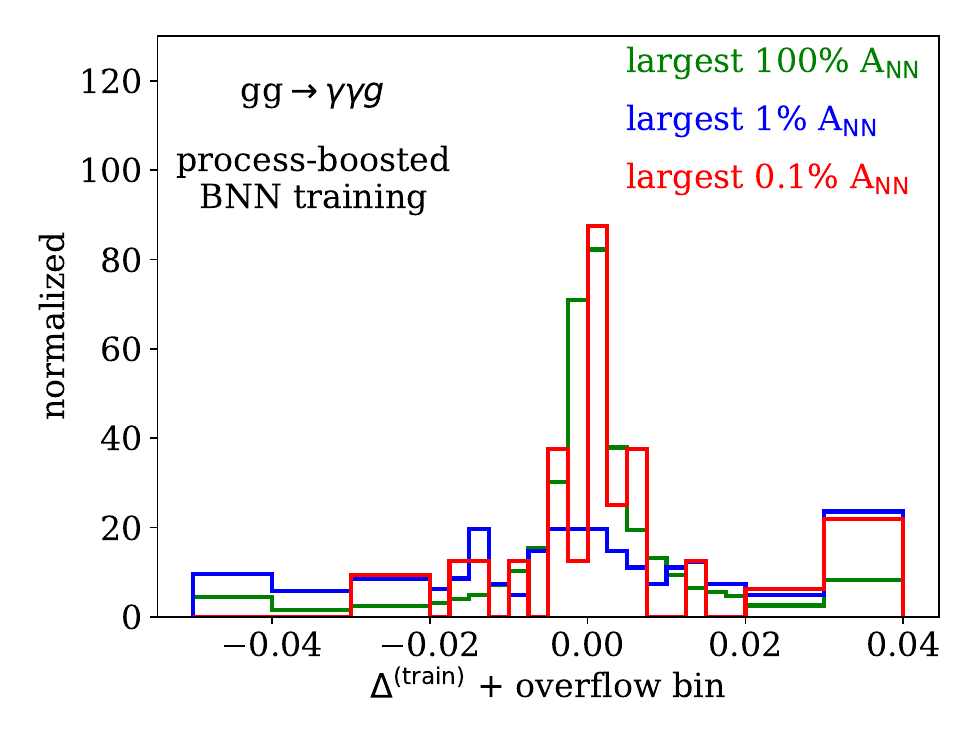}
    \hspace*{0.1\textwidth}
    \includegraphics[width=0.40\textwidth,page=4]{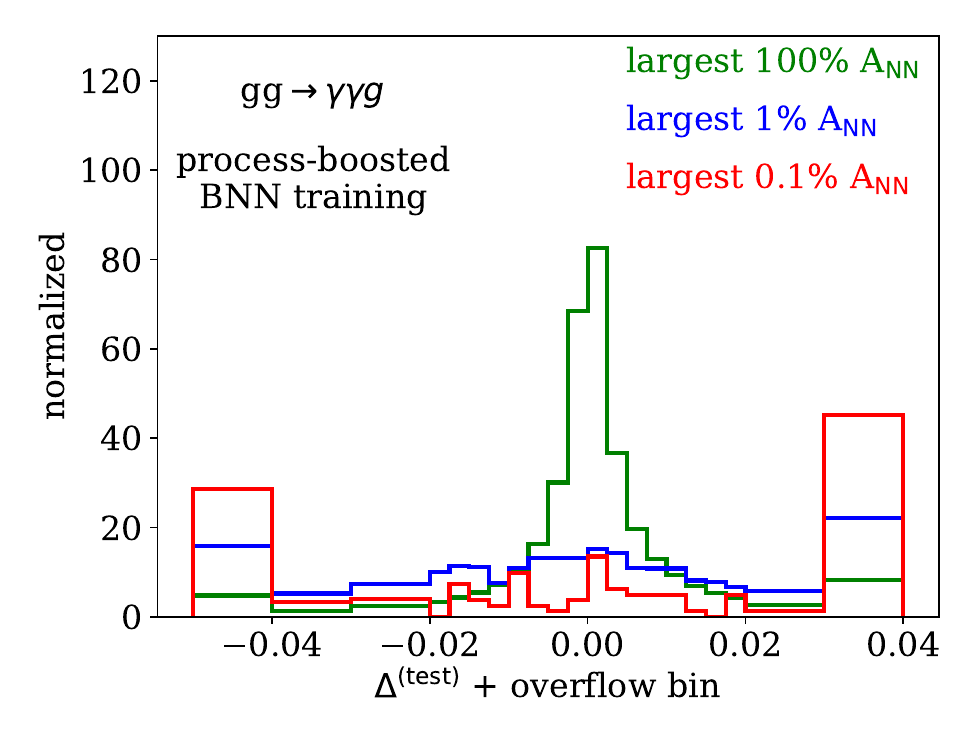}
    \caption{Performance of the 
      process-boosted BNN, in terms of the precision of the generated
      amplitudes, defined in Eq.\eqref{eq:ampl_delta} and evaluated on
      the training (upper) and test datasets. Figures from
      Ref.~\cite{Badger:2022hwf}.}
    \label{fig:amplitudes_boost}
\end{figure}

For the initial BNN run it turns out that the pull distribution indeed
looks like a Gaussian around the peak, but with too large and not
exponentially suppressed tails. We can improve this behavior using an
inspiration from the boosting of a decision tree. We enhance the
impact of critical phase space points through an increased event
weight, leading us to a boosted version of the BNN loss,
\begin{align}
\loss_\text{BBNN}
 &= \int d\theta \; q_{\mu,\sigma}(\theta) \; 
   \sum_\text{points $j$} n_j \times \left[  \frac{\left| \overline{A}_j(\theta) - A_j^\text{truth} \right|^2}{2\sigma_{\text{syst},j}(\theta)^2} + \log \sigma_{\text{syst},j}(\theta)
   \right] \notag \\
  &+ \frac{\sigma_q^2 - \sigma_p^2 + (\mu_q - \mu_p)^2}{2 \sigma_p^2}
  + \log \frac{\sigma_p}{\sigma_q} \; .
\label{eq:loss_bayes_boost}
\end{align}
This boosted or feedback training will improve the network performance
both, in the accuracy of the amplitude prediction and in the learned
uncertainty on the network amplitudes. If we limit ourselves to
self-consistency arguments, we can select the amplitudes with $n_j>1$
through large pull values, as defined in Eq.\eqref{eq:bnn_pull}.  It
turns out that this \underline{loss-based boosting} significantly
improves the uncertainty estimate. Unfortunately, for our amplitude
applications we find that the effect on the large
amplitudes is modest. They still lead to too many outliers
in the network accuracy, and we also tend to systematically
underestimate those large amplitudes.

To further improve the performance of our network we can target the
problematic phase space points directly, by increasing $n_j$ based on
the size of the amplitudes. This \underline{process-specific} boosting
goes beyond the self-consistency of the network and directly improves
a specific task beyond just learning the distribution of amplitudes
over phase space. In Fig.~\ref{fig:amplitudes} we
see that at least for the training data the $\Delta$-distribution
looks the same for small and large amplitudes. 

Going back to
Eq.\eqref{sec:basics_fit}, we can interpret the boosted loss for large
values of $n_j$ as a step towards interpolating the corresponding
amplitudes. While for small amplitudes the network still corresponds
to a fit, we are forcing the network to reproduce the amplitudes at
some phase space points very precisely. Obviously, this boosting will
lead to issues of overtraining. We can see it in the right panel of
Fig.~\ref{fig:amplitudes}, where the improvement through
process boosting for the test dataset does not match the improvement
for the training dataset. However, as long as the performance of the
test dataset improves, even if it is less than for the training
dataset, we improve the network in spite of the overtraining. The
issue with this overtraining is that it becomes harder to control the
uncertainties, which might require some kind of alternating
application of loss-based and process boosting.

This example illustrates three aspects of advanced regression
networks. First, we have seen that a Bayesian network allows us to
construct a likelihood loss even if the training data does not include
an uncertainty estimate. From the pull distribution we know that the 
learned uncertainties are calibrated. 
Second, we can improve the consistency of the
network training by boosting events selected from the pull
distribution. Third, we can further improve the network through
process-specific boosting, where we force the network from a fit to an
interpolation mode based on a specific selection of the input
data. For the latter the benefits do not fully generalize from the
training to the test dataset, but the network performance does improve
on the test dataset, which is all we really require.

\subsubsection{Parton density regression}
\label{sec:basics_regr_nnpdf}

A peculiar and challenging feature of hadron collider physics is that
we can compute scattering rates for quarks and gluons, but not relate
them to protons based on first principles. Lattice simulations might
eventually change this, but for now we instead postulate that all LHC
observables are of the form
\begin{align}
  \sigma(s)= \sum_{\text{partons} \; k,l} \int_0^1 dx_1 \int_0^1 dx_2 \;
  f_k(x_1) f_l(x_2) \; \hat{\sigma}_{kl}(x_1 x_2 s) \; ,
  \label{eq:def_pdf}
\end{align}
where $s$ is the squared energy of the hadronic scattering, and the
two partons carry the longitudinal momentum fractions $x_{1,2}$ of the
incoming proton. The partonic cross section $\hat{\sigma}_{ij}$ is
what we calculate in perturbative QCD. In our brief discussion we omit
the additional dependence on the unphysical renormalization scale and
instead treat the parton densities $f_i(x)$ as mathematical distributions which we
need to compute hadronic cross sections.  There are a few constraints
we have, for example the condition that the momenta of all partons in
the proton have to add to the proton momentum,
\begin{align}
  \int_0^1 dx \; x \; \left[ f_g(x) + \sum_\text{quarks} f_q(x) \right] = 1 \; .
  \label{eq:momentum_sumrule}
\end{align}
We can also relate the fact that the proton consists of three valence
quarks, $(uud)$, to the relativistic parton densities through the sum
rules
\begin{align}
  \int_0^1 dx \; \left[ f_u(x) - f_{\bar{u}}(x) \right] = 2
  \qquad \text{and} \qquad 
  \int_0^1 dx \; \left[ f_d(x) - f_{\bar{d}}(x) \right] = 1 \; .
  \label{eq:valence_sumrules}
\end{align}
Beyond this, the key assumption in extracting and using these parton
densities is that they are universal, which means that in the absence
of a first-principle prediction we can extract them from a
sufficiently large set of measurements, covering different colliders,
final states, and energy ranges.

To extract an expression for the parton densities, for instance the
gluon density, the traditional approach would be a fit to a functional
form. Parton densities have been described by an increasingly complex
set of functions, for example for the gluon in the so-called CTEQ
parametrization, defined at a given reference scale,
\begin{align}
  x f_g(x) &= x^{a_1} \; (1-x)^{a_2} \notag \\
  &\to  a_0 \; x^{a_1} (1-x)^{a_2} \left(1+a_3 x^{a_4} \right) \notag \\
  &\to x^{a_1} (1-x)^{a_2}
  \left[ a_3 (1 - \sqrt{x})^3
    + a_4 \sqrt{x} (1- \sqrt{x})^2
    + (5+2 a_1) x (1-\sqrt{x})
    + x^{3/2}
    \right] \; .
\label{eq:cteq}
\end{align}
We know such fits from Sec.~\ref{sec:basics_fit}, including the loss
function we use to extract the model parameters $a_j$ from data,
including an uncertainty. The problem with fits is that any ansatz
serves as an \underline{implicit bias}, and such an implicit bias
limits the expressivity of the network and leads to an underestimate
of the uncertainties\index{uncertainties}. For instance, when we
describe a gluon density with the first form of Eq.\eqref{eq:cteq} and
extract the parameters $a_{1,2}$ with their respective uncertainty
bands, these uncertainty bands define a set of functions
$f_g(x)$. However, it turns out that if we instead describe the gluon
density with the more complex bottom formula in Eq.\eqref{eq:cteq},
this function will not be covered by the range of allowed versions of
the first ansatz. Even for the common parameters $a_{1,2}$ the uncertainty
bands for the more complex form are likely to increase, because the
better parametrization is more expressive and describes the data with
higher resolution and more potential features.

Because uncertainty estimates are key to all LHC measurements, parton
densities are an example where some implicit bias might be useful, but
too much of it is dangerous. In 2002, \underline{neural network parton
  densities} (NNPDF) were introduced as a non-parametric fit, to allow
for a maximum flexibility and conservative uncertainty estimates.  It
replaces the explicit parametrization of Eq.\eqref{eq:cteq} with
\begin{align}
  xf_g(x) = a_0 \; x^{a_1} (1-x)^{a_2} \; f_\theta(x)  \; ,
  \label{eq:nnpdf}
\end{align}
and similarly for the other partons. NNPDF was and is the first and
leading AI-application to LHC physics~\cite{Forte:2020yip}.

While parton densities are, technically, just another regression
problem, two aspects set it apart from the amplitudes discussed in
Sec.~\ref{sec:basics_regr_amp}. First, the sum in
Eq.\eqref{eq:def_pdf} indicates that densities for different partons
are \underline{strongly correlated}, which means that any set of parton
densities comes with a full correlation or covariance matrix. Going
back to the definition of the $\chi^2$ or Gaussian likelihood loss\index{likelihood loss} 
function in Eq.\eqref{eq:likelihood_loss} we modify it to include
correlations between data points,
\begin{align}
  \boxed{
  \loss_\text{NNPDF}
  = \frac{1}{2} \sum_{i,j} (\mathcal{O} - \mathcal{O}_\theta)_i\,\Sigma_{ij}^{-1} (\mathcal{O} - \mathcal{O}_\theta)_j
  } \; .
  \label{eq:likelihood_corr}
\end{align}
The form of the covariance matrix $\Sigma$ is given as part of the
dataset. Because the inverse of a diagonal matrix is again diagonal,
we see how the diagonal form $\Sigma = \text{diag}(\sigma_j^2)$
reproduces the sum of independent logarithmic Gaussians given in
Eq.\eqref{eq:likelihood_loss}. With the correlated loss function we
should be able to train networks describing the set of parton
densities. The minimization algorithm used for the earlier NNPDF
version is genetic annealing. It has the advantage that it does can
cover different local minima in the loss landscape and is less prone
to implicit bias from the choice of local minima. For the new NNPDF4
approach it has been changed to standard gradient descent.

It is common to supplement a loss function like the NNPDF loss in
Eq.\eqref{eq:likelihood_corr} with additional conditions. First, it
can be shown that using dimensional regularization and the
corresponding $\overline{\text{MS}}$ factorization scheme the parton
densities are positive, a condition that can be added to
Eq.\eqref{eq:likelihood_corr} by penalizing negative values of the
parton densities,
\begin{align}
  \loss_\text{NNPDF}
  \to \loss_\text{NNPDF}
  + \sum_{\text{parton} \; k}
  \sum_{\text{data} \; i}
  \lambda_k \; \text{ELu} \left( -f_k(x_i) \right)
  \qquad \text{with} \qquad
  \text{ELu}(x) =
  \begin{cases}
    \epsilon (e^{-|x|} - 1) \approx - \epsilon |x| & x < 0 \\ x & x > 0 \; , 
  \end{cases}  
  \label{eq:likelihood_nnpdf2}
\end{align}
with $\epsilon \ll 1 $. For $f_k(x_i)<0$ this additional term
increases the loss linearly. Just like a Lagrangian multiplier, finite
values of $\lambda_i$ force the network training to minimize each term
in the combined loss function. A balance between different loss terms
is not foreseen, we will discuss such adversarial loss functions in
Sec.~\ref{sec:gen_gan}, for now we assume $\lambda_i > 0$. The ELu
function can also be used as an activation function, similar to
$\relu$ defined in Eq.\eqref{eq:def_relu}.

\begin{figure}[t]
  \centering
  \includegraphics[width=0.90\textwidth]{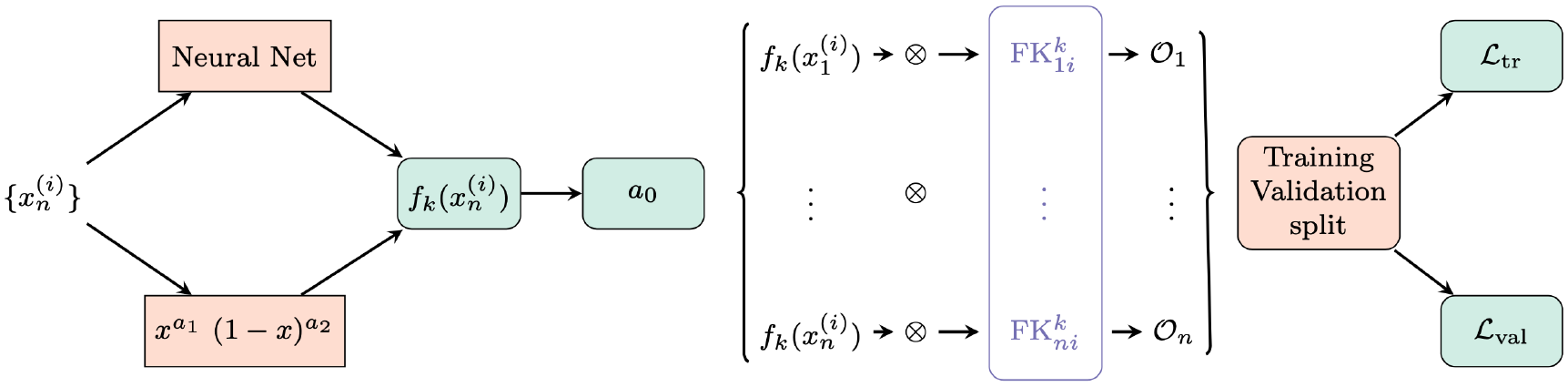}
  \caption{Structure of the NNPDF architecture. Figure slightly
    modified from Ref.~\cite{NNPDF:2021njg}.}
    \label{fig:nnpdf_scheme}
\end{figure}

Another condition arises from the momentum sum rule in
Eq.\eqref{eq:momentum_sumrule}, which requires all densities,
especially the gluon density, to scale like $x^2 f(x) \to 0$ for soft
partons, $x\to 0$.  Similarly, the valence sum rules in
Eq.\eqref{eq:valence_sumrules} require $xf(x) \to 0$ in the same
limit. The condition on the valence sum rule is again included an
additional loss term
\begin{align}
  \loss_\text{NNPDF}
  \to \loss_\text{NNPDF}
  + \sum_{\text{parton} \; k}
  \sum_{\text{soft data} \; i}
  \lambda_k \; \left[ x f_k(x_i) \right]^2 \; .
  \label{eq:likelihood_nnpdf3}
\end{align}
Constructing loss functions with different, independent terms is
standard in machine learning. Whenever possible, we try to avoid this
approach, because the individual coefficients need to be tuned. This
is why we prefer an L2 regularization\index{regularization} as given by the Bayesian
network, Eq.\eqref{eq:bnn_gaussian}, based on a likelihood
loss. Parton densities and the super-resolution networks discussed in
Sec.~\ref{sec:gen_gan_super} are examples where a combined loss
function is necessary and successful.

The architecture of the NNPDF network is illustrated in
Fig.~\ref{fig:nnpdf_scheme}. All parton densities are described by the
same, single network. The quark or gluon nature of the parton is given
to the network as an additional condition, a structural element we
will systematically explore for generative networks in
Sec.\ref{sec:gen_gan_super} and then starting with
Sec.~\ref{sec:gen_inn_events}.  The parametrization describing the parton
densities follows Eq.\eqref{eq:nnpdf}, where the prefactors $a_0$ are
determined through sum rules like the momentum sum rule and the
valence sum rules.  Usually, this parametrization would be considered
preprocessing, with the goal of making it easier for the network to
learn all patterns. In the NNPDF training, the parameters $a_{1,2}$
are varied from instance to instance in the toy dataset, to ensure
that there is no common implicit bias affecting all toy densities in a
correlated manner.  Uniquely to the NNPDF structure, the input is
given to the network through two channels $x$ and $\log x$.  The data
is split into bins $i$ for a given kinematic observable $n$, which has
to be computed using the parton densities encoded in the network,
before we can compute the loss functions for the training and
validation data. FK indicates the tabulated kinematic data.

However, the network architecture is not what makes the NNPDF approach
unique in modern machine learning.  The second aspects that makes
network parton densities special is that they are
\underline{distributions}, similar to but not quite probability
distributions, which means they are only defined stochastically. Even
for arbitrarily precise data $\sigma$ and predictions
$\hat{\sigma}_{ij}$, the best-fit parton densities will fluctuate the
same way that solutions of incomplete inverse problems will
fluctuate. The NNPDF approach distinguishes between \underline{three
  sources of uncertainty} in this situation:
\begin{enumerate}
\item even ignoring uncertainties on the data, we cannot expect that
  the unique minimum in data space translates into a unique minimum in
  the space of of parton densities. This means the extracted parton
  densities will fluctuate between equally good descriptions the
  training data;
\item once we introduce an uncertainty on the data, we can fix the
  mean values in the data distribution to the truth and introduce an
  uncertainty through the toy datasets. Now the the minimum in data
  space is smeared around the true value above, adding to the
  uncertainty in PDF space;
\item finally, the data is not actually distributed around the truth,
  but it is stochastic. This again adds to the uncertainties and it
  adds noise to the set of extracted parton densities.
\end{enumerate}
In the categorization of uncertainties in
Sec.~\ref{sec:basics_regr_amp} we can think of the first uncertainty
as an extreme case of \underline{model} uncertainty, in the sense that
the network is so expressive that the available training data does not
provide a unique solution. This problem might be alleviated when adding
more precise training data, but it does not have to. This is a
standard problem in defining and solving inverse problems, as we will
discuss in Sec.~\ref{sec:cond}. The second uncertainty is
\underline{statistical}\index{statistical uncertainty} in the sense that it would vanish in the limit
of an infinitely large training dataset. The third uncertainty is the
\underline{stochastic} uncertainty introduced before, again not
vanishing for larger training datasets.

Another problem of how to account for all these uncertainties in the
extracted parton densities is that they require a forward simulation,
followed by a comparison of generated kinematic distributions to
data. The solution applied by NNPDF is to replace the dataset and its
uncertainties by a sample of data replicas, with a distribution which
reproduces the actual data, in the Gaussian limit the mean and the
standard deviation for each data point. Each of these toy datasets is
then used to train a NNPDF parton density, and the analysis of the
parton densities, for instance their correlations and uncertainties,
can be performed based on this set of non-parametric NNPDFs. The same
method is applied to other global analyses at the LHC, typically when
uncertainties or measurements are correlated and likelihood
distributions are not Gaussian.  A way to check the uncertainty
treatment is to validate it starting from a known set of pseudo-data
and apply closure tests.

\begin{figure}[t]
  \includegraphics[width=0.325\textwidth]{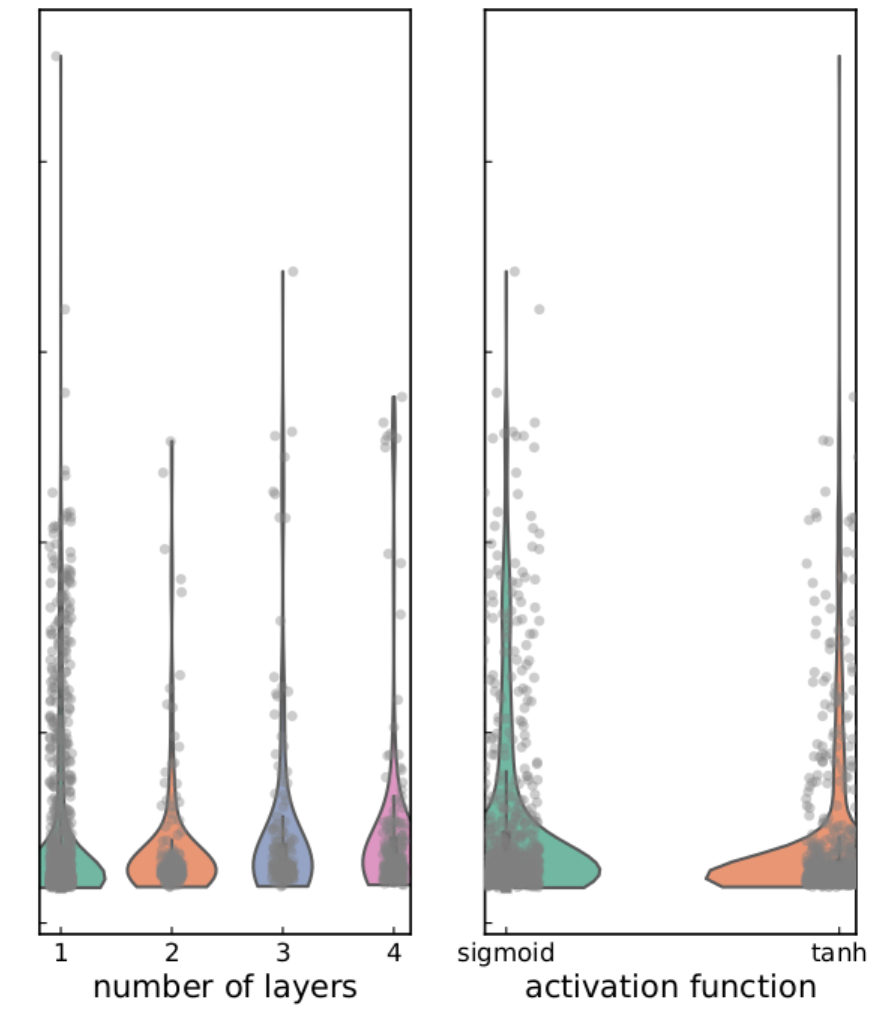}
  \includegraphics[width=0.665\textwidth]{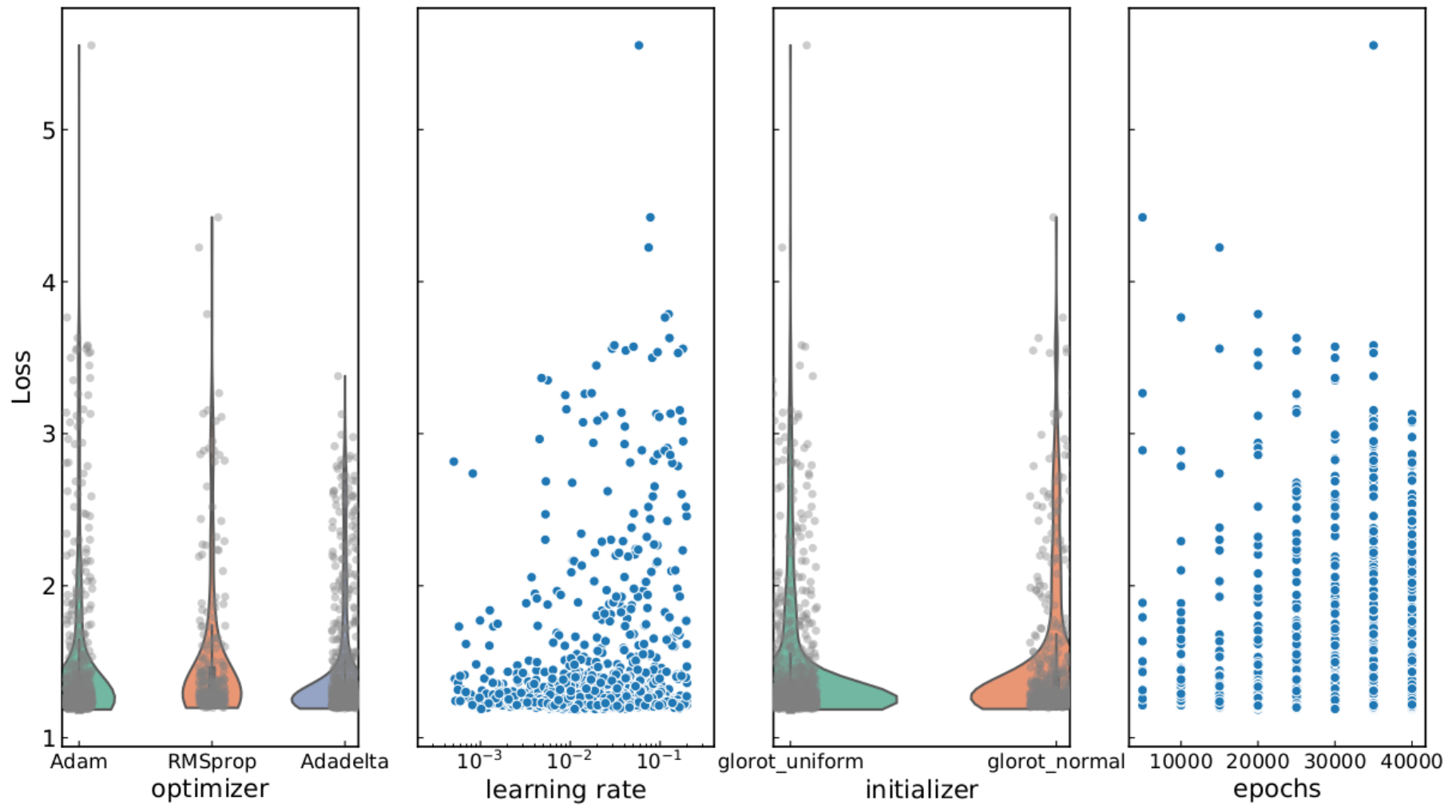}
  \caption{Hyperparameter scan for a test of the NNPDF stability,
    based on data from photon--proton and electron--proton scattering
    only. The $y$-axis shows the average of the test and validations
    losses. Figure from Ref.~\cite{Carrazza:2019mzf}.}
    \label{fig:nnpdf_hyper}
\end{figure}

As the final step of accounting for uncertainties, NNPDF
systematically tests for unwanted systematic biases through the choice
of network architecture and hyperparameters. For a precision
non-parametric fit, where the implicit bias or the smooth
interpolation properties of a network play the key role, it is crucial
that the network parameters do induce an uncontrolled bias. For
instance, a generic $x$-resolution determined by the network
hyper-parameters could allow the network to ignore certain features
and overtrain on others. In Fig.~\ref{fig:nnpdf_hyper} we see the
dependence of the network performance on some network parameters
tested in the NNPDF scan. Each panel shows the loss as a function of
the respective network parameter for 2000 toy densities, specifically
the average of the test and validation losses. These figures are based
on a fit only to data from single-proton interactions with electrons
or (virtual) photons. The violin shape is a visualization of the
density of points as a function of the loss. The different shapes show
how the network details have a sizeable effect on the network
performance.

The final NNPDF network architecture and parameters are determined
through an automatic hyperoptimization. Because the training and
validation datasets are already used in the network training shown in
Fig.~\ref{fig:nnpdf_scheme}, the hyperoptimization requires another
dataset. At the same time, we do not want to exclude relevant data
from the final determination of the parton densities, and the result
of the hyperoptimization will also depend on the dataset used. The
best way out is to use \underline{$k$-folding} to generate these
datasets. Here we divide the dataset into $n_\text{fold}$ partitions,
and for each training we remove one of the folds. The final loss for
each hyperparameter value is then given by the combination of the
losses of the $n_\text{fold}$ individual trainings,
\begin{align}
  \loss_\text{hyperopt} = \frac{1}{n_\text{fold}}
  \sum_k \loss_k \; ,
\end{align}
where the individual losses are given above. Using the sum of the
individual losses turns out to be equivalent to using the maximum of
the individual losses.

\begin{figure}[t]
  \centering
  \includegraphics[width=0.42\textwidth]{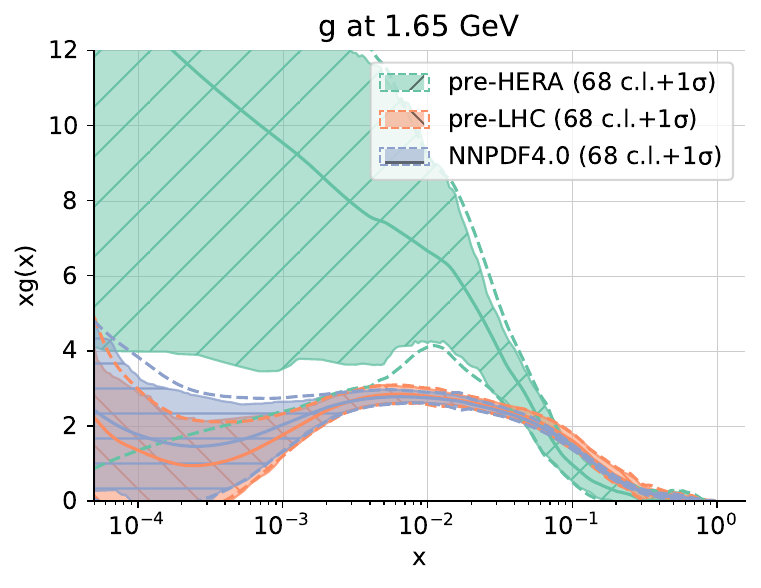}
  \hspace*{0.1\textwidth}
  \includegraphics[width=0.42\textwidth]{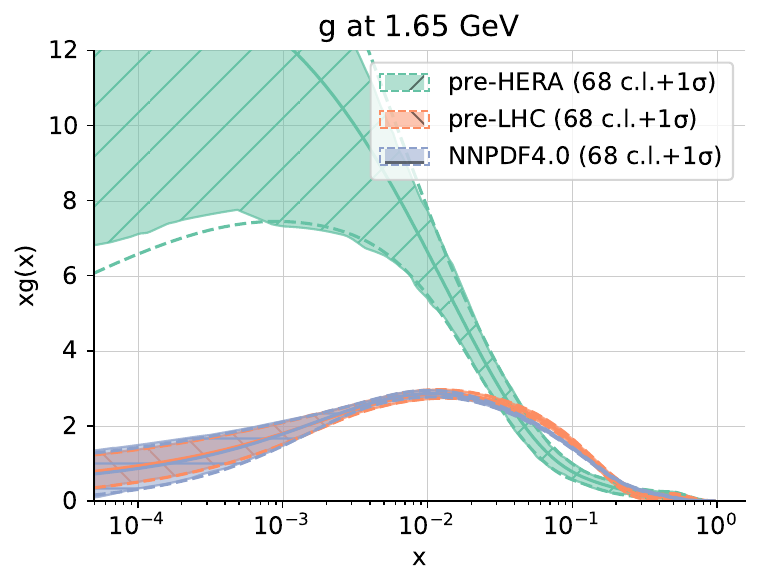}
  \caption{Historic pre-HERA and pre-LHC gluon densities with the
    NNPDF3.1 (left) and the NNPDF4.0 methodologies, compared with the
    full respective datasets of the two methodologies.
    Ref.~\cite{NNPDF:2021njg}.}
    \label{fig:nnpdf_gluon}
\end{figure}

An especially interesting question to ask is what progress we have
made in understanding parton densities over the last decade, both from
a data perspective and based on the generalized fit methodology. The
density to use for this study is the gluon density, because its growth
at small $x$-values is poorly controlled by theory arguments and
assumed fit functions, which means our description of $f_g(x \ll 1)$
is driven by data and an unbiased interpretation of the data. In
Fig.~\ref{fig:nnpdf_gluon} we show this gluon density for three
different datasets: (i) pre-HERA data consists or fixed-target or
beam-on-target measurements of two kinds, photon--proton or
electron--proton, and proton-proton interactions; (ii) Pre-LHC data
adds HERA measurements of electron--proton scattering at small
$x$-values and Tevatron measurements of weak boson production in
proton--antiproton scattering; (iii) NNPDF data then adds a large
range of LHC measurements. The difference between the NNPDF3.1 and
NNPDF4.0 datasets is largely the precision on similar kinematic
distributions.

In the two panels of Fig.~\ref{fig:nnpdf_gluon} we first see that the
NNPDF3.1 and NNPDF4.0 descriptions of pre-HERA data give an unstable
gluon density for $x < 10^{-2}$, where this dataset is simply lacking
information. The gluon density in this regime is largely an
extrapolation, and we already know that neural networks are
interpolation tools and not particularly good at \underline{magic or
  extrapolation}. From an LHC physics perspective, very small $x$-values
are not very relevant, because we can estimate the minimum typical
$x$-range for interesting central processes as
\begin{align}
  x_1 x_2 \sim \frac{m_{W,Z,H}^2}{(14~\tev)^2} \sim \frac{1}{140^2}
  \qquad \Leftrightarrow \qquad
  x_1 \sim x_2 \sim 7 \cdot 10^{-3} \; .
\end{align}
Adding HERA data constrains this phase space region down to $x \sim
10^{-3}$ and pushes the central value of the gluon density to a
reasonable description. Finally, adding LHC data only has mild effects
on the low-$x$ regime, but makes a big difference around $x > 0.05$,
where we measure the top Yukawa coupling in $t\bar{t}H$ production or
test higher-dimensional operators in boosted $t\bar{t}$
production. Comparing the left and right panels of
Fig.~\ref{fig:nnpdf_gluon} shows that the new NNPDF methodology leads
to extremely small uncertainties for the gluon density for the entire
range $x = 10^{-2}~...~1$, where a large number of LHC processes with
their complex correlations make the biggest difference. The quoted
uncertainties account for the historic increase of the dataset
faithfully, except for the pre-HERA guess work. 

As an afterthought, let us briefly think about the difference between
\underline{interpolation} and \underline{extrapolation}. If we want to
encode $f_\theta(x) \approx f(x)$ as a neutral network over $x \in
\mathbb{R}^D$, as defined in Eq.\eqref{eq:nn_mapping}, we rely on the
fact that our training data consists of a sufficiently dense set of
training data point in the space $\mathbb{R}^D$. Compared to a
functional fit, the implicit bias or the assumptions about the
functional form of $f(x)$ are minimal, which means that the network
training works best if for a given point $x_0$ the network can rely on
$x$-values in all directions. This is an assumption, but fairly
obvious. Now we can ask the question how likely it is that we indeed
cover the neighborhood of $x_0$ in $D$ dimensions, and the probability
of finding points in the this neighborhood scales like the volume of
the $D$-dimensional sphere with radius $r$,
\begin{align}
  V_D(r) = \frac{\pi^{D/2}}{\Gamma \left( \frac{D}{2} + 1 \right) } \; r^D
  \qquad \text{with} \qquad 
  \Gamma(n) = (n-1)! \; .
\end{align}
It grows rapidly with $D$, which means that with increasing
dimensionality we are less and less likely to cover the neighborhood
of a given $x_0$. This is a version of the so-called \underline{curse
  of dimensionality}. It is especially true because the relevant
dimensionality is that of the data representation, not of the
underlying physics. The only problem with the general statement that
network training always turns from an interpolation to an
extrapolation problem is that in our language we do not consider a
network an interpolation, but a fit-like approximation.

\subsubsection{Numerical integration}
\label{sec:basics_regr_int}

The last application of a regression network is the numerical
calculation of a $D$-dimensional phase space integral
\begin{align}
  I(s)
  = \int_0^1 dx_1 \dots \int_0^1 dx_D \;f(s;x) \; ,
\label{eq:integral}
\end{align}
where $x_i$ are the integration variables and $s$ is a vector of
additional parameters, not integrated over.  Because the values of the
integrand can span a wide numerical range is useful to normalize the
integrand, for example by its value at the center of the
$x$-hypercube,
\begin{align}
  f(x;s)
  \to \frac{f(s;x)}{f(s; \frac{1}{2},\frac{1}{2},,...,\frac{1}{2})}
  \qquad \Leftrightarrow \qquad 
  I(s)
  \to \frac{I(s)}{f(s;\frac{1}{2},\frac{1}{2},...,\frac{1}{2})} \; .
\end{align}
Without going into details, it is also useful to transform the
integrand into a form which vanishes at the integration boundaries.
Analytically, we would compute the primitive or indefinite integral
$F$,
\begin{align}
  \frac{d^D F(s;x)}{dx_1\dots dx_D}
  = f(s;x) \; ,
\end{align}
and then the integral by evaluating the integration boundaries 
\begin{align}
  I(s)
  &=  \int_0^1dx_1 \dots \int_0^1dx_D \; f(s;x)
  \notag \\
  &=\int_0^1dx_1 \dots \int_0^1dx_D \;\frac{d^{D}F(s;x)}{dx_1\dots dx_D}
  \notag\\
  &=  \int_0^1dx_1 \dots \int_0^1dx_D \;\frac{d}{dx_D}\frac{d^{D-1}F(s;x)}{dx_1\dots dx_{D-1}}
  \notag\\
  &=  \int_0^1dx_1 \dots \int_0^1dx_{D-1} \; \frac{d^{D-1} F(s;x)}{dx_1\dots dx_{D-1}} \Bigg|_{x_D=0}^{x_D=1}
  \notag \\
  &= \sum_{x_1,...,x_D=0,1} (-1)^{D-\sum x_i}F(s;x) \;.
    \label{eq:boundarysum}
\end{align}
In particle physics we really never know the primitive of a phase
space integrand, but we can try to construct it and encode it in a
neural network,
\begin{align}
  F_\theta(s;x) \approx F(s;x) \; .
  \label{eq:train_prim}
\end{align}
On the other hand, we do not have data to train a surrogate network
for $F$ directly. The idea is to instead train an integrand surrogate,
such that its $D$-th derivative matches $f$,
\begin{align}
  \loss_\text{MSE}\left(f(s;x), \frac{dF_\theta(s;x)}{dx_1...dx_D}\right)\,.
\label{eq:loss_int}
\end{align}
If the training on the integrand fixes the network weights such that
the integrand as well as $F$ are determined by the same network, and
$F_\theta$ fulfills Eq.\eqref{eq:train_prim}, we have directly learned
the integral.

\begin{figure}[t]
  \centering
  \includegraphics[width=0.45\textwidth]{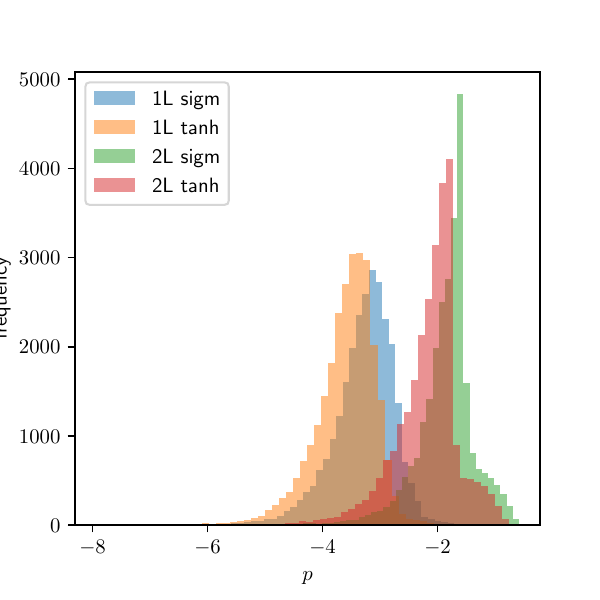}
  \caption{Number of digits accuracy for two integrals, using two
    different activation functions, one of them defined in
    Eq.\eqref{eq:1lbox}. Figure from Ref.~\cite{Maitre:2022xle}.}
    \label{fig:daniel}
\end{figure}

To construct a surrogate which can be differentiated multiple times
with respect to some of its inputs, we need a differentiable
activation function, for example the sigmoid function,
\begin{align}
  \sigmoid(x) = \frac{1}{1+e^{-x}}
  \qquad  \Rightarrow \qquad
  \sigmoid'(x) =  \log( e^x + 1 )
  \qquad \Rightarrow \qquad
  \sigmoid^{(n)} = -\text{Li}_n(-e^x) \; .
\end{align}
There exists a fast sum representation of the di-logarithm
$\text{Li}_2$ for numerical evaluation. The same set of derivative can
be computed for the $\tanh$ activation function.

Next, we need to compute the derivative of this fully connected neural
network.  Following the conventions of Eq.\eqref{eq:forward_pass}, the
input layer for the two input vectors is
\begin{align}
  x^{(0)}_i =
  \begin{cases}
    x_i \qquad & i\leq D \\
    s_{i-D} \qquad & i > D
  \end{cases} \;.  
\end{align}
For the hidden layers we just replace the ReLU activation function in
Eq.\eqref{eq:def_node} with the sigmoid,
\begin{align}
  x^{(n)}_i = \sigmoid
  \left[ W^{(n)}_{ij} x_j^{(n-1)} +b_i^{(n)} \right] \;.
\end{align}
The scalar output of the network with $N$ layers can be
differentiated, for instance, with respect to $x_1$,
\begin{align}
  F_\theta \equiv x^{(N)} &= W^{(N)}_j x_j^{(N-1)}+ b^{(N-1)} \notag \\
  \Rightarrow \qquad
  \frac{d F_\theta}{dx_1}
  &= W^{(N)}_j \; \frac{d x_j^{(N-1)}}{dx_1} \notag \\
  &= \sum_j W^{(N)}_j \;
  \sigmoid' \left[ W^{(N)}_{jk} x_k^{(N-1)} +b_j^{(N)} \right] \; 
  \left[ W^{(N-1)}_{j\ell} \; \frac{d x_\ell^{(N-2)}}{dx_1}\right] \; ,
\end{align}
where we write the sum over $j$ explicitly, while for the other
indices we use the usual summing convention. Next, we differentiate
this expression with respect to $x_2$, altogether $D$ times, to
compute the MSE loss in Eq.\eqref{eq:loss_int}.  The loss can be minimized
with respect to the network parameters $\theta$ using the usual
backpropagation.  Because the integrand is known exactly, there is no
need to regularize the network, but it would also not hurt.  Also, the
numerical generation of integrand values is numerically cheap, which
means $F_\theta$ an be trained using very large numbers of training
data points.

In the original paper, the method is showcased for two integrals, one
of them is
\begin{align}
  I_\text{1L}(s_{12}, s_{14}, m_H^2, m_t^2)
  =& \int_0^1 dx_1\int_0^1 dx_2\int_0^1 dx_3 \frac{1}{F_\text{1L}^2} \notag \\
  \text{with} \qquad
  F_\text{1L}
  =& m_t^2+2 x_3 m_t^2+x_3^2 m_t^2+2 x_2 m_t^2-x_2 s_{14}+2 x_2 x_3 m_t^2
  - x_2 x_3 m_H^2+x_2^2 m_t^2 \notag\\
  &+2 x_1 m_t^2+2 x_1 x_3 m_t^2  - x_1 x_3 s_{12}+2 x_1 x_2 m_t^2-x_1 x_2 m_H^2 +x_1^2 m_t^2\; .
  \label{eq:1lbox}
\end{align}
It is needed to compute the LHC rate for Higgs pair production.

The accuracy of the estimated integral can be measured in
analogy to Eq.\eqref{eq:ampl_delta},
\begin{align}
  p=\log_{10} \left| \frac{I_\text{NN}-I_\text{truth}}{I_\text{truth}}\right| \; ,
\end{align}
giving the effective number of digits the estimates gets right.  The
results for this accuracy are shown in Fig.~\ref{fig:daniel}. It is
based on training an ensemble of eight replicas of the same network,
use their average as the central prediction of the integral, and the
standard deviation as an uncertainty estimate. The entries in the
histogram have different initialisations and are trained on different
training data. The results for the two different activation functions
are similar. The two-loop integral, which we skip in this summary, has
a lower accuracy than the one-loop integral, which is to be expected
given the larger number of integrations.

\clearpage
\section{Classification}
\label{sec:class}

After the short introduction to the simpler regression networks, we
come back to classification as the standard problem in LHC
physics. Whenever we look at a jet or an event, the first question
will be what kind of particles gave us that final state
configuration. This is not trivial, given that jets are complex
objects which can come from a light quark or a gluon, but also from
quarks that decay through the electroweak interaction at the hadron
level, like charm or bottom quarks. They can also come from a tau
lepton, decaying to quarks and a neutrino, or from boosted gauge
bosons or Higgs bosons or top quarks. Even when we are looking for
apparently simple electrons, we need to be sure that they are not one
of the many charged pions which can look like electrons especially in
the forward detector. Similarly, photons are not trivial to separate
from neutral pions when looking at the electromagnetic
calorimeter. Really, the only particle which we can identify fairly
reliably at the LHC are muons.

At the event level we ask the same question again, usually in the
simple form signal vs background. As an example, we want to extract
$t\bar{t}H$ events with an assumed decay $H\to b\bar{b}$, as mentioned
in Eq.\eqref{eq:tth}, from a sample which is dominated by the
$t\bar{t}b\bar{b}$ continuum background and $t\bar{t}jj$ events where
we mis-tagged a light-flavor quark or gluon jet as a $b$-quark. Once
we have identified $t\bar{t}H$ events we can use them to measure for
example the value of the top Yukawa coupling or see if this coupling
respects the CP-symmetry or comes with a complex phase. All of this is
classification, and from our experience with BDTs for jet and event
classification, it is clear that modern neural networks can improve
their performance. Of course, the really interesting part is where we
turn our expertize in these standard classification task into new
ideas, methods, or tools. So the first ML-chapter of these lecture
notes will be on classification with modern neural networks.

We have already introduced many of the underlying concepts and
technical terms for classification tasks for BDTs in
Sec.~\ref{sec:basics_deep_multi}. Let us look at the optimization
behind a classification in a slightly different way. We start with one
event sample $\{ x \}$, distributed according to the true or data
distribution $\pd(x)$. This can be the signal or background sample. We
then construct a model approximating the true distribution in terms of
the network parameters $\theta$, called
\begin{align}
  \pmd(x) \equiv \pmd(x|\theta) \; ,
\label{eq:def_pmd}
\end{align}
where we omit the conditional argument $\theta$. As a function of
$\theta$, the probability distribution $\pmd(x)$ defines a likelihood,
and it should agree with $\pd(x)$.
To define the training goal we use the variational approximation
from in Sec.~\ref{sec:basics_deep_bayes} and 
compare the two probability distributions using the
KL-divergence\index{KL-divergence} defined in Eq.\eqref{eq:def_kl},
\begin{align}
  \kl [\pd,\pmd]
= \XXLangle \log \frac{\pd}{\pmd} \XXRangle_{\pd}
\equiv \int d x \; \pd(x) \; \log \frac{\pd(x)}{\pmd(x)} \; .
\end{align}
Following Eq.\eqref{eq:kl_twofold} we can either evaluate
$\kl [\pd,\pmd]$ or
$\kl [\pmd,\pd]$.
%
The first samples from the data distribution and is called forward
KL-divergence; the second samples from the model and is called reverse
KL-divergence. Since we are working on a well-defined training dataset
we use the first definition to find the best values of $\theta$ and
make sure our trained network approximates the training data well,
\begin{align}
  \kl [\pd,\pmd]
  = \XXLangle  \log \frac{\pd(x)}{\pmd(x)} \XXRangle_{\pd}
  &= \Langle  \log \pd(x) \Rangle_{\pd}
    - \Langle  \log \pmd(x) \Rangle_{\pd} \notag \\
  &= - \Langle  \log \pmd(x) \Rangle_{\pd}
    + \text{const}(\theta) \; .
    \label{eq:def_ce2}
\end{align}
We combine the
KL-divergences of the
signal and a background distributions to a \underline{likelihood-ratio
loss}
\begin{align}
  \loss_\text{class}
  &= \sum_{j=S,B} \kl [\pdj{j},\pmdj{j}] \notag \\
  &= \XLangle \log \pdj{S} - \log \pmdj{S} \XRangle_{\pdj{S}}
  +\XLangle \log \pdj{B} - \log \pmdj{B} \XRangle_{\pdj{B}} \notag \\
  &= - \XLangle \log \pmdj{S} \XRangle_{\pdj{S}}
     - \XLangle \log \pmdj{B} \XRangle_{\pdj{B}} + \text{const}(\theta) \notag \\
     \Rightarrow \qquad 
& \boxed{
     \loss_\text{class}
     = - \sum_{\{x\}} \; \Big[ \pdj{S} \log \pmdj{S} + \pdj{B} \log \pmdj{B} \Big]
     }\; .
     \label{eq:class_loss}
\end{align}
We consistently omit the arguments $x$ and $\theta$ which we included
in Sec.~\ref{sec:basics_deep_multi}.  Comparing this loss to the
definition of the information entropy in
Eq.\eqref{eq:entropy} motivates the alternative name
\textsl{cross entropy} for this loss.  It becomes even simpler when we
take into account that every jet or event has to either signal or
background, $p_S + p_B =1$.  Looking back at Eq.\eqref{eq:kl_twofold}
this simple form of the classification loss tells us that we made the
right choice of KL-divergence.

To mimic the training procedure, we can do variation calculus to
describe the minimization of the loss with respect to $\theta$,
\begin{align}
  \theta_\text{trained} 
  = \argmin_\theta \loss_\text{class}
  = \argmin_\theta  \sum_{j=S,B} \kl [\pdj{j}(x),\pmdj{j}(x|\theta)] \; .
\end{align}
We then replace $p_B = 1 - p_S$ and do a variation with respect to the
$\theta$-dependent model distribution
\begin{align}
  0
  &\really - \frac{\delta}{\delta \pmdj{S}}
  \sum_{\{x\}} \; \Big[
    \pdj{S} \log \pmdj{S}
    + (1-\pdj{S}) \log (1-\pmdj{S}) \Big]  \notag \\
  &= - \sum_{\{x\}} \; \Bigg[
    \frac{\pdj{S}}{\pmdj{S}}
    - \frac{1-\pdj{S}}{1-\pmdj{S}} \Bigg]
  \qqquad \Leftrightarrow \qqquad 
  \boxed{\pdj{S} = \pmdj{S}}
\end{align}
If we work under the assumption that a loss function should be some
kind of log-probability or log-likelihood, we can ask if our minimized
loss function corresponds to some kind of \underline{statistical
  distribution}. Again using $p_B = 1 - p_S$ as the only input in
addition to the definition of Eq.\eqref{eq:class_loss} we find
\begin{align}
  \loss_\text{class}
  &= - \sum_{\{x\}} \; \Big[
      \pdj{S} \log \pmdj{S}
    + (1-\pdj{S}) \log (1 - \pmdj{S}) \Big]  \notag \\
  &= - \sum_{\{x\}} \; 
      \log \Big[ \pmdj{S}^{\pdj{S}} \;
      (1 - \pmdj{S})^{1-\pdj{S}} \Big]  \; .
     \label{eq:class_bernoulli}
\end{align}
We can compare the term in the brackets with the Bernoulli
distribution in Eq.\eqref{eq:def_bernoulli}, which gives the
probability distributions for two discrete outcomes. We find that an
interpretation in terms of the Bernoulli distribution requires for the
outcomes and the expectation value
\begin{align}
  \pdj{S}=x \in \{0,1\}
  \qquad \text{and} \qquad \pmdj{S} = \rho
\end{align}
This means that our learned probability distribution has a Bernoulli
form and works on signal or background jets and events, encoding the
signal vs background expectation value encoded in the trained
network.

If we follow this line of argument, our classification network should
encode and return a signal probability for a given jet or event, which
means the final network layer has to ensure that the network output is
$f_\theta(x) \in [0,1]$. For usual networks this is not the case, but
we can easily enforce this by replacing the ReLU activation function
in the network output layer with a \underline{sigmoid} function,
\begin{align}
  \sigmoid(x) = \frac{1}{1+e^{-x}}
  \qquad \Leftrightarrow \qquad
  \sigmoid^{-1}(x) \equiv \text{Logit}(x) = \log \frac{x}{1-x} \; .
\label{eq:def_sigmoid}
\end{align}
The activation function(s) inside the network only work as a source of
non-linearity and can be considered just another hyper-parameter of
the network. The sigmoid guarantees that the output of the
classification network is automatically constrained to a closed
interval, so it can be interpreted as a probability without the
network having to learn this property. With the loss function of
Eq.\eqref{eq:class_loss} and the sigmoid activation of
Eq.\eqref{eq:def_sigmoid} we are ready to tackle classification
tasks.

This derivaton of the classification loss based on learning
likelihoods will eventually lead us to an interpretation of
classifiers which is very specific to particle physics. In
Sec.~\ref{sec:gen_gan_arch} we will introduce the Neyman-Pearson lemma
which says that the optimal classifier is the likelihood ratio of the
two hypotheses. Based on this mathematical insight we do not consider
classifiers ways to separate two modes in an appropriate
representation, but as a learned ratio of two likelihoods over phase
space.

The final set of naming conventions we need to introduce is the
structure of the training data, for example for classification. If we
train a network to distinguish, for instance, light-flavor QCD jets
from boosted top quarks, the best training data should be jets for
which we know the truth labels. In LHC physics we can produce such
datasets using precision simulations, including full detector
simulations. We call network training on fully labeled data (fully)
\underline{supervised learning}. The problem with supervised learning
at the LHC is that it has to involve Monte Carlo simulations. We will
discuss below how we can define dominantly top jet samples. However,
no sample is ever 100\% pure. This means that training on LHC data
will at best start from a relatively pure signal and background
samples, for which we also know the composition from simulations and a
corresponding analysis. Training a classifier on samples with known
signal and background fractions is called \underline{weakly supervised
  learning}, and whenever we talk about supervised learning on LHC
data we probably mean weakly supervised learning with almost pure
samples. An interesting question is how we would optimize a network
training between purity and statistics of the training data. Of
course, we can find compromises, for instance training a network on a
combination of labeled and unlabeled data. This trick is called
\underline{semi-supervised learning} and can increase the training
statistics, but there seems to be no good example where this helps at
the LHC. One of the reasons might be that training statistics is
usually not a problem in LHC applications. Finally, we can train a
network without any knowledge about the labels, as we will see towards
the end of this section. Here the questions we can ask are different
from the usual classification, and we refer to this as
\underline{unsupervised learning}. This category is exciting, because
it goes beyond the usual LHC analyses.  Playing with unsupervised
learning is the ultimate test of how well we understand a dataset, and
we will discuss some promising methods in Sec.~\ref{sec:auto}.

\subsection{Convolutional networks}
\label{sec:class_cnn}

Modern machine learning, or deep learning, has become a standard
numerical method in essentially all aspects of life and research. Two
applications dominate the applications of modern networks, image
recognition and natural language recognition. It turns out that
particle physics benefits mostly from image recognition research.  The
most active field applying such image-based methods is subjet
physics. It has, for a while, been a driving field for creative
physics and analysis ideas at the LHC.  An established subjet physics
task like identifying the parton leading to an observed jet, is ideal
to develop ML-methods to beat standard methods. The classic,
multivariate approach has two weaknesses. First, it only uses
information preprocessed into theory-inspired and high-level
observables. This can be cured in part by using the more low-level
observables shown in Eq.\eqref{eq:qg_obs}. However, when we just pile
up observables we need to ask how well a BDT can capture the
correlations. Altogether we need to ask the question is we cannot
systematically exploit \underline{low-level information} about a jet
to identify its partonic nature.

\subsubsection{Jet images and top tagging}
\label{sec:class_cnn_tag}

A standard benchmark for jet classification is to separate boosted,
hadronically decaying top quarks from QCD jets. Tops are the only
quarks which decay through their electroweak interactions before can
form hadrons,
\begin{align}
t \to b W^+ \to b \ell^+ \nu_\ell
\qquad \text{or} \qquad
t \to b W^+ \to b u \bar{d} \; .
\end{align}
Most top quarks are produced in pairs and at low transverse
momentum. However, even in the Standard Model a fraction of top quarks
will be produced at large energies. For heavy resonances, like a new
$Z'$ gauge boson decaying to top quarks, the tops will receive
transverse momenta around
\begin{align}
  p_{T,t} \sim p_{T,\bar{t}} \sim \frac{m_{Z'}}{2} - m_t \; .
\label{eq:decay_zprime}
\end{align}
For a 2-prong jet we can compute the typical angular separation
between the two decay jets as a function of the mass and the
transverse momentum, for example in case of a hadronic Higgs decay
\begin{align}
  R_{b\bar{b}} \approx \frac{1}{\sqrt{z(1-z)}} \; \frac{m_H}{p_{T,H}}
  > \frac{2m_H}{p_{T,H}} \; .
  \label{eq:rbb}
\end{align}
The parameter $z$ describes how the energy is divided between the two
decay subjets. Because $R$ is a purely angular separation, it can
become large if one of the two decay products becomes soft. A single
decay jet is most collimated if the energy is split evenly between the
two decay products, which is also the most likely outcome for most
decays. For a 3-prong decay like the top quark we leave it at an
order-of-magnitude estimate of a fat top jet,
\begin{align}
  R_{bjj} \gtrsim \frac{m_t}{p_{T,t}} \; .
  \label{eq:topjet_size}
\end{align}
The inverse dependence can easily be seen when we simulate top decays.
For a standard jet size of $R \sim 0.8$ this relation means we can tag
top jets for $p_{T,t} \gtrsim 300$~GeV, corresponding to decaying
resonances with $m_{Z'} > 1$~TeV. Given the typical reach of the LHC
in resonance searches, this value of $m_{Z'}$ as introduced in
Eq.\eqref{eq:decay_zprime} is actually small, which means that many
resonance searches with hadronic decays nowadays rely on boosted final
states and fat jets from heavy particle decays.

Before we go into the details of ML-based jet tagging\index{jet tagging} we need to
briefly introduce the standard method to define and analyze jets.
Acting on calorimeter output in the usual $\Delta \eta$ vs $\Delta
\phi$ plane, we usually employ QCD-based algorithms to determine the
partonic origin of a given configuration.  Such \underline{jet
  algorithms}\index{jet algorithm} link physical objects, for instance calorimeter towers
or particle flow objects, to more or less physical objects, namely
partons from the hard process. In that sense, jet algorithms invert
the statistical forward processes of QCD splittings\index{QCD splittings}, hadronization,
and hadron decays, shown in Fig.~\ref{fig:simchain}.  Recombination
algorithms try to identify soft or collinear partners amongst the jet
constituents, because we know from QCD that parton splitting are
dominantly soft or collinear. We postpone a more detailed discussion
to Secs.~\ref{sec:auto_dense} and~\ref{sec:cond_inn_bayes} and
instead just mention that in the collinear limit we can describe QCD
splittings again in terms of the energy fraction $z$ of the outgoing
hard parton. For example, a quark radiates a gluon following the
splitting pattern
\begin{align}
  \hat{P}_{q \leftarrow q}(z)
  \propto \frac{1+z^2}{1-z}  \; .
\end{align}
To decide if two subjets come from one parton leaving the hard process
we have to define a geometric measure capturing the collinear and soft
splitting patterns. Such a measure should include the distance
$R_{ij}$ defined in Eq.\eqref{eq:def_r} and the transverse momentum of
one subjet with respect to another or to the beam axis. The three
standard measures are
\begin{alignat}{8}
&k_T \qquad \qquad &
y_{ij} &= \frac{R_{ij}}{R} \;
          \min \left( p_{T,i}, p_{T,j} \right) \qquad  \qquad  \qquad &
y_{iB} &= p_{T,i} \notag \\
&\text{C/A} &
y_{ij} &=\frac{R_{ij}}{R} &
y_{iB} &= 1 \notag \\
&\text{anti-}k_T &
y_{ij} &=\frac{R_{ij}}{R} 
          \min \left( p_{T,i}^{-1}, p_{T,j}^{-1} \right) &
y_{iB} &= p_{T,i}^{-1} \; .
\label{eq:qcd_jet_measure}
\end{alignat}
The parameter $R$ only balances competing jet--jet and jet--beam
distances. In an exclusive jet algorithm\index{jet algorithm} we define two subjets as
coming from one jet if $y_{ij} < y_\text{cut}$, where $y_\text{cut}$
is an input resolution parameter. The jet algorithm then consists of
the steps
\begin{itemize}
\item[] (1) for all combinations of two subjets find $y^\text{min} =
  \text{min}_{ij} (y_{ij},y_{iB})$
\item[] (2a) if $y^\text{min} = y_{ij} < y_\text{cut}$ merge subjets
  $i$ and $j$ and their momenta, keep only the new subjet $i$, go
  back to (1)
\item[] (2b) if $y^\text{min} = y_{iB} < y_\text{cut}$ remove subjet
  $i$, call it beam radiation, go back to (1)
\item[] (2c) if $y^\text{min} > y_\text{cut}$ keep all subjets, call
  them jets, done
\end{itemize}
Alternatively, we can give the algorithm the minimum number of
physical jets and stop there.  As determined by their power dependence
on the transverse momenta in Eq.\eqref{eq:qcd_jet_measure}, three
standard algorithms start with soft constituents ($k_T$), purely
geometric (Cambridge--Aachen), or hard constituents (anti-$k_T$) to
form a jet.  While for the $k_T$ and the $C/A$ algorithms the
clustering history has a physical interpretation and can be associated
with some kind of time, it is not clear what the clustering for the
anti-$k_T$ algorithm means.

\begin{figure}[t]
  \centering
  \includegraphics[width=0.45\textwidth]{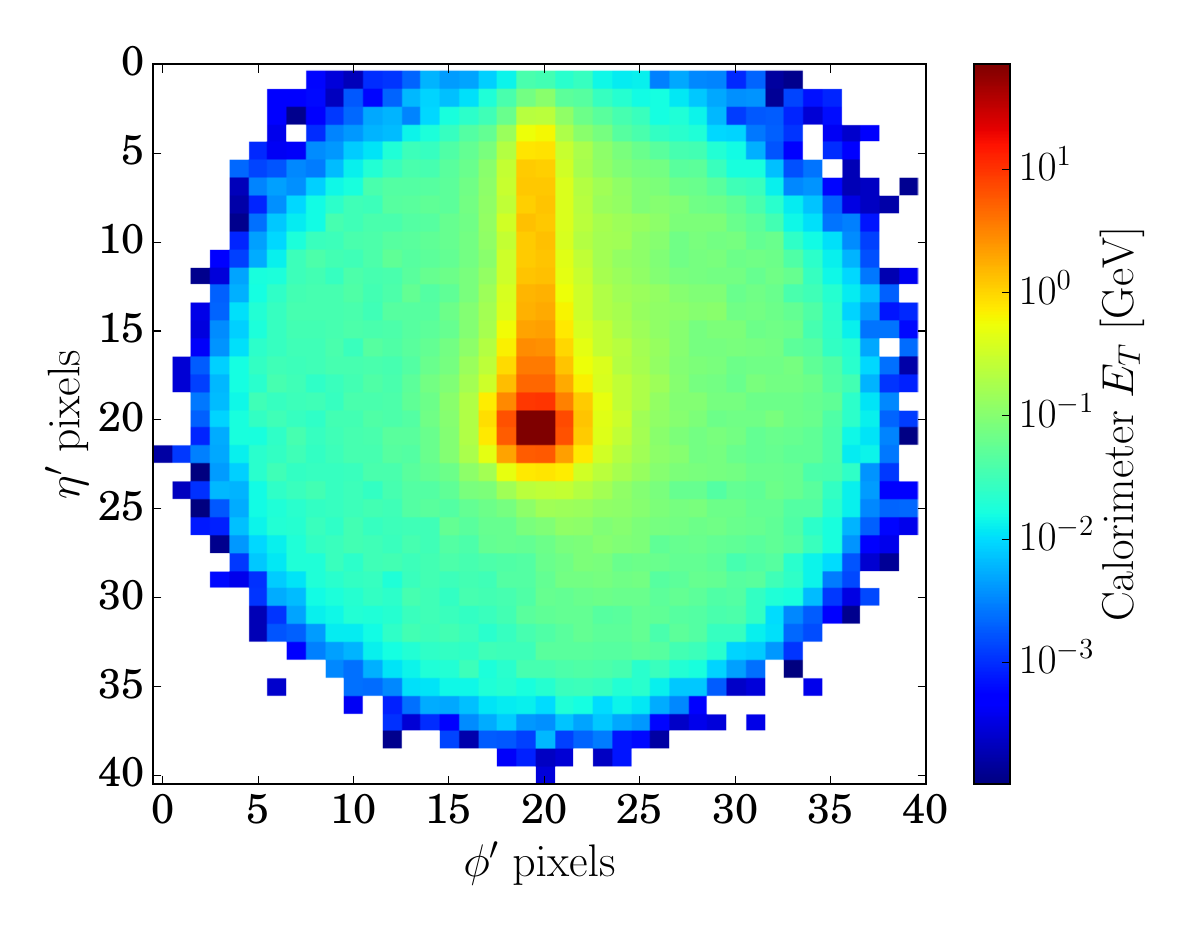}
    \hspace*{0.05\textwidth}
    \includegraphics[width=0.45\textwidth]{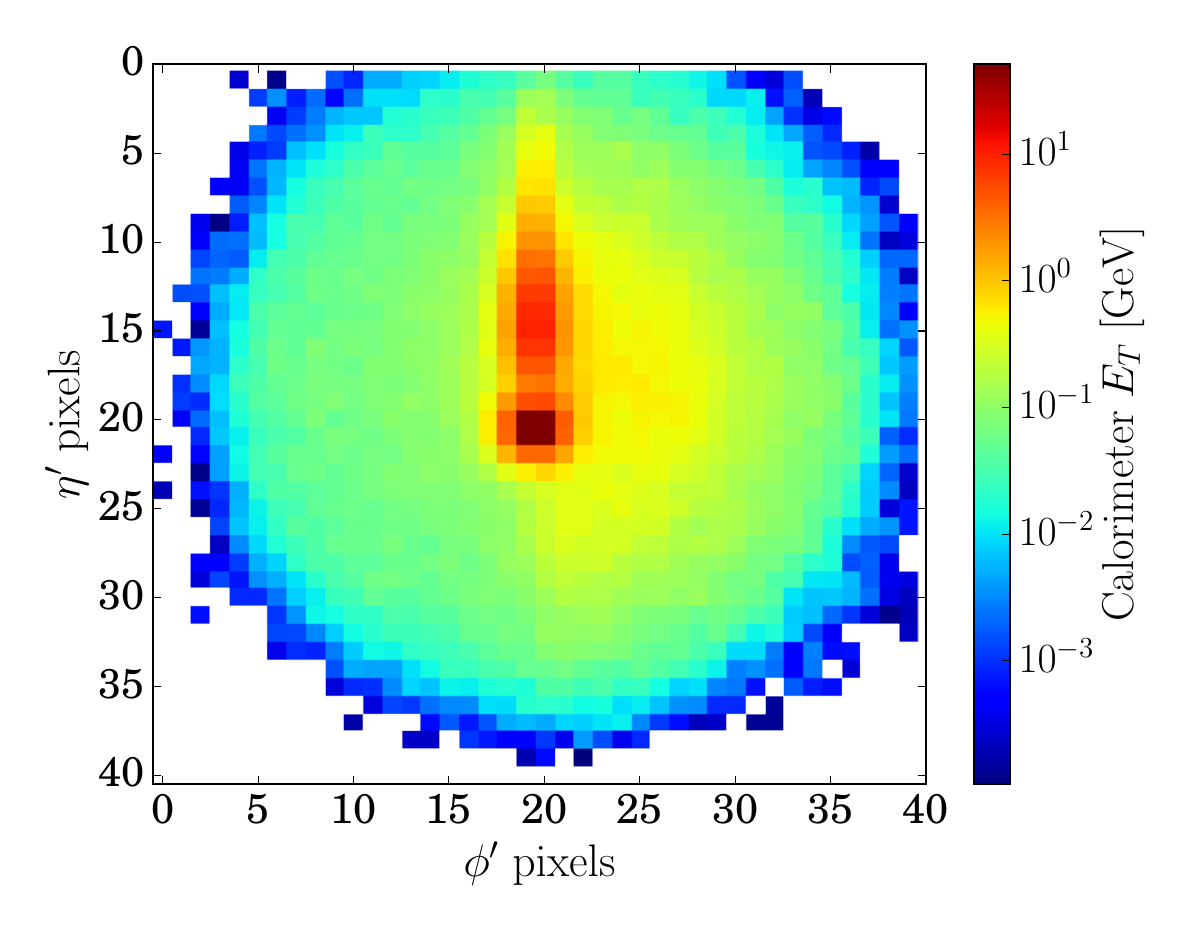}
    \caption{Averaged and preprocessed jet images for a QCD jet
      (left) and a boosted top decay (right) in the rapidity vs
      azimuthal plane plane. The preprocessing steps are introduced
      in the next section. Figures from
      Ref.~\cite{Kasieczka:2017nvn}.}
    \label{fig:jet_images}
\end{figure}

Once we understand the clustering history of a jet, we can try to
determine its partonic content. At this point we are most interested
in finding out if we are looking at the decay jet of a massive
particle or at a regular QCD jets.  For this purpose we start with the
observables given in Eq.\eqref{eq:qg_obs}. For massive decays we can
supplement this set by dedicated tagging observables. For instance, we
construct a proper measure for the number of prongs in a jet in terms
of the $N$-subjettiness variables
\begin{align}
  \tau_N = \frac{1}{R \sum_k p_{T,k}} \sum_k p_{T,k} \; \min \left(
    R _{1,k}, R _{2,k}, \cdots, R_{N,k} \right)^\beta \; .
  \label{eq:tau_N}
\end{align}
It starts with $N$ so-called $k_T$-axes, and a reference distance $R$
combined with a typical power $\beta=1$, and matches the jet
constituents to a given number of axes.  A small value $\tau_N$
indicates consistency with $N$ or less substructure axes, so an
$N$-prong decay returns a small ratio $\tau_N/\tau_{N-1}$, like
$\tau_2/\tau_1$ for $W$-boson or Higgs tagging, or $\tau_3/\tau_2$ for
top tagging.  To identify boosted heavy particles decaying
hadronically we we can also use a mass drop in the jet clustering history,
\begin{align}
  \max(m_1,m_2) < 0.8 \; m_{1+2} \; .
  \label{eq:mod_mass_drop}
\end{align}
If we approximate the full kinematics in the transverse plane, we can
replace the mass drop with a drop in transverse momentum, such that we
search for electroweak decays by requiring $\min(p_{T 1},p_{T 2}) >
p_{T 1+2}$. We can improve the tagging by correlating the two
conditions, on the one hand an enhanced $p_T$-drop and on the other
hand two decay subjets boosted together. This defines the SoftDrop
criterion,
\begin{align}
  \frac{\min(p_{T1},p_{T2})}{p_{T1}+p_{T2}} > 0.2 \left(
  \frac{R_{12}}{R} \right)^\beta \; ,
  \label{eq:soft_drop}
\end{align}
for instance with $\beta=1$.  It allows us to
identify decays through the correlation given in Eq.\eqref{eq:rbb}.
All of these theory-inspired observables, and more, can be applied to
jets and define multivariate subjet analysis tools. Here we typically
use the boosted decision trees introduced in
Sec.~\ref{sec:basics_deep_multi}\index{boosted decision tree}. The question is how modern neutral
networks, like CNNs, working on jet representations like the jet
images\index{jet images} of Fig.~\ref{fig:jet_images}, will perform on such an
established \underline{jet tagging}\index{jet tagging} task.

There are a few reasons why top tagging has become the \textsl{Hello
  World} of classic or ML-based subjet physics. First, as discussed
above, the distinguishing features or top jets are theoretically well
defined. Top decays are described by perturbative QCD, and the
corresponding mass drop and 3-prong structure can be defined in a
theoretically consistent manner without issues with a soft and
collinear QCD description. Second, top tagging is a comparably easy
task, given that we can search two mass drops, three prongs, and an
additional $b$-tag within the top jet. Third, in Sec.~\ref{sec:class}
we mentioned that it is possible to define fairly pure top-jet samples
using the boosted production process
\begin{align}
  pp \to t\bar{t}
  \to (b u \bar{d}) \; (\bar{b} \ell^- \bar{\nu})
  \qquad \text{with} \qquad
  p_{T,t} \sim p_{T,\bar{t}} \gtrsim 500~\gev   \; .
\label{eq:boosted_ttbar}
\end{align}
We trigger on the hard and isolated lepton, reconstruct the
leptonically decaying top using the $W$-mass constraint to replace the
unknown longitudinal momentum of the neutrino, and then work with the
hadronic recoil jet to, for instance, train or calibrate a
classification network on essentially pure samples.

The idea of an image-based top tagger is simple --- if we look at the ATLAS or CMS
calorimeters from the interaction point, they look like the shell of
a barrel, which can be unwrapped into a 2-dimensional plane with the
coordinates rapidity $\eta \approx -4.5~...~4.5$ and azimuthal angle
$\phi = 0~...~2\pi$, with the distance measure $R$ defined in
Eq.\eqref{eq:def_r}. If we encode the transverse energy deposition in
the calorimeter cells as color or grey-scale images, we can use
standard image-recognition techniques to study jets or events. The
network architecture behind the success of ML-image analyses are
convolutional networks, and their application to LHC \underline{jet
  images}\index{jet images} started with Ref.~\cite{deOliveira:2015xxd}. We show a set
of average jet images for QCD jets and boosted top decays in
Fig.~\ref{fig:jet_images}.  The image resolution of, in this case $40
\times 40$ pixels is given by the calorimeter resolution of $0.04
\times 2.25^\circ$ in rapidity vs azimuthal angle.  A single jet image
looks nothing like these average images. For a single jet image, only
20 to 50 of the 1600 pixels have sizeable $p_T$-entries, making even
jet images sparse at the 1\% or 2\% level, not even talking about full
event images. Nevertheless, we will see that standard network
architectures outperform all established methods used in subjet
physics.

\subsubsection{Architecture}
\label{sec:class_cnn_arch}

The basic idea of convolutional networks is to provide correlations
between neighboring pixels, rather than asking the network to learn
that pixels 1 and 41 of a jet image lie next to each other. In
addition, learning every image pixel independently would require a
vast number of network parameters, which typically do not correspond
to the actual content of an image. Alternatively, we can try to learn
structures and patterns in an image under the assumption that a
relatively small number of such patterns encode the information in the
image.  In the simplest case we assume that image features are
\underline{translation-invariant} and define a learnable matrix-like
filter which applies a convolution and replaces every image pixel with
a modified pixel which encodes information about the 2-dimensional
neighborhood. This filter is trained on the entire image, which means
that it will extract the nature and typical distance of 2-dimensional
features. To allow for some self-similarity we can then reduce the
resolution of the image pixels and run filters with a different length
scale. A sizeable number of network parameters is then generated
through so-called feature maps, where we run several filters over the
same image. Of course, this works best if the image is not sparse and
 translation-invariant. For example, an image with diagonal features
will lead to a filter which reflects the diagonal structure.

In Fig.~\ref{fig:cnn_architecture} we illustrate a simple architecture
of a convolutional network (CNN) applied to classify calorimeter
images of LHC jets. The network input is the 2-dimensional jet image,
$(n \times n)$-dimensional matrix-valued inputs $x$ just like in
Eq.\eqref{eq:forward_pass}, and illustrated in Fig.~\ref{fig:jet_images}.
It then uses a set of standard operations:

\begin{itemize}
\item[--] Zero padding $(n\times n) \to (n+1 \times n+1)$: It
  artificially increases the image size by adding zeros to be able to
  use a filter for the pixels on the boundaries
  \begin{align} 
    x_{ij} \; \to  \; 
    \begin{pmatrix} 
      0 & \cdots &0 \\
      \vdots & x_{ij} & \vdots\\
      0 & \cdots & 0
    \end{pmatrix} \; .
  \end{align}

\item[--] Convolution $(n\times n) \to (n \times n)$: To account for
  locality of the images in more than one dimension and to limit the
  number of network parameters, we convolute an input image with a
  learnable filter of size $n_\text{c-size} \cdot
  n_\text{c-size}$. These filters play the role of the nodes in
  Eq.\eqref{eq:single_node},
  \begin{align}
    \boxed{
      x'_{ij} = \sum_{r,s}
    W_{rs}  \;
    x_{i+r,j+s} + b 
    \to \relu (x'_{ij}) } \; .
    \label{eq:def_conv0}
  \end{align}
  As for any network we also apply a non-linear element, for example
  the ReLU activation function defined in Eq.\eqref{eq:def_relu}.

\item[--] Feature maps $n_\text{f-maps} \times (n\times n) \to
  n_\text{f-maps} \times(n \times n)$: Because a single learned filter
  for each convolutional layer defines a small number of network
  parameters and may also be unreliable in capturing the features
  correctly, we introduce a set of filters which turn an image into
  $n_\text{f-maps}$ feature maps. The convolutional layer now returns
  a feature map $x'^{(k)}$ which mixes information from all input maps
  \begin{align}
    x'^{(k)}_{ij} = \sum_{l=0}^{n_\text{f-maps}-1} \quad  \sum_{r,s}
    W^{(kl)}_{rs}  \;
    x^{(l)}_{i+r,j+s} + b^{(k)}
    \qquad \text{for} \quad
    k = 0,...,n_\text{f-maps}-1 \; .
  \label{eq:def_conv}
  \end{align}
  Zero padding and convolutions of a number of feature maps define a
  convolutional layer. We stack $n_\text{c-layer}$ of them.  Each
  $n_\text{c-block}$ block keeps the size of the feature maps, unless
  we use the convolutional layer to slowly reduce their size.

\begin{figure}[t]
  \centering
  \includegraphics[width=\textwidth]{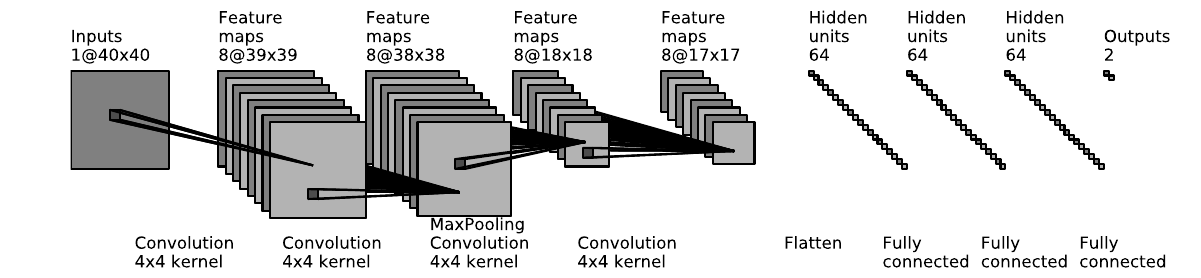} 
  \caption{Simple CNN architecture for the analysis of jet images.
    Figure from Ref.~\cite{Kasieczka:2017nvn}, for a more competitive
    version see Ref.~\cite{Macaluso:2018tck}.}
  \label{fig:cnn_architecture}
\end{figure}

\item[--] Pooling $(n\times n) \to (n/p \times n/p)$: We can reduce
  the size of the feature map through a \underline{downsampling}
  algorithm. For pooling we divide the input into patches of fixed
  size $p \times p$ and assign a single value to each patch, for
  example a maximum or an average value of the pixels. A set of
  pooling steps reduces the dimension of the 2-dimensional image
  representation towards a compact network output. An alternative to
  pooling are stride convolutions, where the center of the moving
  convolutional filter skips pixels. In Sec.~\ref{sec:gen_gan_super}
  we will also study the inverse, upsampling direction.

\item[--] Flattening $(n\times n) \to (n^2 \times 1)$: Because the
  classification task requires, two distinct outputs, we have to
  assume that the 2-dimensional correlations are
  learned and transform the pixel matrix into a 1-dimensional vector,
  \begin{align} 
    x = \left( x_{11},\dots,x_{1n},\dots,x_{n1},\dots,x_{nn} 
        \right) \; .
  \end{align}
  This vector can then be transformed into the usual output of the
  classification network.
  
\item[--] Fully connected layers $n^2 \to n_\text{d-node}$: On
  the pixel vectors we can use a standard fully connected network as
  introduced in Eq.\eqref{eq:forward_pass} with weights, biases, and
  ReLU activation,
  \begin{align}
    x'_i 
        = \relu \left[ \sum_{j=0}^{n^2-1} W_{ij} x_j + b_i \right] \; .
  \label{eq:def_dnn}
  \end{align}
  The deep network part of our classifier comes as a number of fully
  connected layers with a decreasing number of nodes per layer.
  Finally, we use the classification-specific sigmoid activation of
  Eq.\eqref{eq:def_sigmoid} in the last layer, providing a
  2-dimensional output returning the signal and background
  probability for a given jet image.
\end{itemize}
In this CNN structure it is important to remember that the filters are
learned globally, so they do not depend on the position of the central
pixel. This means the size of the CNN does not scale with the number
of pixels in the input image. Second, a CNN with downsampling
automatically encodes different resolutions or fields of vision of the
filters, so we do not have to tune the filter size to the features we
want to extract.  One way to increase the expressivity of the network
is a larger number of feature maps, where each feature map has access
to all feature maps in the previous layer. The number of network
parameters then scales like
\begin{align}
\#_\text{CNN-parameters} \sim n_\text{c-size}^2 \times n_\text{f-maps} \times n_\text{c-layer} \ll n^2 \; .
\label{eq:cnn_scaling}
\end{align}
Of course, CNNs can be defined in any number of dimensions, including
a 1-dimensional time series where the features are symmetric under
time shifts. For a larger number of dimensions the scaling of
Eq.\eqref{eq:cnn_scaling} becomes more and more favorable.

As always, we can speed up the network training through
\underline{preprocessing steps}. They are based on symmetry properties\index{symmetries}
of the jet images\index{jet images}, as we will discuss them in more detail in
Sec.~\ref{sec:class_graph_trans}. For jet images the preprocessing has
already happened in Fig.~\ref{fig:jet_images}.  First, we define a
central reference point, for instance the dominant energy deposition
or some kind of main axis or center of gravity. Second, we can shift
the image such that the main axis is in its center. Third, we use the
rotational symmetry of a single jet by rotating the image such that
the second prong is at 12 o'clock.  Finally, we flip the image to
ensure the third maximum is in the right half-plane.  This is the
preprocessing applied to the averaged jet images shown in
Fig.~\ref{fig:jet_images}. In addition, we can apply the usual
preprocessing steps for the pixel entries from Eq.\eqref{eq:preprocs},
plus a unit normalization of the sum of all pixels in an image.

\begin{figure}[t]
  \includegraphics[width=0.119\textwidth]{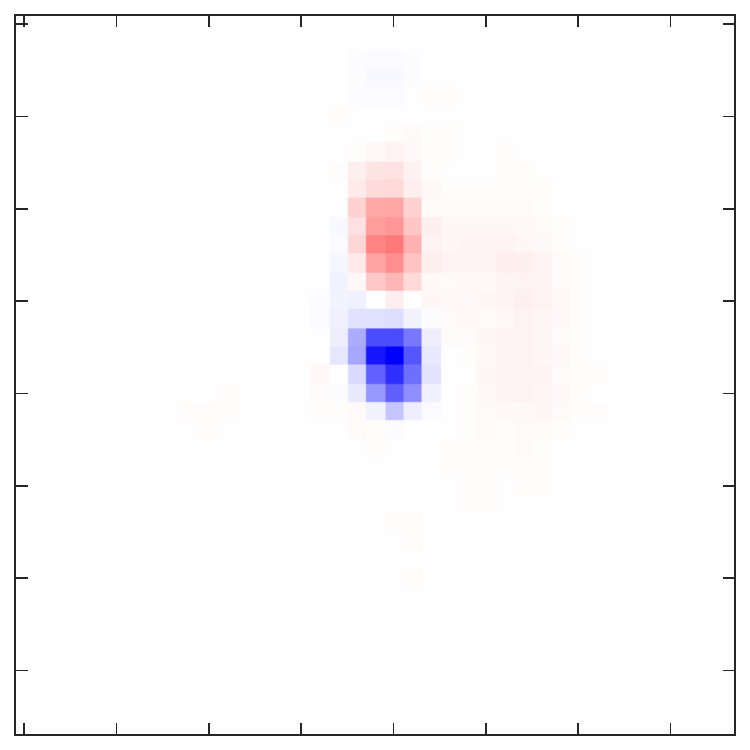}
  \includegraphics[width=0.119\textwidth]{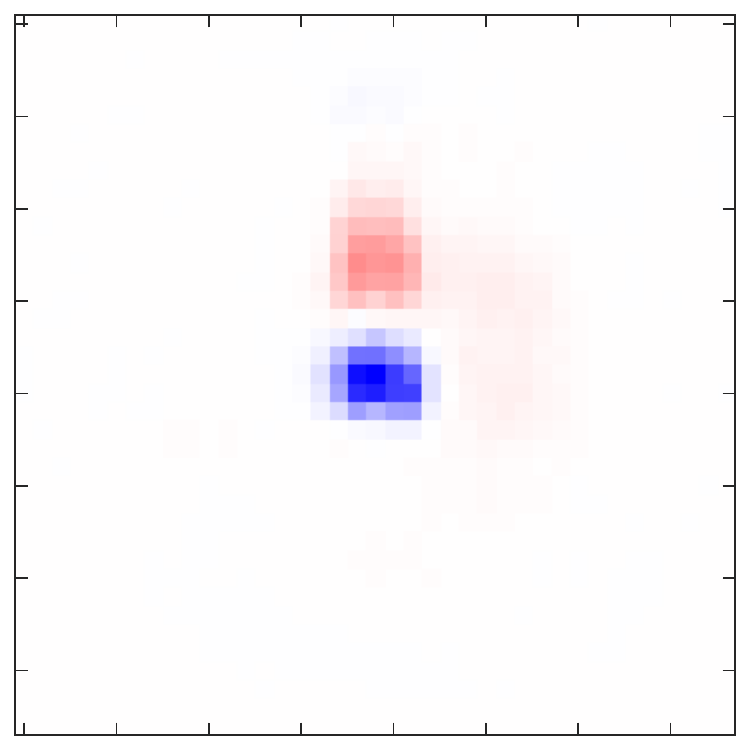}
  \includegraphics[width=0.119\textwidth]{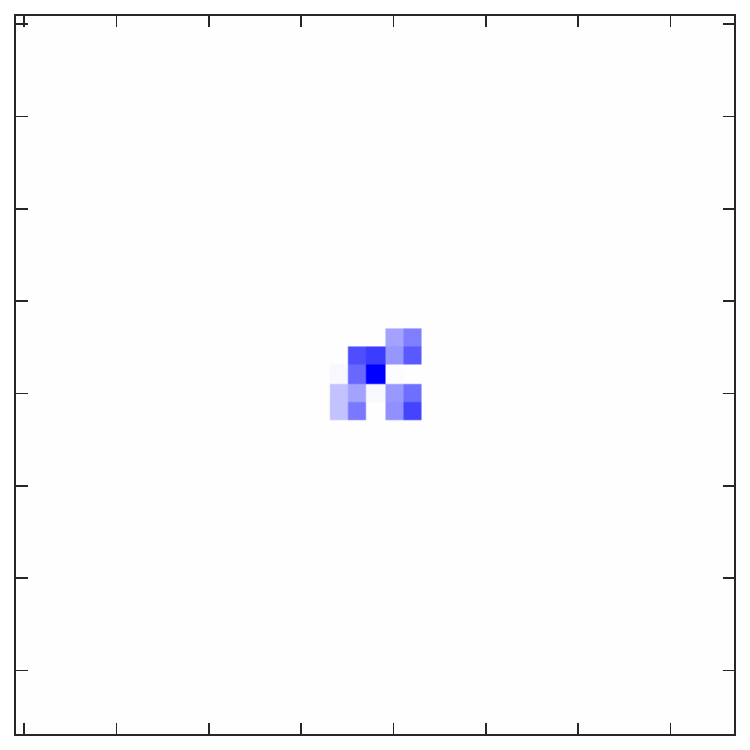}
  \includegraphics[width=0.119\textwidth]{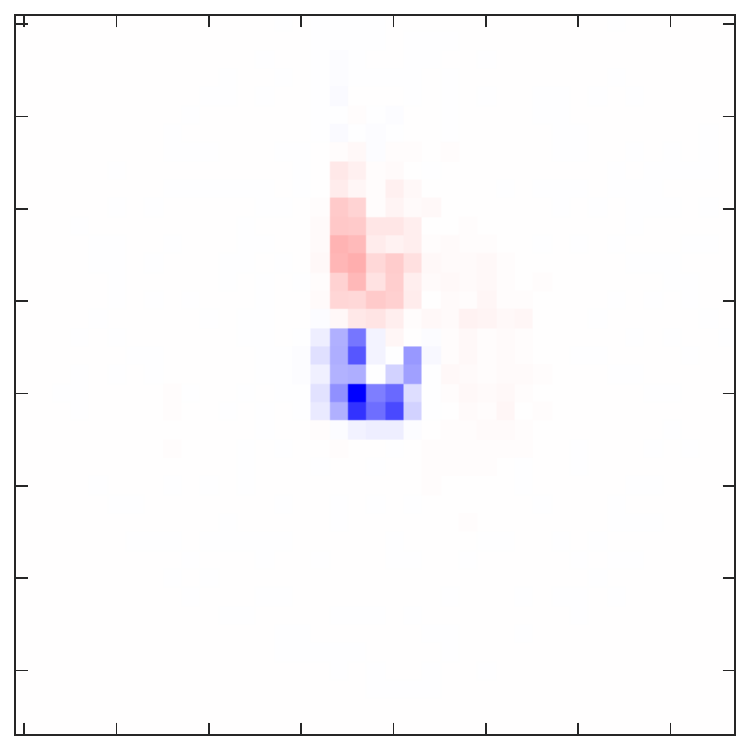}
  \includegraphics[width=0.119\textwidth]{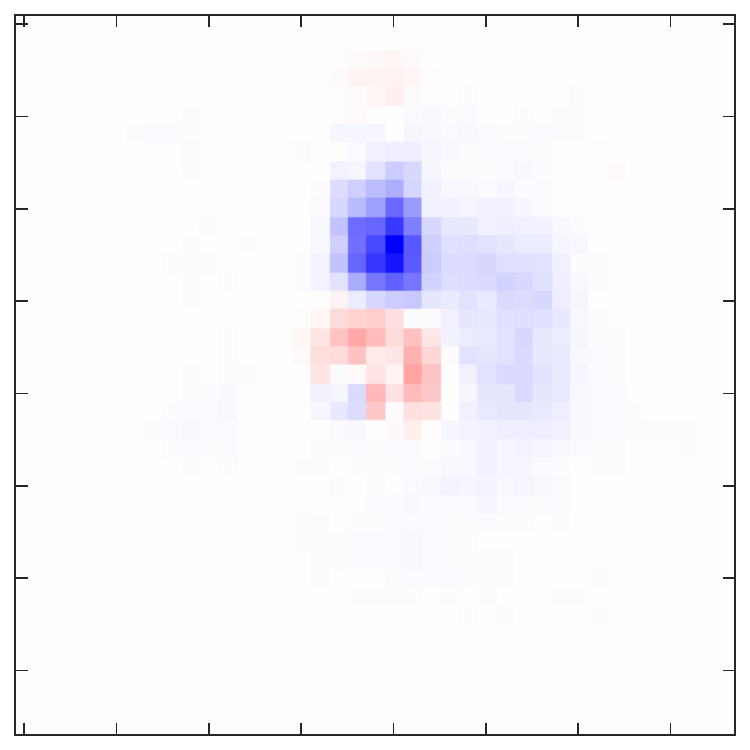}
  \includegraphics[width=0.119\textwidth]{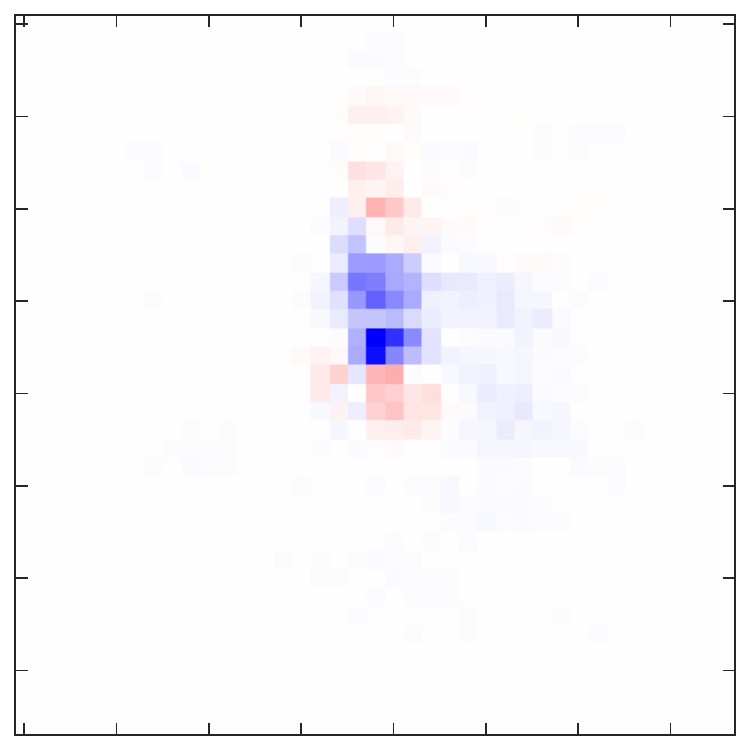} 
  \includegraphics[width=0.119\textwidth]{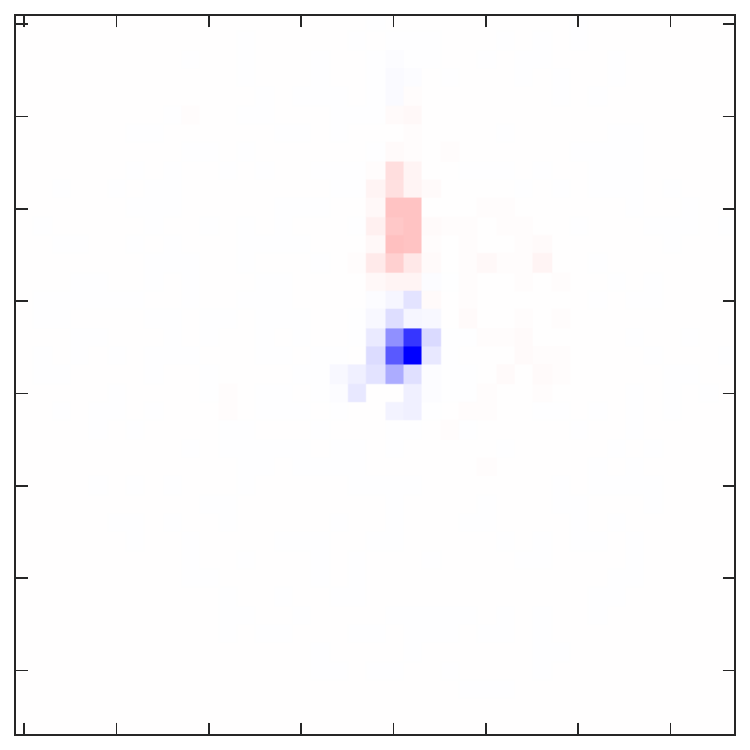} 
  \includegraphics[width=0.119\textwidth]{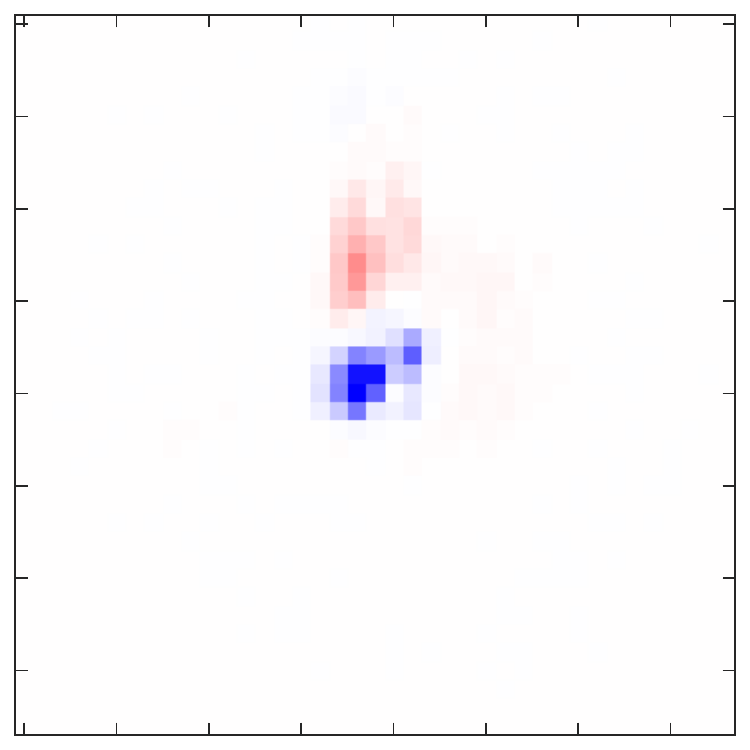} \\
  \includegraphics[width=0.119\textwidth]{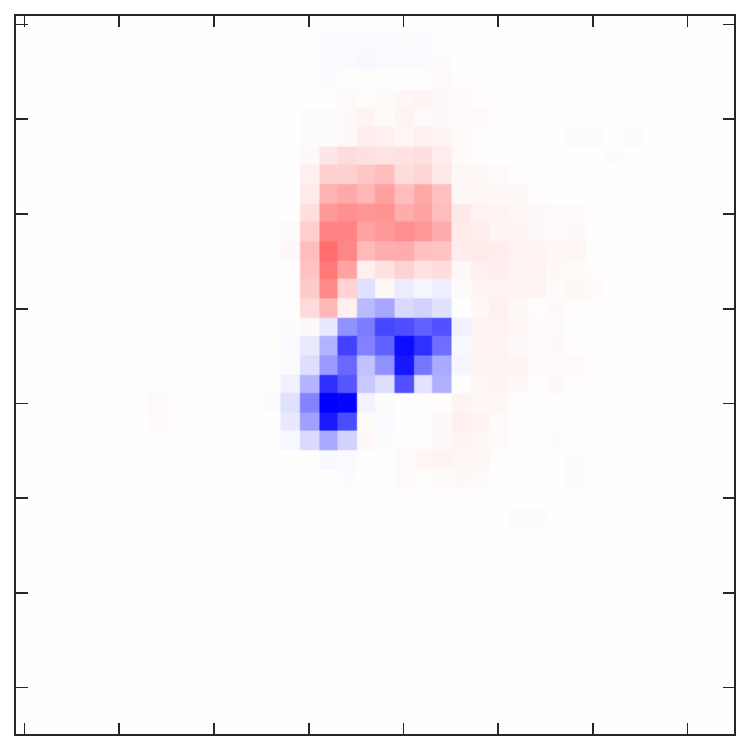}
  \includegraphics[width=0.119\textwidth]{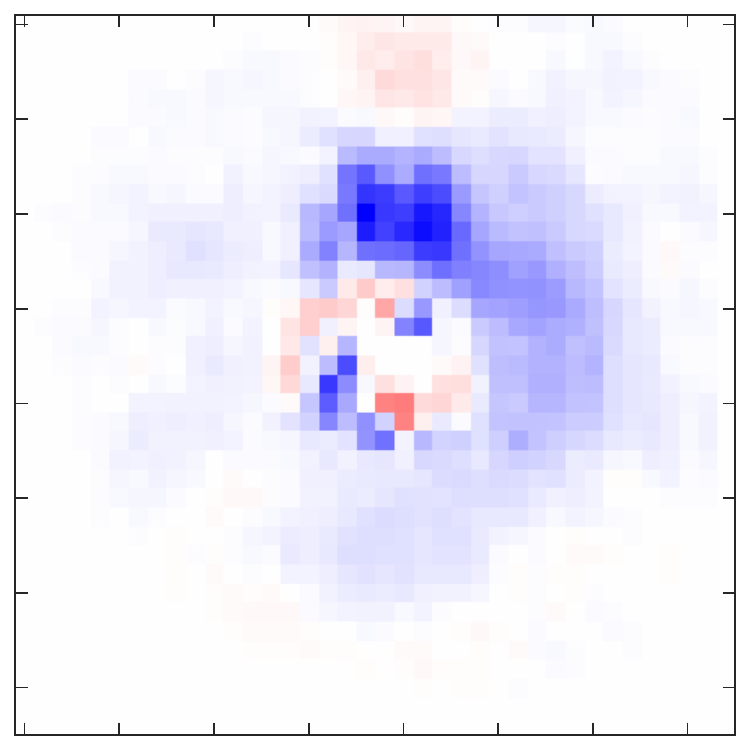}
  \includegraphics[width=0.119\textwidth]{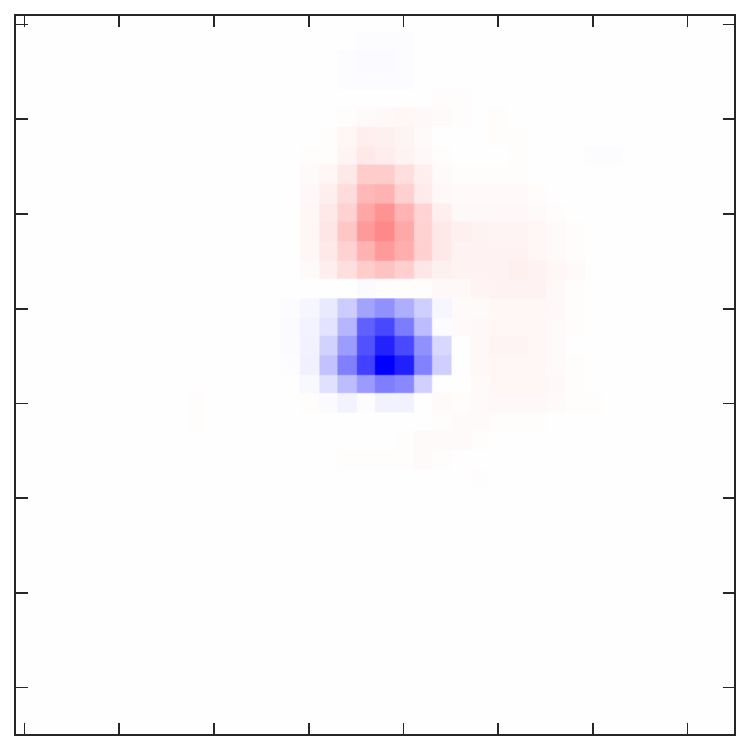}
  \includegraphics[width=0.119\textwidth]{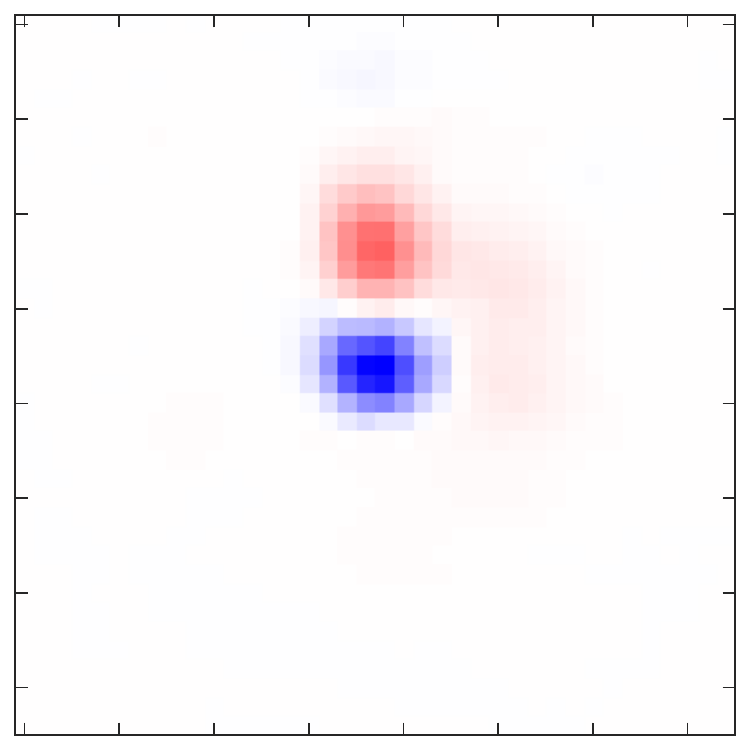}
  \includegraphics[width=0.119\textwidth]{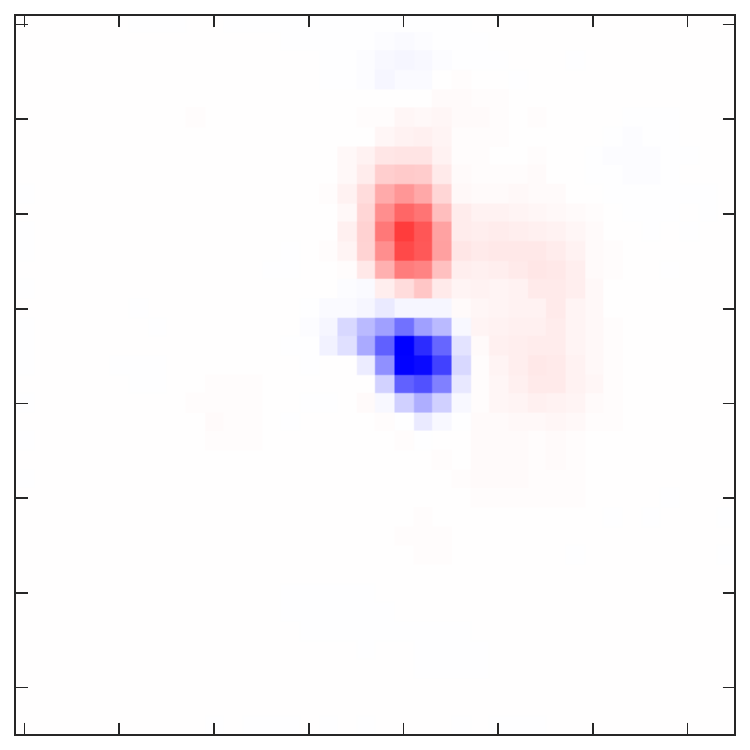}
  \includegraphics[width=0.119\textwidth]{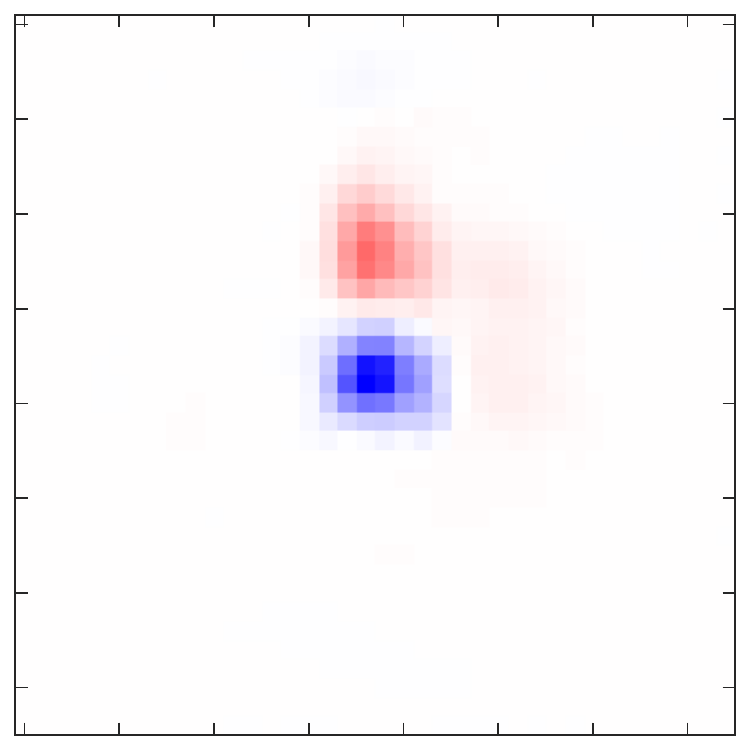} 
  \includegraphics[width=0.119\textwidth]{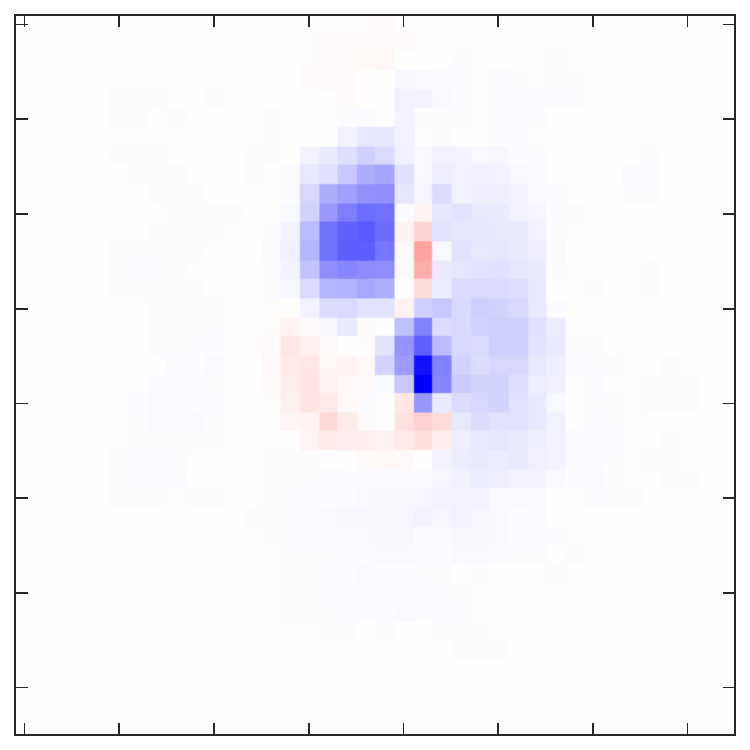} 
  \includegraphics[width=0.119\textwidth]{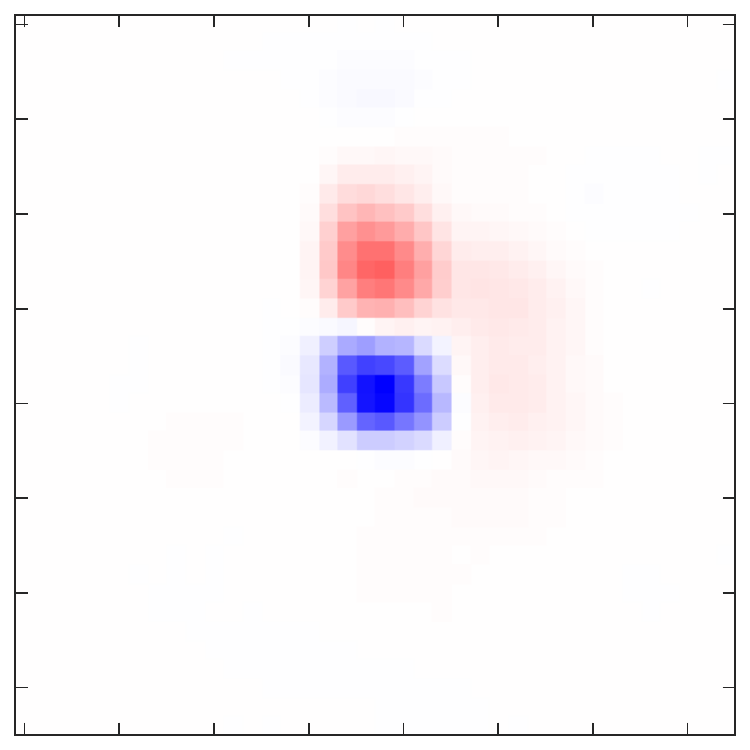} \\
  \includegraphics[width=0.119\textwidth]{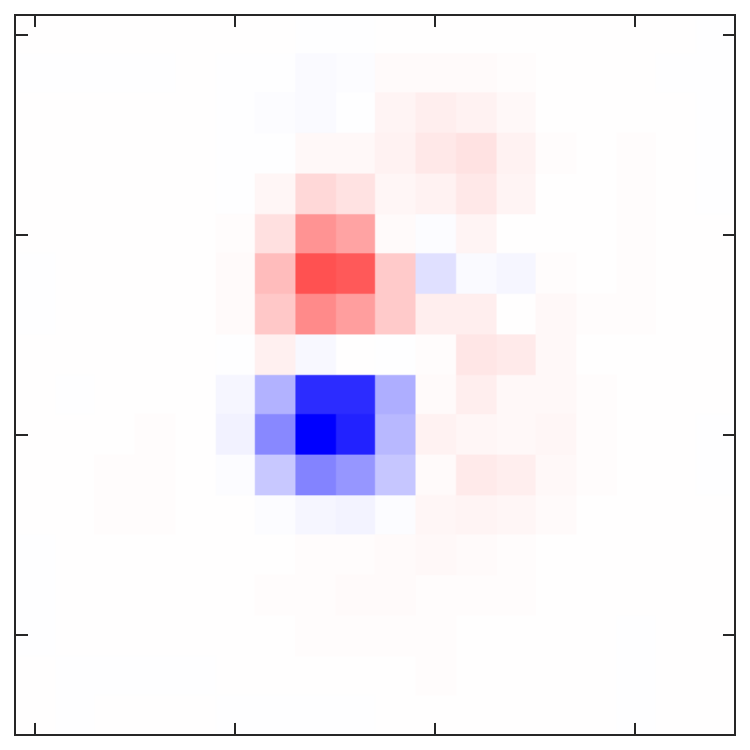}
  \includegraphics[width=0.119\textwidth]{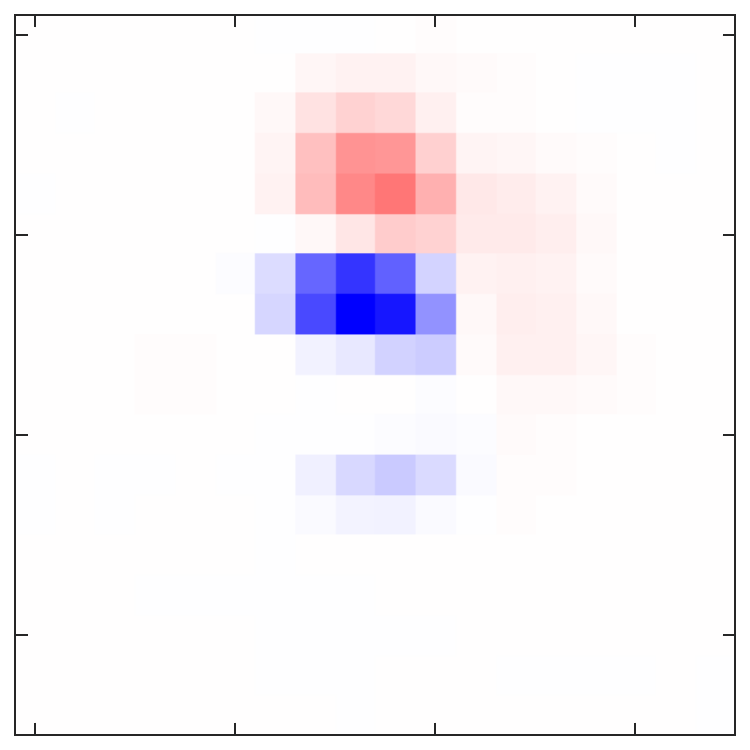}
  \includegraphics[width=0.119\textwidth]{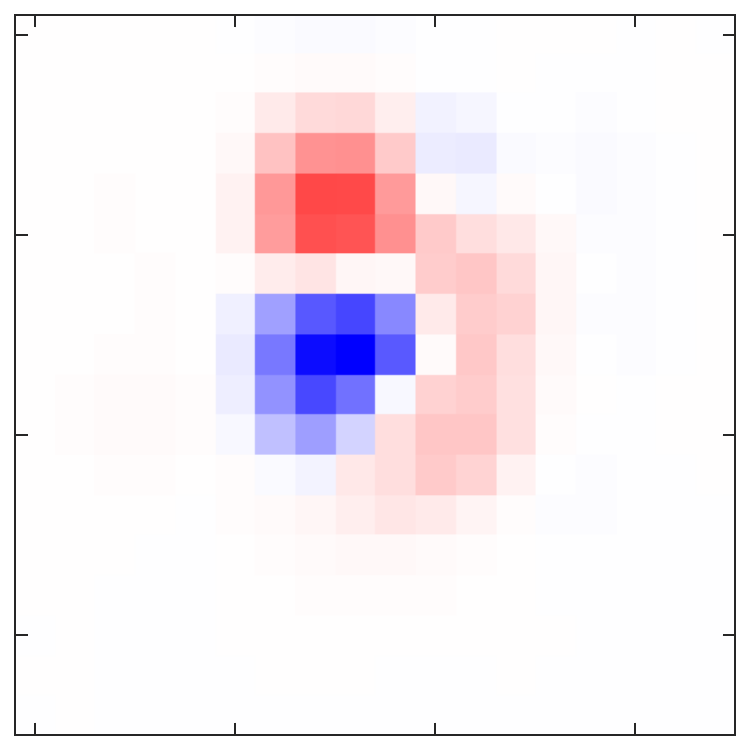}
  \includegraphics[width=0.119\textwidth]{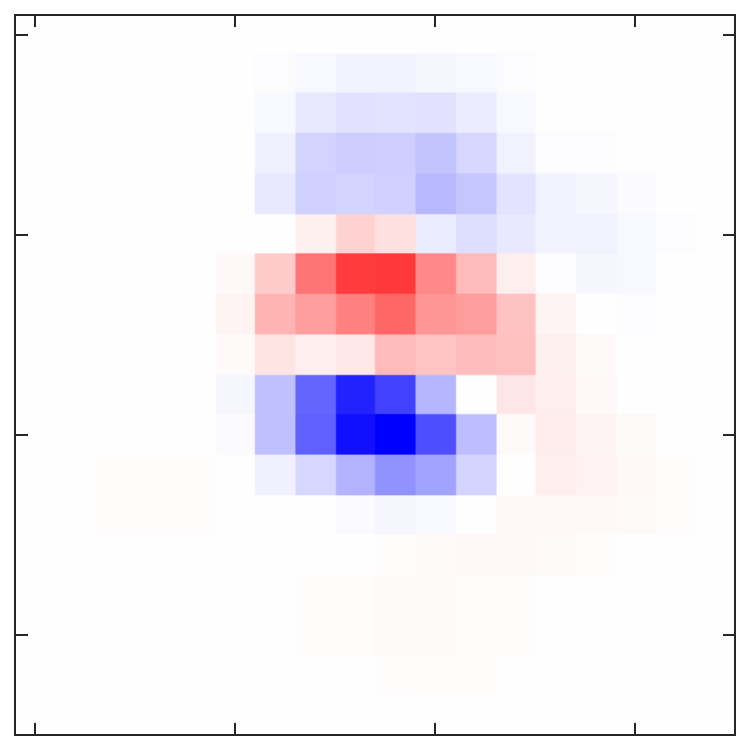}
  \includegraphics[width=0.119\textwidth]{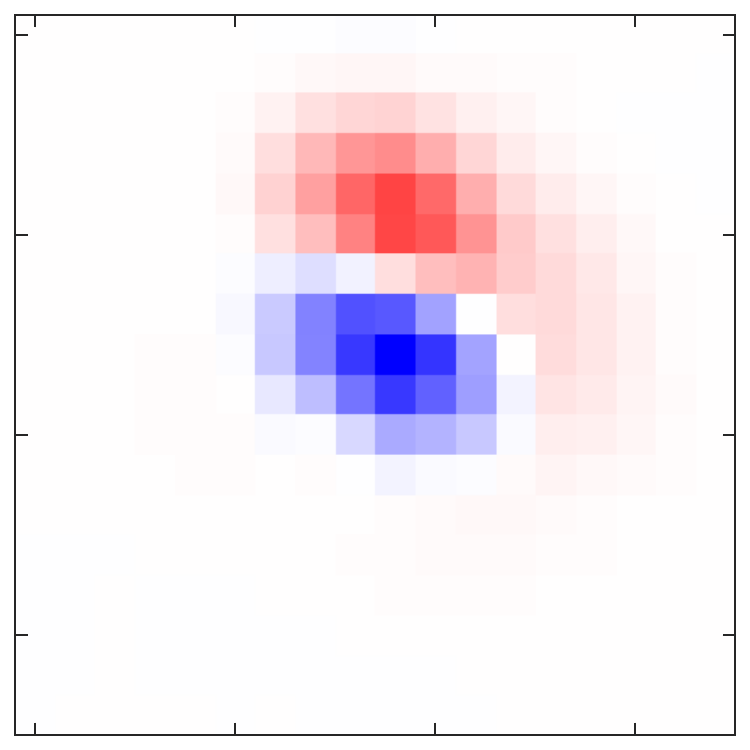}
  \includegraphics[width=0.119\textwidth]{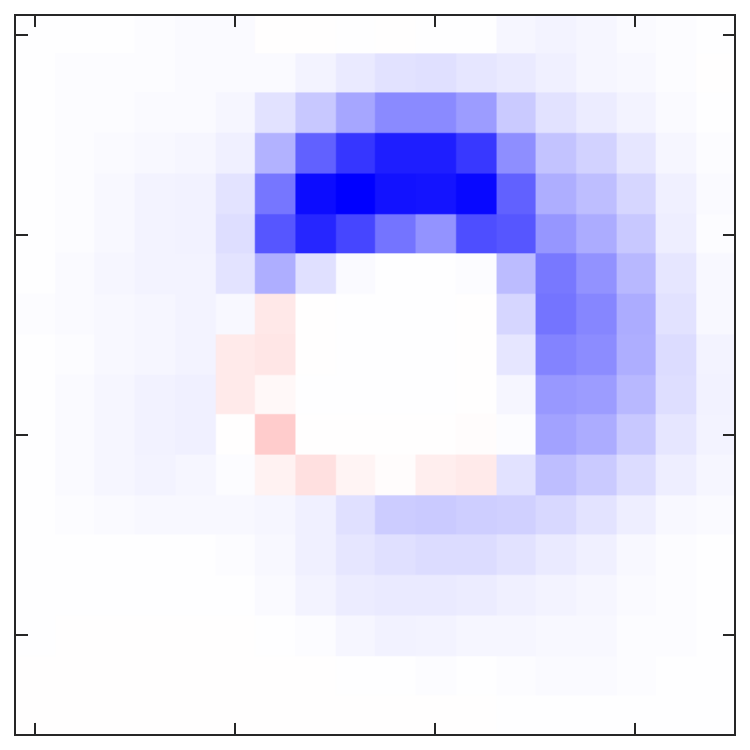} 
  \includegraphics[width=0.119\textwidth]{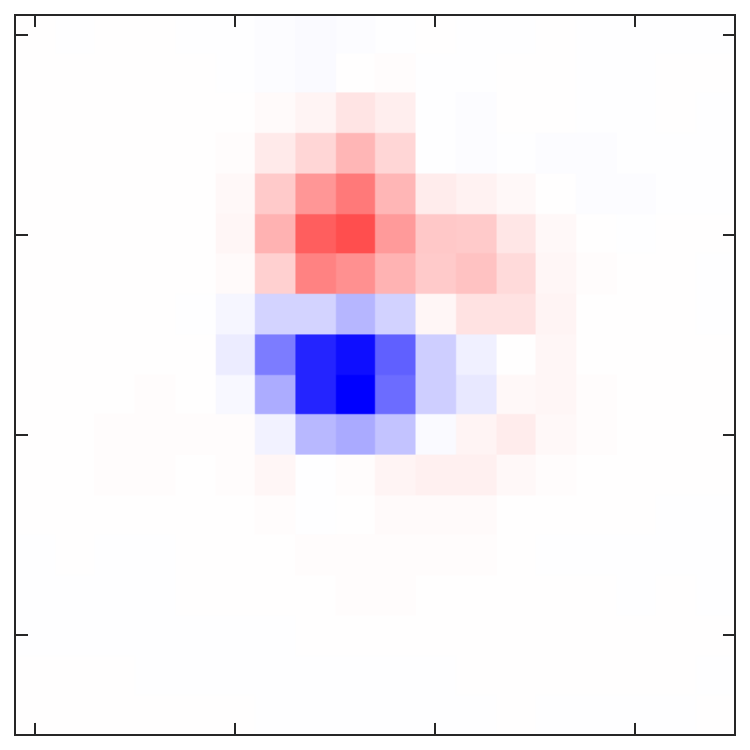} 
  \includegraphics[width=0.119\textwidth]{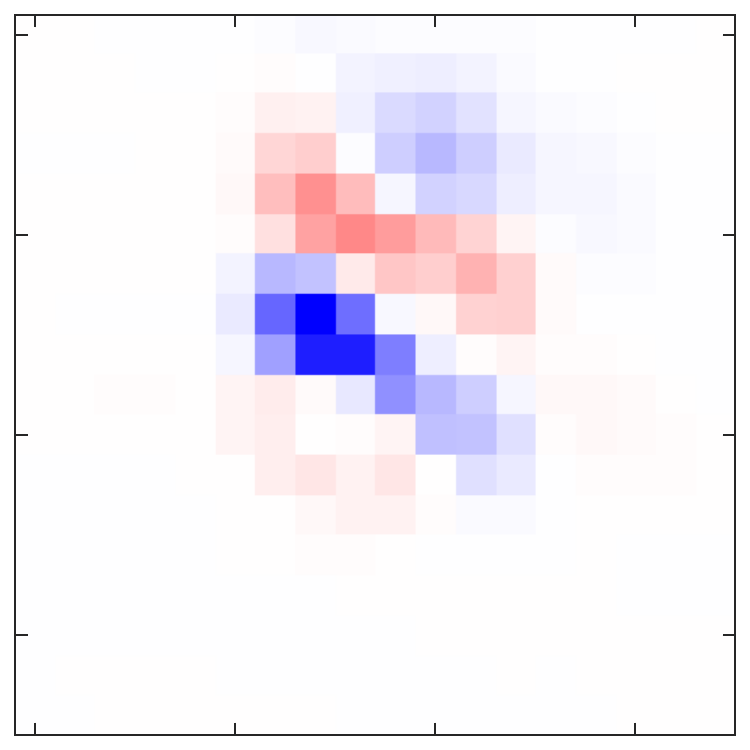} \\
  \includegraphics[width=0.119\textwidth]{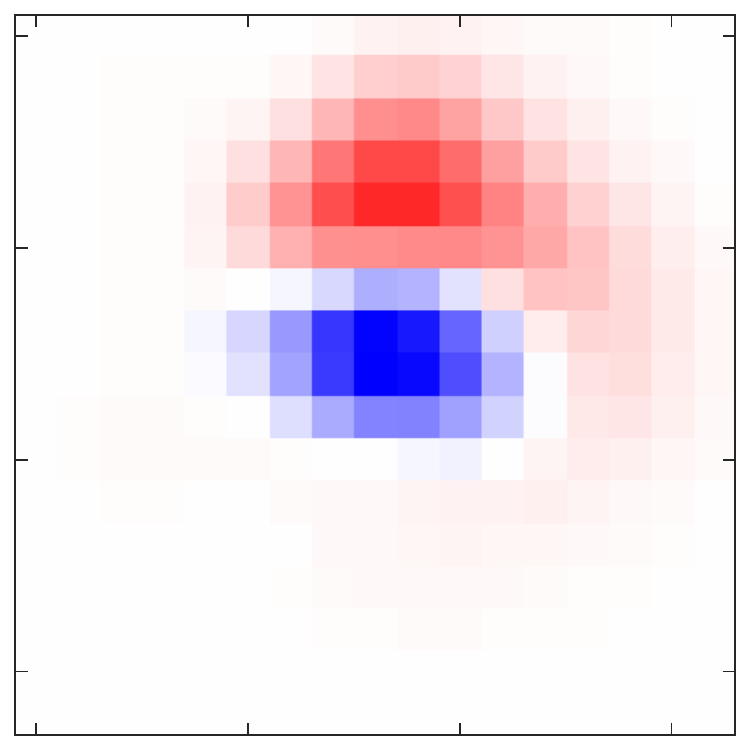}
  \includegraphics[width=0.119\textwidth]{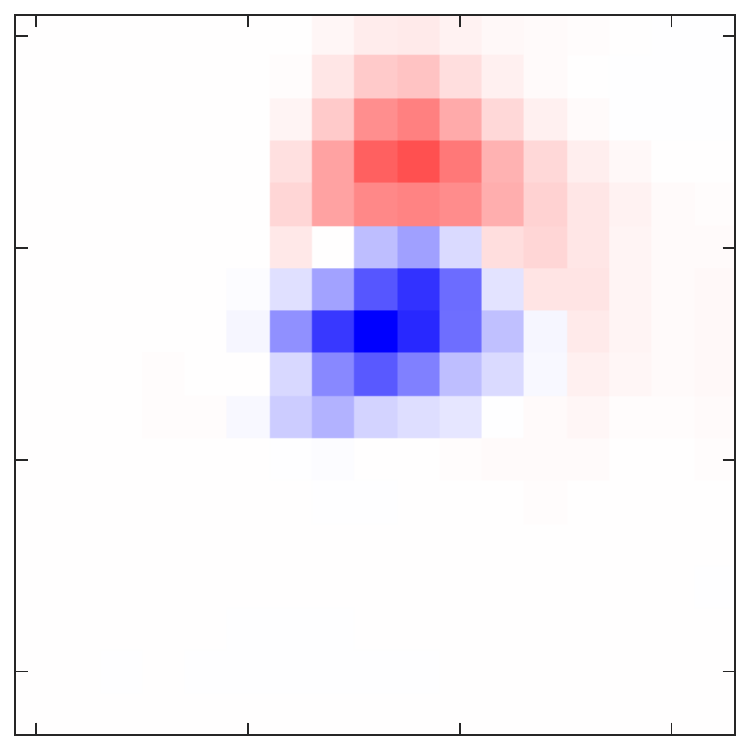}
  \includegraphics[width=0.119\textwidth]{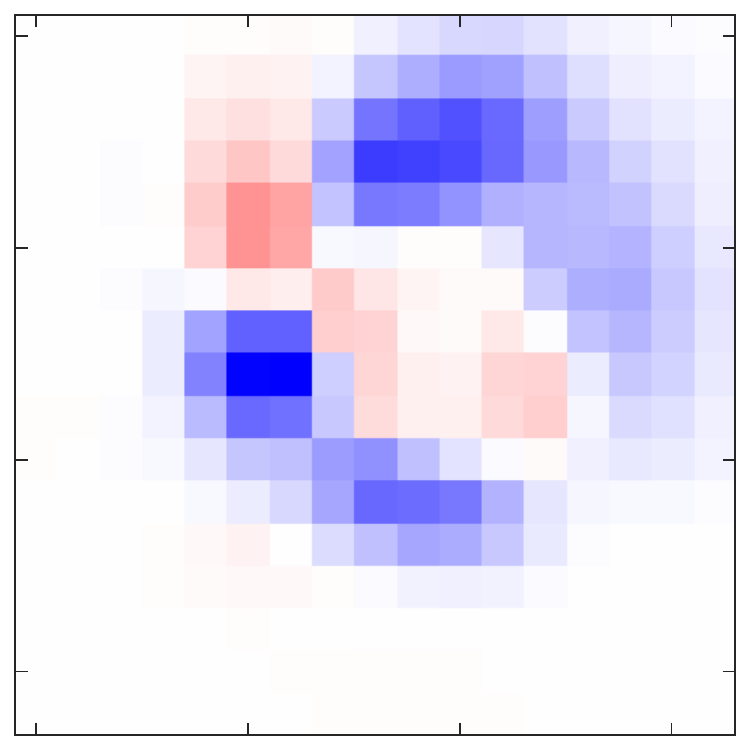}
  \includegraphics[width=0.119\textwidth]{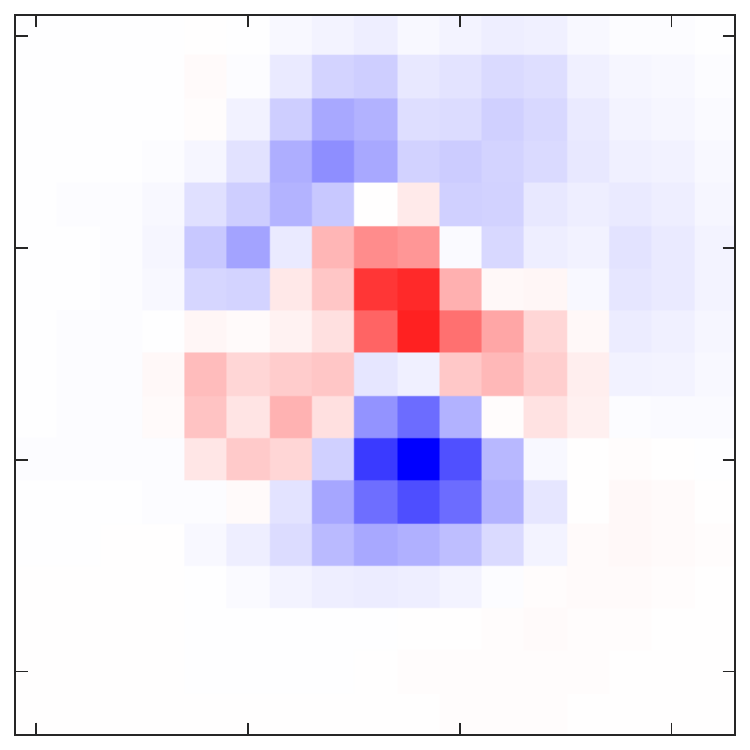}
  \includegraphics[width=0.119\textwidth]{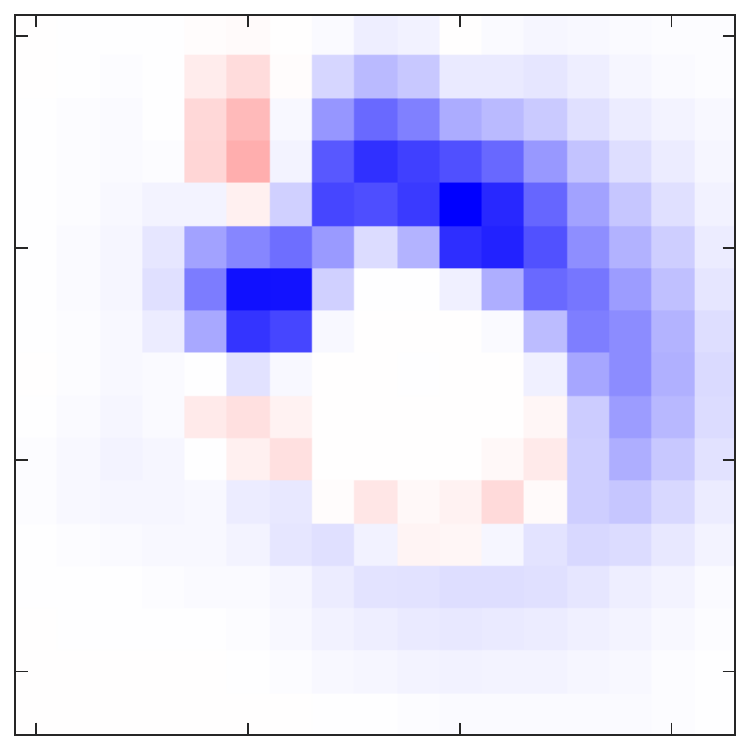}
  \includegraphics[width=0.119\textwidth]{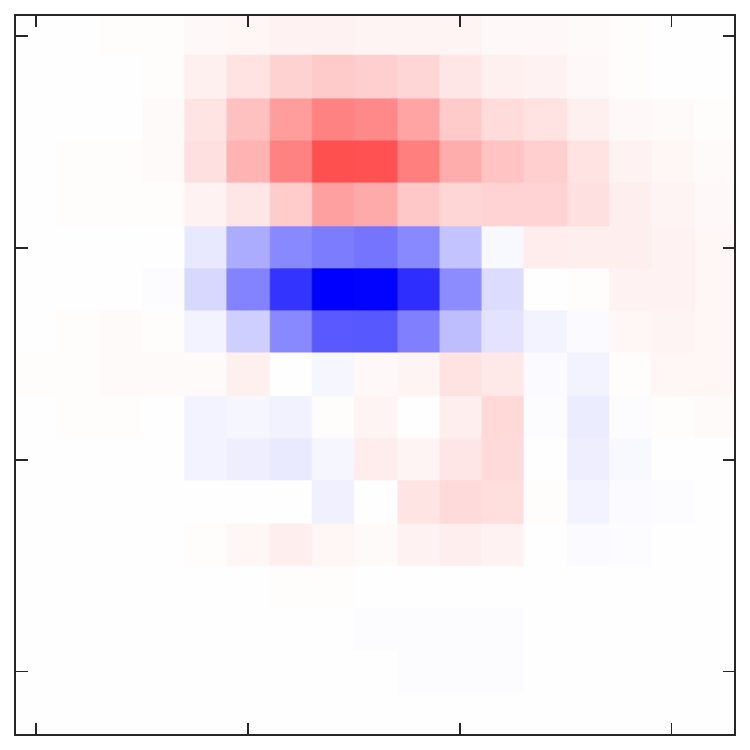}
  \includegraphics[width=0.119\textwidth]{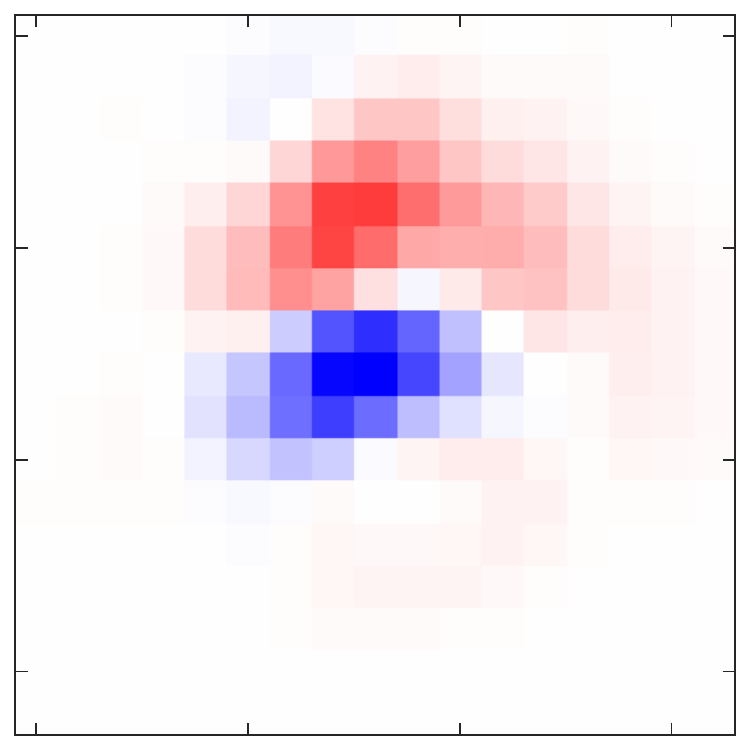}
  \includegraphics[width=0.119\textwidth]{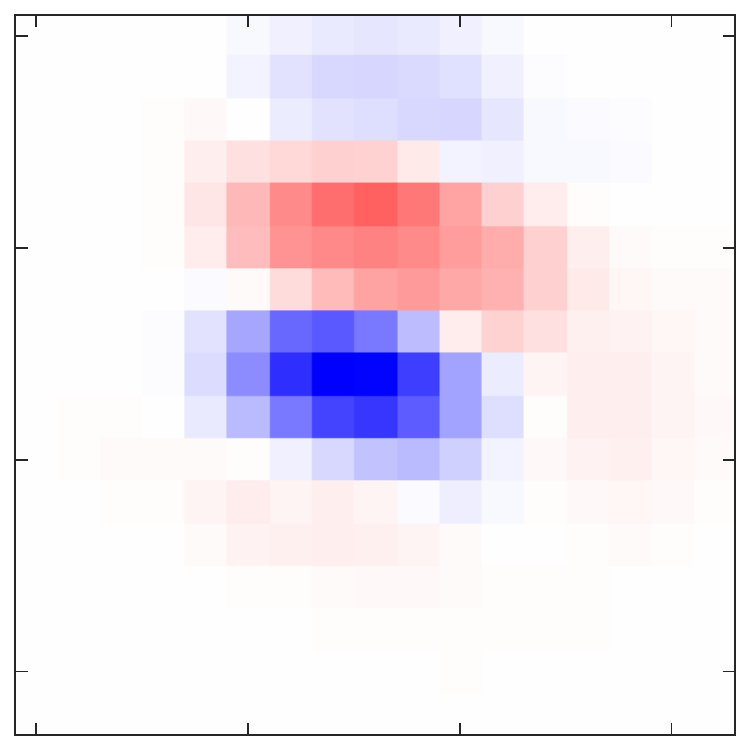}
  \caption{Averaged signal minus background for a simple CNN top tagger.
    The rows correspond to CNN layers, max-pooling reduces the number
    of pixels by roughly a factor four. The columns show different
    feature maps Red areas indicate signal-like regions, blue areas
    indicate background-like regions. Figure from
    Ref.~\cite{Kasieczka:2017nvn}.}
  \label{fig:conv_layers}
\end{figure}

\begin{figure}[t]
\centering
  \includegraphics[width=0.4\textwidth]{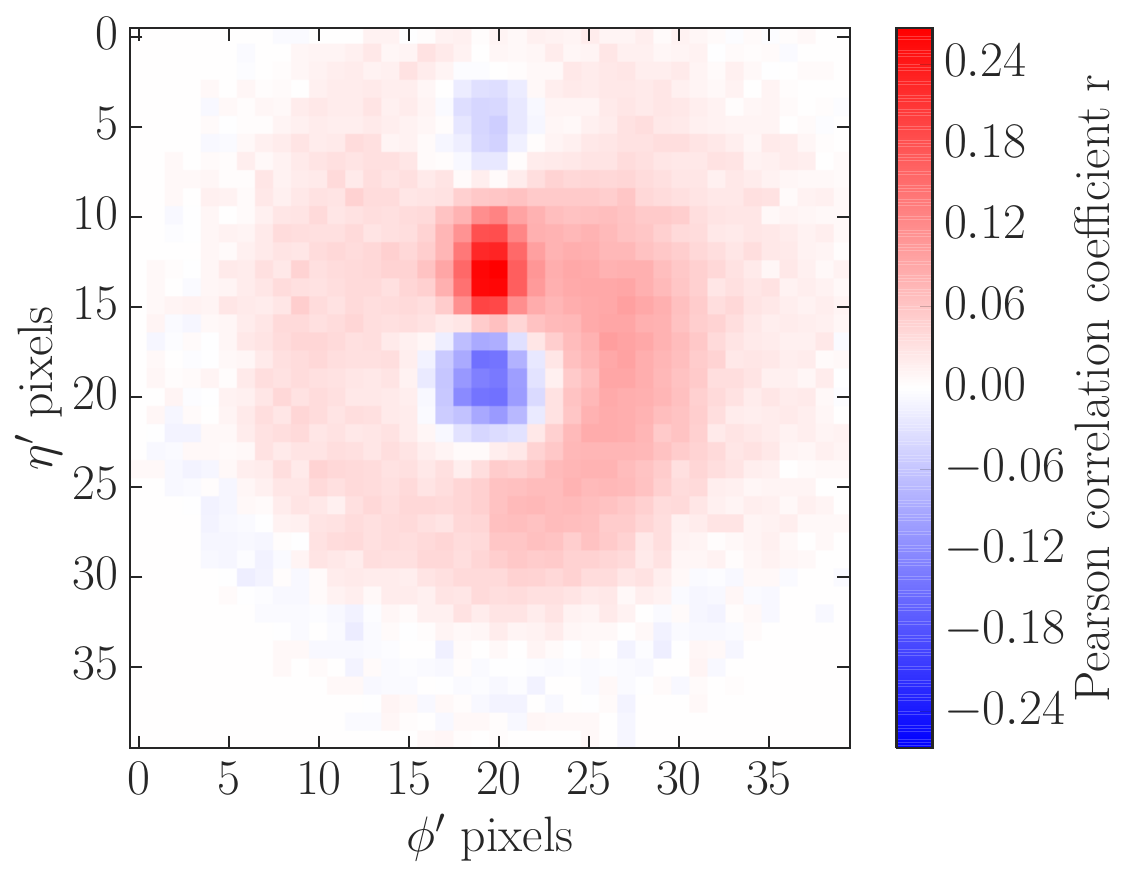}
  \caption{Pearson correlation coefficient for 10,000 signal and
    background images each. The corresponding jet image is illustrated
    in figure~\ref{fig:jet_images}.  Red areas indicate signal-like
    regions, blue areas indicate background-like regions. Figure from
    Ref.~\cite{Kasieczka:2017nvn}.}
  \label{fig:pcc}
\end{figure}

To these jet images we can apply a standard CNN, as illustrated in
Fig.~\ref{fig:cnn_architecture}.  Before we show the performance of a
CNN-based top tagger we can gain some intuition for what is happening
inside the trained CNN by looking at the output of the different
layers in the case of fully preprocessed images. In
Fig.~\ref{fig:conv_layers} we show the difference of the averaged
output for 100 signal-like and 100 background-like images. Each row
illustrates the output of a convolutional layer. Signal-like red areas
are typical for top decays, while blue areas are typical for QCD
jets. The feature maps in the first layer consistently capture a
well-separated second subjet, and some filters of the later layers
also capture a third signal subjet in the right half-plane. While
there is no one-to-one correspondence between the location in feature
maps of later layers and the pixels in the input image, these feature
maps still show that it is possible to see what a CNN learns. One can
try a similar analysis for the fully connected network layers, but it
turns out that we learn nothing.

To measure the impact of the pixels of the preprocessed jet image
impact on the extracted signal vs background label, we can correlate
the deviation of a pixel $x_{ij}$ from its mean value $\bar{x}_{ij}$
with the deviation of the signal probability $y$ from its mean value $\bar{y}$. The
correlation for a given set of combined signal and background images
is given by the \underline{Pearson correlation coefficient}
\begin{align}
  r_{ij} = \frac{\sum_\text{images}\left(x_{ij} - \bar{x}_{ij}\right)
    \left(y - \bar{y}\right)}
  { \sqrt{ \sum_\text{images}\left(x_{ij} -
        \bar{x}_{ij}\right)^2} \sqrt{ \sum_\text{images}\left(y - \bar{y}\right)^2 }
  } \;.
\end{align}
Positive values of $r_{ij}$ indicate signal-like pixels.  In
Fig.~\ref{fig:pcc} we show this correlation coefficient for a simple
CNN.  A large energy deposition in the center leads to classification
as background. A secondary energy deposition at 12 o'clock combined
with additional energy in the right half-plane means top signal,
consistent with Fig.~\ref{fig:jet_images}.

\begin{figure}[b!]
  \centering
  \includegraphics[width=0.505\textwidth]{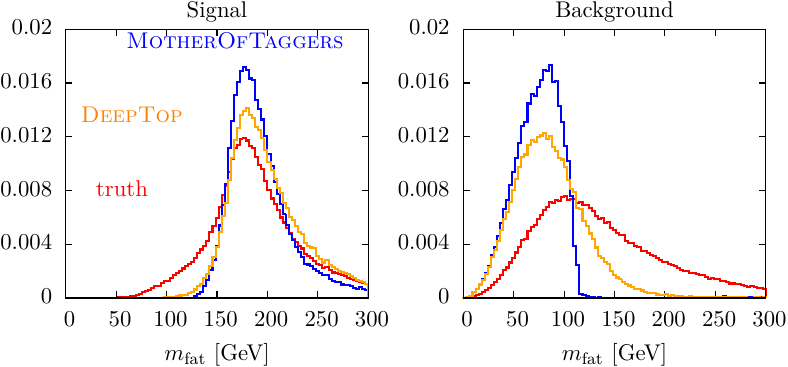}
  \includegraphics[width=0.48\textwidth]{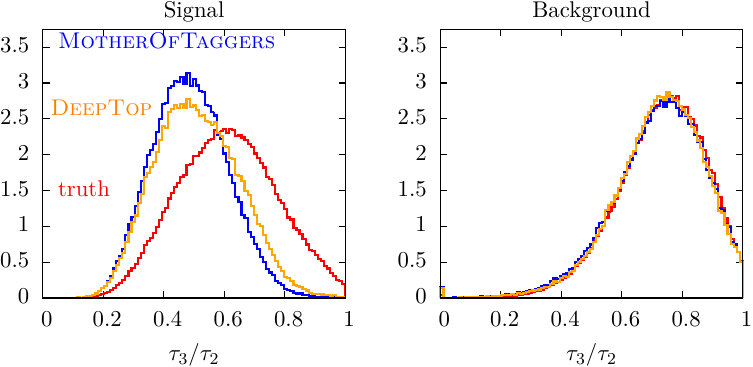}
  \caption{Kinematics observables $m_\text{fat}$ and $\tau_3/\tau_2$
    for events correctly determined to be signal or background by the
    DeepTop CNN and the MotherOfTaggers BDT, as well as Monte Carlo
    truth. Figure from Ref.~\cite{Kasieczka:2017nvn}.}
  \label{fig:inputs}
\end{figure}

Both, for the CNN and for a traditional BDT tagger we can study
signal-like learned patterns in actual signal events by cutting on the
output label $y$.  Similarly, we can use background-like events to
test if the background patterns are learned as expected. In addition,
we can compare the kinematic distributions in both cases to the Monte
Carlo truth. In Fig.~\ref{fig:inputs} we show the distributions for
the fat jet mass $m_\text{fat}$ and $\tau_3/\tau_2$ defined in
Eq.\eqref{eq:tau_N}. The CNN and the classic BDT learn essentially the
same structures. Their results are even more signal-like than the
Monte Carlo truth, because of the stiff cut on $y$. For the CNN and
BDT tagger cases this cut removes events where the signal kinematic
features is less pronounced. The BDT curves for the signal are more
peaked than the CNN curves because these two high-level observables
are BDT inputs, while for the neural network they are derived
quantities.

Going back to the CNN motivation, it turns our that preprocessed jet
images are not translation-invariant, and they are extremely
sparse. This means they are completely different from the kind of
images CNNs were developed to analyze. While the CNNs work very well
for image-based jet classification this raises the question if there
are other, better-suited network architectures for jet taggers.

\subsubsection{Top tagging benchmark}
\label{sec:class_cnn_sample}

\begin{figure}[t]
  \centering
  \includegraphics[width=0.6\textwidth]{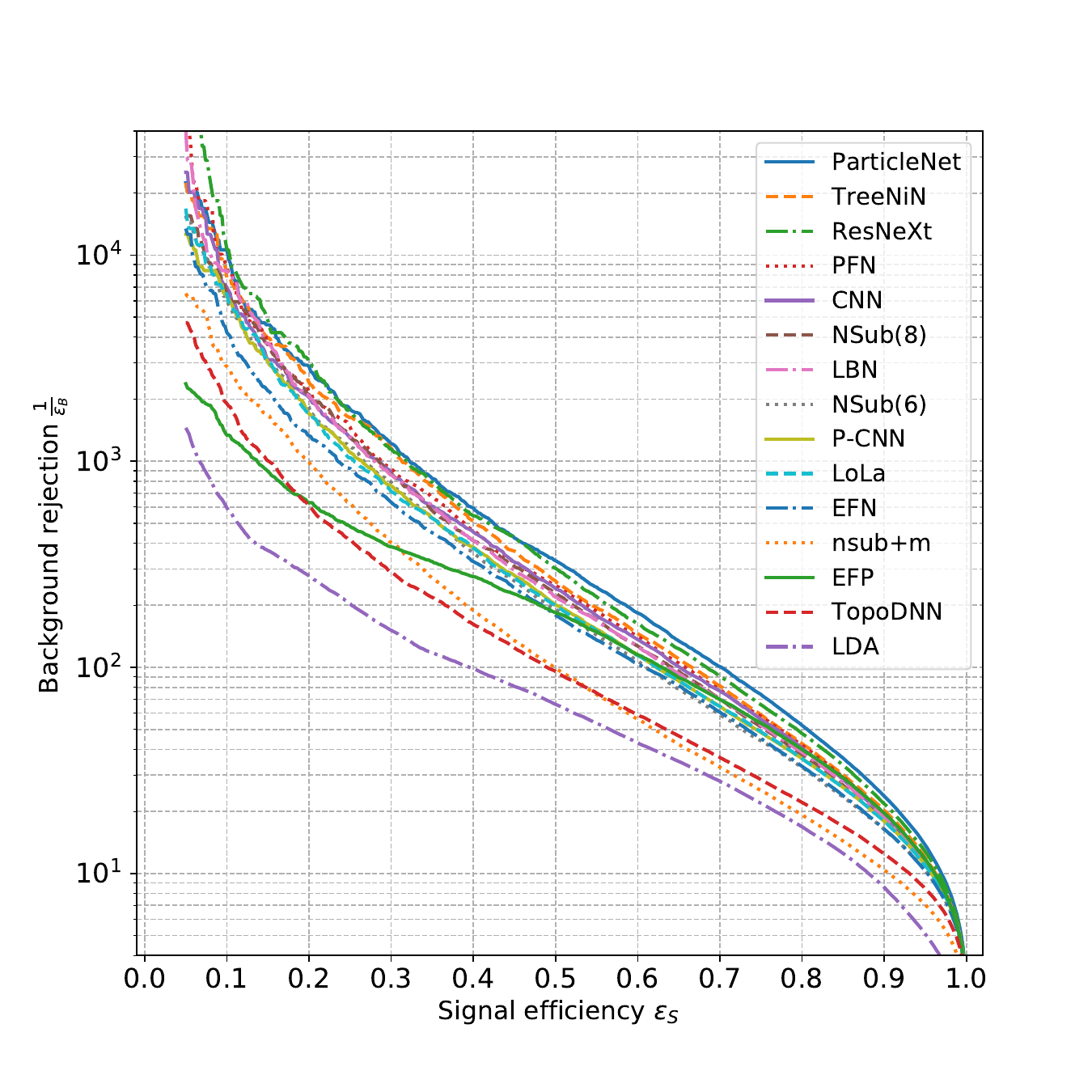}
  \caption{ROC curves for all top-tagging algorithms evaluated on the
    standard test sample. Figure from Ref.~\cite{Kasieczka:2019dbj}.}
  \label{fig:toptagging}
\end{figure}

As common in machine learning, \underline{standard datasets} are
extremely helpful to benchmark the state of the art and develop new
ideas.  For top tagging, the standard dataset consists of 1M top
signal and 1M mixed quark-gluon background jets, produced with the
Pythia8 event generator\index{event generators} and the simplified detector simulation
Delphes. The are divided into 60\% training, 20\% validation, and 20\%
test data. The fat jet is defined through the anti-$k_T$ algorithm
with size $R = 0.8$, which means its boundaries in the jet plane are
smooth. The dataset only uses the leading jet in each $t\bar{t}$ or
di-jet event and requires
\begin{align}
  p_{T,j} = 550~....~650~\gev
  \qquad \text{and} \qquad
  |\eta_j| < 2 \; .
  \label{eq:standard_ptcut}
\end{align}
From Eq.\eqref{eq:topjet_size} we know that we can require a
parton-level top and all its decay partons to be within $\Delta R =
0.8$ of the jet axis for the signal jets. The jet constituents are
extracted through the Delphes energy-flow algorithm, which combines
calorimeter objects with tracking output. The dataset includes the
4-momenta of the leading 200 constituents, which can then be
represented as images for the CNN application.  Particle information
is not included, so $b$-tagging cannot be added as part of the top
tagging, and any quoted performance should not be considered
realistic. The dataset is easily accessible as part of a broader
physics-related reference datasets~\cite{Benato:2021olt}.

Two competitive versions of the image-base taggers were benchmarked on
this dataset, an updated CNN tagger and the standard ResNeXt
network. While the toy CNN shown in Fig.~\ref{fig:cnn_architecture} is
built out of four successive convolutional layers, with eight feature
maps each, and its competitive counterpart comes with four
convolutions layers and 64 feature maps, professional CNNs include 50
or 100 convolutional layers. For networks with this depth a stable
training becomes increasingly hard with the standard convolutions
defined in Eq.\eqref{eq:def_conv}. This issue can be targeted with a
residual network, which is built out of convolutional layers combined
with \underline{skip connections}. In the conventions of
Eq.\eqref{eq:affine} these skip connections come with the additional
term
\begin{align}
  x^{(n-1)} \to x^{(n)} = W^{(n)} x^{(n-1)} + b^{(n)} + x^\text{(earlier layer)}\; ,
  \label{eq:skip_connection}
\end{align}
where the last term can point to any previous layer. Obviously, we can
apply the same structure to the convolutional layer of
Eq.\eqref{eq:def_conv0}
\begin{align}
    x^{(n)}_{ij} = \sum_{r,s=0}^{n_\text{c-size}-1} 
    W^{(n)}_{rs}  \;
    x^{(n-1)}_{i+r,j+s} + b^{(n)} + x_{ij}^\text{(earlier layer)} \; .
\end{align}
Again, we suppress the sum over the feature maps.  These skip
connections are a standard method to improve the training and
stability of very deep networks, and we will come across them again.

In Fig.~\ref{fig:toptagging} we see that the deep ResNeXt is slightly
more powerful than the competitive version of the CNN introduced in
Sec.~\ref{sec:class_cnn_arch}. First, in this study it uses slightly
higher resolution with $64 \times 64$ pixels. In addition, it is much
more complex with 50 layers translating into almost 1.5M parameters,
as compared to the 610k parameters of the CNN. Finally, it uses skip
connections to train this large number of layers. The sizes of the
ResNeXt and the CNN illustrate where some of the power of neural
networks are coming from. They can use up to 1.5M network parameters
to describe a training dataset consisting of 1M signal and background
images each. Each of these sparse calorimeter images includes anything
between 20 and 50 interesting active pixels. Depending on the physics
question we are asking, the leading 10 pixels might encode most of the
information, which translates into 20M training pixels to train 1.5M
network parameters. That is quite a complexity, for instance compared
to standard fits or boosted decision trees. This complexity
also motivates an efficient training, including the back propagation
idea, an appropriately chosen loss function, and numerical GPU
power. Finally, it explains why some people might be sceptical about
the black box nature of neural networks, bringing us back to the
question how we can \underline{control} what networks learn and assign
\underline{uncertainty bands} to their output.

\subsubsection{Bayesian CNN}
\label{sec:class_cnn_bayes}

Since we know from Sec.~\ref{sec:basics_deep_bayes} how to train a
network to not only encode some kind of function $f_\theta(x)$ but
also an uncertainty $\sigma_\theta(x)$. We can apply this method to
our jet classification task, because it makes a difference if a jet
comes with $(60 \pm 20)\%$ or $(60 \pm 1)\%$ signal
probability. Bayesian classification networks provide this information
jet by jet. For jet classification this uncertainty could for example
come from (i) finite, but perfectly labeled training samples; (ii)
uncertainties in the labelling of the training data; and (iii)
systematic differences between the training and test samples.

\begin{figure}[t]
  \centering
  \includegraphics[width=0.49\textwidth]{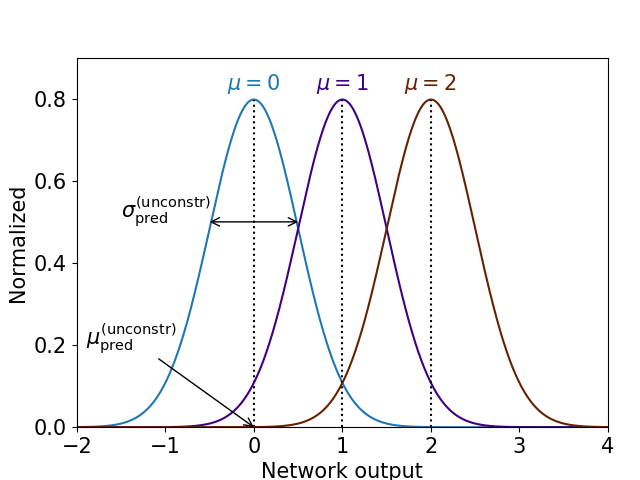}
  \includegraphics[width=0.49\textwidth]{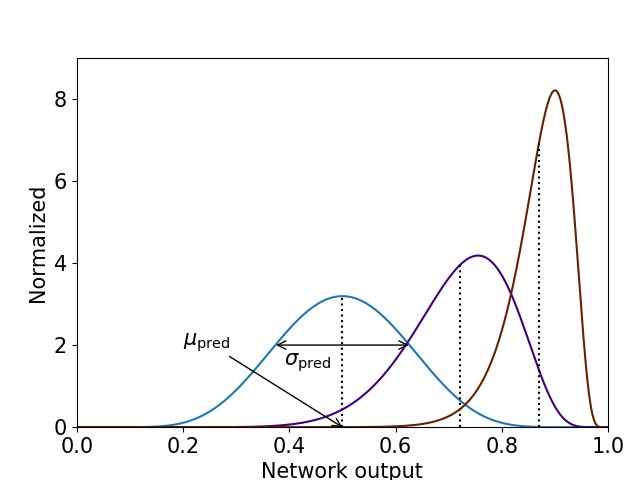} 
  \caption{Effect of the sigmoid transformation on Gaussians with the
    same but different means. Figure from
    Ref.~\cite{Bollweg:2019skg}.}
  \label{fig:bayesian_sigmoid}
\end{figure}

The main difference between a regression network and a classification
network is that the probability outputs requires us to map unbounded
network outputs to the \underline{closed interval} $[ 0,1 ]$ through the final,
sigmoid layer given in Eq.\eqref{eq:def_sigmoid}. Such a sigmoid layer
will change the assumed Gaussian distribution of the Bayesian network
weights. We illustrate this behavior in
Fig.~\ref{fig:bayesian_sigmoid}, where we start from three Gaussians
with the same width but different means and apply a sigmoid
transformation. The results is that the distributions on the closed
interval become asymmetric, to accommodate the fact that even in the
tails of the distributions the functional value can never exceed
one. The Jacobian of the sigmoid transformation is given by
\begin{align}
  \frac{d \; \sigmoid(x)}{dx}
  &= \frac{d}{dx} \big[ 1 + e^{-x} \big]^{-1}
  = - \frac{1}{\big[ 1 + e^{-x} \big]^2} \frac{d}{dx} \big[ 1 + e^{-x} \big]
  = \frac{e^{-x}}{\big[ 1 + e^{-x} \big]^2}\notag \\
  &= \frac{1}{1 + e^{-x}} \frac{1 + e^{-x} - 1}{1+e^{-x}} \notag \\
  &= \sigmoid(x) \Big[ 1- \sigmoid(x) \Big] \; .
\end{align}
%
We can approximate the standard deviation for example of a Gaussian
after the sigmoid transformation assuming a simple linearized form
\begin{align}
\frac{\sigma_\text{stat}^\text{(sigmoid)}}{\sigma_\text{stat}}
\approx \frac{d \sigma_\text{stat}^\text{(sigmoid)}}{d \sigma_\text{stat}}
\approx \frac{d \mu_\text{stat}^\text{(sigmoid)}}{d \mu_\text{stat}}
= \mu_\text{stat}^\text{(sigmoid)} \big[ 1- \mu_\text{stat}^\text{(sigmoid)} \big] \; .
\label{eq:sigmoid_width_approx}
\end{align}
After the sigmoid transformation the uncorrelated parameters $\mu$ and
$\sigma$ turn into a correlated mean and standard deviation; for a
transformed mean $\mu_\text{stat}^\text{(sigmoid)}$ going to zero or
one, the corresponding width $\sigma_\text{stat}^\text{(sigmoid)}$
will vanish. This correlation of the two Bayesian network output is
specific to a classification task. This kind of behavior is not new,
if we remember how we need to replace the Gaussian by a Poisson
distribution, which has the same cutoff feature towards zero count
rates.

\begin{figure}[t]
\includegraphics[width=0.49\textwidth]{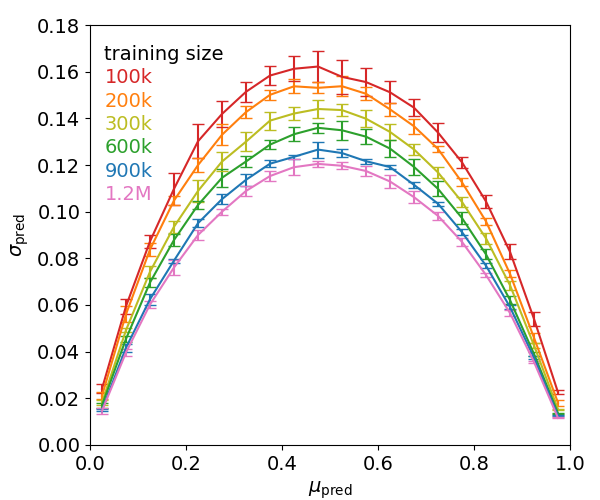}
\includegraphics[width=0.49\textwidth]{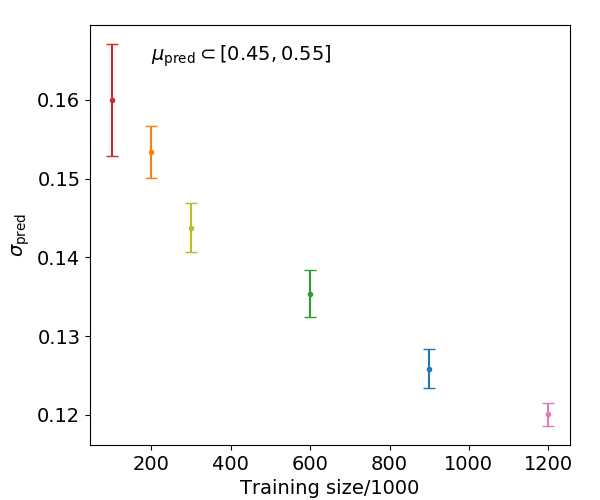} \\
\includegraphics[width=0.24\textwidth]{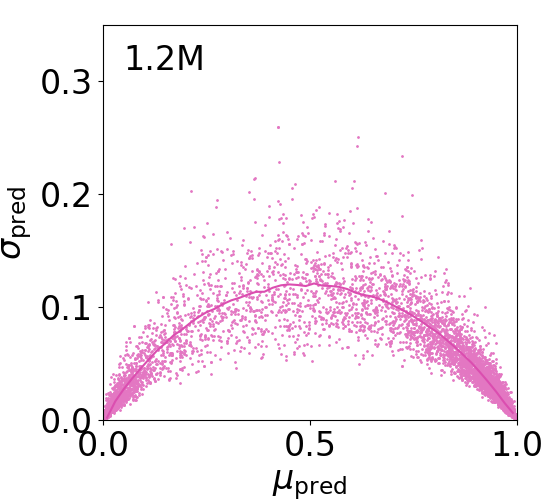}
\includegraphics[width=0.24\textwidth]{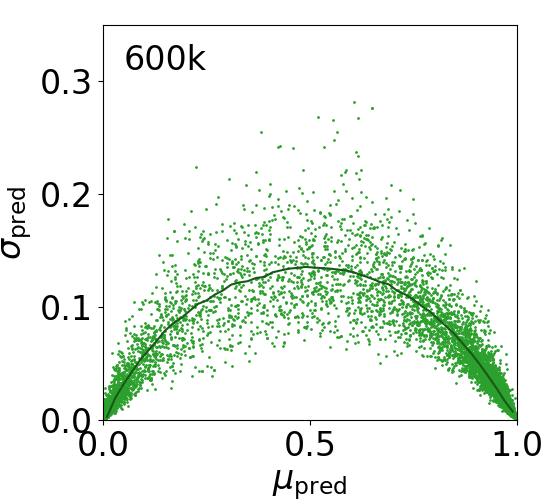}
\includegraphics[width=0.24\textwidth]{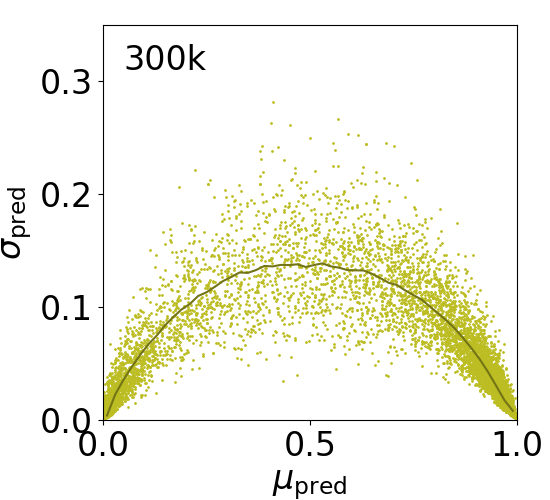}
\includegraphics[width=0.24\textwidth]{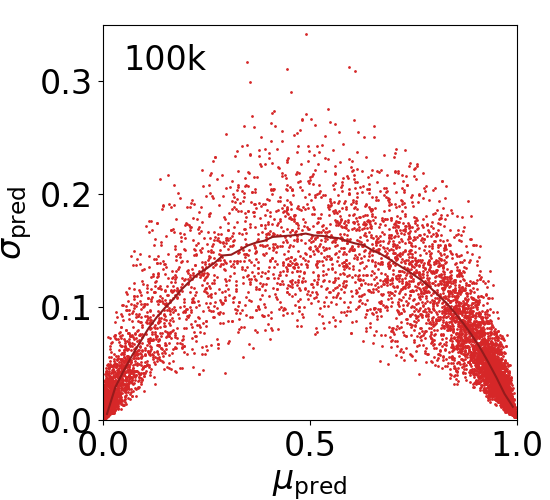}
\caption{Correlation between predictive mean and standard
  deviation. The right panel shows the predictive standard deviation
  for $\mu_\text{stat} = 0.45~...~0.55$ as a function of the size of
  the training sample with the same uncertainty bands from different
  trainings. The lower panels instead show the statistical spread for
  10k jets, signal and background combined. Figure from
  Ref.~\cite{Bollweg:2019skg}.}
\label{fig:mu_sd}
\end{figure}

To see these correlations we can look at a simple source of
statistical\index{statistical uncertainty} uncertainties\index{uncertainties}, a limited number of training jets.  In the
upper left panel of Fig.~\ref{fig:mu_sd} we show the correlation
between the predictive mean and the predictive standard deviation from
the Bayesian CNN\index{Bayesian network}.  For a single correlation curve we evaluate the
network on 10k jets, half signal and half background, and show the
mean values of the 10k jets in slices of $\mu_\text{stat}$, after
confirming that their distributions have the expected Gaussian-like
shape.  The leading feature is the \underline{inverse parabola} shape,
induced by the sigmoid transform, Eq.\eqref{eq:sigmoid_width_approx}.
This is combined with a physics feature, namely that probability
outputs around 0.1 or 0.9 correspond to clear cases of signal and
background jets, where we expect the predictive standard deviation to
be small.  In the upper right panel we illustrate the improvement of
the network output with an increasing amount of training data for the
slice $\mu_\text{stat} = 0.45~...~0.55$.  The improvement is
significant compared to the uncertainty bands, which correspond to different
training and testing samples.  The corresponding spread of these 10k
signal and background jets is illustrated in the four lower panels,
with a matching color code.

\begin{figure}[b!]
  \centering
  \includegraphics[width=0.30\textwidth]{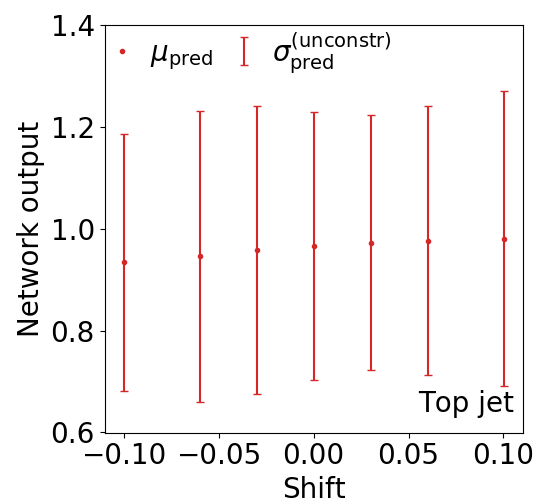}
  \hspace*{0.1\textwidth}
  \includegraphics[width=0.30\textwidth]{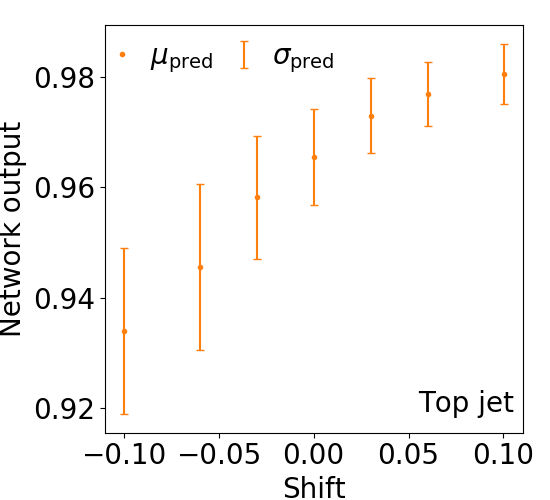} 
\caption{Effect of a shifted energy scale for the hardest constituent
  on the top tagging, showing the network output before the sigmoid
  transformation (left) and the classification output (right). Figure
  from Ref.~\cite{Bollweg:2019skg}.}
\label{fig:adv_attack}
\end{figure}

One of the great advantages of LHC physics is that we can study the
behavior of neural networks on Monte Carlo, before we train or at
least use them on data. This also means that we can extend the
uncertainty treatment of BNNs to also include systematic
uncertainties\index{systematic uncertainty}, as long as we can describe them using \underline{data
  augmentation}. An example is the jet or calorimeter energy scale,
which is determined through reference measurements and then included
in all jet measurements as a function of the detector
geometry. Because hard and soft pixels encode different physics
information and the calibration of soft pixels is strongly correlated
with pile-up removal, we can see what happens for a top tagger when we
change the pixel calibration. As a side remark, for this study we do
not use the standard CNN, because here the pixel entries are usually
normalized. To see an effect on the classification output we shift the
energy of the leading jet constituent by up to 10\%. In the left panel
of Fig.~\ref{fig:adv_attack} we see that this shift has hardly any
effect on the network output before we apply the sigmoid
activation. In the right panel we see what happens after the sigmoid
activation: depending on the sign of the systematic shift the network
is systematically less or more sure that a top jets corresponds to a
signal. From a physics perspective this is expected, because top jets
are more hierarchical, owed to the weak decays and the mass
drop. Changing the calibration of the hard constituent(s) then acts
like an \underline{adversarial attack} on the classification network
--- a change exactly in the feature that dominates the classification
output.

Finally, as mentioned above, using neural networks beyond black-box mode
requires to control if the network has captured the underling
feature(s) correctly and then assign an uncertainty band to the network
output. These two aspects have to be separated, because networks fall
into the same trap as people --- not knowing anything they tend to
underestimate their uncertainty by ignoring unknown unknowns. For
Bayesian classification we can tackle this problem: first, we use the
correlation of the two outputs $(\mu_\text{stat},\sigma_\text{stat})$
of the Bayesian classification network for \underline{control}. A
poorly trained network will not reproduce the quadratic correlation of
Eq.\eqref{eq:sigmoid_width_approx} and reveal its fundamental
ignorance. Once we have convinced ourselves that the network behaves
as expected, we can use the predictive jet-wise
\underline{uncertainty} as an input to the actual analysis.

\subsubsection{Capsules}
\label{sec:class_caps}

As we have seen, CNNs are great tools to analyze jet images at the
LHC, even though jet images do not look at all like the kind of images
these networks were developed for. However, at the LHC we are only
interested in jets as an, admittedly extremely interesting, part of a
collision event. This leads to the general question is how we can
combine information at the event level with the subjet information
encoded in each jet. For images this means we move from already sparse
jet images to extremely sparse events. For this task, a natural
extension of CNNs are capsule networks or CapsNets.  They allow us to
analyze structures of objects and their geometric layout
simultaneously. At the LHC we would like them to combine subjet
information with the event-level kinematics of jets and other
particles.

The idea of capsules as a generalization of CNNs follows from the
observation that CNNs rely on a 1-dimensional scalar representation of
images.  The idea behind capsules is to represent the entries of the
feature maps as vectors in signal or background feature space,
depending on which a given capsule describes.  Only the absolute value
of the capsule encodes the signal vs background classification.  The
direction of the vectors can track the actual geometric position and
orientation of objects, which is useful for images containing multiple
different objects.  In particle physics, an event image is a perfect
example of such a problem, so let us see what we can do when we
replaces this single number with vectors extracted from the feature
maps.


\begin{figure}[b!]
\centering
\includegraphics[width=0.5\textwidth]{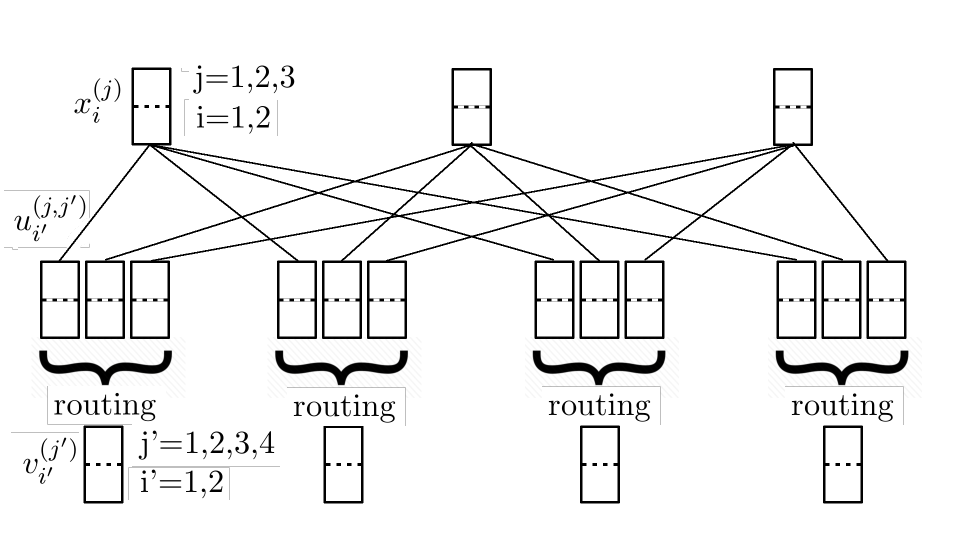}
\caption{Sketch of a CapsNet module with two simple capsule
  layers. Figure from Ref.~\cite{Diefenbacher:2019ezd}.}
\label{fig:Capsnet}
\end{figure}

Just like a scalar CNN, a CapsNet starts with a pixelized image, for
instance the calorimeter image of a complete event with $180 \times
180$ pixels. It is analyzed with a convolutional filter, combined with
pooling or stride convolutions to reduce the size of the feature
maps. The CapsNet's convolutional layers are identical to a scalar
CNN.  The new idea is to transform the feature maps after the
convolution into \underline{pixel-wise vectors}.  Each layer then
consists of a number of capsule vectors, for example 24 feature maps
with $40 \times 40$ entries each can be represented as 1600
capsule-vectors of dimension 24, or 3200 capsules of dimension 12, or
4800 capsules of dimension 8, etc.

The capsules have to transfer information matching their vector
property.  In Fig.~\ref{fig:Capsnet} we illustrate a small, 2-layer
CapsNet with three initial 2-dimensional capsules $\vec{x}^{(j)}$
linked through \underline{routing by agreement} to four 2-dimensional
capsules $\vec{v}^{(j')}$.  For deeper networks the dimensionality of
the resulting capsule vector can, and should, be larger than the
incoming capsule vector. We can write the complete matrix
transformation from the input $x$ to the output $v$ as
\begin{align}
  v_{i'}^{(j')}
  = \sum_{j=1,2,3} \sum_{i=1,2} \left( C \; W\right)_{i' i}^{(j' j)} x_i^{(j)} \; .
\end{align}
The connecting matrix has two sets of indices and size,$2\times 2$ in
$i$ and $3 \times 4$ in $j$. We can reduce the number of parameters by
factorizing the two steps.  To get from three to four capsules we
first define four combinations of the three initial capsules with the
entries $u^{(j,j')}_{i'}$, related to the initial capsule vectors
$\vec{x}^{(j)}$ through trainable weight matrices,
\begin{align}
u^{(j,j')}_{i'} = \sum_{i=1,2} W^{(j,j')}_{i' i} \; x^{(j)}_i 
\qqquad \text{for $j=1,2,3$ and $j'=1,2,3,4$} \; .
\label{eq:weight_u}
\end{align}
Next, we contract the original index $j$ to define the four outgoing
capsules using another set of trainable weights,
\begin{align}
  v_{i'}^{(j')} = \sum_{j=1,2,3} C^{(j',j)} \; u^{(j,j')}_{i'}
  \qquad \text{with} \qquad
    \sum_{j'=1,2,3,4} C^{(j,j')} &= 1 \quad \forall j \; .
\label{eq:output_capsule}
\end{align}
The normalization ensures that the contributions from one capsule in
the former to each capsule in the current layer add up to one.
Furthermore, a squashing step after each capsule layer ensures
that the length of every capsule vector remains between 0 and 1,
\begin{align}
\vec v \to \vec v' &= \frac{|\vec v|}{\sqrt{1+ |\vec v|^2}} \; \hat{v} \; ,
\label{eq:squash2}
\end{align}
with $\hat{v}$ defined as the unit vector in $\vec v$-direction.

\begin{figure}[b!]	
\centering
\includegraphics[width=0.2\textwidth]{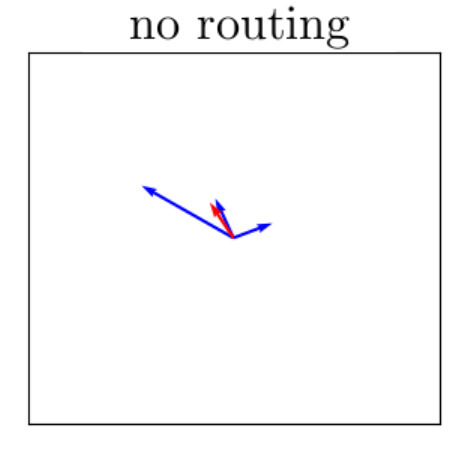}
\includegraphics[width=0.2\textwidth]{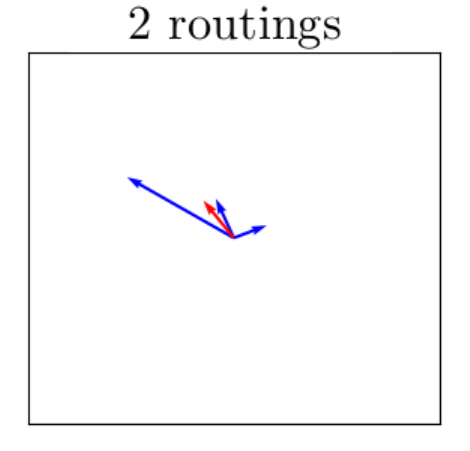}
\includegraphics[width=0.2\textwidth]{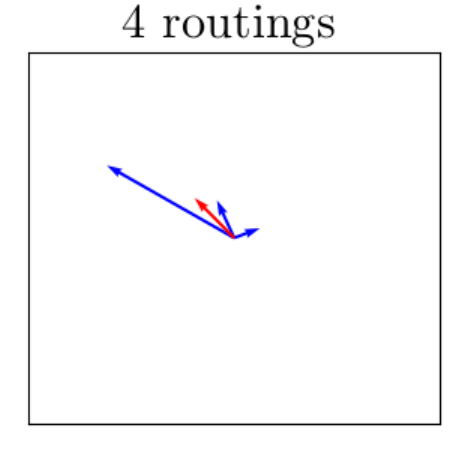}
\includegraphics[width=0.2\textwidth]{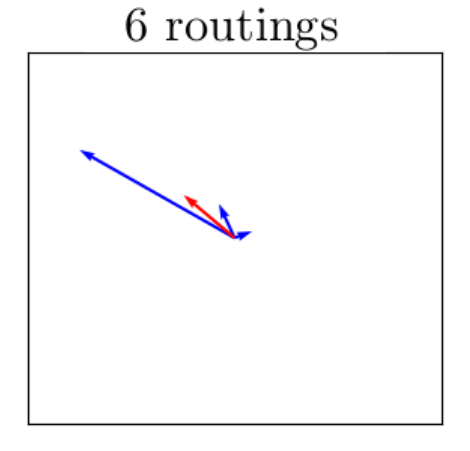}
\caption{Effects of the routing/squashing combination. In blue we show
  the intermediate vectors, in red we show the output vector after
  squashing. Figure from Ref.~\cite{Diefenbacher:2019ezd}.}
\label{fig:routing_illus}
\end{figure}

Up to now we have constructed a set of four capsules from a set of
three capsules through a number of trainable weights, but not enforced
any kind of connection between the two sets of capsule vectors.  We
can extend the transformation in $j$-space,
Eq.\eqref{eq:output_capsule}, to consecutively align the vectors
$\vec{u}^{(j,j')}$ and $\vec{v}^{(j')}$ through a re-definition of the
weights $C^{(j,j')}$. This means we compute the scalar product between
the vector $\vec{u}^{(j,j')}$ and the squashed vector $\vec{v}^{(j')}$
and replace in Eq.\eqref{eq:output_capsule}
\begin{align}
C^{(j,j')} \longrightarrow C^{(j,j')} + \vec{u}^{(j,j')} \cdot \vec{v}^{(j')} \; .
\end{align}
We can iterate this additional condition and construct a series of
vectors $v^{(j')}$, which has converged once $\vec{u}^{(j,j')}$ and
$\vec{v}^{(j')}$ are parallel. It is called \underline{routing by
  agreement} and is illustrated in Fig.~\ref{fig:routing_illus}, where
the blue vectors represent the three $\vec{u}^{(j,j')}$ in each set
and the red vector is the output $\vec{v}^{(j')}$.  With each routing
iteration the vectors parallel to $\vec{v}^{(j')}$ become longer while
the others get shorter.

Unlike the CNN, the CapsNet can now encode information in the length
and the direction of the output vectors.  We typically train the
network such that the length of the output vectors provide the
classification. Just like for the scalar CNN we differentiate between
signal and background images using two output capsules. The more
likely the image is to be signal or background, the longer the
corresponding capsule vector will be.  For simple classification the
capsule-specific part of the loss function consists of a 2-terms
\underline{margin loss}
\begin{align}
  \boxed{
  \loss_\text{CapsNet} = \max \left( 0,m_+ - | \vec{v}^{(1)} | \right)^2
  + \lambda \max \left( 0, | \vec{v}^{(2)} |- m_- \right)^2 } \; .
\end{align}
The first term vanishes if the length of the signal vectors $\vec
v^{(1)}$ exceeds $m_+$. The second term vanishes for background
vectors $\vec v^{(2)}$ shorter than $m_-$.  Typical target numbers
$m_+ = 0.9$ and $m_- = 0.1$ can sum up to one, nothing forces the
actual length of all capsules in a prediction to do the same.


If we want to use CapsNets to analyze full events including subjet
information, we can apply them to extract the signal process
\begin{align}
pp \to Z' \to t \bar{t} \to (b jj) \; (\bar{b} jj)
\label{eq:signal}
\end{align}
with $m_{Z'}= 1$~TeV from \underline{two backgrounds}
\begin{align}
  pp \to t\bar{t}
  \qqquad \text{and} \qqquad 
  pp \to jj \; ,
\end{align}
all with $p_{T,j} > 350$~GeV and $|\eta_j| < 2.0$. As usual, we
transform the calorimeter hits into a 2-dimensional image, now
including with $180 \times 180$ pixels to covering the entire detector
with $|\eta| < 2.5$ and $\phi = 0~...~2\pi$.

If we want to extract a signal from two backgrounds we can train the
classifier network either on two or on three classes. It often depends
on the details of the network architecture and performance what the
best choice is. In the case of capsules it turns out that the QCD
background rejection benefits from the 3-class setup, because a first
capsule can focus on separating the signal from the $t\bar{t}$
continuum background while a dedicated QCD capsule extracts the subjet
features. On the other hand, we need to remember is that training
multi-class networks require more data to learn all relevant features
reliably.

\begin{figure}[t]	
\centering
\includegraphics[width=0.40\textwidth]{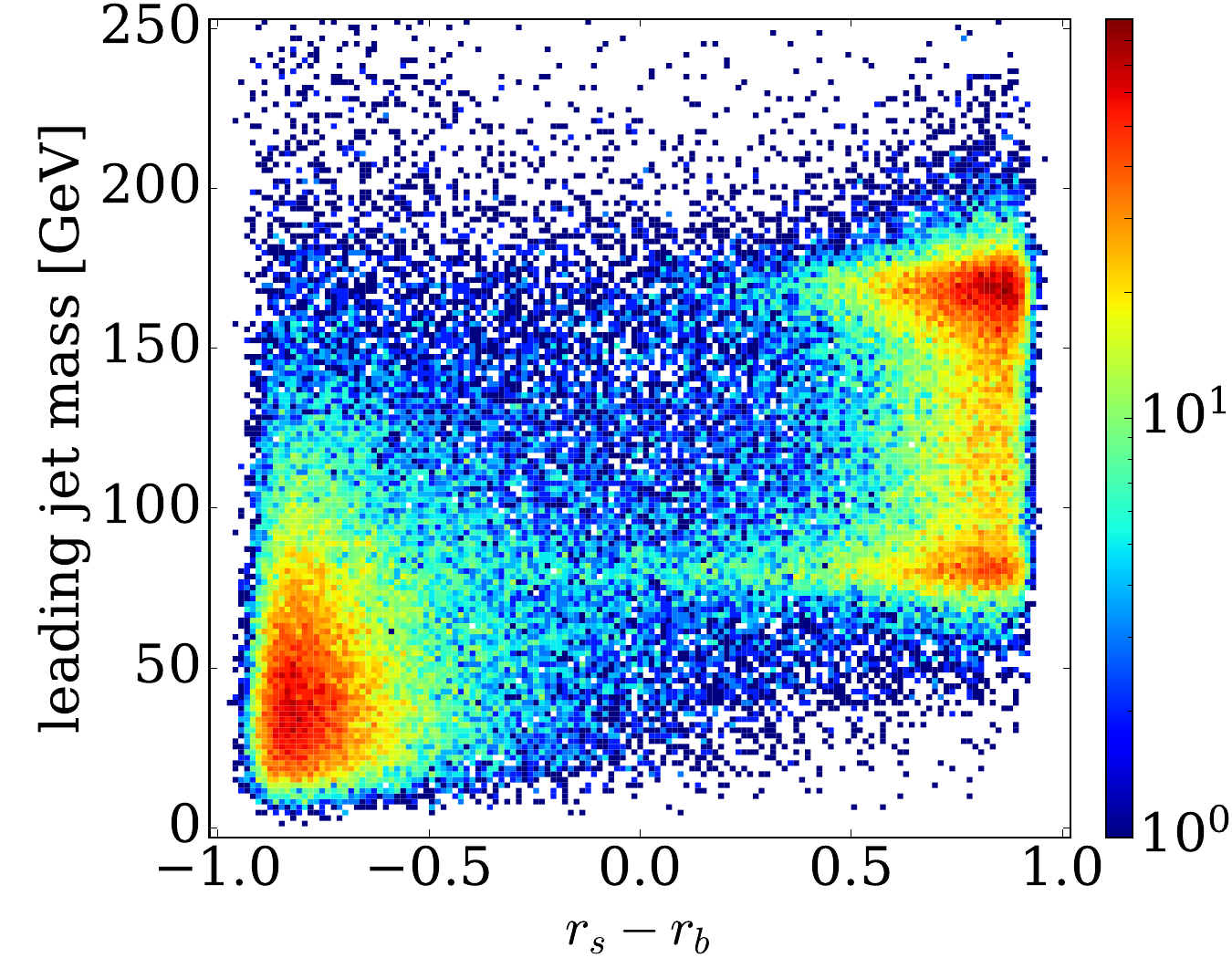}
\hspace*{0.01\textwidth}
\includegraphics[width=0.40\textwidth]{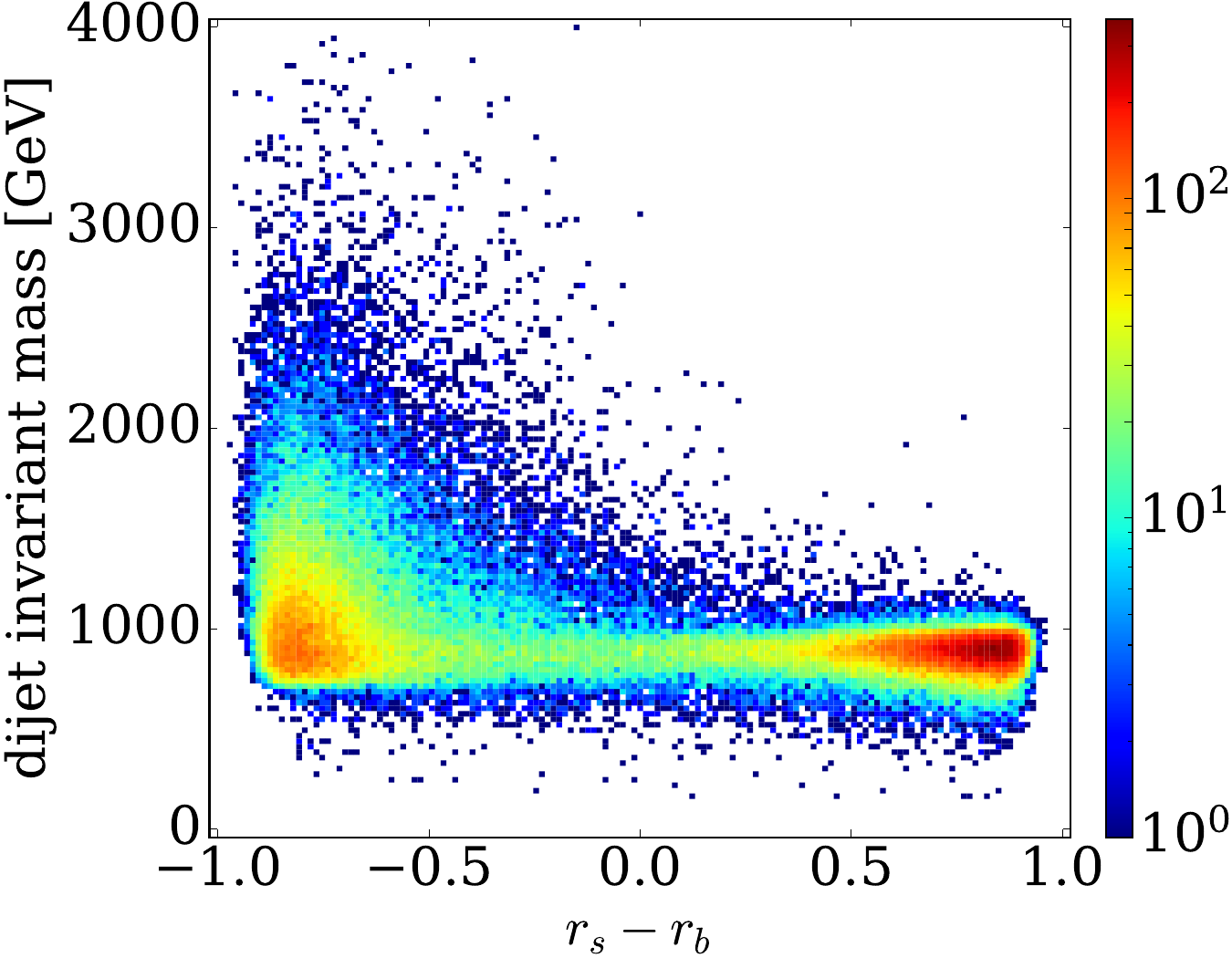} 
\caption{Correlation between capsule outputs $r_S-r_B$ and the leading
  jet mass, the di-jet $m_{jj}$, and $\Delta \eta_{jj}$ for true
  signal and background events.  Finally we show the correlation of
  the signal $\varphi$ vs the the mean $\eta_j$ for true signal
  events.  Figure from Ref.~\cite{Diefenbacher:2019ezd}.}
\label{fig:correlations}
\end{figure}

As one advantage of capsules we will see that they can combine
\underline{jet tagging with event kinematics}\index{jet tagging}, a problem for regular
CNNs.  To simplify our study we train the CapsNet to only separate the
$Z' (\to t\bar{t})$ signal from QCD di-jet events, so the signal and
background differs in event-level kinematics and in jet substructure.
We then define a signal-background discriminator
\begin{align}
  r_S - r_B \equiv |\vec{v}^{(S)}|-|\vec{v}^{(B)}|
  =
  \begin{cases}
    +1 & \text{signal events} \\
    -1 & \text{background events} \;    
  \end{cases}
\end{align}
Confronting this value with two key observables in
Fig.~\ref{fig:correlations}, we first confirm that the network
associates a large jet mass with the top signal, where the secondary
peak in the leading jet mass arises from cases where the jet image\index{jet images}
only includes two of the three top decay jets and learns either $m_W$
or the leading $m_{jb} \approx m_W$. We also see that the capsules
learn to identify the peak in the di-jet invariant mass at
approximately 1~TeV as a signal feature, different from the falling
spectrum for background-like events.

\begin{figure}[t]	
\centering
\includegraphics[width=0.65\textwidth]{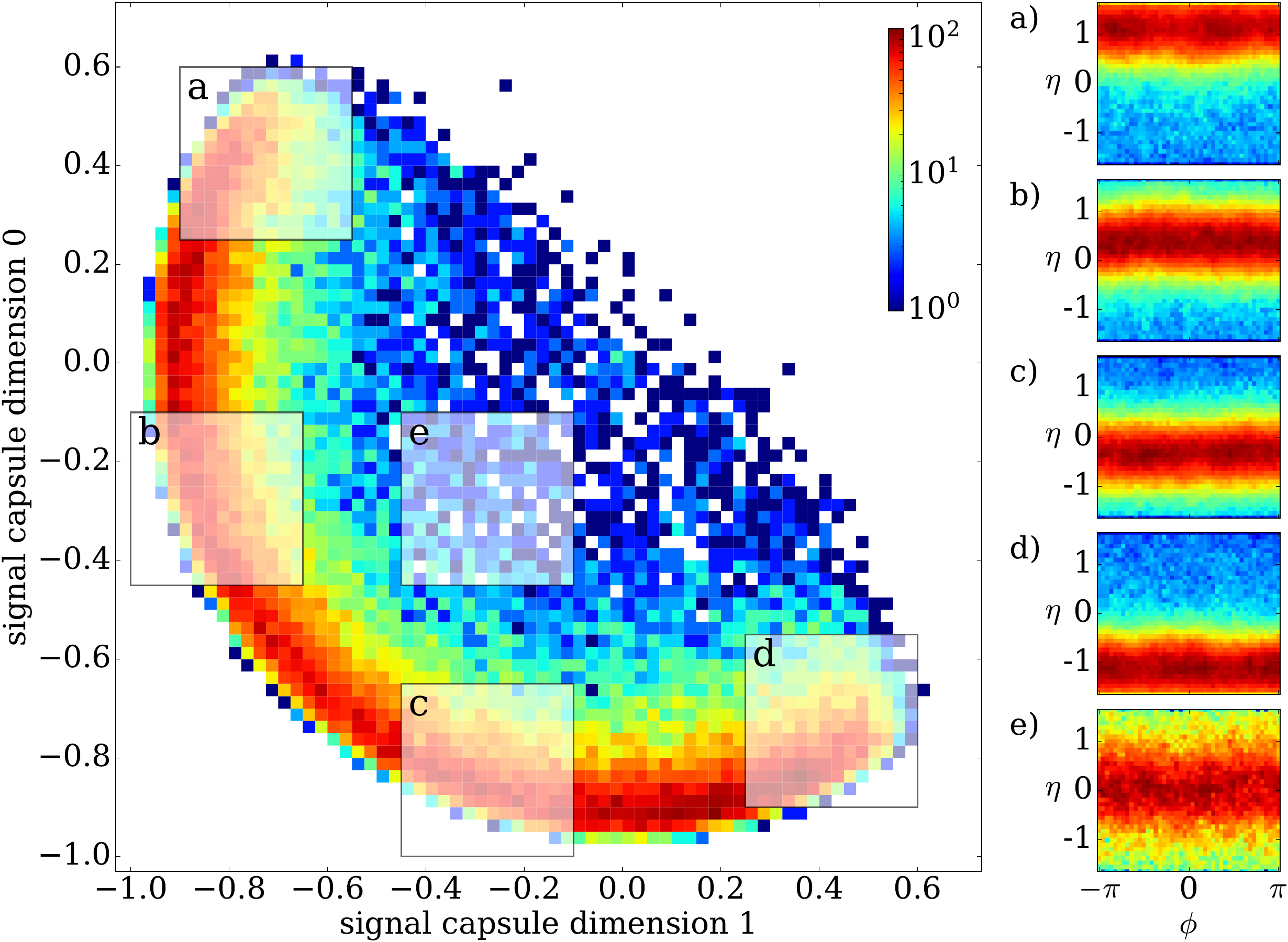}
\caption{Distribution of the two entries in the 2-dimensional signal
  capsule for signal events. Right: average event images in the
  $\eta-\phi$ plane. Figure from Ref.~\cite{Diefenbacher:2019ezd}.}
\label{fig:SigCaps}
\end{figure}

The second advantage of capsules is that they organize features in
their vector structure, so we can understand what the CapsNet has
learned. Only a certain combination of the vector entries is required
to separate signal and background, the rest of them is free to learn a
convenient \underline{representation space}.  This representation can
cover patterns which affect the classification output, or not.  Again,
the two output capsules correspond to the $Z'$ signal and the
light-flavor QCD background, with two dimensions each, making it easy
to visualize the output capsules. The classification output is then
mapped back into the image space for visualization purposes.

In Fig.~\ref{fig:SigCaps} we show the density of the two output
entries in the 2-dimensional signal capsule for true signal
events. Each event corresponds to a point in the 2-dimensional
plane. Because the classification output is proportional to the length
of the capsule vector, it corresponds to the distance of each point
from the origin.  Correctly identified signal events sit on the
boundary of the circle segment.  The rotation of the circle segment is
not fixed a priori, and nothing forces the network to fill the full
circle In this 2-dimensional capsule plane we select five
representative regions indicated by semi-transparent squares. For each
region we identify the contributing events and super-impose their
detector images in the $\eta - \phi$ plane in the right panels of
Fig.~\ref{fig:SigCaps}. For our signal events we observe bands in
rapidity, smeared out in the azimuthal angle. This indicates that the
network learns an event-level correlation in the two $\eta_j$ as an
identifying feature of the signal.

As we can see, the CapsNet architecture is a logical and interesting
extension of the CNN, especially when we are interested in combining
event-level and jet-level information and in understanding the
classification outcome. Capsules define a representation or latent
space, which we will study systematically in Sec.~\ref{sec:gen_rl}. All of
this means that CapsNets are extremely interesting conceptually. The
problem in particle physics applications is still, that extremely
sparse jet or calorimeter images are not the ideal representation, and
in Fig.~\ref{fig:toptagging} we have seen that other architectures are
more promising. So we will leave CapsNets behind, but keep in mind
their structural advantages.

\subsection{Representing point clouds}
\label{sec:class_graph}

In the last section we have argued the case to move from jet images to
full event information. Event images work for this purpose, but their
increasingly sparse structure cuts into their original motivation. The
actual data format behind LHC jets and events are not images, but a
set of 4-vectors with additional information on the particle
content. These 4-vectors include energy measurements from the
calorimeter and momentum measurements from the tracker. The difference
between the two is that calorimeters observe neutral and charges
particles, while tracking provides information on the charged
particles with extremely high angular resolution. Calorimeter and
tracking information are combined through dedicated
\underline{particle flow} algorithms, which is probably the better
option than combining sparse jet images of vastly different
resolution. Now, we could switch all the way from image recognition to
natural language recognition networks, but those do not reflect the
main symmetry\index{symmetries} of 4-vectors or other objects describing LHC collisions,
the permutation symmetry. Instead, we will see there are image-based
concepts which work extremely well with the LHC data format.

\subsubsection{4-Vectors}
\label{sec:class_graph_4vec}

The basic constituents entering any LHC analysis are a set of $C$
measured 4-vectors sorted by $p_T$, for example organized as the
matrix
\begin{align}
( k_{\mu,i} ) = 
\begin{pmatrix}
k_{0,1} &  k_{0,2} & \cdots \\
k_{1,1} &  k_{1,2} & \cdots \\ 
k_{2,1} &  k_{2,2} & \cdots \\ 
k_{3,1} &  k_{3,2} & \cdots   
\end{pmatrix} \; ,
\label{eq:def_input}
\end{align}
where for now we ignore additional information on the particle
identification.  Such a high-dimensional data representation in a
general and often unknown space is called a \underline{point cloud}.
If we assume that all constituents are approximately massless, a
typical jet image\index{jet images} would encode the relative phase space position to
the jet axis and the transverse momentum of the constituent,
\begin{align}
  k_{\mu,i} \quad \to \quad
  \begin{pmatrix}
    \Delta \eta_i \\
    \Delta \phi_i \\
    p_{T,i} 
  \end{pmatrix} \; .
\label{eq:vec_image}
\end{align}
For the generative networks we will introduce in Sec.~\ref{sec:gen} we
implement this transformation as preprocessing, but for the simpler
classification network it turns out that we can just work with the
4-vectors given in Eq.\eqref{eq:def_input}.

\begin{figure}[t]
  \centering
  \includegraphics[width=0.40\textwidth]{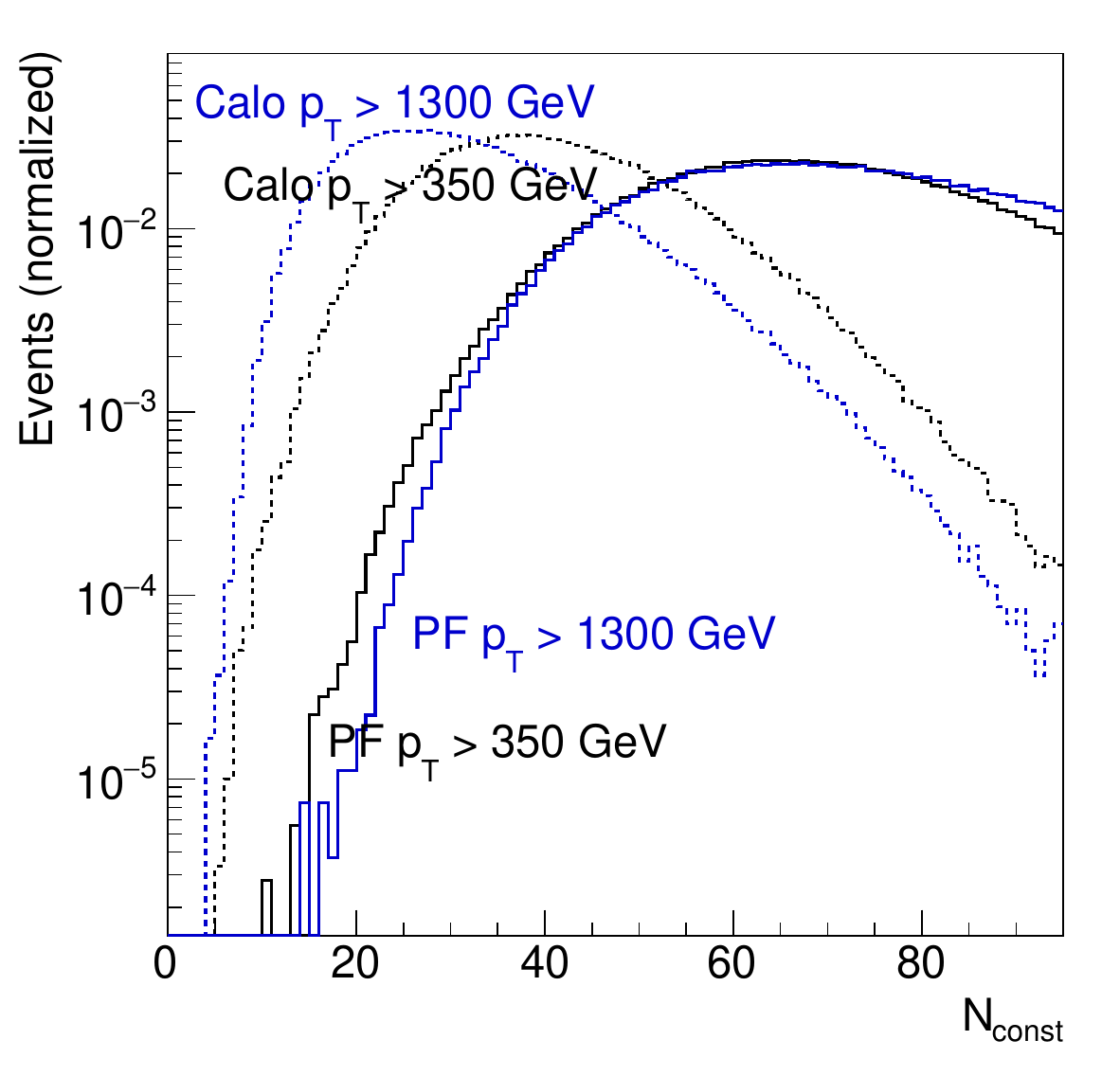}
  \hspace*{0.1\textwidth}
  \includegraphics[width=0.40\textwidth]{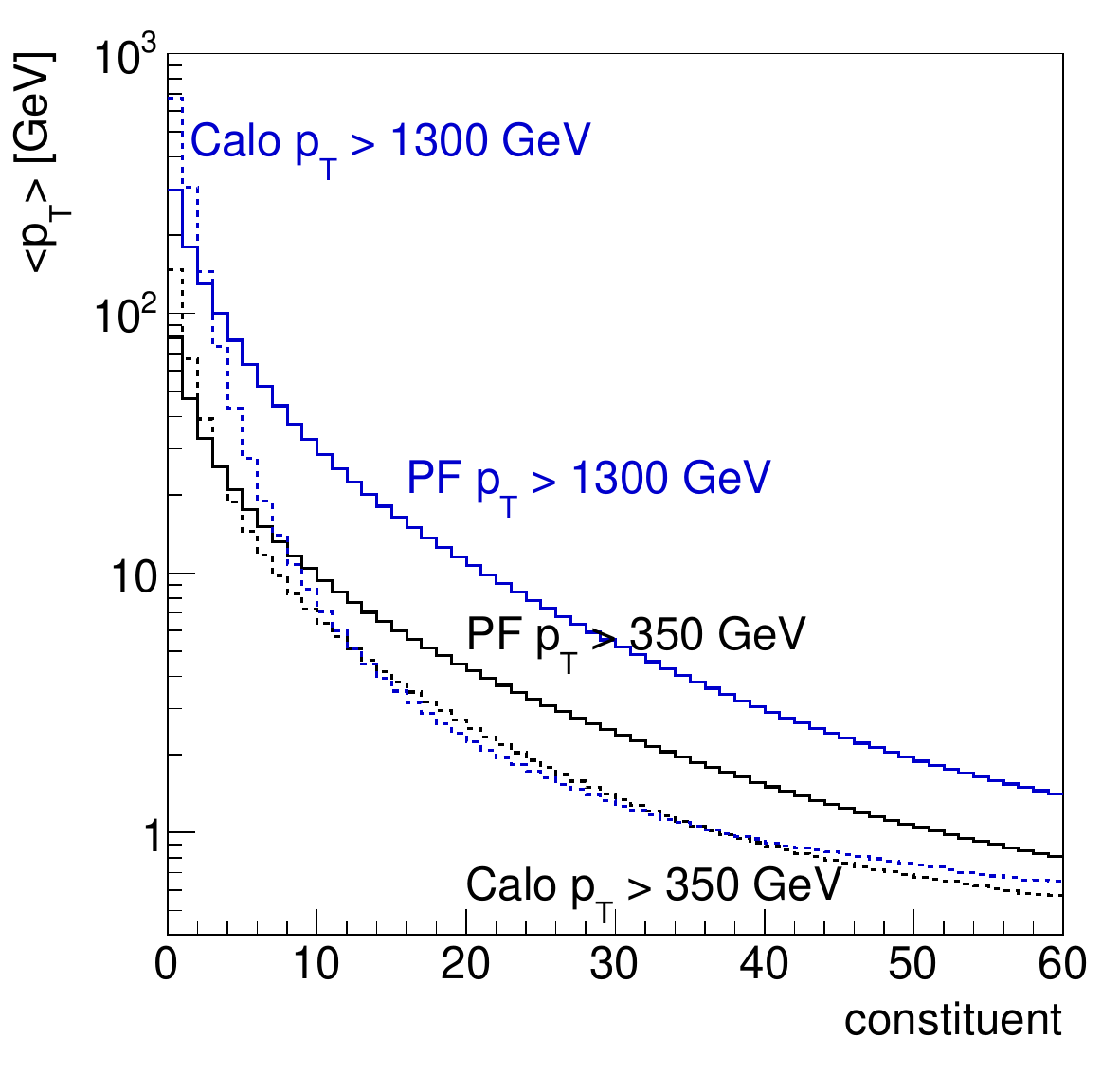}
  \caption{Number of top jet constituents (left) and mean of the
    transverse momentum (right) of the ranked constituent 4-vectors in
    Eq.\eqref{eq:def_input}. We show information from jet images
    (dashed) and from the combined through particle flow
    (solid). Figure from Ref.~\cite{Butter:2017cot}.}
\label{fig:nconst}
\end{figure}

To illustrate the difference between the image representation and the
4-vector representation we replace our standard top tagging dataset of
Eq.\eqref{eq:standard_ptcut} with two distinct samples, corresponding
to moderately boosted tops from Standard Model processes and highly
boosted tops from resonance searches,
\begin{align}
  p_{T,j} = 350~...~450~\gev
  \qquad \text{and} \qquad 
  p_{T,j} = 1300~...~1400~\gev \; .
\label{eq:lola_pt_range}
\end{align}
In the left panel of Fig.~\ref{fig:nconst} we show the number of
calorimeter-based and particle-flow 4-vectors $k_{\mu,i}$ available
for our analysis, $N_\text{const}$.  We see that including tracking
information roughly doubles the number of available 4-vectors and
reduces the degradation towards higher boost.  In the right panel we
show the mean transverse momentum of the $p_T$-ordered 4-vectors,
indicating that the momentum fraction carried by the charged
constituents is sizeable. The fact that the calorimeter-based
constituents do not get harder for higher boost indicates a serious
limitation from their resolution.  In practice, we know that the
hardest 40 constituents tend to saturate tagging performances, while
the remaining entries will typically be much softer than the top decay
products and hence carry little signal or background information from
the hard process. Comparing this number to the calorimeter-based on
particle flow distributions motivates us to go beyond calorimeter
images.

As a starting point, we introduce a simple constituent-based tagger
which incorporates some basic \underline{physics structure}.  First,
we mimic a jet algorithm\index{jet algorithm} and multiply the 4-vectors from
Eq.\eqref{eq:def_input} with a matrix $C_{ij}$ and return a set of combined
4-vectors $\tilde{k}_j$ as linear combinations of the input
4-vectors,
\begin{align}
k_{\mu,i} \stackrel{\text{CoLa}}{\longrightarrow}
\widetilde{k}_{\mu,j} 
= k_{\mu,i} \; C_{ij} \; .
\label{eq:def_4vectors}
\end{align}
The explicit form of the matrix $C$ defining this combination layer
(CoLa) ensures that the $\tilde{k}_j$ include each original momentum
$k_i$ as well as a trainable set of $M-C$ linear combinations.  These
$\tilde{k}_j$ could be analyzed by a standard, fully connected or dense network.

However, we already known that the relevant distance measure between
two substructure objects, or any two 4-vectors sufficiently far from
the closest black hole, is the \underline{Minkowski metric}.  This
motivates a Lorentz layer, which transforms the $\tilde{k}_j$ into the
same number of measurement-motivated invariants $\hat{k}_j$,
\begin{align}
\tilde{k}_j 
\stackrel{\text{LoLa}}{\longrightarrow}
\hat{k}_j = 
 \begin{pmatrix}
  m^2(\tilde k_j)\\ 
  p_T(\tilde k_j)\\ 
  \Box_m w^{(E)}_{jm} \,E(\tilde k_m)\\ 
  \Box_m w^{(d)}_{jm} \, d^2_{jm}\\ 
 \end{pmatrix} \; .
\label{eq:lola}
\end{align}
The first two $\hat{k}_j$ map individual $\tilde{k}_j$ onto their
invariant mass and transverse momentum, using the Minkowski distance
between two four-momenta,
\begin{align}
d^2_{jm} &= (\tilde k_j - \tilde k_m)_{\mu} \; g^{\mu \nu} \; (\tilde k_j - \tilde k_m)_{\nu} \; .
\label{eq:minkowski}
\end{align}
In case the invariant masses and transverse momenta are not sufficient
to optimize the classification network, the additional weights
$w_{jm}$ are trainable. The third entry in Eq.\eqref{eq:lola}
constructs a linear combination of all energies, evaluated with one of
several possible \underline{aggregation functions}
\begin{align}
  \Box \in \left\{ \text{max}, \text{sum}, \text{mean}, \cdots \right\} \; .
\label{eq:aggregation}
\end{align}
Similarly, the fourth entry combines all Minkowski distances of
$\tilde{k}_m$ with a fixed $\tilde{k}_j$. Again, we can sum over or
minimize over the internal index $m$ while keeping the external index
$j$ fixed.

A technical challenge related to the Minkowski metric for example in a
\underline{graph convolutional network} (GCN) language is that it
combines two different features: two subjets are Minkowski-close if
they are collinear or when one of them is soft ($k_{i,0} \to
0$). Because these two scenarios correspond to different, but possibly
overlapping phase space regions, they are hard to learn for the
network. To see how the network does and what kind of structures drive
the network output, we turn the problem around and ask the question if
the Minkowski metric is really the feature distinguishing top decays
and QCD jets. This means we define the invariant mass $m(\tilde{k}_j)$
and the distance $d_{jm}^2$ in Eq.\eqref{eq:lola} with a trainable
diagonal metric and find
\begin{align}
g = \text{diag} ( &\quad 0.99 \pm 0.02, 
\notag \\ 
&-1.01 \pm 0.01, -1.01 \pm 0.02, -0.99 \pm 0.02 ) \; , 
\label{eq:learned_minkowski}
\end{align}
where the uncertainties are given by five independently trained copies. This
means that for top tagging the appropriate space to relate the
4-vector data of Eq.\eqref{eq:def_4vectors} is defined by the
Minkowski metric. Obviously, this is not going to be true for all
analysis aspects. For instance, at the event level the rapidities or
the scattering angles include valuable information on decay products
from heavy resonance compared to the continuum background induced by
the form of the parton densities. Still, the LoLa tagger motivates the
question how we can combine a data representation as 4-vectors with an
appropriate metric for these objects.

In Fig.~\ref{fig:toptagging} we see that the CoLa-LoLa network does
not provide the leading performance. One might speculate that its
weakness is that it is a little over-constructed with too much physics
bias, with the positive side effect that the network only needs 127k
parameters.

The fact that we can extract the Minkowski metric as the relevant
metric for top-tagging based on 4-vectors leads us to the concept of
graphs. Here we assume that our point-cloud data populates
a space for which we can extract some kind of metric or geometry. The
optimal metric in this space depends on the task we are training the
network to solve. This becomes obvious when we extend the 4-vectors of
jet constituents or event objects with entries encoding details from
the tracker, like displaced vertices, or particle identification. The
question is how we can transform such a point cloud into a structure
which allows us to perform the kind of operations we have seen for the
CoLa-LoLa tagger, but in an abstract space.

\subsubsection{Graph convolutional network}
\label{sec:class_graph_arch}

The first problem with the input 4-vectors given in
Eq.\eqref{eq:def_input} is that we do not know which space
they live in.  The structure that generalizes our CoLa-LoLa approach
to an abstract space is, first of all, based on defining
\underline{nodes}, in our case the vectors describing a jet
constituent. These nodes have to be connected in some way, defining an
\underline{edge} between each pair of nodes. The LoLa ansatz in
Eq.\eqref{eq:lola} already assumes that these edges have to go beyond
the naive Minkowski metric. The object defined by a set of nodes and
their edges is called a \underline{graph}.

\begin{figure}[t]
  \centering
  \includegraphics[width=0.22\textwidth]{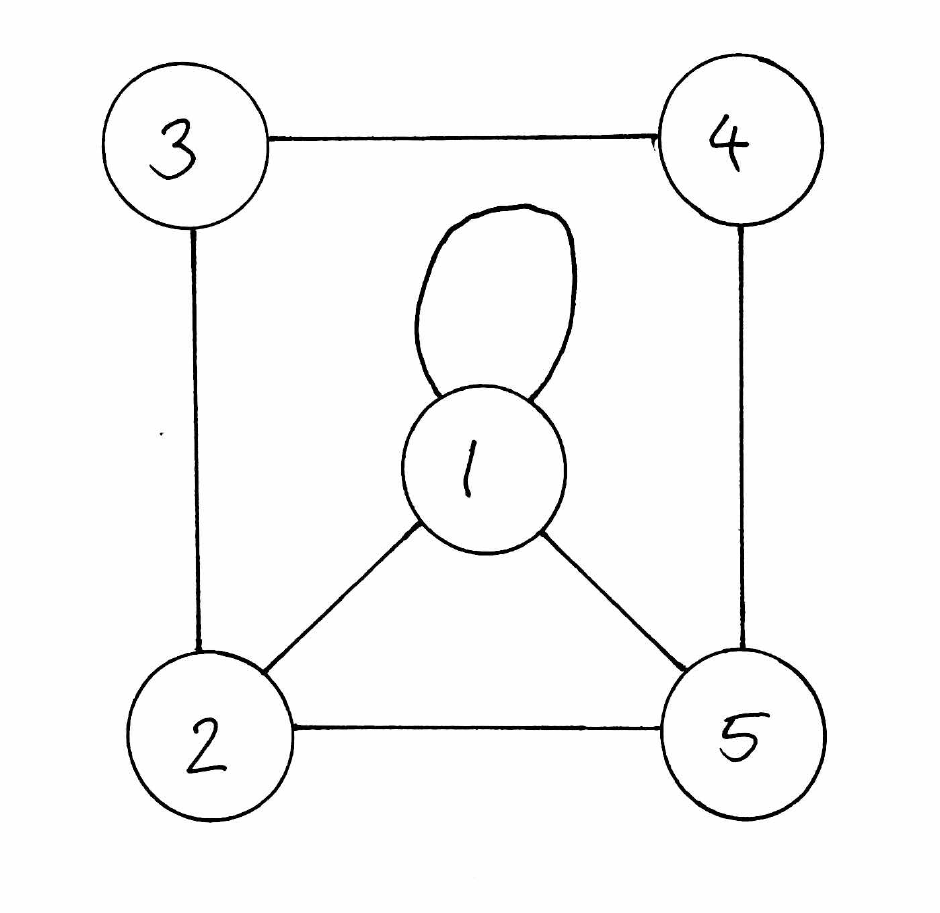}
  \hspace*{0.15\textwidth}
  \includegraphics[width=0.22\textwidth]{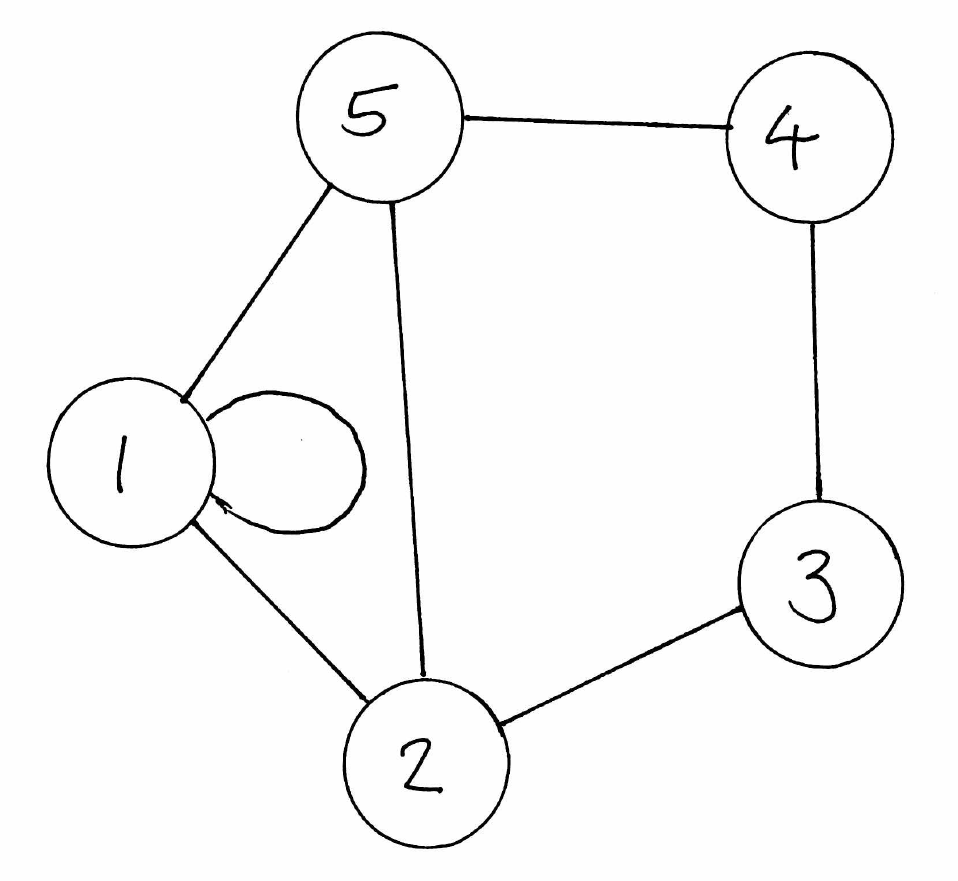}
  \caption{Example for a simple graph.}
  \label{fig:graph}
\end{figure}

The basic object for analyzing a graph with $C$ nodes is the $C \times
C$ adjacency matrix. It encodes the $C^2$ edges, where we allow for
self-interactions of nodes. In the simplest case where we are just
interested in the question if two nodes actually define a relevant
edge, the adjacency matrix includes zeros and ones. In the graph
language such an adjacency matrix defines an undirected --- edges do
not depend on their direction between two nodes --- and
unweighted. Even for this simple case this matrix can be useful. Let
us look at an example of $C=5$ nodes with the six edges
\begin{align}
A =   \begin{pmatrix}
    2 & 1 & 0 & 0 & 1 \\
    1 & 0 & 1 & 0 & 1 \\
    0 & 1 & 0 & 1 & 0 \\
    0 & 0 & 1 & 0 & 1 \\
    1 & 1 & 0 & 1 & 0 
  \end{pmatrix} \; .
\label{eq:adjacency}
\end{align}
This graph is illustrated in Fig.~\ref{fig:graph}, skipping the 6th
node in the image.  Because the graph is undirected, the adjacency
matrix is symmetric. Only the first node has a self-interaction, which
we count twice because it can be used in two directions. Now we can
compute powers of the adjacency matrix, like
\begin{align}
  A^2
  =
  \begin{pmatrix}
    6 & 3 & 1 & 1 & 3 \\
    3 & 3 & 0 & 2 & 1 \\
    1 & 0 & 2 & 0 & 2 \\
    1 & 2 & 0 & 2 & 0 \\
    3 & 1 & 2 & 0 & 3 
  \end{pmatrix} 
%
%
  \qquad \text{and} \qquad 
  A^3 =
  \begin{pmatrix}
    18 & 10 & 4 & 4 & 10 \\
    10 &  4 & 5 & 1 &  8 \\
     4 &  5 & 0 & 4 &  1 \\
     4 &  1 & 4 & 0 &  5 \\
    10 &  8 & 1 & 5 &  4 
  \end{pmatrix} \; .
\end{align}
The matrix $A^n$ encodes the number of different paths of length $n$
which we can take between the two nodes given by the matrix entry. For
instance, we can define four length-3 connections between the node and
and itself, two different loops each with two directions.  Once we
have defined a set of edges we can use the existence of an edge, or a
non-zero entry in $A$, to define neighboring nodes, and neighboring
nodes is what we need for operations like filter convolutions, the
basis of a \underline{graph-convolutional network}.

Once we have defined our set of nodes and their adjacency matrix, the
simplest way to use a filter is to go over all nodes and train a
universal filter for their respective neighbors.  We already know that
our nodes are not just a simple number, but a collection of different
features. In that case we can define each node with a feature vector
$x_i^{(k)}$ for the nodes $i=1~...~C$. In analogy to the feature maps of
Eq.\eqref{eq:def_conv} we can then define a filter $W_j^{(kl)}$ where
$j$ goes through the neighboring nodes of $i$ and matrix entries match the
size of the central and neighboring feature vectors.  This means
neighboring pixels of Eq.\eqref{eq:def_conv} become nodes with edges,
and feature maps turn into feature vectors.  A convolutional network
working on one node now returns
\begin{align}
  x'^{(k)}_i
  = \sum_{\text{features} \; l} \sum_{\text{neighbors}\; j} 
  W^{(kl)}_{ij} \; x_j^{(l)}
  \equiv \sum_{\text{features} \; l} \sum_{\text{nodes} \; j} W^{(kl)}_{ij} A_{ij} \; x_j^{(l)} \; ,
\label{eq:graph_conv}
\end{align}
where in the second form we have used the adjacency matrix to define
the neighbors.  When using such graph convolutions we can add a
normalization factor to this adjacency matrix. Obviously, we also need
to apply the usual non-linear activation unction, so the network will
for instance return $\relu (x'_i)$. Finally, for the definition of the
universal filter it will matter how we order the neighbors of each
node.

Instead of following the convolution explicitly, as in
Eq.\eqref{eq:graph_conv}, we can define a more general
transformation than in Eq.\eqref{eq:def_conv}, namely a vector-valued
function of two feature vectors $x_{i,j}$, and with the same number of
dimensions as the feature vector $x_i$,
\begin{align}
  x'_i
  = \sum_{\text{neighbors}\; j}  W_\theta( x_i,x_j)
  \label{eq:edge_conv1}
\end{align}
The sum runs over the neighboring nodes, and we omit the explicit sum
in feature space. If we consider the strict form of
Eq.\eqref{eq:graph_conv} a convolutional prescription, and its
extension to $W^{(kl)}_{ij} \to W^{(kl)}_{ij}(x_i,x_j)$ as an
attention-inspired generalization, the form in
Eq.\eqref{eq:edge_conv1} is often referred as the most general
\underline{message passing}.  Looking at this prescription and
comparing it for instance with Eq.\eqref{eq:lola}, it is not clear why
we should sum over the neighboring nodes, so we can define more
generally
\begin{align}
  \boxed{
  x'_i
  = \Box_j \; W_\theta(x_i,x_j)
  } \; ,
  \label{eq:graph_conv2}
\end{align}
with the \underline{aggregation functions} $\Box_j$ defined in
Eq.\eqref{eq:aggregation}. The corresponding convolutional layer is
called an \underline{edge convolution}. Just like a convolutional
filter, the function $W$ is independent of the node position $i$,
which means it will scale as economically as the CNN. The form of $W$
allows us to implement symmetries\index{symmetries} like
\begin{align}
W_\theta(x_i,x_j) &= W_\theta(x_i-x_j)     &\qquad & \text{translation symmetry} \notag \\
W_\theta(x_i,x_j) &= W_\theta(|x_i-x_j|)   &\qquad & \text{rotation symmetry} \notag \\
W_\theta(x_i,x_j) &= W_\theta(x_i,x_i-x_j) &\qquad & \text{translation symmetry conditional on center.}
\label{eq:gcn_filters}
\end{align}
The most important symmetry for applications in particle physics is
the \underline{permutation symmetry} of the constituents in a jet or
the jets in an event. The edge convolution is symmetric as long as
$W_\theta$ does not spoil such a symmetry and the aggregation function
is chosen like in Eq.\eqref{eq:graph_conv2}.

\begin{figure}[t]
  \centering
  \includegraphics[width=0.27\textwidth]{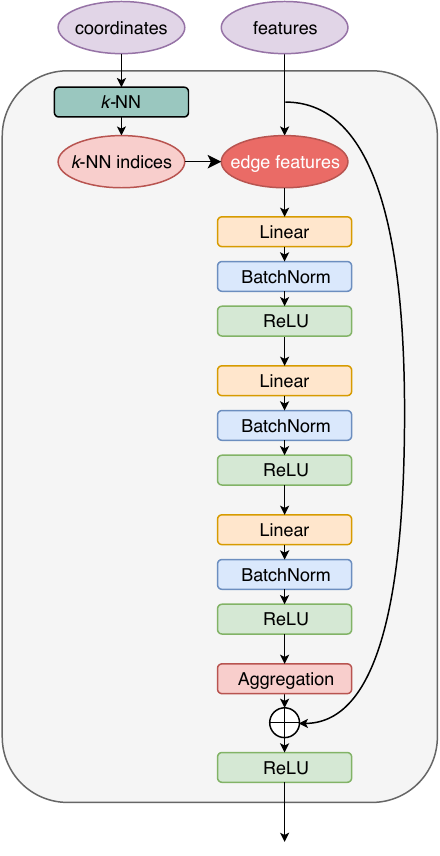}
  \hspace*{0.2\textwidth}
  \includegraphics[width=0.20\textwidth]{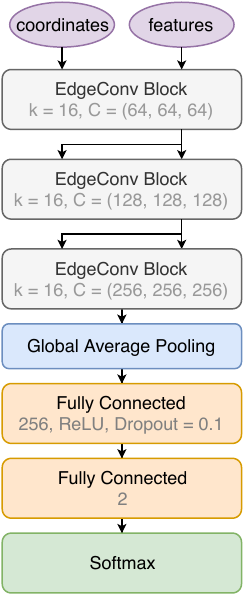}
\caption{Architectures of an edge convolution block (left) and the
  ParticleNet implementation for jet tagging. In the right panel, $k$
  is the number of nearest neighbors considered and $C$ the number of
  channels per edge convolution layer. Figure from
  Ref.~\cite{Qu:2019gqs}.}
\label{fig:particle_net}
\end{figure}

Going back to the point clouds describing LHC jets or events, we first
need to transform the set of possibly extended 4-vectors into a graph
with nodes and edges. Obviously, each extended 4-vector of an input
jet can become a node described by a feature vector. Edges are defined
in terms of an adjacency matrix. We already know that we will modify
the adjacency matrix as the initial form of the edges through edge
convolutions, so we can just define a reasonably first set of
neighbors for each node. We can do that using standard
nearest-neighbor algorithms, defining the input to the first
edge-convolution layer. When stacking edge convolutions, each layer
produces a new set of feature vectors or nodes. This leads to a
re-definition of the adjacency matrix after each edge convolution,
removing the dependence on the ad-hoc first choice. This architecture
is referred to as a Dynamic Graph Convolutional Neural Network
(DGCNN). The structure of the edge convolution is shown in the left
panel of Fig.~\ref{fig:particle_net}. It combines the nearest-neighbor
definition for the graph input with a series of linear edge
convolutions and a skip connection as introduced in
Eq.\eqref{eq:skip_connection}.

One reason to introduce the dynamic GCN is that it provides the best
jet tagging results in Fig.~\ref{fig:toptagging}. The ParticleNet
tagger is based on the third ansatz for the filter function in
Eq.\eqref{eq:gcn_filters} with the simple linear combination
\begin{align}
  \Box_j \; W_\theta(x_i,x_j)
  = \text{mean}_j \Big[ \theta_\text{diff} \cdot (x_i - x_j)  
  + \theta_\text{local} \cdot x_i \Big]
\end{align}
This is the form of the linear convolution referred to in the left
panel of Fig.~\ref{fig:particle_net}. Batch normalization is a way to
improve the training of deep networks in practice. It evaluates the
inputs to a given network layer for a minibatch defined in
Eq.\eqref{eq:derivative_loss} and changes their normalization to mean
zero and standard deviation one. It is known to improve the network
training, even though there seems to be no good physics or other
reason for that improvement.

In the right panel of Fig.~\ref{fig:particle_net} we show the
architecture of the network. After the edge convolutions we need to
collect all information in a single vector, which in this case is
constructed by average-pooling over all nodes for each of the
channels. The softmax activation function of the last layer is the
multi-dimensional version of the sigmoid defined in
Eq.\eqref{eq:def_sigmoid}, required for a classification network. It
defines a vector of the same size as its input, but such that all
entries are positive and sum to one,
\begin{align}
  \softmax_i (x) = \frac{e^{x_i}}{\sum_j e^{x_j}} \; .
  \label{eq:def_softmax}
\end{align}
If we only look at our standard classification setup with two network
outputs, where the first gives a signal probability, the softmax
function becomes a scalar sigmoid
\begin{align}
  \softmax_1 (x)
  = \frac{e^{x_1}}{e^{x_1} + e^{x_2}}
  = \frac{1}{1 + e^{x_2- x_1}}
  \equiv \sigmoid ( x_2-x_1 ) \; ,
\end{align}
for the difference of the inputs, as defined in
Eq.\eqref{eq:def_sigmoid}. Just as we can use the cross entropy
combined with a sigmoid layer to train a network to give us the
probability of a binary classification, we can use the multi-class
cross entropy combined with the softmax function to train a network
for a multi-label classification. Maybe I will show this in an updated
version of these notes, but not in the first run.

The fact that for top tagging the GCN outperforms all its competitors
indicates that graphs with their permutation invariance are a better
representation of jet than images. The number of network parameters is
500k, similar in size to the CNN used for the top-tagging challenge,
but leading to the best tagging performance of all architectures.

\subsubsection{Transformer}
\label{sec:class_graph_trans}

Now that the graph network can construct an appropriate task-dependent
space for neighboring nodes, we can tackle the second problem with the
input 4-vectors given in Eq.\eqref{eq:def_input}, namely that we do
not know how to order permutation-invariant nodes. We can use graphs
to solve this problem by removing the adjacency matrix and instead
relating all nodes to all nodes, constructing a fully connected graph.
A modern alternative, which ensures permutation invariance and can be
applied to point clouds, are transformers. Their origin is language
analysis, where they provide a structure to analyze how words compose
sentences in different languages, where the notion of neighboring
works does not really mean anything. Their defining building block is
attention, or more specifically,
\underline{self-attention}\index{self-attention}.  It is the natural
extension of an adjacency matrix, as shown in Eq.\eqref{eq:adjacency},
where a zero means that no information from that node will enter the
graph analysis. Self-attention allows an element to assign learned
weights to other elements, or, mathematically, a square matrix with an
appropriate normalization. These weights define how much `attention'
is placed on those elements whenever our ML-task requires us to define
a relation between them.

We motivate the construction of self-attention using a toy model in
representation space.  The goal of our construction is to describe and
then learn some kind of relation between two entries of an input
vector or sequence $x$, and then construct a representation of $x_i$
using these relations.

First, we define an input entry $x_i$ and represent it by a normalized
\underline{query} entry
\begin{align}
  x_i \; \longrightarrow \; 
  q = \frac{x_i}{|x|} \; .
\label{eq:toy_transformer_wq}
\end{align}
It represents the input $x_i$ as a unit vector in a compact latent
space.  Next, we need a set of vectors for which we analyse the
relations to $q$, the so-called \underline{value} vectors $v$.  If
they form an orthonormal basis, we can write our input vector as
\begin{align}
  q = \sum_j (q \cdot v_j) \; v_j
  \equiv \sum_j a_j v_j
  \qquad \text{with} \qquad 
  a_j = (q \cdot v_j) \; .
  \label{eq:toy_transformer_a}
\end{align}
The scalar product $a_j$ represents the strength of the connection
between the query vector and a given value vector $v_j$. However, when
implementing this method as a trained network we should not require a
normalized basis. Instead, we replace the value basis by another basis
of so-called \underline{key} vectors,
\begin{align}
  q = \sum_j (q \cdot k_j) \; v_j
  \qquad \text{or} \qquad 
  a_j = (q \cdot k_j) \; .
  \label{eq:toy_transformer_ax}
\end{align}
As an example, the keys for an orthogonal, but not normalized basis
are
\begin{align}
  k_j = \frac{v_j}{v^2} \; .
\end{align}
Using our freedom in choosing the keys, we then transform the input
$x_i$ into a representation $z_i$, defined as as
\begin{align}
  \boxed{
  x_i \; \longrightarrow  \;
  z_i = \sum_j (q \cdot k_j) \; v_j } \; .
  \label{eq:toy_transformer_z}
\end{align}
This representation takes into account the relation to all other
entries of our input vector or sequence. Because of the sum over the
value basis, it is explicitly permutation invariant. The
transformer-encoder is only trained together with the respective
network, but we can assume that the network training will construct an
orthogonal basis of key and value vectors to make optimal use of the
information encoded in the data.

\begin{figure}[t]
  \centering
  \includegraphics[width=0.55\textwidth]{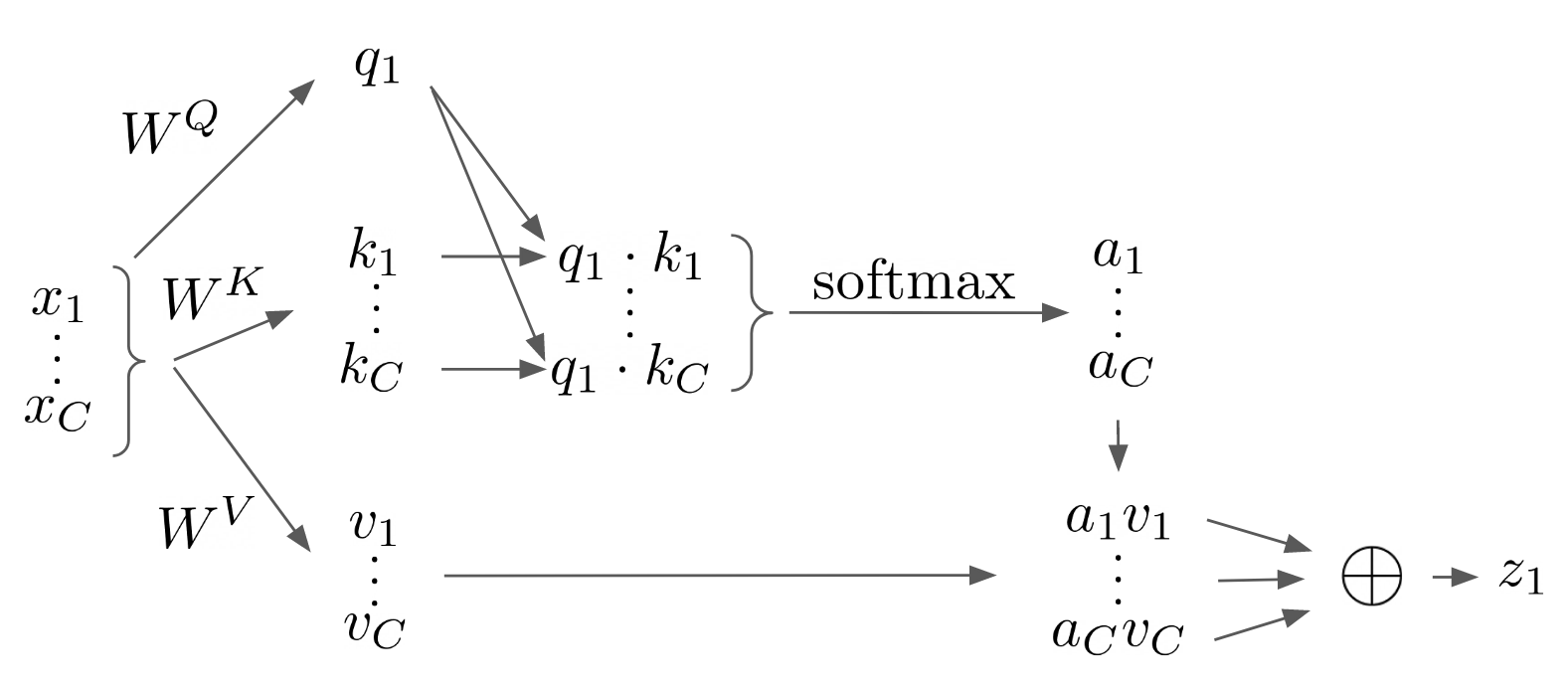}
  \caption{Illustration of single-headed self-attention. Figure from
    Ref.~\cite{Dillon:2021gag}.}
  \label{fig:attn}
\end{figure}

Let us move on to the proper single-headed
self-attention illustrated in
Fig.~\ref{fig:attn}. Let us assume that we are again analyzing $C$ jet
constituents, so the vector $x_1$ describes the phase space position
for the first constituent in an unordered list. If we stick to an
image-like representation, the complete vector $x$ will be $C$ copies
of the 3-dimensional phase space vector given in
Eq.\eqref{eq:vec_image}. First, we generalize
Eq.\eqref{eq:toy_transformer_wq} and define a proper query
representation of $x_1$ in a general latent space through a learned
\underline{weight matrix} $W^Q$,
\begin{align}
  q_1 = W^Q x_1 \; .
\label{eq:def_wq}
\end{align}
If we cover all constituents, $W^Q$ becomes a block-diagonal matrix of
size $3C$.  To relate all constituents $x_1~...~x_C$ to $x_1$ we
define a learned \underline{key matrix}
\begin{align}
  \begin{pmatrix} k_1 \\ \vdots \\ k_C \end{pmatrix}
  = W^K 
  \begin{pmatrix} x_1 \\ \vdots \\ x_C \end{pmatrix}
  \label{eq:transformer_wk}
\end{align}
We use a scalar product to project all keys $k_1~...~k_C$ onto
$q_1$. Modulo some details this defines a set of 3-dimensional vectors
in analogy to Eq.\eqref{eq:toy_transformer_a},
\begin{align}
  \begin{pmatrix} a_1^{(1)} \\ \vdots \\ a_C^{(1)} \end{pmatrix}
  = \softmax  \begin{pmatrix} (q_1 \cdot k_1) \\ \vdots \\ (q_1 \cdot k_C) \end{pmatrix}
  \qquad \text{with} \quad a_j^{(1)} \in [0,1] \qquad \sum_j a_j^{(1)} = 1 \; ,
\label{eq:transformer_a}
\end{align}
all in reference to $x_1$. It gives us a quadratic
attention matrix, similar to the adjacency matrix
Eq.\eqref{eq:adjacency} of a graph,
\begin{align}
  \softmax (q_i \cdot k_j)
  \equiv a_j^{(i)}
  \ne   a_i^{(j)}
  \equiv \softmax (q_j \cdot k_i) \; .
  \label{eq:att_matrix}
\end{align}
Finally, we transform the complete set of inputs $x_1~...~x_C$ into a
latent \underline{value representation}, in analogy to the constrained
query form of Eq.\eqref{eq:def_wq}, but allowing for full
correlations,
\begin{align}
  \begin{pmatrix} v_1 \\ \vdots \\ v_C \end{pmatrix}
  = W^V 
  \begin{pmatrix} x_1 \\ \vdots \\ x_C \end{pmatrix} \; ,
  \label{eq:transformer_wv}
\end{align}
through the learned matrix $W^V$. Generalizing from $x_1$ to $x_j$
gives us the output vector for the transformer-encoder layer in
analogy to Eq.\eqref{eq:toy_transformer_z}
\begin{align}
\boxed{   z_i
  = \sum_{j=1}^{3C} a_j^{(i)} v_j 
  = \sum_j \softmax_j \Big[ (W^Q x_i) \cdot (W^K x_j) \Big] \; (W^V x)_j }
  \; . 
  \label{eq:transformer_all}
\end{align}
In this formula that the matrices $W^Q$ and $W^K$ do not have to be
quadratic and can define internal representations $W x$ with any
number of dimensions. The size of $W^V$ defines the dimension of the
output vector $z$.

A practical problem with the self-attention described above is that
each element tends to attend dominantly to itself, which means that in
Eq.\eqref{eq:transformer_a} the diagonal entries $a_j^{(j)}$ dominate.
This numerical problem can be cured by extending the network to
multiple heads, which means we perform several self-attention
operations in parallel, each with separate learned weight matrices,
and concatenate the outputs before applying a final linear layer. This
might seem not efficient in computing time, but in practice we can do
the full calculation for all constituents, all attention heads, and an
entire batch in parallel with tensor operations, so it pays off.

Because a transformer can be viewed as a preprocessor for the jet
constituents, enforcing permutation invariance before any kind of NN
application, we can combine with it any other preprocessing step. For
instance, we know that constituents with small $p_T$ just represent
noise, either from QCD or from pileup, and they should have no effect
on the physics of the subjet constituents. We can implement this
constraint in a \underline{IR-safe transformer}, where we add a
correction factor to Eq.\eqref{eq:transformer_a},
\begin{align}
  a_j^{(i)} = \softmax [ (q_i \cdot k_j) + \beta \log p_T ] \; .
  \label{eq:transformer_ir}
\end{align}
For small $p_T$ the second contribution ensures that constituents with
$p_T \to 0$ do not contribute to the attention weights $a_j$.  In
addition, we replace $z_j \to p_{T,j} z_j$ in the transformer
output.

\begin{figure}[t]
  \centering
  \includegraphics[width=0.45\textwidth]{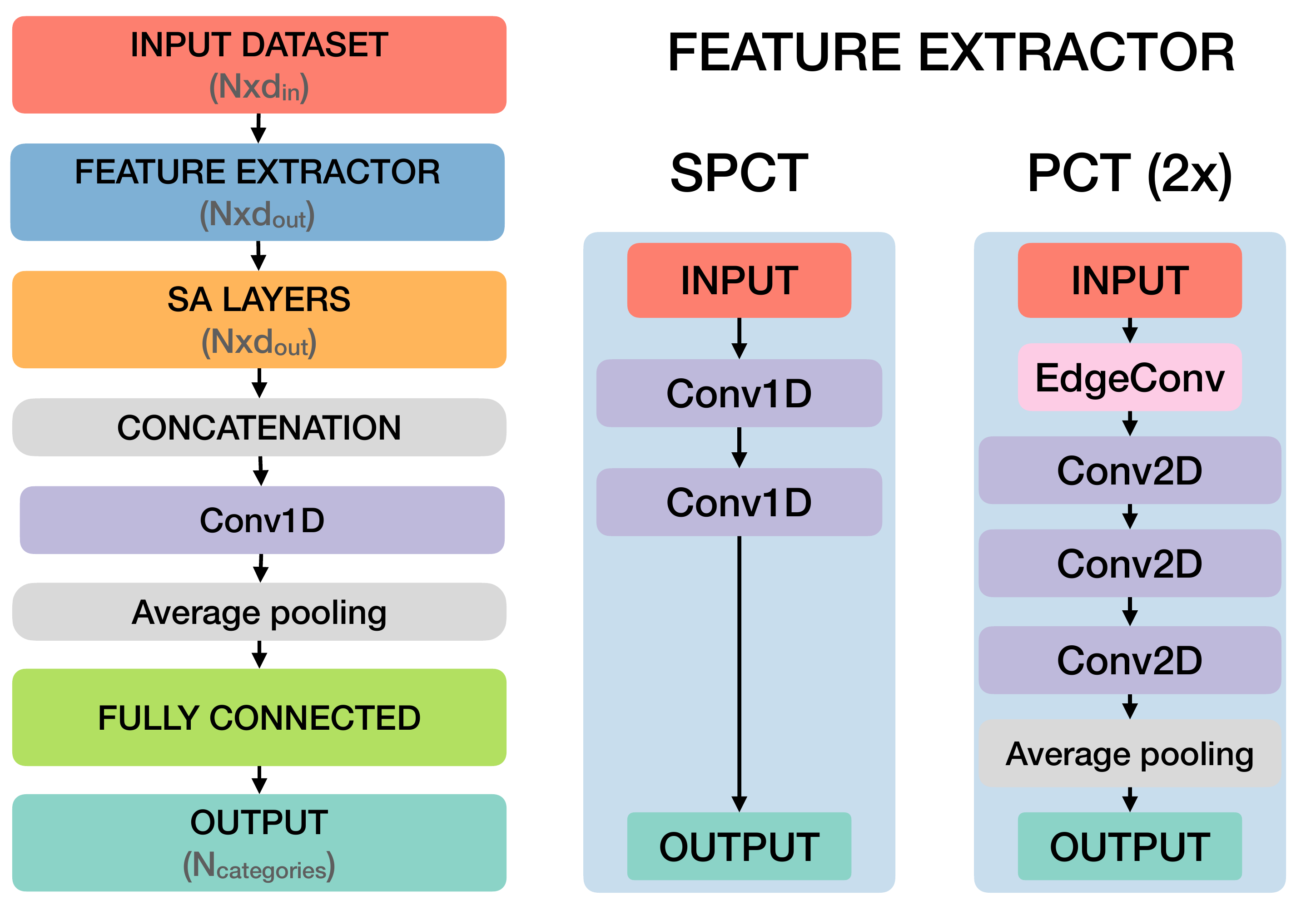}
  \caption{Network architecture and feature extractors for the top
    tagging application of the point cloud transformer. Figure from
    Ref.~\cite{Mikuni:2021pou}.}
  \label{fig:pct}
\end{figure}

The transformer-encoder layer can then be used as part of different
network architectures, for example to analyze jets. The output $z$ of
a stack of transformer layers can be fed into a classification network
directly, or it can be combined with the input features $x$, in the
spirit of the skip connections discussed in
Sec.~\ref{sec:class_cnn_sample}, to define an offset-attention.  In
Fig.~\ref{fig:pct} we illustrate the combination of the transformer
with two different feature extractors, a set of 1-dimensional
convolutional layers running over the features $x$ of each constituent
(SPCT), and an edge convolution as introduced in
Sec.~\ref{sec:class_graph_arch} (PCT). The edge convolution uses a
large number of $k=20$ nearest neighbors and is followed by
2-dimensional convolutions. The output of the transformer layers is
then concatenated to an expanded feature dimension and fed through a
fully connected network to provide the usual classification
output. This way the SPCT is an transformer-enhanced fully connected
network and the PCT combines a simplified GCN structure with a
transformer-encoder. The performance of the simple SCPT network
matches roughly the standard CNN or LoLa results shown in
Fig.~\ref{fig:toptagging}, but with only 7k network parameters. The
PCT performs almost as well as the leading ParticleNet architecture,
but with only 200k instead of almost 500k network parameters. So while
transformers with their different learned matrices appear
parameter-intensive, they are actually efficient in reducing the size
of the standard networks while ensuring permutation invariance as the
key ingredient to successful jet tagging. The challenge of the
transformer preprocessing is their long training time.

\subsubsection{Deep sets}
\label{sec:class_graph_sets}

Motivated by the same argument of permutation invariance as the
transformers, another approach to analyzing LHC jets and events is
based on the mathematical observation that we can approximate any
observable of 4-vectors using a combination of per-particle mappings
and a continuous pooling function,
\begin{align}
\{ k_{\mu,i} \} \to  F_\theta \left[ \sum_i \phi_\theta(k_{\mu, i}) \right]
\label{eq:deepsets}
\end{align}
where $\phi \in \mathbb{R}^\ell$ is a \underline{latent space}
representation of each input 4-vector or extended particle
information. Latent spaces are abstract, intermediate spaces
constructed by neural networks. By some kind of dedicated requirement,
they organize the relevant information and form the basis for example
of all generative networks discussed in Sec.~\ref{sec:gen}.  As a side
remark, the observable $F$ can be turned infrared and collinear safe
by replacing $\phi(k_{\mu, i}) \to p_{T,i} \phi$, where $\phi$ now
only depends on the angular information of the $k_\mu$. This is the
same strategy as the IR-safe transformer in
Eq.\eqref{eq:transformer_ir}. Such an IR-safe \underline{energy flow
  network} (EFN) representation is an additional restriction, which
means it will weaken the distinguishing power of the discriminative
observable, but it will also make it consistent with perturbative
QFT. In Tab.~\ref{tab:sumobs} we show a few such representations,
including subjet observables from Eq.\eqref{eq:qg_obs}.

A \underline{particle flow network} (PFN) implementation of the deep
sets architecture is, arguably, the simplest way to analyze point
clouds by using two networks. The first network constructs the latent
representations respecting the permutation symmetry of the inputs.  It
is a simple, fully connected network relating the 4-vectors and the
particle-ID, for the PFN-ID version, to a per-vector
$\ell$-dimensional latent representation $\phi_\theta(k_\mu) \in
\mathbb{R}^\ell$.  The second, also fully connected network, sums the
$\phi_\theta$ for all 4-vectors just like the graph aggregation
function in Eq.\eqref{eq:graph_conv2}, and feeds them through a fully
connected classification network with a softmax or sigmoid activation
function as its last layer. The entire classification network is
trained though a cross-entropy loss.

For the competitive top tagging results shown in
Fig.~\ref{fig:toptagging} the energy flow network (EFN) and the
particle flow network (PFN) only require 82k parameters in the two
fully connected networks, with a 256-dimensional latent space. The
suppression of QCD jets for a given top tagging efficiency is roughly
20\% smaller when we require soft-collinear safety from the EFN rather
than using the full information in the PFN.

\begin{table}[t]
  \centering
  \begin{small} \begin{tabular}{ll|ll}
\toprule
& & $\phi$ & $F$ \\
\midrule
mass & $m$  & $p^\mu$ & $F(x^\mu) = \sqrt{x^\mu x_\mu}$ \\ 
multiplicity & $n_\text{PF}$ & $1$ & $F(x) = x$ \\
momentum dispersion & $p_TD$ & $(p_{T}, p_{T}^2)$ & $F(x,y) = \sqrt{y}/x $\\
\bottomrule
\end{tabular} \end{small}
\caption{Example for observables decomposed into per-particle maps
  $\phi$ and functions $F$ according to Eq.\eqref{eq:deepsets}.  In
  the last column, the arguments of $F$ are placeholders for the
  summed output of $\phi$. Table from Ref.~\cite{Komiske:2018cqr}.}
\label{tab:sumobs}
\end{table}

Just like for the graph network in Sec.~\ref{sec:class_graph_arch} it
is also clear how one would add information on the identified
particles in a jet or an event.  The question is how to best add
additional entries to the $k_\mu$ introduced in
Eq.\eqref{eq:def_input}. In analogy to the preprocessing of jet
images, described in Sec.~\ref{sec:class_cnn_arch}, we would likely
use the $\eta$ and $\phi$ coordinates relative to the jet axis,
combined with the (normalized) transverse momentum $p_T$. A fourth
entry could then be the mass of the observed particle, if non-zero.
The charge could simply be a fifth entry added to $k_\mu$. Particle
identification would then tell us if a jet constituent is a
\begin{align}
\gamma, e^\pm, \mu^\pm, \pi^0, \pi^\pm, K_L, K^\pm, n, p \cdots  
\end{align}
One way to encode such \underline{categorical data} would be to assign
a number between zero and one for each of these particles and add a
sixth entry to the 4-momentum. This is equivalent to assigning the
particle code an integer number and then normalizing this entry of the
feature vector. To learn the particle-ID the network then learns the
ordering in the corresponding direction.

The problem with this encoding becomes obvious when we remind
ourselves that a loss function forms a scalar number out of the
feature vector, so the network needs to learn some kind of filter
function to extract this information. This means combining different
categories into one number is not helping the network. Another problem
is a possible bias from the network architecture or loss function,
leading to an enhanced sensitivity of the network to larger values of
the particle-ID vector.  Some ranges in the ID-directions might be
preferred by the network, bringing us back to permutation invariance,
the theme of this section. Instead, we can encode the particle-ID in a
permutation-invariant manner, such that a simple unit vector in all
directions can extract the information. Using \underline{one-hot
  encoding} the phase space vector of Eq.\eqref{eq:vec_image} becomes
\begin{align}
  k_\mu \quad \to \quad
  \begin{pmatrix}
    \Delta \eta \\
    \Delta \phi \\
    p_T \\
    m \\
    \delta_{\text{ID}= \gamma} \\
    \delta_{\text{ID}= e} \\
    \vdots
  \end{pmatrix} \; .
  \label{eq:vec_graph}
\end{align}
The additional dimensions can only have entries zero and one. This
method looks like a waste of dimensions, until we remind ourselves
that a high-dimensional feature space is actually a strength of neural
networks and that this kind of information is particularly easy to
extract and de-correlate from the feature space.

\subsubsection{CNNs to transformers and more}
\label{sec:class_graph_struc}

After introducing a whole set of network architectures, developed for
image or language applications, we can illustrate their differences
slightly more systematically. By now, we know to consider our input
data as elements of a point cloud $x_{i,j}$, which can be represented
as nodes of a graph or similar construction. It is convenient to
divide their properties into node features and edge features
$e_{ij}$. A neural network is a trainable function $F_\theta$ or
$\phi_\theta$. The symmetry\index{symmetries} properties of a network are determined by
an aggregation function $\Box$ as introduced in
Eq.\eqref{eq:aggregation}, typically a sum or a modification of a
sum. Note that a product is just a sum in logarithmic space. Finally,
we denote a generic activation function as $\relu$.

\underline{Convolutional networks}, including graph-convolutional
networks, are defined by Eq.\eqref{eq:def_conv0}. We can write this
transformation as
\begin{align}
  x_i' = \relu \; F_\theta \left[ x_i, \Box_{j \in N} c_{ij} \phi_\theta(x_j) \right]  \; ,
  \label{eq:link_cnn}
\end{align}
where constant values $c_{ij}$ imply the crucial weight sharing. The
aggregation combines the central node with a neighborhood $N$. In
physics application we often choose this neighborhood small, because
physics effects are usually local in an appropriate space.

For self-attention\index{self-attention}, the basis of \underline{transformers} defined in
Sec.~\ref{sec:class_graph_trans}, we replace the $c_{ij}$ by as a
general link function of $x_i$ and $x_j$, with access to the feature
structure
\begin{align}
  x_i' = \relu \; F_\theta \left[ x_i, \Box_{j \in N} a(x_i,x_j,e_{ij}) \phi_\theta(x_j) \right] \; .
  \label{eq:link_att}
\end{align}
Many transformers cover all nodes instead of a local neighborhood, but
this approach is expensive to train and not always required. We refer
to them as masked transformers.

Even more generally, we can avoid the factorization into linking
relation and the node features and replace it with a general learned
function. This defines \underline{message passing networks},
introduced in Eq.\eqref{eq:edge_conv1},
\begin{align}
  x_i' = \relu \; F_\theta \left[ x_i, \Box_{j \in N} \phi_\theta(x_i,x_j,e_{ij}) \right] \; .
  \label{eq:link_mp}
\end{align}
Finally, we can also understand the efficiency gain of the
\underline{deep sets} architecture in Eq.\eqref{eq:deepsets} in this
form,
\begin{align}
  x_i' = \relu \;  F_\theta \left[ \Box_j \phi_\theta(x_j) \right] \; ,
\end{align}
Here they are not edges, which means nodes have to be encoded without
any notion of an underlying space or metric, and the function
$F_\theta$ operates on the aggregation of the latent representations
of the nodes.

\clearpage
\section{Non-supervised classification}
\label{sec:auto}

Searches for BSM physics at the LHC traditionally start with a theory
hypothesis, such that we can compare the expected signature with the
Standard Model background prediction for a given phase space region
using likelihood methods. The background hypothesis might be defined
through simulation or through an extrapolation from a background into
a signal region. This traditional approach has two fundamental
problems which we will talk about in this section and which will take
us towards a more modern interpretation of LHC searches.

First, we can generalize classification, for example of LHC events, to
the situation where our training data is measured data and therefore
does not come with event-wise labels. However, a standard assumption
of essentially any experimental analysis is that signal features are
localized in phase space, which means we can define background
regions, where we assume that there is no signal, and signal regions,
where there still are background, but accompanied by a sizeable signal
fraction. This leads us to classification based on \underline{weakly
  supervised learning}.

Second, any searches based on hypothesis testing does not generalize
well in model space, because we can never be sure that our model
searches actually cover an existing anomaly or sign of physics beyond
the Standard Model. We can of course argue that we are performing such
a large number of analyses that it is very unlikely that we will miss
an anomaly, but this approach is at the very least extremely
inefficient. We also need to remind ourselves that ruling out some
parameter space in a pre-defined model is not really a lasting result.
This means we should find ways to identify for example anomalous jets
or events in the most model-independent way. Such a method can be
purely data-driven or rely on simulations, but in either case we only
work with background data to extract an unknown signal, a method
refereed to as \underline{unsupervised learning}.

\subsection{Classification without labels}
\label{sec:auto_dense}

Until now we have trained classification networks on labelled, pure
datasets. Following the example of top tagging in
Sec.~\ref{sec:class_cnn_tag}, such training data can be simulations or
actual data which we understand particularly well. The problem is that
in most cases we do not understand a LHC dataset well enough to
consider it fully labelled. What is much easier is to determine the
relative composition of such a dataset with the help of simulations,
for instance 80\% top jets combined with 20\% QCD jets on the one hand
and 10\% top jets combined with 90\% QCD jets on the other.

Let us go back to Sec.~\ref{sec:basics_deep_multi} where we looked at
phase space distributions of signal and background jets or events
$p_{S,B}(x)$. What we observe are not labelled signal and background
samples, but two mixed samples with global \underline{signal
  fractions} $f_{1,2}$ and \underline{background fractions}
$1-f_{1,2}$ in our two training datasets. The mixed phase space
densities $p_{1,2}$ are related to the pure densities as
\begin{align}
  \begin{pmatrix}
    p_1(x) \\ p_2(x)
  \end{pmatrix}
  &= 
  \begin{pmatrix}
    f_1 & 1-f_1 \\ f_2 & 1-f_2 
  \end{pmatrix}
  \begin{pmatrix}
    p_S(x) \\ p_B(x) 
  \end{pmatrix} \notag \\
  \Leftrightarrow \qquad 
  \begin{pmatrix}
    p_S(x) \\ p_B(x)
  \end{pmatrix}
  &= 
  \frac{1}{f_1 - f_2}
  \begin{pmatrix}
    1-f_2 & f_1-1 \\ -f_2 & f_1
  \end{pmatrix}
  \begin{pmatrix}
    p_1(x) \\ p_2(x)
  \end{pmatrix} \notag \\
  \Rightarrow \qquad \;
  \frac{p_S(x)}{p_B(x)} &= \frac{(1-f_2) p_1(x) + (f_1-1)p_2(x)}{-f_2 p_1(x) + f_1p_2(x)} \; .
\label{eq:fractions}
\end{align}
The last line implies that, if we know $f_{1,2}$ and can also extract
the mixed densities $p_{1,2}(x)$ of the two training datasets, we can
compute the signal and background distributions and with them the
likelihood ratio as the optimal test statistics.

Next, we can ask the question how the likelihood ratios for signal vs
background classification, $p_S/p_B$, and the separation of the two
mixed samples, $p_1/p_2$, are related. This will lead us to a shortcut
in our classification task,
\begin{align}
  \frac{p_1(x)}{p_2(x)}
  &= \frac{f_1 p_S(x) + (1-f_1) p_B(x)}{f_2 p_S(x)  + (1-f_2) p_B(x)} 
  = \frac{f_1 \dfrac{p_S(x)}{p_B(x)} + 1-f_1}
         {f_2 \dfrac{p_S(x)}{p_B(x)} + 1-f_2}
  \notag \\
  \frac{d}{d (p_S/p_B)} \frac{p_1(x)}{p_2(x)}
  &= \frac{f_1 \left[ f_2 \dfrac{p_S(x)}{p_B(x)} + 1 - f_2 \right] - f_2 \left[ f_1 \dfrac{p_S(x)}{p_B(x)} + 1-f_1 \right]}
       {\left[ f_2 \dfrac{p_S(x)}{p_B(x)} + 1- f_2 \right]^2} 
   = \frac{f_1 - f_2}
              {\left[ f_2 \dfrac{p_S(x)}{p_B(x)} + 1 - f_2 \right]^2} \; .
\label{eq:cwola}
\end{align}
The sign of this derivative is a global $\sign (f_1 - f_2)$ and does
not change if we vary the likelihood ratios. This means the two
likelihood ratios are linked through a monotonous function, which
means that we can exchange them as a test statistics at no cost. In
other words, if we are interested in an optimal classifier we can skip
the translation into $p_S/p_B$ and just use the classifier between the
two mixed samples, $p_1/p_2$, instead. This is an attractive option,
because it means that we do not need to know $f_{1,2}$ if we are just
interested in the likelihood ratio.

\begin{figure}[t]
  \centering
  \includegraphics[width=0.40\textwidth]{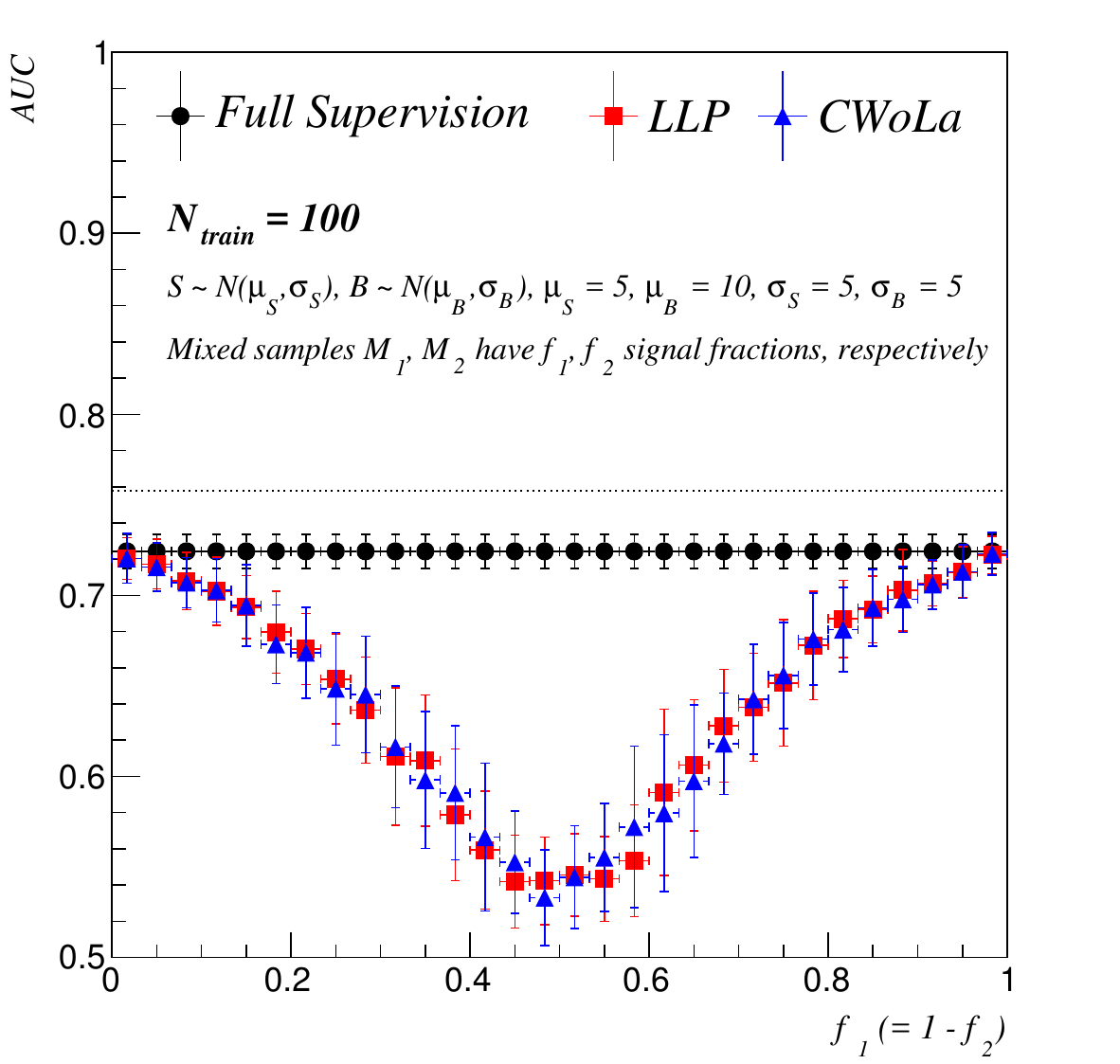}
  \hspace*{0.1\textwidth}
  \includegraphics[width=0.40\textwidth]{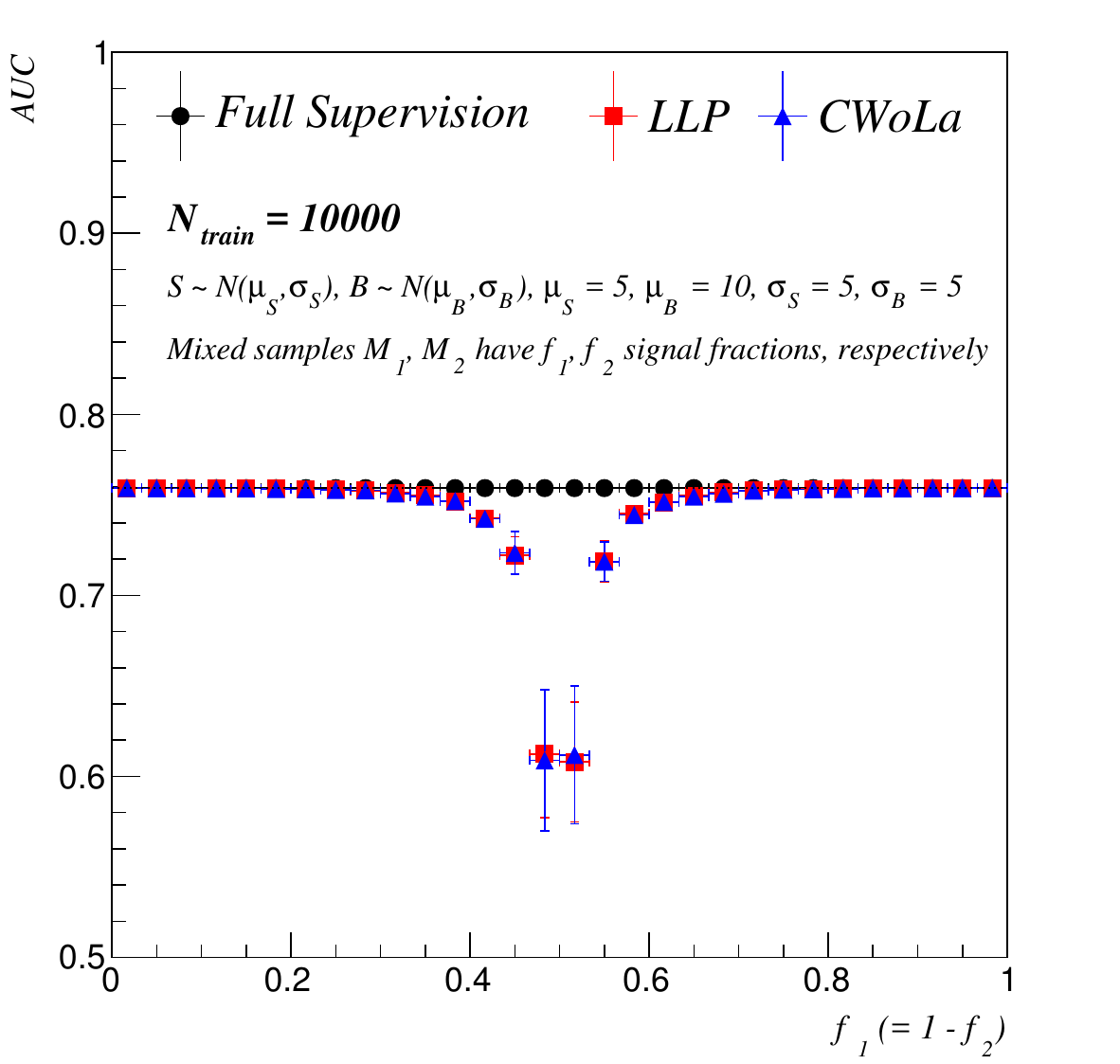}
  \caption{Performance of the CWoLa method for the double-Gaussian toy
    example as a function of the signal fraction in one of the
    training datasets for 100 (left) and 10000 (right) training
    points. Figure from Ref.~\cite{Metodiev:2017vrx}.}
  \label{fig:cwola1}
\end{figure}

While this classification without labels (CWoLa)\index{CWoLa} does not require use
to know the signal and background fractions and is therefore, strictly
speaking, an unsupervised method, we always work under the assumption
that we have a background-dominated and a signal-dominated dataset.
Moreover, any such analysis tool needs to be calibrated. If the
classification outcome is not a signal or background probability, as
discussed in Sec.~\ref{sec:class}, we need to define a working point
for our classifier and determine its signal and background
efficiencies. From Eq.\eqref{eq:fractions} we see that we only need
two samples with known signal and background fractions to extract
$p_S(x)$ and $p_B(x)$ for any given working point.

We illustrate the unsupervised method using the original toy model of
a 1-dimensional, binned observable $x$ and Gaussian signal and
background distributions $p_{S,B}(x) = \normal(x)$. We have three
options to train a classifier on this dataset, which means we can
\begin{enumerate}
\item compute the full supervised likelihood ratio $p_S/p_B(x)$ from
  the truth distributions;
\item use Eq.\eqref{eq:fractions} to compute the likelihood ratio from
  $p_{1,2}(x)$ and known \underline{label proportions} $f_{1,2}$
  (LLP);
\item follow the CWoLa method and use $p_1/p_2(x)$ to separate
  signal and background instead of the two samples.
\end{enumerate}
The AUC values for the three methods are shown in
Fig.~\ref{fig:cwola1} as a function of the signal fraction of one of
the two samples, chosen the same as the background fraction for the
second sample. The horizontal dashed line indicates the
fully-supervised AUC with infinite training statistics. By
construction, the AUC for full supervision is independent of
$f_1$. The weakly supervised and unsupervised methods start coinciding
with the fully supervised method as long as we stay away from
\begin{align}
  f_1 \approx f_2 \approx \frac{1}{2} \; .
\end{align}
We can see the problem with this parameter point in
Eq.\eqref{eq:fractions}, where a matrix inversion is not
possible. Similarly, in Eq.\eqref{eq:cwola} the linear dependence of
the two likelihood ratios vanishes in the same parameter
point.

For jet tagging\index{jet tagging} the CWoLa method becomes relevant when we cannot
construct pure training samples for either experimental or theoretical
reasons. As discussed above, the standard top tagging benchmark is
chosen to be well defined theoretically and experimentally. This is
different for quark vs gluon tagging.  However,
\underline{quark--gluon tagging} is experimentally extremely
attractive, because it would for instance allow us to suppress
backgrounds to the weak-boson-fusion processes discussed in
Sec.~\ref{sec:basics_deep_multi}.  On the theory side, we know that
partons split into pairs of collinear daughter partons with a
probability described by the unregularized splitting kernels\index{QCD splittings}
\begin{align}
  \hat{P}_{g \leftarrow g}(z)
  &= N_c \; \left[ \frac{z}{1-z} + \frac{1-z}{z} + z (1-z) \right]
  &\qquad 
  \hat{P}_{q \leftarrow g}(z)
  &= \frac{1}{2} \; \big[ z^2 + (1-z)^2 \big] \notag \\
  \hat{P}_{g \leftarrow q}(z)
  &= \frac{N_c^2-1}{2N_c} \; \frac{1+(1-z)^2}{z}
  &\qquad 
  \hat{P}_{q \leftarrow q}(z)
  &= \frac{N_c^2-1}{2N_c} \; \frac{1+z^2}{1-z} 
\label{eq:qcd_splittings}
\end{align}
where $z$ is the energy fraction carried by the harder of the two
daughter partons which also describes the splitting $\hat{P}_{j
  \leftarrow i}$. The off-diagonal splitting probabilities imply that
partonic quarks and gluons are only defined probabilistically, which
means any kind of quark-gluon tagging at the LHC will at most be
weakly supervised. Another theoretical complication is that quarks
pairs coming for example from a large-angle gluon splitting $g \to
q\bar{q}$ and from an electroweak decay $Z \to q\bar{q}$ have
different color-correlations, so they are not really identical at the
single-quark level.  Experimentally, it is also impossible to define
pure quark vs gluon jets samples. While most LHC jets at moderate
energies are gluons, we can use $W$ and $Z$ decays to collect quark
jets, but some of those decays actually include three or more jets, $Z
\to q\bar{q}g$, which again leads us to learn from mixed samples.

\begin{figure}[t]
  \centering
  \includegraphics[width=0.39\textwidth]{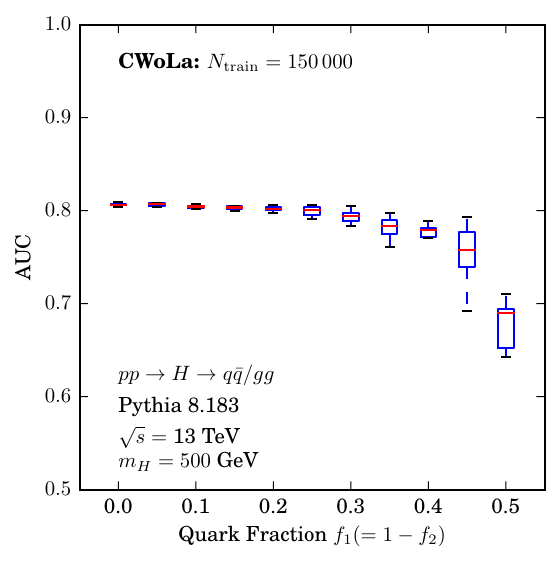}
  \hspace*{0.1\textwidth}
  \includegraphics[width=0.42\textwidth]{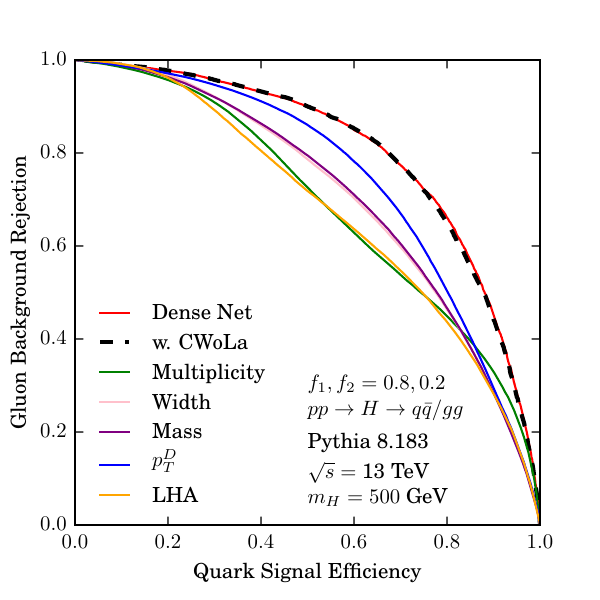}
  \caption{Performance of the CWoLa method for quark-gluon tagging as
    a function of the signal fraction (left), and the corresponding
    ROC curve (right). Figure from Ref.~\cite{Metodiev:2017vrx}.}
  \label{fig:cwola2}
\end{figure}

We can apply CWoLa to quark-gluon tagging, using a very simple 
classification network based on five substructure observables like
those shown in Eq.\eqref{eq:qg_obs}. To simplify things, the jets are
generated from hypothetical Higgs decays with $m_H = 500$~GeV
\begin{align}
pp \to H \to qq \; / \; q\bar{q} \; .
\end{align}
Again we train the network on two mixed samples and show the results
in Fig.~\ref{fig:cwola2}. In this example, we do not extract the
likelihood ratio, but train a standard classifier to separate the two
training samples by minimizing the cross entropy. We then apply the
classifier to split signal and background, so we rely on the fact that
our network knows the Neyman-Pearson lemma\index{Neyman-Pearson lemma}.  For the exact relation
between a trained discriminator and the likelihood ratio we refer to
Sec.~\ref{sec:gen_gan_arch}.  In the left panel we see that the AUC is
again stable as long as we stay away from equally mixed training
samples. In the right panel we see that the performance of the network
is as good as the same network trained on labelled data.

Another application of CWoLa would be event classification, where we
associate a signal with a specific phase space configuration which can
also be generated by background processes. This is an example for LHC
analyses not based on detailed hypothesis tests. A search with a
minimal model dependence is a \underline{bump hunt} in the invariant
mass of a given pair of particles. Looking for a resonance decaying to
two muons or jets we get away with minimum additional assumptions, for
example on the underlying production process or correlations with
other particles in the final state, by letting a mass window slide
through the invariant mass distribution and searching for a feature in
this smooth observable. Technically, we can analyse a binned
distribution by fitting a background model, taking out individual
bins, and checking if $\chi^2$ changes. The reason why this method
works is that we have many data points in the background regions and a
localized signal region, which can be covered by the background model
through interpolation. This analysis idea is precisely what we have in
mind for an enhanced classification network with weakly supervised or
unsupervised training.

Looking at our usual reference process, there exists no point in the
signal phase space for the process
\begin{align}
  pp \to t\bar{t}H \to t\bar{t} \; (b\bar{b})
\end{align}
which is not also covered by the $t\bar{t} b\bar{b}$ continuum
background. The only difference is that in background regions like
$m_{bb} = 50~...~100$~GeV or $m_{bb} > 150$~GeV there will be hardly
any signal contamination while under the Higgs mass peak $m_{bb} \sim
125$~GeV the signal-to-background ratio will be sizeable. This
separation might appear trivial, but the mass selection will
propagate into the entire phase space.  This means we can use
sidebands to extract the number of background events expected in the
signal region and subtract the background from the combined $m_{bb}$
distribution. If we want to subtract the background event by event
over the entire phase space, we need to include additional information
from the remaining phase space directions, ideally encoded in an
event-by-event classifier.

For a signal-background separation using the kind of likelihood-based
classifier described above we need two mixed training samples. We can
use the $m_{bb}$ distribution to define two event samples, one with
almost only background events and one with an enhanced signal
fraction.  The challenge of such a construction is that over the
entire phase space the mixed samples need to be based on the same
underlying signal and background distributions $p_{S,B}(x)$, just with
different signal fractions $f_{1,2}$, as introduced in
Eq.\eqref{eq:fractions}. This means that if we assume that the
distribution of all phase space features $x$, with the exception of
$m_{bb}$, are the same for signal and background, a CWoLa classifier
will detect the signal correctly.  This is a strong assumption, and we
will revisit this challenge in Sec.~\ref{sec:gen_cathode} in the
context of density estimation.

Finally, on a slightly philosophical note we can argue that it is
difficult to separate signal from background events in a situation
where both processes are defined though transition amplitudes over
phase space and will interfere. Still, in this case we can define a
background label for all events which remain in the limit of zero
signal strength and a signal label to the pure signal and the
interference with the background. Note that the interference can be
constructive or destructive for a given phase space point.

\subsection{Anomaly searches}
\label{sec:auto_ano}

The main goal of the LHC is to search for new and interesting effects
which would require us to modify our underlying theory.  If BSM
physics is accessible to the LHC, but more elusive than expected, we
should complement our hypothesis-based search strategies with more
general approaches. For example, we can search for jets or events
which stick out by some measure, turning them into prime candidates
for a BSM physics signature. If we assume that essentially all events
or jets at the LHC are described well by the Standard Model and the
corresponding simulations, such an outlier search is equivalent to
searching for the most non-SM instances. The difference between these
two statements is that the first is based on unlabeled data, while the
second refers to simulations and hence pure SM samples. They are
equivalent if the data-based approach effectively ignores the outliers
in the definition of the background-like sample and its implicitly
underlying phase space density.

\subsubsection{(Variational) autoencoders}
\label{sec:auto_ano_ae}

\begin{figure}[t]
\begin{subfigure}[t]{0.42\textwidth}
\raisebox{0.9mm}{ \includegraphics[width=\textwidth]{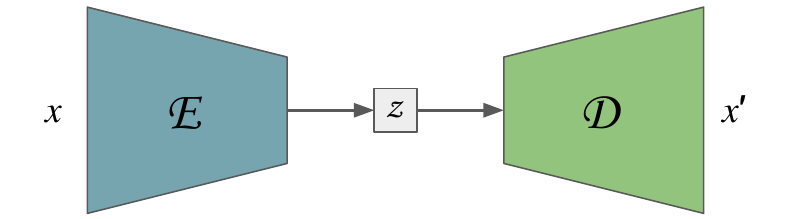} }
\end{subfigure}
\begin{subfigure}[t]{0.54\textwidth}
\includegraphics[width=\textwidth]{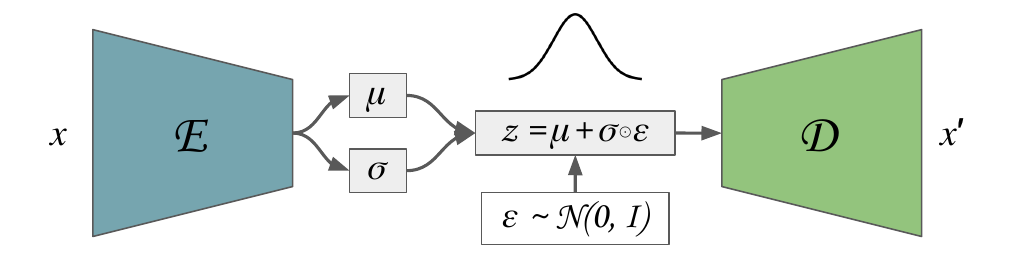} 
\end{subfigure} 
\par \bigskip
\begin{subfigure}[b]{\textwidth}
\includegraphics[width=\textwidth]{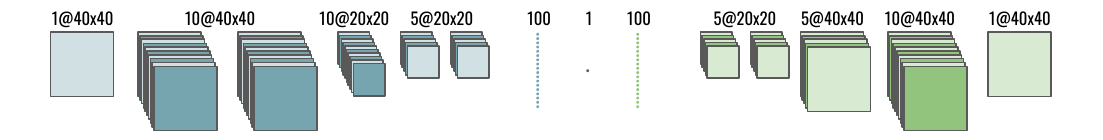}
\caption{Encoder and decoder}
\label{fig:aevae_archs_c}
\end{subfigure}
\caption{Architectures used for AE (left) and VAE (right) networks.
  All convolutions use a 5x5 filter size and all layers use PReLU
  activations. Downsampling from 40x40 to 20x20 is achieved by average
  pooling, while upsampling is nearest neighbor interpolation. Figure
  from Ref.~\cite{Dillon:2021nxw}.}
\label{fig:aevae_archs}
\end{figure}

For the practical task of anomaly searches\index{anomaly detection}, autoencoders (AEs) are the
simplest unsupervised ML-tool.  In the AE architecture an encoder
compresses the input data into a \underline{bottleneck}, encoding each instance of
the dataset in an abstract space with a dimension much smaller than
the dimension of the input data.  A decoder then attempts to
reconstruct the input data from the information encoded in the
bottleneck.  This architecture is illustrated in the upper left panel
of Fig.~\ref{fig:aevae_archs} and works for jets using an image or
4-vector representation. Without the bottleneck the AE could just
construct an identity mapping of a jet on itself; with the bottleneck this
is not possible, so the network needs to construct a compressed representation
of the input. This should work well if the ambient or apparent dimensionality of
our data representation is larger than the intrinsic or physical dimensionality
of the underlying physics.

The loss function for such an AE can be the MSE defined in
Eq.\eqref{eq:mse_loss}, quantifying the agreement of the pixels in an
input jet image\index{jet images} $x$ and the average jet image output $x'$, summed over
all pixels,
\begin{align}
\loss_\text{MSE} = \frac{1}{N} \sum_{i=1}^N \; \left| x_i - x_i' \right|^2 \; .
\label{eq:aeloss1}
\end{align}
This sum is just an expectation value over a batch of jet images sampled from
the usual data distribution $\pd(x)$,
\begin{align}
\boxed{
  \loss_\text{MSE} = \XLangle \left| x - x' \right|^2 \XRangle_{\pd} } \; .
\label{eq:aeloss2}
\end{align}
The idea behind the anomaly search is that the AE learns to compress
and reconstruct the training data very well, but different test data
passed through the AE results in a large loss.

Because AEs do not induce a structure in latent space, we have no
choice but to use this reconstruction uncertainty or loss also as the
anomaly score.  This corresponds to a definition of anomalies as an
unspecific kind of outliers. Using the latent loss as an anomaly score
leads to a conceptual weakness when we switch the standard and
anomalous physics hypothesis. For example, QCD jets with a limited
underlying physics content of the massless parton splittings given in
Eq.\eqref{eq:qcd_splittings} can be described by a small
bottleneck. An AE trained on QCD jets will not be able to describe
top-decay jets with their three prongs and massive decays. Turning the
problem around, we can train the AE on top jets, in which case it will
be able to describe multi-prong topologies as well as frequently
occurring single-prong topologies in the top sample. QCD jets are now
just particularly simple top jets, which means that they will not lead
to a large anomaly score. This bias towards identifying more complex
data is also what we expect from the standard use of a bottleneck for
data compression.

Moving beyond purely reconstruction-based autoencoders,
\underline{variational autoencoders} (VAEs) add structure to the
latent bottleneck space, again illustrated in
Fig.~\ref{fig:aevae_archs}. In the encoding step, a high-dimensional
data representation is mapped to a low-dimensional latent
distribution, from which the decoder learns to generate the original,
high-dimensional objects. The latent bottleneck space then contains
structured information which might not be apparent in the
high-dimensional input representation. This means the VAE architecture
consists of a learnable encoder with the output distribution $z \sim
p^\text{E}_\theta(z|x)$ mapping the phase space $x$ to the latent space $z$,
and a learnable decoder with the output distribution $x \sim
p^\text{D}_\theta(x|z)$. The loss combines two terms
\begin{align}
  \boxed{
  \loss_\text{VAE}
  = \XLangle - \Langle \log p^\text{D}_\theta(x|z) \Rangle_{p^\text{E}_\theta(z|x)}
  + \beta_\text{KL} \kl [p^\text{E}_\theta(z|x), \pl(z)] \XRangle_{\pd}
  } \; .
\label{eq:vaeloss}
\end{align}
The first terms is the reconstruction loss, where we compute the
likelihood of the output of the decoder $p^\text{D}_\theta(x|z)$ and given the
encoder $p^\text{E}_\theta(z|x)$, evaluated on batches sampled from $\pd$.  We
get back the MSE version serving as the AE reconstruction loss in
Eq.\eqref{eq:aeloss2} when we approximate the decoder output
$p^\text{D}_\theta(x|z)$ as a Gaussian with a constant width. The second term
is a latent loss, comparing the latent-space distribution from the
encoder to a \underline{prior} $\pl(z)$, which defines the structure
of the latent space. A more systematic deviation of the VAE loss as a
likelihood loss will follow in Sec.~\ref{sec:gen_vae}.

For a Gaussian prior we use the so-called
\underline{re-parametrization trick} to pretend to sample from any
multi-dimensional Gaussian with mean $\mu$ and standard deviation
$\sigma$ by instead sampling from a standard Gaussian
\begin{align}
  z = \mu + \sigma \; \epsilon
  \qquad \text{with} \qquad \epsilon \sim \normal(\mu=0,\sigma=1) \; .
\end{align}
For such a standard Gaussian prior $\pl(z)$ and a Gaussian encoder
output $p^\text{E}_\theta(z|x)$ the KL-divergence\index{KL-divergence} defined in
Eq.\eqref{eq:def_kl} turns into the form of Eq.\eqref{eq:kl_gaussian}
\begin{align}
\kl [p^\text{E}_\theta(z|x),\normal(0,1)] = \frac{1}{2n} \sum_{i=1}^n \left(\sigma_i^2 + \mu_i^2 - 1 - 2 \log \sigma_i \right),
\label{eq:klvae}
\end{align}
where $\sigma_i$ and $\mu_i$ replace the usual $n$-dimensional encoder
output $z$ for a given $x$.  Combining the simple MSE-reconstruction
loss of Eq.\eqref{eq:aeloss1} with this form of the latent loss gives
us an explicit form of the VAE loss for a batch of events or jet
images.

\begin{figure}[t]
\includegraphics[width=0.32\textwidth]{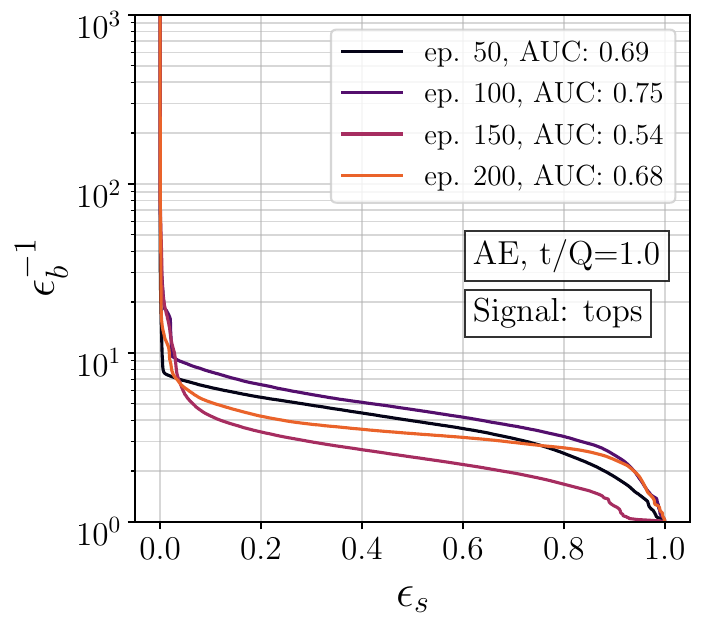}
\includegraphics[width=0.32\textwidth]{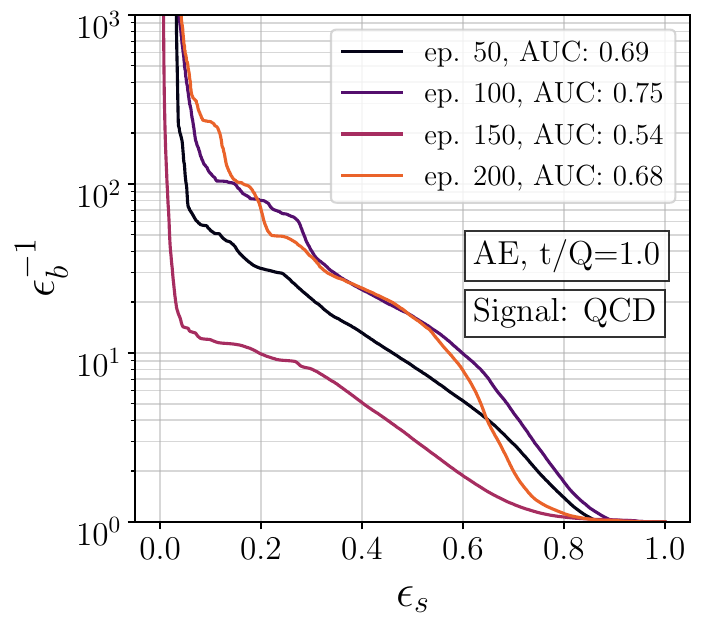}
\includegraphics[width=0.32\textwidth]{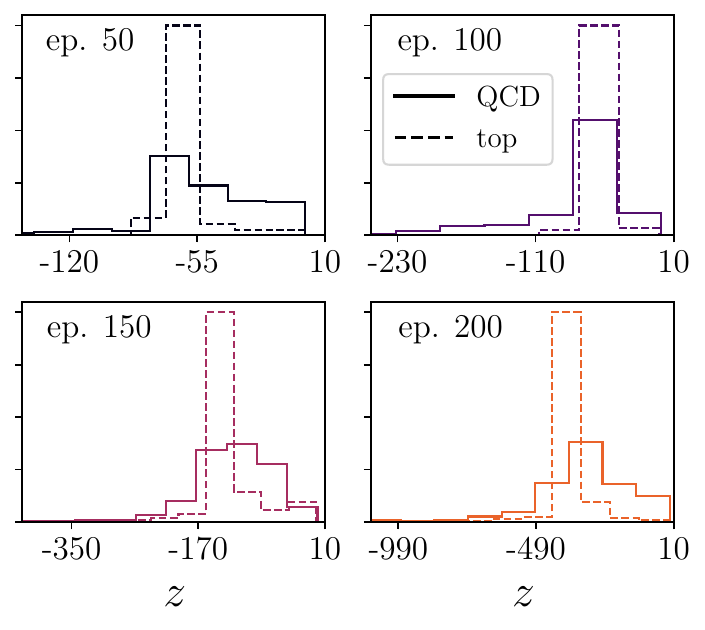} \\
\includegraphics[width=0.32\textwidth]{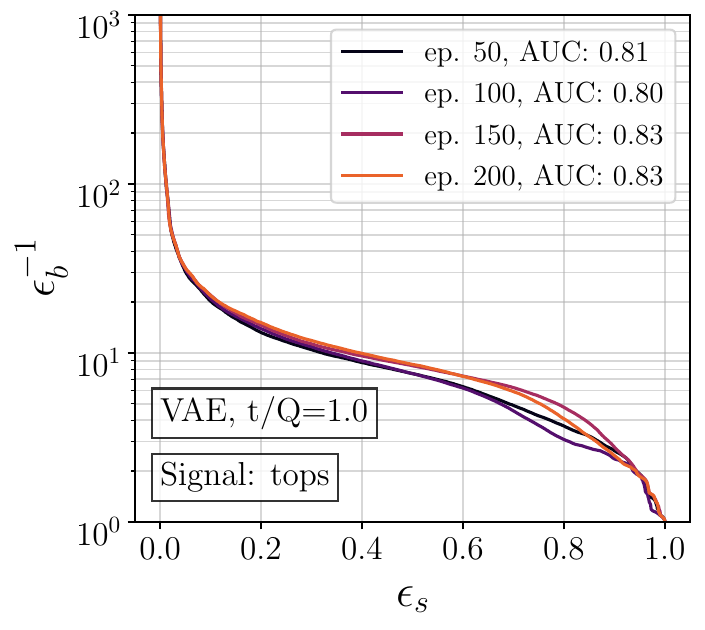}
\includegraphics[width=0.32\textwidth]{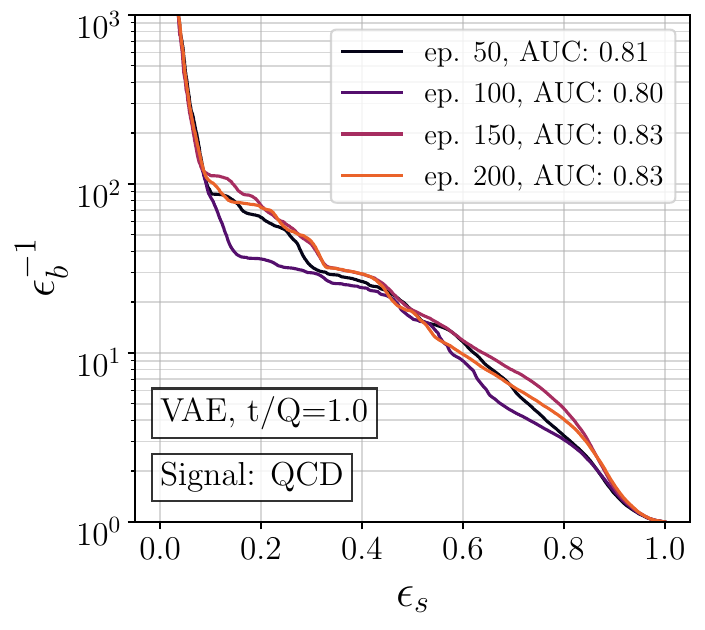}
\includegraphics[width=0.32\textwidth]{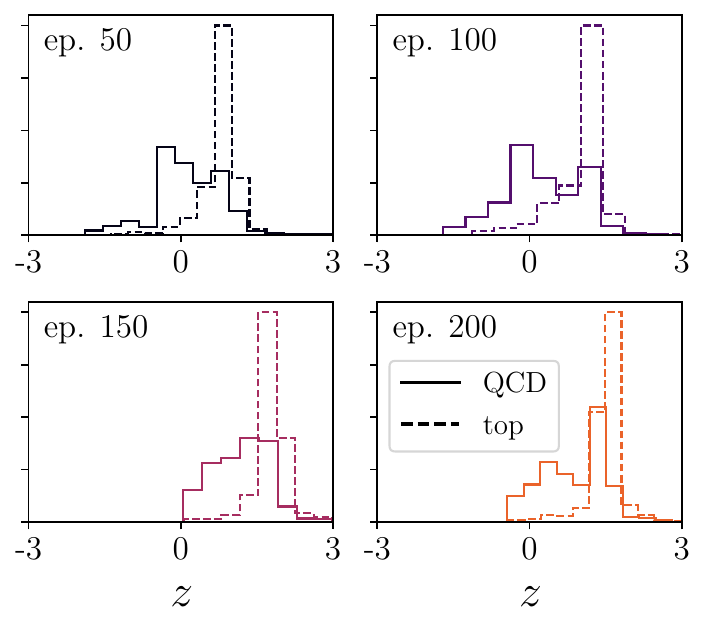}
\caption{Symmetric performance of the toy-AE (upper) and toy-VAE
  (lower).  In the large panels we show the ROC curves for tagging top
  and QCD as signals. In the small panels we show the distributions of
  top and QCD jets in the 1-dimensional latent space. Figure from
  Ref.\cite{Dillon:2021nxw}.}
\label{fig:aevae}
\end{figure}

Because of the structured bottleneck we now have a choice of anomaly
scores, either based on the reconstruction or the latent space. This
way the VAE can avoid the drawbacks of the AE by using an alternative
anomaly score to the reconstruction uncertainty. We can compare the
alternative choices using the standard QCD and top jet-images with a
simple preprocessing described in Sec.~\ref{sec:class_cnn_arch}.  To
force the network to learn the class the jet belongs to and to be able
to visualize this information, we restrict the bottleneck size to one
dimension. In the VAE case this gives us a useful probabilistic
interpretation, since the mapping to the encoded space is performed by
a probability distribution $p^\text{E}_\theta(z|x)$.  To simplify the training
of the classifier, we train on a sample with equal numbers of top and
QCD jets.  We are interested in three aspects of the VAE classifiers:
(i) performance as a top-tagger, (ii) performance as a QCD-tagger,
(iii) stable encoding in the latent space.  Some results are shown in
Fig.~\ref{fig:aevae}. Compared to the AE results, the regularisation
of the VAE latent space generates as relatively stable and structured
latent representation, even converging to a representation in which
the top jets are clustered slightly away from the QCD jets. On the
other hand, the separation in the VAE latent space is clearly not
sufficient to provide a competitive anomaly score.

For an unsupervised signal vs background classification a natural
assumption for the latent space would be a bi-modal or
\underline{multi-modal} structure. This is not possible for a VAE with
a unimodal Gaussian prior, but we can try imposing an uncorrelated
Gaussian mixture prior on the $n$-dimensional latent space.  The GMVAE
then minimizes the same loss as the VAE given in
Eq.\eqref{eq:vaeloss}.  However, for the regularization\index{regularization} term we cannot
calculate the KL-divergence analytically anymore. Instead, we estimate
it using Monte Carlo integration and start by re-writing it according
to its definition in Eq.\eqref{eq:def_kl}
\begin{align}
  \kl [p^\text{E}_\theta(z|x),\pl(z)]
  = \Langle \log p^\text{E}_\theta(z|x) \Rangle_{p^\text{E}_\theta(z|x)}
  - \Langle \log \pl(z) \Rangle_{p^\text{E}_\theta(z|x)} \; .
\end{align}
For an arbitrary number of Gaussians the combined likelihood prior is
given by
\begin{align}
  \pl(z) &= \sum_r p(z;r) \; p_r \notag \\
  \text{with} \qquad 
  p(z;r) &= \prod_{i=1}^n \; \frac{1}{\sqrt{2\pi} \sigma_{r,i}} \;
  \exp \left( -\frac{(z_i - \mu_{r,i})^2}{2\sigma^2_{r,i}} \right)
\end{align}
where $p_r$ is the weight of the respective mixture component
$\mu_{r,i}$ is the mean of mixture component $r = 1~...~R$ in
dimension $i$, and $\sigma^2_{r,i}$ is the corresponding variance.
All means, variances, and mixture weights are learned network
parameters.

The problem with the GMVAE is that it does not work well for
unsupervised classification.  When training a 2-component mixture
prior on our jet dataset the mixture components tend to collapse onto
a single mode. To prevent this, we need to add a \underline{repulsive
  force} between the modes, calculated as a function of the
\underline{Ashman distance} between two Gaussian distributions in one
dimension,
\begin{align}
D^2 = \frac{2(\mu_1 - \mu_2)^2}{\sigma^2_1 + \sigma^2_2} \; .
\end{align}
A value of $D > 2$ indicates a clear mode separation.  We encourage
bi-modality in our latent space by adding the loss term
\begin{align}
\loss_\text{GMVAE} = \loss_\text{VAE} -\lambda \tanh \frac{D}{2} \; .
\end{align}
The tanh function is meant to eventually saturate and stop pushing
apart the modes.  To cut the long story short, the Gaussian mixture
prior will shape the latent space for the top vs QCD application and
lead to a stable training. However, there is no increase in
performance in going from the VAE to the GMVAE.  The reason for this
is that while the top jets occupy just one mode in latent space, the
QCD jets occupy both.  The mode assignment is mostly based on the
amount of pixel activity within the jet, rather than on specific jet
features. This means we need to go beyond the GMVAE in that we need a
network which distributes the signal and background into the two
different modes.

\subsubsection{Dirichlet-VAE}
\label{sec:auto_ano_dvae}

The motivation, but ultimate failure of the GMVAE leads us to another
way of defining a VAE with a geometry that leads to a mode separation.
We can use the Dirichlet distribution, a family of continuous
multivariate probability distributions, as the latent space prior,
\begin{align}
  \mathcal{D}_\alpha(r)
  = \frac{ \Gamma( \sum_i \alpha_i ) }{ \prod_i \Gamma( \alpha_i ) } \; 
  \prod_i r_i^{\alpha_i-1},
  \qquad \text{with} \; i=1~...~R \; .
\end{align}
This $R$-dimensional Dirichlet structure is defined on an
$R$-dimensional simplex, which means that all but one of the $r$
vector components are independent and $\sum_i r_i =1$.  In our simple
example with $R=2$ the latent space is described by $r_1\in[0,1]$,
with $r_0 = 1-r_1$.  The weights $\alpha_i > 0$ can be related to the
expectation values for the sampled vector components,
\begin{align}
  \langle r_i \rangle = \frac{\alpha_i}{\sum_j \alpha_j} \; ,
\end{align}
which means that the Dirichlet prior it will create a hierarchy among
different mixture components.  An efficient way to generate numbers
following a Dirichlet distributions is through a softmax-Gaussian
approximation,
\begin{align}
  r \sim \text{softmax} \, \normal(z;\tilde{\mu},\tilde{\sigma}) &\approx \mathcal{D}_\alpha(r) \notag \\
  \text{with} \quad 
\tilde{\mu}_i & = \log\alpha_i - \frac{1}{R}\sum_{i}\log\alpha_i		\notag  \\
\tilde{\sigma}_i & = \frac{1}{\alpha_i}\left(1-\frac{2}{R}\right) + \frac{1}{R^2}\sum_i\frac{1}{\alpha_i} \; .
\label{eq:softdir}
\end{align}
The loss function of the Dirichlet-VAE (DVAE) includes the usual
reconstruction loss and latent loss.  For the higher-dimensional
reconstruction loss we can use the cross-entropy between the inputs
and the outputs, while the latent loss is given by the KL-divergence\index{KL-divergence}
between the per-jet latent space representation and the Dirichlet
prior.  This is easily calculated for the Gaussians in the softmax
approximation of the Dirichlet distribution and the Gaussians defined
by the encoder output 
\begin{align}
  \boxed{ 
    \loss_\text{DVAE}  = \XLangle - \Langle \log p^\text{D}_\theta(x|r) \Rangle_{p^\text{E}_\theta(r|x)} + \beta_\text{KL} \kl [p^\text{E}_\theta(r|x),\mathcal{D}_\alpha(r)] \XRangle_{\pd}
    } \; ,
\label{eq:dvae_loss}
\end{align}
where the KL-divergence can be evaluated using
Eq.\eqref{eq:kl_gaussian}. This way, training the DVAE is essentially
equivalent to the standard VAE with its loss given in
Eq.\eqref{eq:vaeloss}.

\begin{figure}[t]
\centering
\includegraphics[width=0.8\textwidth]{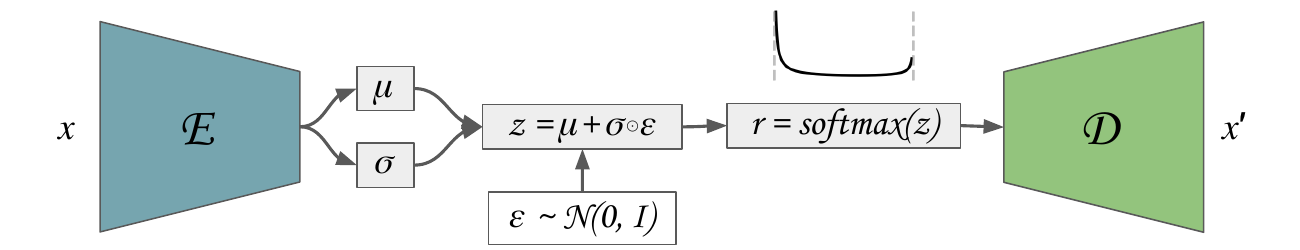}
\caption{Architecture for the Dirichlet VAE.  The approximate
  Dirichlet prior is indicated by the softmax step and the bi-modal
  distribution shown above it. Figure from Ref.\cite{Dillon:2021nxw}.}
\label{fig:dvae_arch}
\end{figure}

With a Dirichlet latent space the sampled $r_i$ can be interpreted as
mixture weights of mixture components describing the jets.  If we view
the DVAE as a classification tool for a multinomial mixture model,
these mixture components correspond to probability distributions over
the image pixels. These distributions enter into the likelihood
function of the model, which is parameterised by the decoder network.
For our 2-dimensional example, the pure component distributions are
given by $p^\text{D}_\theta(x|r_1 = 0)$ and $p^\text{D}_\theta(x|r_1 = 1)$, and any
output for $r_1 \in[0,1]$ is a combination of these two
distributions. Our simple decoder architecture exactly mimics this
scenario.

\begin{figure}[b!]
\includegraphics[width=0.32\textwidth]{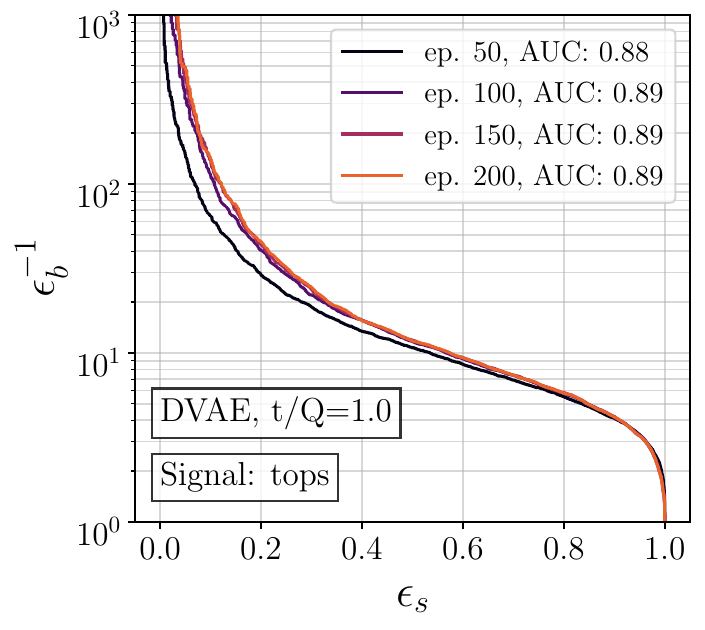}
\includegraphics[width=0.32\textwidth]{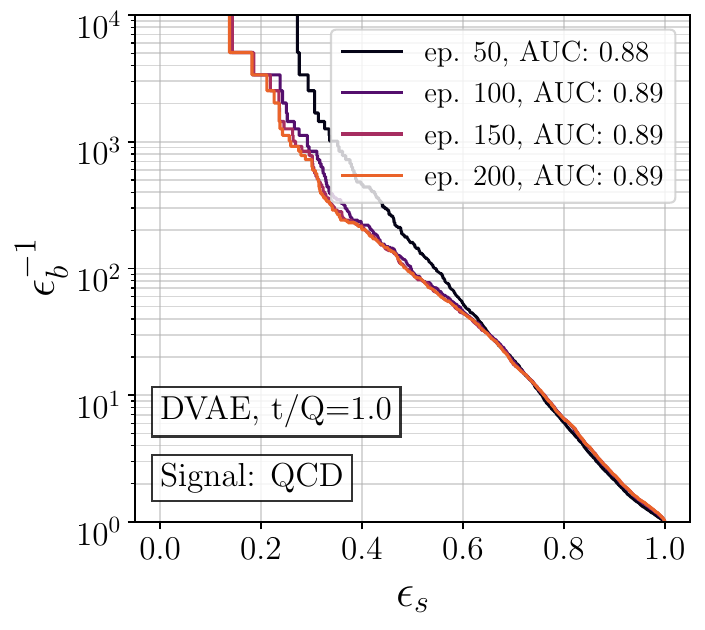}
\includegraphics[width=0.32\textwidth]{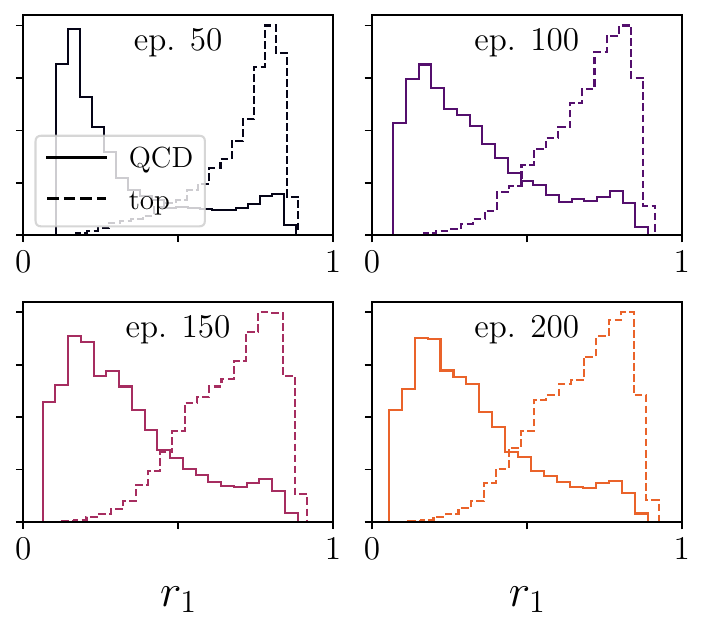}
\caption{Results for the DVAE with $\alpha_{1,2} = 1$. In the large
  panels we show the ROC curves for tagging top and QCD as signal. In
  the small panels we show the distributions of the top and QCD jets
  in the latent space developed over the training. Figure from
  Ref.\cite{Dillon:2021nxw}.}
\label{fig:dvaeplot}
\end{figure}

To compare with the AE and VAE studies in Fig.~\ref{fig:aevae} we can
again train the DVAE on a mixture of QCD and top jets. We look for
symmetric patterns in QCD and top tagging and show the latent space
distributions in Fig.~\ref{fig:dvaeplot}.  First of all, the DVAE
combined with its latent-space anomaly score identifies anomalies in
both directions. The reason for this can be seen in the latent space
distributions, which quickly settles into a bi-modal pattern with
equal importance given to both modes.  The QCD mode peaks at $r_1=0$,
indicating that the $p^\text{D}_\theta(x|r_1=0)$ mixture describes QCD jets,
while the top mode peaks at $r_1 = 1$, indicating that the
$p^\text{D}_\theta(x|r_1 = 1)$ mixture describes top jets. This means that a
DVAE indeed solves the fundamental problem of detecting anomalies
symmetrically and without the complexity bias of the standard AE
architecture.

\subsubsection{Normalized autoencoder}
\label{sec:auto_ano_norm}

The problem with AE and VAE applications to anomaly searches leads us
to the question what it means for a jet to be anomalous. While the AE
only relies on a bottleneck and the compressibility of the jet
features, the DVAE adds the notion of a properly defined latent
space. However, Eq.\eqref{eq:dvae_loss} still combines an MSE-like
reconstruction loss with a shaped latent space, falling short of the
kind of likelihood losses or probabilistically interpretable losses we
have learned to appreciate.  A normalised autoencoder (NAE) goes one
step further and constructs a statistically interpretable latent
space, a bridge from an out-of-distribution definition of anomalies to
density-based anomalies.

We start by introducing \underline{energy-based models} (EBMs)\index{energy-based network}, a
class of models which can estimate probability densities especially 
well. They are defined through a normalizable energy function,
which is minimized during training. The energy function can be chosen
as any non-linear mapping of a phase space point to a scalar value,
\begin{align}
  E_\theta(x):\;  \mathbb{R}^D \rightarrow \mathbb{R} \; ,
  \label{eq:energy}
\end{align}
where $D$ is the dimensionality of the phase space. This kind of
mapping of a complex and often noisy distribution to a single system
energy can be motivated from statistical physics.

The EBM uses this energy function to define the loss based on a
Boltzmann distribution describing the probability density over phase
space
\begin{align}
p_\theta(x) = \frac{e^{-E_\theta(x)}}{Z_\theta}
\qquad \text{with} \qquad 
Z_\theta = \int_x dx e^{-E_\theta(x)} \; ,
\label{eq:ebm}
\end{align}
with the partition function $Z_\theta$.  The main practical feature of
a Boltzmann distribution is that low-energy states have the highest
probability.  Furthermore, the Boltzmann distribution has two
advantages: first, it can easily be integrated, which means we can at
least hope to compute $Z_\theta$. Second, it can be singled out as the
probability distribution $p_\theta(x)$ which gives the largest entropy
\begin{align}
S = - \int dx \; p_\theta(x) \log p_\theta(x) \; ,
\end{align}
translating into a large model flexibility. Going beyond our usual
likelihood loss, the loss of our normalized AE is the
\underline{negative logarithmic probability}
\begin{align}
  \boxed{
  \loss_\text{NAE}
  = - \Langle \log  p_\theta(x) \Rangle_{\pd} 
  = \Langle E_\theta(x) + \log Z_\theta \Rangle_{\pd}
  } \; ,
\label{eq:ebm_loss}
\end{align}
where we define the total loss as the expectation value over the
per-sample loss. Unlike for the VAE, the loss really minimizes the
posterior of the data distribution to be correct, as a function of the
network parameters $\theta$; this is nothing but a likelihood
loss\index{likelihood loss}.  The difference between the EBM and
typical implementations of likelihood-ratio losses is that we do not
use Bayes' theorem\index{Bayes' theorem} and a prior, so the
normalization term depends on $\theta$ and becomes part of the
training.

To train the network we want to minimize the loss in
Eq.\eqref{eq:ebm_loss}, so we have to compute the gradient of the full
probability,
\begin{align}
- \nabla_\theta \log p_\theta(x) 
&= \nabla_\theta E_\theta(x)
+ \nabla_\theta \log Z_\theta \notag \\
  &=  \nabla_\theta E_\theta(x)
  + \frac{1}{Z_\theta} \nabla_\theta \int_x dx e^{-E_\theta(x)} \notag \\
  &=  \nabla_\theta E_\theta(x)
  -  \int_x dx \frac{e^{-E_\theta(x)}}{Z_\theta} \nabla_\theta E_\theta(x) \notag \\
&=  \nabla_\theta E_\theta(x)
-  \XLangle \nabla_\theta E_\theta(x) \XRangle_{p_\theta} \; .
\label{eq:dec}
\end{align}
The first term in this expression can be obtained using automatic
differentiation from the training sample, while the second term is
intractable and must be approximated.

In the minimization of the loss we evaluate the gradient of the loss
as the expectation value over $\pd(x)$. This allows us to rewrite the
gradient of the loss as the difference of two energy gradients
\begin{align}
 \XLangle - \nabla_\theta \log p_\theta(x)\XRangle_{\pd} 
=  \XLangle \nabla_\theta E_\theta(x) \XRangle_{\pd}
- \XLangle \nabla_\theta E_\theta(x) \XRangle_{p_\theta} \; .
\label{eq:diff}
\end{align}
The first term samples from the training data, the second from the
model.  According to the sign of the energy in the loss function, the
contribution from the training dataset is referred to as positive
energy and the contribution from the model as negative energy. There
are three ways to to look at the second term, induced by the
normalization constant $Z_\theta$, in the loss gradient: (i) as a
normalization which ensures that the loss vanishes for $p_\theta (x) =
\pd(x)$, similar to the usual likelihood ratio loss of
Eq.\eqref{eq:def_ce2}; (ii) as a sampling covering both, the data
distribution and the model distribution, a little like a forward and
reverse KL-divergence; and (iii) as a \underline{background
  structure} into the minimization of the likelihood minimization.

Looking at the loss in Eq.\eqref{eq:diff} we can identify the training
as a minmax problem, where we minimize the energy (or MSE) of the
training samples and maximize the energy (or MSE) of the modelled
samples. While the energy of training data points is pushed downwards,
the energy of points sampled from the model distribution will be
pushed upwards.  For instance, if $p_\theta(x)$ reproduces $\pd(x)$
over most of the phase space $x$, but $p_\theta(x)$ includes an
additional mode, the phase space region corresponding to this extra
mode will be assigned large $E_\theta(x)$ through the minimization of
the loss.  This process of adjusting the energy continues until the
model reaches the equilibrium in which the model distribution is
identical to the training data distribution.

\begin{figure}[t]
  \includegraphics[width=0.4\textwidth]{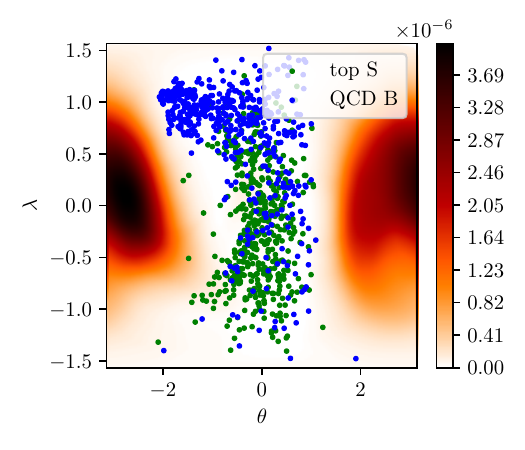}
  \hspace*{0.1\textwidth}
  \includegraphics[width=0.4\textwidth]{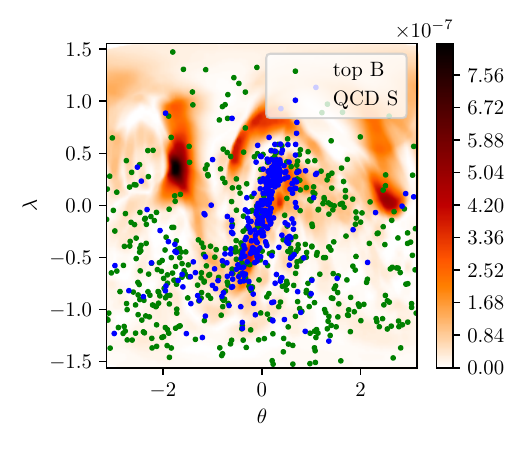} \\
  \includegraphics[width=0.4\textwidth]{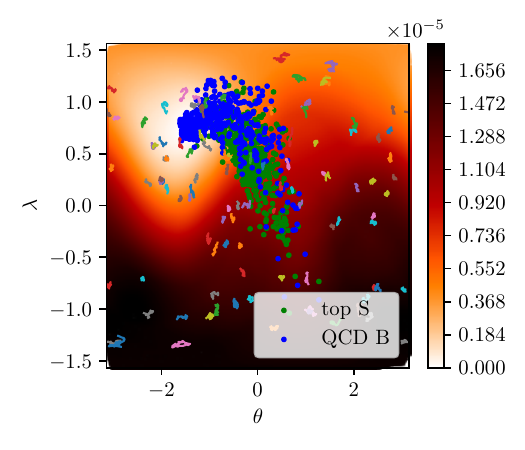}
  \hspace*{0.1\textwidth}
  \includegraphics[width=0.4\textwidth]{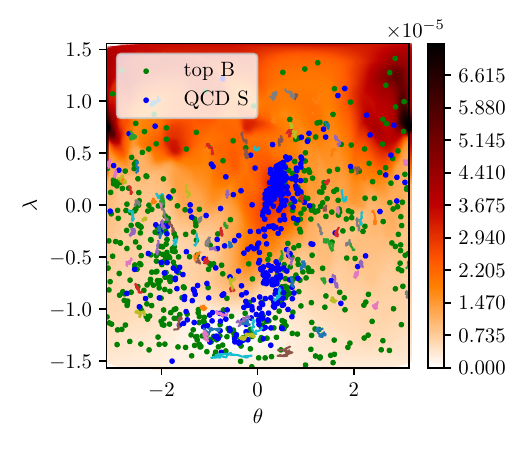}
  \caption{Equirectangular projection of the latent space after
    pre-training (upper) and after NAE training (lower). The $x$- and
    $y$-axis are the longitude and latitude on the latent sphere. We
    train on QCD jets (left) and on top jets (right). The lines
    represent the path of the LMCs in the current iteration. Figure
    from Ref.~\cite{Dillon:2022mkq}.}
  \label{fig:top_lat}
\end{figure}

One practical way of sampling from $p_\theta(x)$ is to use
\underline{Markov-Chain Monte Carlo} (MCMC).  The NAE uses Langevin
Markov Chains, where the steps are defined by drifting a random walk
towards high probability points according to
\begin{align}
  x_{t+1} = x_t + \lambda_x \nabla_x \log p_\theta(x) + \sigma_x \epsilon_t
  \qquad \text{with} \qquad
  \epsilon_t &\sim \normal(0,1) \; .
\label{eq:lmc}
\end{align}
Here, $\lambda$ is the step size and $\sigma$ the noise standard
deviation. When $2\lambda = \sigma^2$ the equation resembles Brownian
motion and gives exact samples from $p_\theta(x)$ in the limit of
$t\rightarrow +\infty$ and $\sigma \rightarrow 0$.

For ML applications working on images, the high dimensionality of the
data makes it difficult to cover the entire physics space $x$ with
Markov chains of reasonable length. For this reason, it is common to
use shorter chains and to choose $\lambda$ and $\sigma$ to place more
weight on the gradient term than on the noise term. If $2\lambda \neq
\sigma^2$, this is equivalent to sampling from the distribution at a
different temperature
\begin{align}
  T = \frac{\sigma^2}{2\lambda} \; .
\end{align}
By upweighting $\lambda$ or downweighting $\sigma$ we are effectively
sampling from the distribution at a low temperature, thereby
converging more quickly to the modes of the distribution.

Despite the well-defined algorithm, training EBMs is difficult due to
instabilities arising from (i) the minmax optimization, with similar
dynamics to balancing a generator and discriminator in a GAN; (ii)
potentially biased sampling from the MCMC due to a low effective
temperature; and (iii) instabilities in the LMC chains.  Altogether,
stabilizing the training during its different phases requires serious
effort.

Because the EBM only constructs a normalized probability loss based on
whatever energy function we give it, we can upgrade a standard AE with
the encoder-decoder structure,
\begin{align}
  f_\theta(x): \;
  \mathbb{R}^D \rightarrow \mathbb{R}^{D_z} \rightarrow \mathbb{R}^D \; .
\end{align}
The training minimizes the per-pixel difference between the original
input and its mapping, so we upgrade the AE to a probabilistic NAE by
using the MSE as the energy function in Eq.\eqref{eq:energy}
\begin{align}
  \boxed{
    E_\theta(x) = \text{MSE} = \left| x - f_\theta(x) \right|^2
    } \; .
\end{align}
By using the reconstruction uncertainty as the energy, the model will learn
to poorly reconstruct inputs not in the training distribution. This
way it guarantees the behavior of the model all over phase space,
especially in the region close to but not in the training data
distribution. We cannot give such a guarantee for a standard
AE, which only sees the training distribution and could
assign arbitrary reconstruction scores to data outside this
distribution.

For the training of the NAE, specifically the estimation of the
normalization $Z_\theta$ we complement the standard MCMC in phase
space with the mapping between latent and phase spaces provided by the
AE.  If we accept that different initializations of the MCMC defined
in Eq.\eqref{eq:lmc} lead to different results, we can tune
$\lambda_x$ and $\sigma_x$ in such a way that we can use a sizeable
number of short, non-overlapping Markov chains. Next, we apply
On-Manifold Initialization (OMI). It is motivated by the observation
that sampling the full data space is inefficient due to its high
dimensionality, but the training data lies close to a low-dimensional
manifold embedded in the data space $x$. All we need to do is to
sample close to this manifold. Since we are using an AE this manifold
is defined implicitly as the image of the decoder network, meaning
that any point in the latent space $z$ passed through the decoder will
lie on the manifold. This means we can first focus on the manifold by
taking samples from a suitably defined distribution in the
low-dimensional latent space, and then map these samples into data
space via the decoder. After that, we perform a series of MCMC steps
in the full ambient data space to allow the Markov chains to minimize
the loss around the manifold.  During the OMI it is crucial that we
cover the entire latent space, thus a \underline{compact latent space}
is preferable.  For this purpose we normalize the latent vectors so
that they lie on the surface of a hypersphere $\mathbb{S}^{D_z-1}$,
allowing for a uniform sampling of the initial batch in the latent
space.

As usual, we apply the NAE to unsupervised top tagging\index{jet tagging}, after training
on QCD jets only, and vice versa.  For the latent space dimension we
use $D_z = 3$, which allows us to visualize the latent space nicely.
Before starting the NAE training we pre-train the network using the
standard AE procedure with the standard MSE loss.  In the upper panels
of Fig.~\ref{fig:top_lat} we show a projection of the latent space
after training the usual AE.  In the left panels we train on the
simpler QCD background, which means that the latent space has a simple
structure. The QCD jets are distributed widely over the low-energy
region, while the anomalous top jets cluster slightly away from the
QCD jets. This changes when we train on the more complex top jets, as
shown in the right panels. The latent MSE-landscape reflects this
complex structure with many minima, and top jets spread over most of
the sphere. After the NAE training, only the regions populated by
training data have a low MSE. The sampling procedure has shaped the
decoder manifold to correctly reconstruct only training jet
images. For both training directions, the Markov chains move from a
uniform distribution to mostly cover the region with low MSE, leading
to an improved separation of the respective backgrounds and
signals.

\begin{figure}[t]
  \centering
  \includegraphics[width=0.4\textwidth]{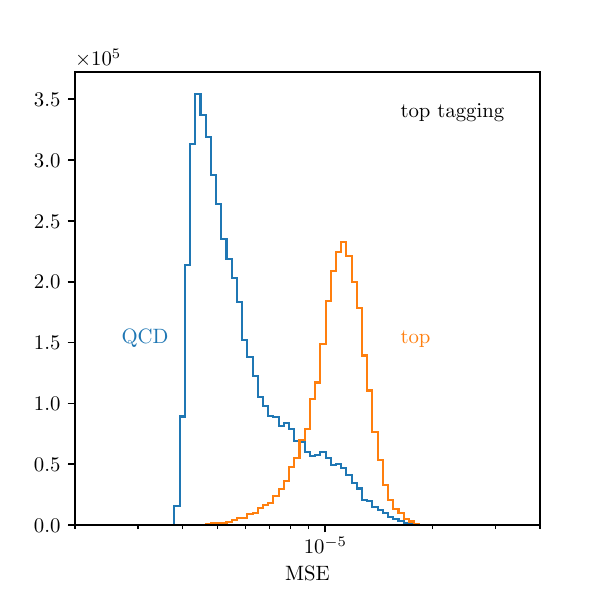}
  \hspace*{0.1\textwidth}
  \includegraphics[width=0.4\textwidth]{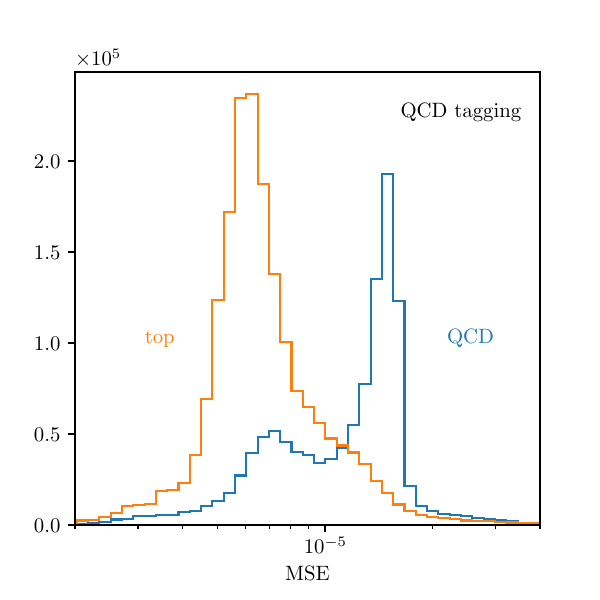}
  \caption{Distribution of the MSE after training on QCD jets (left)
    and on top jets (right). We show the MSE for QCD jets (blue) and
    top jets (orange) in both cases. Figure from
    Ref.~\cite{Dillon:2022mkq}.}
  \label{fig:top_hist}
\end{figure}
 
To show how the NAE works \underline{symmetrically} for anomalous tops
and for anomalous QCD jets, we can look at the respective MSE
distributions.  In the left panel of Fig.~\ref{fig:top_hist} we first
see the result after training the NAE on QCD jets. The MSE values for
the background are peaked strongly, cut off below $4 \cdot 10^{-5}$
and with a smooth tail towards larger MSE values. The MSE distribution
for top jets is peaked at larger values, and again with an
unstructured tail into the QCD region.  Alternatively, we see what
happens when we train on top jets and search for the simpler QCD jets
as an anomaly. In the right panel of Fig.~\ref{fig:top_hist} the
background MSE is much broader, with a significant tail also towards
small MSE values. The QCD distribution develops two distinct peaks, an
expected peak in the tail of the top distribution and an additional
peak under the top peak. The fact that the NAE manages to push the QCD
jets towards larger MSE values indicates that the NAE works beyond the
compressibility ordering of the simple AE. However, the second peak
shows that a fraction of QCD jets look just like top jets to the NAE.

The success of the NAE shows that anomaly searches\index{anomaly detection} at the LHC are
possible, but we need to think about the definition of anomalous jets
and are naturally lead to phase space densities. This is why we will
leave the topic of anomaly searches for now, work on methods for
density estimation, and return to the topic in
Sec.~\ref{sec:gen_cathode}.

\clearpage
\section{Generation and simulation}
\label{sec:gen}

In the previous chapters we have discussed mainly classification using
modern machine learning, using all kinds of supervised and
unsupervised training. We have seen how this allows us to extract more
complete information and significantly improve LHC analyses. However,
classification is not the same as modern analyses in the sense that
particle physics analyses have to be related to some kind of
fundamental physics question. This means we can either measure a
fundamental parameters of the Standard Model Lagrangian, or we can
search for physics beyond the Standard Model. Measuring a Wilson
coefficient or coupling of the effective field theory version of the
Standard Model is the modern way to unify these two approaches. To
interpret LHC measurements in such a theory framework the central tool
are simulations --- how do we get from a Lagrangian to a prediction
for observed LHC events in one of the detectors?

Simulations, or \underline{event generation}\index{event generators} based on perturbative QFT
is where modern machine learning benefits the theory side of particle
physics most. Their input to our simulations is a Lagrangian, from
which we can extract Feynman rules, which describe the interactions of
the particles we want to produce at the LHC. Using these Feynman rules
we then compute the transition amplitudes for the partonic LHC process
at a given order in perturbation theory.  We have learned how to
approximate these amplitudes with NN-surrogates in
Sec.~\ref{sec:basics_regr_amp}.  As illustrated in
Fig.~\ref{fig:simchain} this includes the production and decays, if we
want to separate them, as well as the additional jet radiation from
the partons in the initial and final states. Just as the parton
splittings forming jets, discussed in Sec.~\ref{sec:class_cnn_tag},
this part of the simulation is described by the QCD parton
splittings. Fragmentation, of the formation and decay of hadrons out
of partons is, admittedly, the weak spot of perturbative QCD when it
comes to LHC predictions. Finally, we need to describe the detector
response through a precision simulation, which is historically in the
hands of the experimental collaborations. For theory studies we rely
on fast detector simulations, like Delphes, as mentioned in our
discussion of the top tagging dataset in
Sec.~\ref{sec:class_cnn_sample}. It turns out that from an
ML-perspective event generation and detector simulations are similar
tasks, requiring the same kind of generative networks introduced in
this section.

The concept which allows us to apply modern machine learning to LHC
event generation and simulations is generative networks. To train
them, we start with a dataset which implicitly encodes a probability
density over a physics or phase space. A generative network learns
this underlying density and maps it to a latent space from which we
can then sample using for example flat or Gaussian random
numbers,
\begin{align}
  r \sim \pl(r)
  \quad \to \quad
  x = f_\theta(r) \sim \pmd(x) \approx \pd(x)  \; .
\label{eq:generative_basic}
\end{align}
The last step represents the network training, for instance in terms
of a variational approximation. A typical latent distribution is the
standard multi-dimensional Gaussian,
\begin{align}
    \pl(r) = \normal(0,1) \; .
\end{align} 
Generative networks allow us to produce samples following a learned
distribution. The generated data should then have the same form as the
training data, in which case the generative network will produce
statistically independent samples reproducing the implicit underlying
structures of the training data. Because we train on a distribution of
events and there are no labels or any other truth information about
the learned phase space density, generative network training is
considered unsupervised.

If the network is trained to learn a phase space density, we expect
generative networks to require us to compare different distributions,
for instance the training distributions $\pd(x)$ and the encoded
density $\pmd(x)$.  We already know one way to compare the actual and
the generated phase space densities from
Eq.\eqref{eq:kl_twofold}. However, the
KL-divergence\index{KL-divergence} is only one way to compare such a
probability distributions and part of a much bigger field called
\underline{optimal transport}. We remind ourselves of the definition
of the KL-divergence,
\begin{align}
  \kl [\pd,\pmd]
= \XXLangle \log \frac{\pd(x)}{\pmd(x)} \XXRangle_{\pd}  
= \int d x \; \pd(x) \; \log \frac{\pd(x)}{\pmd(x)} \; ,
\label{eq:def_kl3}
\end{align}
The KL-divergence between two identical distributions is zero. A
disadvantages of this measure is that it is not symmetric, which in
the above form means that phase space regions where we do not have
data will not contribute to the comparison of the two probability
distributions. Two distributions with zero overlap have infinite
KL-divergence. If the asymmetric form of the KL-divergence turns out
to be a problem, we can easily repair it by introducing the
\underline{Jensen-Shannon divergence}
\begin{align}
  D_\text{JS}[\pd,\pmd]
  &= \frac{1}{2} \left[
       \kl \left[ \pd,\frac{\pd+\pmd}{2} \right]
    +  \kl \left[ \pmd,\frac{\pd+\pmd}{2} \right] \right] \notag \\
  &= \frac{1}{2} \left[
      \int d x \; \pd(x) \; \log \frac{2\pd(x)}{\pd(x)+\pmd(x)} 
    + \int d x \; \pmd(x) \; \log \frac{2\pmd(x)}{\pd(x)+\pmd(x)} 
    \right] \notag \\
  &= \frac{1}{2} 
  \int d x \;
  \left[ \pd(x) \; \log \frac{\pd(x)}{\pd(x)+\pmd(x)} 
       + \pmd(x) \; \log \frac{\pmd(x)}{\pd(x)+\pmd(x)} 
    \right] + \log 2 \notag \\
  &\equiv
  \frac{1}{2} \XXLangle 
  \log \frac{\pd(x)}{\pd(x)+\pmd(x)} 
  \XXRangle_{\pd}
  + \frac{1}{2} \XXLangle
  \log \frac{\pmd(x)}{\pd(x)+\pmd(x)} 
  \XXRangle_{\pmd}
  + \log 2 \; .
  \label{eq:def_js}
\end{align}
The JS-divergence between two identical distributions also vanishes,
but because it samples from both, $\pd$ and $\pmd$, it will not
explode when the distributions have zero overlap.  Instead, we find in
this limit
\begin{align}
  D_\text{JS}[\pd,\pmd]
  \to \frac{1}{2} 
  \int d x \;
  \left[ \pd(x) \; \log \frac{\pd(x)}{\pd(x)} 
       + \pmd(x) \; \log \frac{\pmd(x)}{\pmd(x)} 
    \right] + \log 2 
  = \log 2 \; .
\end{align}
On thing the KL-divergence and the JS-divergence have in common is
that they calculate the difference between two distributions based on
the log-ratio of two functional values. Consequently, they reach their
maximal values for two distributions with no overlap in $x$, no matter
how the two distributions look. This is counter-intuitive, because a
distance measure should notice the difference between two identical
and two very different distributions with vanishing overlap, for
instance a two Gaussians vs a Gaussian and a double-Gaussian. This
brings us to the next distance measure, which is meant to work
horizontally in the sense that it guarantees for example
\begin{align}
  W[\pd(x),\pmd(x) = \pd(x-a)] \approx a \; .
  \label{eq:wasserstein_shift}
\end{align}
This \underline{Wasserstein distance} or earth mover distance can be
most easily defined for weighted sets of points defining each of the
two distributions $\pd(x)$ and $\pmd(y)$ in a discretized description
\begin{align}
  M_\text{data} = \sum_{i=1}^{N_1} \pdj{i} \delta_{x_i}
  \qquad \text{and} \qquad 
  M_\text{model} = \sum_{j=1}^{N_2} \pmdj{j} \delta_{y_j}  \; .
\end{align}
We then define a preserving transport strategy as a matrix relating
the two sets, namely
\begin{align}
  \pi_{ij} \ge 0
  \qquad \text{with} \qquad 
  \frac{1}{N_2} \sum_j \pi_{ij} = \pdj{i} \qquad
  \frac{1}{N_1} \sum_i \pi_{ij} = \pmdj{j} \; .
\end{align}
The first normalization condition ensures that all entries in the model
distributions $j$ combined with the data entry $i$ reproduce the full
data distribution, the second normalization condition works the other
way around.  We define the distance between the two represented
distributions as
\begin{align}
W[\pd,\pmd] = \min_\pi \frac{1}{N_1 N_2} \sum_{i,j} \vert x_i-y_j \vert \; \pi_{ij} \; ,
\end{align}
where the minimum condition implies that we choose the best transport
strategy. For our example from Eq.\eqref{eq:wasserstein_shift} with
two identical functions this strategy would give $x_i - y_j = a$.
Alternatively, we can write the Wasserstein distance as an expectation
value of the distance between two points. This distance has to be
sampled over the combined probability distributions and then minimized
over the so-called transport plan,
\begin{align}
  W[\pd,\pmd] = \min \XLangle \vert x - y \vert \XRangle_{\pd(x),\pmd(y)}
\end{align}
From the algorithmic definition we see that computing the Wasserstein
distance is expensive and scales poorly with the number of points in
our samples. We will see that these three different distances between
distributions can be used for different generative networks, to learn
and then sample from an underlying phase space distribution.

Finally, if we want to test the performance of a generative network a
classifier or discriminator trained to distinguish training data and
generated data seems an obvious choice. As a matter of fact, this kind
of comparison can already be part of the network training, as we will
see in Secs.~\ref{sec:gen_gan} and~\ref{sec:gen_inn_events}. the reason
why we are mentioning this here is that the Neyman-Pearson lemma tells
us that in this case the discriminator has to learn the likelihood
ratio or a simple variation of it. For a generative network over phase
space this means we can extract the scalar field $\pmd(x)/\pd(x)$ as
the unsupervised counterpart to the agreement between a regression
network and its training data from Eq.\eqref{eq:ampl_delta}.

\subsection{Variational autoencoders}
\label{sec:gen_vae}

We studied autoencoders and variational autoencoders already in
Sec.~\ref{sec:auto_ano_ae}, with the idea to map a physics space onto
itself with an additional bottleneck (AE) and an induced latent space
structure in this bottleneck (VAE). Looking at their network
architecture from a generative network point of view, we can also
sample from this latent space with corresponding random numbers, for
instance with a multi-dimensional Gaussian distribution.  In that case
the decoder part of the VAE will generate events corresponding to the
properties translated from the input phase space to the latent space
through the encoder. This means we already know one simple generative
network.

In Sec.~\ref{sec:auto_ano_ae} we introduced the two-term VAE loss
function somewhat ad hoc and without any reference to a probability
distribution or a stochastic justification. Let us now tackle it a
little more systematically. We start by assuming that we know the data
$x$ and implicitly its probability distribution $\pd(x)$. We also want
to enforce a given latent distributions $\pl(r)$. In that case the the
\underline{encoder} generates according to the conditional probability
$\pmd(r|x)$, while the \underline{generator} is described by
$\pmd(x|r)$.

We start with the encoder training, which should approximate something
like $\pmd(r|x) \sim \pl(r)$. However, this condition cannot be the
final word, since it missed the conditional structure. Instead, we
need to construct the reference distribution $p(r|x)$ from Bayes'
theorem
\begin{align}
  p(r|x) \; \pd(x)
  = \pmd(x|r) \; \pl(r) \; .
\end{align}
On the left side we have to complete the reference distribution by the
data-defined $\pd(x)$, while on the right side we combine the
conditional generator with the known latent distribution.

To train the encoder we resort to the variational approximation from
Sec.~\ref{sec:basics_deep_bayes}, specifically
Eq.\eqref{eq:first_bayes}.  The goal is to construct a network
function $\pmd(r|x)$ which approximates $p(r|x)$, just like in
Eq.\eqref{eq:first_bayes}. As before, we construct this approximation
using the KL-divergence\index{KL-divergence} of Eq.\eqref{eq:def_kl},
\begin{align}
\kl [\pmd(r|x),p(r|x)]
  &= \XXLangle  \log \frac{\pmd(r|x)}{p(r|x)} \XXRangle_{\pmd(r|x)} \notag \\
  &= \XXLangle  \log \pmd(r|x)  - \log \frac{\pmd(x|r) \pl(r)}{\pd(x)} \XXRangle_{\pmd(r|x)} \notag \\
  &= - \XLangle \log \pmd(x|r) \XRangle_{\pmd(r|x)}
    + \XLangle \log \pmd(r|x) - \log  \pl(r) \XRangle_{\pmd(r|x)} + \log \pd(x) \notag \\
  &= - \XLangle \log \pmd(x|r) \XRangle_{\pmd(r|x)}
    + \kl [ \pmd(r|x), \pl(r)]  + \log \pd(x) \; .
\end{align}
As before, the evidence $\pd$ is independent of our network training,
which means we can use $\kl [\pmd(r|x),p(r|x)]$ modulo the last term as
the VAE loss function. Also denoting that everything is always
evaluated on batches of events from the training dataset we reproduce
Eq.\eqref{eq:vaeloss}, namely
\begin{align}
  \boxed{
    \loss_\text{VAE}
    = \XLangle - \Langle \log \pmd(x|r) \Rangle_{\pmd(r|x)} + \beta_\text{KL} \kl [\pmd(r|x), \pl(r)] \XRangle_{\pd}
    } \; .
\label{eq:vaeloss2}
\end{align}
We introduce the parameter $\beta_\text{KL}$ to allow for a little
more flexibility in balancing the two training tasks which are
otherwise linked through Bayes' theorem.  Just like the Bayesian
network, the VAE loss function derived through the variational
approximation includes a KL-divergence as a regularization.

As mentioned before, the structure of the latent space $r$, from which
we sample, is introduced by the prior $\pl(r) = \normal(0,1)$ If
we control the $r$-distribution, we can consider the conditional
decoder $\pmd(x|r)$ a generative network producing events with the
probability we desire. The problem with variational AEs in particle
physics is that they rely on the assumption that all features in the
data can be compressed into a low-dimensional and limited latent
space. In the mix of expressivity and achievable accuracy VAEs are
usually not competitive with the other generative network
architectures we will discuss next. An exception might be detector
simulations, where the underlying physics of calorimeter showers is
simple enough to be encoded in a low-dimensional latent space, while
the space of detector output channels is huge.

\subsection{Generative Adversarial Networks}
\label{sec:gen_gan}

In our discussion of the VAE we have seen that generative
networks are structurally more complex then regression or
classification networks. In the language of probability distributions
and likelihood losses, we need a generator or decoder
network which relies on a conditional probability $\pmd(x|r)$ for the
target phase space distribution $x$ given the distribution of the
incoming random numbers $r$. For the VAE we used the variational
inference trick to construct this latent representation.

An alternative way to learn the underlying density is to combine two
networks with \underline{adversarial training}. Adversarial training
means we combine two loss functions
\begin{align}
  \boxed{
    \loss_\text{adv} = \loss_1 - \lambda \loss_2
    } \; ,
\label{eq:def_adv}
\end{align}
where the first loss can for example train a classifier and the second
term can compute an observable we want to decorrelate. The second
network uses the information from the first, classifier
network. Because of the negative sign the two sub-networks will now
play with each other to find a combined minimum of the loss
function. An excellent classifier with small $\loss_1$ will use and
correctly reproduce the to-be-decorrelation variable, implying also a
small $\loss_2$. However, for large enough $\lambda$ the two networks
can also work towards a smaller combined loss, where the classifier
becomes less ambitious, indicated by a finite $\loss_1$, but
compensated by an even larger $\loss_2$. This balanced gain works
best if the classifier is trained as well as possible, but leaving out
precisely the aspects which allow for large values of $\loss_2$. The
two networks playing against each other will then find a compromise
where variables entering $\loss_2$ are ignored in the classifier
training represented by $\loss_1$.

Mathematically, the constructive balance or compromise of two players
is called a \underline{Nash equilibrium}. Varying the coupling
$\lambda$ we can strengthen and weaken either of the two sub-networks,
a stable Nash equilibrium means that without much tuning of $\lambda$
the two networks settle into a combined minimum. This does not have to
be the case, a combination of two networks can of course be unstable
in the sense that depending on the size of $\lambda$ either $\loss_1$
or $\loss_2$ wins. Another danger in adversarial training is that the
adversary network might force the original network to construct
nonsense solutions, which we have not thought about, but which
formally minimize the combined loss.  In the beginning of
Sec.~\ref{sec:gen} we have observed a mechanism which can lead to such
poor solution, where the KL-divergence is insensitive to a massive
disagreement between data and network, as long as the distribution we
sample from in Eq.\eqref{eq:def_kl3} vanishes. In this section we will
use adversarial training to construct a generative network.

\subsubsection{Architecture}
\label{sec:gen_gan_arch}

Similar to the VAE structure, the first element of a generative
adversarial network (GAN) is the learned generator, just like the
VAE decoder
\begin{align}
  \pmd(x|r) \Bigg|_{\pl(r)=\normal(0,1)} \; .
\label{eq:gan_random}
\end{align}
Now the latent space $r$ is replaced by a random number generator
for $r$, following some simple Gaussian or flat distribution. The argument
$x$ is the physical phase space of a jet, a scattering process at the
LHC, or a detector output. We remind ourselves that (unweighted)
events are nothing but positions in phase space. The difference to the
VAE is that we do not train the generator as an inversion or
encoder, but use an adversarial loss function like the one shown in
Eq.\eqref{eq:def_adv}. The GAN architecture is illustrated in
Fig.~\ref{fig:arch_gan1}.

\begin{figure}[b!]
\centering
\begin{tikzpicture}[node distance=2cm]

\node (generator) [generator] {Generator};
\node (random) [process,left of=generator, xshift=-1cm] {$\{ r \}$};
\draw [arrow, color=black] (random) -- (generator);

\node (in1) [io, right of=generator, xshift=2cm] {$\{ x_G \}$};
\draw [arrow,color=Gcolor] (generator) -- (in1);

\node (in2) [io, right of=in1] {$\{ x_T \}$};
\node (data) [process, right of=in2, xshift=2cm] {MC Data};
\draw [arrow,color=black] (data) -- ( in2);

\node (discriminator) [discriminator, below of=in2,xshift=1cm] {Discriminator};

\draw [arrow, color=Gcolor] (in1.325) -- (discriminator);

\draw [arrow, color=black] (in2) -- (discriminator);

\node (gloss) [decision, below of=in1] {$\loss_G$};
\node (dloss) [decision, below of=discriminator] {$\loss_D$};
\draw [arrow,color=Dcolor] (discriminator) -- (dloss);
\draw [arrow,color=Gcolor] (discriminator) -- (gloss);

\draw [arrow,dashed, color=Gcolor] (gloss) -| (generator);
\coordinate[right of=dloss, xshift = 1cm](a);
\coordinate[above of=a](b);

\draw[thick,dashed,color=Dcolor] (dloss) -- (a);
\draw[thick,dashed,color=Dcolor] (a) -- (b);
\draw[arrow,dashed, color=Dcolor] (b) -- (discriminator);

\end{tikzpicture}
\caption{Schematic diagram for a GAN.  The input $\{r\}$ describes a
  batch of random numbers, $\{ x \}$ denotes a batch of phase space
  points sampled either from the generator or the training data.}
\label{fig:arch_gan1}
\end{figure}

We know from Sec.~\ref{sec:class} that it is not hard to train a
classification network for jets or events, which means that given a
reference dataset $\pd(x)$ and a generated dataset $\pmd(x)$ we can
train a discriminator or classification network to tell apart the true
data and the generated data phase space point by phase space point.
This discriminator network is trained to give
\begin{align}
  D(x) =
  \begin{cases}
    0 & \qquad \text{generated data} \\
    1 & \qquad \text{true data} \\
  \end{cases}
\end{align}
and values in between otherwise. If our discriminator is set up as a
proper classification network, its output can be interpreted as the
probability of an event being true data. Given a true dataset and a
generated dataset, we can train the discriminator to minimize any
combination of
\begin{align}
  \Langle 1 - D(x) \Rangle_{\pd}
  \qquad \text{and} \qquad 
  \Langle D(x) \Rangle_{\pmd} \; .
  \label{eq:gan_naive}
\end{align}
For a perfectly trained discriminator both terms will vanish.  On the
other hand, we know from Sec.~\ref{sec:class} that the loss
function for such classification task should be the cross entropy,
which motivates the discriminator loss
\begin{align}
  \loss_D
  &= 
\Langle -\log D(x) \Rangle_{\pd} 
+ \Langle - \log [1-D(x)] \Rangle_{\pmd} \notag \\
  &= -
  \int dx \Big[ \pd(x) \log D(x) + \pmd(x) \log (1-D(x)) \Big] \; .
\label{eq:gan_dloss}
\end{align}
Comparing this form to the two objectives in Eq.\eqref{eq:gan_naive}
we simply enhance the sensitivity by replacing $1 - D \to -\log
D$. The loss is always positive, and a perfect discriminator will
produce zeros for both contributions.  From the discriminator loss, we
can compute the \underline{optimal discriminator} output
\begin{align}
  \frac{\delta}{\delta D} \Big[ \pd(x) \log D + \pmd(x) \log (1-D) \Big]
  = \frac{\pd(x)}{D} - \frac{\pmd(x)}{1-D} 
  & = 0 \notag \\
  \Leftrightarrow \qquad
  D_\text{opt}(x) \; \pmd(x) &= (1-D_\text{opt}(x)) \; \pd(x) \notag \\
  \Leftrightarrow \qquad
  D_\text{opt}(x) &= \frac{\pd(x)}{\pd(x) + \pmd(x)} \; ,
\label{eq:gan_dopt}
\end{align}
assuming that the maximum of the integrand also maximizes the integral
because of the positive probability distributions.  In general, this
formula says that the optimal discriminator is given by the ratio of
the two likelihoods. This results is much more generally known as the
\underline{Neyman-Pearson lemma}\index{Neyman-Pearson lemma}.  It
tells us that the ratio of the two likelihoods is the most powerful
test statistic to distinguish the two underlying hypotheses, defined
as the smallest false negative error for a given false positive rate.
This provides us with a statistically deep and extremely useful link
between classification networks and density estimation for instance
through generative networks. Using a classifier to extract a
likelihood ratio is called the \underline{likelihood ratio trick}.

To train the generator network $\pmd(x|r)$ we now use our adversarial
idea. The trained discriminator encodes the agreement of the true and
generated datasets, and all we need to do is evaluate it on the
generated dataset
\begin{align}
  \boxed{
    \loss_G= \Langle - \log D(x) \Rangle_{\pmd}
    } \; .
\label{eq:gan_gloss}
\end{align}
This loss will vanish when the discriminator (wrongly) identifies all
generated events as true events with $D=1$. 

In our GAN application this discriminator network gets successively
re-trained for a fixed true dataset and evolving generated data.  In
combination, training the discriminator and generator network based on
the losses of Eq.\eqref{eq:gan_dloss} and~\eqref{eq:gan_gloss} in an
alternating fashion forms an adversarial problem which the two
networks can solve amicably. The Nash equilibrium between the losses
implies, just like in Eq.\eqref{eq:def_adv}, that a perfectly trained
discriminator cannot tell apart the true and generated samples.

To match the literature, we can merge the two GAN losses
Eq.\eqref{eq:gan_dloss} and Eq.\eqref{eq:gan_gloss} into one
formula after replacing the sampling $x \sim \pmd(x)$ with a sampling
$r \sim \pl(r)$ and $x = G(r)$, 
\begin{align}
  \loss_D &
  = \Langle -\log D(x) \Rangle_{\pd} 
  + \Langle - \log [1-D(G(r))] \Rangle_{\pl} \notag \\
  \loss_G &
  = \Langle - \log D(G(r)) \Rangle_{\pl}
  \; \sim \; \Langle \log [ 1- D(G(r))] \Rangle_{\pl} \; .
  \label{eq:gan_combined}
\end{align}
For the generator loss we use the fact that minimizing $-\log D$ is
the same as maximizing $\log D$, which is again the same as minimizing
$\log (1-D)$ in the range $D \in [0,1]$. The $\sim$ indicates that the
two functions will lead to the same result in the minimization, but we
will see later that they differ by a finite amount.  After modifying
the generator loss we can write the two optimizations for the
discriminator and generator training as a \underline{min-max game}
\begin{align}
  \boxed{
    \min_G \max_D \;
  \Langle \log D(x) \Rangle_{\pd}
  +   \Langle \log [1-D(G(r))] \Rangle_{\pl}
  } \; .
  \label{eq:gan_minmax}
\end{align}
Finally, we can evaluate the discriminator and generator losses in the
limit of the optimally trained discriminator given in
Eq.\eqref{eq:gan_dopt},
\begin{align}
  \loss_D
  &\to 
  - \Langle \log D_\text{opt}(x) \Rangle_{\pd} 
- \Langle  \log [1-D_\text{opt}(x)] \Rangle_{\pmd} \notag \\
  &= 
  - \XXLangle \log \frac{\pd(x)}{\pd(x) + \pmd(x)} \XXRangle_{\pd} 
  - \XXLangle \log \frac{\pmd(x)}{\pd(x) + \pmd(x)} \XXRangle_{\pmd} \notag \\
  &\equiv - 2 D_\text{JS}[\pd,\pmd] + 2 \log 2 \; ,
\end{align}
just inserting the definition in Eq.\eqref{eq:def_js}. It shows where
the GAN will be superior to the KL-divergence-based VAE, because the
JS-divergence is more efficient at detecting a mismatch between
generated and training data.  For the same optimal discriminator the
modified generator loss from Eq.\eqref{eq:gan_combined} becomes
\begin{align}
  \loss_G
  &\to \Langle \log (1- D_\text{opt}(x)) \Rangle_{\pmd} \notag \\
  &= \XXLangle \log \frac{2\pmd(x)}{\pd(x) + \pmd(x)} \XXRangle_{\pmd} - \log 2 \notag \\
  &\equiv \kl \left[ \pmd,\frac{\pd+\pmd}{2} \right] - \log 2  \; .
  \label{eq:gan_gloss2}
\end{align}
For a perfectly trained discriminator and generator the Nash
equilibrium is given by
\begin{align}
  \pd(x) = \pmd(x)
  \qquad \Rightarrow \qquad 
  \loss_D = 2 \log 2
  \qquad \text{and} \qquad
  \loss_G = - \log 2 \; 
\end{align}
We can find the same result from the original definitions of
Eq.\eqref{eq:gan_dloss} and~\eqref{eq:gan_gloss}, using our correct
guess that the perfect discriminator in the Nash equilibrium is
constant, namely
\begin{align}
  D(x) = \frac{1}{2}
  \qquad \Rightarrow \qquad 
  \loss_D &= - \log \frac{1}{2} - \log \frac{1}{2} = 2 \log 2 \notag \\
  \loss_G &= - \log \frac{1}{2} = \log 2 \; .
\end{align}
The difference in the value for the generator loss correspond to the
respective definitions in Eq.\eqref{eq:gan_combined}.

The fact that the GAN training searches for a generator minimum given
a trained discriminator, which is different from the generator-alone
training, leads to the so-called \underline{mode collapse}. Starting
from the generator loss, we see in Eq.\eqref{eq:gan_gloss} that it
only depends on the discriminator output evaluated for generated
data. This means the generator can happily stick to a small number of
images or events which look fine to a poorly trained discriminator. In
the discriminator loss in Eq.\eqref{eq:gan_dloss} the second term
will, by definition, be happy with this generator output as well. From
our discussion of the KL-divergence we know that the first term in the
discriminator loss will also be fine if large gradients of $\log D(x)$
only appear in regions where the sampling through the training dataset $\pd(x)$ is
poor, which means for example unphysical regions.

After noticing that the JS-divergence of the GAN discriminator loss
improves over the KL-divergence-base VAE, we can go one step further
and use the Wasserstein distance between the distributions $\pd$ and
$\pmd$. The Wasserstein distance of two non-intersecting distributions
grows roughly linearly with their relative distance, leading to a
stable gradient. According to the Kantorovich-Rubinstein duality, the
Wasserstein distance between the training and generated distributions
is given by
\begin{align}
W(\pd,\pmd) = \max_D 
\left[ \Langle  D(x) \Rangle_{\pd}
-  \Langle D(x) \Rangle_{\pmd} \right]  \; .
\end{align}
For the WGAN the discriminator is also called critic.  The definition
of the Wasserstein distance involves a maximization in discriminator
space, so the discriminator has to be trained multiple times for each
generator update. A 1-Lipschitz condition can be enforced through a
maximum value of the discriminator weights. It can be be replaced by a
gradient penalty, as it is used for regular GANs.

\subsubsection{Event generation}
\label{sec:gen_gan_events}

\begin{figure}[b!]
\centering
\begin{tikzpicture}[node distance=2cm]

\node (generator) [generator] {Generator};
\node (random) [process,left of=generator, xshift=-1cm] {$\{ r \}, \{ m \}$};
\draw [arrow, color=black] (random) -- (generator);

\node (in1) [io, right of=generator, xshift=2cm] {$\{ x_G \}$};
\draw [arrow,color=Gcolor] (generator) -- (in1);

\node (in2) [io, right of=in1] {$\{ x_T \}$};
\node (data) [process, right of=in2, xshift=2cm] {MC Data};
\draw [arrow,color=black] (data) -- ( in2);

\node (discriminator) [discriminator, below of=in2,xshift=1cm] {Discriminator};
\node (mmd) [process, below of=in1,xshift=-1cm] {MMD$^2$};

\draw [arrow, color=Gcolor] (in1) -- (mmd);
\draw [arrow] (in2) -- (mmd);
\draw [arrow, color=Gcolor] (in1.325) -- (discriminator);

\draw [arrow, color=black] (in2) -- (discriminator);

\node (gloss) [decision, below of=mmd] {$L_G$};
\draw [arrow,color=Gcolor] (mmd) -- (gloss);
\node (dloss) [decision, below of=discriminator] {$L_D$};
\draw [arrow,color=Dcolor] (discriminator) -- (dloss);
\draw [arrow,color=Gcolor] (discriminator) -- (gloss);

\draw [arrow,dashed, color=Gcolor] (gloss) -| (generator);
\coordinate[right of=dloss, xshift = 1cm](a);
\coordinate[above of=a](b);

\draw[thick,dashed,color=Dcolor] (dloss) -- (a);
\draw[thick,dashed,color=Dcolor] (a) -- (b);
\draw[arrow,dashed, color=Dcolor] (b) -- (discriminator);

\end{tikzpicture}
\caption{Schematic diagram for an event-generation GAN. It corresponds
  to the generic GAN architecture in Fig.~\ref{fig:arch_gan1}, but
  adds the external masses to the input and the MMD loss defined in
  Eq.\eqref{eq:mmd}.}
\label{fig:arch_gan2}
\end{figure}

As discussed in Sec.~\ref{sec:basics_particle_sim}, event generation\index{event generators}
is at the heart of LHC theory. The standard approach is Monte Carlo
simulation, as indicated in Fig.~\ref{fig:simchain}, and in this
section we will describe how it can be supplemented by a generative
network. There are several motivations for training a generative
network on events: (i) we can use such a network to efficiently encode
and ship standard \underline{event samples}, rather than re-generating them every
time a group needs them; (ii) we will see in Sec~\ref{sec:gen_gan_amp}
that we can typically produce several times as many events using a
generative network than used for the training; (iii) understanding
generative networks for events allows us to test different ML-aspects
which can be used for phase space integration and generation; (iv) we
can train generative network flexibly at the parton level or at the
jet level; (v) finally, we will describe potential applications in the
following sections and use generative networks to construct inverse
networks.

The training dataset for event-generation networks are unweighted
events, in other words phase space points whose density represents a
probability distribution over phase space. One of the standard
reference processes is top pair production including decays,
\begin{align}
  pp
  \to t^* \bar{t}^*
  \to (b W^{+ *}) \; (\bar{b} W^{- *})
  \to (b u \bar{d}) \; (\bar{b} \bar{u} d)
\end{align}
The star indicates \underline{on-shell intermediate particles}, described by a
Breit-Wigner propagator and shown in the Feynman diagrams
\begin{center}
\begin{fmfgraph*}(120,90)
\fmfset{arrow_len}{2mm}
\fmfset{curly_len}{3mm}
\fmfset{wiggly_len}{3mm}
\fmfstraight
\fmfleft{i1,i2}
\fmfright{oj11,oj12,ob1,ob2,oj21,oj22}
\fmf{fermion,tension=1.0,width=0.6,label=$q$,lab.side=right}{i1,vp1}
\fmf{fermion,tension=1.0,width=0.6,label=$\bar q$,lab.side=right}{vp1,i2}
\fmf{gluon,tension=2.0,width=0.5}{vp1,vp2}
\fmf{fermion,tension=1.0,width=0.6}{ob1,vd11}
\fmf{fermion,tension=1.5,width=0.6,label=$t$,lab.side=left}{vd11,vp2}
\fmf{fermion,tension=1.5,width=0.6,label=$t$,lab.side=left}{vp2,vd21}
\fmf{fermion,tension=1.0,width=0.6}{vd21,ob2}
\fmf{photon,tension=1.0,width=0.6,label=$W$,lab.side=right}{vd11,vd12}		
\fmf{photon,tension=1.0,width=0.6,label=$W$,lab.side=left}{vd21,vd22}		
\fmf{fermion,tension=0.5,width=0.6}{oj12,vd12,oj11}
\fmf{fermion,tension=0.5,width=0.6}{oj22,vd22,oj21}
\fmflabel{$\bar d$}{oj11}
\fmflabel{$u$}{oj12}
\fmflabel{$b$}{ob1}
\fmflabel{$b$}{ob2}
\fmflabel{$\bar u$}{oj21}
\fmflabel{$d$}{oj22}
\end{fmfgraph*}
\hspace*{0.1\textwidth}
\begin{fmfgraph*}(130,100)
\fmfset{arrow_len}{2mm}
\fmfset{curly_len}{3mm}
\fmfset{wiggly_len}{3mm}
\fmfstraight
\fmfleft{i1,i2}
\fmfright{oj11,oj12,ob1,ob2,oj21,oj22}
\fmf{gluon,tension=1.0,width=0.6,label=$g$,lab.side=right}{i1,vp1}
\fmf{gluon,tension=1.0,width=0.6,label=$g$,lab.side=right}{vp2,i2}
\fmf{fermion,tension=0.7,width=0.5}{vp1,vp2}
\fmf{fermion,tension=1.0,width=0.6}{ob1,vd11}
\fmf{fermion,tension=1.5,width=0.6,label=$t$,lab.side=left}{vd11,vp1}
\fmf{fermion,tension=1.5,width=0.6,label=$t$,lab.side=left}{vp2,vd21}
\fmf{fermion,tension=1.0,width=0.6}{vd21,ob2}
\fmf{photon,tension=1.0,width=0.6,label=$W$,lab.side=right}{vd11,vd12}		
\fmf{photon,tension=1.0,width=0.6,label=$W$,lab.side=left}{vd21,vd22}		
\fmf{fermion,tension=0.5,width=0.6}{oj12,vd12,oj11}
\fmf{fermion,tension=0.5,width=0.6}{oj22,vd22,oj21}
\fmflabel{$\bar d$}{oj11}
\fmflabel{$u$}{oj12}
\fmflabel{$X$}{ob1}
\fmflabel{$b$}{ob2}
\fmflabel{$\bar u$}{oj21}
\fmflabel{$d$}{oj22}
\end{fmfgraph*}

\end{center}
For intermediate on-shell particles the denominator of the respective
propagator is regularized by extending it into the complex plane, with
a finite imaginary part~\cite{Plehn:2009nd}
\begin{align}
  \left|
  \frac{1}{s - m^2 + i m \Gamma}
  \right| ^2
  = \frac{1}{(s-m^2)^2 + m^2 \Gamma^2} \; .
\label{eq:breit_wigner}
\end{align}
By cutting the corresponding self-energy diagrams, $\Gamma$ can be
related to the decay width of the intermediate particle. In the limit
$\Gamma \ll m$ we reproduce the factorization into production rate and
branching ratio, combined with the on-shell phase space condition
\begin{align}
  \lim _{\Gamma \to 0}
  \frac{\Gamma_\text{part}}{(s-m^2)^2 + m^2 \Gamma^2}
  = \Gamma_\text{part} \; \frac{\pi}{\Gamma} \; \delta (s - m^2)
  = \pi \text{BR}_\text{part}  \; \delta (s - m^2)  
\end{align}
For a decay coupling $g$ the width scales like $\Gamma \sim m g^2$, so
weak-scale electroweak particles have widths in the GeV-range, which
means means the Breit-Wigner propagators for top pair production
define four sharp features in phase space.

As a first step we ignore additional jet radiation, so the phase space
dimensionality of the final state is constant. Each particle in the
final state is described by a 4-vector, which means the $t\bar{t}$
phase space has $6 \times 4 = 24$ dimensions. If we are only
interested in generating events, we can ignore the detailed kinematics
of the initial state, as it will be encoded in the training
data. However, we can simplify the phase space because all external
particles are on their mass shells and we can compute their energies
from their momenta,
\begin{align}
  p^2 = E^2 - m^2
  \qquad \Leftrightarrow \qquad
  E = \sqrt{p^2 + m^2} \; ,
\end{align}
leaving us with an 18-dimensional phase space and six final-state
masses as constant input to the network training. Another possible
simplification would be to use transverse momentum conservation
combined with the fact that the incoming partons have no momentum in
the azimuthal plane. We will now use this additional condition in the
network training and instead use it to test the accuracy of the
network. A symmetry\index{symmetries} we could use is the global azimuthal angle of the
process and replacing the azimuthal angles of all final state particle
with an azimuthal angle difference to one reference particle.

The main challenge of training a GAN to learn and produce $t\bar{t}$
events is that the Breit-Wigner propagators strongly constrain four of
the 18 phase space dimensions, but those directions are hard to
extract from a generic parametrization of the final state. To
construct the invariant mass of each of the tops the discriminator and
generator have to probe a 9-dimensional part of the phase space, where
each direction covers several 100~GeV to reproduce a top mass peak
with its width $\Gamma_t=1.5$~GeV. For a given LHC process and its
Feynman diagrams we know which external momenta form a resonance, so
we can construct the corresponding invariant mass and give it to the
neural network to streamline the comparison between true and generated
data. This is much less information than we usually use in Monte Carlo
simulations, where we define an efficient phase space mapping from the
known masses and widths of every intermediate resonance.

One way to focus the network on a low-dimensional part of the phase
space is the \underline{maximum mean discrepancy} (MMD) combined with
a kernel-based method to compare two samples drawn from different
distributions. Using one batch of training data points and one batch
of generated data points, it computes a distance between the
distributions as
\begin{align}
\text{MMD}^2(\pd,\pmd)
&= 
\Langle  k(x, x') \Rangle_{x, x' \sim \pd}
+ \Langle  k(y, y') \Rangle_{y, y' \sim \pmd}
-2 \Langle  k(x, y) \Rangle_{x \sim \pd, y  \sim \pmd} \; ,
\label{eq:mmd}
\end{align}
where $k(x,y)$ can be any positive definite, narrow kernel
function. Two identical distributions lead to $\text{MMD}(p,p)=0$
given enough statistics. Inversely, if $\text{MMD}(\pd,\pmd)=0$ for
randomly sampled batches, the two distributions have to be identical
$\pd(x) = \pmd(x)$. The shape of the kernels determines how local the
comparison between the two distributions is evaluated, for instance
though a Gaussian kernel with exponentially suppressed tails or a
Breit-Wigner with larger tails. The kernel width becomes a resolution
hyperparameter of the combined network. We can include the MMD loss to
the generator loss of Eq.\eqref{eq:gan_gloss},
\begin{align}
\loss_G \to 
\loss_G  + \lambda_\text{MMD} \, \text{MMD}^2 \; ,
\label{eq:mmdloss}
\end{align}
with a properly chosen coupling $\lambda$, similar to the parton
density loss introduced in Sec.~\ref{sec:basics_regr_nnpdf}. The
modified GAN setup for event generation is illustrated in
Fig.~\ref{fig:arch_gan2}.

\begin{figure}[t]
  \centering
  \includegraphics[width=0.45\textwidth]{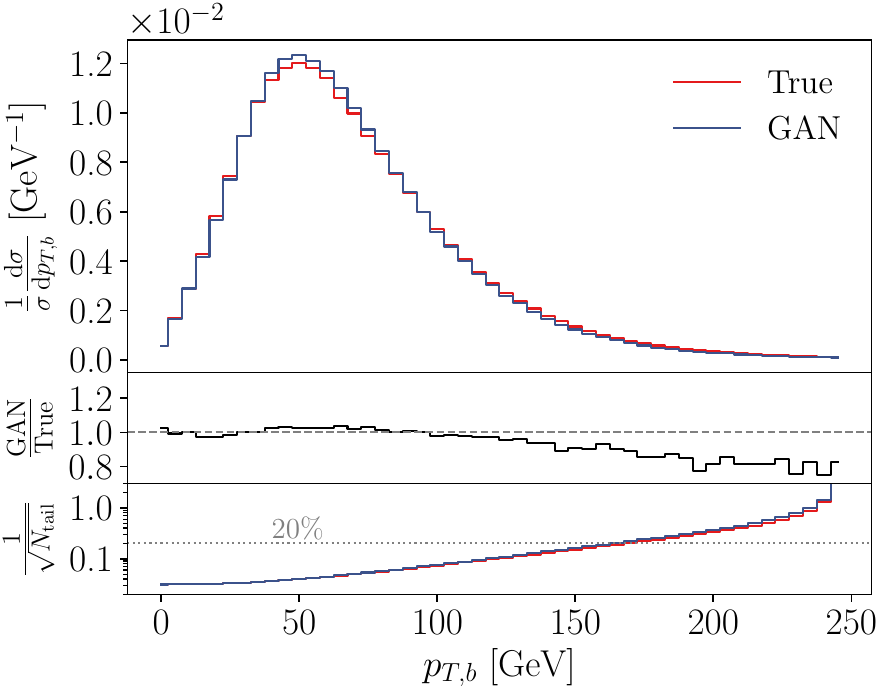}
  \hspace*{0.05\textwidth}
  \includegraphics[width=0.45\textwidth]{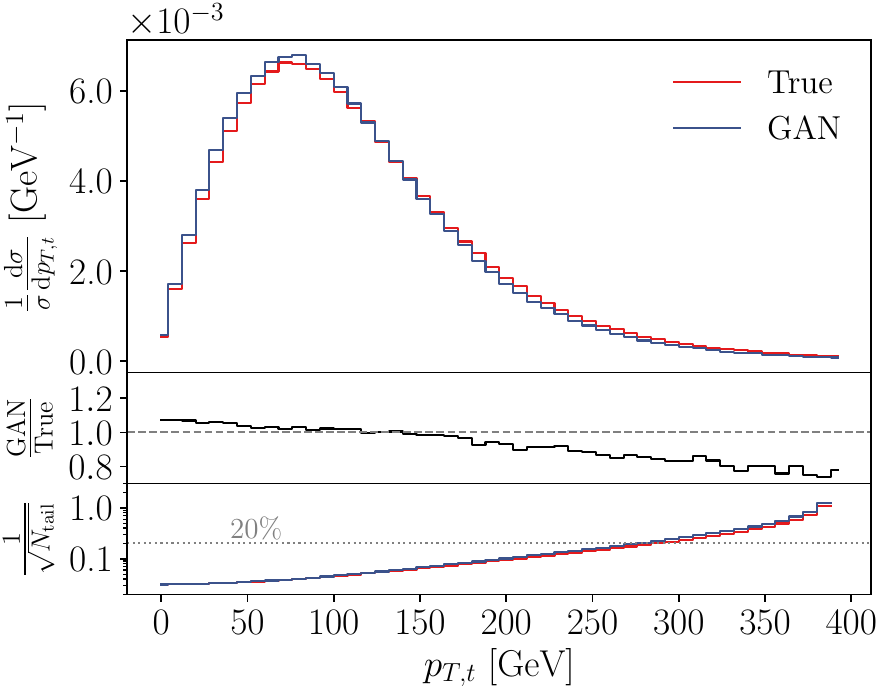} \\
  \includegraphics[width=0.45\textwidth]{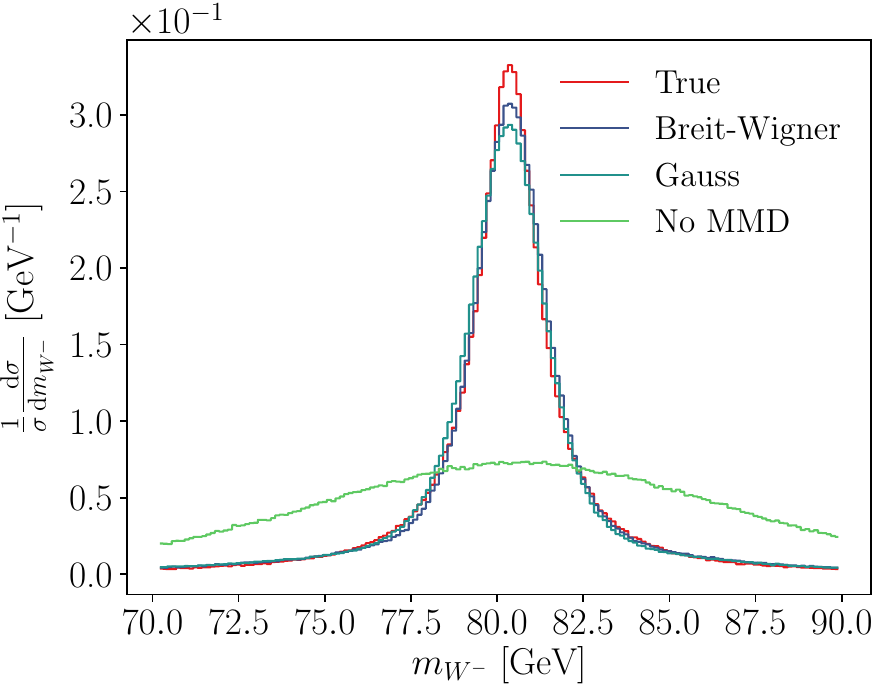}
  \hspace*{0.05\textwidth}
  \includegraphics[width=0.45\textwidth]{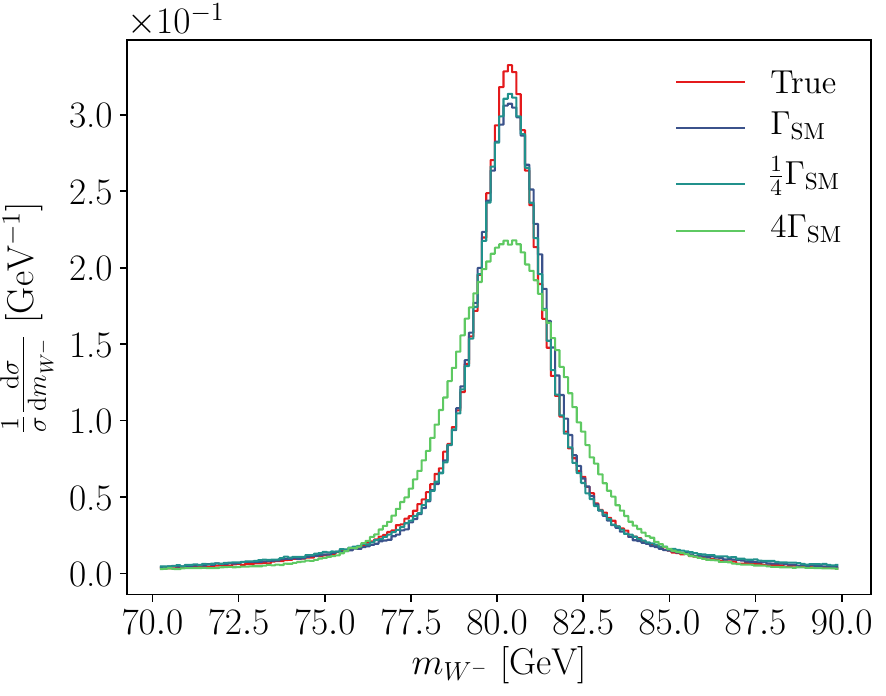}
  \caption{Upper: transverse momentum distributions of the final-state
    $b$-quark and the decaying top quark for MC truth and the GAN. The
    lower panels give the bin-wise ratios and the relative statistic
    uncertainty on the cumulative number of events in the tail of the
    distribution for our training batch size. Lower: comparison of
    different kernel functions and varying widths for reconstructing
    the invariant $W$-mass. Figure from Ref.\cite{Butter:2019cae}.}
\label{fig:gan_distris}
\end{figure}

To begin with, we can look at relatively flat distributions like
energies, transverse momenta, or angular correlations. In
Fig.~\ref{fig:gan_distris} we see that they are learned equally well
for final-state and intermediate particles.  In the kinematic tails we
see that the bin-wise difference of the two distributions increases to
around 20\%.  To understand this effect we estimate the impact of
limited training statistics per 1024-event batch through the relative
statistical\index{statistical uncertainty} uncertainty\index{uncertainties} on the number of events $N_\text{tail}(p_T)$
in the tail above the quoted $p_T$ value.  For the
$p_{T,b}$-distribution the GAN starts deviating at the 10\% level
around 150~GeV. Above this value we expect around 25 events per batch,
leading to a relative statistical uncertainty of 20\%. The top
kinematics is slightly harder to reconstruct, leading to a stronger
impact from low statistics.

Next, we can look at sharply peaked kinematic distributions,
specifically the invariant masses which we enhance using the MMD
loss. In the lower panels of Fig.~\ref{fig:gan_distris} we show the
effect of the additional MMD loss on learning the invariant $W$-mass
distribution.  Without the MMD in the loss, the GAN barely learns the
correct mass value.  Adding the MMD loss with default kernel widths of
the Standard Model decay widths drastically improves the results.  We
can also check the sensitivity on the kernel form and width and find
hardly any effect from decreasing the kernel width.  Increasing the
width reduces the resolution and leads to too broad mass peaks.

In the following sections we will first discuss three aspects of
generative networks and statistical limitations in the training
data. First, in Sec.~\ref{sec:gen_gan_amp} we study how GANs add
physics information to a problem similar to a fit to a small number of
training data points. Second, in Sec.~\ref{sec:gen_gan_subtract} we
apply generative networks to subtracting event samples, a problem
where the statistical uncertainties\index{statistical uncertainty} scales poorly.  Third, in
Sec.~\ref{sec:gen_gan_unweight} we use a GAN to unweight events, again
a problem where the standard method is known to be extremely
inefficient. Finally, in Sec.~\ref{sec:gen_gan_super} we use GANs to
enhance the resolution of jet images\index{jet images}, again making use of an implicit
bias orthogonal to the partonic nature of the jets.

\subsubsection{GANplification}
\label{sec:gen_gan_amp}

\begin{figure}[t]
  \centering
  \includegraphics[width=0.40\textwidth]{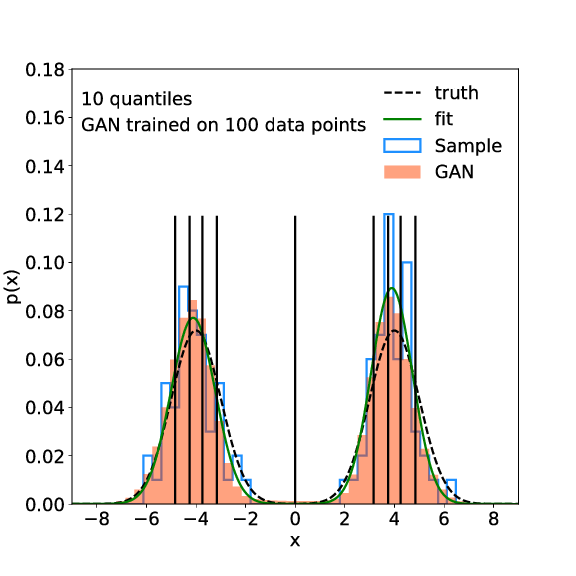}
  \hspace*{0.1\textwidth}
  \includegraphics[width=0.37\textwidth]{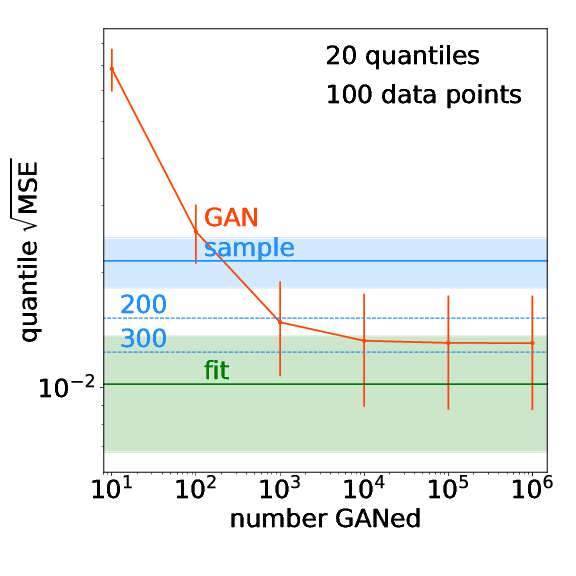}
  \caption{Left: 1-dimensional camel back function, we show the true
    distribution (black), a histogram with 100 sample points (blue), a
    fit to the samples data (green), and a high-statistics GAN sample
    (orange). Right: quantile error for sampling (blue), 5-parameter
    fit (green), and GAN (orange), shown for 20 quantiles. Figure from
    Ref.~\cite{Butter:2020qhk}.}
\label{fig:camel}
\end{figure}

An interesting question for neural networks in general, and generative
networks in particular, is how much physics information the networks
include in addition to the information from a statistically limited
training sample. For a qualitative answer we can go back to our
interpretation of the network as a non-parametric fit. For a fit,
nobody would ever ask if a function fitted to a small number of
training points can be used to generate a much larger number of
points. Also for a network it is clear that the network setup adds
information. This is a positive effect of an \underline{implicit bias}.  For neural
networks applied to regression tasks we use such an implicit bias,
namely that the relevant functions $f(x)$ which we want to approximate
are smooth and do not include features below a certain
$x$-resolution. Such a smoothness argument also applies to generative
networks and the underlying phase space density, the question becomes
how much this implicit bias accounts for in terms of events we can
generate, compared to the number of training events.

Because it is impossible to generate information out of nothing, we need to ask 
ourselves which aspects of our knowledge GANplification exchanges for an
increased number of valid events. The information included in a dataset 
is the information entropy defined in Eq.\eqref{eq:entropy}.
It can also be defined for a continuous probability distribution and
reaches its maximum for a constant probability,
\begin{align}
  H 
  = - \int_0^c dx p(x) \ln p(x) 
  \qquad \text{with} \qquad 
  \int_0^c dx p(x) = 1  \; .
\end{align}
It can be approximated by a histogram with $B$ bins with equal widths $\Delta = c/B$,
\begin{align}
  H
  \approx - \Delta \sum_i p(x_i) \ln p(x_i)
  \qquad \text{with} \qquad 
  \Delta \sum_i p(x_i) =1 \; .
\end{align}
Alternatively, we can evaluate the information entropy in a discrete histogram 
with $n_i$ counts per bin as 
\begin{align}
  H_B(N) 
  = - \sum_i \frac{n_i}{N} \ln \frac{n_i}{N}
  \qquad \text{with} \qquad 
  \sum_i \frac{n_i}{N} = 1 \; .
\end{align}
A sampled distribution cannot include more information 
that the underlying truth, which means
\begin{align}
  H > H_B(N) \; .
\label{eq:two_entropies1}
\end{align}   
To relate the approximate formula for $H$ and $H_B(N)$ we first use the
normalization condition and then find 
\begin{align}
   \frac{n_i}{N} \equiv p(x_i) \Delta 
   \qquad \Rightarrow \qquad 
   H_B(N) 
   &= - \sum_i \left[ p(x_i) \Delta \right] \ln \left[ p(x_i) \Delta \right]  \notag \\
   &= - \Delta \sum_i p(x_i) \ln p(x_i) 
      - \Delta \; \ln \Delta \sum_i p(x_i) \notag \\
   &= - \Delta \sum_i p(x_i) \ln p(x_i) 
      - \ln \Delta  \notag \\
   &= H - \ln \Delta 
\label{eq:two_entropies2}
\end{align}
It looks like minimizing $\Delta \to 0$ we can increase $H_B$ to exceed 
$H$, which contradicts Eq.\eqref{eq:two_entropies1}. 
This is resolve by the fact that more and finer bins only add 
information if there are enough entries per bin, such that the statistical 
fluctuations do not wash out the bin-wise information. 
This means Eq.\eqref{eq:two_entropies2} defines the optimal
bin width to evaluate a continuous underlying density without loss of information
\begin{align}
 \Delta = e^{H - H_B(N)} \; .
\label{eq:width_opt1}
\end{align}
This formula relates the resolution $\Delta$ to the binned entropy $H_B(N)$,
provided we know the full entropy $H$. In the next step, we want to 
replace $H_B(N)$ with $N$, using the general relation
\begin{align}
  H_B(N) = \frac{1}{M} \ln N 
      = \ln N^{1/M} 
  \qquad \text{with} \qquad M \ge 1 \; ,
\label{eq:bins_magic}
\end{align}
where the maximum information entropy and hence minimal $M =1$ is reached for a
constant probability. In this limit we know
\begin{alignat}{7}
&& H   &= - \int_0^c dx \frac{1}{c} \ln \frac{1}{c}
       =  \ln c 
       =  \ln B + \ln \Delta \notag \\
 &\Rightarrow &\qquad 
 H_B &= \ln B \equiv \ln N^{1/M} \notag \\
 &\Leftrightarrow &\qquad 
 B &= N^{1/M} \; .
\end{alignat}
If we want to stay away from too small event counts per bin, the range 
$M>2$ or $B < \sqrt{N}$ is reasonable.
For non-trivial probability distributions, Eq.\eqref{eq:width_opt1} rewritten 
in terms of $N$ becomes
\begin{align}
  \Delta 
  = e^H \; e^{\ln N^{1/M}} 
  = e^H N^{1/M} \; .
\label{eq:width_opt2}
\end{align}
Using Eq.\eqref{eq:width_opt2} we can argue that for constant $H_B$ we can 
describe a function with different event statistics $N$, leading to different 
$M$-values or bin counts. According to Eq.\eqref{eq:width_opt1} the same 
binned information entropy implies a universal resolution 
$\Delta$ for the training data and the generated events,
\begin{align}
  \Delta 
  = e^H N_\text{train}^{1/M_\text{train}}
  &\equiv e^H N_\text{gen}^{1/M_\text{gen}} \notag \\
  \Leftrightarrow \qquad 
  \frac{1}{M_\text{train}} \log N_\text{train} 
  &=   \frac{1}{M_\text{gen}} \log N_\text{gen} \notag \\
  \Leftrightarrow \qquad \qquad \quad 
  \frac{M_\text{gen}}{M_\text{train}} 
  &= \frac{\log N_\text{gen}}{\log N_\text{train}}
\end{align}
We assume a reasonable binning for the training dataset with 
$M_\text{train} \approx 2$ and 
then avoid overfitting by requiring 
\begin{align}
 M_\text{gen} < 3
 \qquad \Leftrightarrow \qquad 
 \frac{\log N_\text{gen}}{\log N_\text{train}} 
 < \frac{3}{2} 
\end{align}
This formula gives a maximum value for the amplification factor 
$N_\text{gen}/N_\text{train}$, assuming that we increase $N$ without
really increasing the resolution of the learned function beyond 
the training data.


A simple, but instructive toy example is a one-dimensional camel back
function, two Gaussians defined by two means, two widths, and a
relative normalization, shown in the left panel of
Fig.~\ref{fig:camel}. Reflecting the fixed resolution $\Delta$, we
divide the $x$-axis is into $n_\text{quant}$ quantiles, which means
that for each bin $j$ we expect the same number of events
$\bar{x}_j$. To quantify the amount of information for this fixed
resolution, for instance in the training data, we compute the average
quantile error
\begin{align}
\text{MSE}
= \frac{1}{n_\text{quant}} \; 
  \sum_{j=1}^{n_\text{quant}} \left( x_j - \bar{x}_j \right)^2 
= \frac{1}{n_\text{quant}} \; 
  \sum_{j=1}^{n_\text{quant}} \left( x_j - \frac{1}{n_\text{quant}} \right)^2 \; ,
\label{eq:quantile_error}
\end{align}
corresponding to the MSE defined in Eq.\eqref{eq:mse_loss}. Here $x_j$
is the estimated probability for an event to end up in each of the
$n_\text{quant}$ quantiles, and $\bar{x}_j = 1/n_\text{quant}$ is the
constant reference value. In the right panel of Fig.~\ref{fig:camel}
we show this MSE for 100 training points statistically distributed
over 20 quantiles. The uncertainty indicated by the shaded region
corresponds to the standard deviation from 100 statistically
independent sample.

Next, we apply a simple 5-parameter fit to the two means, the two
standard deviations, and the relative normalization of the camel
back. As expected, the MSE for the fit is much smaller than the MSE
for the training data, because the fit function defines a significant
implicit bias and is solidly over-constrained by the 100 data
points. Again, the error bar corresponds to 100 independent
fits. Quantitatively, we find that the fitted function is
statistically worth around 500 events instead of the 100-event
sample. This means, the fit leads to a \underline{statistical
  amplification}\index{statistical amplification} by a factor five. If we define an amplification
factor for matched MSEs we can write our result as
\begin{align}
  \boxed{
  A_\text{fit}
  = \frac{N_\text{sampled}(\text{MSE}=\text{MSE}_\text{fit})}{N_\text{train}}
  \approx 5
    }  \; .
\end{align}
Finally, we train a very simple GAN for the 1-dimensional target space
on the same 100 data points which we used for the fit. For the first
100 GANned events we find that the MSE corresponds to the 100 training
events. This means that the training and generated samples of the same
size include the same kind of information. This shows that the
properties of the training data are correctly encoded in the
network. However, we can use the trained GAN to generate many more
events, and in Fig.~\ref{fig:camel} we see that the MSE improves up to
$10^4$ events. After that the MSE reaches a plateau and does not
benefit from the additional statistics. This curve reflects two
effects. First, while the first 100 generated events carry as much
information, per event, as the training data, the next 900 events only
carry the same amount of information as the next 100 training
events. The full information encoded in the network is less than 300
training events, which means additional generated events carry less
and less information. Second, the GAN does include roughly as much
information as 300 training events, implying an amplification factor
\begin{align}
  A_\text{GAN}
  = \frac{N_\text{sampled}(\text{MSE}=\text{MSE}_\text{GAN})}{N_\text{train}}
  \approx 3 \; ,
\end{align}
surprisingly close to the parametric fit. This confirms our initial
picture of a neural network as a non-parametric fit also for the
underlying density learned by a generative network.

This kind of behavior can be observed generally for sparsely populated
and high-dimensional phase space, and the amplification factor
increases with sparseness.  While a quantitative result on achievable
amplification factors of generative networks in LHC simulations will
depend on many aspects and parameters, this result indicates that
using generative networks in LHC simulations can lead to an increase
in precision.

\subsubsection{Subtraction}
\label{sec:gen_gan_subtract}

The basis of all generative network is that we can encode the density,
for instance over phase space, implicitly in a set of events. For
particle physics simulations, we can learn the density of a given
signal or background process from Monte Carlo simulations, possibly
enhanced or augmented by data or in some other way. Most LHC searches
include a signal and several background processes, so we have to train
a generative network on the combination of background samples. This is
not a problem, because the combined samples will describe the sum of
the two phase space densities. 

The problem becomes more interesting when we instead want to train a
generative network to describe the difference of two phase space
densities, both given in terms of event samples.  There are at least
two instances, where subtracting event samples becomes
relevant. First, we might want to study signal events based on one
sample that includes signal and background and one sample that
includes background only. For kinematic distributions many analyses
subtract a background distribution from the combined signal plus
background distribution. Obviously, this is not possible for
individual events, where we have to resort to event weights
representing the probability that a given event is signal.  Second, in
perturbative QCD not all contributions to a cross section prediction
are positive. For instance, we need to subtract contributions included
in the definition of parton densities from the scattering process in
the collinear phase space regions. We also might want to subtract
on-shell contributions described by a higher-order simulation code
from a more complex but lower-order continuum production, for example
in top pair production. In both cases, we need to find a way to train
a generative network on two event samples, such that the resulting
events follow the \underline{difference} of their individual phase space
densities.

Following the GANplification argument from Sec.~\ref{sec:gen_gan_amp},
extracting a smooth phase space density for the difference of two
samples also has a statistical advantage. If we subtract two samples
using histograms we are not only limited in the number of phase space
dimensions, we also generate large statistical uncertainties\index{statistical uncertainty}. Let us
start with $S + B$ events and subtract $B \gg S$ statistically
independent events. The uncertainty on the resulting event number per
bin is then
\begin{align}
\Delta_S
&= \sqrt{ \Delta_{S+B}^2 + \Delta_B^2 } \notag \\
&= \sqrt{ (S+B) + B } 
\approx \sqrt{2 B} \gg \sqrt{S} \; .
\label{eq:error}
\end{align}
The hope is that the uncertainty on learned signal density turns out
smaller than this statistical uncertainty, because the neural network
with its implicit bias constructs smooth distributions for $S+B$ and
for $B$ before subtracting them.

\begin{figure}[b!]
\centering
\definecolor{Gcolor}{HTML}{2c7fb8}
\definecolor{Dcolor}{HTML}{f03b20}

\tikzstyle{generator} = [thick, rectangle, rounded corners, minimum width=1.8cm, minimum height=1cm,text centered, draw=Gcolor]
\tikzstyle{discriminator} = [thick, rectangle, rounded corners, minimum width=1.8cm, minimum height=1cm,text centered, draw=Dcolor]
\tikzstyle{io} = [thick,circle, trapezium left angle=70, trapezium right angle=110, minimum width=1.2cm, minimum height=1cm, text centered, draw=black]

\tikzstyle{process} = [thick, rectangle, minimum width=1cm, minimum height=1cm, text centered, draw=black]

\tikzstyle{xG} = [thick,rectangle, minimum width=2.2cm, minimum height=3cm, text depth= 2.2cm, draw=black]
\tikzstyle{s0} = [thick,rectangle, minimum width=2cm, minimum height=3cm, text centered]
\tikzstyle{s1} = [thick, dotted, rectangle, minimum width=1.6cm, minimum height=1.1cm, text centered, draw=black]

\tikzstyle{decision} = [thick,rectangle, minimum width=1cm, minimum height=1cm, text centered, draw=black]

\tikzstyle{dots} = [circle, minimum size=2pt, inner sep=0pt,outer sep=0pt, draw=Dcolor, fill = Dcolor]

\tikzstyle{arrow} = [thick,->,>=stealth]

\begin{tikzpicture}[node distance=2cm]

\node (generator) [generator] {$G$};
\node (random) [io, left of=generator, xshift=-0.2cm, yshift=0cm] {$\{ r \}$};
\draw [arrow, color=black] (random) -- (generator);

\node (xG) [xG, right of=generator, xshift=1.5cm, yshift=0cm] {$c\in\mathcal{C}_S \cup\mathcal{C}_B$};
\node (xs0) [right of=generator, xshift=1.5cm, yshift=0.2cm] {$\{ x_G,c\}$};
\node (xs1) [s1, right of=generator, xshift=1.5cm, yshift=-0.8cm] {$c\in\mathcal{C}_B$};
\draw [arrow, color=Gcolor] (generator) -- (xG);

\node (d1) [discriminator, right of = xG, xshift=1.0cm, yshift=2cm] {$D_{S+B}$};
\node (d2) [discriminator, right of = xG, xshift=1.0cm, yshift=-2cm] {$D_B$};
\node (x1) [io, above of = xG, xshift=0cm, yshift=0.5cm] {$\{x_{S+B} \}$};
\node (x2) [io, below of = xG, xshift=0cm, yshift=-0.5cm] {$\{x_B \}$};

\draw [arrow, color=Gcolor] (xG) -- (d1);
\draw [arrow, color=Gcolor] (xs1) -- (d2);
\draw [arrow, color=black] (x1) -- (d1);
\draw [arrow, color=black] (x2) -- (d2);

\node (data1) [process, left of=x1, xshift=-0.2cm, yshift=0cm] {Data $S+B$};
\node (data2) [process, left of=x2, xshift=-0.2cm, yshift=0cm] {Data $B$};
\draw [arrow, color=black] (data1) -- (x1);
\draw [arrow, color=black] (data2) -- (x2);

\node (dloss1) [process, right of=d1, xshift=0.5cm, yshift=0cm] {$\loss_{D_{S+B}}$};
\node (dloss2) [process, right of=d2, xshift=0.5cm, yshift=0cm] {$\loss_{D_B}$};
\node (gloss) [process, right of=xG, xshift=1.0cm, yshift=0cm] {$\loss_G$};
\draw [arrow, color=Gcolor] (d1) -- (gloss);
\draw [arrow, color=Gcolor] (d2) -- (gloss);
\draw [arrow, color=Gcolor] (xG) -- (gloss);
\draw [arrow, color=Dcolor] (d1) -- (dloss1);
\draw [arrow, color=Dcolor] (d2) -- (dloss2);
\draw[arrow, dashed, color=Dcolor] (dloss1) [out=-120, in=-60] to (d1);
\draw[arrow, dashed, color=Dcolor] (dloss2) [out=120, in=60] to (d2);

\coordinate[ right of = gloss, xshift=1.5cm, yshift=0.0cm] (out1);
\coordinate[ below of = out1, xshift=0cm, yshift=-1.5cm] (out2);
\coordinate[ below of = generator, xshift=0cm, yshift=-1.5cm] (out3);
\draw[thick, dashed, color=Gcolor] (gloss) -- (out1);
\draw[thick, dashed, color=Gcolor] (out1) -- (out2);
\draw[thick, dashed, color=Gcolor] (out2) -- (out3);
\draw[arrow, dashed, color=Gcolor] (out3) -- (generator);

\end{tikzpicture}
\caption{Structure of the subtraction GAN.  The training data is given
  by labelled events $\{ x_{S,B} \}$ and $\{ x_B \}$. The label $c$
  encodes the category of the generated events. Figure from
  Ref.~\cite{Butter:2019eyo}.}
\label{fig:GANsimple}
\end{figure}

We start with a simple 1-dimensional toy model, \ie events which
are described by a single real number $x$. We define a combined
distribution $p_{S+B}$ and a subtraction distribution $p_B$ as
\begin{align}
p_{S+B}(x)=\frac{1}{x}+0.1
\qquad \text{and} \qquad 
p_B(x)=\frac{1}{x} \;.
\label{eq:diff_sub1a}
\end{align}
The distribution we want to extract is
\begin{align}
p_S = 0.1 \; .
\label{eq:diff_sub1b}
\end{align}
The subtraction GAN is trained to reproduce the labelled training
datasets $\{x_{S+B} \}$ and $\{x_B \}$ simultaneously.  The
architecture is shown in Fig.~\ref{fig:GANsimple}. It consists of one
generator and two independent discriminators.  The losses for the
generator and the two discriminators follow the standard GAN setup in
Eq.\eqref{eq:gan_combined}.  The generator maps random numbers $\{
r\}$ to samples $\{x_G, c\}$, where $x_G$ stands for an event and $c = S,B$
for a label.  To encode the class label $c$ there are different
options. First, we can assign integer values, for instance $c=0$ for
background and $c=1$ for signal. The problem with an integer or real
label is that the network has to understand this label relative to a
background metric. Such a metric will not be symmetric for the two
points $c=0,1$. Instead, we can use \underline{one-hot encoding} by
assigning 2-dimensional vectors, just as in Eq.\eqref{eq:vec_graph}
\begin{align}
  c^\text{one-hot}_S =
  \begin{pmatrix} 1 \\ 0 \end{pmatrix}
  \qquad \text{and} \qquad 
  c^\text{one-hot}_B =
  \begin{pmatrix} 0 \\ 1 \end{pmatrix}
\label{eq:one_hot}
\end{align}
This representation can be generalized to many labels, again without
inducing an ordering and breaking the symmetry of the encoding.

In Fig.~\ref{fig:GANsimple} we see that for the class $\mathcal{C}_B$
and the union of $\mathcal{C}_S$ with $\mathcal{C}_B$ we train two
discriminators to distinguish between events from the respective input
samples and the labelled generated events.  This training forces the
events from class $\mathcal{C}_B$ to reproduce $p_B$ and all events to
reproduce $p_{S+B}$. If we then normalize all samples correctly, the
events labelled as $\mathcal{C}_S$ will follow $p_S$.  The additional
normalization first requires us to assign labels to each event and
then to encode them into a counting function based on the one-hot
label encoding $I(c)$.  This counting function allows us to define a
loss term which ensures the correct relative weights of signal and
background events,
\begin{align}
  \loss_G \to  \loss_G +
  \lambda_\text{norm} 
  \left( 
 \dfrac{\sum_{c \in \mathcal{C}_i} I(c)}
       {\sum_{c \in \mathcal{C}_{S+B}} I(c)}
       -\dfrac{\sigma_i}{\sigma_{S+B}}
       \right)^2\; .
  \label{eq:gloss_norm}
\end{align}
The individual rates $\sigma_i$ are input parameters for example from
Monte Carlo simulations.

\begin{figure}[t]
  \centering
  \includegraphics[page=1, width=0.45\textwidth]{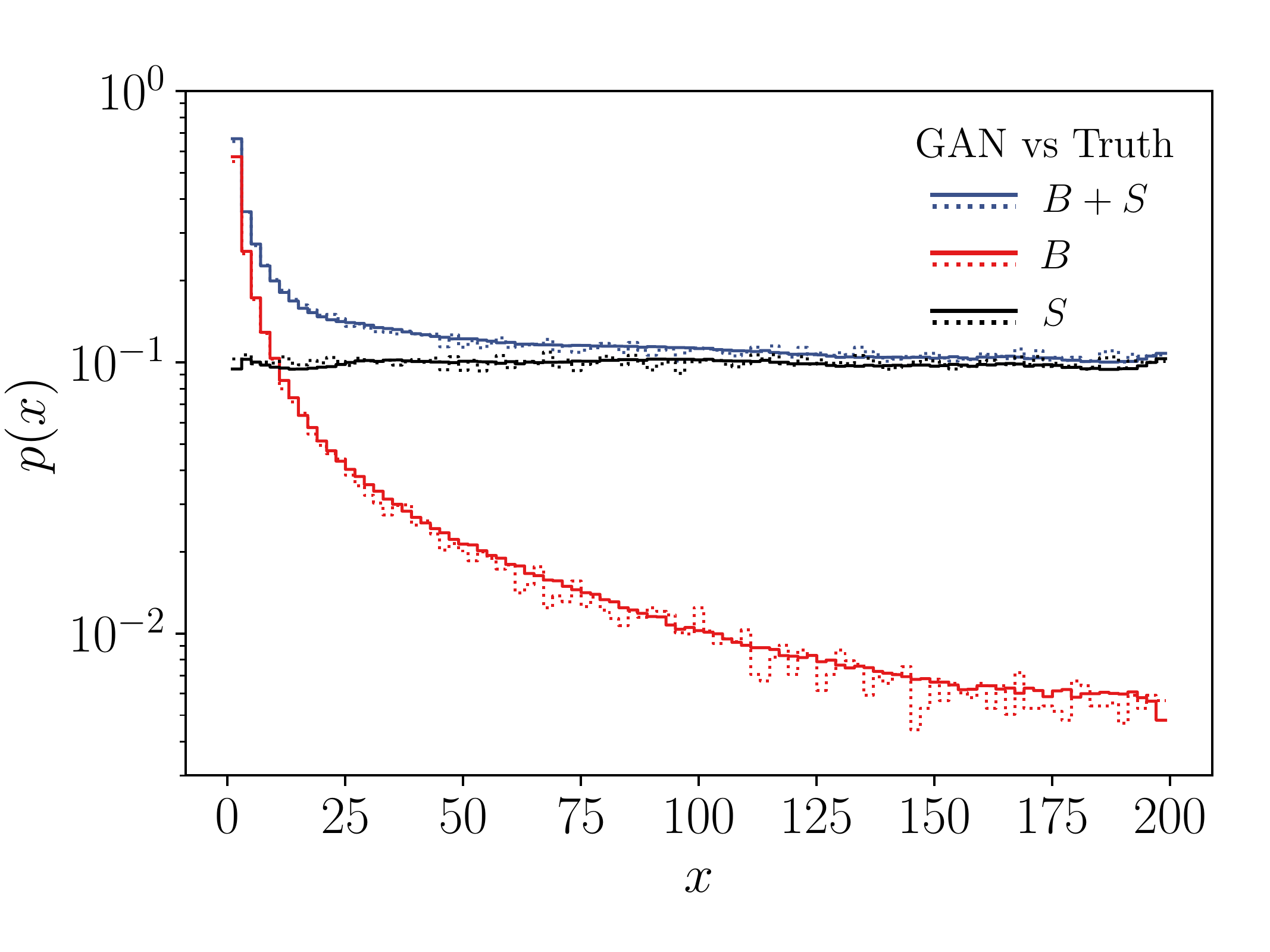}
  \hspace*{0.05\textwidth}
  \includegraphics[page=2, width=0.45\textwidth]{toy_subgan} \\
  \includegraphics[page=1, width=0.45\textwidth]{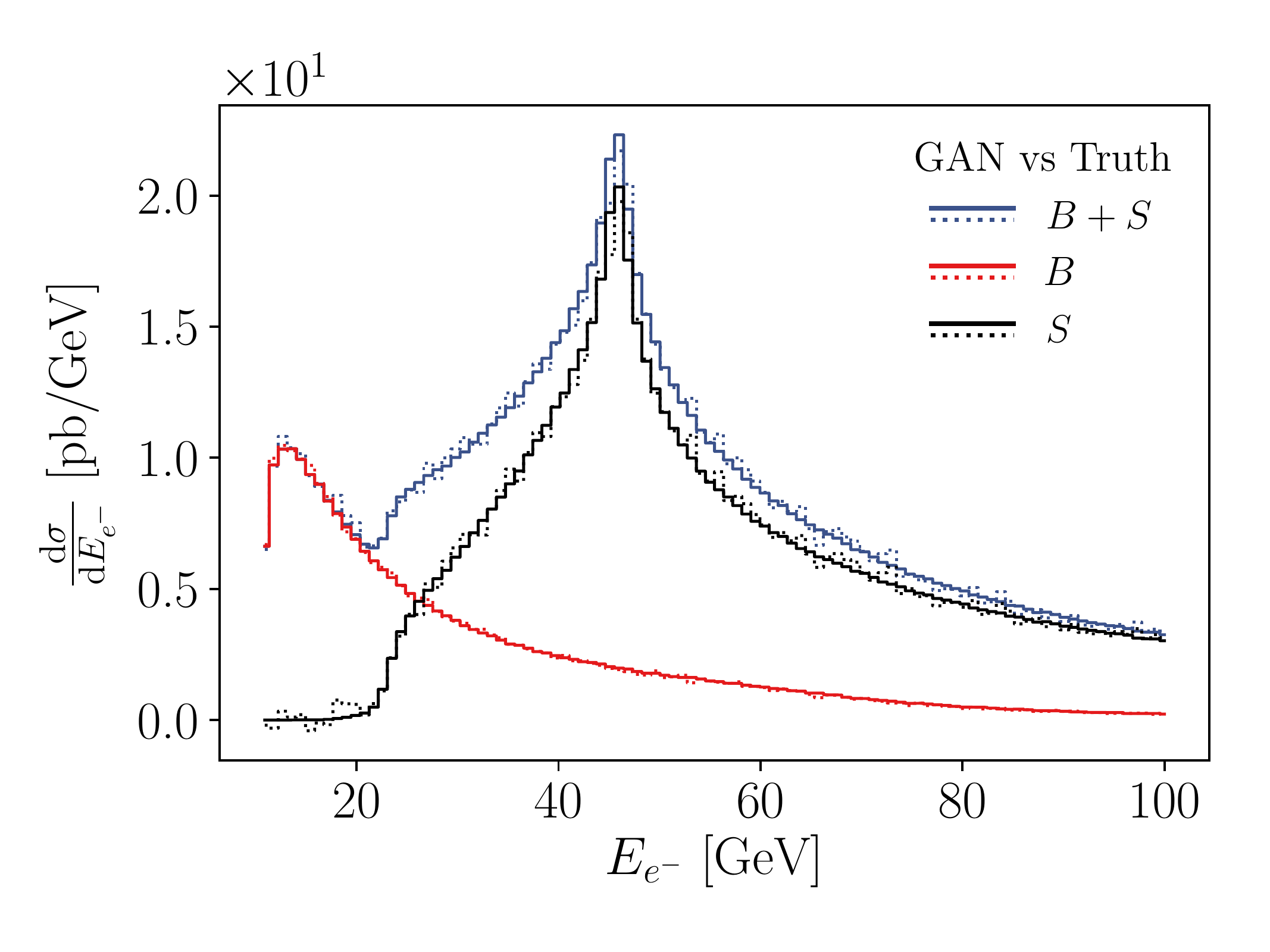}
  \hspace*{0.05\textwidth}
  \includegraphics[page=2, width=0.45\textwidth]{lhc_subgan}
  \caption{Illustration of the subtraction GAN for 1-dimensional toy
    events (top) and Drell--Yan events at the LHC.  Left: Generated
    (solid) and true (dashed) events for the two input distributions
    and the subtracted output. Right: distribution of the subtracted
    events, true and generated, including the uncertainty envelope
    propagated from the input statistics. Figures from
    Ref.~\cite{Butter:2019eyo}.}
  \label{fig:subGAN}
\end{figure}

In the left panel of Fig.~\ref{fig:subGAN} we show the input
distributions of Eq.\eqref{eq:diff_sub1a}, as well as the true and
generated subtracted distribution. The dotted lines show the training
dataset, the full lines show the generated distributions.  This
comparison confirms that the GAN learns the input information
correctly. We also see that the generated subtracted or signal events
follow Eq.\eqref{eq:diff_sub1b}. In the right panel we zoom into the
subtracted sample to compare the statistical uncertainties from the
input data with the generated signal events\index{uncertainties}.  The truth uncertainty
for the subtracted sample is computed from Eq.\eqref{eq:error}.
Indeed, the GAN delivers more stable results than what we expect from
the bin-by-bin statistical noise of the training data.  The GANned
distribution shows systematic deviations, but also at a visibly
smaller level than the statistical fluctuation of the input
data.

After illustrating ML-subtraction of event samples on a toy model, we
need to show how this method works in a particle physics context with
4-momenta of external particles as unweighted events.  A simple LHC
example is the Drell--Yan process, with a \underline{continuum photon
  contributions} and a $Z$-peak. The task is to subtract the photon
background from the full process and generate events only for the
on-shell $Z$-exchange and the interference with the background,
\begin{align}
S+B: \qquad pp &\to  \mu^+ \mu^- \notag \\
B: \qquad pp &\to \gamma \to \mu^+ \mu^- \; .
\end{align}
Aside from the increased dimension of the phase space the subtraction
GAN has exactly the same structure as shown in
Fig.\eqref{fig:GANsimple}.  In the lower panels of
Fig.~\ref{fig:subGAN} we see how the subtraction clearly extracts the
$Z$-mass peak in the lepton energy of the full sample, compared with
the feature-less photon continuum in the subtraction sample.  The
subtracted curve should describe the on-shell $Z$-pole and its
interference with the continuum. It vanishes for small lepton
energies, where the interference is negligible. In contrast, above the
Jacobian peak from the on-shell decay a finite interference remains as
a high-energy tail. In the right panel we again show the subtracted
curve including the statistical uncertainties from the input samples.
Obviously, our subtraction of the background to a di-electron
resonance is not a state-of-the-art problem in LHC physics, but it
illustrates how neural networks can be used to circumvent conceptual
and statistical limitations.

\subsubsection{Unweighting}
\label{sec:gen_gan_unweight}

A big technical problem in LHC simulations is how to get from, for
example, cross section predictions over phase space to predicted
events. Both methods can describe a probability density over phase space,
but using different ways of encoding this information:
\begin{enumerate}
\item Differential cross sections are usually encoded as values of the
  probability density for given phase space points. The points
  themselves follow a distribution, but this distribution has no
  relevance for the density and only ensures that all features of the
  distribution are encoded with the required resolution.
\item Events are just phase space points without weights, and the
  phase space density is encoded in the density of these unweighted
  events.
\end{enumerate}
Obviously, it is possible to combine these two extreme choices and
encode a phase space density using weighted events for which the phase
space distribution matters.

If we want to compare simulated with measured data, we typically rely
on unweighted events on both sides. There is a standard approach to
transform a set of \underline{weighted events} to a set of unweighted
events. Let us consider an integrated cross section of the form
\begin{align}
  \sigma
  = \int d x\,\frac{d \sigma}{d x}
  \equiv \int d x\, w(x) \; .
\label{eq:event_generation}
\end{align}
The event weight $w(x)$ is equivalent to the probability for a single
event $x$. To compute this integral numerically we draw events $\{x\}$
and evaluate the expectation value. If we sample with a flat
distribution in $x$ this means
\begin{align}
  \sigma
  \approx \left\langle \frac{d\sigma}{d x} \right\rangle_\text{flat}
  \equiv \Langle w(x) \Rangle_\text{flat} \; .
\end{align}
For flat sampling, the information on the cross section is again
included in the event weights alone.  We can then transform the
weighted events $\left\{x, w \right\}$ into unweighted events
$\left\{x\right\}$ using a hit-or-miss algorithm.
Here, we rescale the weight $w$ into a probability to keep
or reject the event $x$,
\begin{align}
w_\text{rel} =\frac{w}{w_\text{max}} < 1 \; ,
\end{align}
and then use a random number $r \in[0,1]$ such that the event is kept
if $w_\text{rel}> r$. A shortcoming of this method is that we lose
many events, for a given event sample the unweighting efficiency is
\begin{align}
\epsilon_\text{uw}=\frac{\langle w \rangle}{w_\text{max}} \ll 1 \;.
\label{eq:uw_efficiency}
\end{align}
We can improve sampling and integration with a suitable coordinate
transformations
\begin{align}
  x \to x'
  \qquad \Rightarrow \qquad 
  \sigma
  = \int dx\,w(x)
  = \int dx' \,\left|\frac{\partial x}{\partial x'}\right|\,w(x')
  \equiv \int dx' \; \tilde{w}(x') \;.
\label{eq:coord_trafo}
\end{align}
Ideally, the new integrand $\tilde{w}(x')$ is nearly constant and the
structures in $w(x)$ are fully absorbed by the Jacobian. In this case
the unweighting efficiency becomes
\begin{align}
  \tilde\epsilon_\text{uw}
  =\frac{\langle\tilde w\rangle}{\tilde{w}_\text{max}} \approx 1 \; .
\end{align}
This method of choosing an adequate coordinate transformation is
called importance sampling, and the standard tool in particle physics
is Vegas.

An alternative approach is to train a generative network to produce
unweighted events after training the network on weighted events.  We
start with the standard GAN setup and loss defined in
Eq.\eqref{eq:gan_combined}.  For weighted training events, the
information in the true distribution factorizes into the distribution
of sampled events $\pd$ and their weights $w(x)$. To capture this
combined information we replace the expectation values from sampling
$\pd$ with weighted means,
\begin{align}
  \boxed{
    \loss_D =  \frac{\Langle - w(x)\,\log D(x)\Rangle_{\pd}}{\langle w(x)\rangle_{\pd}}
    + \Langle -\,\log [ 1-D(G(r))] \Rangle_{\pl}
  } \; .
  \label{eq:uwGAN1}
\end{align}
The generator loss is not affected by this change,
\begin{align}
\loss_G = \Langle - \log D(G(r)) \Rangle_{\pl} \; ,
  \label{eq:uwGAN2}
\end{align}
and because it still produces unweighted events, their weighted means
reduce to the standard expectation value of
Eq.\eqref{eq:gan_combined}.

\begin{figure}[t]
  \centering
  \includegraphics[page=7, width=.45\textwidth]{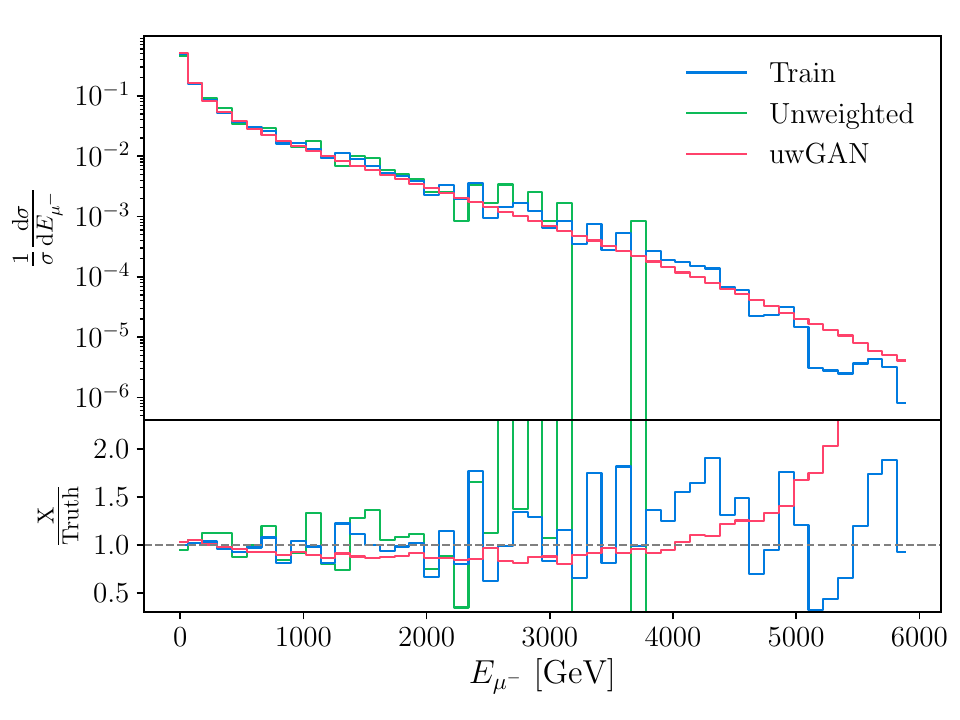} 
  \hspace*{0.05\textwidth}
  \includegraphics[page=4, width=.45\textwidth]{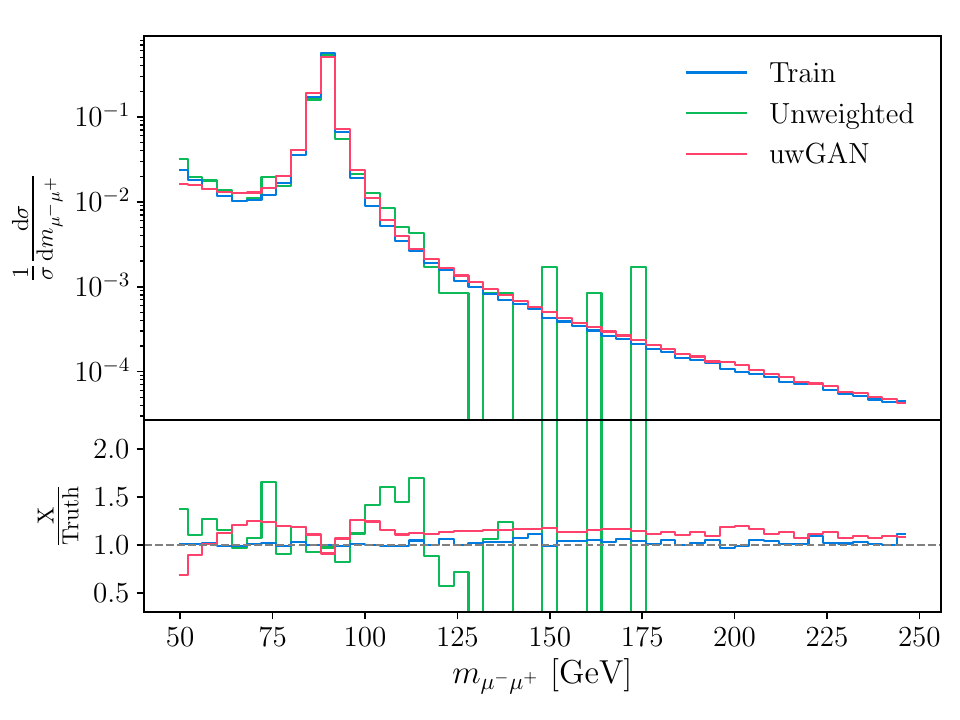}
  \caption{Kinematic distributions for the Drell-Yan process based on
    500k weighted training events, 1k unweighted events using standard
    unweighting, and 30M unweighted events generated with the
    GAN. Figure from Ref.\cite{Backes:2020vka}.}
  \label{fig:uw_drellyan}
\end{figure}

As in Sec.~\ref{sec:gen_gan_subtract}, we use weights describing the
Drell--Yan process
\begin{align}
p p  \to \mu^+ \mu^- \; ,
\end{align}
now with a minimal acceptance cut
\begin{align}
  m_{\mu\mu} > 50~\gev \; .
\label{eq:cuts}
\end{align}
We can generate weighted events with a naive phase space mapping and
then apply the unweighting GAN to those events.  For a training
dataset of 500k events the weights range from $10^{-30}$ to $10^{-4}$,
even if we are willing to ignore more than 0.1\% of the generated
events. Effects contributing to this vast range are the $Z$-peak, the
strongly dropping $p_T$-distributions, and our deliberately poor phase
space mapping. The classic unweighting efficiency defined by
Eq.\eqref{eq:uw_efficiency} is 0.22\%, a high value for
state-of-the-art tools applied to LHC processes.

In Fig.~\ref{fig:uw_drellyan} we show a set of kinematic
distributions, including the deviation from a high-precision truth
sample. First, the training dataset describes $E_\mu$ all the way to
6~TeV and $m_{\mu \mu}$ beyond 250~GeV with deviations below 5\%,
albeit with statistical limitation in the low-statistics tail of the
$m_{\mu \mu}$ distribution.  The sample after hit-and-miss unweighting
is limited to 1000 events. Correspondingly, these events only cover
$E_\mu$ to 1~TeV and $m_{\mu\mu}$ to 110~GeV.  In contrast, the
unweighting GAN distributions reproduce the truth information, if
anything, better than the fluctuating training data. Unlike for most
GAN applications, we now see a slight overestimate of the phase space
density in the sparsely populated kinematic regions.  For illustration
purpose we can translate the reduced loss of information into a
corresponding size of a hypothetical hit-and-miss training sample, for
instance in terms of rate and required event numbers, and find up to
an enhancement factor around 100. While it is not clear that an
additional GAN-unweighting is the most efficient way of accelerating
LHC simulations, it illustrates that generative networks can bridge
the gap between unweighted and weighted events. We will further
discuss this property for normalizing flow networks in
Sec.~\ref{sec:gen_inn_events}.

\subsubsection{Super-resolution}
\label{sec:gen_gan_super}

One of the most exciting applications of modern machine learning in
image analysis is super-resolution networks. The simple question is if
we can train a network to enhance low-resolution images to higher
resolution images, just exploiting general features of the objects in
the images.  While super-resolution of LHC objects is ill-posed in a
deterministic sense, it is well-defined in a statistical sense and
therefore completely consistent with the standard analysis techniques
at the LHC.  Looking at a given analysis object or jet, we would not
expect to enhance the information stored in a low-dimensional image by
generating a corresponding high-resolution image. However, if we
imaging the problem of having to combine two images at different
resolution, it should be beneficial to first apply a super-resolution
network and then combine the images at high resolution, rather than
downsampling the sharper image, losing information, and then combining
the two images. Beyond this obvious application, we could ask the
question if implicit knowledge embedded in the architecture of the
super-resolution network can contribute information in a manner
similar as we have seen in Sec.~\ref{sec:gen_gan_amp}.

Related to the combination of images with different resolution,
super-resolution networks will be used in next-generation particle
flow algorithms~\cite{DiBello:2020bas}. A related question on the
calorimeter level alone is the consistent combination of different
calorimeter layers or to ensure an optimal combination of calorimeter
and tracking information for charged and neutral aspects of an event.
Such approaches are especially promising when both sides of the
up-sampling, for instance low-resolution calorimeter data and
high-resolution tracking data, are available from data rather than
simulations.

We can apply super-resolution to jet images\index{jet images} and use the top-tagging
dataset described in Sec.~\ref{sec:class_cnn_sample} to test the model
dependence. The task is then to generate a high-resolution (HR),
super-resolved (SR) version of a given low-resolution (LR) image. This
kind of question points towards \underline{conditional generative
  networks}, which use their sampling functionality to generate events
or jets for or under the condition of a fixed event or jet starting
point. As before, our QCD and top jet images are generated in the
boosted and central kinematic regime with
\begin{align}
  p_{T,j}=550~...~650~\gev
  \qquad \text{and} \qquad
  |\eta_j|<2 \; ,
\end{align}
including approximate detector effects. The jet images consist of
roughly 50 active pixels, which means a sparsity of 99.8\% for
$160\times160$ images.  For each jet image we include two
representations, one encoding $p_T$ in the pixels and one including a
\underline{power-rescaled $p_T^{0.3}$}. The first image will
efficiently encode the hard pixels, while the second image gains
sensitivity for softer pixels especially for QCD jets.  This allows
the network to cover peaked patterns as well as more global
information.  The training dataset consists of paired LR/HR images,
generated by down-sampling the HR image with down-scaling factors 2,
4, and 8.

\begin{figure}[t]
    \centering
    \includegraphics[width=.35\textwidth]{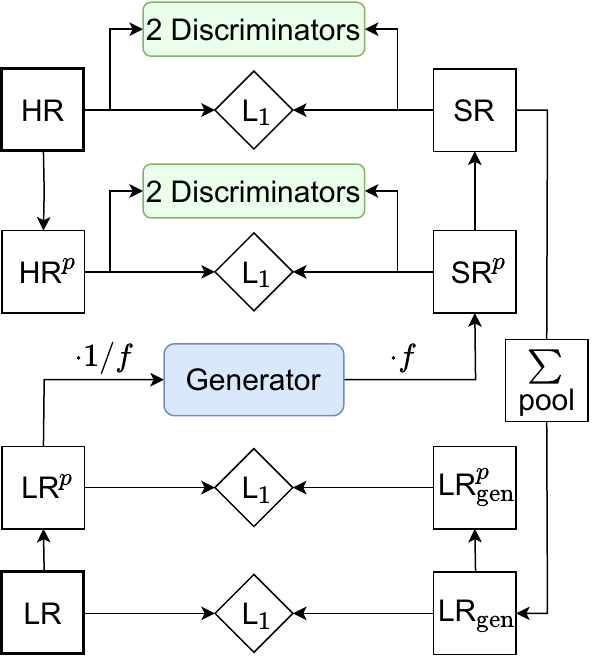}
    \caption{Training process for super-resolution jet images. Figure
      from Ref.~\cite{Baldi:2020hjm}.}
    \label{fig:super_net}
\end{figure}

The main building block of the super-resolution network, illustrated
in Fig.~\ref{fig:super_net}, is an image generator, following the
enhanced super-resolution GAN (ESRGAN) architecture.  It converts a LR
image into a SR image using a convolutional network.
\underline{Upsampling} a 2-dimensional image works in complete analogy
to the downsampling using a convolutional filter, described in
Sec.~\ref{sec:class_cnn}. For simplicity, let us assume that we want
to triple the size of an image using a ($3\times 3$)-filter, which is
globally trained. We can then replace every LR-pixel by $3 \times 3$
SR pixels multiplying the original pixel with the filter, referred to
as a patch. If we want to use the same filter to upsample only by a
factor two, we just sum all SR pixel contributions form the LR
image. This method is called transposed convolutions, and it can
incorporate the same aspects like padding or strides as the regular
convolution. An alternative upsampling method is pixel-shuffle. It
uses the, in our case 64 feature maps. To double the resolution two
dimensions we combine four feature maps and replace each LR pixel with
$2 \times 2$ SR pixels, one from each feature map.  For jet images it
turns out that up to three steps with an upsampling factor of two
works best if we alternate between pixel-shuffle and transposed
convolutions.

The discriminator network is a simple convolutional network.  It
measures how close the generated SR dataset is to the HR training data
and is trained through the usual loss function from
Eq.\eqref{eq:gan_combined}.
\begin{align}
  \loss_D =
  \langle - \log D(x) \rangle_{\pd} 
  + \langle - \log[ 1-D(G(r)) ] \rangle_{\pl} \; . 
\end{align}
To improve the sensitivity of the discriminator, we also include two
versions of it, one trained continuously and one where we erase the
memory by resetting all parameters after a certain number of batches.

\begin{figure}[t]
\centering 
\includegraphics[width=0.30\textwidth,page=2]{qcd_M2}
\includegraphics[width=0.30\textwidth,page=3 ]{qcd_M2}
\includegraphics[width=0.30\textwidth,page=1]{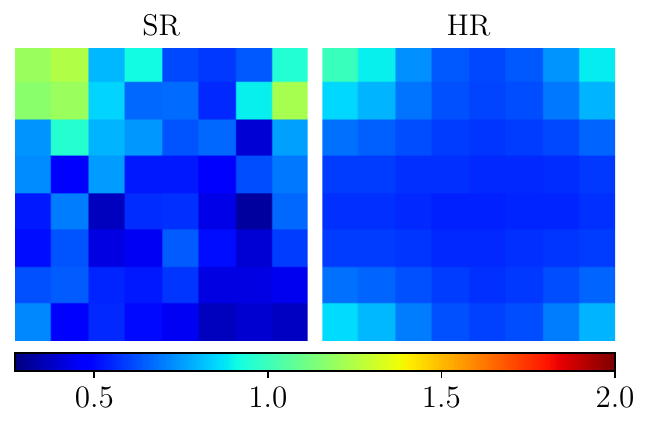} \\
\includegraphics[width=0.30\textwidth,page=4 ]{qcd_M2}
\includegraphics[width=0.30\textwidth,page=5]{qcd_M2}
\includegraphics[width=0.30\textwidth,page=9 ]{qcd_M2}\\
\includegraphics[width=0.30\textwidth,page=11 ]{qcd_M2}
\includegraphics[width=0.30\textwidth,page=14 ]{qcd_M2}
\includegraphics[width=0.30\textwidth,page=15 ]{qcd_M2}\\
\includegraphics[width=0.30\textwidth,page=11 ]{QCDev_top_M2}
\includegraphics[width=0.30\textwidth,page=14 ]{QCDev_top_M2}
\includegraphics[width=0.30\textwidth,page=15 ]{QCDev_top_M2}
\caption{Performance of a super-resolution network trained on QCD jets
  and applied to QCD jets. We show averaged HR and SR images, average
  patches for the SR and the HR images, pixel energies, and high-level
  observables. In the bottom row we show results after training on top
  jets. Figure from Ref.~\cite{Baldi:2020hjm}.}
\label{fig:qcd-qcd}
\end{figure}

The super-resolution generator loss include some additional
functionalities. We can start with the usual generator loss from
Eq.\eqref{eq:gan_combined}.  To ensure that the generated SR images
and the true HR images really resemble one another, we aid the
discriminator by adding a specific term $\loss_\text{HR}
(\text{SR},\text{HR} )$ to the generator loss. Similarly, we can
downsample the generates SR images and compare this $\lrgen$ image to
the true LR image pixel by pixel. The corresponding contribution to
the generator loss is $\loss_\text{LR} (\lrgen,\text{LR})$. Finally,
when upsampling the LR image we need to distribute each LR pixel
energy over the appropriate number of SR pixels, a so-called patch. We
force the network to spread the LR pixel energy such that the number
of active pixels corresponds to the HR truth through the loss
contribution $\loss_\text{patch}
(\text{patch(SR)},\,\text{patch(HR)})$, to avoid artifacts.  The
combined generator loss over the standard and reweighted jet images
is then
\begin{align}
  \boxed{
  \loss_G  \to \sum_{p=0.3,1} 
    \lambda_p \left(\lambda_\text{HR}\, \loss_\text{HR}
    + \lambda_\text{LR} \, \loss_\text{LR}
    + \lambda_G \, \loss_G
    + \lambda_\text{patch} \, \loss_\text{patch} \right)
    } \; ,
    \label{eq:totalloss}
\end{align}
This loss adds a sizeable number of hyperparameters to the network,
which we can tune for example by looking at the set of controlled
subjet observables given in Eq.\eqref{eq:qg_obs}.

\begin{figure}[t]
\centering
\includegraphics[width=0.30\textwidth,page=2]{top_M2}
\includegraphics[width=0.30\textwidth,page=3 ]{top_M2}
\includegraphics[width=0.30\textwidth,page=1]{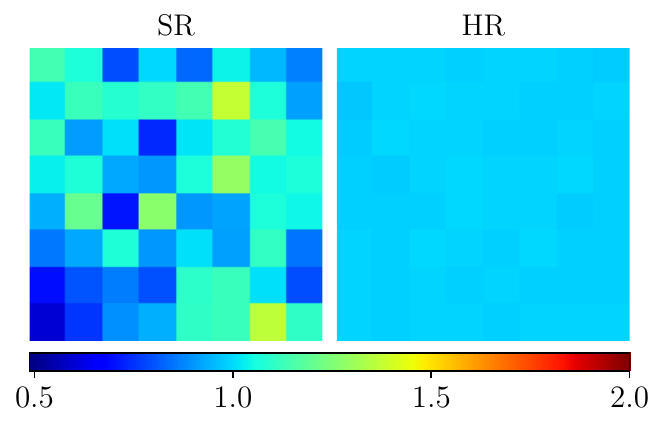}\\
\includegraphics[width=0.30\textwidth,page=4 ]{top_M2}
\includegraphics[width=0.30\textwidth,page=5]{top_M2}
\includegraphics[width=0.30\textwidth,page=9 ]{top_M2}\\
\includegraphics[width=0.30\textwidth,page=11 ]{top_M2}
\includegraphics[width=0.30\textwidth,page=14 ]{top_M2}
\includegraphics[width=0.30\textwidth,page=15 ]{top_M2}\\
\includegraphics[width=0.30\textwidth,page=11 ]{TOPev_qcd_M2}
\includegraphics[width=0.30\textwidth,page=14 ]{TOPev_qcd_M2}
\includegraphics[width=0.30\textwidth,page=15 ]{TOPev_qcd_M2}
\caption{Performance of a super-resolution network trained on top jets
  and applied to top jets. We show averaged HR and SR images, average
  patches for the SR and the HR images, pixel energies, and high-level
  observables. In the bottom row we show results after training on QCD
  jets. Figure from Ref.~\cite{Baldi:2020hjm}.}
\label{fig:top-top}
\end{figure}

In a first test, we train and test the super-resolution network on QCD
jets, characterized by a few central pixels.  In
Fig.~\ref{fig:qcd-qcd} we compare the HR and SR images for QCD jets,
as well as the true LR image with their generated $\lrgen$
counterpart. In addition to average SR and HR images and the relevant
patches, we show some pixel energy spectra and high-level observables
defined in Eq.\eqref{eq:qg_obs}.  In the pixel distributions we see
how the LR image resolution reaches its limits when, like for QCD
jets, the leading pixel carries most of the energy.  For the 10th
leading pixel we see how the QCD jet largely features soft noise. This
transition between hard structures and noise is the weak spot of the
SR network.  Finally, we show some of the jet substructure observables
defined in Eq.\eqref{eq:qg_obs}. The jet mass peaks around the
expected 50~GeV, for the LR and for the HR-jet alike, and the
agreement between LR and $\lrgen$ on the one hand and between HR and
SR on the other is better than the agreement between the LR and HR
images. A similar picture emerges for the $p_T$-weighted girth
$w_\text{PF}$, which describes the extension of the hard pixels.  The
pixel-to-pixel correlation $C_{0.2}$ also shows little deviation
between HR and SR on the one hand and LR and $\lrgen$ on the other.

The situation is different for top jets, which are dominated by two
electroweak decay steps. Comparing Figs.~\ref{fig:top-top}
with~\ref{fig:qcd-qcd} we see that the top jets are much wider and
their energy is distributed among more pixels. From a SR point of
view, this simplifies the task, because the network can work with more
LR-structures. Technically, the original generator loss becomes more
important, and we can balance the performance on top-quark jets vs QCD
jets using $\lambda_G$.  Looking at the ordered constituents,
the mass drop structure is learned very well. The leading four
constituents typically cover the three hard decay sub-jets, and they
are described better than in the QCD case. Starting with the 4th
constituent, the relative position of the LR and HR peaks changes
towards a more QCD-like structure, so the network starts splitting one
hard LR-constituent into hard HR-constituents. This is consistent
with the top-quark jet consisting of three well-separated patterns,
where the QCD jets only show this pattern for one leading constituent.
Among the high-level observables, the SR network
shifts the jet mass peak by about 10~GeV and does well on the girth
$w_\text{PF}$, aided by the fact that the jet resolution has hardly
any effect on the jet size. As for QCD-jets, $C_{0.2}$ is no challenge
for the up-sampling.

The ultimate goal for jet super-resolution is to learn general jet
structures, such that SR images can be used to improve multi-jet
analyses. In practice, such a network would be trained on any
representative jet sample and applied to QCD and top jets the
same. This means we need to test the \underline{model dependence} by
training and testing our network on the respective other samples. In
the bottom panels of Fig.~\ref{fig:qcd-qcd} and~\ref{fig:top-top} we
see that this cross-application works almost as well as the consistent
training and testing. This means that the way the image pixels are
distributed over patches is universal and hardly depends on the
partonic nature of the jet.

\subsection{Normalizing flows and invertible networks}
\label{sec:gen_inn}

After discussing the generative VAE and GAN architectures we remind
ourselves that controlling networks and uncertainty estimation are
really important for regression and classification networks. So the
question is if we can apply the recipes from
Sec.~\ref{sec:basics_deep_bayes} to capture \underline{statistical or
  systematic limitations of the training data} for generative
networks.  In LHC applications we would want to know the uncertainties\index{uncertainties}
on phase space distributions, for example when we rely on simulations
for the background $p_T$-distribution in mono-jet searches for dark
matter.  If we generate large numbers of events, saturating the
GANplification\index{statistical amplification} effect from Sec.~\ref{sec:gen_gan_amp}, uncertainties
on generated LHC distributions are uncertainties on the accuracy with
which our generative network has learned the underlying phase space
distribution it then samples from.  There exist, at least, three
sources of uncertainty. First, $\sigma_\text{stat}(x)$ arises from
statistical limitations of the training data.  Two additional terms,
$\sigma_\text{sys}(x)$ and $\sigma_\text{th}(x)$ reflect our ignorance
of aspects of the training data, which do not decrease when we
increase the amount of training data.  If we train on data, a
systematic uncertainty\index{systematic uncertainty} could come from a poor calibration of particle
energy in certain phase space regions. If we train on Monte Carlo, a
theory uncertainty\index{theory uncertainty} will arise from the treatment of large Sudakov
logarithms of the kind $\log (E/m)$ for boosted phase space
configurations.  Once we know these uncertainties as a function of
phase space, we can include them in the network output as additional
event entries, for instance supplementing
\begin{align}
  \text{ev}  = \begin{pmatrix} \{ x_{\mu,j} \} \\ \{ p_{\mu,j} \} \end{pmatrix}
  \quad \longrightarrow \quad 
  \begin{pmatrix} \sigma_\text{stat}/p \\ \sigma_\text{syst}/p \\ \sigma_\text{th}/p \\ \{ x_{\mu,j} \} \\ \{ p_{\mu,j} \} \end{pmatrix} \; ,
  \qquad \text{for each particle $j$.} 
  \label{eq:ext_evt}
\end{align}
The first challenge is to extract $\sigma_\text{stat}$ without
binning, which leads us to introduce normalizing flows directly in the
Bayesian setup, in analogy to our first regression network in
Secs.~\ref{sec:basics_deep_bayes} and \ref{sec:basics_regr_amp}.

\subsubsection{Architecture}
\label{sec:gen_inn_arch}

To model complex densities precisely and sample from them in a
controlled manner, we would like to modify the VAE architecture such
that the latent space can encode all phase space correlations. We can
choose the dimensionality of the latent vector $r$ the same as the
dimensionality of the phase space vector $x$. This gives us the
opportunity to define the encoder and decoder as \underline{bijective
  mappings}, which means that the encoder and the decoder are really
the same network evaluated in opposite directions. This architecture
is called a normalizing flow or an invertible neural network (INN),
\begin{align}
\text{latent $r \sim \pl$} 
\quad
\stackrel[\leftarrow \; \overline{G}_\theta(x)]{G_\theta(r) \rightarrow}{\xleftrightarrow{\hspace*{1.5cm}}}
\quad
\text{phase space $x \sim \pd$} \; ,
\label{eq:inn_mapping}
\end{align}
where $\overline{G}_\theta(x)$ denotes the inverse transformation to
$G_\theta(r)$.  Given a sample $r$ from the latent distribution, we
can use $G$ to generate a sample from the target
distribution. Alternatively, we can use a sample $x$ from the target
distribution to compute its density using the inverse direction.  In
terms of the network $G_\theta(r)$ the physical phase space density
and the latent density are related as
\begin{alignat}{7}
&&  dx \; \pmd(x) &= dr \; \pl(r) \notag \\
&\Leftrightarrow& \qquad 
  \pl(r) 
  &= \pmd(x) \left| \frac{\partial G_\theta(r)}{\partial r}\right|
  = \pmd \big(G_\theta(r)\big)\left| \frac{\partial G_\theta(r)}{\partial r}\right| \notag \\
&\Leftrightarrow& \qquad 
  \pmd(x) 
  &= \pl(r) \left| \frac{\partial G_\theta(r)}{\partial r}\right|^{-1}
  = \pl \big(\overline{G}_\theta(x)\big)\left| \frac{\partial \overline{G}_\theta(x)}{\partial x}\right| 
  \label{eq:cov}
\end{alignat}

For an INN we require the latent distribution $\pl$ to be known and
simple enough to allow for efficient sample generation, $G_\theta$ to
be flexible enough for a non-trivial transformation, and its Jacobian
determinant to be efficiently computable. We start by choosing a
multivariate Gaussian with mean zero and an identity matrix as the
covariance at the distribution $\pl$ in the unbounded latent space.
The INN we will use is a special variant of a normalizing flow
network, inspired by the RealNVP architecture, which guarantees a
\begin{itemize}
\item bijective mapping between latent space and physics space;
\item equally fast evaluation in both direction;
\item tractable Jacobian, also in both directions.
\end{itemize}
The construction of $G_\theta$ relies on the usual assumption that a
chain of simple invertible nonlinear maps gives us a complex map. This
means we transform the latent space into phase space with several
transformation layers, for which we need to know the Jacobians. For
instance, we can use \underline{affine coupling layers} as building
blocks. Here, the input vector $r$ is split in half, $r = (r_1,r_2)$,
allowing us to compute the output $x=(x_1,x_2)$ of the layer as
\begin{align}
\begin{pmatrix} x_1 \\ x_2 \end{pmatrix} =
\begin{pmatrix}
r_1 \odot e^{s_2(r_2)} + t_2(r_2) \\
r_2 \odot e^{s_1(x_1)} + t_1(x_1)
\end{pmatrix}
\qquad \Leftrightarrow \qquad 
\begin{pmatrix} r_1 \\ r_2 \end{pmatrix} =
\begin{pmatrix}
(x_1 - t_2(r_2)) \odot e^{-s_2(r_2)} \\
(x_2 - t_1(x_1)) \odot e^{-s_1(x_1)}
\end{pmatrix} \; .
\label{eq:layer1},
\end{align}
where $s_i, t_i$ ($i=1,2$) are arbitrary functions, and $\odot$ is the
element-wise product. In practice each of them will be a small
multi-layer network.  The Jacobian of the transformation $G$ is an
upper triangular matrix, and its determinant is just the product of
the diagonal entries.
\begin{align}
  \frac{\partial G_\theta(r)}{\partial r}
  &= 
  \begin{pmatrix} 
    \partial x_1/\partial r_1 &  \partial x_1/\partial r_2 \\
    \partial x_2/\partial r_1 &  \partial x_2/\partial r_2
  \end{pmatrix}
  = 
  \begin{pmatrix} 
    \text{diag}\left(e^{s_2(r_2)}\right) & \text{finite} \\ 
    0 & \text{diag}\left(e^{s_1(x_1)}\right) 
  \end{pmatrix} \notag \\
  \qquad \Rightarrow \qquad
  \left| \frac{\partial G_\theta(r)}{\partial r} \right|
  &= \prod e^{s_2(r_2)} \;  \prod e^{s_1(x_1)}
\; .
\end{align}
Such a Jacobian determinant is computationally inexpensive and still
allows for complex transformations. We refer to the sequence of
coupling layers as $G_\theta(r)$, collecting the parameters of the
individual nets $s$, $t$ into a joint $\theta$.

Given the invertible architecture we proceed to train our network via
a \underline{likelihood loss}\index{likelihood loss}, which we already used as the
first term of the VAE loss in Eq.\eqref{eq:vaeloss2}.  It relies on
the assumption that we have access to a dataset which encodes the
intractable phase space distribution $\pd(x)$ and want to fit our
model distribution $\pmd(x)$ via $G_\theta$. The likelihood
loss\index{likelihood loss} for the INN is
\begin{align}
  \boxed{
  \loss_\text{INN}
  = - \XLangle \log \pmd(x) \XRangle_{\pd} 
  =- \XXLangle \log \pl\big(\overline{G}_\theta(x)\big) + \log \left| \frac{\partial \overline{G}_\theta(x)}{\partial x}\right|
  \XXRangle_{\pd}  } \; .
  \label{eq:MLE}
\end{align}
The first of the two terms ensures that the latent representation
remains, for instance, Gaussian, while the second term constructs the
correct transformation to the phase space distribution. Given the
structure of $\overline{G}_\theta(x)$ and the latent distribution
$\pl$, both terms can be computed efficiently.  As in
Eq.\eqref{eq:def_kl}, one can view this maximum likelihood approach as
minimizing the KL-divergence between the true but unknown phase space
distribution $\pd(x)$ and our approximating distribution
$\pmd(x)$.

While the INN provides us with a powerful generative model of the
underlying data distribution, it does not account for an uncertainty
in the network parameters $\theta$. However, because of its bijective
nature with the known Jacobian, the INN allows us to meaningfully
switch from deterministic sub-networks $s_{1,2}$ and $t_{1,2}$ to
their Bayesian counterparts. Here we follow exactly the same setup as
in Sec.~\ref{sec:basics_deep_bayes} and recall that we can write the
BNN loss function of Eq.\eqref{eq:loss_bayes1}\index{Bayesian network} as
\begin{align}
  \loss_\text{BNN} = - \XLangle \log \pmd(x) \XRangle_{\theta \sim q}
  + \kl [q(\theta),p(\theta)] \; .
\end{align}
We now approximate the intractable posterior $p(\theta|x_\text{train})$ with a
mean-field Gaussian as the variational posterior $q(\theta)$ and then
apply Bayes' theorem\index{Bayes' theorem} to train the network on the usual \underline{ELBO
  loss}, now for event samples
\begin{align}
  \loss_\text{B-INN}
  &= -  \XLangle \log \pmd(x) \XRangle_{\theta\sim q, x \sim \pd}
  + \kl [q(\theta), p(\theta)] \notag \\
  &= - \XXLangle \log \pl\big(\overline{G}_\theta(x)\big)
  + \log \left| \frac{\partial \overline{G}_\theta(x)}{\partial x}\right| \XXRangle_{\theta\sim q, x \sim \pd}
  + \kl [q(\theta), p(\theta)] 
  \label{eq:loss_binn}
\end{align}
By design, the likelihood, the Jacobian, and the KL-divergence can be
computed easily.

To generate events using this model and with statistical
uncertainties\index{statistical uncertainty}, we remind ourselves how Bayesian network sample over
weight space in Eq.\eqref{eq:sig_pred}, but how to predict a phase
space density with a local uncertainty map. In terms of the BNN
network outputs analogous to Eq.\eqref{eq:bnn_output} this means
\begin{align}
p(x) &= \int d\theta \; q(\theta) \; \pmd(x)  \notag \\
\sigma_\text{stat}^2(x) &= \int d\theta \; q(\theta) \left[ \pmd(x | \theta) - p(x) \right]^2 \; .
\label{eq:sigma_pred}
\end{align}
Here $x$ denotes the initial phase space vector from
Eq.\eqref{eq:ext_evt}, and the predictive uncertainty can be
identified with the corresponding relative statistical uncertainty
$\sigma_\text{stat}$.

Let us illustrate \underline{Bayesian normalizing flows} using a set
of 2-dimensional toy models.  First, we look a simple 2-dimensional
ramp distribution, linear in one direction and flat in the other,
\begin{align}
  p(x, y) = 2 x \; .
\label{eq:linear_dens}
\end{align}
The factor two ensures that $p(x,y)$ is normalized. 
The network input and output consist of unweighted
events in the 2-dimensional parameters space, $(x,y)$.

\begin{figure}[t]
\includegraphics[width=0.32\textwidth, page=1]{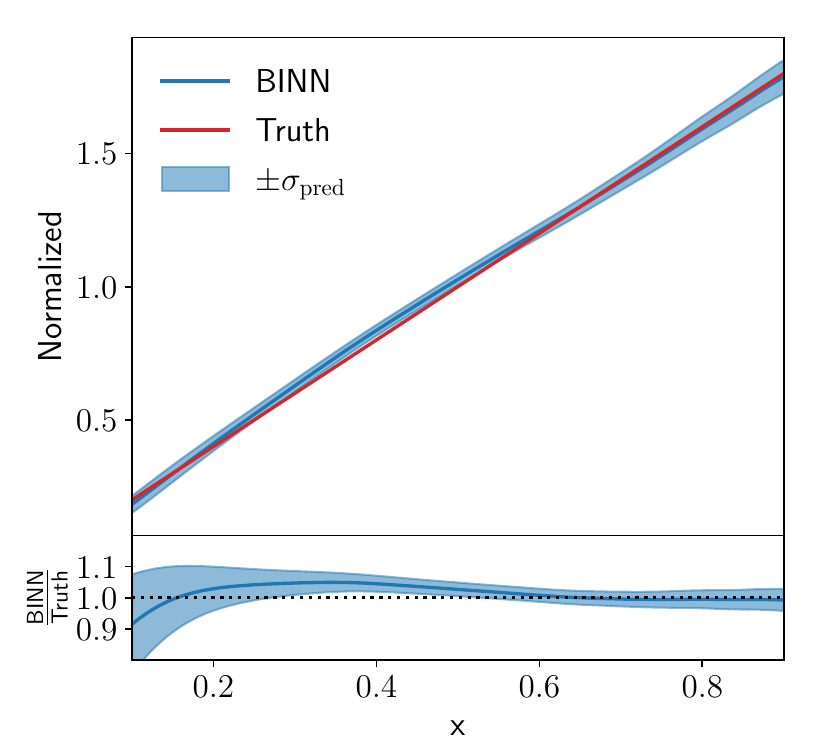}
\includegraphics[width=0.32\textwidth, page=3]{linear_1dplots}
\includegraphics[width=0.32\textwidth, page=2]{linear_1dplots}
\caption{Density and predictive uncertainty distribution for a linear
  wedge ramp using a B-INN. The uncertainty on $\sigma_\text{stat}$ is
  given by its $y$-variation.  The green curve represents a
  2-parameter fit to Eq.\eqref{eq:fit_wedge}. Figure from
  Ref.~\cite{Bellagente:2021yyh}.}
  \label{fig:linear_unc}
\end{figure}

In Fig.~\ref{fig:linear_unc} we show the network prediction, including
the \underline{predictive uncertainty} on the density. Both, the phase
space density and the uncertainty are scalar fields defined over phase
space, where the phase space density has to be extracted from the
distribution of events and the uncertainty is explicitly given for
each event or phase space point. For both fields we can trivially
average the flat $y$-distribution.  In the left panel we indicate the
predictive uncertainty as an error bar around the density estimate,
covering the deviation from the true distribution well.

In the central and right panels of Fig.~\ref{fig:linear_unc} we show
the relative and absolute predictive uncertainties. The relative
uncertainty decreases towards larger $x$. However, the absolute
uncertainty shows a distinctive minimum around $x \approx 0.45$.
To understand this minimum we focus on the non-trivial $x$-coordinate
with the linear form
\begin{align}
  p(x) = a  x + b
  \qquad \text{with} \qquad x \in [0,1] \; .
  \label{eq:def_2d_ramp}
\end{align}
Because the network learns a density, we can remove $b$ by
fixing the normalization,
\begin{align}
  1 = \int_0^1 dx \; (ax + b) = \frac{a}{2} + b
  \qquad \Rightarrow \qquad 
  p(x) = a \left( x - \frac{1}{2} \right) + 1 \; .
\end{align}
Let us now assume that the network acts like a one-parameter fit of
$a$ to the density, so we can propagate the uncertainty on the density
into an uncertainty on $a$,
\begin{align}
\sigma_\text{stat} \equiv \Delta p \approx \left| x - \frac{1}{2} \right| \; \Delta a \; .
\label{eq:simple_wedge}
\end{align}
The absolute value appears because the uncertainties are defined to be
positive, as encoded in the usual quadratic error propagation. The
minimum at $x=1/2$ explains the pattern we see in
Fig.~\ref{fig:linear_unc}. What this simple approximation cannot
explain is that the predictive uncertainty is not symmetric and does
not reach zero. However. we can modify our simple ansatz to vary the
hard-to-model boundaries and find
\begin{align}
  p(x) =& a  x + b
  \qquad \text{with} \qquad x \in [x_\text{min},x_\text{max}] \notag \\
  \Rightarrow \qquad 
  p(x)
  =& a x
  +  \frac{ 1 - \dfrac{a}{2}(x_\text{max}^2 - x_\text{min}^2) }{ x_\text{max} - x_\text{min} } \; .
\end{align}
For the corresponding 3-parameter fit we find
\begin{align}
\sigma_\text{stat}^2 \equiv (\Delta p)^2 =
    \left( x - \frac{1}{2} \right)^2 (\Delta a)^2
    + \left(1 + \frac{a}{2} \right)^2 (\Delta x_\text{max} )^2
    + \left(1 - \frac{a}{2} \right)^2 (\Delta x_\text{min} )^2 \; .
\label{eq:fit_wedge}
\end{align}
While the slight shift of the minimum is not explained by this form,
it does lift the minimum uncertainty to a finite value. If we evaluate
the uncertainty as a function of $x$ we cannot separate the effects
from the two boundaries.  The green line in Fig.~\ref{fig:linear_unc}
gives a 2-parameter fit of $\Delta a$ and $\Delta x_\text{max}$ to the
$\sigma_\text{stat}$ distribution from the Bayesian INN.

\begin{figure}[t]
\includegraphics[width=0.32\textwidth, page=1]{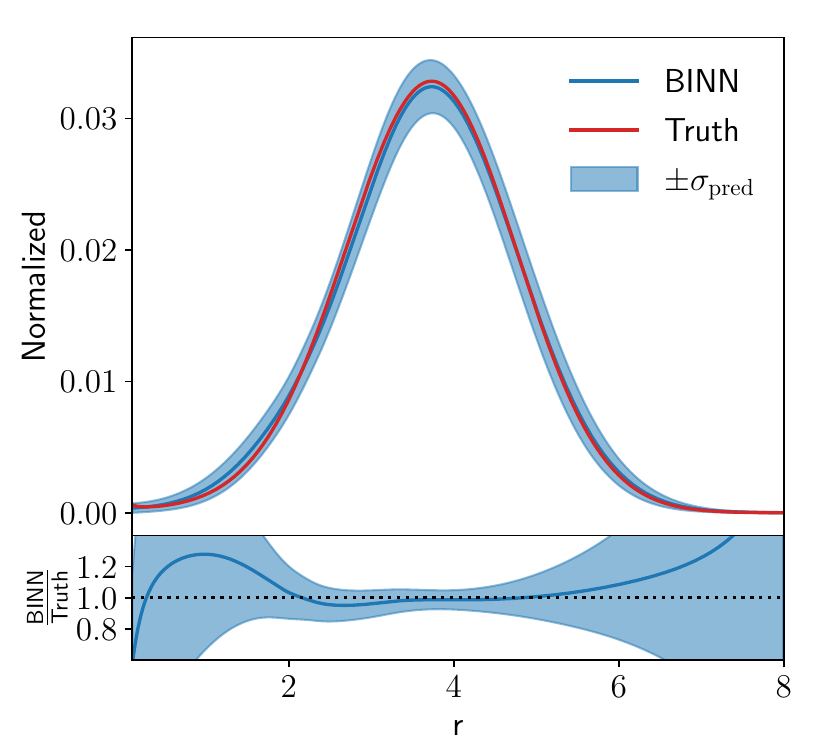}
\includegraphics[width=0.32\textwidth, page=3]{gauss_ring_1dplots}
\includegraphics[width=0.32\textwidth, page=2]{gauss_ring_1dplots} 
\caption{Density and predictive uncertainty distribution for the
  Gaussian ring. The uncertainty band on $\sigma_\text{stat}$ is given
  by different radial directions. The green curve represents a
  2-parameter fit to Eq.\eqref{eq:gauss_fit}. Figure from
  Ref.~\cite{Bellagente:2021yyh}.}
\label{fig:gauss_unc}
\end{figure}

While the linear ramp example could describe how an INN learns a
smoothly falling distribution, like $p_T$ of one of the particle in
the final state, the more dangerous phase space features are sharp
intermediate mass peaks. The corresponding toy model is a
2-dimensional Gaussian ring in terms of polar coordinates,
\begin{align}
  p(r, \phi) &= \normal(r;\mu=4, \sigma=1)
  \qquad \text{with} \quad \phi \in [0, \pi] \notag \\
  \Leftrightarrow \qquad 
  p(x, y) &= \normal(\sqrt{x^2+y^2};\mu=4,\sigma=1) \times \frac{1}{\sqrt{x^2 + y^2}} \; .
\end{align}
where the Jacobian $1/r$ ensures that both probability distributions
are correctly normalized. We train the Bayesian INN on Cartesian
coordinates, just like for the ramp discussed above.  In
Fig.~\ref{fig:gauss_unc} we show the Cartesian density, evaluated on a
line of constant angle. This form includes the Jacobian and leads to a
shifted maximum.  Again, the uncertainty covers the deviation of the
learned from the true density.

Also in Fig.~\ref{fig:gauss_unc} we see that the absolute predictive
uncertainty shows a dip at the peak position.  As before, this leads
us to the interpretation in terms of appropriate fit parameters. For
the Gaussian radial density we use the mean $\mu$ and the width
$\sigma$, and the corresponding variations of the Cartesian density
give us
\begin{align}
  \sigma_\text{stat}
  =  \Delta p &\supset
  \left| \dfrac{d}{d \mu} \; p(x, y) \right| \; \Delta \mu \notag  \\
  &= 
  \frac{1}{r} \; \frac{1}{\sqrt{2 \pi}\sigma} \left| \dfrac{d}{d \mu} \; e^{-(r - \mu)^2/(2\sigma^2)} \right| \; \Delta \mu   \\
  &= 
  \frac{p(x,y)}{r} \left| \frac{2(r-\mu)}{2\sigma^2} \right| \; \Delta \mu  
  = \frac{p(x,y)}{r} \; \frac{|r-\mu|}{\sigma^2} \; \Delta \mu  \; .
\label{eq:gauss_fit}
\end{align}
This turns out the dominant uncertainty, and it explains the local
minimum at the peak position. Away from the peak, the uncertainty is
dominated by the exponential behavior of the Gaussian.

The patterns we observe for the Bayesian INN indicate that our
bilinear mapping is constructed very much like a fit. The network
first identifies the family of functions which describe the underlying
phase space density, and then it adjusts the relevant parameters, like
the derivative of a falling function or the peak position of the
Gaussian. We emphasize that this result is extracted from a joint
network training on the density and the uncertainty on the density,
not from a visualization. It also applies to normalizing flows
only. Without a tractable Jacobian it is not clear how it can be
generalized for example to GANs, even though the results on
super-resolution with jets trained on QCD jets and applied to top jets
in Fig.~\ref{fig:top-top} suggest that also the GANs first learn
general aspects before focusing on details of their model.

\subsubsection{Event generation with uncertainties}
\label{sec:gen_inn_events}

When we want to use normalizing flows to describe complex phase space
distributions we can replace the simple affine coupling layer of
Eq.\eqref{eq:layer1} with more powerful transformations. One example
are splines, smooth rational functions which we typically use for
interpolations.  Such a spline is spanned by a set of rational
polynomials interpolating between a given number of knots.  The
positions of the knots and the derivatives of the target function at
the knots are parameterized by the network as bin widths $\theta_x^j$,
bin heights $\theta_y^j$, and knot derivatives $\theta_d^j$,
illustrated in Fig.~\ref{fig:spline}.

\begin{figure}[t]
  \centering
  \input{incl_splineflow}
  \caption{Visualisation of a spline layer for normalizing flows.  The
    network positions of the spline knots and the derivative at each
    knot. Splines linking the knots define a monotonic
    transformation. Figure from Ref.~\cite{Stienen:2020gns}.}
  \label{fig:spline}
\end{figure}

We already know from Sec.~\ref{sec:gen_gan_unweight} that it is easy
to translate weighted into unweighted event samples through
hit-and-miss unweighting.  However, we have also seen that this
unweighting is computationally expensive, because the range of weights
is large.  Furthermore, we know from Sec.~\ref{sec:gen_gan_subtract}
that LHC predictions can include events with negative weights, and in
this case we cannot easily unweight the sample.  What we can do is
train a generative network on weighted events, using the modified loss
from Eq.\eqref{eq:MLE}
\begin{align} 
  \boxed{
  \loss_\text{INN, weighted} 
  = - \XLangle w(x) \log \pmd(x) \XRangle_{\pd}
  } \; .
  \label{eq:weighted_loss}
\end{align}
This loss defines the correct optimization of the generative network for
all kinds of event samples, including negative weights. Because this
loss function only changes the way the network learns the phase space
density, the generator will still produce unweighted events.

We postpone the discussion of standard event generation to
Sec.~\ref{sec:gen_inn_events}, where we also have a look at different sources
of uncertainties for generative networks\index{uncertainties}. Instead, we will show how
normalizing flows can be trained on events with negative weights,
specifically the process
\begin{align}
p p \rightarrow t \bar{t}  \; .
\end{align}
If we generate events for this process to \underline{NLO in QCD}, for
instance using the MadGraph\index{event generators} event generator. For the default setup
$23.9\%$ of all events have negative weights.

\begin{figure}[t]
  \centering
  \includegraphics[width=0.45\textwidth]{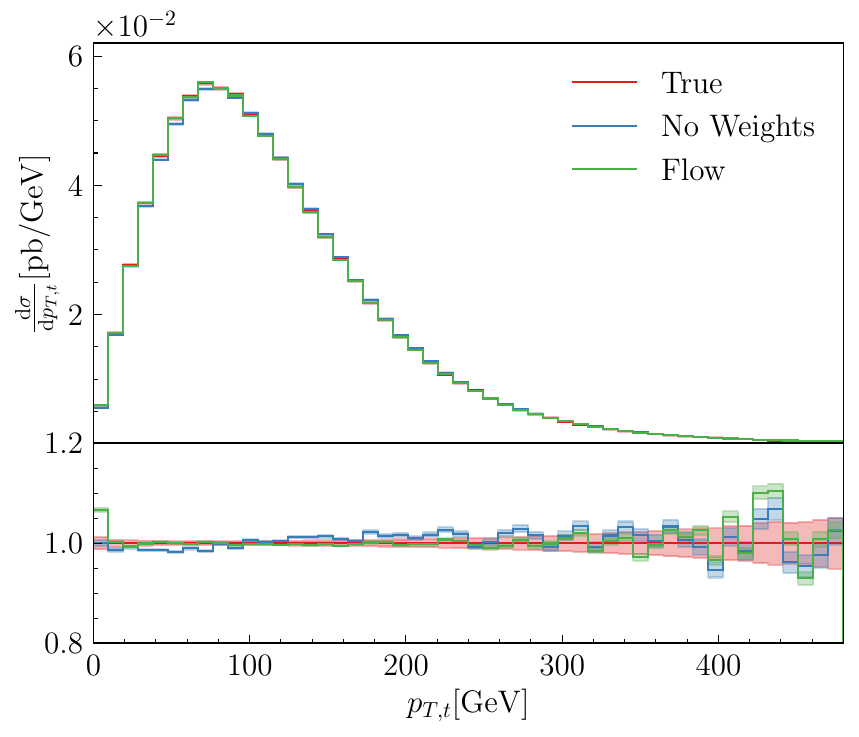}
  \hspace*{0.05\textwidth}
  \includegraphics[width=0.48\textwidth]{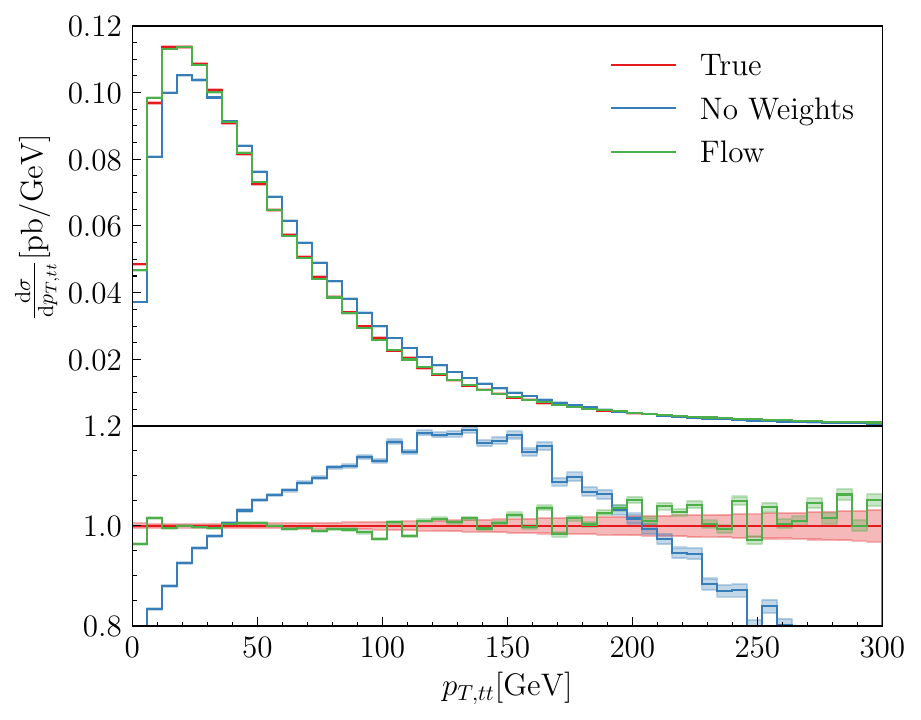}
  \caption{Generated events for $t\bar{t}$ production at NLO,
    including negative weights. We show the MC truth, the generative
    normalizing flow, and the MC truth ignoring the sign of the event
    weights, the latter to illustrate the effects of the negative
    weights.  Figure from Ref.~\cite{Stienen:2020gns}.}
\label{fig:flow_unweight}
\end{figure}

In Fig.~\ref{fig:flow_unweight} we show two example distributions
generated from the unweighting flow-network.  The distributions
without event weights are to show that the network indeed learns the
effect of the events with negative weights. We see that $p_{T,t}$ is
essentially unaffected by the weights, because this distribution is
already defined by the LO-kinematics, and negative weights only induce
to a slight bias for small values of $p_{T,t}$.  As always, the
generative network runs out of precision in the kinematic tails of the
distributions. The picture changes for the $p_{T,tt}$ distribution,
which is zero at leading order and only get generated through
real-emission corrections, which in turn are described by a mix of
positive-weight and negative-weight events. The generative network
learns the distribution correctly over the entire phase space. The
normalizing flow results in Fig.~\ref{fig:flow_unweight} can be
compared to the GANned events in Fig.~\ref{fig:gan_distris}, and we
see that the agreement between the normalizing flow and the true phase
space density is significantly better than for GANs, which means that
normalizing flows are better suited for this kind of precision
simulations.

After confirming that normalizing flows can generate events, let us go
back to the Bayesian setup introduced in Sec.~\ref{sec:gen_inn} and
analyse a little more systematically how we can extract a controlled
and precise prediction from an INN.  In Sec.~\ref{sec:class_cnn_bayes}
we have already introduced the strategy we will now follow for
precision generative networks. In two steps we first need to
\underline{control} that our network has captured all features of the
phase space density and only then can we compute the different
\underline{uncertainties}\index{uncertainties} on the encoded density.

\begin{figure}[t]
    \centering
    \includegraphics[page = 7,width=0.80\textwidth]{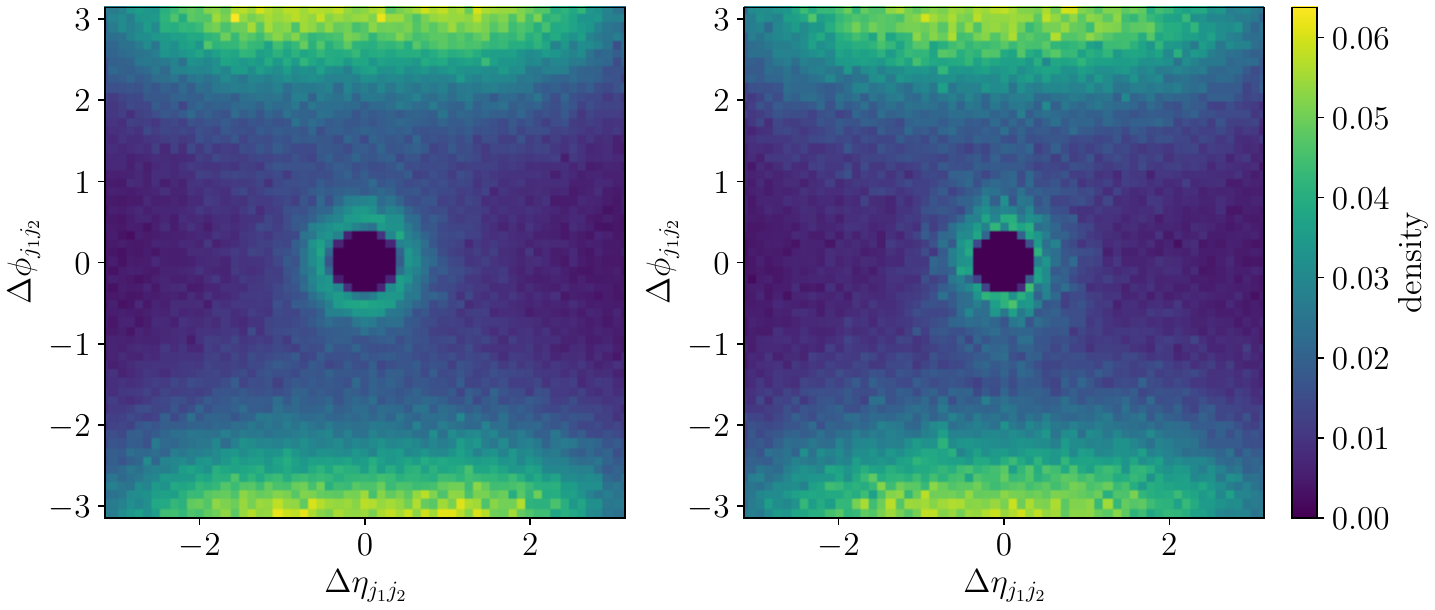}
    \caption{Jet-jet correlations for events with two jets. We show
      truth (left) and INN-generated events (right). Figure from
      Ref.~\cite{Butter:2021csz}.}
    \label{fig:inn_2d}
\end{figure}

As for most NN-based event generators we now use unweighted LO events
as training data, excluding detector effects because they soften sharp
phase space features.  The production process
\begin{align}
pp \to Z_{\mu \mu} + \{ 1,2,3 \}~\text{jets} 
\label{eq:ref_proc_z}
\end{align}
is a challenge for generative networks, because it combines a sharp
$Z$-resonance with the geometric separation of the jets and a variable
phase space dimensions.  If we assume that the muons are on-shell, but
the jets come with a finite invariant mass, the phase space has $6 +
n_\text{jets} \times 4$ dimensions. Standard cuts for reconstructed
jets at the LHC are, as usual
\begin{align}
  p_{T,j} > 20~\gev
  \qqquad \text{and} \qqquad
  \Delta R_{jj} >  0.4 \; .
\label{eq:jet_cuts}
\end{align}
In Fig.~\ref{fig:inn_2d} we show an example correlation between two
jets, with the central hole for $\Delta R_{jj} < 0.4$. Strictly
speaking, such a hole changes the topology of the phase space and
leads to a fundamental mismatch between the latent and phase
spaces. However, for small holes we can trust the network to
interpolate through the hole and then enforce the hole at the
generation stage.

Since we have learned that preprocessing will make it easier for our
network to learn the phase space density with high accuracy, we
represent each final-state particle by
\begin{align}
    \{ \; p_T, \eta, \Delta \phi, m \; \} \; ,
\label{eq:inn_obs}
\end{align}
where we choose the harder of the two muons as the reference for the
azimuthal angle and replace the azimuthal angle difference by
$\text{atanh} ( \Delta\phi/\pi )$, to create an approximately Gaussian
distribution.  We then use the jet cuts in Eq.\eqref{eq:jet_cuts} to
re-define the transverse momenta as $\tilde{p}_T = \log ( p_T -
p_{T,\text{min}})$, giving us another approximately Gaussian phase
space distribution. Next, we apply a centralization and
normalization
\begin{align}
  \tilde{q}_i = \frac{q_i - \overline{q_i}}{\sigma(q_i)} 
\end{align}
for all phase space variables $q$. Finally, we would like each phase
space variable to be uncorrelated with all other variables, so we
apply a linear decorrelation or whitening transformation separately
for each jet multiplicity.

After all of these preprocessing steps, we are left with the
challenge of accommodating the \underline{variable jet multiplicity}.
While we will need individual generative networks for each final state
multiplicity, we want to keep these networks from having to learning
the basic features of their common hard process $pp \to Zj$.  Just as
for simulating jet radiation, we assume that the kinematics of the
hard process depends little on the additional jets. This means our
base network gets the one-hot encoded number of jets as
\underline{condition}. Each of the small, additional networks is
conditioned on the training observables of the previous networks and
on the number of jets. This means we define a likelihood loss\index{likelihood loss} for a
\underline{conditional INN} just like Eq.\eqref{eq:MLE},
\begin{align}
  \pmd(x) \; \to \; \pmd(x|c,\theta)
  \qquad \Rightarrow \qquad 
  \loss_\text{cINN}
  &= - \XLangle \log \pmd(x|c,\theta) \XRangle_{\pd} \; ,
  \label{eq:cond_inn_1}
\end{align}
where the vector $c$ includes the conditional jet number, for
example. While the three networks for the jet multiplicities are
trained separately, they form one generator with a given fraction of
$n$-jet events.

\begin{figure}[t]
  \centering
  \includegraphics[page =  3, width=0.45\textwidth]{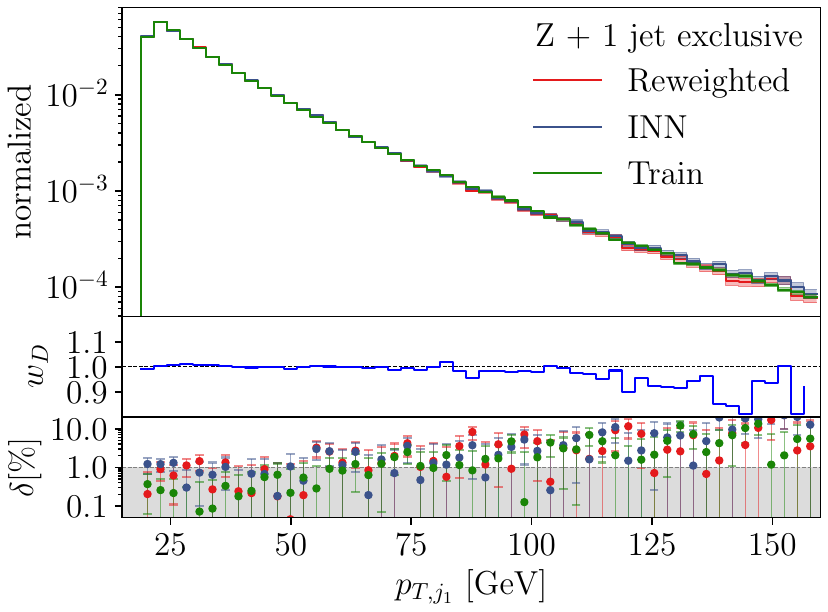}
  \hspace*{0.05\textwidth}
  \includegraphics[page = 34, width=0.45\textwidth]{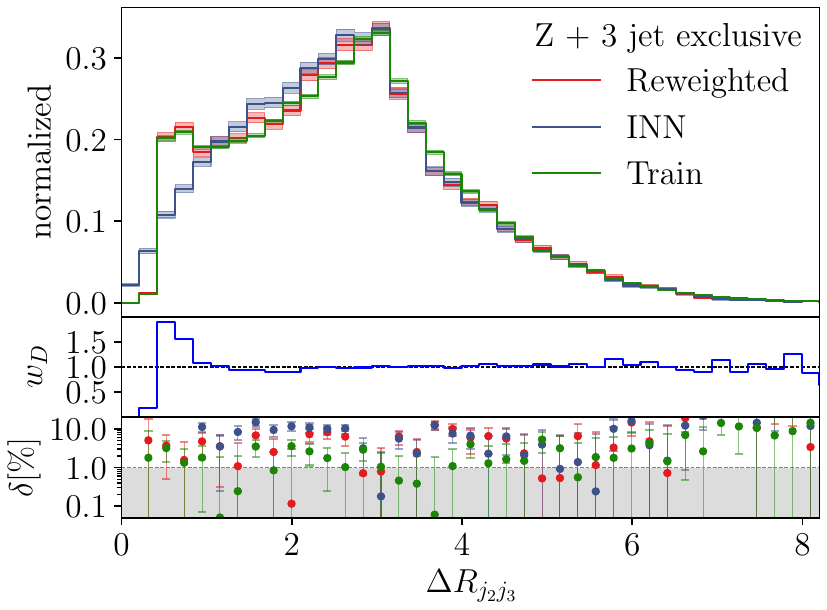}
  \caption{Discriminator-reweighted INN distributions for $Z+$jets
    production.  The bottom panels show the average correction factor
    obtained from the discriminator output.  Figure from
    Ref.~\cite{Butter:2021csz}.}
  \label{fig:reweight}
\end{figure}

To make our INN more expressive with a limited number of layers, we
replace the affine coupling blocks of Eq.\eqref{eq:layer1} with
cubic-spline coupling blocks.  The coupling
layers are combined with random but fixed rotations to ensure
interaction between all input variables.  In Fig.~\ref{fig:reweight}
we show a set of kinematic distributions from the high-statistics
1-jet process to the more challenging 3-jet process. Without any
modification, we see that the $m_{\mu \mu}$ distribution as well as
the $\Delta R_{jj}$ distributions require additional work.

One way to systemically improve and control a precision INN-generator
is to combine it with a discriminator. This way we can try to exploit
different \underline{implicit biases} of the discriminator and
generator to improve our results. The simplest approach is to train
the two networks independently and reweight all events with the
discriminator output. This only requires that our discriminator output
can be transformed into a probabilistic correction, so we train it by
minimizing a cross-entropy loss in Eq.\eqref{eq:gan_combined} to
extract a probability
\begin{align}
    D(x_i) \to \begin{cases} 0 \qquad & \text{generator} \\ 1 & \text{truth/data}
    \end{cases}
\end{align}
for each event $x_i$. For a perfectly generated sample we should get
$D(x_i) = 0.5$. The input to the three discriminators, one for each
jet multiplicity, are the kinematic observables given in
Eq.\eqref{eq:inn_obs}.  In addition, we include a set of challenging
kinematic correlations, so the discriminator gets the generated and
training events in the form
\begin{align}
  x_i = \{p_{T,j},\eta_j,\phi_j, M_j \}
  \cup \{M_{\mu \mu}\}
  \cup\{\Delta R_{2,3}\}
  \cup \{\Delta R_{2,4}, \Delta R_{3,4}\} \; .
\end{align}
If the discriminator output has a \underline{probabilistic interpretation} we
can compute an event weight
\begin{align}
  w_D(x_i) = \frac{D(x_i)}{1-D(x_i)} \; \to \; \frac{\pd(x_i)}{\pmd(x_i)} \; .
\label{eq:reweight}
\end{align}
We can see the effect of the additional discriminator in
Fig.~\ref{fig:reweight}.  The deviation from truth is defined through
a high-statistics version of the training dataset.  The reweighted
events are post-processed INN events with the average weight per bin
shown in the second panel. While for some of the shown distribution a
flat dependence $w_D = 1$ indicates that the generator has learned to
reproduce the training data as well as the discriminator can tell, our
more challenging distributions are significantly improved by the
discriminator.

While the discriminator reweighting provides us with an architecture
that learns complex LHC events at the percent level, it comes with the
disadvantage of generating weighted events and it does not use the
opportunity for the generator and discriminator to improve each other.
If it is possible to train the discriminator and generator networks
jointly, we could benefit from a GAN-like setup. The problem is that
we have not been able to reach the required Nash equilibrium in an
adversarial training for the INN generator.  As an alternative
approach we can include the discriminator information in the
appropriately normalized generator loss of Eq.\eqref{eq:MLE} through
an additional event weight
\begin{align}
  \boxed{
    \loss_\text{DiscFlow}
  = - \XXLangle w_D(x)^\alpha \log \frac{\pmd(x)}{\pd(x)}
  \XXRangle_{\pd} } \; ,
\end{align}
in complete analogy to the weighted INN loss in
Eq.\eqref{eq:weighted_loss}. In the limit of a well-trained
discriminator this becomes
\begin{align}
  \loss_\text{DiscFlow}
  &= - \XXLangle \left( \frac{\pd(x)}{\pmd(x)} \right)^\alpha \log \frac{\pmd(x)}{\pd(x)}
  \XXRangle_{\pd} \notag \\
  &= - \XXLangle \left( \frac{\pd(x)}{\pmd(x)} \right)^{\alpha} \log \pmd(x) \XXRangle_{\pd}
  + \XXLangle \left( \frac{\pd(x)}{\pmd(x)} \right)^\alpha \log \pd(x) \XXRangle_{\pd} \; .
  \label{eq:discflow1}
\end{align}
Because in our simple DiscFlow setup the discriminator weights
$\omega_D$ approximating $\pd(x)/\pmd(x)$ do not have gradients with
respect to the generative network parameters this simplifies to
\begin{align}
  \loss_\text{DiscFlow}
  &= - \XXLangle \left( \frac{\pd(x)}{\pmd(x)} \right)^\alpha \log \pmd(x) \XXRangle_{\pd} 
  =
  - \int dx \; \left( \frac{\pd(x)}{\pmd(x)} \right)^\alpha \pd(x) \;
  \log \pmd(x)  \; .
  \label{eq:discflow2}
\end{align}
The hyperparameter $\alpha$ determines the impact of the discriminator
output and can be scheduled.  The loss in Eq.\eqref{eq:discflow2} shows
that we are, effectively, training on a shifted reference
distribution. If the generator model populates a phase space region
too densely, $\omega_D <1$ reduces the weight of the training events;
if a region is too sparsely populated by the model, $\omega_D > 1$
amplifies the impact of the training data. As the generator converges
to the training data, the discriminator will give $w_D(x) \rightarrow
1$, and the generator loss approaches the standard INN form.  Unlike
the GAN setup from Sec.~\ref{sec:gen_gan} this
discriminator--generator coupling does not require a Nash equilibrium
between two competing networks.

\begin{figure}[t]
  \centering
  \includegraphics[page =  3, width=0.45\textwidth]{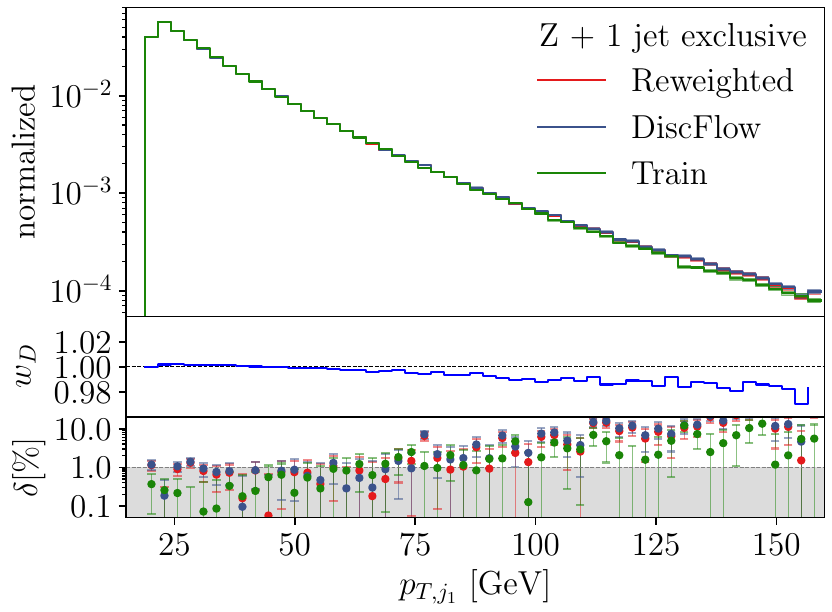}
  \hspace*{0.05\textwidth}
  \includegraphics[page = 34, width=0.45\textwidth]{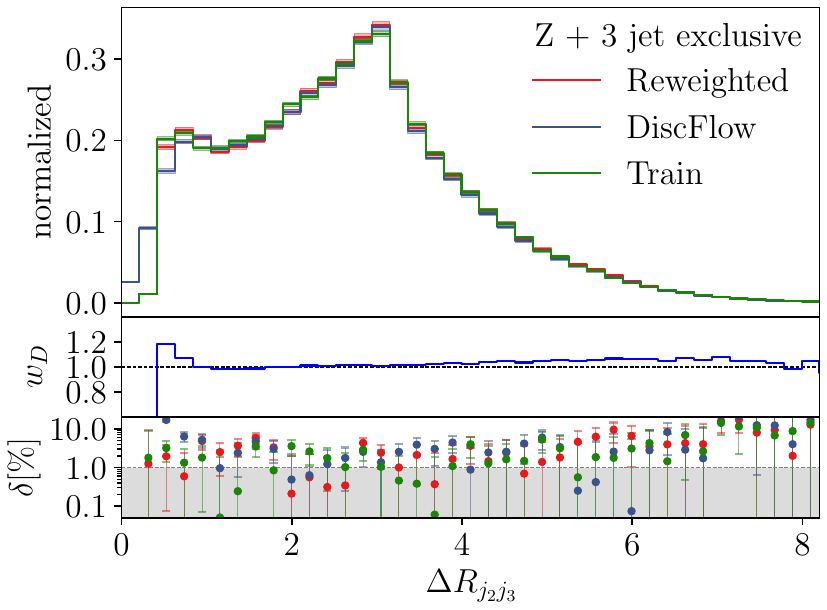}
  \caption{Discriminator-reweighted DiscFlow distributions for
    $Z+$~jets production. Figure from Ref.~\cite{Butter:2021csz}.}
  \label{fig:discflow_re}
\end{figure}

The \underline{joint discriminator-generator training} has the
advantage over the discriminator reweighting that it produces
unweighted events. In addition, we can keep training the discriminator
after decoupling the generator and reweight the events following
Eq.\eqref{eq:reweight}.  In Fig.~\ref{fig:discflow_re} we show some
sample distributions for $Z$+jets production. The DiscFlow generator
produces better results than the standard INN generator shown in
Fig.~\ref{fig:reweight}, but it still benefits from a reweighting
postprocessing.  For $\Delta R_{jj} < 0.4$ we also see that the
DiscFlow training only improves the situation in phase space regions
where we actually have training data to construct $\omega_D(x)$.

\begin{figure}[t]
  \centering
  \includegraphics[width=0.495\textwidth, page=3]{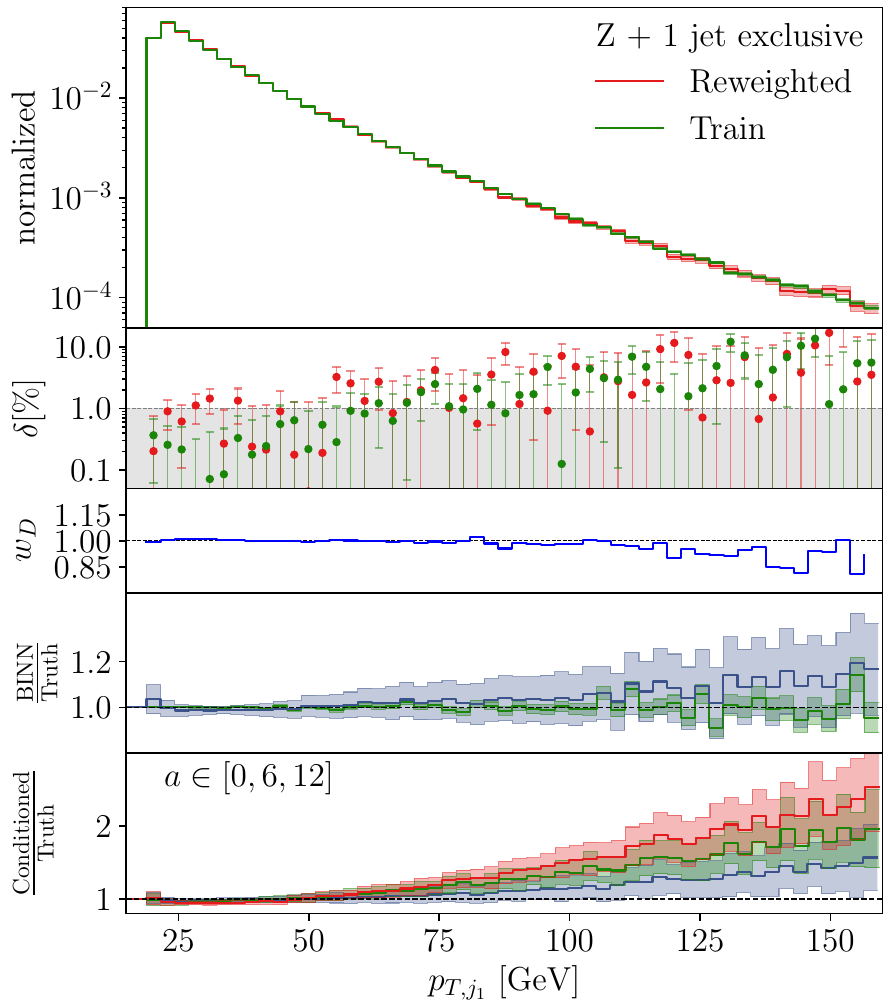}
  \caption{Illustration of uncertainty-controlled INN-generator. We
    show the reweighted $p_{T,j_1}$-distribution for the inclusive
    $Z$+jets sample, combined with the discriminator $D$, the B-INN
    uncertainty, and the sampled systematic uncertainty defined
    through the data augmentation of Eq.\eqref{eq:augment}. Figure
    from Ref.~\cite{Butter:2021csz}.}
  \label{fig:inn_final}
\end{figure}

Following our earlier argument, the improved accuracy of the
INN-generator and the control through the discriminator naturally lead
us to ask what the uncertainty on the INN prediction is. We already
know that the uncertainty estimate from the BNN automatically covers
limitations in the training data and training statistics. However, for
the generative network we have not discussed systematic uncertainties\index{systematic uncertainty}
and data augmentations, like we applied them for instance in
Sec.~\ref{sec:class_cnn_bayes}.

In the top panel of Fig.~\ref{fig:inn_final} we show the
$p_{T,j}$-distribution for $Z+1$~jet production. We choose the simple
2-body final state to maximize the training statistics and the
network's accuracy. In the second panel we show the relative
deviation of the reweighted INN-generator and the training data from
the high-statistics truth. While the network does not exactly match
the precision of the training data, the two are not far apart,
especially in the tails where training statistics becomes an issue. In
the third panel we show the discriminator weight $w_D$ defined in
Eq.\eqref{eq:reweight}. In the tails of the distribution the generator
systematically overestimates the true phase space density, leading to
a correction $w_D(x) < 1$, before in the really poorly covered phase
space regions the network predictions starts to fluctuate.

The fourth panel of Fig.~\ref{fig:inn_final} shows the uncertainty
reported by the Bayesian version of the INN-generator, using the
architecture introduced in Sec.~\ref{sec:gen_inn_arch}.  Like its
deterministic counterpart, the B-INN overestimates the phase space
density in the poorly populated tail, but the learned uncertainty on
the phase space density covers this discrepancy reliably.

Moving on to systematic or \underline{theory uncertainties}\index{theory uncertainty}, the
problem with generative networks and their unsupervised training is
that we do not have access to the true phase space density. The
network extracts the density itself, which means that any
augmentation, like additional noise, will define either a different
density or make the same density harder to learn. One way to include
data augmentations is by turning the network training from
unsupervised to supervised through an the augmentation by a parameter
known to the network. This means we augment our data via a nuisance
parameter describing the nature and size of a systematic\index{systematic uncertainty} or theory
uncertainty, and train the network conditionally on this
parameter. The nuisance parameter is added to the event format given
in Eq.\eqref{eq:ext_evt}.  As a simple example, we can introduce a
theory uncertainty proportional to a transverse momentum, inspired by
an electroweak Sudakov logarithm.  As a function of the parameter $a$
we shift the unit weights of the training events to
\begin{align}
w = 1 + a \;  \left( \frac{p_{T, j_1} -15~\gev}{100~\gev} \right)^2 \; ,
\label{eq:augment}
\end{align}
where the transverse momentum is given in GeV, we account for a
threshold at 15~GeV, and we choose a quadratic scaling to enhance the
effects of this augmentation. We train the Bayesian INN conditionally
on a set of values $a = 0~...~30$, just extending the conditioning of
Eq.\eqref{eq:cond_inn_1}
\begin{align}
  \pmd(x) \; \to \; \pmd(x|a,c,\theta)
\label{eq:cond_inn_2}
\end{align}
In the bottom panel of Fig.~\ref{fig:inn_final} we show generated
distributions for different values of $a$.  To incorporate the
uncertainty described by the nuisance parameter in the event
generation incorporating we sample the $a$-values for example using a
standard Gaussian.  In combination, we can cover a whole range of
statistical, systematic, and theoretical uncertainties by using
precise normalizing flows as generative networks. It is not clear if
these flows are the final word on LHC simulations and event
generation, but for precision simulations they appear to be the
leading generative network architecture.

\subsubsection{Phase space generation}
\label{sec:gen_inn_phase}

\begin{figure}[t]
  \includegraphics[width=0.45\textwidth]{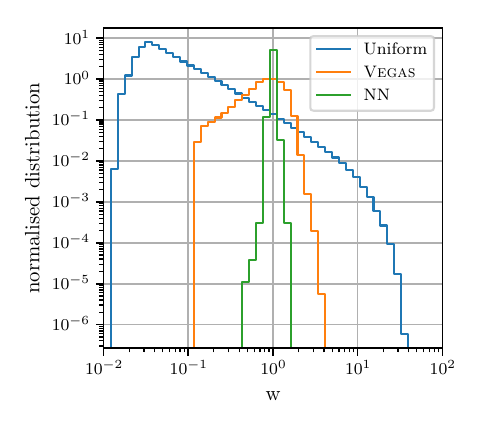}
  \hspace*{0.05\textwidth}
  \includegraphics[width=0.45\textwidth]{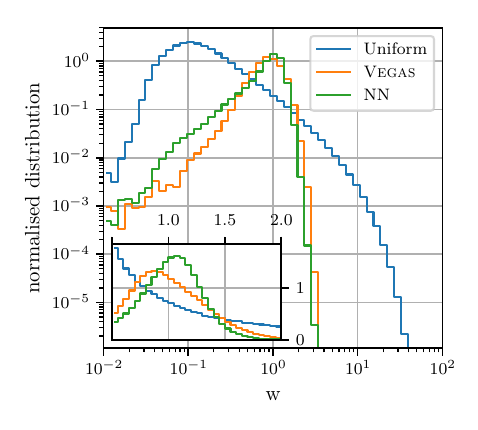}  
  \caption{Event weight distributions for sampling the total cross
    section for $gg \to 3$~jets (left) and 4~jets (right) for
    $\sqrt{s}=1~\text{TeV}$ with $N=10^6$ points, comparing Vegas
    optimization, INN-based optimization and an unoptimized
    (``Uniform'') distribution. The inset in the right panel shows the
    peak region on a linear scale. Figure from
    Ref.~\cite{Bothmann:2020ywa}.}
  \label{fig:inn_weights}
\end{figure}

In addition to learning the phase space distribution of events from a
set of unweighted events, parton-level events can also be learned from
the differential cross section directly. Mathematically speaking, the
problem can be formulated as sampling random numbers according to a
given functional form. An ML-framework for this is not just useful for
LHC event generation\index{event generators}, is can be used for any kind of numerical
Monte-Carlo integration in high dimensions.

Numerical integration with Monte-Carlo techniques and importance
sampling are already discussed in
Sec.~\ref{sec:gen_gan_unweight}. Now, we want the coordinate
transformation of Eq.\eqref{eq:coord_trafo} to be learned by a
normalizing flow.  In the language of LHC rates of events, we have to
transform the \underline{differential cross section}, $d\sigma$, into a
properly normalized distribution,
\begin{align}
  \pcs = \frac{d\sigma}{\int d\sigma} \; .
  \label{eq:def_cross_sec}
\end{align}
Then, we can use the loss functions discussed at the beginning of
Sec.~\ref{sec:gen}, like the KL-divergence, to learn the
distribution. Looking back at Eq.\eqref{eq:def_kl3}, we have with the
identification of $\pd = \pcs$
\begin{align}
  \loss_\text{phase Space}
  = \kl [\pcs,\pmd] = \int d x \; \pcs(x) \; \log \frac{\pcs(x)}{\pmd(x)} 
  \; .
\label{eq:KL_integration}
\end{align}
The main difference to Eq.\eqref{eq:def_kl3} is that now we do not
have training samples from $\pd$, because generating samples according
to $\pcs$ is the problem we want to solve. However, we do have samples
distributed according to $\pmd$, generated from our model. We can use
them for training, provided we correct their weights by introducing a
factor $\pmd/\pmd $ into Eq.\eqref{eq:KL_integration} and then
interpreting it as an expectation value over $\pmd$,
\begin{align}
  \loss_\text{phase Space}
  = \XXLangle \frac{\pcs(x)}{\pmd(x)} \log \frac{\pcs(x)}{\pmd(x)} \XXRangle_{\pmd} \; .
\label{eq:KL_integration.numerical}    
\end{align}
However, we have to be careful when computing the gradients of the
loss of Eq.\eqref{eq:KL_integration.numerical} with respect to the
network weights. These enter, in principle, every quantity in
Eq.\eqref{eq:KL_integration.numerical}, namely
\begin{itemize}
\item $\pcs$ in Eq.\eqref{eq:def_cross_sec} is computed by estimating
  the integral in the denominator with a Monte-Carlo estimate, which
  can be improved using importance sampling and samples from $\pmd$.
  \begin{align}
    \int d\sigma = \frac{1}{N}\sum d\sigma = \XXLangle \frac{d\sigma}{\pmd}\XXRangle_{\pmd} \; .
  \end{align} 
  The estimate, however, should just be a constant normalization
  factor, so the dependence on the network parameters is spurious and
  a gradient with respect to them will not help the optimization task.
\item $\pmd$ in the denominator of the prefactor corrects for the fact
  that we have only samples from the model, not according to
  $\pcs$. These can come from the current state of the model, or from
  a previous one when we recycle previous evaluations. But since this
  distribution is external to the optimization objective, we should
  not take gradients of this prefactor.
\item $\pmd$ in the denominator of the logarithm is the one we want to
  optimize, so we need its gradient.
\end{itemize}

The figure of merit, as introduced in Sec.~\ref{sec:gen_gan_unweight},
is the unweighting efficiency. It tells us what fraction of events
survives a hit-and-miss unweighting as given in
Eq.\eqref{eq:uw_efficiency}. Figure~\ref{fig:inn_weights} compares the
unweighting efficiency for uniformly sampled, Vegas sampled, and INN
sampled points in $gg\to 3$ jets and $gg\to 4$ jets. We see that the
normalizing flow beats the standard Monte Carlo method by a large
margin for the simpler case, but loses its advantage in the right
panel. This points to a problem in the scaling of neural network
performance with the number of phase space dimensions, as compared to
the logarithmic scaling of established Monte Carlo methods.

\subsubsection{Calorimeter shower generation}
\label{sec:gen_inn_calo}

\begin{figure}[t]
  \centering
  \includegraphics[width=0.60\textwidth]{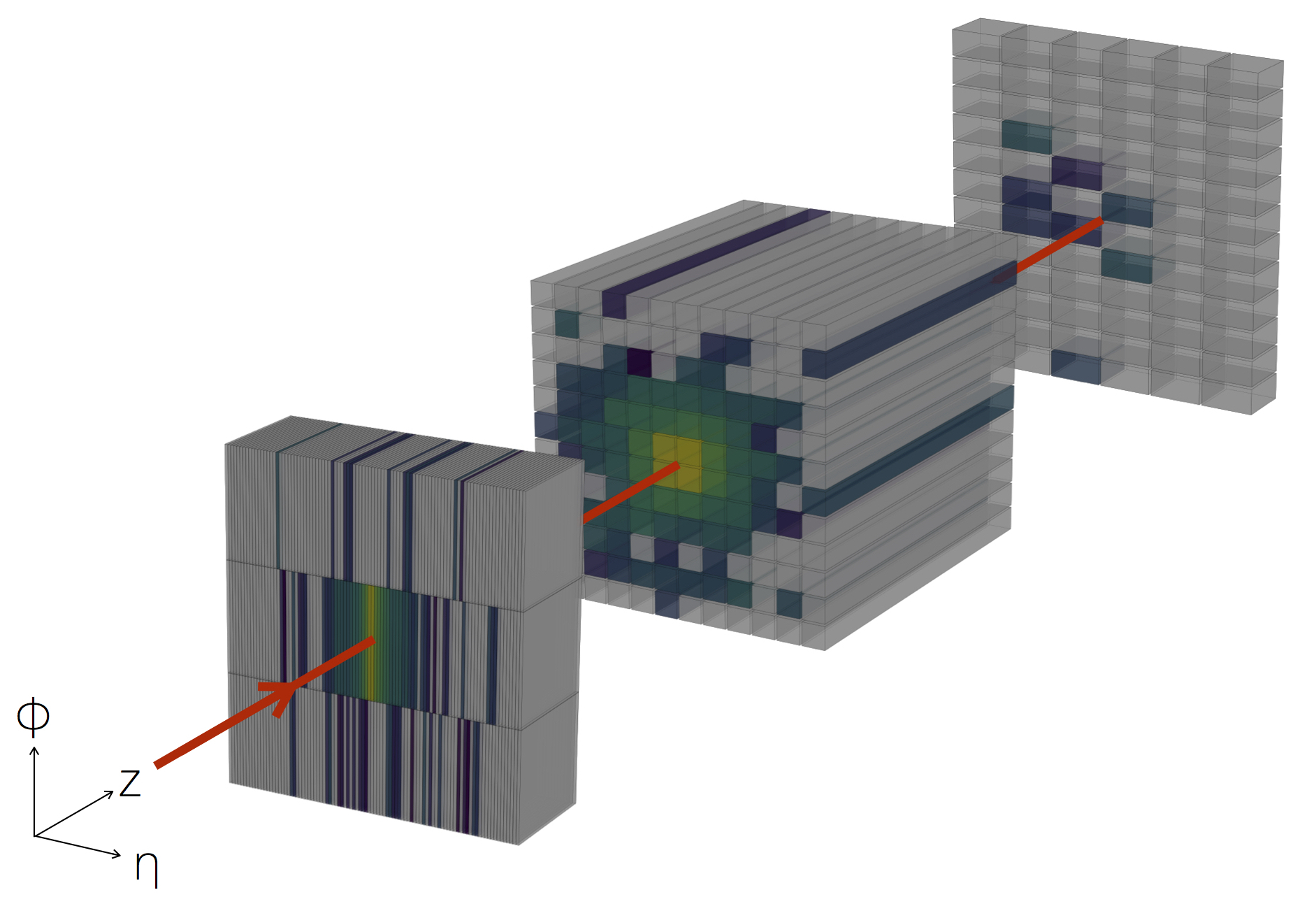}
  \caption{3-dimensional view of calorimeter shower produced by an
    incoming $e^+$ with $E_{\text{inc}}=10~$GeV, simulated with
    GEANT4. Figure from Ref.~\cite{calogan2}. }
  \label{fig:CaloGAN_shower_3d}
\end{figure}

Another application of INNs as generative networks in LHC physics is the
simulation of the detector response to incoming particles. Such
interactions are stochastic and implicitly defines a phase space distribution
\begin{align}
  p(x_\text{shower}|\text{initial state}) \; .
\end{align}
The initial state is
characterized by the incoming particle type, energy, position,
and angle to the detector surface. For simplicity, we focus on
showers originating from the center of the detector volume, and always
perpendicular to the surface. We also train a different INN for each
particle type, so the learning task is simplified
to
\begin{align}
  p(x_\text{shower}|E_{\text{inc}}) \; .
\end{align}
 
As ground truth, we use 100,000 showers simulated with GEANT4, a
simulation framework based on first principles. Simulating all LHC
events with GEANT4 is computationally extremely expensive.  Especially
high-energy showers can take extremely long to simulate, orders of
magnitude longer than low-energy showers due to the larger number of
secondary particles needing to be produced and tracked.  Whenever only
some features of the showers and not all details on energy depositions
are needed, fast simulations can be used.  In these frameworks, a few
high-level features are simulated instead of the full shower
information, trading shower fidelity for evaluation speed. Generative
networks, such as GANs or INNs, instead allow us to generate showers
with many more details in the high-dimensional low-level feature space
and with the same speed for all incident energies.

\begin{figure}[b!]
  \centering
  \includegraphics[width=\textwidth]{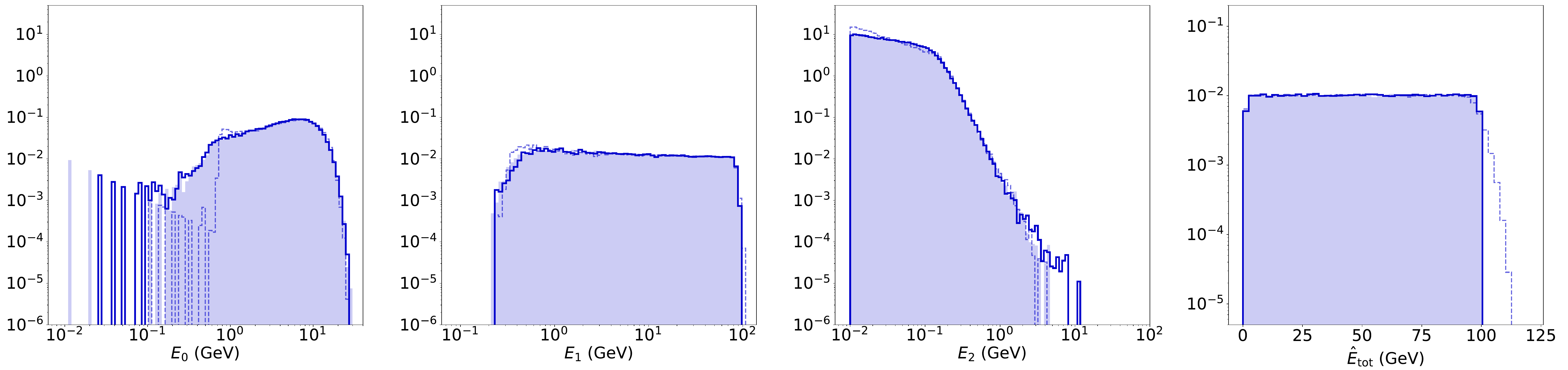}
  \includegraphics[width=0.50\textwidth]{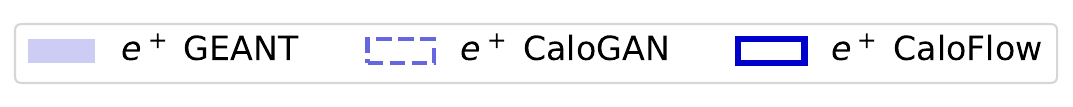}
  \caption{Energy deposition per layer and total energy
    deposition of $e^+$ showers.
    Figure from Ref.~\cite{Krause:2021ilc}. }
  \label{fig:CaloFlow_hists}
\end{figure}

The GANplification effect of Sec.~\ref{sec:gen_gan_amp} also applies
to calorimeter showers\index{statistical amplification}, but there are two more effects that work in
our favor and allow us to generate many more events than are in the
training sample. The use the fact that calorimeter showers are
independent of the hard scattering and independent of each other. Even
for a fixed number of training showers the combinatorial factors of
combining them with the particles produced in the hard scattering will
increase the number of statistically independent events. Second,
depending on the chosen voxelization, the generated showers can be
rotated before they are combined with an LHC event.

As an example, we consider $e^+, \gamma$, and $\pi^+$ showers in a
simplified version of the ATLAS electromagnetic calorimeter. Showers
are digitized into three layers of voxels, with varying size per
layer. In total, the dataset contains 504 voxels. The incident energy
will be sampled uniformly in $[1,
  100]$~GeV. Figure~\ref{fig:CaloGAN_shower_3d} shows such a
shower. Characteristic for calorimeter datasets, we see a high degree
of sparsity and energy depositions ranging over several orders of
magnitude.  Since the three layers vary in their physical length, the
energy deposited on average per layer also differs. To avoid being
dominated by the central layer in training, we normalize all showers
such that they sum to one in each layer. This normalization can be
inverted in the generation when we know the layer energies, so we also
have to learn the probability $p(E_0, E_1, E_2|E_{\text{inc}})$. This
can either be done by an independent generative network, or by
augmenting the normalized voxel energies with the layer energy
information.

When training normalizing flows for calorimeter showers, the high
degree of sparsity tends to overwhelm the likelihood loss. The network
focuses on learning the many zero entries and does not do well on
single voxels with the dominant energy depositions. To ameliorate
this, we add noise below the read-out threshold of the detector to the
voxels. After generation, we then remove all entries below that
threshold and recover the correct sparsity.  In
Fig.~\ref{fig:CaloFlow_hists} we show the deposited energies in the
three layers and overall, for $e^+$ showers from GEANT4, a GAN, and a
normalizing flow.

The main advantage of generative networks over GEANT4 is speed. The
first flow-based approach, however, was based on an autoregressive
architecture, which is fast to train but a factor 500 (given by the
dimensionality of the voxel space) slower in generation. This means
that for a batch size of 10k showers, a GPU can produce a single
shower in 36ms. Even though this is significantly faster than GEANT4
(1772ms), it is much slower than the GAN (0.07ms). This difference can
be addressed in two ways. One would be to use an alternative
normalizing flow architecture. The second one is to train a second
autoregressive flow with the inverse speed preference: fast sample
generation and slow evaluation of the likelihood loss. This is referred to as 
probability density distillation or
teacher-student training. Starting from a flow trained with the
likelihood loss (CaloFlow v1), we freeze its weights and train a
second flow to match the output of the first flow. Using an MSE loss
after every transformation and also to match the output of each of the
NNs, we ensure that this new flow becomes a copy of the first one, but
with a fast generative direction.

\subsubsection{Multi-channel importance sampling}
\label{sec:madnis}

In Sec.~\ref{sec:gen_inn_phase}, we have introduced the concept of
importance sampling and how to construct a better sampling density
using normalizing flows. A recurring feature of LHC scattering
processes is multi-modal phase-space structures. It is generally
difficult, if not impossible, to find and parametrize a single density
$g$ that describes $d\sigma$ across the entire phase space. Instead,
we decompose the phase-space integral into $n_c$ different
channels
\begin{align}
  \sigma
  = \int dx\frac{d\sigma}{dx}
  &\equiv\int dx\,w(x) \notag \\
  &= \sum_i^{n_c} \int dx \,\alpha_i(x)\,w(x)
  = \sum_i^{n_c} \sigma_i\; ,
  \label{eq:multi-channel.integral}    
\end{align}
by introducing normalized \underline{channel weights}
\begin{align}
  \sum_i^{n_c} \alpha_i(x) = 1
  \qquad \text{with} \qquad \alpha_i\in[0,1] \;.
    \label{eq:multi-channel.weights}  
\end{align}
For each channel we define a coordinate transformation
\begin{align}
x \;
\xleftrightarrow[\quad \leftarrow \overline{G}_i(z)\quad]{G_i(x)\rightarrow} 
\; z  \; ,
\label{eq:multi-channel.ps_mapping}
\end{align}
which induces properly normalized densities
\begin{align}
   g_i(x)=\left\vert\frac{\partial G_i(x)}{\partial x}\right\vert
   \qquad \text{with} \quad 
   \int dx\,g_i(x)=1  \; .
   \label{eq:multi-channel.channel_densities}
\end{align}
This turns the phase-space integral into
\begin{align}
    \boxed{ \sigma=\sum_i^{n_c} \int dx \,\alpha_i(x)w(x)
    =\sum_i^{n_c} \int dz\left.\alpha_i(x)\frac{w(x)}{g_i(x)}\right\vert_{x=\overline{G}_i(z)} } \; ,
    \label{eq:multi-channel.transform} 
\end{align}
Normally, the construction of the channels is physics-inspired and
employs analytic mappings which are then further refined by an
adaptive algorithm, for instance Vegas.  While the channel cross
sections $\sigma_i$ are unchanged under the mappings $G_i$, the
variance
\begin{align}
  \text{var}_i^2\equiv\text{var}_i^2\left[\frac{\alpha_i w}{g_i}\right]
  &=\int dz\left(\left.\frac{\alpha_i(x) w(x)}{g_i(x)}\right\vert_{x\equiv x(z)}-\sigma_i\right)^2 \notag \\
  &=\int dx\,g_i(x) \left(\frac{\alpha_i(x) w(x)}{g_i(x)}-\sigma_i\right)^2
\label{eq:multi-channel.def_sigma}
\end{align}
is minimized by the optimal mapping
\begin{align}
g_i(x)\Bigg|_\text{opt}=\frac{\alpha_i(x)w(x)}{\sigma_i}\; .
\label{eq:multi-channel.optimal}
\end{align}
Now, the goal of \underline{MadGraph Neural Importance Sampling}
(MadNIS)\index{MadNIS} is to parametrize and learn both the channel weights
$\alpha_i(x)$ and the coordinate transformation $ G_i(x)$ with neural
networks.

MadNIS implements the channel weights as simple
regression network,
\begin{align}
   \alpha_{i}(x)\equiv\alpha_{i\xi}(x) \; ,
  \label{eq:multi-channel.trained_weights}
\end{align}
with network parameters $\xi$ and the number of output nodes equal to
the number of channels. The normalization condition from
Eq.\eqref{eq:multi-channel.weights} is guaranteed by the last layer
\begin{align}
    \softmax \alpha_{i\xi}(x)=\frac{e^{\alpha_{i\xi}(x)}}{\sum_j e^{\alpha_{j\xi}(x)}} \;.
\end{align}
It is often beneficial to use prior knowledge when training a neural
network. While the channel weight network might converge to the
optimal channel weights with a sufficiently large training dataset, we
can use physics to define a good starting point. For
instance, in MadGraph two well-motivated and normalized priors are
\begin{align}
    \alpha_i^\text{MG}(x)&= \frac{|\mathcal{M}_i(x)|^2}{\sum_j |\mathcal{M}_j(x)|^2} \notag \\
    \alpha_i^\text{MG}(x) &= \frac{P_i(x)}{\sum_j P_j(x)} 
  \quad \text{with} \quad
  P_i(x) =  \prod_{k \in \text{prop}} \frac{1}{|p_k(x)^2-m_k^2 -\mathrm{i} m_k\Gamma_k|^2} \; ,
\label{eq:multi-channel.mg_prior}
\end{align}
defining the single-diagram enhanced multi-channel method.  Relative
to either of them we learn a correction
\begin{align}
  \alpha_{i\xi} (x) &= \log \alpha^{\text{MG}}_i(x)+ \Delta_{i\xi}(x) \notag \\
  \softmax \alpha_{i\xi}(x) &= \frac{\alpha^{\text{MG}}_i(x) \; e^{\Delta_{i\xi}(x)}}
           {\sum_j \left[ \alpha^{\text{MG}}_j(x) \; e^{\Delta_{j\xi}(x)} \right] } \; ,
\label{eq:delta_channel}
\end{align}
initialized as $\Delta_{i\xi}(x)=0$.

Next, MadNIS implements the channel-mappings as a combination of a
physics-inspired analytic mappping and an INN, as introduced in
Sec.~\ref{sec:gen_inn}
\begin{align}
x  
\xleftrightarrow[\hspace{1.1cm}]{\text{analytic}} 
y
\xleftrightarrow[\hspace{1.1cm}]{\text{INN}}
z\; .
\label{eq:multi-channel.def_inn}
\end{align}
This chain effectively replaces Vegas with an INN. MadNIS improves the
physics-inspired phase-space mappings by training an INN as
\begin{align}
  z = G_{i\theta}(x)
  \qquad \text{or} \qquad
  x = \overline{G}_{i \theta}(z) \; ,
  \label{eq:multi-channel.trained_mappings}
\end{align} 
with the network weights $\theta$.  Similar to the multi-channel
weights, we use physics knowledge to simplify the task of the
normalizing flow and improve its efficiency.

Combining the two networks, MadNIS simultaneously optimizes both the
channel weights and the mappings to reduce the variance of the
integral.  In contrast to Sec.~\ref{sec:gen_inn_phase}, it does not
normalize the integrands of each channel to probability distributions,
%
%
because the channel integrals 
anymore but are now functions of the trainable channel weights
$\alpha_{i\xi}$. The loss function has to be constructed such that the
relative channel contributions are preserved. MadNIS uses the weighted
squared sum of the variances of the channels from
Eq.\eqref{eq:multi-channel.def_sigma} to minimize the complete
variance
\begin{align}
\loss_\text{variance}
=\sum^{n_c}_{i=1} \frac{N}{N_i}\text{var}^2_i
&=\sum^{n_c}_{i=1} \frac{N}{N_i}\int dx\,g_{i\theta}(x) \left(\frac{\alpha_{i\xi}(x) w(x)}{g_{i\theta}(x)}-\sigma_{i\xi}\right)^2 \notag \\
&=\sum^{n_c}_{i=1} \frac{N}{N_i}\int dx\, \left(
\frac{\alpha_{i\xi}(x)^2 w(x)^2}{g_{i\theta}(x)} 
- 2 \alpha_{i\xi}(x) w(x) \sigma_{i\xi}
+ \sigma_{i\xi}^2 \right) \notag \\
&=\sum^{n_c}_{i=1} \frac{N}{N_i}\left(\int dx \,\frac{\alpha_{i\xi}(x)^2 w(x)^2}{g_{i\theta}(x)}-\left[\int dx \,\alpha_{i\xi}(x) w(x)\right]^2\right) \; .
\end{align}
In the last line we use the definition of $\sigma_{i\xi}$ as the
learned approximation of Eq.\eqref{eq:multi-channel.integral}. To
evaluate the $x$-integration numerically we introduced a new sampling
density $q_i(x)$ which is independent of the model density
$g_{i\theta}(x)$,
\begin{align}
\loss_\text{variance}
&=\sum^{n_c}_{i=1}\frac{N}{N_i}\left(
    \left\langle \frac{\alpha_{i\xi}(x)^2 w(x)^2}{g_{i\theta}(x)\,q_i(x)} \right\rangle_{x\sim q_i(x)}
      - \left\langle \frac{\alpha_{i\xi}(x) w(x)}{q_i(x)} \right\rangle_{x\sim q_i(x)}^2 \right) \; .
\label{eq:multi-channel.varloss}
\end{align}
The reason for choosing an independent sampling density is twofold:
(i) potentially intractable integrand gradients, and (ii) buffered
training.

First, to optimize the normalizing flow and learn $g_{i\theta}$, we
have to compute the gradients $\nabla_\theta \loss_\text{variance}$.
f the sampling density and model density coincide, $q_i=g_{i\theta}$,
the sampled events $x\equiv x_\theta$ depend on the network parameters
$\theta$. Computing the gradients requires gradients of all terms of
the integrand, including $\nabla_\theta w(x_\theta)$. In general, the
weights are not differentiable. By choosing a sampling density $q$
independent of the network parameters $\theta$, we only need the
gradient of $g_{i\theta}$ which is always tractable.

Second, to train a normalizing flow in the standard importance
sampling setup requires continuously generating new training events
and train online.  For expensive integrands, like higher-order or
high-multiplicity matrix elements, this limits the training
capabilities. In that case it helps to buffer generated events and
their underlying sampling density and train on these events while
continuously replacing the buffer with newly generated samples. To
stabilize the combined \underline{online and buffered
  training}\index{buffered vs online training} and to compute the loss
with high precision, we choose a sampling density $q_i(x)\simeq
g_{i\theta}(x)$.

A critical hyperparameter in the variance loss in
Eq.\eqref{eq:multi-channel.varloss} is the numbers of points per
channel, $N_i$, during training and integral evaluation. Its optimal
choice depends on the $\text{var}_i$ and can be computed
analytically by minimizing the loss with respect to $N_i$ under the condition
$N=\sum_i N_i$. It gives
\begin{align}
  N_i
 = N \frac{\text{var}_{i}}{\sum_j \text{var}_{j}} 
 \qquad \Leftrightarrow \qquad 
 \frac{N}{N_i} = \frac{\sum_j \text{var}_j}{\text{var}_i}  \; .
 \label{eq:multi-channel.stratified-sampling}
\end{align}
This known result from stratified sampling defines the MadNIS loss
\begin{align}
    \boxed{\loss_{\text{MadNIS}} 
    = \left[\sum_{i=1}^{n_c} \text{var}_i\right]^2
    = \left[\sum_{i=1}^{n_c} \left(
    \left\langle \frac{\alpha_{i\xi}(x)^2 w(x)^2}{g_{i\theta}(x)q_i(x)} \right\rangle_{x\sim q_i(x)}
      - \left\langle \frac{\alpha_{i\xi}(x) w(x)}{q_i(x)} \right\rangle_{x\sim q_i(x)}^2 \right)^{1/2} \right]^2 }
    \;.
    \label{eq:multi-channel.madnis_loss}
\end{align}
The variance of a channel usually scales with its cross
section. Moreover, even if different channels contribute similarly to
the total cross section, their variances can still be very
imbalanced. The MadNIS loss is always dominated by channels with a
large variance, which can lead to unstable training if the number of
points in these channels is too small. We can further improve the
MadNIS training by two additional tricks:
Because the variance of the channels changes during the training of
the channel weights and during the training of the importance
sampling, we track running means of the channel variances and use them
to adjust the number of points per channel during online training. To
ensure a stable training we first distribute a fraction of points
evenly among channels. The remaining events are distributed according
to the channel variances, following
Eq.\eqref{eq:multi-channel.stratified-sampling}.

For most LHC applications, MadNIS captures the relevant the phase
space with a few channels. We can exploit this and drop channels with
negligible contribution to the total cross section.  If we drop a
channel, we also remove its contribution to the buffered sample and
adjust the normalization of the channel weights. This way no training
time is invested into dropped channels and the result remains
unbiased.

\begin{figure}[t]
  \includegraphics[width=0.45\textwidth]{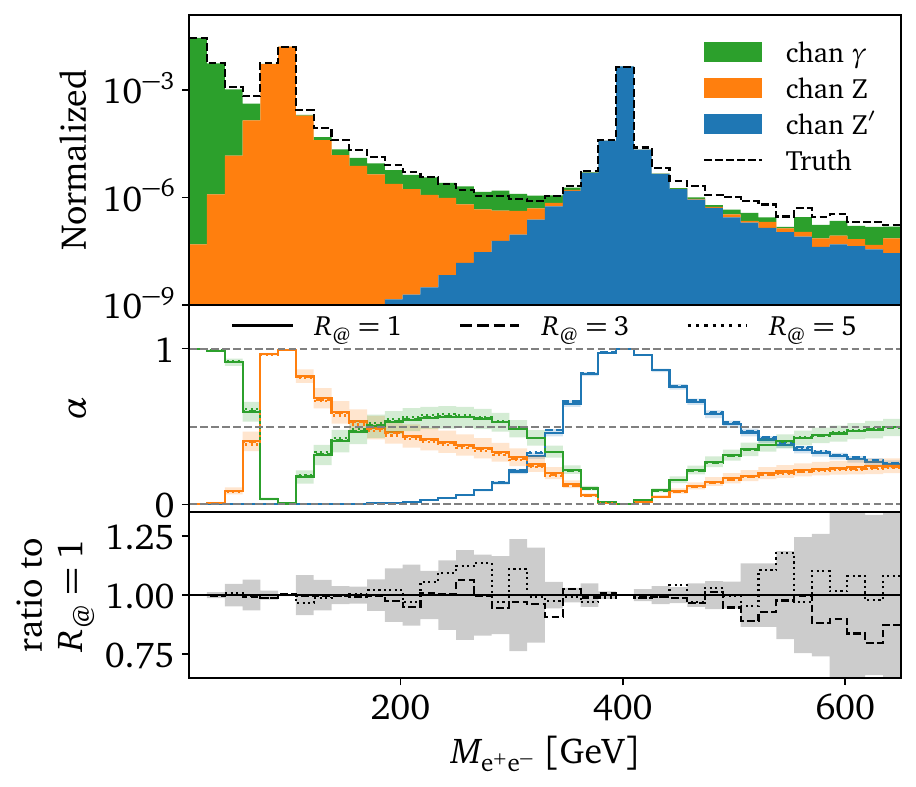}  
  \hspace*{0.05\textwidth}
  \includegraphics[width=0.45\textwidth]{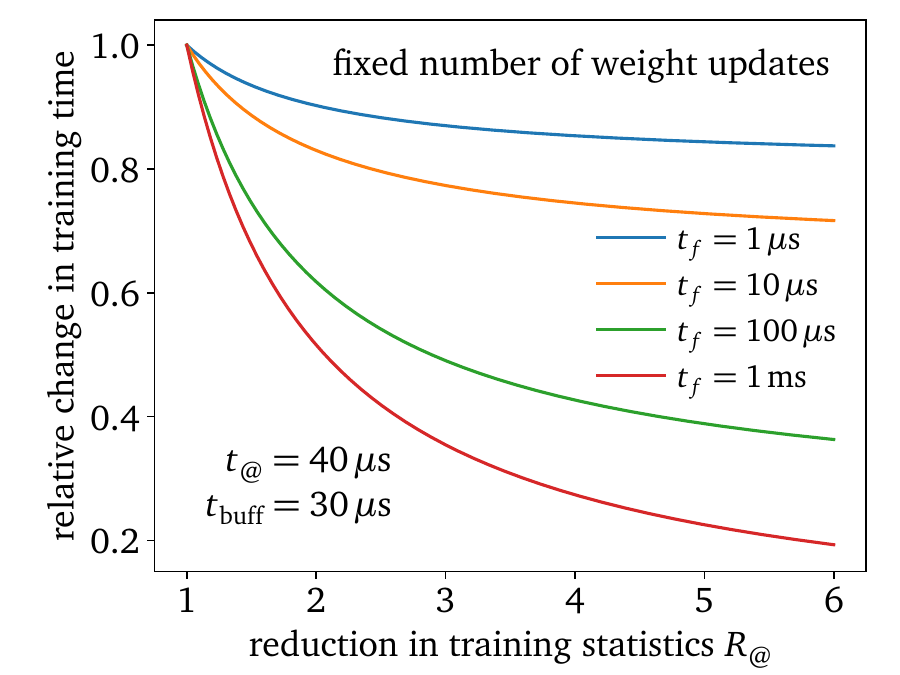} 
  \caption{Left: learned $M_{e^{+}e^{-}}$ distribution where the
    middle panel shows the learned channel weights, and the lower
    panel shows the ratio of the combined distribution to pure online
    training for reduction factors in training statistics. Right:
    change in training time as a function of the reduction in training
    statistics $R_\text{@}$ for integrands with different
    computational costs. Figures from Ref.~\cite{Heimel:2022wyj}.}
  \label{fig:madnis_results}
\end{figure}

We can benchmark the performance of MadNIS for different LHC
processes. First, we consider the Drell-Yan process with an additional
$Z'$-resonance,
\begin{align}
     p p  \to \gamma, Z^*, {Z'}^* \to e^+ e^-\; ,
\label{eq:drell-yan}
\end{align}
assuming $M_{Z'} = 400.0$~GeV and $\Gamma_{Z'} = 0.5$~GeV.  The left
plot in Fig.~\ref{fig:madnis_results} shows how the learned channels
map out the different Breit-Wigner peaks and simplify the INN
tasks. Each learned channel weight $\alpha_i$ dominates in a specific
phase space region. In the bottom panel and in the right plot, the
effect of reduced training statistics induced by the buffered training
is compared to the traditional online training. For the same relative
uncertainty the buffered training reduces the CPU time by up to 80\%.

Targeting actual LHC challenges, we can implement MadNIS in the
MadGraph framework, to give is access to scattering matrix elements,
parton densities and analytic phase space mappings. We can then apply
it to the cutting-edge scattering processes
\begin{alignat}{4}
W+\text{jets}&  \qquad  &
g g &\to W^+  d \bar{u} \qquad &
g g &\to W^+  d \bar{u} g \qquad &
g g &\to W^+  d \bar{u} gg
\notag \\
\text{$t\bar{t}$+jets}&  \qquad  & 
g g &\to t\bar{t}+g \qquad &
g g &\to t\bar{t}+gg \qquad &
g g &\to t\bar{t}+ggg\; ,
\label{eq:proc}  
\end{alignat}
The number of Feynman diagrams at leading order ranges from 8 for
$Wd\bar{u}$ production to 1240 for the $t\bar{t}ggg$ process.

\begin{figure}[t]
    \includegraphics[width=0.495\textwidth]{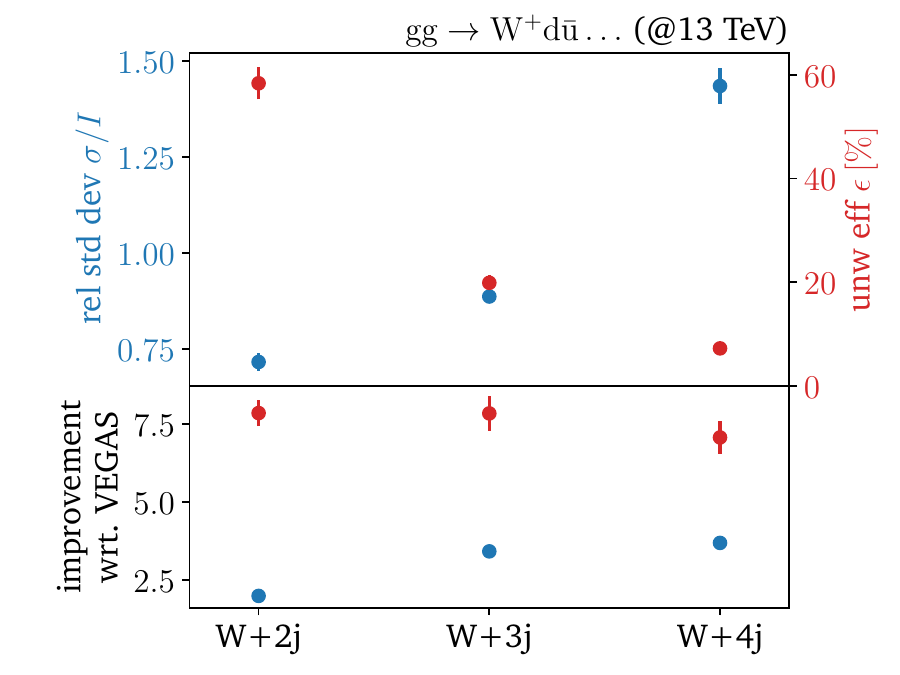}
    \includegraphics[width=0.495\textwidth]{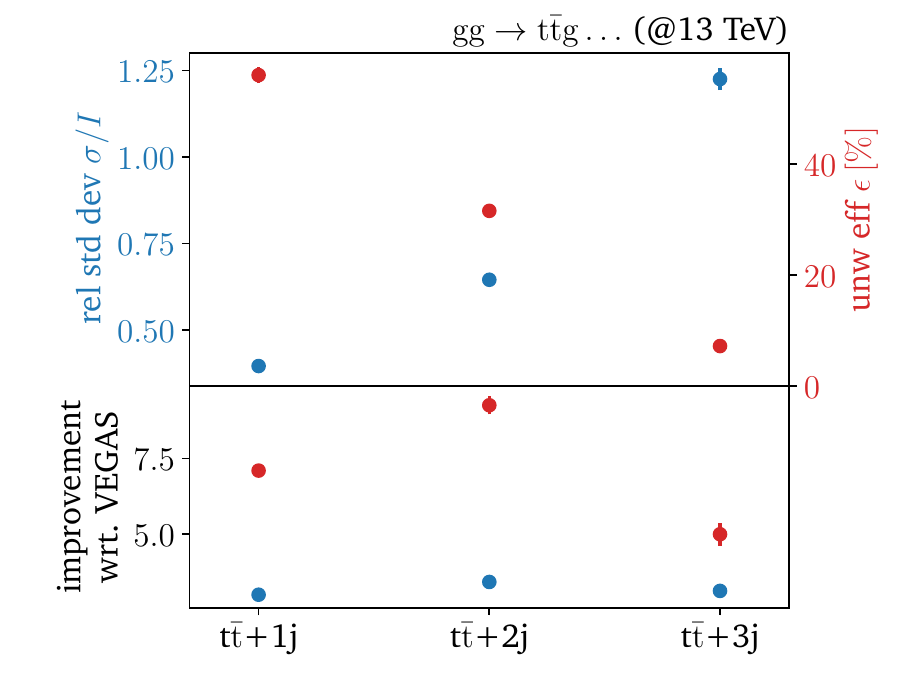}
    \caption{
      Relative standard deviation and unweighting
      efficiency for W+jets and $t\bar{t}$+jets with different numbers
      of gluons in the final state. The performance gain is
      illustrated in the lower panels. Figure from
      Ref.\cite{Heimel:2023ngj}.}
    \label{fig:process_comparison}
\end{figure}

In Fig.~\ref{fig:process_comparison}, we show the performance gain of
MadNIS compared to Madgraph using Vegas only. As quality measure, we
show two different metrics: the relative standard deviation
$\text{var}_\sigma / \sigma$, as is it independent from the sampling
statistics, and the unweighting efficiency $\epsilon$.
Looking at the scaling with the number of additional gluons in the
final state, the unweighting efficiency decreases and standard
deviation increases towards higher multiplicities, and the relative
gain of MadNIS over Vegas remains roughly constant for W+jets
production. For the even more challenging $t\bar{t}$+jets process the
gain decreases for the 3-gluon case.  This can partially be explained
by an suboptimal implementation for many channels (945 for
$t\bar{t}+3\text{jets}$), leaving work for the actual implementation
in upcoming Madgraph releases.

\subsection{Diffusion networks}
\label{sec:gen_diff}

Diffusion networks are a new brand of generative networks which are
similar to normalizing flows, but unlike the INN they are not fully
bijective and symmetric. On the positive side, they tend to be more
expressive that the relatively constrained normalizing flows. We will
look at two distinctly different setups, one based on a discrete time
series in Sec.~\ref{sec:gen_diff_ddpm} and another one based on
differential equations in Sec.~\ref{sec:gen_diff_cfm}.

\subsubsection{Denoising diffusion probabilistic model}
\label{sec:gen_diff_ddpm}

Looking at the INN mapping illustrated in Eq.\eqref{eq:inn_mapping},
we can interpret this basic relation as a time evolution from the phase
space distribution to the latent distribution or back,
\begin{align}
  \pmd(x_0)
  \quad 
  \stackrel[\leftarrow \text{backward}]{\text{forward} \rightarrow}{\xleftrightarrow{\hspace*{1.5cm}}}
  \quad
  \pl(x_T) \; ,
\label{eq:time_series}
\end{align}
identifying the original parameters $x \to x_0$ and $r \to x_T$ in the
spirit of a discrete time series with $t = 0~...~T$. The step-wise
adding of Gaussian noise defines the forward direction of Denoising
Diffusion Probabilistic Models (DDPMs).  The task of the reverse,
generative process is to to denoise the diffused data.

The forward process, by definition, turns a phase space distribution
into Gaussian noise. The multi-dimensional distribution is factorized
into independent steps,
\begin{align}
    p(x_1,...,x_T|x_0) = \prod_{t=1}^T p(x_t | x_{t-1}) 
    \qquad \text{with} \qquad 
    p(x_t | x_{t-1}) = \normal(x_t;\sqrt{1 - \beta_t}x_{t-1}, \beta_t) \; .
    \label{eq:qxtt-1}
\end{align}
Each step describes a conditional probability and adds Gaussian noise
with an appropriately chosen variance $\beta_t$ and mean $\sqrt{1 -
  \beta_t} x_{t-1}$ to generate $x_t$. Following the original paper we
can choose a linear scaling $\beta_t \sim 2 \cdot 10^{-2} (t-1)/T$.
Naively, we would add noise with mean $x_{t-1}$, but in that case each
step would broaden the distribution, since independent noise sources
add their widths in quadrature. To compensate, we add noise with
scaled mean $\sqrt{1 - \beta_t} x_{t-1}$. In that case we can combine
all Gaussian convolutions and arrive at
\begin{align}
p(x_t|x_0) &= \int dx_1...dx_{t-1} \; \prod_{i=1}^t \; p(x_{i} | x_{i-1}) \notag \\
& = \normal(x_t; \sqrt{1 - \Bar{\beta}_t} x_0, \bar{\beta}_t)
\qquad \text{with} \qquad 
1 - \Bar{\beta}_t = \prod_{i=1}^t ( 1- \beta_i ) \; .
\label{eq:samplet}
\end{align}
%

The reverse process in \eqref{eq:time_series} starts with a Gaussian
and gradually transforms it into the phase-space distribution through
the same discrete steps as Eq.\eqref{eq:qxtt-1}.  The corresponding
generative network approximates each forward step and should produce
the correct phase-space distribution
\begin{align}
    \pmd(x_0) 
    &= \int dx_1...dx_T \; 
    \pl(x_T) \prod_{t=1}^T p_\theta(x_{t-1}|x_t) \notag \\
    \text{with} \qquad 
    p_\theta(x_{t-1}|x_t) &= \normal(x_{t-1}; \mu_\theta(x_t,t),\sigma^2_\theta(x_t,t)) \; .
\label{eq:pthetajoint}
\end{align}
Here, $\mu_\theta$ and $\sigma_\theta$ are learnable parameters
describing the individual conditional probability slices $x_t \to
x_{t-1}$. It turns out that in numerical practice we can simplify it
to $\sigma^2_\theta(x_t,t) \to \sigma^2_t$. 

To link the forward and
reverse directions, we first apply Bayes' theorem\index{Bayes' theorem} on each slice defined in
Eq.\eqref{eq:qxtt-1} to give us the reverse $p(x_{t-1}|x_t)$.
%
%
The problem with this inversion is that the full probability
distribution $p(x_1,...,x_T|x_0)$ in Eq.\eqref{eq:qxtt-1} is conditioned on $x_0$. 
With this dependence we can compute the conditioned forward
posterior as a Gaussian with an $x_0$-dependent mean,
\begin{align}
  p(x_{t-1}|x_t, x_0)
  &= \frac{p(x_t|x_{t-1}) p(x_{t-1}|x_0)}{p(x_t|x_0)}
   =\normal(x_{t-1}; \hat{\mu}(x_t,x_0), \hat{\beta}_t) \notag \\
   \text{with} \quad
   \hat{\mu}(x_t,x_0) &= \frac{\sqrt{1 - \bar{\beta}_{t-1}} \beta_t}{\bar{\beta}_t} x_0 + \frac{\sqrt{1-\beta}_t \Bar{\beta}_{t-1}}{\bar{\beta}_t} x_t 
   \quad \text{and} \quad
   \hat{\beta}_t = \frac{\bar{\beta}_{t-1}}{\Bar{\beta}_t}\beta_t \; .
   \label{eq:forward_bayes}
\end{align}
For training the DDPM network we need to just approximate a set of
Gaussians, Eq.\eqref{eq:forward_bayes}, with their learned
counterparts in Eq.\eqref{eq:pthetajoint}.

The loss function of the diffusion network is the same sampled
likelihood\index{likelihood loss} as for the INN, given in Eq.\eqref{eq:MLE}, which we can
simplify by inserting and then dividing by $p(x_1,...,x_T|x_0)$
following Eq.\eqref{eq:qxtt-1}
\begin{align}
    - \XLangle \log \pmd(x_0) \XRangle_{\pd} 
    &= -\int dx_0 \; \pd(x_0) \; \log \left( \int dx_1...dx_T \; \pl(x_T) \prod_{t=1}^T p_\theta(x_{t-1}|x_t) \right) \notag \\
    &= -\int dx_0 \; \pd(x_0) \; \log \left( \int dx_1...dx_T \; \pl(x_T) p(x_1,...,x_T|x_0) \prod_{t=1}^T \frac{p_\theta(x_{t-1}|x_t)}{p(x_t|x_{t-1})} \right) \notag \\
    & = -\int dx_0 \; \pd(x_0) \; \log \XXLangle \pl(x_T) \prod_{t=1}^T \frac{p_\theta(x_{t-1}|x_t)}{p(x_t|x_{t-1})} \XXRangle_{p(x_1,...,x_T|x_0)}
\end{align}
This expression includes a logarithm of an expectation value. There is
a standard relation, Jensen's inequality, for convex functions,
\begin{align}
  f(\langle x \rangle) \leq \left \langle f(x) \right\rangle \; .
  \label{eq:jensen}
\end{align}
Convex means that if we linearly interpolate between two points of the
function, the interpolation lies above the function. This is obviously
true for a function around the minimum, where the Taylor series gives
a quadratic (or even-power) approximation. We use this inequality for
our negative log-likelihood around the minimum. It provides an upper
limit to the negative log-likelihood, which we minimize instead,
\begin{align}
    - \XLangle \log \pmd(x_0) \XRangle_{\pd} 
    &\leq -\int dx_0 \; \pd(x_0) \; \XXLangle \log \left(  \pl(x_T) \prod_{t=1}^T \frac{p_\theta(x_{t-1}|x_t)}{p(x_t|x_{t-1})}\right) \XXRangle_{p(x_1,...,x_T|x_0)} \notag \\ 
    &= -\int dx_0...dx_T \; \pd(x_0) \; p(x_1,...,x_T|x_0) \; \log \left( \pl(x_T) \prod_{t=1}^T \frac{p_\theta(x_{t-1}|x_t)}{p(x_t|x_{t-1})} \right) \notag\\ 
    &\equiv -\int dx_0...dx_T \; p(x_0,...,x_T) \; \log \left( \pl(x_T) \prod_{t=1}^T \frac{p_\theta(x_{t-1}|x_t)}{p(x_t|x_{t-1})} \right) \notag\\
    &= \XXXLangle -\log \pl(x_T) - \sum_{t=1}^T \log \frac{p_\theta(x_{t-1}|x_t)}{p(x_t|x_{t-1})} \XXXRangle_{p(x_0,...,x_T)} 
\end{align}
Now we use Bayes' theorem for the individual slices $p_\theta(x_{t-1}|x_t)$ and compare them with the reference form from Eq.\eqref{eq:forward_bayes},
\begin{align}
    - \XLangle \log \pmd(x_0) \XRangle_{\pd} 
    &\le \XXLangle -\log \pl(x_T) - \sum_{t=2}^T \log \frac{p_\theta(x_{t-1}|x_t)}{p(x_t|x_{t-1})} - \log \frac{p_\theta(x_{0}|x_1)}{p(x_1|x_0)}\XXRangle_{p(x_0,...,x_T)} \notag\\
    &= \XXLangle -\log \pl(x_T) - \sum_{t=2}^T \log \frac{p_\theta(x_{t-1}|x_t) p(x_{t-1}|x_0)}{p(x_{t-1}|x_t, x_0) p(x_t|x_0)} - \log \frac{p_\theta(x_0|x_1)}{p(x_1|x_0)}\XXRangle_{p(x_0,...,x_T)}\notag \\
    &= \XXLangle -\log \pl(x_T) - \sum_{t=2}^T \log \frac{p_\theta(x_{t-1}|x_t)}{p(x_{t-1}|x_t,x_0)} -\log \frac{p(x_1|x_0)}{p(x_T|x_0)} - \log \frac{p_\theta(x_0|x_1)}{p(x_1|x_0)}\XXRangle_{p(x_0,...,x_T)} \notag\\
    &= \XXLangle -\log \frac{\pl(x_T)}{p(x_T|x_0)} - \sum_{t=2}^T \log \frac{p_\theta(x_{t-1}|x_t)}{p(x_{t-1}|x_t,x_0)} - \log p_\theta(x_0|x_1)\XXRangle_{p(x_0,...,x_T)}
\end{align}
As usual, we ignore terms which do not depend on the network weights $\theta$,
\begin{align}
    \XLangle \log \pmd(x_0) \XRangle_{\pd} 
    &\le \sum_{t=2}^T  \XXLangle \log \frac{p(x_{t-1}|x_t,x_0)}{p_\theta(x_{t-1}|x_t)} \XXRangle_{p(x_0,...,x_T)}
    - \XLangle \log p_\theta(x_0|x_1)\XRangle_{p(x_0,...,x_T)} + \text{const} \notag\\
    &= \sum_{t=2}^T  \int dx_0...dx_T \; p(x_0,...,x_T) \log \frac{p(x_{t-1}|x_t,x_0)}{p_\theta(x_{t-1}|x_t)} 
    - \XLangle \log p_\theta(x_0|x_1)\XRangle_{p(x_0,...,x_T)} + \text{const} \notag\\
    &= \sum_{t=2}^T \XLangle \kl [p(x_{t-1}|x_t,x_0),p_\theta(x_{t-1}|x_t)]\XRangle_{p(x_0,x_t)} 
    - \XLangle \log p_\theta(x_0|x_1)\XRangle_{p(x_0,...,x_T)} + \text{const} \notag\\
    &\approx \sum_{t=2}^T \XLangle \kl [p(x_{t-1}|x_t,x_0),p_\theta(x_{t-1}|x_t)]\XRangle_{p(x_0,x_t)} 
    \label{eq:DKL0}
\end{align}
The sampling follows $p(x_0,x_t) = p(x_t|x_0)$ $\pd(x_0)$.  The second
sampled term will be numerically negligible compared to the first
$T-1$ terms.  The KL-divergence\index{KL-divergence} compares the two Gaussians from
Eq.\eqref{eq:forward_bayes} and Eq.\eqref{eq:pthetajoint}, with the
two means $\mu_\theta(x_t,t)$ and $\hat{\mu}(x_t,x_0)$ and the two
standard deviations $\sigma_t^2$ and $\hat{\beta}_t$,
\begin{align}
  \boxed{
  \loss_\text{DDPM} 
  = \sum_{t=2}^T \XXLangle \frac{1}{2\sigma_t^2} \left|\hat{\mu}-\mu_\theta\right|^2 \XXRangle_{p(x_0,x_t)} } \; .
  \label{eq:loss_ddpm}
\end{align}
To provide $\hat{\mu}$ we use the reparametrization
trick on $x_t(x_0,\epsilon)$ as given in Eq.\eqref{eq:samplet}
\begin{align}
  x_t(x_0,\epsilon) &= \sqrt{1 - \bar{\beta}_t} x_0 + \sqrt{\Bar{\beta}_t}\epsilon 
  \qquad \text{with} \qquad 
  \epsilon \sim \normal(0,1)  \notag \\
  \Leftrightarrow \qquad
  x_0(x_t,\epsilon) &= \frac{1}{\sqrt{1 - \Bar{\beta}_t}}\left( x_t -\sqrt{\Bar{\beta}_t}\epsilon \right) \; ,
\end{align}
This form we insert into Eq.\eqref{eq:forward_bayes} to find, after a few simple steps.
\begin{align}
  \hat{\mu}(x_t,\epsilon) 
  &= \frac{1}{\sqrt{1 - \beta_t}} \left( x_t(x_0,\epsilon)- \frac{\beta_t}{\sqrt{\Bar{\beta}_t}} \epsilon \right) \; .
    \label{eq:repara}
\end{align}
The same method can be applied to provide $\mu_\theta (x_t,t) \equiv
\hat{\mu}(x_t,\epsilon_\theta)$, in terms of the trained network
regression $\epsilon_\theta$.

\begin{figure}[t]
    \centering
    \begin{tikzpicture}[node distance=2cm, scale=0.5, every node/.style={transform shape}]

\node (xT) [txt] {$x_T \sim \mathcal{N}(0,1)$};
\node (modelT_b) [cinn_black, below of=xT, yshift=-0.5cm] {};
\node (modelT) [cinn, below of=xT, yshift=-0.5cm, fill=Bcolor] {DDPM};
\node (T) [txt, left of=modelT_b, xshift=-0.5cm]{$t=T$};
\node (eps) [expr, right of = modelT_b, xshift=1cm]{$\epsilon_\theta$};
\node (xTmO) [xts, right of =eps, xshift=4cm]{$x_{T-1} = \frac{1}{\sqrt{1 - \beta_T}} \left(x_T - \frac{\beta_T}{\sqrt{\Bar{\beta}_T}} \epsilon_\theta\right) + \sigma_T z_T$};
\node (zT) [txt, above of=xTmO, yshift=0.5cm,xshift=-1.5cm] {$z_{T-1} \sim \mathcal{N}(0,1)$};

\node (modelTmO_b) [cinn_black, below of=xTmO, yshift=-0.8cm,xshift=-1.5cm] {};
\node (modelTmO) [cinn, below of=xTmO, yshift=-0.8cm, fill=Bcolor,xshift=-1.5cm] {DDPM};
\node (TmO) [txt, left of=modelTmO_b, xshift=-1cm]{$t=T-1$};

\node (epsZ) [expr, right of = modelTmO_b, xshift=1cm]{$\epsilon_\theta$};
\node (ppp) [txt, right of=epsZ, xshift=0.5cm, font=\Huge]{...};
\node (x1) [xt, right of=ppp, xshift=3cm]{$x_1 = \frac{1}{\sqrt{1 -\beta_2}}\left(x_2  - \frac{\beta_2}{\sqrt{\Bar{\beta}_2}} \epsilon_\theta \right) + \sigma_2 z_2$};
\node (z1) [txt, above of=x1, yshift=0.5cm,xshift=-1.5cm] {$z_1 \sim \mathcal{N}(0,1)$};
\node (modelO_b) [cinn_black, below of=x1, yshift=-0.8cm,xshift=-1.5cm] {};
\node (modelO) [cinn, below of=x1, yshift=-0.8cm, fill=Bcolor,xshift=-1.5cm] {DDPM};
\node (O) [txt, left of=modelO_b, xshift=-0.5cm]{$t=1$};
\node (eps0) [expr, right of=modelO_b, xshift=0.8cm]{$\epsilon_\theta$};
\node (x0) [xt, below of=eps0, yshift=-0.5cm]{$x_0 = \frac{1}{\sqrt{1 - \beta_1}} \left(x_1  - \frac{\beta_1}{\sqrt{\Bar{\beta}_1}} \epsilon_\theta\right)$};

\draw [arrow, color=black] (zT.south) -- ([xshift=-1.5cm]xTmO.north);
\draw [arrow, color=black] (xT.south) -- (modelT_b.north);
\draw [arrow, color=black] (modelT_b.east) -- (eps.west);
\draw [arrow, color=black] (T.east) -- (modelT_b.west);
\draw [arrow, color=black] (eps.east) -- (xTmO.west);

\draw [arrow, color=black] ([xshift=-1.5cm]xTmO.south) -- (modelTmO_b.north);
\draw [arrow, color=black] (modelTmO_b.east) -- (epsZ.west);
\draw [arrow, color=black] (TmO.east) -- (modelTmO_b.west);
\draw [arrow, color=black] (epsZ.east) -- (ppp.west);

\draw [arrow, color=black] (ppp.east) -- (x1.west);
\draw [arrow, color=black] (z1.south) -- ([xshift=-1.5cm]x1.north);

\draw [arrow, color=black] ([xshift=-1.5cm]x1.south) -- (modelO_b.north);
\draw [arrow, color=black] (modelO_b.east) -- (eps0.west);
\draw [arrow, color=black] (O.east) -- (modelO_b.west);
\draw [arrow, color=black] (eps0.south) -- (x0.north);

\end{tikzpicture}
\caption{DDPM sampling algorithm, Figure from Ref.~\cite{Butter:2023fov}.}
    \label{fig:sampling_ddpm}
\end{figure}

The derivation of the Bayesian INN in Sec.~\ref{sec:gen_inn_arch} can
just be copied to define a Bayesian DDPM\index{Bayesian network}. Its loss follows from 
Eqs.\eqref{eq:loss_ddpm} and~\eqref{eq:loss_binn}, with a sampling
over network parameters $\theta \sim q(\theta)$ and the regularization
term\index{regularization},
\begin{align}
\loss_\text{B-DDPM} 
&= \XLangle \loss_\text{DDPM} \XRangle_{\theta \sim q} 
+ \kl[q(\theta),p(\theta)] \; .
\label{eq:BDDPM_loss}
\end{align}
We turn the deterministic DDPM into the B-DDPM through two steps, (i)
swapping the deterministic layers to the corresponding Bayesian
layers, and (ii) adding the regularization term to the loss.  To
evaluate the Bayesian network sample over the network weight
distribution.

For the DDPM training we start with a phase-space point $x_0 \sim
\pd(x_0)$ drawn from the true phase space distribution and also draw the
time step $t$ from a uniform distribution and $\epsilon$ from a standard 
Gaussian. We then use Eq.\eqref{eq:repara} to compute the 
diffused data point after $t$ time steps, $x_t$. This means the
uses many different time steps $t$
for many different phase-space points $x_0$ to learn the
step-wise reversed process. , which is why we use a relatively
simple residual dense network architecture, which is trained over many
epochs.

The (reverse) DDPM sampling is illustrated in
Fig.~\ref{fig:sampling_ddpm}. We start with $x_T \sim \pl(x_T)$, drawn
from the Gaussian latent space distribution. Combining the learned
$\epsilon_\theta$ and the drawn Gaussian noise $z_{T-1} \sim
\normal(0,1)$ we calculate $x_{T-1}$, assumed to be a slightly less
diffused version of $x_T$. We repeat this sampling until we reach the
phase space distribution of $x_0$.  Because the network needs to predict
$\epsilon_\theta$ $T$ times, it is slower than classic generative
networks like VAEs, GANs, or INNs.


\begin{figure}[t]
\includegraphics[width=0.33\textwidth, page=1]{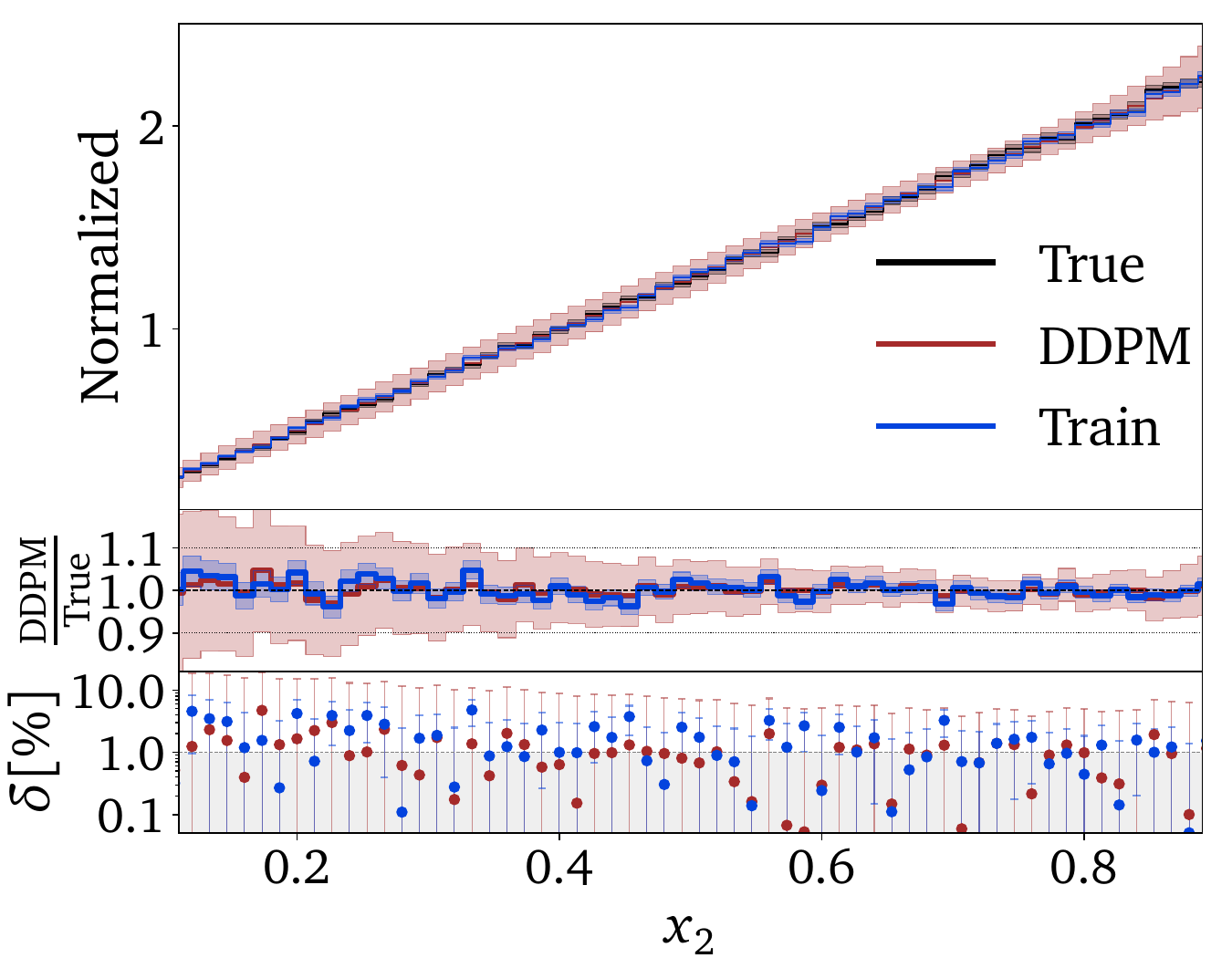}
\includegraphics[width=0.33\textwidth, page=2]{DDPM_ramp}
\includegraphics[width=0.33\textwidth, page=2]{CFM_ramp}
\caption{Density and predictive uncertainty distribution for a linear
  wedge ramp using diffusion networks. We show DDPM results (left and
  center) and CFM results (right). The uncertainty on
  $\sigma_\text{stat}$ is given by its $x_1$-variation.  Figure from
  Ref.~\cite{Butter:2023fov}.}
  \label{fig:diff_ramp}
\end{figure}

From Sec.~\ref{sec:gen_inn_arch} we know that we can illustrate the
training of a generative network using a simple toy model, for example
the linear ramp defined in Eq.\eqref{eq:def_2d_ramp}. In
Fig.~\ref{fig:diff_ramp} we show the corresponding results to the
B-INN results from Fig.~\ref{fig:linear_unc}. The key results is that
the DDPM learns the simple probability distribution with high
precision and without a bias. Moreover, the absolute predictive
uncertainty on the estimated density has a minimum around $x_2 \sim
0.7$, albeit less pronounced than for the B-INN. This suggests that
the DDPM has some similarity to the INN, but its implicit bias on the
fitted function and its expressivity are different.

\subsubsection{Conditional flow matching}
\label{sec:gen_diff_cfm}

An alternative approach to diffusion networks is Conditional Flow
Matching~(CFM).  Like the DDPM, it uses a time evolution to transform
a phase space distributions into Gaussian noise, but instead of a
discrete chain of conditional probabilities it constructs and solves a
continuous \underline{ordinary differential equation} (ODE)
\begin{align}
\boxed{
  \frac{d x(t)}{dt} = v(x(t), t) } \; ,
\label{eq:sample_ODE}
\end{align}
where $v(x(t), t)$ is called the velocity field.  We will learn the
velocity field to generate samples by integrating this ODE from $t=1$
to $t=0$.  The ODE can be linked to a probability density $p(x,t)$
through the \underline{continuity equation}
\begin{align}
\boxed{
  \frac{\partial p(x,t)}{\partial t} + \nabla_x \left[ p(x,t) v(x,t) \right] = 0
  } \; .
\label{eq:continuity}
\end{align} 
These two equations are equivalent in that for a given probability
density path $p(x,t)$ any velocity field $v(x,t)$ describing the
sample-wise evolution Eq.\eqref{eq:sample_ODE} will be a solution of
Eq.\eqref{eq:continuity}, and vice versa. We will use the continuity
equation to learn the velocity field $v_\theta(x,t) \approx v(x,t)$ in
the network training and then use the ODE with $v_\theta(x,t)$ as a
generator.

To realize the diffusion model ansatz of Eq.\eqref{eq:time_series} we
need to express the velocity field in terms of our known density $p(x,t)$ with the boundary conditions
\begin{align}
 p(x,t) \to 
 \begin{cases}
  \pd(x) \qqquad & t \to 0 \\
  \pl(x) = \normal(x;0,1) \qqquad & t \to 1  \; ,
\end{cases} 
\label{eq:fm_limits}
\end{align}
where we use $x$ as the argument of $\pl$, rather than $r$, to
illustrate that it is an evolved form of the phase space distribution
$x$. In the forward, diffusion direction, the time evolution goes from
a phase space point $x_0$ to the latent standard Gaussian. In a
conditional form we can write a simple linear interpolation
\begin{align}
    x(t|x_0) 
    &= (1-t) x_0 + t r 
    \to \begin{cases}
      x_0 \qqquad & t \to 0 \\
      r \sim \normal(0,1) \qqquad & t \to 1 
    \end{cases} \notag \\
\Leftrightarrow \qquad 
p(x,t|x_0) &= \normal(x ; (1-t) x_0, t ) \; .
\label{eq:conditional_path}
\end{align}
This conditional time evolution is similar to the DDPM case in
Eq.\eqref{eq:samplet}. Formally, we can compute the full probability
path from it, ensuring that it fulfills both boundary conditions of
Eq.\eqref{eq:fm_limits},
\begin{align}
    p(x,t) &= \int d x_0 \; p(x,t|x_0) \; \pd (x_0) \notag \\
    \text{with} \qquad
    p(x,0) &= \int d x_0 \; p(x,0|x_0) \; \pd(x_0) 
    = \int d x_0 \; \delta(x-x_0) \; \pd(x_0) = \pd(x) \notag \\
    \text{and} \qquad
    p(x,1) &= \int d x_0 \; p(x,1|x_0) \; \pd(x_0) 
    = \normal(x; 0,1) \int d x_0 \; \pd(x_0) = \normal(x; 0,1) \; .
\label{eq:marginal_path_01}
\end{align}

For a generative model we need the velocity corresponding to this
probability density path. We start with the conditional velocity,
associated with $p(x,t|x_0)$, and combine Eq.\eqref{eq:sample_ODE}
and~\eqref{eq:conditional_path} to
\begin{align}
   v(x(t|x_0),t| x_0) 
   &= \frac{d x(t|x_0)}{dt}\notag \\
   &= \frac{d}{dt} \left[ (1-t) x_0 + t r \right] = - x_0 + r \; .
\label{eq:conditional_velocity}
\end{align}
Our linear interpolation leads to a time-constant velocity, which
solves the continuity equation for $p(x,t|x_0)$ because we construct
it as a solution to the ODE
\begin{align}
\frac{\partial p(x,t|x_0)}{\partial t} + \nabla_x \left[ p(x,t|x_0) v(x,t|x_0) \right] = 0 \; .
\label{eq:continuity2}
\end{align} 
Just like for the probability paths, Eq.\eqref{eq:marginal_path_01},
we can compute the unconditional velocity from its conditional
counterpart,
\begin{align}
    \frac{\partial p(x,t) }{\partial t}
    &= \frac{\partial}{\partial t} \int d x_0 \; p(x,t|x_0)\pd(x_0) \notag \\
    &= \int dx_0 \; \frac{\partial p(x,t|x_0)}{\partial t} \; \pd(x_0) \notag \\
    &= -\int dx_0 \; \nabla_x \left[ p(x,t|x_0) v(x,t|x_0)\right] \; \pd(x_0) \notag \\
    &= -\nabla_x \left[ p(x,t) \int dx_0 \; \frac{p(x,t|x_0)v(x,t|x_0)\pd(x_0)}{p(x,t)} \right] 
    \equiv -\nabla_x \left[ p(x,t)v(x,t)\right] \notag \\
\Leftrightarrow \qquad 
    v(x,t) &= \int dx_0 \; \frac{p(x,t|x_0)v(x,t|x_0)\pd(x_0)}{p(x,t)} \; .
    \label{eq:velocity}
\end{align}
While the conditional velocity in Eq.\eqref{eq:conditional_velocity}
describes a trajectory between a normal distributed and a phase space
sample $x_0$ that is specified in advance, the full velocity in
Eq.\eqref{eq:velocity} evolves samples from $\pd$ to $\pl$ and vice
versa.

Training the CFM means learning the velocity field in
Eq.\eqref{eq:velocity}, a simple regression task, $v(x,t) \approx
v_\theta(x,t)$. The straightforward choice is the MSE-loss,
\begin{align}
    \loss_\text{FM} 
    &= \XLangle \left[ v_\theta(x,t) - v(x,t) \right]^2 \XRangle_{t, x\sim p(x,t)} \notag \\
    &= \XLangle v_\theta(x,t)^2 \XRangle_{t, x\sim p(x,t)}
    - \XLangle 2v_\theta(x,t)v(x,t) \XRangle_{t, x\sim p(x,t)} + \text{const} \; , 
\label{eq:FMloss}
\end{align}
where the time is sampled uniformly over $t \in [0,1]$.  Again, we can
start with the conditional path in Eq.\eqref{eq:conditional_path} and
calculate the conditional velocity in
Eq.\eqref{eq:conditional_velocity} for the MSE loss. We rewrite the
above loss in terms of the conditional quantities, so the first term
becomes
\begin{align}
\XLangle v_\theta(x,t)^2 \XRangle_{t, x\sim p(x,t)} 
    &= \XXLangle \int dx \; p(x,t)  v_\theta(x,t)^2  \XXRangle_t \notag \\
    &= \XXLangle \int dx v_\theta(x,t)^2 \int dx_0 \; p(x,t|x_0)\pd(x_0)  \XXRangle_t \notag \\
    &= \XLangle v_\theta(x,t)^2 \XRangle_{t,x_0\sim \pd, x \sim p(x,t|x_0)} \notag \\
    &= \XLangle v_\theta(x(t|x_0),t)^2 \XRangle_{t,x_0\sim \pd, r}
\end{align}
In the last term we use the simple form of $x(t|x_0)$ given in
Eq.\eqref{eq:conditional_path}, which needs to be sampled over
$r$.  Similarly, we rewrite the second loss term as
\begin{align}
 -2 \XLangle v_\theta(x,t)v(x,t) \XRangle_{t, x\sim p(x,t)} 
    &= -2\XXLangle \int dx \;  p(x,t) v_\theta(x,t) \; \frac{\int dx_0 p(x,t|x_0)v(x,t|x_0)\pd(x_0)}{p(x,t)} \; \XXRangle_t  \notag \\
    &= -2\XXLangle \int dx dx_0 \; v_\theta(x,t) \; v(x,t|x_0) \; p(x,t|x_0) \; \pd(x_0) \XXRangle_t \notag \\
    &= -2\XLangle v_\theta(x,t) \; v(x,t|x_0)\XRangle_{t,x_0\sim \pd, x \sim p(x,t|x_0)} \notag \\
    &= -2\XLangle v_\theta(x(t|x_0),t) \; v(x(t|x_0),t|x_0)\XRangle_{t,x_0\sim \pd, r} \; .
\end{align}
The (conditional) Flow Matching loss of Eq.\eqref{eq:FMloss} then becomes
\begin{align}
  \boxed{
    \loss_\text{CFM} = \XLangle \left[ v_\theta(x(t|x_0),t) - v(x(t|x_0),t|x_0) \right]^2 \XRangle_{t,x_0\sim \pd, r}
  } \; .
\label{eq:CFM_loss}
\end{align}
We can compute it using the linear ansatz from
Eq.\eqref{eq:conditional_path} as
\begin{align}
\loss_\text{CFM}
= \XXLangle \left[ v_\theta(x(t|x_0),t) - \frac{d x(t|x_0))}{dt} \right]^2 \XXRangle_{t,x_0\sim \pd, r} 
= \XLangle \left[ v_\theta((1-t)x_0+tr,t) - (r - x_0)\right]^2 \XRangle_{t,x_0\sim \pd, r}  \; .
\label{eq:CFM_loss2}
\end{align}
As usually, we can turn the CFM into a Bayesian generative network.  For
the Bayesian INN or the Bayesian DDPM the loss is a sum of the
likelihood loss\index{likelihood loss}  and a KL-divergence regularization, shown in
Eqs.\eqref{eq:loss_binn} and~\eqref{eq:BDDPM_loss}. Unfortunately, the
CFM loss in Eq.\eqref{eq:CFM_loss} is not a likelihood loss. To mimic
the usual setup we still modify the CFM loss by switching to Bayesian
network\index{Bayesian network} layers and adding a KL-regularization\index{regularization},
\begin{align}
    \loss_\text{B-CFM} &= \XLangle \loss_\text{CFM}\XRangle_{\theta \sim q(\theta)} + c  \kl[q(\theta),p(\theta)].
\label{eq:BCFM_objective}
\end{align}
While for a likelihood loss the factor $c$ is fixed by Bayes' theorem,
in this case it is a free hyperparameter. However, we find that the
network predictions and their associated uncertainties are very stable
when varying it over several orders of magnitude\index{uncertainties}.

To train the CFM we sample a data point $x_0 \sim \pd(x_0)$ and
$r \sim \normal(0,1)$ as the starting and end points of a
trajectory, as well as a time from a uniform distribution. We first
compute $x(t|x_0)$ according to Eq.\eqref{eq:conditional_path} and
then $v(x(t|x_0),t|x_0)$ following
Eq.\eqref{eq:conditional_velocity}. The point $x(t|x_0)$ and the time
$t$ are passed to a neural network which encodes the conditional
velocity field
\begin{align}
  v_\theta(x(t|x_0),t) \approx v(x,t|x_0) \; .
\end{align}
One property of the training algorithm is that the same network input,
a time $t$ and a position $x(t|x_0)$, can be produced by many
different trajectories with different conditional velocities. While
the network training is based on a wide range of possible
trajectories, the CFM loss in Eq.\eqref{eq:CFM_loss} ensures that
sampling over many trajectories returns a well-defined velocity field.

Once the CFM network is trained, the generation of new samples is
straightforward. We start by drawing a sample from the latent
distribution $r \sim \pl = \normal(0,1)$ and calculate its time
evolution by \underline{numerically solving the ODE} backwards in time
from $t=1$ to $t=0$
\begin{align}
\frac{d}{dt}x(t) &= v_\theta(x(t),t) 
\qquad \text{with} \quad r = x(t=1) \notag \\
\Rightarrow \qquad 
x_0 &= r - \int_0^1 v_\theta (x,t) dt \equiv G_\theta(r)
\; ,
\label{eq:ODE_solution}
\end{align}
This generation is fast, as is the CFM training, because we can rely
on established ODE solvers.  Under mild regularity assumptions this
solution defines a bijective transformation between the latent space
sample and the phase space sample $G_\theta(x_1)$, similar to its
definition in the INN case, Eq.\eqref{eq:inn_mapping}.

Like the INN and unlike the DDPM, the CFM network also allows us to
calculate phase space likelihoods. Making use of the continuity
equation, Eq.\eqref{eq:continuity}, we can write
\begin{align}
    \frac{d p(x,t)}{dt} 
    &= \frac{\partial p(x,t)}{\partial t} + \left[ \nabla_x p(x,t) \right] \; v(x,t) \notag \\
    &= \frac{\partial p(x,t)}{\partial t} + \nabla_x \left[ p(x,t) v(x,t)\right] - p(x,t) \left[ \nabla_x v(x,t) \right] \notag \\
    &= - p(x,t) \; \nabla_x v(x,t) \; .
    \label{eq:CFM:likelihood_ode}
\end{align} 
Its solution can be cast in the INN notation from Eq.\eqref{eq:cov},
\begin{align}
    \frac{p(r,t=1)}{p(x_0,t=0)} 
    & = \exp \left(  - \int_0^1 dt \; \nabla_x v(x(t),t) \right) \notag \\
    & \equiv \frac{\pl(G_\theta^{-1}(x_0))}{\pmd(x_0)} 
    = \left| \text{det}\frac{\partial G_\theta^{-1}(x_0)}{\partial x_0} \right|^{-1} 
    \notag \\
    \Leftrightarrow \qquad 
    \left| \text{det}\frac{\partial G_\theta^{-1}(x_0)}{\partial x_0} \right| 
    & = \exp \left( \int_0^1 dt  \; \nabla_x v_\theta(x(t),t) \right) \; . 
\label{eq:CFM:jacobian}
\end{align}
Calculating the Jacobian requires integrating over the divergence of
the learned velocity field, which is fast if we use automatic
differentiation.

To understand how the generative network learn, we can again use the
2-dimensional linear wedge combined with the Bayesian setup. In the
right panel of Fig.~\ref{fig:diff_ramp} we show the corresponding
distribution. First, we see that the scale of the uncertainty is a
factor five below the DDPM uncertainty, which means that the CFM is
easily and reliably trained for a small number of dimensions. In
addition, we see that the minimum in the absolute uncertainty is even
flatter and at $x_2 \sim 0.3$. Again, this is reminiscent of the
fit-like INN behavior, but much less pronounced.

An interesting aspect of the CFM is that, unlike the other generative
models we have discussed, nothing in our derivation requires the
latent distribution to be a Gaussian. Instead of
Eq.\eqref{eq:inn_mapping} or Eq,\eqref{eq:time_series} we can link any
two distributions, sample from one and into the second.  An
interesting LHC application is the generation of events from a narrow
phase space into a wider phase space. This kind of problem occurs for
collinear or soft or on-shell subtraction terms, where we need to
model cancellations between phase spaces with different numbers of
particles in the final state. Similarly, we can try to generate
off-shell decay configurations from an on-shell or Breit-Wigner
approximation. Such off-shell effects are expensive to simulate and
numerically suppressed. The advantage of training a generative network
to modify the on-shell distributions as compared to training a
generative network to simulate off-shell events is that the on-shell
events already include the full correlations, so the network only has
to learns small modifications rather than the correlations themselves.

\begin{figure}[t]
  \centering
  \includegraphics[width=0.4\textwidth, page=88]{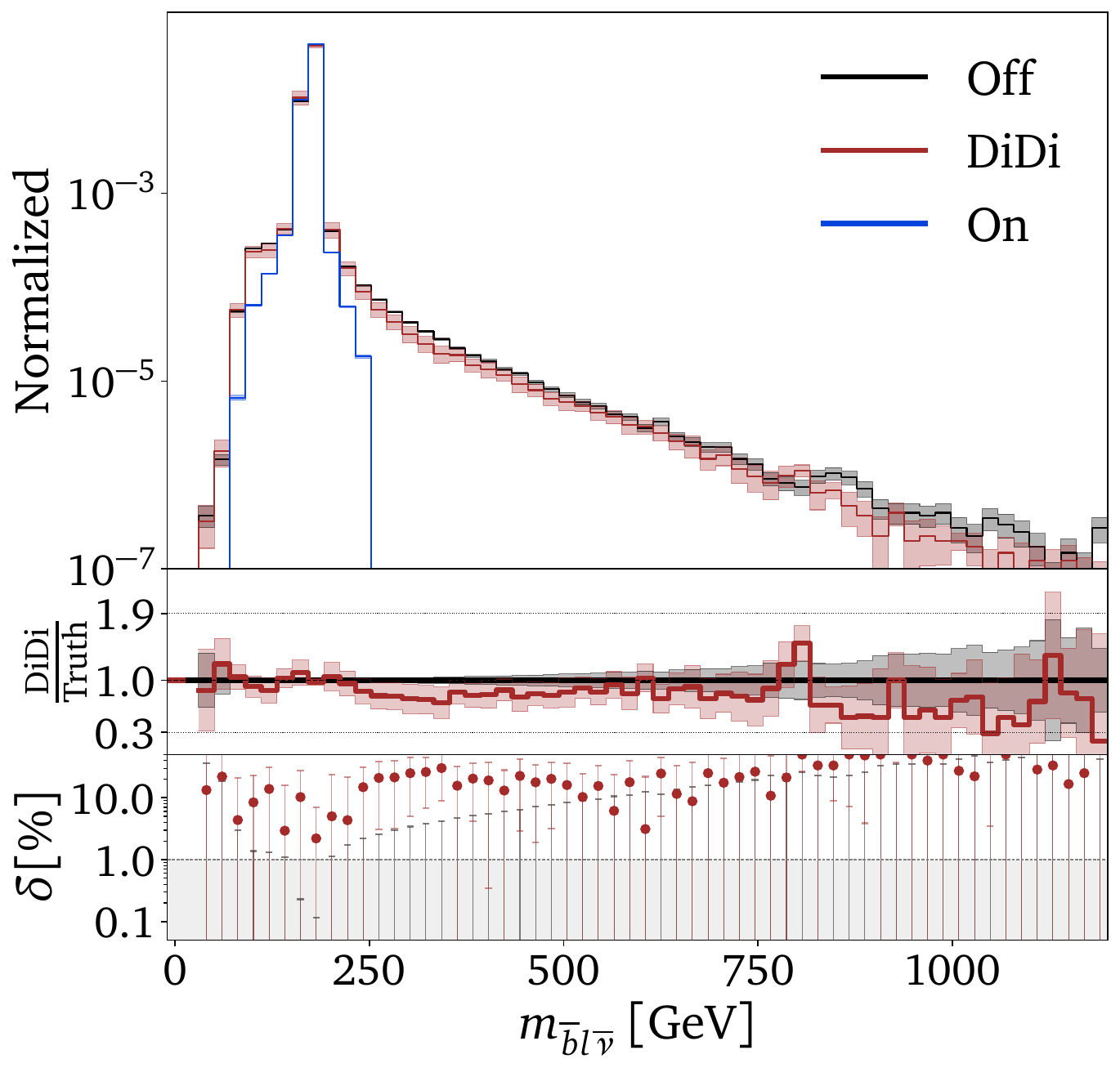}
  \hspace*{0.1\textwidth}
  \includegraphics[width=0.42\textwidth, page=2]{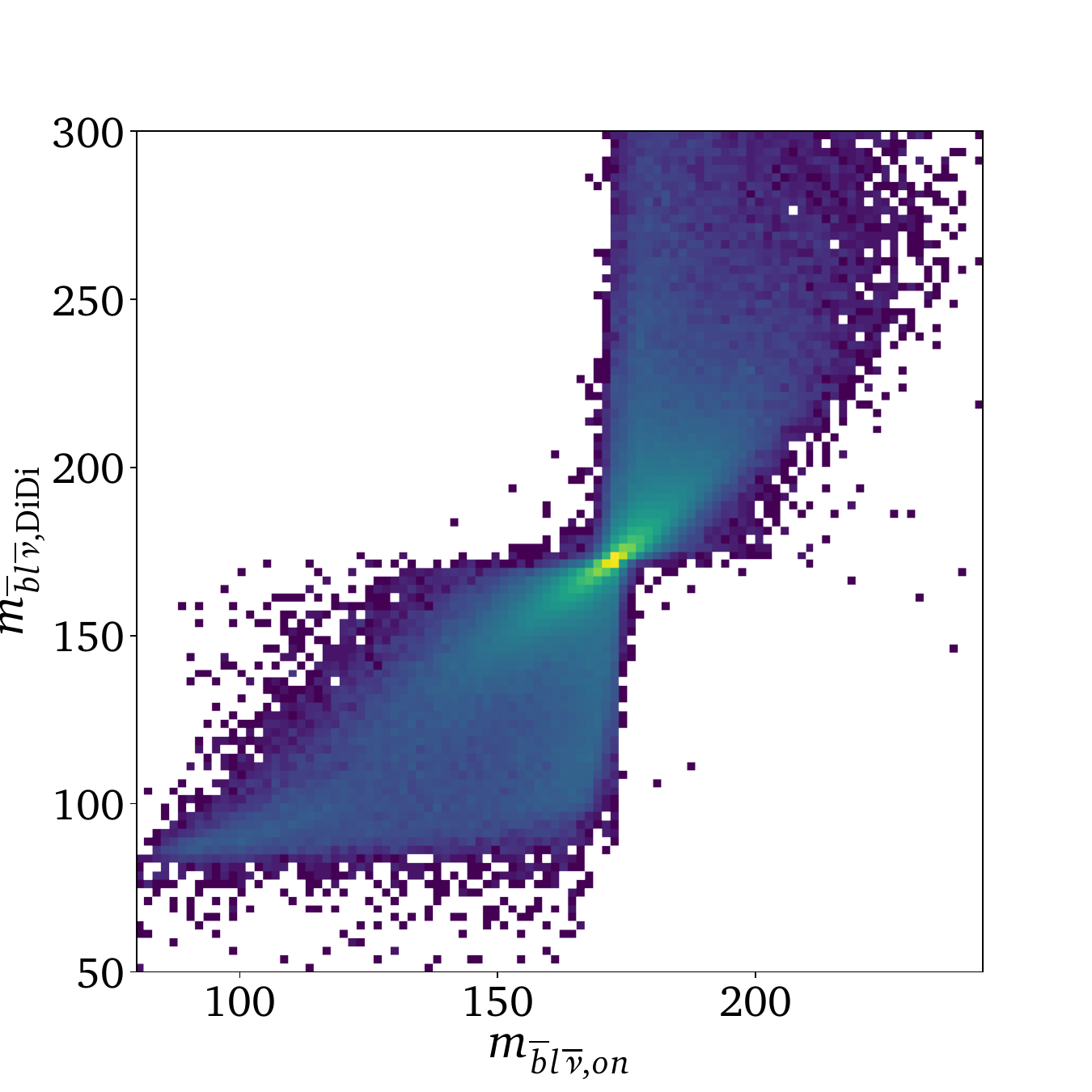}
  \caption{Left: kinematic distributions of on-shell vs off-shell top
    pair production, compared with the generated off-shell
    distributions (DiDi). Right: migration plot between on-shell and
    generated off-shell configurations. Figure from Ref.~\cite{Butter:2023ira}.}
  \label{fig:didi}
\end{figure}

Let us look at the production of off-shell top quarks
\begin{align}
 p p \to ( b e^+ \nu_e ) \; ( \bar{b} \mu^- \nu_\mu ) \; .
\label{eq:off-shell}
\end{align}
The propagators of massive and unstable intermediate particles with
mass $m$ include the particle width $\Gamma$ and have the form
\begin{align}
  \frac{\Gamma_\text{part}}{(s-m^2)^2 + m^2 \Gamma_\text{tot}^2}
  \; \stackrel{\Gamma \to 0}{\longrightarrow} \; 
  \Gamma_\text{part} \; \frac{\pi}{\Gamma_\text{tot}} \; \delta(s - m^2)
  \equiv \pi \; \br_\text{part} \; \delta(s-m^2) \; . 
\end{align}
For vanishing widths, the decay propagator leads to a phase space
constraint and the branching ratio $\br_\text{part}$ into the partial
decay channel given by the off-shell process. For small, but finite
width we can use this so-called Breit-Wigner propagator to simulate
off-shell effects. Far away from the on-shell pole this description is
wrong, because Eq.\eqref{eq:off-shell} does not actually require top
quarks to appear in the Feynman diagrams. In Fig.~\ref{fig:didi} we
show the on-shell and off-shell distributions for the reconstructed
top mass.

Generalizing Eq.\eqref{eq:inn_mapping} the generative task becomes
\begin{align}
  \text{on-shell $x \sim p_{\text{on}}(x)$}
\quad 
\xleftrightarrow{\hspace*{1.5cm}}
\quad 
  \text{off-shell $x \sim \pmd(x|\theta) \approx p_\text{off}(x)$}  \; .
\label{eq:generative}
\end{align}
The time-dependent probability distributions of
Eq.\eqref{eq:fm_limits} become
\begin{align}
 p(x,t) \to 
 \begin{cases}
  p_\text{off}(x) \quad & t \to 0 \\
  p_\text{on}(x)  \quad & t \to 1  \;,
\end{cases} 
\end{align}
and the loss function from Eq.\eqref{eq:CFM_loss} is samples from the
on-shell and off-shell events,
\begin{align}
    \loss_\text{CFM} &= \XLangle \left[ v_\theta((1-t)x_0+t x_1,t) - (x_1 - x_0)\right]^2 \XRangle_{t,x_0\sim p_\text{off}, x_1 \sim p_\text{on}}  \; .
\label{eq:CFM_loss3}
\end{align}
For the training, we can use paired events, if available, but we do
not have to. This kind of mapping can also be achieved using the
Flows-For-Flows setup or the so-called Schr\"odinger bridge, but as a
generalization of the CFM model to \underline{Direct Diffusion} (DiDi)
it is particularly obvious.

In the left panel of Fig.~\ref{fig:didi} we compare the performance of
a simple direct diffusion network to the on-shell base distribution
and the off-shell target distribution. The generated off-shell events
reproduce the correct distribution in the, apparently, extrapolated
phase space region with reasonable precision. Deviations from the
target distributions are covered by the uncertainty bands of the Bayesian
version. As always, the performance of the DiDi generator can be
improved through classifier reweighting.

Given that the off-shell extrapolation does not follow a well-defined
simulation model, the DiDi training on unmatched event samples has to
construct some kind of optimal transport prescription. This is already
discussed in some of the original literature on CFM. We can illustrate
the learned optimal transport prescription through the correlations
between each starting phase space point $x_0$ and the generated
off-shell phase space point $x_1$. In the right panel of
Fig.~\ref{fig:didi} we show this correlation, for the transverse
momentum of the anti-bottom quark. As one might expect, the optimal
transport keeps most events close to where they are and avoids moving
event from one side of the top mass peak to the other.

\subsection{Autoregressive transformer}
\label{sec:gen_at}

Another recent generative network we can use for LHC physics is known as
ChatGPT. This stands for generative pre-trained transformer, the same
transformer we have introduced as a permutation-invariant
preprocessing in Sec.~\ref{sec:class_graph_trans}. It needs
pre-training for large language models, because we already know that
the transformer learns the relations or links between objects without
assuming locality. For our purpose we combine our JetGPT with a
Gaussian mixture model for the density estimation, to cover the
partonic or reconstruction-level phase space of LHC events.


A common problem of all network architectures introduced until now is
the scaling of the network performance, in terms of precision,
training time, training dataset, with the phase space
dimensionality. Because they learn all correlations in all phase space
directions simultaneously, we automatically see a power-law scaling.
The \underline{autoregressive setup} can alleviate this by interpreting a phase
space vector $x = (x_1,...x_n)$ as a sequence of directions $x_i$, with
factorizing conditional probabilities
\begin{align}
  \pmd(x)
  &= \prod_{i=1}^n p_\theta(x_i|x_1,...,x_{i-1}) \notag \\
  &= \prod_{i=1}^n p_\theta(x_i|\omega^{(i-1)}) 
  \; ,
  \label{eq:def_autoreg}
\end{align}
where the parameters $\omega^{(i-1)}$ encode the conditional
dependence on $x_1,...x_{i-1}$. For LHC applications the $x_i$ are the
standard phase space directions, like energies or transverse momenta
or angles. The autoregressive setup improves the scaling with the
phase space dimensionality in two ways. First, each distribution
$p_\theta(x_i|x_1,...x_{i-1})$ is easier to learn than a distribution
conditional on the full phase space vector $x$. Second, we can use our
physics knowledge to group challenging phase space directions early in
the sequence $x_1,...,x_n$.

From our earlier discussion we know that generative networks for the
LHC can be understood as \underline{density estimation} over an
interpretable phase space, from which the network then samples. We
need a way to encode the phase space probability for our transformer.
A naive choice are binned probabilities $w_j^{(i-1)}$ in each phase
space direction.  If we want our autoregressive transformer to scale
better with dimensionality, a better approach is a Gaussian mixture
with learnable means and widths,
\begin{align}
  p_\theta(x_i|\omega^{(i-1)} )= \sum_\text{Gaussian $j$} w_j^{(i-1)} \normal (x_i; \mu_j^{(i-1)}, \sigma_j^{(i-1)} ) \; .
\label{eq:at_gaussians}
\end{align} 
The network architecture for the transformer generator of LHC events
follows the Generative Pretrained Transformer (GPT) models. The input
data is a sequence of $x_i$, followed by a linear layer to map each
value $x_i$ in the latent space.  Next follow a series of blocks,
which combine the self-attention\index{self-attention} layer of
Sec.~\ref{sec:class_graph_trans} with a standard feed-forward network.
The self-attention constructs correlations between the phase space
directions.  Finally, another linear layer leads to the representation
$\omega^{(i-1)}$ given in Eq.\eqref{eq:def_autoreg}.  The only
difference to the definition of the self-attention matrix $a_i^{(j)}$
in Eq.\eqref{eq:att_matrix} has to reflect the autoregressive ansatz,
\begin{align}
  a_j^{(i)} = 0
        \qquad \text{for} \quad j>i \; .
\end{align}
Moreover, because the standard self-attention leads to permutation
invariance in the phase space components, we need to break it by
adding positional information to the latent representation of $x_i$
through a linear layer using the one-hot encoded phase space position
$i$.

To train the autoregressive transformer we evaluate the chain of
conditional likelihoods for the realized values $x_i$, providing
$\pmd(x)$ for the usual likelihood loss\index{likelihood loss} 
\begin{align}
  \boxed{
  \loss_\text{AT} 
  = - \XLangle \log \pmd(x)\XRangle_{x\sim\pd} 
  = - \sum_{i=1}^n \XLangle \log p (x_i|\omega^{(i-1)})\XRangle_{x\sim \pd}
  } \; .
\label{eq:transformerloss}
\end{align}
As any generative network, we bayesianize the transformer by drawing
its weights from a set of Gaussians. In addition, we need to add the
KL-regularization\index{regularization} to the likelihood loss, giving us
\begin{align}
    \loss_\text{B-AT} = \XLangle \loss_\text{AT}\XRangle_{\theta \sim q(\theta)} +   \kl[q(\theta),p(\theta)].
\end{align}
For large generative networks, we encounter the problem that too many
Bayesian weights destabilize the network training. While a
deterministic network can switch of unused weights by just setting
them to zero, a Bayesian network\index{Bayesian network} can only set the mean to zero In that
case its width will approach the prior $p(\theta)$, so excess weights
contribute noise to the training. This problem can be solved either by
adjusting the prior hyperparameter or by only bayesianizing a fraction
of the network weights. A standard, extreme choice is be to
bayesianize only the last layer. In any case it is crucial to confirm
that the uncertainty estimate from the network is on a stable plateau
of the prior hyperparameter\index{uncertainties}.

The transformer generation is illustrated in
Fig.~\ref{fig:ALPs_sampling}. For each component, $\omega^{(i-1)}$
encodes the dependence on the previous components $x_1,..., x_{i-1}$,
and correspondingly we sample from $p (x_i|\omega^{(i-1)})$. The
parameters $\omega^{(0)},...,\omega^{(i-2)}$ from the sampling of
previous components are re-generated in each step, but not used
further. This means event generation is less efficient than the
likelihood evaluation during training, because it cannot be
parallelized.

\begin{figure}[t]
    \centering
    \begin{tikzpicture}[node distance=2cm, scale=0.5, every node/.style={transform shape}, on top/.style={preaction={draw=white,-,line width=#1}}, on top/.default=4pt]
\node (x0) [font=\LARGE] {$x_0=0$};
\node (model0_b) [cinn_black, right of=x0, xshift=3.5cm] {};
\node (model0) [cinn, fill=Bcolor, right of=x0, xshift=3.5cm] {AT};
\node (w0) [font=\LARGE, right of=model0, xshift=0.5cm] {$\omega^{(0)}$};
\node (cpdf0) [rectangle, font=\LARGE, fill=G2color, right of=w0, xshift=1.3cm, minimum width=3.5cm, minimum height=1.2cm, rounded corners] {$p (x_1|\omega^{(0)})$};

\draw [arrow] (x0.east) -- (model0.west);
\draw [arrow] (model0.east) -- (w0.west);
\draw [arrow] (w0.east) -- (cpdf0.west);

\node (xlist) [rectangle, fill=Rcolor, below of=x0, yshift=-4.9cm, minimum width=1.5cm, minimum height=8.5cm, rounded corners] {};

\node (x1) [font=\LARGE, below of=x0, yshift=-1.3cm] {$x_1$};
\node (model1_b) [cinn_black, right of=x1, xshift=6cm, yshift=0.2cm] {};
\node (model1) [cinn, fill=Bcolor, right of=x1, xshift=6cm, yshift=0.2cm] {AT};
\node (w1) [right of=model1, align=center, xshift=0.5cm, font=\LARGE] {$\omega^{(0)}$\\$\omega^{(1)}$};
\node (cpdf1) [rectangle, right of=w1, xshift=1.3cm, yshift=-0.3cm, minimum width=3.5cm, minimum height=1.2cm, fill=G2color, font=\LARGE, rounded corners] {$p (x_2|\omega^{(1)})$};

\draw [arrow] (cpdf0.south) -- ++ (0cm, -1cm) -- ++ (-13cm, 0cm) |- ([xshift=-0.3cm]x1.west);
\draw [arrow] ([xshift=0.3cm]x1.east) -- ([yshift=-0.2cm]model1.west);
\draw [arrow] (x0.east) -- ++ (2.5cm, 0cm) -- ++ (0cm, -2cm) |- ([yshift=0.5cm]model1.west);
\draw [arrow] ([yshift=0.5cm]model1.east) -- ([yshift=0.5cm]w1.west);
\draw [arrow] ([yshift=-0.3cm]model1.east) -- ([yshift=-0.3cm]w1.west);
\draw [arrow] ([yshift=-0.3cm]w1.east) -- (cpdf1.west);

\node (xn) [font=\LARGE, below of=x1, yshift=-3cm, align=center] {$x_2$\\$\vdots$\\$x_{n-1}$};
\draw [arrow] (cpdf1.south) -- ++ (0cm, -0.7cm) -- ++ (-15.5cm, 0cm) |- ([yshift=0.7cm]xn.west);

\node (ddots) [font=\LARGE, below of=w1, xshift=-0.1cm, yshift=-0.5cm, align=center, rotate=-10] {$\ddots$};

\node (modeln_b) [cinn_black, right of=xn, xshift=10cm, yshift=0.5cm] {};
\node (modeln) [cinn, fill=Bcolor, right of=xn, xshift=10cm, yshift=0.5cm] {AT};
\node (wn) [right of=modeln, align=center, xshift=1cm, font=\LARGE] {$\omega^{(0)}$\\$\vdots$\\$\omega^{(n-1)}$};
\node (cpdfn) [rectangle, right of=wn, xshift=1.8cm, yshift=-0.8cm, minimum width=4cm, minimum height=1.2cm, fill=G2color, font=\LARGE, rounded corners] {$p (x_n|\omega^{(n-1)})$};
\node (vdots) [font=\LARGE, left of=xn, xshift=0.3cm, yshift=0.2cm] {$\vdots$};
\draw [arrow] ([xshift=-1cm, yshift=-0.4cm]xn.west) |- ([yshift=-0.8cm]xn.west);

\draw [arrow] (x0.east) -- ++ (2.5cm, 0cm) -- ++ (0cm, -6cm) |- ([yshift=0.8cm]modeln.west);
\draw [arrow] ([xshift=1.0cm]x1.east) -- ++ (0.3cm, 0cm) -- ++ (0cm, -4cm) |- ([yshift=0.5cm]modeln.west);
\draw [arrow] ([yshift=0.7cm]xn.east) -- ([yshift=0.2cm]modeln.west);
\draw [arrow] ([yshift=-0.3cm]xn.east) -- ([yshift=-0.8cm]modeln.west);
\draw [arrow] ([yshift=0.8cm]modeln.east) -- ([yshift=0.8cm]wn.west);
\draw [arrow] ([yshift=-0.8cm]modeln.east) -- ([yshift=-0.8cm]wn.west);
\draw [arrow] ([yshift=-0.8cm]wn.east) -- (cpdfn.west);
\node (vdots2) [font=\LARGE, left of=modeln, xshift=0.6cm, yshift=-0.2cm] {$\vdots$};

\node (xlast) [font=\LARGE, below of=xn, yshift=-0.2cm, align=center] {$x_n$};
\draw [arrow] (cpdfn.south) -- ++ (0cm, -0.5cm) -- ++ (-20.5cm, 0cm) |- ([xshift=-0.3cm]xlast.west);

\draw [line, on top] ([xshift=2.5cm, yshift=-1.5cm]x0.east) -- ([xshift=2.5cm, yshift=-1.7cm]x0.east);
\draw [line, on top] ([xshift=2.5cm, yshift=-3.2cm]x0.east) -- ([xshift=2.5cm, yshift=-3.4cm]x0.east);
\draw [line, on top] ([xshift=2.5cm, yshift=-4.9cm]x0.east) -- ([xshift=2.5cm, yshift=-5.1cm]x0.east);
\draw [line, on top] ([xshift=1.3cm, yshift=-1.6cm]x1.east) -- ([xshift=1.3cm, yshift=-1.8cm]x1.east);

\end{tikzpicture}
    \caption{Sampling algorithm for the autoregressive transformer.}
    \label{fig:ALPs_sampling}
\end{figure}


\begin{figure}[b!]
\includegraphics[width=.33\textwidth,page=2]{paper_binned_ramp6} 
\includegraphics[width=.33\textwidth,page=2]{paper_GMM_ramp6} 
\includegraphics[width=.33\textwidth,page=1]{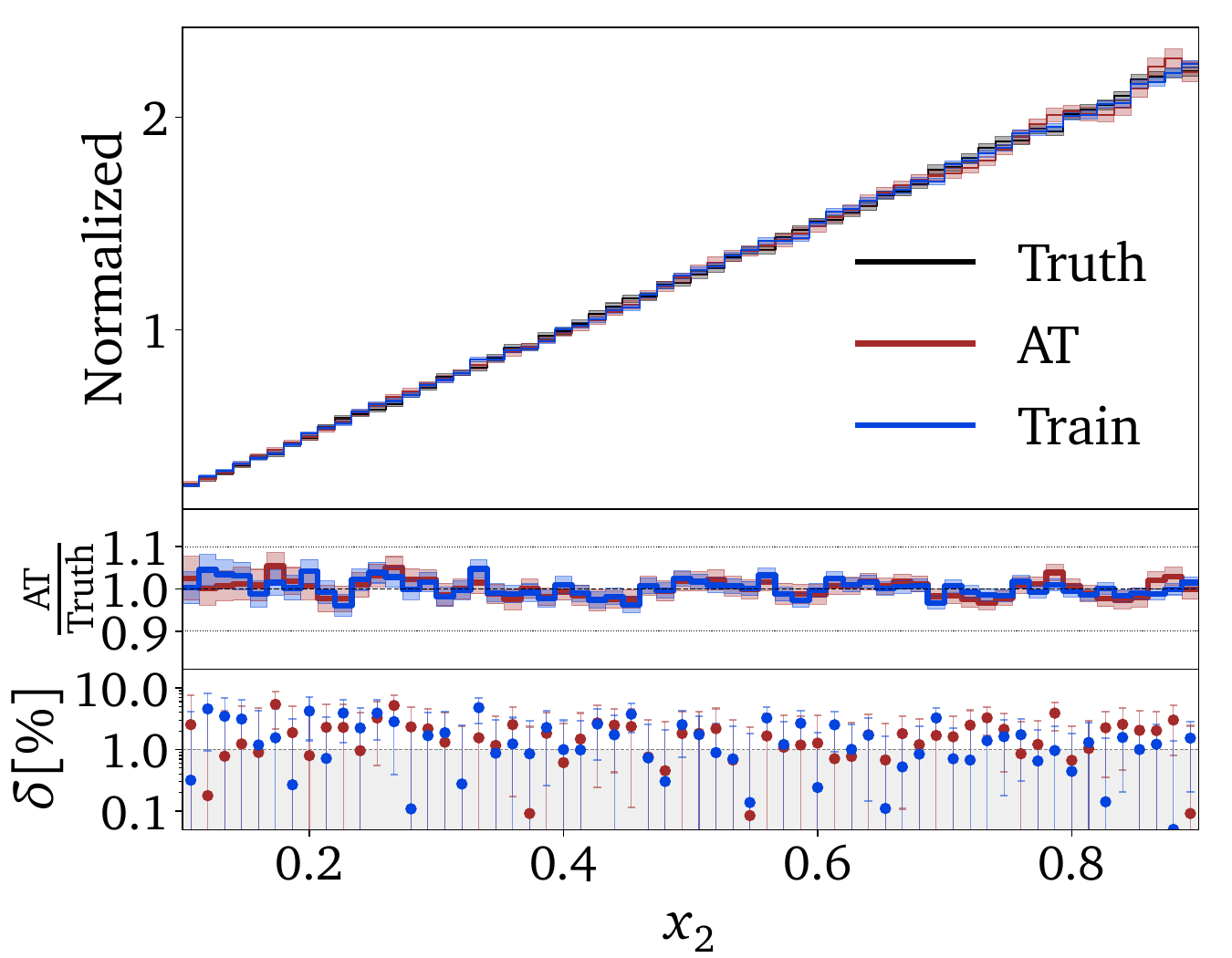} 
\caption{Density and predictive uncertainty distribution for a linear
  wedge ramp using a regressive transformer. We show results from
  binned density encoding (left) and from a Gaussian mixture model
  (center and right). The uncertainty on $\sigma_\text{pred} \equiv
  \sigma_\text{stat}$ is given by its $x_1$-variation. The blue line
  gives the statistical uncertainty of the training data. Figure from
  Ref.~\cite{Butter:2023fov}.}
\label{fig:ATramp}
\end{figure}

To benchmark the Bayesian autoregressive transformer, or JetGPT, we
target the same two-dimensional ramp which we also used for the B-INN
and the two diffusion models, Eq.\eqref{eq:linear_dens}.  In
Fig.~\ref{fig:ATramp} we show results from two representation of the
phase space density. In general, the density is described
accurately. Unlike for the B-INN, Fig.~\ref{fig:linear_unc}, and the
Bayesian diffusion networks, Fig.~\ref{fig:diff_ramp}, we do not observe
any structure in the absolute or relative uncertainties. This means
that the training of the transformer does indeed not benefit from the
maximum of available correlations in the middle of the phase
space. Instead, the absolute uncertainty grows with the density and
the relative uncertainty decreases with the density, as we expect
from a bin-wise counting experiment. The autoregressive transformer
cannot just be interpreted as a simple fit of a class of functions.

In the left panel of Fig.~\ref{fig:ATramp} we see that a naive, binned
density encoding leads to small uncertainties. In the center panel we
show the same results for a mixture of 21 Gaussians, leading to a much
larger uncertainty. While for two dimensions the advantage over the
binned distribution is not obvious, it is clear that we need such a
representation for LHC phase spaces. The main problem can be seen in
the right panel, at the upper edge of the ramp. Here, we have enough
training data to determine a well-suited model, but the Gaussian
mixture model cannot reproduce the flat growth towards the sharp upper
edge. Instead, it introduces an artifact, just covered by the
uncertainty. Because the transformer does no construct a fitting
function with a beneficial implicit bias, we also compare its
predictive uncertainty with the statistical uncertainty of the
training data. As one would hope, the uncertainty of the generative
network conservatively covers the limitations of the training data.


To judge the promise of the diffusion networks and the JetGPT
transformer for LHC applications, we can test them on the same LHC process
as we did for the normalizing flows, Sec.~\ref{sec:gen_inn_events},
\begin{align}
pp \to Z_{\mu \mu} + \{ 1,2,3 \}~\text{jets} \; .
\label{eq:ref_proc_z2}
\end{align}
We already know that the main challenge lies in the variable number of
jets, the $Z$-peak, and the pairwise $R$-separation of the jets. The
phase space dimensionality is three per muon and four per jet, \ie 10,
14, and 18 dimensions altogether.  Each particle is represented by
\begin{align}
    \{ \; p_T, \eta, \phi, m \; \} \; ,
\label{eq:def_obs}
\end{align}
and $\log (p_T - p_{T,\text{min}})$ provides us with approximately
Gaussian shape. All azimuthal angles are given relative to the leading
muon, and the jet mass is encoded as $\log m$.  Momentum conservation
is not guaranteed and can be used to test the network.

\begin{figure}[t!]
\includegraphics[width=.425\textwidth, page=2]{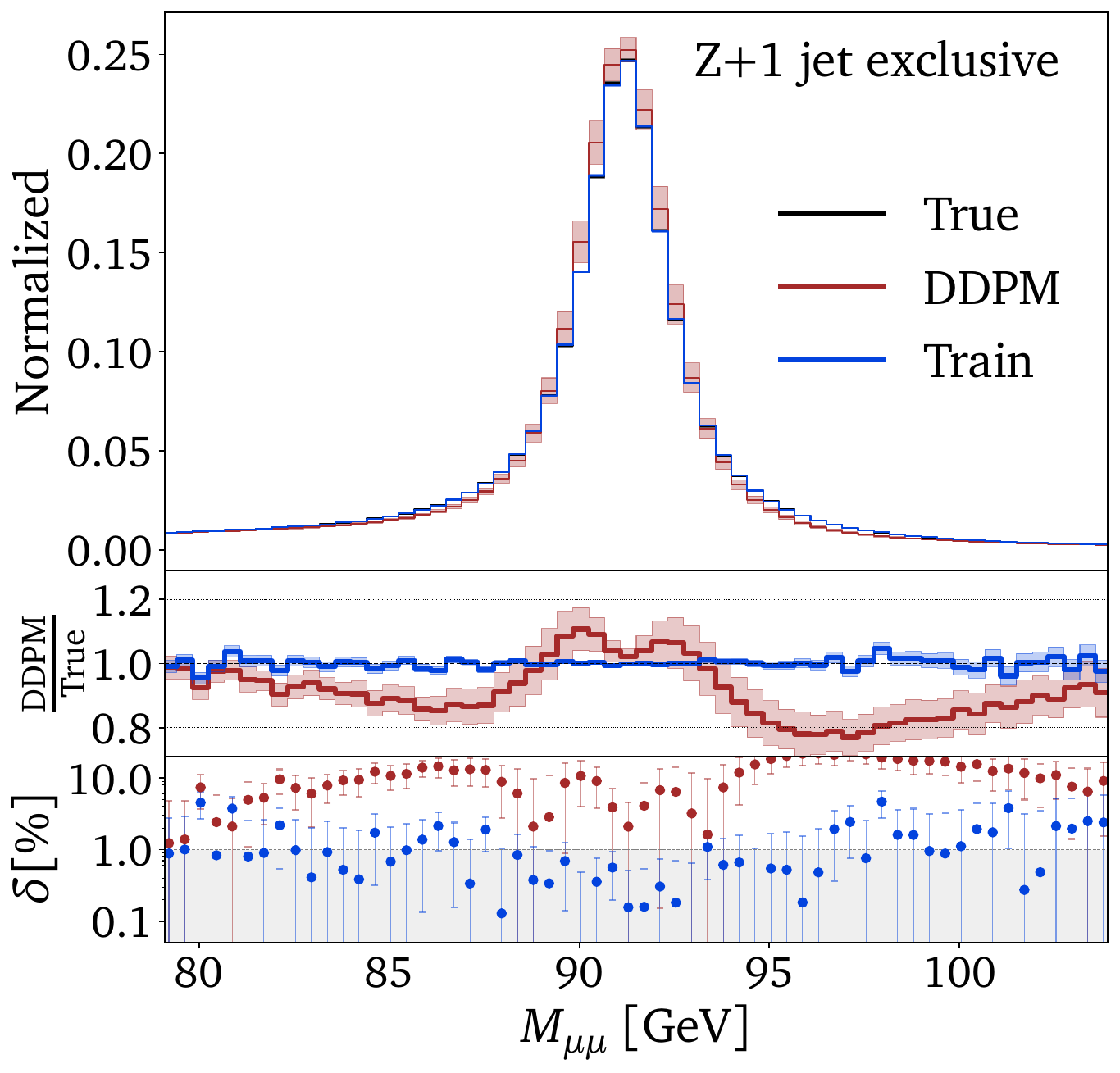} \hspace*{0.13\textwidth}
\includegraphics[width=.425\textwidth, page=13]{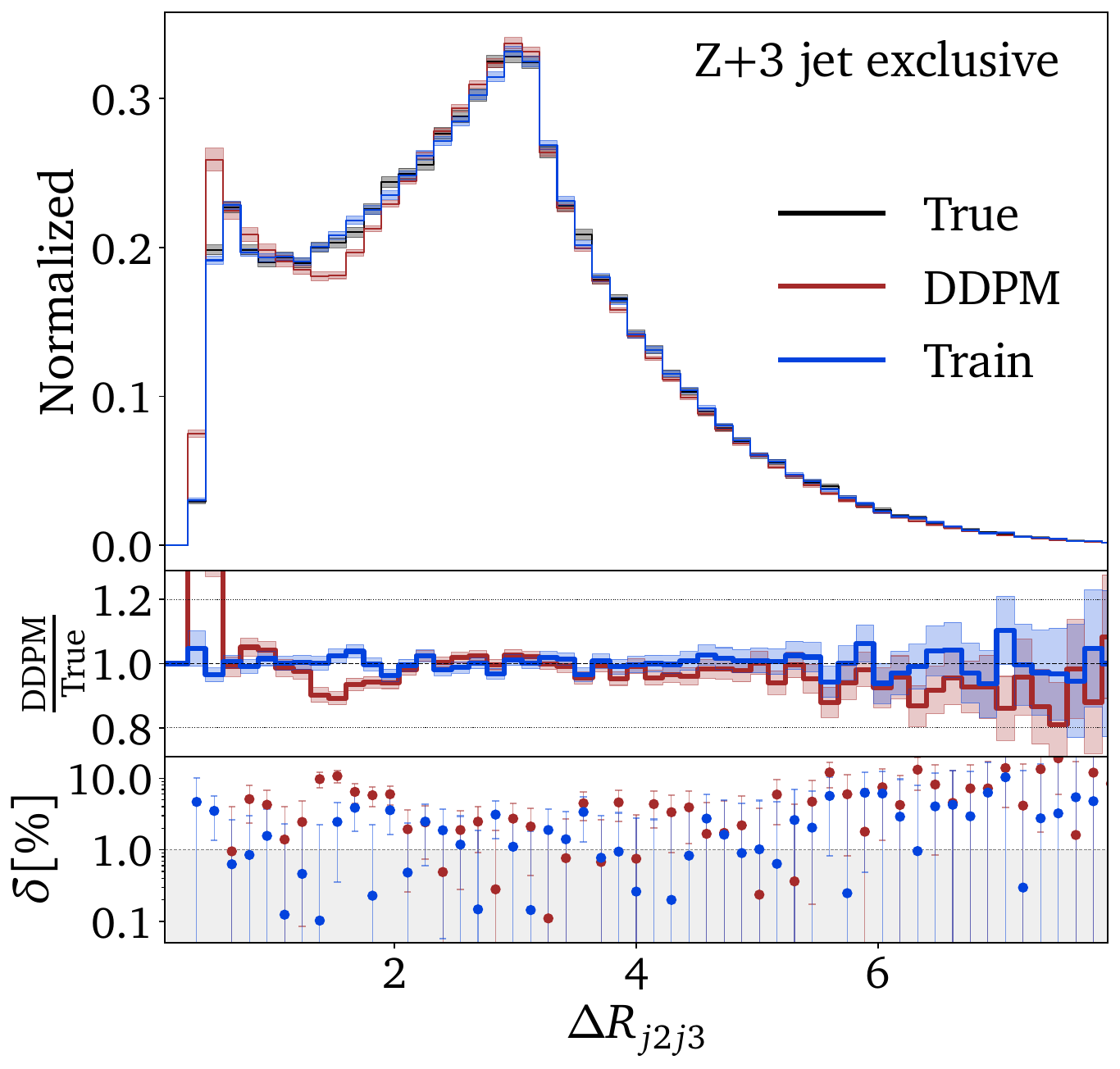} \\
\includegraphics[width=.425\textwidth, page=2]{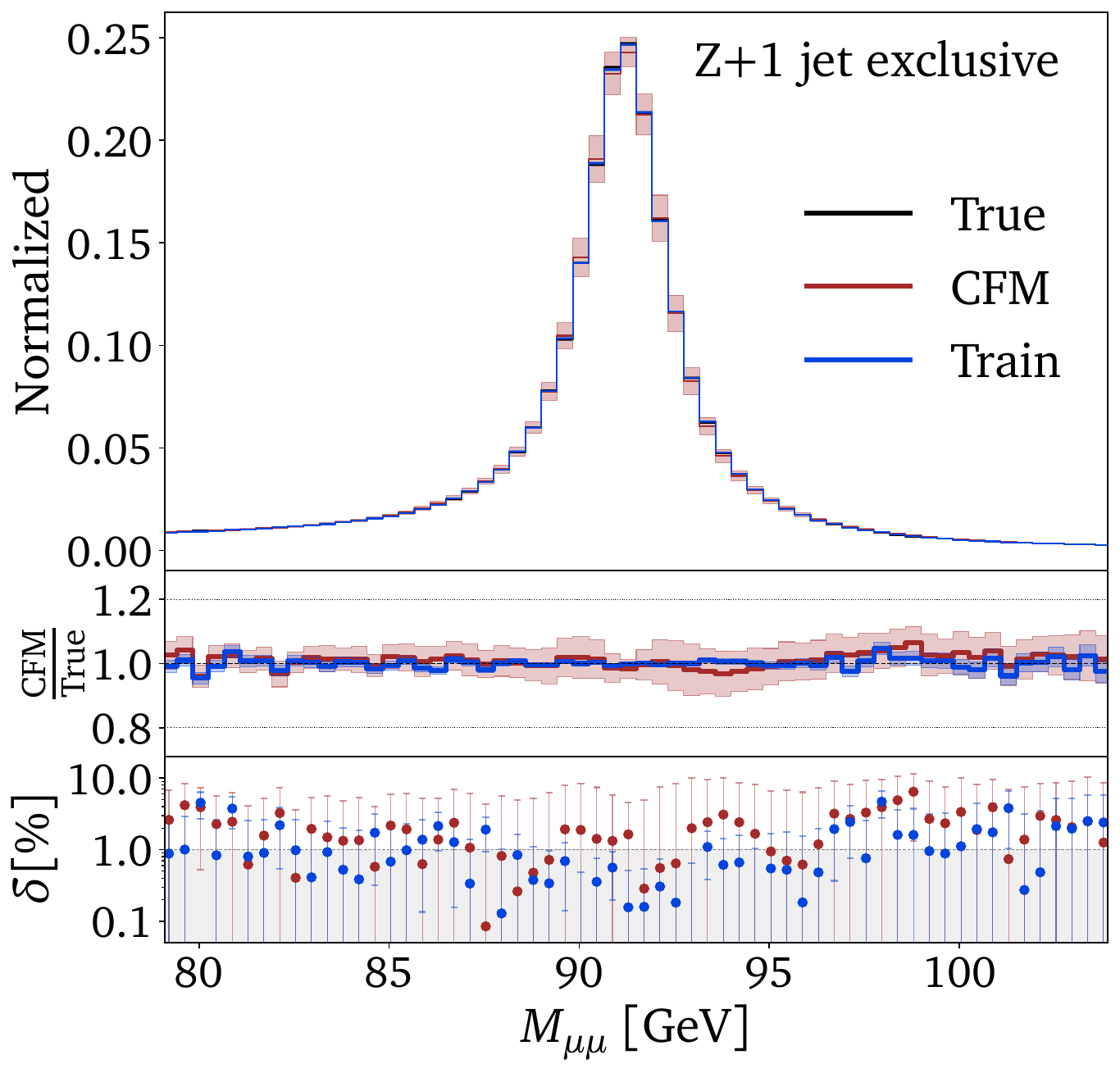} \hspace*{0.13\textwidth}
\includegraphics[width=.425\textwidth, page=13]{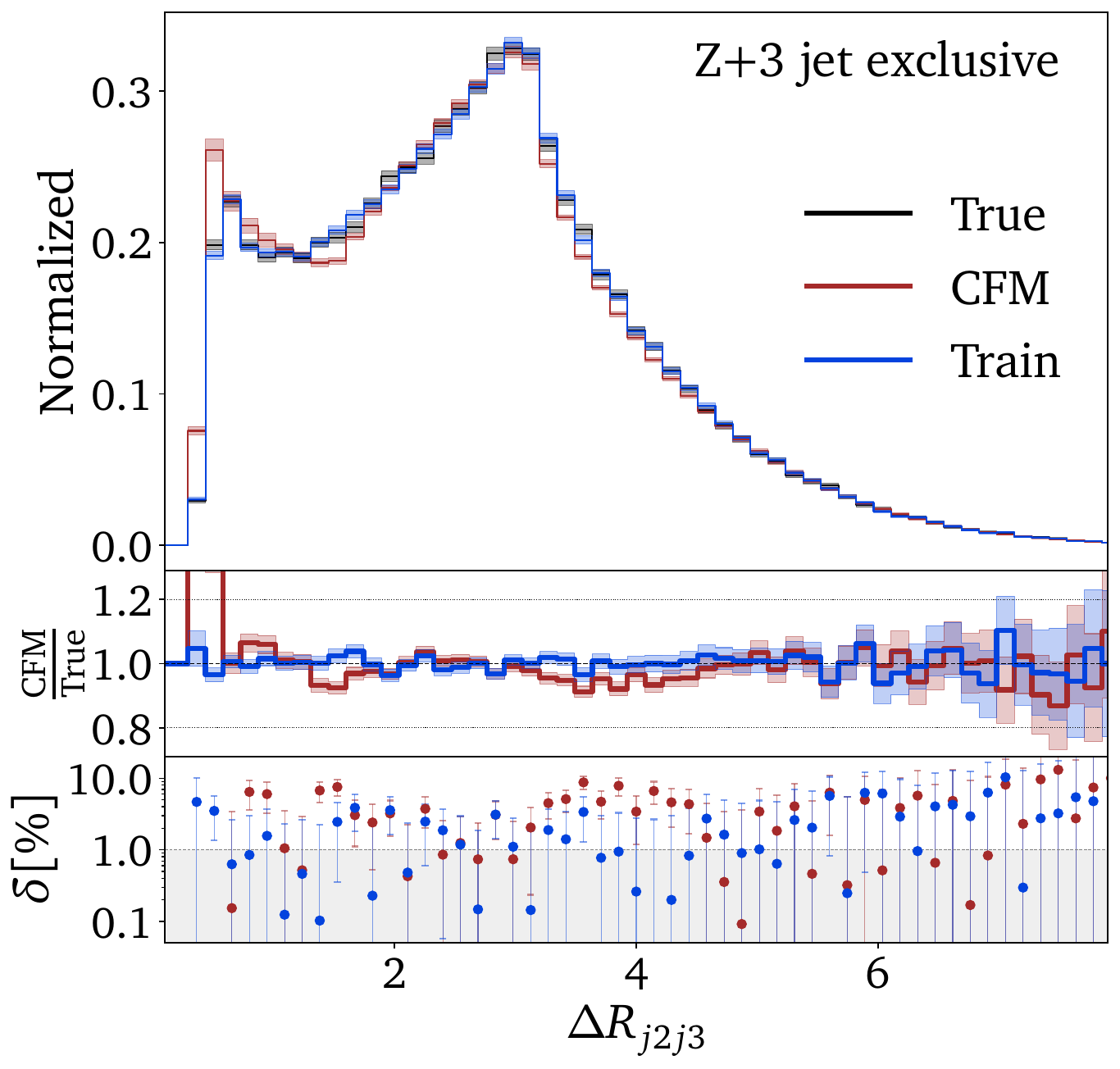} \\
\includegraphics[width=.425\textwidth, page=2]{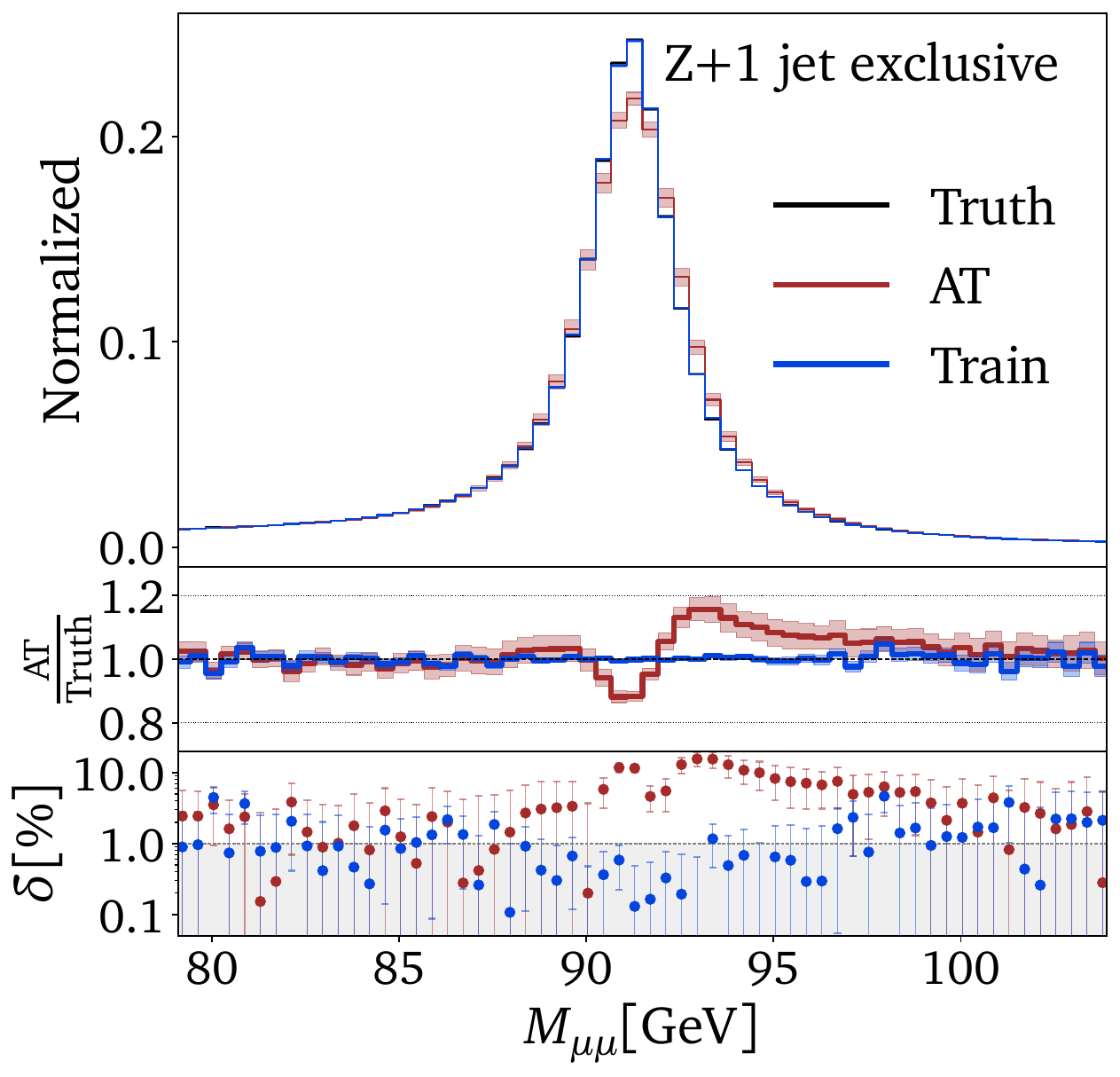} \hspace*{0.13\textwidth}
\includegraphics[width=.425\textwidth, page=7]{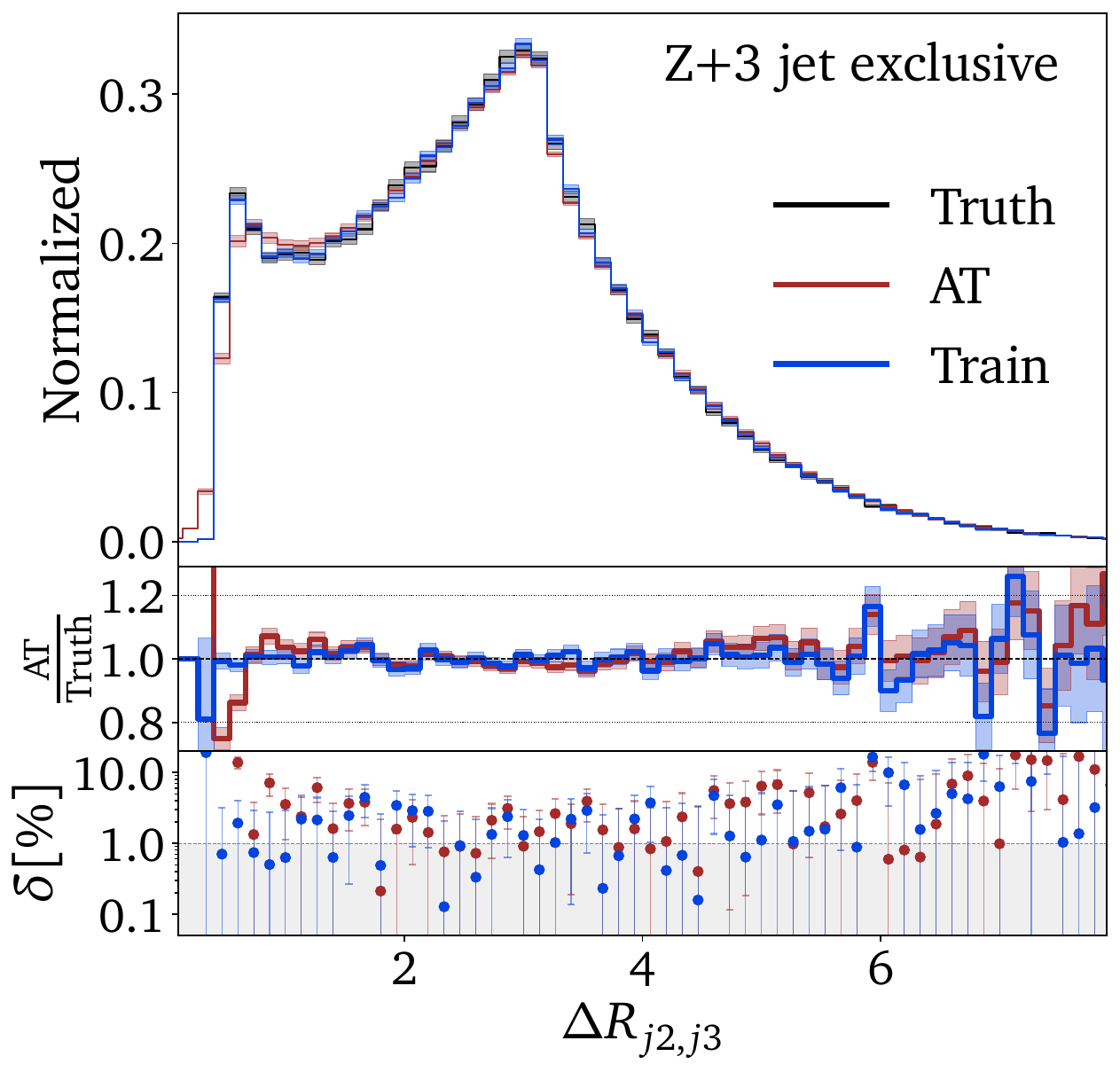}
\caption{Kinematic distributions form the DDPM diffusion network, the
  CFM diffusion network, and the JetGPT autoregressive
  transformer. Shown are the critical $\Delta R_{j_2 j_3}$ and $M_{\mu
    \mu}$ distributions. Figure from Ref.~\cite{Butter:2023fov}.}
\label{fig:jetgpt}
\end{figure}

In the top panels of Fig.~\ref{fig:jetgpt} we show the two key
distributions from the B-DDPM network. In the left panel we see that
the network has learned the sharp $Z$-peak well, albeit not
perfectly. The distinctive shape of the ratio indicates that the DDPM
models the $Z$-peak as slightly too wide. The $R$-separation between
the second and third jets has a global peak around $R_{j_2 j_3} =
\pi$, and the subleading feature around the hard cut $R_{jj} >
0.4$. This feature shows the onset of the collinear divergence, and
comes out too strong for the DDPM. This result can be compared to the
B-INN performance in Fig.~\ref{fig:discflow_re}, which instead washes
out the same distribution. This implies that diffusion networks have a
strong implicit bias in the way they fit a function to the data, but
that it is different from the normalizing flow case.

Moving on to the CFM diffusion networks, we see that it learns the
$Z$-peak perfectly, and much better than the predictive uncertainty
would suggest. For the $R_{j_2 j_3}$ distribution the two diffusion
networks show exactly the same limitations.

For the autoregressive transformer we make use of our freedom to order
the phase space directions, allowing it to focus on the angular
correlations.
\begin{align}
\left( \;
(\phi, \eta )_{j_{1,2,3}}, \; 
(p_T, \eta)_{\mu_1}, \;
(p_T, \phi, \eta)_{\mu_2}, \;
(p_T, m)_{j_{1,2,3}} \; 
\right) \; .
\label{eq:trans_lhc}
\end{align}
In the bottom row of Fig.~\ref{fig:jetgpt} we see that this strategy
leads to a significant improvement of the $R_{jj}$ distributions, but
at the expense of the learned $Z$-peak. The latter now comes with an
increased width and a shift of the central value. These differences
illustrate that, also in comparison to the B-INN, the diffusion
networks and the autoregressive transformer come with distinct
advantages and disadvantages, which suggests that for applications of
generative networks to LHC phase space we need to keep an open mind.

\clearpage
\section{Symmetries and representation learning}
\label{sec:gen_rl}

In particle physics, the feature space of neural network applications
is usually some kind of phase space. This phase space is, to
phycisists, interpretable, and we know many exact or approximate
symmetries. If we assume that a suitable data representation is
important for the performance of a realistic network, trained in a
finite time on a finite amount of data, we need to find and implement
such a representation.

Since Emmy Noether we know that
symmetries are the most basic structure in physics, especially in
particle physics.
LHC physics is defined by an extremely complex
symmetry structure, starting with LHC data, the detector geometry, to
the relativistic space-time symmetries and local gauge symmetries
defining the underlying QFT. If we want to use machine learning we
need to embrace these symmetries.

The most obvious symmetry is permutation invariance of particles for
instance in a jet, as discussed in Sec.~\ref{sec:class_graph_struc}.
The jet images\index{jet images} introduced in
Sec.~\ref{sec:class_cnn} are defined in rapidity vs azimuthal angle,
and their preprocessing exploits the rotation symmetry around the jet
axis.  Theoretical invariances under infrared transformations
motivated the IR-safe transformer and the energy flow networks.  The
most relevant and interesting symmetry in LHC physics the Lorentz
symmetry of relativistic particles. We discussed it in
Sec.~\ref{sec:class_graph_4vec} and gave an example for the learned
underlying Minkowski metric in
Eq.\eqref{eq:learned_minkowski}. Indeed, the question is how many
networks in particle physics have to spend GPU time on learning
permutation symmetry or the Minkowski metric for a set of particle
4-vectors.

From a structural perspective, there are two ways symmetries can
affect neural network functions. First, we call a network
\underline{equivariant} or covariant if for a symmetry operation $S$ and a network
output $f_\theta(x)$ we have
\begin{align}
\boxed{  f_\theta(S(x)) = S(f_\theta(x)) } \; .
\label{eq:def_equivariant}
\end{align}
This means we can recover a symmetry operation $S$ on the data $x$ as a
symmetry operation of an equivariant network output.  For example, if
we shift all pixels in an input image to a CNN in one direction, the
training of the convolutional filters will not change and the feature
maps inside the network will just be shifted as well, so the CNN is
equivariant under translations.  An equivalent definition of an
equivariant network is that for two different inputs with the same
output, two transformed inputs also give the same output,
\begin{align}
  f_\theta(x_1) = f_\theta(x_2)
  \qquad \Rightarrow \qquad 
  f_\theta(S(x_1)) = S(f_\theta(x_1)) =S(f_\theta(x_2)) = f_\theta(S(x_2)) \; .
\end{align}

A stronger symmetry requirement is an \underline{invariant} network,
namely
\begin{align}
\boxed{  f_\theta(S(x)) = f_\theta(x) } \; .  
\label{eq:def_invariant}
\end{align}

Following this line of thought, we want to implement a general
representation for LHC data which is based on a set of 4-vectors, but allows
the networks to use the benefits of symmetry-aware representations,
and works for many network tasks. There are, at least, three ways to
achieve this:
\begin{enumerate}
\item hard-code a
data representation which is, for instance, equivariant under Lorentz
transformations;
\item learn a, for instance, symmetric data representation in
  a self-supervised manner, through contrastive learning;
\item learn such a data representation very generally, as what is
  usually advertized as pre-training or a foundation model.  
\end{enumerate}

\subsection{Lorentz-equivariance}
\label{sec:gen_rl_eq}

To generally simplify network tasks on LHC phase spaces, we can
guarantee Lorentz equivariance for different networks tasks. From
Sec.~\ref{sec:class_graph_trans} we know that transformers provide
flexible and powerful data representations, and from many applications
in physics and beyond we also know that these data representations
lead to great performance, at least when we have enough training
data. This means we could try to develop a universal
Lorentz-equivariant transformer.

One way of constructing an equivariant transformer uses geometric
algebra (L-GATr).  Such a geometric algebra is an extension of a
vector space with an additional the \underline{geometric product}. For
two vectors it is decomposed into a symmetric and an antisymmetric
contribution,
\begin{align}
    xy = \frac{\{x,y \}}{2} + \frac{[x,y]}{2}  \; ,
    \label{eq:xy_gen}
\end{align}
where the anti-commutator is the usual inner product and the
commutator is a new outer product.  Neither of them is part of the
original vector space.

We are interested in the spacetime algebra in terms of a 4-vector
space is $\mathbb{R}^{4}$ and the Minkowski metric. As its basis we
choose a set of four real vectors $\gamma^\mu$, which satisfy
\begin{align}
    \left\{\gamma^{\mu}, \gamma^{\nu} \right\} = 2 g^{\mu\nu} \; .
    \label{eq:gamma-matrices}
\end{align}
This inner product establishes the basis elements as a set of
orthogonal vectors and fixes their normalization. The difference to
the similar Dirac algebra and spinors is that they are defined over a
complex rather than real vector space.  Next, we construct new
elements of the algebra. All higher-order elements can be
characterized as antisymmetric products of $\gamma^\mu$, organized in
terms of grade or the number of $\gamma^\mu$ they require. For
instance, the antisymmetric tensor $\sigma^{\mu\nu}$ is generated from
the geometric product of two $\gamma^\mu$ and consequently has grade
two,
\begin{align}
    \gamma^{\mu} \gamma^{\nu} 
    = \frac{\{ \gamma^\mu, \gamma^\nu \}}{2} + \frac{[ \gamma^\mu, \gamma^\nu]}{2} 
    \equiv g^{\mu\nu} + \sigma^{\mu \nu}  \; .
\end{align}
The commutator $\sigma^{\mu\nu}$ is a bivector, which can be
interpreted as the plane in Minkowski space. The symmetric term in the
geometric product reduces the grade, while the antisymmetric term
increases it. The whole product $\gamma^\mu \gamma^\nu$ is a sum of
grade zero (scalar) and grade two. A generic element of the algebra
that mixes grade information is called a multivector.

Next, the geometric product of three vectors contains the
antisymmetric tensor
$\epsilon_{\mu\nu\rho\sigma}\gamma^{\mu}\gamma^{\nu}\gamma^{\rho}$ and
forms an axial vector. The product of all four vectors defines the
pseudoscalar
\begin{align}
    \gamma^5
    =\gamma^0\gamma^1\gamma^2\gamma^3 
    \equiv \frac{1}{4!}\epsilon_{\mu\nu\rho\sigma}\gamma^\mu\gamma^\nu\gamma^\rho\gamma^\sigma.
\end{align}
It acts as parity reversal and can be used to write axial vectors as
$\gamma^\mu\gamma^5$.  The missing factor $i$ compared to the usual
definition of $\gamma^5$ reflects the difference between the complex
Dirac algebra and our real spacetime algebra.  Geometric products with
more than four $\gamma^\mu$ can be reduced to lower-grade structures.

Combining all these elements, we can express any \underline{multivector} of the
algebra as
\begin{align}
    x= x^S \; 1 
    + x^V_\mu  \; \gamma^\mu 
    + x^B_{\mu\nu} \; \sigma^{\mu\nu} 
    + x^A_\mu \; \gamma^\mu \gamma^5 
    + x^P \; \gamma^5 
    \qquad \text{with} \qquad 
    \begin{pmatrix}
    x^S \\ x^V_\mu \\ x^B_{\mu\nu} \\ x^A_\mu \\ x^P 
    \end{pmatrix} 
    \in\mathbb{R}^{16} \; .
\label{eq:multivector}
\end{align}
In this representation, we only include the nonzero and independent
entries in the antisymmetric bivector.  Multivectors can be used to
represent both spacetime objects and Lorentz transformations. For
instance, particles are characterized by their particle
identification and their 4-momentum
\begin{align}
    x^S = \text{PID} 
    \qqquad x^V_{\mu} = p_{\mu}
    \qqquad x^T_{\mu\nu}=x^A_\mu=x^P=0 \; . 
    \label{eq:embedding}
\end{align}
Similarly, we can represent objects like transition amplitudes as a
function of 4-momentum multivectors. A matrix element as a function of
4-momenta can be decomposed into parity-even and parity-odd terms and
then gives us two scalar function and one pseudo-scalar function
\begin{align}
  |\mathcal{M}|^2
  &= |\mathcal{M}_E|^2 + |\mathcal{M}_O|^2 + 2\, \text{Re}\left(\mathcal{M}_E^* \mathcal{M}_O \right) \notag \\
  &= x^S1 + x^P \gamma^5 
    \qqquad \text{with} \qquad 
    x^S1 = |\mathcal{M}_E|^2 + |\mathcal{M}_O|^2
    \quad \text{and} \quad 
    x^P \gamma^5 = 2 \text{Re}\left(\mathcal{M}_E^* \mathcal{M}_O \right) \; . 
\end{align}

Finally, the geometric algebra also describe symmetry operations on
objects. In this framework Lorentz transformations act as
\begin{align}
    \Lambda_v(x) = vxv^{-1} \; , 
    \label{eq:lorentz_law}
\end{align}
where $v$ is a multivector representing an element of the Lorentz
group acting on the algebra element $x$. The representation $v$ is
built by a simple rule: a multivector encoding an object that is
invariant under a Lorentz transformation will also represent the
transformation itself. This gives a dual interpretation to spacetime
algebra elements as geometric objects and as Lorentz transformations.
This way, the algebra representation allows us to apply this boost on
any object in the geometric algebra, such that Lorentz transformations
will never mix grades. Each algebra grade transforms under a separate
sub-representation of the Lorentz group.

Based on the multivector representation, we construct an equivariant
transformer network L-GATr. In the form of
Eq.\eqref{eq:def_equivariant} its Lorentz equivariance implies
\begin{align}
  \boxed{
    \text{L-GATr}\Big(\Lambda(x)\Big) = \Lambda \Big(\text{L-GATr}(x)\Big)
    } \; .
\end{align}

L-GATr follows the standard transformer architecture introduced in
Sec.~\ref{sec:class_graph_trans}, including standard operations like
linear layer, attention, layer normalization, or an activation
function. The only difference is that the feature or phase space is
multivectors $x$ as defined in Eq.\eqref{eq:multivector}. The only
additional structure is the geometric product defined in
Eq.\eqref{eq:xy_gen}.

\begin{figure}[t]
    \includegraphics[width=0.495\textwidth]{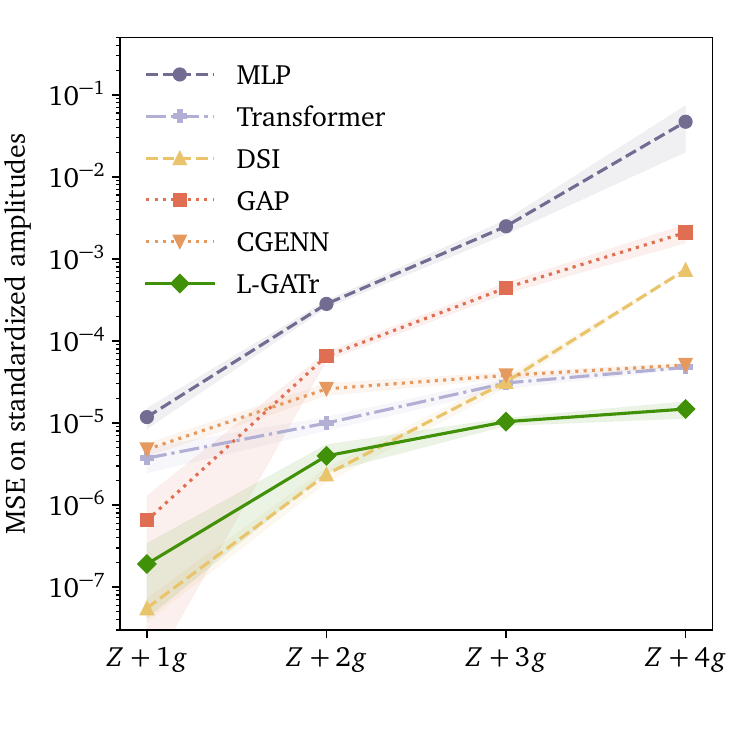}
    \includegraphics[width=0.495\textwidth]{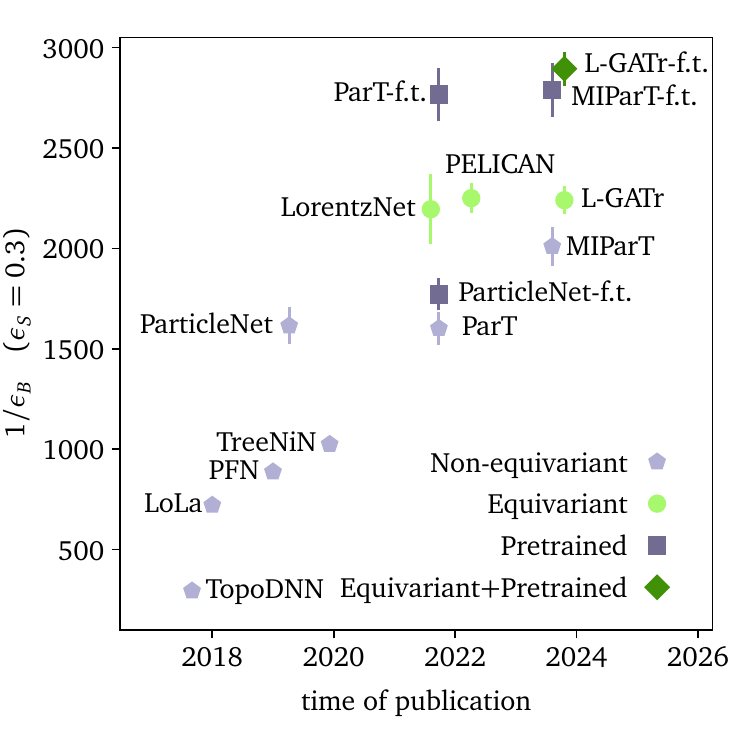}
    \caption{Left: prediction error from L-GATr and several baselines
      for $Z+ng$ amplitudes with increasing particle multiplicity.
      size. Right: history of top taggers since
      Fig.~\ref{fig:toptagging}. Figure from
      Ref.~\cite{Brehmer:2024yqw}.}
    \label{fig:lgatr_amplitudes}
\end{figure}

The main claim of representation learning is that we should be able to
use the L-GATr data representation for different LHC tasks, from
amplitude regression to jet tagging and even generation. One caveat is
that for some LHC tasks Lorentz symmetry is not exact. In that case we
have to allow L-GATr to break the symmetry in a flexible manner, for
example through hard-breaking reference multivectors as additional
inputs and without altering the network architecture. The network now
has the option to tune out the reference vectors when they are not
needed.  An example is the LHC beam direction, which breaks the
Lorentz group to the subgroup of rotations around and boosts along the
beam axis. The natural reference vector is this beam direction
$x^V_\pm = (0,0,0,\pm 1)$, or alternative the bivector representing
the $x-y$ plane, $x^B_{12} = 1$. The fact that for some tasks the
Lorentz symmetry is broken implies that we will have to at least
fine-tune the universal L-GATr representation:
\begin{enumerate}

\item Our first L-GATr case study is on amplitude regression. Because we
will see that one of the key advantages is an improved scaling of the
network accuracy with the number of particles in the final state, we
switch from the $\gamma \gamma g(g)$ final state from
Sec.~\ref{sec:basics_regr_amp} to a simpler, tree level process
\begin{align}
 q \bar{q} \to Z + n  \, g, 
 \qqqquad n=1 \ldots 5 \; .
 \label{eq:zplusjets} 
\end{align}
To avoid soft or collinear divergences, we require 
\begin{align}
 p_T > 20~\gev
 \qquad \text{and}\qquad
 \Delta R > 0.4 
\end{align}
for all final-state objects.  The input data is still the initial and
final state 4-momenta combined with standardized amplitudes. The
standard MSE loss can easily be expanded to a heteroskedastic loss or
even to a BNN extension. The improved scaling towards up to six
particles in the final state is due to the permutation symmetry of
transformers, as discussed in Sec.~\ref{sec:class_graph_trans}.

The MSE or absolute accuracy of the amplitude network is shown in the
left panels of Fig.~\ref{fig:lgatr_amplitudes}. The benchmarks are a
standard MLP; a non-equivariant transformer; the deep-sets invariant
network from Sec.~\ref{sec:basics_regr_amp}; an equivariant MLP using
the geometric algebra (GAP); and an equivariant graph network using
the geometric algebra (CGENN).  We first see that transformer and
graph networks scale better with the number of external
particles. L-GATr is roughly on par with the leading DSI network for
two gluons, but its improved scaling gives it the lead for three or
more gluons. Going to five gluons confirms the improved L-GATr
scaling, in spite of a reduced number of training amplitudes
reflecting the event generation CPU limitations.

\begin{table}[b!]
    \centering
    \begin{small}
    \begin{tabular}{l l c llll}
        \toprule
        & Network && Accuracy & AUC & $1/\epsilon_B$ ($\epsilon_S=0.5$) & $1/\epsilon_B$ ($\epsilon_S=0.3$) \\
        \midrule
        \multirow{4}{*}{architectures}
        &ATLAS TopoDNN  && 0.916\dz{} & 0.972\dz{} & -- & \dz{}\result{295}{\dz{}\dz{}5} \\
        &LoLa (Sec.~\ref{sec:class_graph_4vec})     && 0.929\dz{} & 0.980\dz{} & -- & \dz{}\result{722}{\dz{}17} \\
        &ParticleNet (Sec.~\ref{sec:class_graph_arch}) && 0.940\dz{} & 0.9858 & \result{397}{\dz{}7} & \result{1615}{\dz{}93} \\
        &ParT (Sec.~\ref{sec:class_graph_trans})       && 0.940\dz{} & 0.9858 & \result{413}{16} & \result{1602}{\dz{}81} \\
        \midrule
        \multirow{3}{*}{equivariant}
        &LorentzNet  && 0.942\dz{} & 0.9868 & \result{498}{18} & \result{2195}{173} \\
        &PELICAN     && \result{0.9426}{0.0002} & \result{0.9870}{0.0001} & -- & \result{2250}{\dz{}75} \\
        &L-GATr      && \result{0.9423}{0.0002} & \result{0.9870}{0.0001} & \result{540}{20} & \result{2240}{\dz{}70} \\
        \midrule
        \multirow{3}{*}{pre-trained}
        &ParticleNet-f.t. && 0.942\dz{} & 0.9866 & \result{487}{\dz{}9} & \result{1771}{\dz{}80} \\
        &ParT-f.t.        && 0.944\dz{} & 0.9877 & \result{691}{15} & \result{2766}{130} \\
        &L-GATr-f.t.      && \result{0.9446}{0.0002} & \result{0.98793}{0.00001}  & \result{651}{11} & \result{2894}{84}  \\
        \bottomrule
    \end{tabular}
    \end{small}
    \caption{Top tagging accuracy, AUC, and background rejection
      $1/\epsilon_B$ for the top tagging dataset described in
      Sec.~\ref{sec:class_cnn_tag}. The background suppression is the
      same as in the right panel of
      Fig.~\ref{fig:lgatr_amplitudes}. The error bars are based on the
      mean and standard deviation of five random seeds. Table from
      Ref.~\cite{Brehmer:2024yqw}.}
    \label{tab:lgatr_toptagging}
\end{table}

\item Next, we see what happens for jet tagging, the LHC task impacted
  most by modern ML.  The specific task is given by tagging boosted
  top jets, as introduced in Sec.~\ref{sec:class_cnn_tag}.  The
  performance of the L-GATr tagger is shown in the right panel of
  Fig.~\ref{fig:lgatr_amplitudes}, as a function of time. The
  performance measure is the background suppression at a fixed signal
  efficiency of $\epsilon_s = 30\%$, shown in
  Tab.~\label{tab:lgatr_toptagging}. For some of the older taggers the
  corresponding ROC curves are shown in Fig.~\ref{fig:toptagging}.

The classic baselines are ParticleNet, a convolutional graph network
discussed in Sec.~\ref{sec:class_graph_arch} and the
Particle Transformer, a fully implemented tagger similar to the
transformer introduced in Sec.~\ref{sec:class_graph_trans}. They
define the cutting edge of standard architectures with a background
suppression around 1600, more than a factor 5 better than the first
ATLAS DNN-tagger and all tools from the pre-ML era.

A systematic performance boost comes from Lorentz-equivariant
architectures, for instance LorentzNet as an equivariant version of
ParticleNet, PELICAN as an alternative with a focus on equivariant
aggregation functions, and L-GATr as the first equivariant
transformer. All three show a background rejection around 2200.

The final improvement comes from pre-training the data-intensive
transformers on another, much larger jet dataset. For this purpose, we
use a standard set of 100M jets from different partons, split into 10
different jet classes and referred to as the JetClass dataset. After
that, the transformers are fine-tuned by switching the last layer of
the network to map to a single output channel and re-initializing its
weights.  All three transformers benefit from pre-training, where the
equivariant and pre-trained L-GATr takes the lead with a background
suppression around 2900, a factor ten better than the pre-DNN taggers.

\begin{figure}[t!]
    \includegraphics[width=0.49\linewidth]{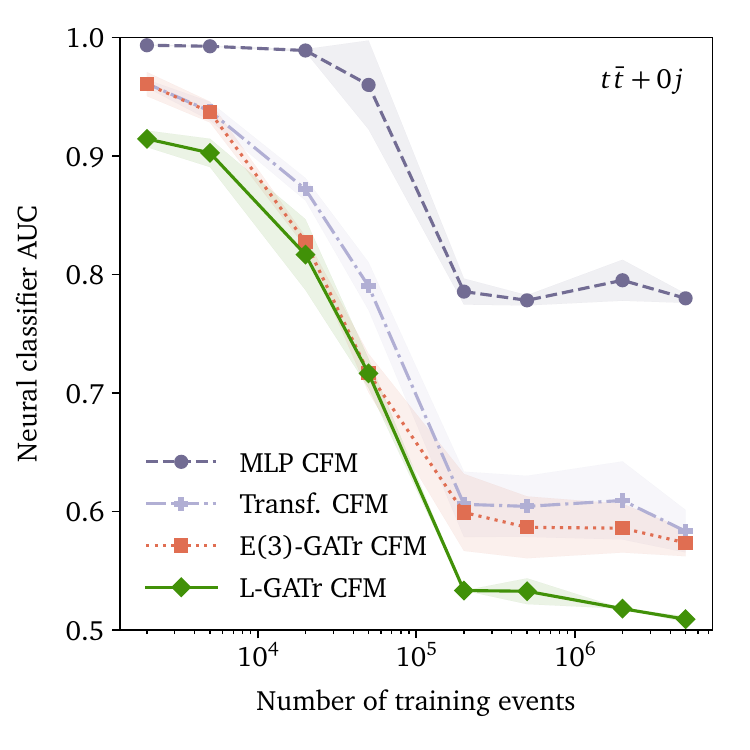} 
    \includegraphics[width=0.49\linewidth]{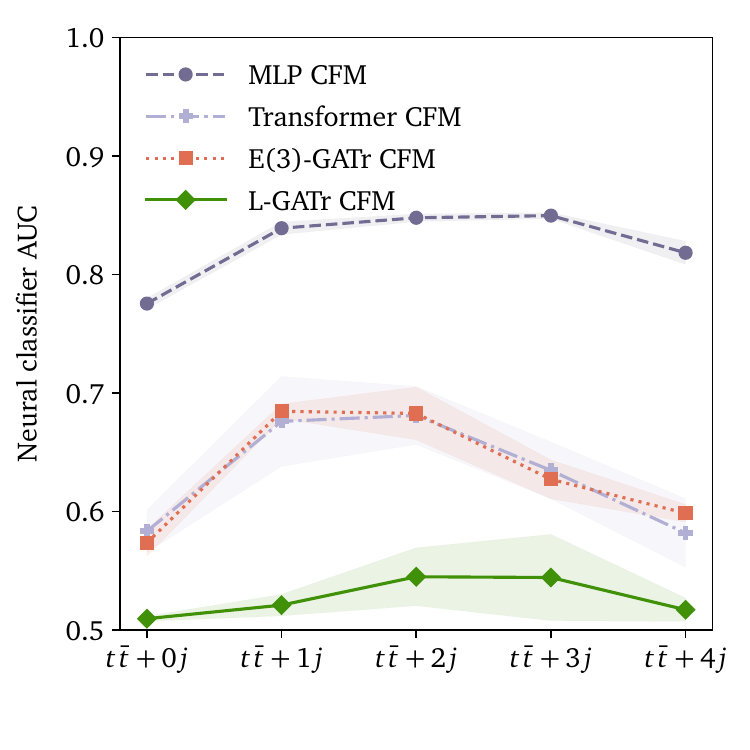}
    \caption{Performance of the LGATr generator in terms of a trained classifier AUC. We show the scaling with the size of the training dataset size (left) and with the number of particles in the final state. Figure from
      Ref.~\cite{Brehmer:2024yqw}.}
    \label{fig:lgatr_cfm_scaling}
\end{figure}

\item The third benchmark application is event generation. From
  Sec.~\ref{sec:gen_gan_events} we know that top pair production is a
  challenge because of the intermediate on-shell tops and
  $W$-bosons. We suspect from the amplitude regression that an
  equivariant transformer will scale well with more particles in the
  final state, so we choose a more challenging all-hadronic partonic
  process
\begin{align}
    pp\to t_h \bar t_h + n\, j , 
    \qqqquad n=0\dots 4 \; ,
\end{align}
Again, to avoid soft and collinear divergences we require
\begin{align}
 p_{T,j} > 22~\gev 
 \qquad 
 \Delta R_{jj} > 0.5 
 \qquad |\eta_j| < 5 \; ,
\end{align} 
and two $b$-tagged jets. The sizes of the $t\bar t+n\; j$ datasets
reflect the frequency of the respective processes, resulting in 9.8M,
7.2M, 3.7M, 1.5M and 480k events for $n=0 \ldots 4$. We train separate
networks for each multiplicity, to allow for a direct comparison
between different architectures, but we know that transformers
can also be trained jointly on all
multiplicities.

As the generative network architecture we employ the CFM from
Sec.~\ref{sec:gen_diff_cfm}.
The phase space parametrization for which we require straight
trajectories is crucial for the generator performance. A
standard MLP and transformer CFMs work on $x$ defined as
\begin{align}
p =
\begin{pmatrix} 
 E\\p_x\\p_y\\p_z 
\end{pmatrix} 
\quad \to  \quad
f^{-1}(p) = x =  
\begin{pmatrix} 
 x_p  \\ x_m \\ x_\eta \\ x_\phi
\end{pmatrix} 
\equiv
\begin{pmatrix} 
 \log( p_T-p_T^\text{min}) \\ \log m^2 \\ \eta \\ \phi
 \end{pmatrix} \; ,
\label{eq:momentum_rep}
\end{align}
to encode $v(x(t),t)$.  For the Lorentz-equivariant generator, we
choose a base distribution $\pl$ that is invariant under the symmetry
group and use Gaussians in $(p_x, p_y, p_z, \log m^2)$ with mean and
standard deviation fitted to the phase space distribution $\pd $.
L-GATr starts with $p$ and transforms $x$ into the corresponding
4-momenta $p=f(x)$. For them, L-GATr encodes the velocity $v(p(t),t)
\equiv \text{L-GATr}(p)$.  Finally, we transform this velocity back
into $x$, using the jacobian of the backwards transformation.  As a
test, we also define E(3)-GATr by encoding $(p_x, p_y, p_z)$ as a
vector and $x_m$ as a scalar.

To analyze the scaling, we can use the AUC of a classifier trained to
tell apart generated from training events. In the left panel of
Fig.~\ref{fig:lgatr_cfm_scaling} we find a clear performance increase
with increasing symmetry awareness. The transformer architecture does
not have a visible effect on the required training dataset. In the
right panel we see that the scaling as a function of the numbers of
addional jets is essentially flat. Here is is important to keep in
mind that the challenge of the event generation is already in the hard
$t\bar{t}$ process, which might explain why additional jets hardly
have an effect.

\end{enumerate}

Altogether, we see that permutation invariance and Lorentz-equivariant
encoded in the L-GATr data representation does improve regression,
classification, and generative networks. However, this improvement
requires us to know these symmtries and to encode them such that the
network is allowed to break them, if needed.

\subsection{Contrastive learning}
\label{sec:gen_rl_clr}

Especially when encoding symmetries in network architectures is hard,
we would like a training procedure and loss function which ensures
that the latent or representation space is invariant under symmetries
or invariances defined by symmetric training data or symmetric data
augmentations. A way to achieve this is \underline{contrastive
  learning} of representations (CLR).  The goal of such a network is
to map, for instance, jets $x_i$ described by their constituents to a
latent or representation space,
\begin{align}
  f_\theta: \quad x_i \rightarrow z_i \;,
\label{eq:mapping}
\end{align}
which is invariant to symmetries and theory-driven augmentations, and
remains discriminative for the training and test datasets.

\begin{figure}[b!]
  \centering
  \includegraphics[width=0.6\textwidth]{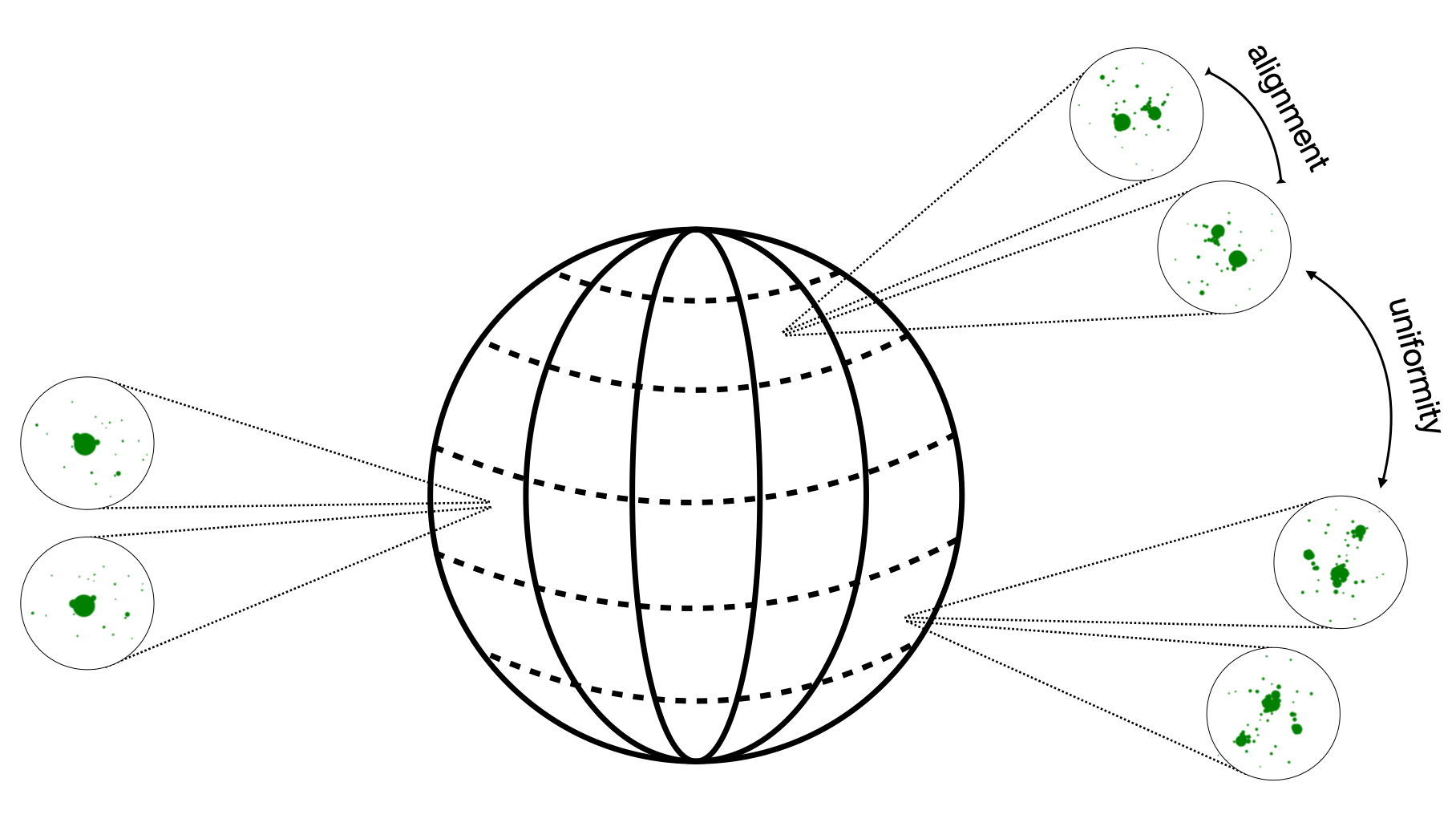}
  \caption{Illustration of the uniformity and alignment concepts
    behind the contrastive learning. Figure from
    Ref.~\cite{Dillon:2021gag}.}
  \label{fig:sphere}
\end{figure}

As usually, our jets $x_i$ are described by $n_C$ massless
constituents and their phase space coordinates given in
Eq.\eqref{eq:vec_image}, so the jet phase space is
$3n_C$-dimensional. We first take a batch of jets $\{x_i\}$ from the
dataset and apply one or more symmetry-inspired augmentations to each
jet. This generate an augmented batch $\{x_i^\prime\}$. We then pair
the original and augmented jets into two datasets
\begin{alignat}{7}
  \text{positive pairs:} &\qqquad \{(x_i,x_i^\prime)\} \notag \\
  \text{negative pairs:} &\qqquad \{(x_i,x_j)\} \cup \{(x_i,x_j^\prime)\}
  \quad \text{for} \quad i \neq j \; .
\end{alignat}
Positive pairs are symmetry-related and negative are not.  The goal of
the network training is to map positive pairs as close together in
representation space as possible, while keeping negative pairs far
apart. Simply pushing apart the negative pairs allows our network to
encode any kind of information in their actual position in the latent
space.  Labels, for example indicating if the jets are QCD or top, are
not used in this training strategy, which is referred to as
\underline{self-supervised} training. The trick to achieve this split
in the representation space is to replace vectors $z_i$, which the
network outputs, by their normalized counterparts,
\begin{align}
  f_\theta(x_i) = \frac{z_i}{|z_i|}
  \qquad \text{and} \qquad 
  f_\theta(x_i^\prime) = \frac{z_i^\prime}{|z_i^\prime|} \; .
\end{align}
This way the jets are represented on a \underline{compact
  hypersphere}, just as we already used it for the normalized
autoencoder in Sec.~\ref{sec:auto_ano_norm}.  On this space we can
define the similarity between two jets as
\begin{align}
s(z_i,z_j)=\frac{z_i\cdot z_j}{|z_i||z_j|} \in [-1,1] \; ,
\label{eq:cosine}
\end{align}
which is just the cosine of the angle between the jets in the latent
space.  This similarity is not a proper distance metric, but we could
instead define an angular distance in terms of the cosine, such that
it satisfies the triangle inequality.  Based on this similarity we
construct a \underline{contrastive loss}. It can be understood in
terms of alignment versus uniformity on the unit hypersphere,
illustrated in Fig.~\ref{fig:sphere}.  Starting with negative pairs,
a loss term like
\begin{align}
  \boxed{
  \loss_\text{CLR}
  \supset \sum_{j \neq i \in\text{batch}} \left[ e^{s(z_i,z_j)/\tau} + e^{s(z_i,z_j^\prime)/\tau} \right] }
\end{align}
will be push them apart, preferring $s \to -1$, but on the compact
hypersphere they cannot be pushed infinitely far apart. This means our
loss will be minimal when the unmatched jets are uniformly
distributed.  To map the jets to such a uniform distribution in a
high-dimensional space, the mapping will identify features to
discriminate between them and map them to different points. We have
seen this self-organization effect for the capsule networks in
Sec.~\ref{sec:class_caps}. Next, we want the loss to become minimal
when for the positive pairs all jets and their respective augmented
counterparts are aligned in the same point, $s \to 1$
\begin{align}
  \loss_\text{CLR}
  \supset - \sum_{i\in\text{batch}} s(z_i,z_i^\prime)
\end{align}
This
additional condition induces the invariance with respect to
augmentations and symmetries.  The contrastive loss is given by a
sum of the two corresponding conditions
\begin{align}
  \loss_\text{CLR}
  &= - \sum_{i\in\text{batch}} \frac{s(z_i,z_i^\prime)}{\tau}
  + \sum_{i\in\text{batch}} \log \sum_{j \neq i \in\text{batch}} \left[ e^{s(z_i,z_j)/\tau} + e^{s(z_i,z_j^\prime)/\tau} \right]  \notag \\
  &= - \sum_{i\in\text{batch}} \log \frac{ e^{s(z_i,z_i^\prime)/\tau} }{ \sum_{j \neq i \in\text{batch}} \left[ e^{s(z_i,z_j)/\tau} + e^{s(z_i,z_j^\prime)/\tau} \right]}  \; ,
\label{eq:clloss}
\end{align}
The so-called temperature $\tau>0$ controls the relative influence of
positive and negative pairs.  The first term sums over all positive
pairs and reaches its minimum in the alignment limit.  The negative
pairs contribute to the second term, and the expression in brackets is
summed over all negative-pair partners of a given jet. If such a
solution were possible, the loss would force all individual distances
to their maximum. For the spherical latent space the best the network
can achieve is the smallest average $s$-value for a uniform
distribution.

Next, we can apply contrastive learning to LHC jets, to see if the
network learns an invariant representation and defines some kind of
structure through the uniformity requirement.  Before applying
symmetry transformations and augmentations we start preprocessing the
jets as described in Sec.\ref{sec:class_cnn_arch} and ensure that the
$p_T$-weighted centroid is at the origin in the $\eta - \phi$
plane. Now, rotations around the jet axes are a very efficient
symmetry we can impose on our representations.  We apply them to a
batch of jets by rotating each jet by angles sampled from
$0~...~2\pi$. Such rotations in the $\eta - \phi$ plane are not
Lorentz transformations and do not preserve the jet mass, but for
narrow jets with $R\lesssim1$ the corrections to the jet mass can be
neglected.  As a second symmetry we implement translations in the
$\eta - \phi$ plane. Here, all constituents in a jet are shifted by
the same random distance, where shifts in each direction are limited
to $-1~...~1$.

In addition to (approximate) symmetries, we can also use
\underline{theory-inspired augmentations}. QFT tells us that soft
gluon radiation is universal and factorizes from the hard physics in
the jet splittings\index{QCD splittings}.  To encode this invariance in the latent
representation, we augment our jets by smearing the positions of the
soft constituents, in $\eta$ and $\phi$ using from a Gaussian
distribution centered on the original coordinates
\begin{align}
  \eta^\prime \sim \normal \left( \eta, \frac{ \Lambda_\text{soft}}{ p_T } \right)
  \qquad \text{and} \qquad
  \phi^\prime \sim \normal \left( \phi, \frac{ \Lambda_\text{soft}}{ p_T } \right) \; ,
\end{align}
with a $p_T$-suppression in the variance relative to
$\Lambda_\text{soft}=100$~MeV.  Secondly, collinear splittings lead to
divergences in perturbative QFT. In practice. they are removed through
the finite angular resolution of a detector, which cannot resolve two
constituents with $p_{T,a}$ and $p_{T,b}$ at vanishing $\Delta
R_{ab}\ll1$. We introduce collinear augmentations by splitting
individual constituents such that the total $p_T$ in an infinitesimal
region of the detector is unchanged,
\begin{align}
  p_T \to p_{T,a}+p_{T,b}
  \qquad \text{with} \qquad 
  \eta_a = \eta_b = \eta \qquad 
  \phi_a = \phi_b = \phi \; .
\end{align}
These soft and collinear augmentations will enforce a learned
IR-safety in the jet representation, unlike the modified versions of
the transformer or the EFPs.

Finally, we include the \underline{permutation symmetry} among the
constituents, for instance through a transformer-encoder. The
combination of contrastive loss and a permutation-invariant network
architecture defines the JetCLR approach.

\begin{figure}[t]
  \centering
  \includegraphics[width=0.40\textwidth]{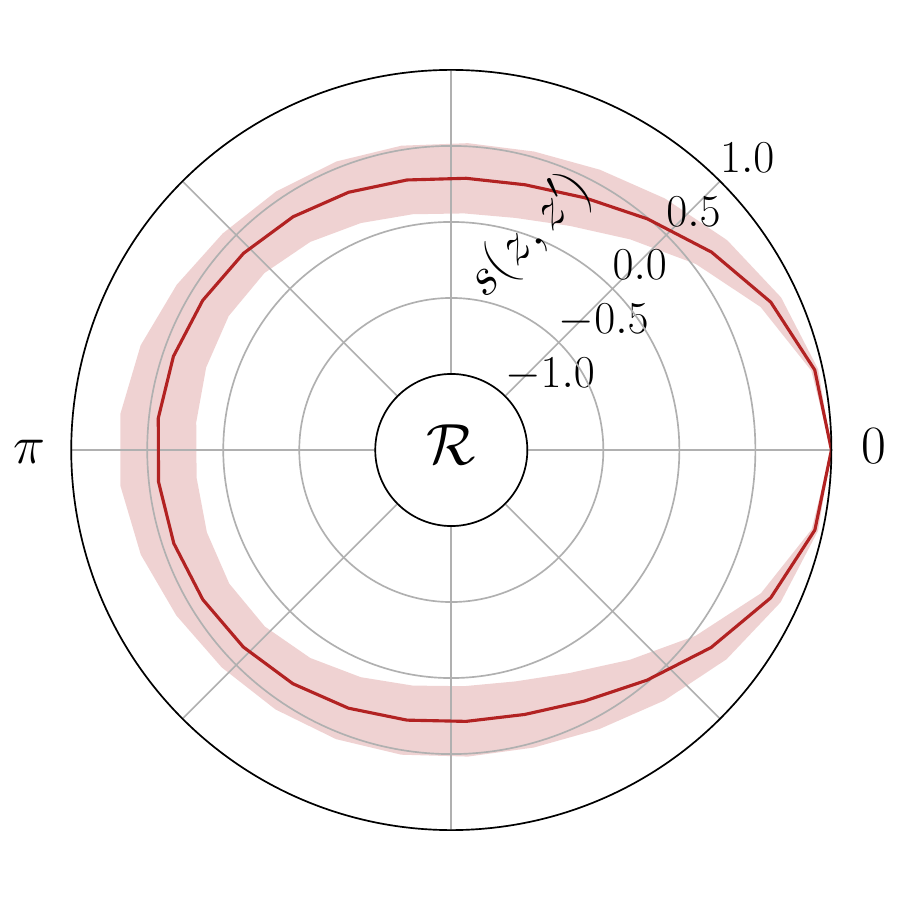}
  \hspace*{0.05\textwidth}
  \includegraphics[width=0.40\textwidth]{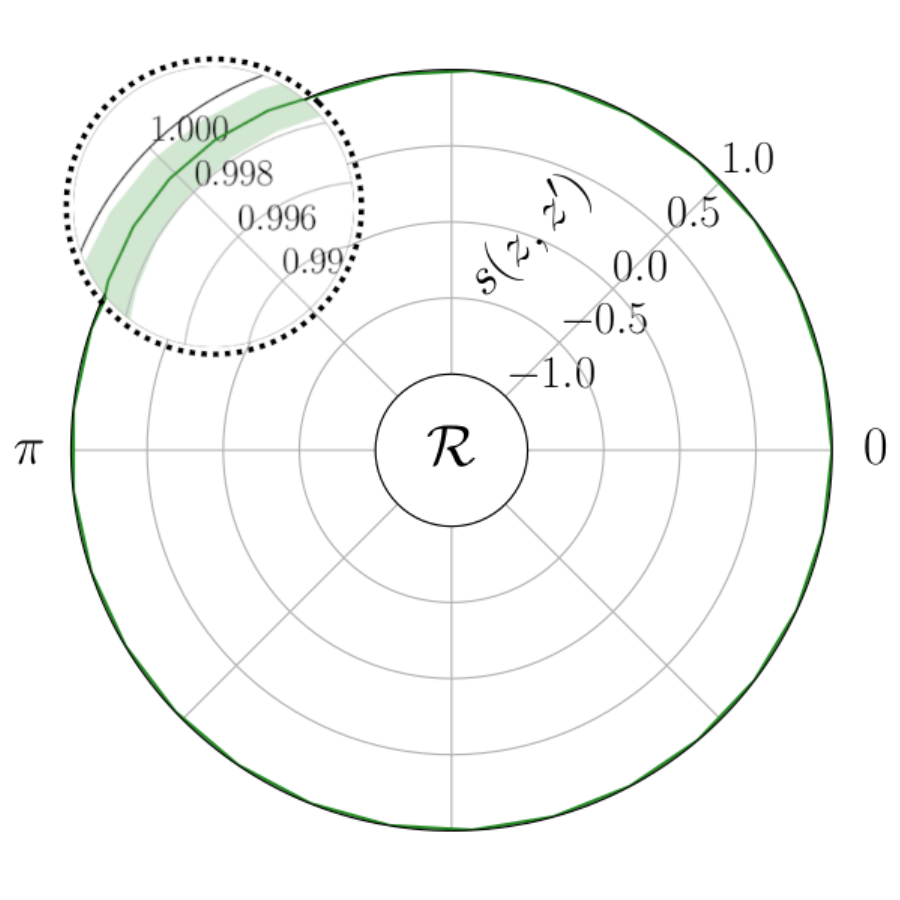}
  \caption{Visualization of the rotational invariance in
    representation space, where $s(z,z') = 1$ indicates identical
    representations. We compare JetCLR representations trained without
    (left) and with (right) rotational transformations. Figure from
    Ref.~\cite{Dillon:2021gag}.}
  \label{fig:rot-radial}
\end{figure}

\begin{table}[b!]
\centering
\begin{small}\begin{tabular}{l|cc}
\toprule
Augmentation 					& $\epsilon^{-1}(\epsilon_s\!=\!0.5)$  		& AUC					  		\\
 \midrule
 none 						& 15 								& 0.905 						\\
 translations					& 19 								& 0.916						\\
 rotations						& 21 								& 0.930						\\
 soft+collinear					& 89 								& 0.970						\\
 combined 					& 181							& 0.979						\\
\bottomrule
\end{tabular} \end{small}
\caption{JetCLR classification results for different symmetries and
  augmentations and $S/B\!=\!1$. The combined setup includes
  translation and rotation symmetries, combined with soft and
  collinear augmentations. Table from Ref.~\cite{Dillon:2021gag}.}
\label{tab:aug}
\end{table}

For a test sample of jets we can check if our JetCLR network indeed
encodes symmetries.  To illustrate the encoded rotation symmetry we
show how the representation is invariant to actual rotations of jets.
We start with a batch of 100 jets, and produce a set of rotated copies
for each jet, with rotation angles evenly spaced in $0~...~2\pi$.  We
then pass each jet and its rotated copy through the JetCLR network,
and calculate their similarity in the latent representation,
Eq.\eqref{eq:cosine}.  In Fig.~\ref{fig:rot-radial} we show the mean
and standard deviation of the similarity as a function of the rotation
angle without and with the rotational symmetry included in the JetCLR
training. In the left panel the similarity varies between $0.5$ and
$1.0$ as a function of the rotation angle, while in the right panel
the JetCLR representation is indeed rotationally invariant.  From the
scale of the radial axis $s(z,z')$ we see that the representations
obtained by training JetCLR with rotations are very similar to the
original jets.

Before using the JetCLR construction for an explicit task, we can can
analyze the effect of the different symmetries and theory
augmentations using a linear classifier test (LCT). For this test we
train a linear neural network with a binary cross-entropy loss to
distinguish top and QCD jets, while our JetCLR training does not know
about these labels. This means the LCT tells us if the uniformity
condition has encoded some kind of feature which we assume to be
correlated with the difference between QCD and top jets. A high AUC
from the LCT points to a well-structured latent representation. From
first principles, it is not clear which symmetries and augmentations
work best for learning representations.  In Tab.~\ref{tab:aug} we
summarize the results after applying rotational and translational
symmetry transformations and soft+collinear augmentations.  It turns
our that, individually, the soft+collinear augmentation works best.
Translations and rotations are less powerful individually, but the
combination defines by far the best-ordered representations.

\subsection{Transfer learning and foundation models}
\label{sec:gen_rl_fm}

In the right panel of Fig.~\ref{fig:lgatr_amplitudes} we have seen
that one way to improve, especially, the performance of transformers
is fine-tuning. The reason for this improvement is that the training
dataset for top taggers introduced in Sec.~\ref{sec:class_cnn_sample}
includes only 2M training jets. For CNNs or graph networks this is
sufficient, but for transformers the size of the training sample
becomes a limiting factor. The solution is \underline{pre-training} a
top tagger on a larger dataset, for instance the JetClass dataset with
100M jets from many partons (including tops), and then fine-tune the
transformers on the actual top tagging dataset. If the task for the
pre-training and the fine-tuning are sufficiently different, this
procedure is called \underline{transfer learning}.  The natural
question is, how specific does pre-training have to be and how far
away from the original dataset or network task does it still help us.
If we can generalize pre-training to many LHC applications, such a
pre-trained network will serve as a \underline{foundation model}.

Foundation models outside LHC physics are usually trained using
self-supervision, for instance masking data and learn to fill in the
blanks. This allows the network to represent the underlying structure
of the data, for instance the underlying symmetries.  If we assume
that we can in principle train a perfect networks in terms of
expressivity and accuracy, the goal of a foundation model is to make
the training on a limited dataset more efficient. This goal is defined
for given task, like amplitude regression, jet or event
classification, anomaly searches, generation, or conditional
genetation. If this is our goal, a foundation model for LHC can be
trained on supervised tasks. In this section we will look at a
foundation model for many jet physics tasks at the LHC.

\begin{figure}[b!]
    \centering
        \includegraphics[width=.99\textwidth]{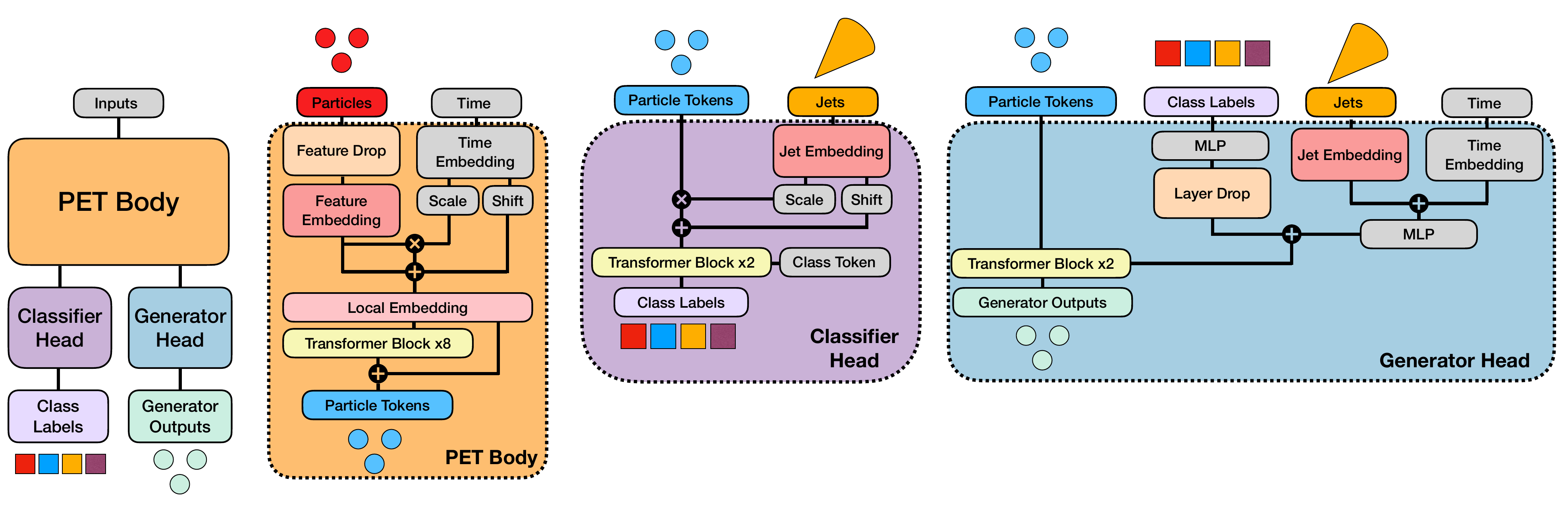}
    \caption{Architecture of the \textsc{OmniLearn} foundation
      model. The global structure is shown in the left, the individual
      blocks in the right. Figure from Ref.~\cite{Mikuni:2025tar}.}
    \label{fig:pet}
\end{figure}

The dual architecture of the foundation model is illustrated in
Fig.~\ref{fig:pet}. The input to the central \textsc{PET} body are the
particles in the jet, as know from Sec.~\ref{sec:class_graph_trans},
conditioned on time parameter in the generative diffusion network. The
time information is encoded to a higher dimensional space using a time
embedding layer.  Contrary to other diffusion networks, the time
embedding includes a multiplication of the output of the Fourier
features by the time parameter, such that the output of the time
embedding is zero when the input time is also zero. This way the time
embedding is turned off when the network is evaluated in classifier
mode.  Because not all jet datasets include particle identification
(PID), the network training uses feature drop, where with a given
probability the PID information is replaced by zeros. This encourages
the network to learn a useful representation both in the presence and
absence of these features.  The outputs of the feature embedding are
combined with the time information though a shift and scaling
operation. Before the transformer block the network uses a positional
token for the geometrical information of the neighborhood surrounding
each particle. Even though transformers can learn general correlations
between particles, local information for instance in terms of the
dynamic graph convolution and its edges introduced in
Sec.~\ref{sec:class_graph} often improves performance.  The particle
embedding equipped with local information is then passed through
multiple transformer blocks introduced in
Sec.~\ref{sec:class_graph_trans}.

The classification of jet types takes as inputs the particles from the
the \textsc{PET} body, supplemented with high-level kinematic
observables like the jet mass, transverse momentum, pseudorapidity,
and particle multiplicity.  It helps the network converge faster when
evaluated over datasets covering different fiducial regions than the
ones used during training. The jet information is embedded in a
higher-dimensional space using a jet embedding layer.  A trainable
class token is then used to summarize the information of the particle
embeddings before the classification output. It is essentially
interpreted as an additional particle, concatenated to the true
particle inputs. Inside the transformer block, the outputs of the
\textsc{PET} body are not updated but only the class token is allowed
to change at the end of each transformer block.

The generator head also takes as inputs the particle embeddings and
the kinematic kinematics. Additionally, it uses the time information
and the set of class labels to condition the generator over the jet
types to be simulated. The time and jet information are embedded in a
higher dimensional space using the same encoding blocks as in the
\textsc{PET} body and the classifier head.  The outputs are passed
through a layer drop operation, because when the network is used during
downstream tasks, the classes used to condition the \textsc{PET}
architecture are hardly going to be the same as the ones used during
training. Randomly ignoring the class labels encourages the entire
architecture to learn both a general and specialized representation.
The results of the layer drop operation are added to the outputs of
the combined jet and time embeddings, to serve as a diffusion token.
Similarly to the classification token, it summarizes the particle
embedding information inside the transformer block. However, while the
classification token is interpreted as an additional particle, the
diffusion token is a conditional shift of the particle embeddings
produced by the \textsc{PET} body. Initially, all particles are
simultaneously shifted by the diffusion token created from the
combined class labels, time, and jet information. The diffusion tokens
are then updated with every transformer layer. The diffusion
prediction is then the sum of the original \textsc{PET} body outputs
with the learned diffusion tokens.The \textsc{PET} body model has 1.3M
trainable weights, while the classifier and generator heads have 268k
and 416k trainable parameters, respectively.
 
\begin{table}[b!]
    \centering
    \begin{small}
    \begin{tabular}{lccccc}
    \noalign{\smallskip}\hline
          &  Acc &AUC & \multicolumn{2}{c}{1/$\epsilon_B$} \\
          \cline{4-5}
          & & & $\epsilon_S = 0.5$ & $\epsilon_S = 0.3$ \\
            \hline
            P-CNN & 0.930 & 0.9803 & 201 $\pm$ 4 & 759 $\pm$ 24 \\
            PFN & - & 0.9819 & 247 $\pm$ 3 & 888 $\pm$ 17 \\
            ParticleNet & 0.940 & 0.9858 & 397 $\pm$ 7 & 1615 $\pm$ 93\\
            ParT & 0.940 & 0.9858 & 413 $\pm$ 16 &  1602 $\pm$ 81 \\
            \hline
            LorentzNet & 0.942 & 0.9868 & 498 $\pm$ 18 & 2195 $\pm$ 173 \\
            PELICAN & 0.9425 & 0.9869 & - & 2289 $\pm$ 204  \\
            L-GATr & 0.9423 & 0.9870 & 540 $\pm$ 20 & 2240 $\pm$ 70 \\
            \hline
            ParticleNet-f.t. & 0.942 & 0.9866 & 487 $\pm$ 9 & 1771 $\pm$ 80 \\ 
            ParT-f.t. & 0.944 &  0.9877 & 691 $\pm$ 15 & 2766 $\pm$ 130 \\
            L-GATr-f.t. & 0.9446 & 0.98793 & 651 $\pm$ 11 & 2894 $\pm$ 84 \\
            \hline
            \textsc{PET} Classifier & 0.938 & 0.9848 & 340 $\pm$ 12 & 1318 $\pm$ 39\\
            \textsc{OmniLearn} & 0.942 & 0.9872 & 568 $\pm$ 9 & 2647 $\pm$ 192 \\
        \hline
	\noalign{\smallskip}
	\end{tabular}
        \end{small}
    \caption{Performance of the \textsc{OmniLearn} foundation model
      for top tagging. The uncertainty quoted
      corresponds to the standard deviation of five trainings with
      different random weight initialization. If the uncertainty is
      not quoted then the variation is negligible compared to the
      expected value. Table from Ref.\cite{Mikuni:2025tar}, with
      information added from Tab.~\ref{tab:lgatr_toptagging}.}
    \label{tab:omnilearn_top}
\end{table}

The loss function of the foundation model
combines the classification and generation tasks,
\begin{align}
  \loss &= \loss_\text{class} + \loss_\text{gen} + \loss_\text{class smear} \notag \\
        &= \loss_\text{class}(y,y_\theta) +  \left[ v_\theta - v \right]^2 + \alpha^2(t)\loss_\text{class} (y,y_\theta;t) \; .
\label{eq:loss_fm}
\end{align}
We denote the input data as $x$. For the supervised \textsc{PET}
classifier we use the standard cross entropy loss from
Eq.\eqref{eq:class_loss} for the classification output
$y_{\text{pred}}$ and true labels $y$. The loss for the \textsc{PET}
is the usual learned velocity which we know from the CFM in
Eq.\eqref{eq:CFM_loss}. The third contribution to the loss comes from
an additional classification along the CFM trajectories parametrized
by $t$. From Eq.\eqref{eq:fm_limits} we know the probability
distributions at the two endpoints, $\pd$ for $t=0$ and $\pl$ for $t =
1$. We chose a linear trajectory in Eq.\eqref{eq:conditional_path},
while for \textsc{PET} the general trajectories are written as
\begin{align}
  x(t) = \alpha(t) x + \sigma(t) r \; .
\end{align}
The weight in front of the $t$-dependent classifier ensures that for
$t=0$ or $\alpha(0) = 1$ and $\sigma(0) = 0$ the classifier uses the
unperturbed data. For $t=1$, with $\alpha(1) = 0$ and $\sigma(1) = 1$,
the pure noise data cannot have any impact on the classifier training.
The foundation model is trained on the same 100M jets as the
fine-tuned transformers in Sec.~\ref{sec:gen_rl_eq}.  Up to 150
particles are saved per jet to be used during training.  For the
fine-tuning the learning rate of the \textsc{PET} body is reduced by a
factor 10 compared to the rest of the \textsc{PET} architecture.

From the nine applications of the \textsc{OmniLearn} foundation model
mentioned in the original paper, we only look at two. The first is the
usual top tagging task. The results are shown in
Tab.~\ref{tab:omnilearn_top}, and they are directly comparable to the
fine-tuning results for standard and equivariant transformers in
Tab.~\ref{tab:lgatr_toptagging}. While the \textsc{PET} classifier is
not meant to be competitive with the best-performing jet classifiers,
the foundation model trained on 100M jets and then fine-tuned on the
top tagging dataset performs at a level consistent with the best
fine-tuned transformers. Given that the datasets and the transformer
architectures are the same, this is not surprising. The question is
how this performance generalizes to different jet physics tasks.

\begin{figure}[t]
        \includegraphics[width=.49\textwidth]{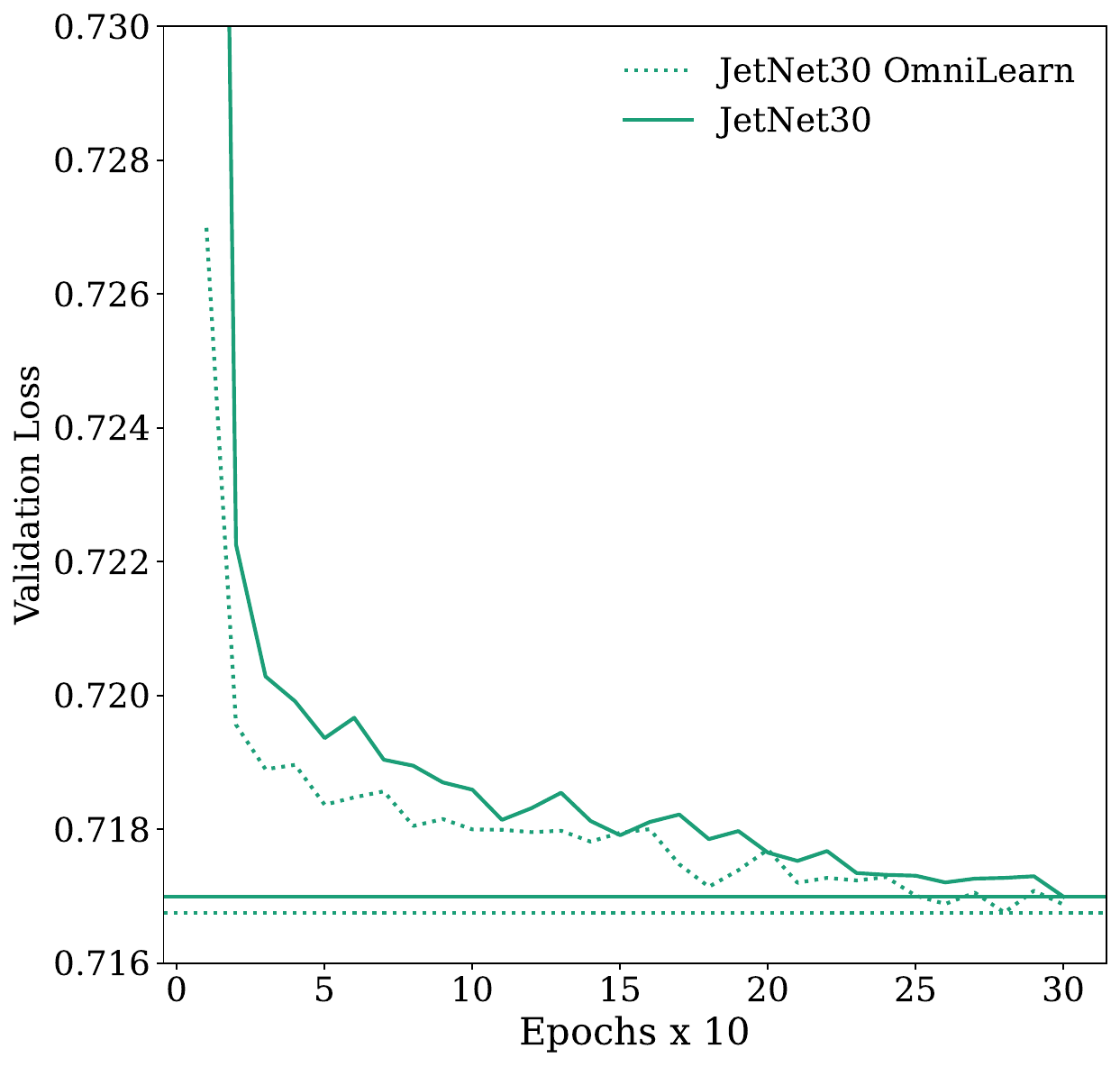}
        \includegraphics[width=.49\textwidth]{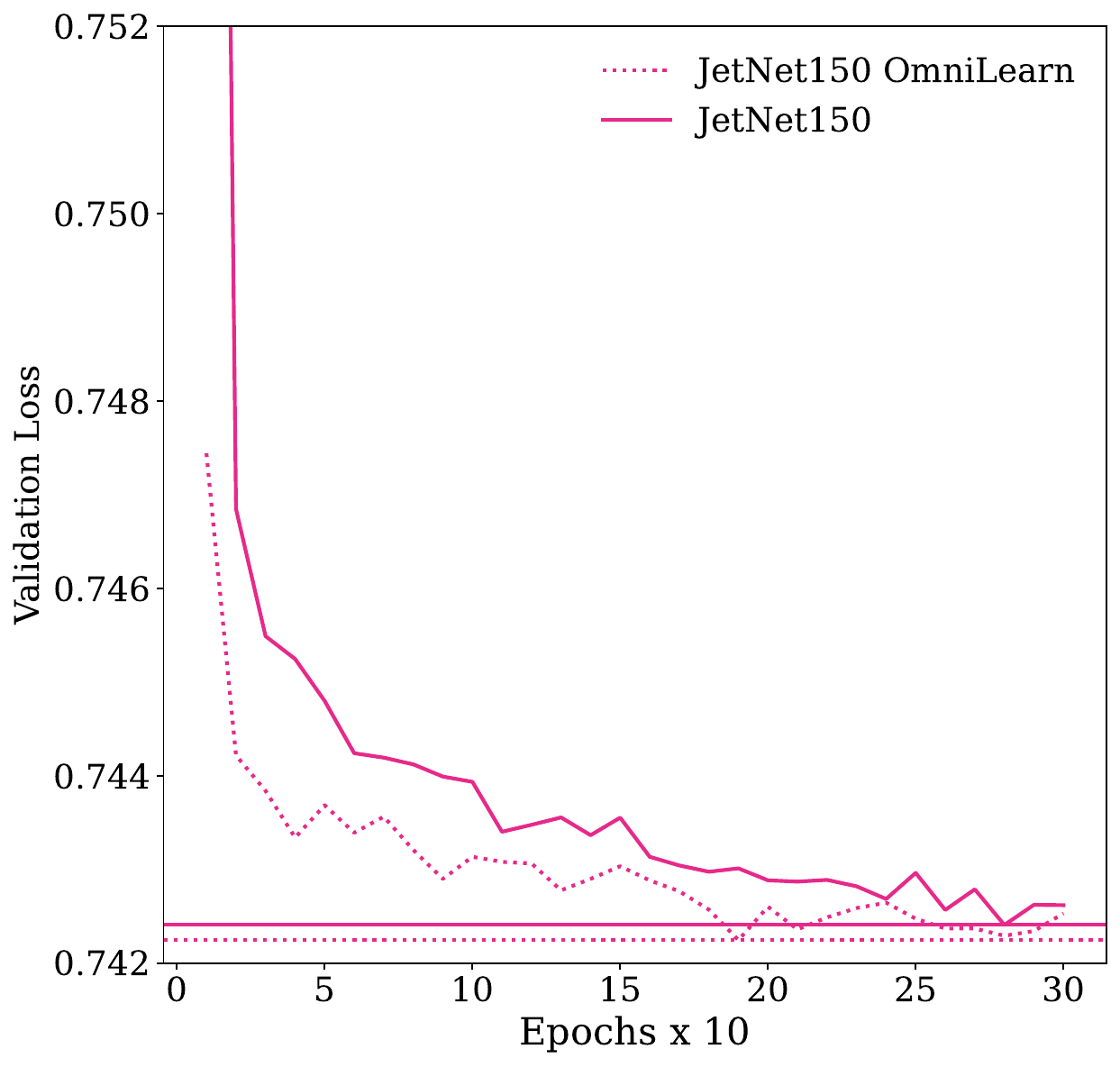}
    \caption{Validation loss curves obtained in the JetNet dataset
      with 30 (left) and 150 (right) particles. The \textsc{OmniLearn}
      validation loss is compared with the \textsc{PET} classifier
      trained from scratch. Figure from Ref.\cite{Mikuni:2025tar}.}
    \label{fig:loss_jetnet}
\end{figure}

As a second task, \textsc{OmniLearn} can be evaluated, for instance,
on precision generation of very hard jets, with $p_T \sim 1$~TeV and
consisting of up to either up to 30 or up to 150 particles.  For each
jet, the four-momentum of the jet is provided together with the
particle multiplicity. One network learns to generate the jet
kinematics, and the diffusion network of the foundation model uses
this information as a condition to generate the set of particle
4-momenta.  In Fig.~\ref{fig:loss_jetnet}, the validation loss of the
\textsc{PET} generator and the \textsc{OmniLearn} training shows how
the foundation model starts from a smaller loss and converges quicker,
requiring around 30$\%$ fewer epochs than a generator trained from
scratch.

\begin{table}[t!]
\centering
\begin{small}
\begin{tabular}{cl|ccccccccccc}
\hline
Parton origin & Model & $W^\mathrm{PM}$ ($\times 10^{-3}$) & $W^\mathrm{P}$ ($\times 10^{-3}$) & $W^\mathrm{PEFP}$ ($\times 10^{-5}$) & Coverage $\uparrow$ & MMD \\
\hline
          & FPCD & 0.44 $\pm$ 0.11 & 0.28 $\pm$ 0.05 & 0.91 $\pm$ 0.16 & 0.56 & 0.03 \\
          & \textsc{EPiC-GAN} & 0.4 $\pm$ 0.1 & 3.2 $\pm$ 0.2 & 1.1 $\pm$ 0.7 & -& -\\
 Gluon    & \textsc{PET} generator & 0.32 $\pm$ 0.09 & 0.34 $\pm$ 0.07 & 1.19 $\pm$ 0.26 & 0.56 & 0.02 \\
          & \textsc{OmniLearn} & 0.47 $\pm$ 0.18 & 0.31 $\pm$ 0.11 & 1.05 $\pm$ 0.23 & 0.56 & 0.02 \\
\hline
          & FPCD &  0.46 $\pm$ 0.05 & 0.24 $\pm$ 0.02 & 0.43 $\pm$ 0.09 & 0.54 & 0.02 \\
          & \textsc{EPiC-GAN} & 0.4 $\pm$ 0.1 & 3.9 $\pm$ 0.3 & 0.7 $\pm$ 0.4 & -& -\\
Light Quark & \textsc{PET} generator & 0.41 $\pm$ 0.04 & 0.34 $\pm$ 0.08 & 0.74 $\pm$ 0.18 & 0.55 & 0.02 \\
          & \textsc{OmniLearn} & 0.46 $\pm$ 0.13 & 0.39 $\pm$ 0.11 & 0.54 $\pm$ 0.14 & 0.53 & 0.02 \\
\hline
          & FPCD & 0.40 $\pm$ 0.07 & 0.30 $\pm$ 0.03 & 2.23 $\pm$ 0.16 & 0.58 & 0.05 \\
          & \textsc{EPiC-GAN} & 0.6 $\pm$ 0.1 & 3.7 $\pm$ 0.3 & 2.8 $\pm$ 0.7 & -& -\\
 Top Quark& \textsc{PET} generator & 0.40 $\pm$ 0.08 & 0.28 $\pm$ 0.08 & 1.81 $\pm$ 0.33 & 0.57 & 0.04 \\
          & \textsc{OmniLearn} & 0.38 $\pm$ 0.05 & 0.30 $\pm$ 0.07 & 1.84 $\pm$ 0.30 & 0.57 & 0.04 \\
          
\hline
          & FPCD & 0.29 $\pm$ 0.02 & 0.23 $\pm$ 0.02 & 0.22 $\pm$ 0.04 & 0.55 & 0.02 \\
 $W$ Boson & \textsc{PET} generator & 0.15 $\pm$ 0.02 & 0.27 $\pm$ 0.07 & 0.12 $\pm$ 0.03 & 0.55 & 0.02 \\
          & \textsc{OmniLearn} & 0.18 $\pm$ 0.01 & 0.27 $\pm$ 0.05 & 0.14 $\pm$ 0.04 & 0.56 & 0.02 \\

\hline
          & FPCD & 0.28 $\pm$ 0.05 & 0.22 $\pm$ 0.03 & 0.23 $\pm$ 0.03 & 0.55 & 0.02 \\
 $Z$ Boson & \textsc{PET} generator & 0.24 $\pm$ 0.06 & 0.35 $\pm$ 0.06 & 0.20 $\pm$ 0.04 & 0.55 & 0.02 \\
          & \textsc{OmniLearn} & 0.19 $\pm$ 0.05 & 0.38 $\pm$ 0.11 & 0.19 $\pm$ 0.03 & 0.56 & 0.02 \\
\hline
\end{tabular}
\end{small}
\caption{Comparison of different generative networks used for jet
  generation with 150 particles.  Lower is better for all metrics
  except for Coverage. Table from Ref.\cite{Mikuni:2025tar}.}
\label{tab:jetnet150}
\end{table}

After training, we can check the performance of the jet generation for
different generating networks. As benchmarks, one can use a point
cloud diffusion network (FPCD) trained from scratch and a taylormade
point-cloud EPiC-GAN. The first three metrics are using the
Wasserstein distance $W$ introduced in Sec.~\ref{sec:gen}. The
Wasserstein distance can be evaluated on the particle momenta $(p_T,
\eta, \phi)$ (P), adding the jet mass (PM), and a set of five leading
coefficients of so-called energy flow polynomials (PEFP). Two more
metrics are based on the closest point cloud in two sets of jets, for
instance constructed by the smallest Wasserstein distance. Once we
have such pairings, we can measure the diversity in the generated
sample through the ratio of generated jets matched to one jet in a
reference dataset. Ideally, this co-called coverage is one. The
quality of the generated jets is based on the distance between the
closest matched jet and the reference jet. It can be expressed as the
miminum matching distance (MMD) defined in Eq.\eqref{eq:mmd}.  We show
the results from the \textsc{PET} generator and the \textsc{OmniLearn}
model in Tab.~\ref{tab:jetnet150}.  In all metrics \textsc{OmniLearn}
shows at least as good performance as the benchmark networks.

Interestingly, for LHC physics tasks the foundation model does not
outperform networks trained from scratch for a given task. The reason
is that all benchmark networks are fully trained on a sufficiently
large training dataset. However, the foundation model is significantly
more efficient in training time. Unlike for instance large language
models such a physics foundation model does not have to especially
large, even a moderately large transformer architecture gives it
sufficient expressivity. The common tasks a foundation model trained
in a supervised manner includes jet classification, jet generation,
likelihood estimation, and anomaly detection.

\clearpage
\section{Inverse problems and inference}
\label{sec:cond}

If we use our \underline{simulation chain} in Fig.~\ref{fig:simchain}
in the forward direction, the typical LHC analysis starts with a new,
theory-inspired hypothesis encoded in a Lagrangian as new particles
and couplings. For every point in the BSM parameter space we simulate
events, compare them to the measured data using likelihood methods,
and discard the BSM physics hypothesis. This approach is inefficient
for a variety of reasons:
\begin{enumerate}
\item BSM physics hypotheses have free model parameters like masses or
  couplings, even if an analysis happens to be independent of them.
  If the predicted event rates follow a simple scaling, like for a
  truncated effective theory, this is simple, but usually we need to
  simulate events for each point in model space.
\item There is a limit in electroweak or QCD precision to which we can
  reasonably include predictions in our standard simulation
  tools. Beyond this limit we can, for instance, only compute a
  limited set of kinematic distributions, which excludes these
  precision prediction from standard analyses.
\item Without a major effort it is impossible for model builders to
  derive competitive limits on a new model by recasting an existing
  analysis, if the analysis requires the full experimental and
  theoretical simulation chains.
\end{enumerate}
These three shortcomings point to the same task: we need to
\underline{invert the simulation chain}, apply this inversion to the
measured data, and compare hypotheses at the level of the hard
scattering. Detector unfolding\index{unfolding} is a known, but non-standard
application. For hadronization and fragmentation an approximate
inversion is standard in that we always apply jet algorithms\index{jet algorithm} to
extract simple parton properties from the complex QCD jets. Removing
QCD jets from the hard process is a standard task in any analysis
using jets for the hard process and leads to nasty combinatorial
backgrounds. Unfolding to parton level is being applied in top
physics, assuming that the top decays are correctly described by the
Standard Model. Finally, unfolding all the way to the parton-level is
called the matrix element method and has been applied to Tevatron
signatures, for example single top production. All of these inverse
simulation tasks have been tackled with classical methods, and we will
see how the can benefit from modern neutral networks.

Inverse problems in particle physics can be illustrated most easily
for the case of detector effects. As a one-dimensional binned toy
example we look at a parton-level distribution $f_{\text{parton},j}$
which gets transformed into $f_{\text{reco},j}$ at detector or
reconstruction level. We can model these detector effects as part of
the forward simulation through a \underline{response matrix} $G_{ij}$,
\begin{align}
  f_{\text{reco},i} = \sum_{j=1}^N G_{ij} \; f_{\text{parton},j} \; .
\label{eq:inv_matrix1}
\end{align}
We postulate the existence of an inversion with 
$\overline{G}$ through 
\begin{align}
  f_{\text{parton},k}
  = \sum_{i=1}^N \overline{G}_{ki} f_{\text{reco},i} 
  = \sum_{j=1}^N \left( \sum_{i=1}^N \overline{G}_{ki} G_{ij} \right) f_{\text{parton},j} 
  \quad \text{with} \quad
  \sum_{i=1}^N \overline{G}_{ki} G_{ij} = \delta_{kj} \; .
\label{eq:inv_matrix2}
\end{align}
If we assume that we know the $N^2$ entries of $G$, this
form gives us the $N^2$ conditions to compute its inverse $\overline{G}$.
We illustrate this one-dimensional binned case with a toy-smearing
matrix
\begin{align}
  G
  &=
  \begin{pmatrix}
  1 - \epsilon & \epsilon & 0 \\ \epsilon & 1-2\epsilon & \epsilon \\ 0 & \epsilon & 1-\epsilon
  \end{pmatrix} \; .
\label{eq:gtoy}
\end{align}
We can assume $\epsilon \ll 1$, but we do not have to.  We look at two input
vectors, keeping in mind that in an unfolding problem we typically
only have one kinematic distribution to determine the inverse matrix
$\overline{G}$,
\begin{alignat}{7}
  f_\text{parton} &= n \begin{pmatrix} 1 \\ 1 \\ 1 \end{pmatrix}
  &\quad &\Rightarrow \quad
  f_\text{reco} = f_\text{parton} \; , \notag \\
  f_\text{parton} &= \begin{pmatrix} 1 \\ n \\ 0 \end{pmatrix}
  &\quad &\Rightarrow \quad
  f_\text{reco} 
  = f_\text{parton} + \epsilon \begin{pmatrix} n-1 \\ -2n + 1 \\ n \end{pmatrix} \; .
\end{alignat}
The first example shows how for a symmetric smearing matrix a flat
distribution removes all information about the detector effects. This
implies that we might end up with a choice of reference
process and phase space such that we cannot extract the
detector effects from the available data. The second example
illustrates that for bin migration from a dominant peak the
information from the original $f_\text{parton}$ gets overwhelmed
easily. We can also compute the inverse of the smearing matrix in
Eq.\eqref{eq:gtoy} and find
\begin{align}
  \overline{G}
  \approx \frac{1}{1-4\epsilon} 
  \begin{pmatrix}
  1-3\epsilon & -\epsilon & \epsilon^2 \\ -\epsilon & 1-2\epsilon & -\epsilon \\ \epsilon^2 & -\epsilon & 1-3\epsilon
  \end{pmatrix} \; ,
\end{align}
where we neglect the sub-leading $\epsilon^2$-terms whenever there is a
linear term as well. The unfolding matrix extends beyond the nearest
neighbor bins, which means that local detector effects lead to a
\underline{global unfolding matrix} and unfolding only works well if
we understand our entire dataset. The reliance on useful kinematic
distributions and the global dependence of the unfolding define the
main challenges once we attempt to unfold the full phase space of an
LHC process.

Current unfolding methods follow essentially this simple approach and
face three challenges.  First, the required binning into histograms is
chosen before the analysis and ad-hoc, so it is not optimal. Second,
the matrix structure behind correlated observables implies that we can
only unfold two or three observables simultaneously.  Finally,
detector unfolding often misses some features which affect the
detector response.


To illustrate how ML-methods enable unfolding, we
define the problem using 
four \underline{unbinned and high-dimensional} phase space densities. For the training we 
always rely on simulated predictions $\psim$, giving us paired 
events at the
parton level $\xp$ and at the reconstruction level $\xr$. 
This forward simulation, for 
example of detector effects, models the probability that a given configuration
$\xp$ turns into $\xr$, it encodes the 
conditional probability $p(\xr|\xp)$.
Classical simulation-based inference then compares forward-simulated and 
measured events over the phase space $\xr$. 
Unfolding will then be applied to the measured $\pd (\xr)$ and 
unfolds the corresponding events to follow $\punf(\xp)$, 
\begin{alignat}{9}
  & \psim(\xp)
  \quad \xleftrightarrow{\text{unfolding inference}} \quad 
  && \punf(\xp)
  \notag \\
  & \hspace*{-9mm} {\scriptstyle p(\xr|\xp)} \Bigg\downarrow
  && \hspace*{+6mm} \Bigg\uparrow {\scriptstyle p(\xp|\xr)}
  \notag \\
  & \psim(\xr) 
  \quad \xleftrightarrow{\text{\; forward inference \;}} \quad 
  && \pd(\xr)
\label{eq:schematic}
\end{alignat}
Because our forward simulation is \underline{statistical}, if not
quantum, we can define an inverse simulation in the same
non-deterministic sense as $p(\xp|\xp)$.  Inference based on unfolded
data works in $\xp$, which means that we only need to unfold, for
instance, the detector once and can then do inference using an event
generator without detector simulation. Because the forward simulation
in Fig.~\ref{fig:simchain} always adds phase space dimensions and
complexity, unfoldig inference should be easier than forward
inference, and depending on the physics of the tested theory
hypothesis it might still be optimal.

If we rely on paired simulated events to define the unfolding, we can
use \underline{Bayes' theorem}\index{Bayes' theorem} to relate the
forward and inverse simulations,
\begin{align}
  p(\xp|\xr)
  = p(\xr|\xp)  \; \frac{\psim(\xp)}{\psim(\xr)} \; .
  \label{eq:bayes_unfold}
\end{align}
In the forward direction, the paired events describe $p(\xr|\xp)$,
where the initial distribution of the events is given by $p(\xp)$.
In the  unfolding direction the paired events give us $p(\xp|\xr)$,
but with a different prior $p(\xr)$. 
If we are lucky, the simulation agrees with data perfectly,
$\psim(\xr) = \pd(\xr)$, so we can transfer unfolding probability from
the simulation to the data blindly.  In reality, this will not be the
case, so we need to apply some kind of correction to remove any
unwanted network dependence or prior, as discussed below.

\subsection{Unfolding by reweighting}
\label{sec:cond_omni}

OmniFold is an ML-based technique that iteratively
unfolds\index{unfolding} detector effects. It works on
high-dimensional phase spaces and does not require histograms or
binning.  Its \underline{reweighting} technique is illustrated in
analogy to Eq.\eqref{eq:schematic},
\begin{alignat}{9}
  & \psim(\xp)
  \quad \xrightarrow{\text{\; classifier weights (3) \;}} \quad 
  && \punf(\xp)
  \notag \\
  & \hspace*{-16mm} \text{\footnotesize pull (2)/push weights(4)} \Bigg\updownarrow
  && 
  \notag \\
  & \psim(\xr) 
  \quad \; \xleftrightarrow{\text{\; classifier weights (1) \;}} \quad 
  && \pd(\xr)
\end{alignat}
It start from
reconstruction-level data and predict the parton-level configuration
for a measured configuration. It avoids a direct mapping from
reconstruction-level events to parton level events and instead employs
an iterative reweighting based on \underline{simulated pairs} of
parton-level and reconstruction-level configurations,
\begin{align}
   ( \xp, \xr) \; .
\end{align}
These simulated events need to be reweighted at the reconstruction
level, to reproduce the measured data. Because of the pairing, this
reweighting can be pushed to the parton level.

As a starting point, we assume that the simulated event pairs have finite
weights, while the data starts with unit weights.
The Omnifold algorithm consists of four steps:
\begin{enumerate}
\item First, it learns weights $w(\xr)$, such that $w(\xr) \psim(\xr) \approx \pd(\xr)$. 
\item It uses the paired simulation to pull the
  weights $w(\xr)$ to the corresponding $w(\xp)$. 
  While $w(\xr)$ is well defined,
  $w(\xp)$ might not be, for instance when one $\xp$ corresponds
  to several $\xr$ with different weights. 
\item If
    $w(\xp) \psim(\xp)$ is a good candidate for $\punf(\xp)$,
    we learn $w'(\xp)$ re-defining 
    $\punf(\xp) = w'(\xp) p(\xp)$, where $p(\xp)$ only
    has unit weights. Formally, this turns the unfolded distribution
    into a function of $\xp$ only.
\item Again, we use the paired simulation to push the improved weights
  from $w'(\xp)$ to $w'(\xr)$.  The change from the $w(\xr)
  \psim(\xr)$ to $w'(\xr) \psim(\xr)$ should be small.
\item Now starting from $w'(\xp) \psim(\xr)$ instead of $\psim(\xr)$ we 
    iterate these steps until they have converged and 
    \begin{align}
        \punf(\xp) = w'_\infty(\xp) \; \psim(\xp) \; .
    \end{align} 
\end{enumerate}
Because the iteration steps do not require additional simulations and work on
fixed, paired datasets, this method is fast and efficient. The ultimate
question is how well it describes correlations between phase space
points.

\begin{figure}[t]
  \centering
  \includegraphics[width=0.45\textwidth]{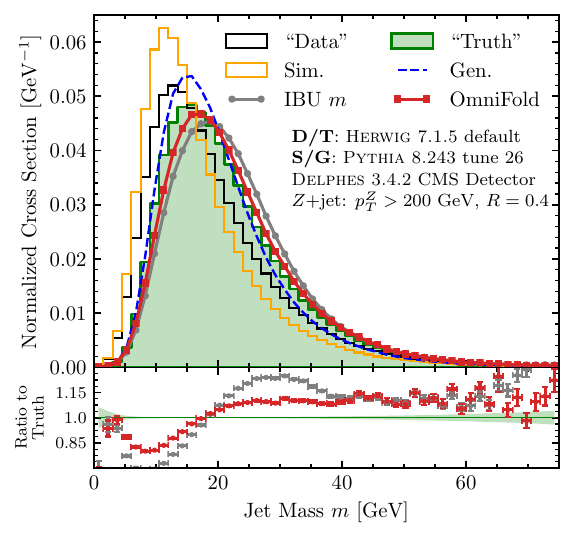}
  \hspace*{0.05\textwidth}
  \includegraphics[width=0.45\textwidth]{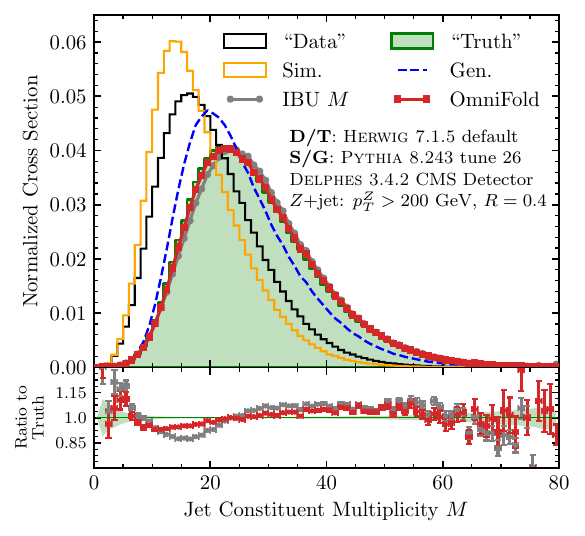}
  \caption{Unfolding results for sample substructure observables,
    using Herwig jets as data and using Pythia for the simulation.
    The lower panels include statistical uncertainties on the
    1-dimensional distributions. The jet multiplicity $M$ corresponds
    to $n_\text{PF}$ defined in Eq.\eqref{eq:qg_obs}. Figure from
    Ref.~\cite{Andreassen:2019cjw}.}
    \label{fig:unfoldmultiple}
\end{figure}

To illustrate the performance of Omnifold, we look at jets and
represent the actual data, $\pd(\xr)$, with the Herwig event
generator\index{event generators}. For the simulation of paired jets, $\psim(\xp)$
and $\psim(\xr)$, we use the Pythia event generator. We
then use Omnifold can unfold the Herwig data to the parton level, and
compare these results with the true Herwig distributions at the parton level
\begin{align}
  \punf(\xp)
  \; \leftrightarrow \; 
  \pd(\xp) \; .
\end{align}

While the event information can be fed into the network in many ways,
the Omnifold authors choose to use the \underline{deep sets}
representation of point clouds as introduced in
Sec.~\ref{sec:class_graph_sets}.  For a quantitative test we show a set
of jet substructure observables, defined in Eq.\eqref{eq:qg_obs}.  The
detector effects, described by the fast Delphes simulation can be seen
in the difference between the generation curves,
$\sim(\xp)$, and the simulation curves,
$\sim(\xr)$. We see that the jet mass and the constituent
multiplicity are significantly reduced by the detector resolution and
thresholds. The data distributions from Herwig represent
$\pd(\xr)$, and we observe a significant difference to the
simulation, where Herwig jets have more constituents than Pythia jets.

If we now use the Pythia-trained Omnifold algorithm to unfold the mock
data we can compare the results to the green truth curves. The ratio
$\punf(\xp)/\pd(\xp)$ is shown in
the lower panels. We see that, as any network, this deviation is
systematic in the bulk and becomes noisy in the kinematic tails. The
same is true for the classical, bin-wise iterative Bayesian unfolding
(IBU), and the unbinned and multi-dimensional Omnifold method clearly
beats the bin-wise method.

\subsection{Conditional generative networks}
\label{sec:cond_inn}

Instead of constructing a backwards mapping for instance from the
detector level to the parton level using classifier reweighting, we
can also tackle this inverse problem with generative networks,
specifically a conditional normalizing flow or cINN. In this section
we will describe two applications of this versatile network
architecture, one for unfolding and one for inference or measuring
model parameters.

\subsubsection{Generative unfolding}
\label{sec:cond_inn_unfold}

\begin{figure}[t]
  \centering
  \includegraphics[width=0.7\textwidth]{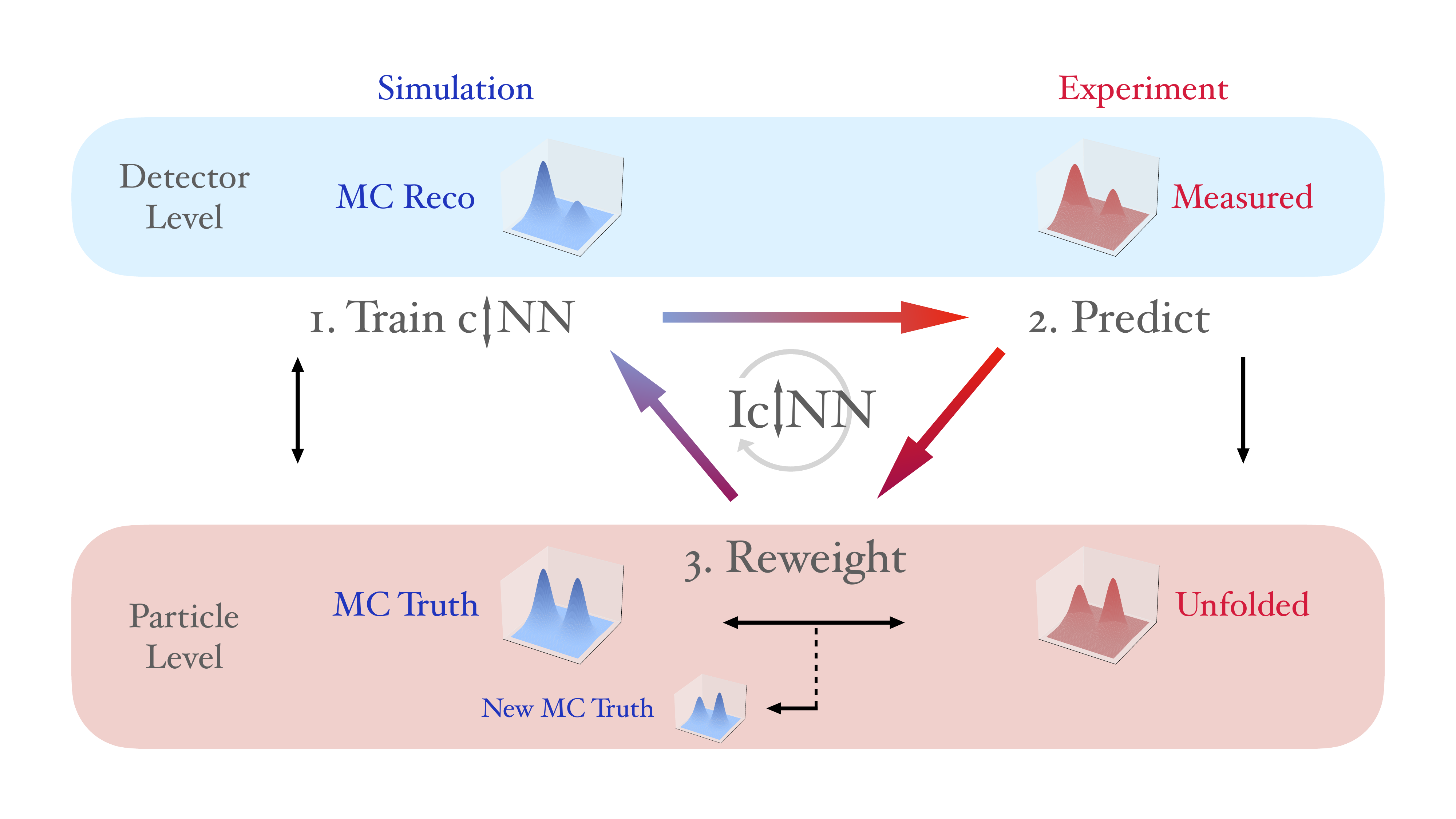}
  \caption{Illustration of the iterative cINN unfolding. In the text,
  we use $\xp$ instead of `Truth'. Figure from
    Ref.~\cite{Backes:2022vmn}.}
    \label{fig:it_cINN}
\end{figure}

The idea of using conditional generative networks for
unfolding\index{unfolding} is to encode the conditional probability
$p(\xp|\xr)$ in a conditional generative network. Using the scheme of
Eq.\eqref{eq:schematic} this means
\begin{alignat}{9}
  & \psim(\xp)
  && \punf(\xp)
  \notag \\
  & \hspace*{-9mm} \text{\footnotesize paired data} \Bigg\updownarrow
  && \hspace*{+6mm} \Bigg\uparrow {\scriptstyle \pmd(\xp|\xr)}
  \notag \\
  & \psim(\xr) 
  \quad \xleftrightarrow{\text{\; correspondence \;}} \quad 
  && \pd(\xr)
\label{eq:unfold_pic_1}
\end{alignat}
Just as Omnifold, the inverse simulation is trained on
pairs of simulated events. 
Technically, we start from a simple latent distribution, with the
generative network transforms into the required phase space
distribution,
\begin{align}
  z \sim \pl(z)
  \quad 
  \longrightarrow
  \quad
  \xp \sim \pmd(\xp|\xr) \; .
  \label{eq:generative_unfolding}
\end{align}
In terms of the conditional probability, the phase space distribution of 
an unfolded dataset can be derived from Bayes' theorem in 
Eq.\eqref{eq:bayes_unfold},
\begin{align}
  p(\xr|\xp) \psim(\xp) &= p(\xp|\xr) \psim(\xr) \notag \\
  \psim(\xp) \equiv 
  \psim(\xp) \int d\xr \; p(\xr|\xp) 
  &= \int d\xr \; p(\xp|\xr) \psim(\xr) \; ,
\end{align}
where we are using the normalization condition of the 
conditional probability in the first argument. If we 
encode the conditional probability into a network  with the 
learned $\pmd$ and apply it to the actual data, the 
unfolded phase space distribution becomes
\begin{align}
  \punf(\xp) = \int d \xr \; \pmd(\xp|\xr) \; \pd(\xr) \; .
\label{eq:unfold_model}
\end{align}
Technically, we can evaluate this integral over $\xr$ by sampling
the learned conditional unfolding probability from $\pd(xr)$ and
ignore the argument $\xr$ of the originally joint distribution. 
This form uses prior distributions for individual events, which
means that we can also take single measured event and unfold them 
any number of times. The problem is that we train the network 
on simulated data, and we do not know how well the simulations
agree with the actual data. This difference cannot be absorbed
into a reweighting, but has to involve re-training the network. 
We will discuss an iterative re-training in Fig.~\ref{fig:it_cINN}.

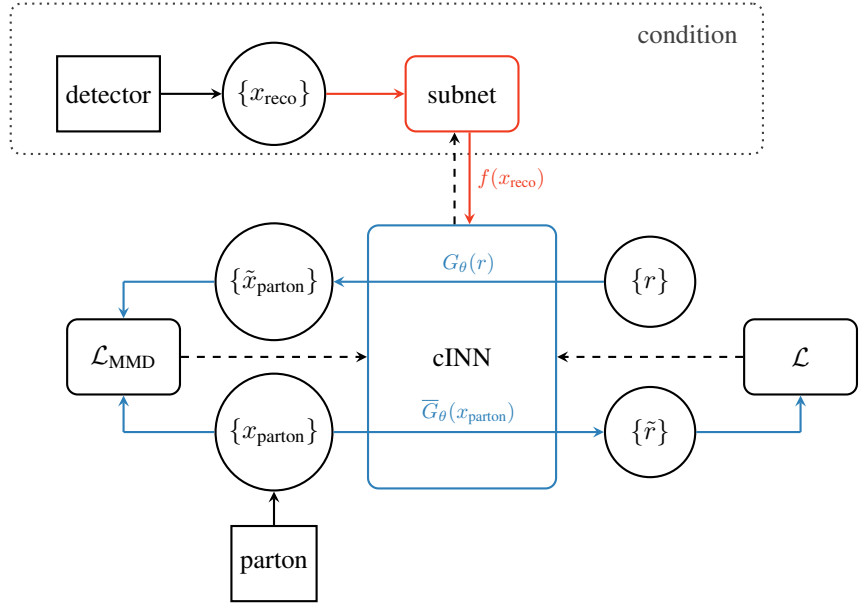
\begin{figure}[t]
\centering

\definecolor{Gcolor}{HTML}{2c7fb8}
\definecolor{Dcolor}{HTML}{f03b20}

\tikzstyle{cINN} = [thick, rectangle, rounded corners, minimum width=2.5cm, minimum height=3.5cm,text centered, draw=Gcolor]
\tikzstyle{preprocessor} = [thick, rectangle, rounded corners, minimum width=1.5cm, minimum height=1cm,text centered, draw=Dcolor]
\tikzstyle{mmd} = [thick, rectangle, rounded corners, minimum width=1.5cm, minimum height=1cm,text centered, draw=black]
\tikzstyle{io} = [thick,circle, minimum width=1.2cm, minimum height=1cm, text centered, draw=black]

\tikzstyle{cond} = [thick, rectangle, dotted, rounded corners, minimum width=10.0cm, minimum height=2cm,text centered, draw=gray!50!black]

\tikzstyle{iodotted} = [thick, circle, minimum width=1.2cm, minimum height=1cm, text centered, draw=black, dotted]

\tikzstyle{process} = [thick, rectangle, minimum width=1cm, minimum height=1cm, text centered, draw=black]

\tikzstyle{xG} = [thick,rectangle, minimum width=2.2cm, minimum height=3cm, text depth= 2.2cm, draw=black]
\tikzstyle{s0} = [thick,rectangle, minimum width=2cm, minimum height=3cm, text centered]
\tikzstyle{s1} = [thick, dotted, rectangle, minimum width=1.6cm, minimum height=1.1cm, text centered, draw=black]

\tikzstyle{decision} = [thick,rectangle, minimum width=1cm, minimum height=1cm, text centered, draw=black]

\tikzstyle{dots} = [circle, minimum size=2pt, inner sep=0pt,outer sep=0pt, draw=Dcolor, fill = Dcolor]

\tikzstyle{arrow} = [thick,->,>=stealth]

\begin{tikzpicture}[node distance=2cm]

\node (cINN) [cINN] {cINN};

\node (xG) [io, left of = cINN, xshift=-0.5cm, yshift=1cm] {$\{\tilde{x}_\text{parton}\}$};
\node (rG) [io, right of = cINN, xshift=0.5cm, yshift=-1cm] {$\{ \tilde{r} \}$};

\node (xp) [io, left of = cINN, xshift=-0.5cm, yshift=-1cm] {$\{x_\text{parton}\}$};
\node (parton) [process, below of=xp, xshift=0cm, yshift=0.25cm] {parton};
\node (mmd) [mmd, left of=xp, xshift=0.0cm, yshift=1cm] {$\loss_\text{MMD}$};

\node (random) [io, right of=cINN, xshift=0.5cm, yshift=1cm] {$\{ r \}$};
\node (gauss) [mmd, right of=rG, xshift=0.0cm, yshift=1cm] {$\loss$};

\node (cond) [cond, above of = xG, xshift=1.5cm, yshift=0.7cm] {};
\node (condi) [above of = xG, xshift=5.5cm, yshift=1.3cm, color=gray!50!black] {condition};

\node (preprocessor) [preprocessor, above of=cINN, xshift=0cm, yshift=1.5cm] {subnet};
\node (xd) [io, left of = preprocessor, xshift=-0.5cm, yshift=0cm] {$\{x_\text{reco}\}$};
\node (detector) [process, left of=xd, xshift=-0.2cm, yshift=0cm] {detector};

\coordinate[ right of = rG, xshift=0cm, yshift=0cm] (Gin1);
\coordinate[ right of = random, xshift=0cm, yshift=0cm] (Gin2);

\coordinate[ left of = xG, xshift=0cm, yshift=0cm] (MMDin1);
\coordinate[ left of = xp, xshift=0cm, yshift=0cm] (MMDin2);


\draw [arrow, color=black] ([yshift=0em]parton.north) -- ([yshift=0em]xp.south);

\draw [thick, color=Gcolor] ([yshift=0em]xp.west) -- ([yshift=0em]MMDin2);
\draw [arrow, color=Gcolor] ([yshift=0em]MMDin2) -- ([yshift=0em]mmd.south);
\draw [thick, color=Gcolor] ([yshift=0em]xG.west) -- ([yshift=0em]MMDin1);
\draw [arrow, color=Gcolor] ([yshift=0em]MMDin1) -- ([yshift=0em]mmd.north);

\draw [thick, color=Gcolor] ([yshift=0em]rG.east) -- ([yshift=0em]Gin1);
\draw [arrow, color=Gcolor] ([yshift=0em]Gin1) -- ([yshift=0em]gauss.south);

\draw [arrow, color=black] ([yshift=0em]detector.east) -- ([yshift=0em]xd.west);
\draw [arrow, color=Dcolor] ([yshift=0em]xd.east) -- ([yshift=0em]preprocessor.west);
\draw [arrow, color=Dcolor] ([yshift=0em, xshift=1mm]preprocessor.south) -- node[scale=0.8, anchor=center, right, color=Dcolor]{$f(x_\text{reco})$} ([yshift=0em,xshift=1mm]cINN.north);
\draw [arrow, dashed, color=black] ([yshift=0em, xshift=-1mm]cINN.north) -- ([yshift=0em, xshift=-1mm]preprocessor.south);

\draw[arrow, thick, color=Gcolor] (random.west) -- node[scale=0.8, sloped, anchor=center, above, color=Gcolor]{$G_\theta(r)$} (xG.east);
\draw[arrow, thick, color=Gcolor] (xp.east) -- node[scale=0.8, sloped, anchor=center, above, color=Gcolor]{$\overline{G}_\theta(x_\text{parton})$} (rG.west);

\draw[arrow, thick, dashed, color=black] (mmd) -- (cINN.west);
\draw[arrow, thick, dashed, color=black] (gauss) -- (cINN.east);

\end{tikzpicture}
\caption{Illustration of the cINN setup. Figure from
  Ref.~\cite{Bellagente:2020piv}.}
\label{fig:cinn}
\end{figure}

A key ingredient to unfolding with generative
networks~\cite{Bellagente:2019uyp} is to train the network with a
likelihood loss~\cite{Bellagente:2020piv} or ensuring that its learns to
generate a probability like for the CFM.
We employ an INN to map an set of random
numbers to a parton-level phase space with the corresponding
dimensionality.  To capture the information from the
reconstruction-level events we condition the INN on such an event.
Trained on a given process the network will now generate probability
distributions for parton-level configurations given a
reconstruction-level event and an unfolding model. The cINN is still
invertible in the sense that it includes a bi-directional training
from Gaussian random numbers to parton-level events and back, but the
invertible nature is not what we use to invert the detector
simulation.  We will eventually need show how the conditional INN
retains a proper statistical notion of the inversion to parton level
phase space.  This avoids a major weakness of standard unfolding
methods, namely that they only work on large enough event samples
condensed to one-dimensional or two-dimensional kinematic
distributions, such as a missing transverse energy distribution
in mono-jet searches or the rapidities and transverse momenta in top
pair production.

The structure of the conditional INN (cINN) is illustrated in
Fig.~\ref{fig:cinn}.  We first preprocess the reconstruction-level
data by a small subnet, $\xr\to f(\xr)$ and omit
this additional step below.  After this preprocessing, the detector
information is passed to the functions $s_i$ and $t_i$ in
Eq.\eqref{eq:layer1}, which now depend on the input, the output, and
on the fixed condition.  The cINN is trained a loss similar to
Eq.\eqref{eq:MLE}, but now maximizing the \underline{probability
  distribution} for the network weights $\theta$, conditional on
$\xp$ and $\xr$ always sampled in pairs
\begin{align}
\boxed{
  \loss_\text{cINN}
  = -  \XLangle \log \pmd(\theta |\xp,\xr)
  \XRangle_{p_\text{parton} \sim p_\text{reco}}
  } \; .
\label{eq:cinn_loss0}
\end{align}
Next, we need to turn the posterior for the network parameters into a
likelihood for the network parameters and evaluate the probability
distribution for the event configurations. Because we sample the
parton-level and reconstruction-level events as pairs, it does not
matter which of the two we consider for the probability. We use Bayes'
theorem to turn the probability for $\theta$ into a likelihood\index{likelihood loss}, with
$\xp$ as the argument,
\begin{align}
  \loss_\text{cINN}
  &= -  \XXLangle \log \frac{\pmd(\xp |\xr, \theta) \pmd(\theta|\xr)}{\pmd(\xp|\xr)}
        \XXRangle_{p_\text{parton} \sim p_\text{reco}} \notag \\
  &= - \XLangle \log \pmd(\xp |\xr,\theta)
             \XRangle_{p_\text{parton} \sim p_\text{reco}}
        - \log \pmd(\theta) + \text{const} \; .
\label{eq:cinn_loss1}
\end{align}
As before, we  ignore all terms irrelevant for
the minimization. The second term is a simple weight regularization\index{regularization},
which we also drop in the following. We now apply the usual coordinate
transformation, defined in Fig.~\ref{fig:cinn}, to introduce the
trainable Jacobian,
\begin{align}
  \loss_\text{cINN}
  &= - \XLangle \log \pmd(\xp |\xr,\theta)
             \XRangle_{p_\text{parton} \sim p_\text{reco}} \notag \\
  &= - \XXLangle \log \pl(\overline{G}_\theta(\xp|\xr))
  + \log \left| \frac{\partial \overline{G}_\theta(\xp|\xr)}{\partial \xp} \right| \XXRangle_{p_\text{parton} \sim p_\text{reco}} \; .
\label{eq:cinn_loss2}
\end{align}
This is the usual likelihood loss, but conditional on
reconstruction-level events and trained on event pairs. As before, we
assume that we want to map the parton-level phase space to Gaussian
random numbers. In that case the first term becomes
\begin{align}
  \log \pl(\overline{G}_\theta(\xp)) = -\frac{|| \overline{G}_\theta(\xp))||_2^2}{2} \; .
\end{align}
We can briefly discuss the symmetry of the problem. If we think of the
forward simulation as generating a set of reconstruction-level events
for a given detector-level event and using Gaussian random numbers for
the sampling, the forward simulation and the unfolding are completely
equivalent. The reason for this symmetry is that in the entire
argument we never consider individual events, but phase space
densities, so our inversion is stochastic and not
deterministic.

\begin{figure}[t]
  \centering
  \includegraphics[page = 9, width=0.455\textwidth]{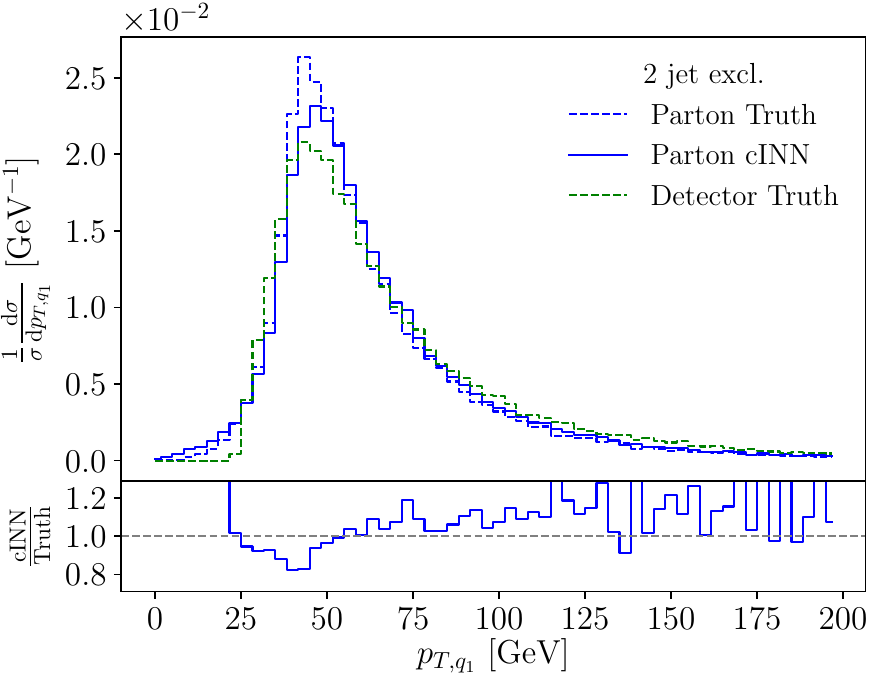} 
  \hspace*{0.05\textwidth}
  \includegraphics[page =10, width=0.455\textwidth]{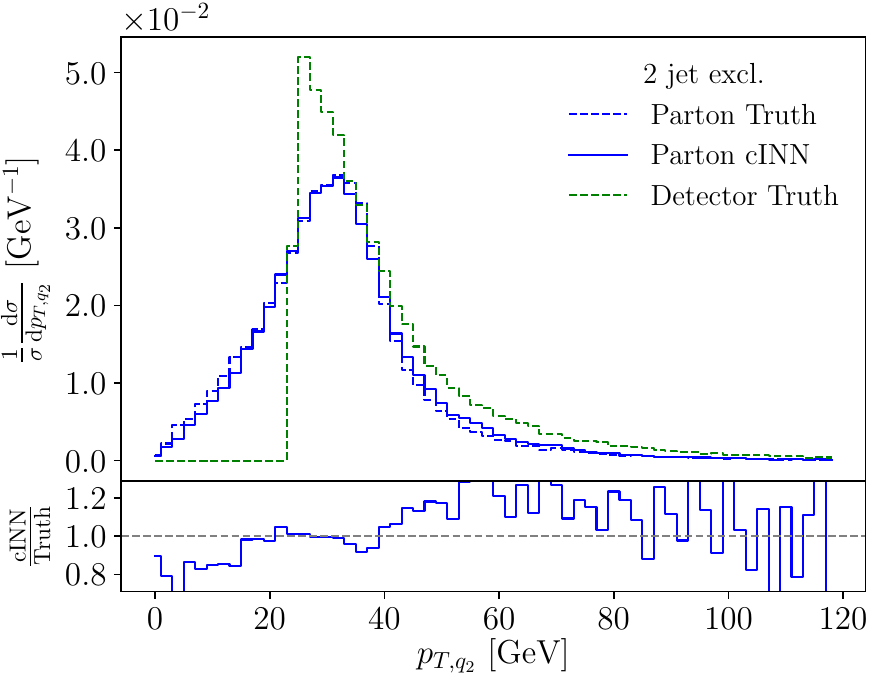} 
\caption{cINNed $p_{T,q}$ distributions.  Training and testing events
  include exactly two jets. The lower panels give the ratio of cINNed
  to parton-level truth. Figure from Ref.~\cite{Bellagente:2020piv}.}
\label{fig:2j}
\end{figure}

To see what we can achieve with cINN unfolding we use the hard
reference process
\begin{align}
q \bar{q} 
\to ZW^\pm
\to (\ell^- \ell^+) \; (j j ) \; .
\end{align}
We could use this partonically defined process to test the cINN
\underline{detector unfolding}. However, we know from
Sec.~\ref{sec:gen_inn_events} that once we include incoming protons we
also need to include QCD jet radiation, so our actual final state is
given by the hadronic process
\begin{align}
pp
\to 
(\ell^- \ell^+) \; (j j ) + \{ 0,1,2 \}~\text{jets} \; .
\label{eq:proc_jets}
\end{align}
All jets are required to pass the basic acceptance cuts
\begin{align}
p_{T, j} > 25~\gev 
\qquad \text{and} \qquad 
|\eta_j| < 2.5 \; .
\label{eq:jetcond}
\end{align}
These cuts regularize the soft and collinear divergences at
fixed-order perturbation theory.

The second task of our cINN unfolding will be to determine which of
the final-state jets come from the $W$-decay and which arise from
initial-state QCD radiation.  Since ISR can lead to harder jets than
the $W$-decay jets, an assignment by $p_{T,j}$ will not solve the
\underline{jet combinatorics}.  This unfolding requires us to define a
specific hard process with a given number of external jets. We can
illustrate this choice using two examples. First, a di-lepton
resonance search typically probes the hard process $pp \to \mu^+
\mu^-+X$, where $X$ denotes any number of additional,
analysis-irrelevant jets. We would invert this measurements to the
partonic process $pp \to \mu^+ \mu^-$. A similar mono-jet analysis
instead probes the process $pp \to Z' j (j) +X$, where $Z'$ is a dark
matter mediator decaying to two invisible dark matter
candidate. Depending on the analysis, the relevant background process
to invert is $pp \to Z_{\nu \nu} j$ or $pp \to Z' jj$, where the
missing transverse momentum recoils against one or two hard
jets. Because our inversion network in trained on matched pairs of
simulated events, we implicitly define the appropriate hard process
when generating the training data.

\begin{figure}[t]
  \centering
  \includegraphics[page = 9, width=0.455\textwidth]{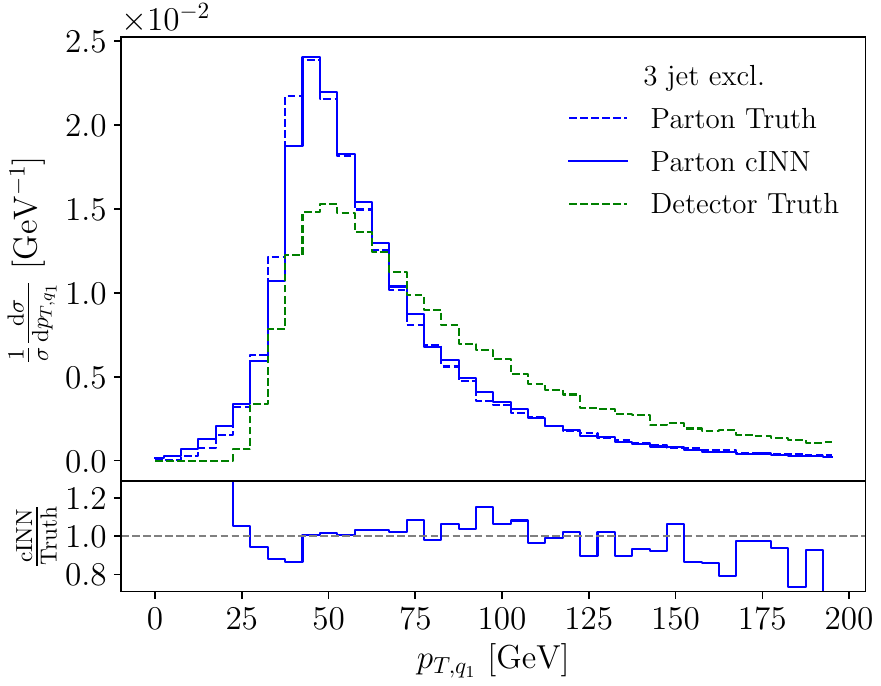}
  \hspace*{0.05\textwidth}
  \includegraphics[page =10, width=0.455\textwidth]{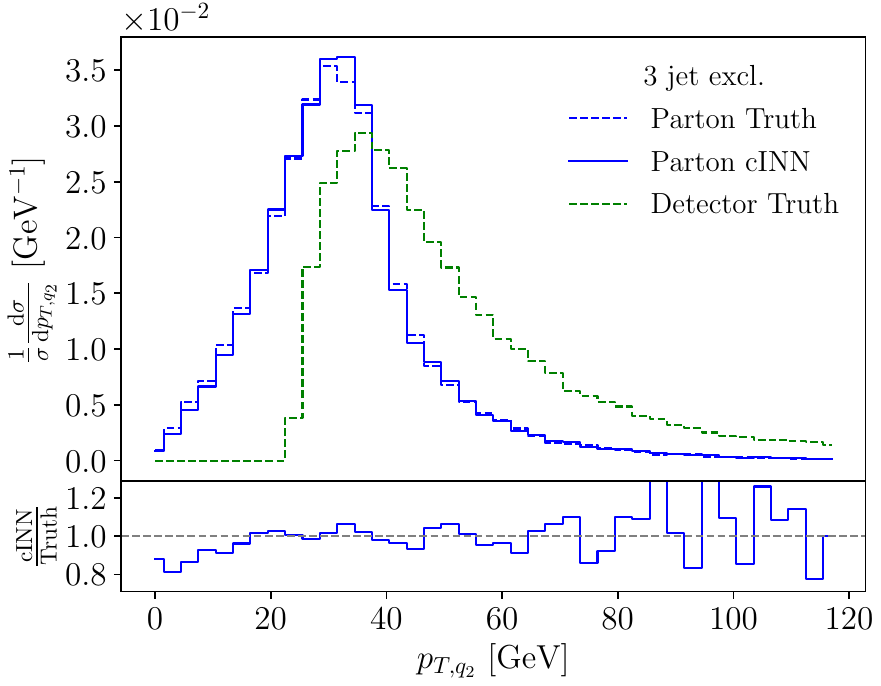} \\
  \includegraphics[page =13, width=0.455\textwidth]{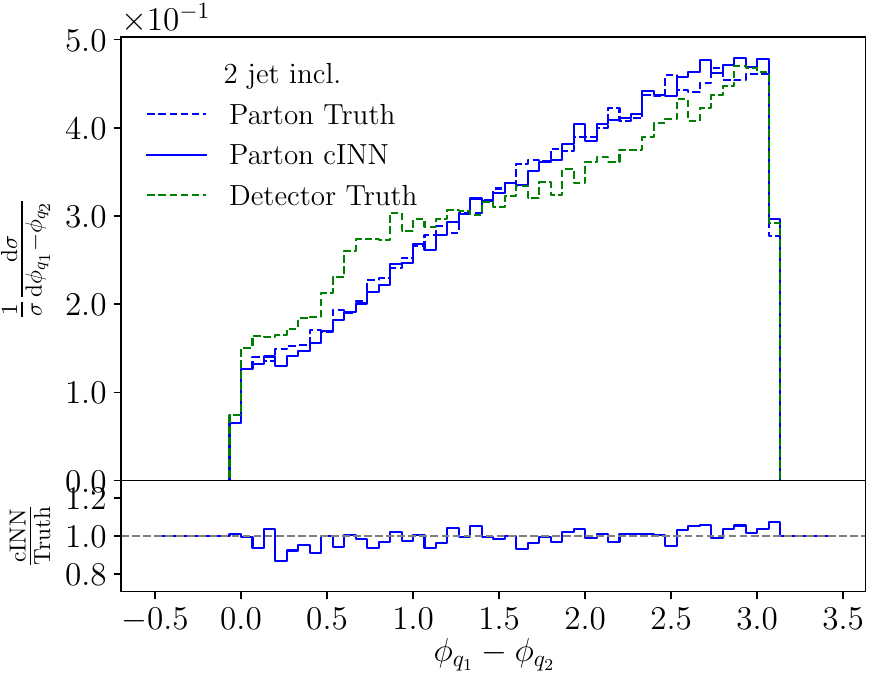}
  \hspace*{0.05\textwidth}
  \includegraphics[page =19, width=0.455\textwidth]{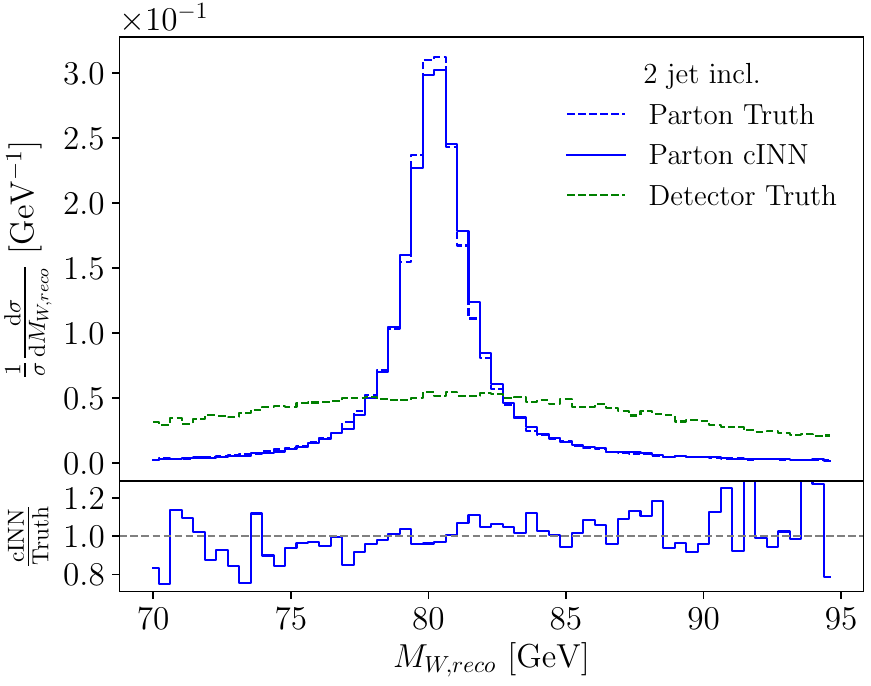}
\caption{cINNed $p_{T,q}$ and $m_{W,\text{reco}}$ distributions.  Training and
  testing events include exactly three (left) and four (right) jets
  from the dataset including ISR. Figure from
  Ref.~\cite{Bellagente:2020piv}.}
\label{fig:allj}
\end{figure}

In Fig.~\ref{fig:2j} we show the unfolding performance of the cINN,
trained and tested on exclusive 2-jet events.  The two jets are mapped
on the two hard quarks from the $W$-decay, ordered by $p_T$. In the
left panel we see that the detector effects on the harder decay jets
are mild and the unfolding is not particularly challenging. On the
right we the effect of the minimum-$p_T$ cut and how the network is
able to remove this effect when unfolding to the decay quarks.  The
crucial new feature of this cINN output is that it provides
probability distribution in parton-level phase space for a given
reconstruction-level event. By definition of the loss function in
Eq.\eqref{eq:cinn_loss1} we can feed a single reconstruction-level
event into the network and obtain a probability distribution over
parton-level phase space for this single event. Obviously, this
guarantees that any kinematic distribution and any correlation between
events is unfolded correctly at the sample level.

Next, we can demonstrate how the cINN unfolds a sample of events with
a variable number of jets from the hard process and from ISR.  First,
we show the unfolding results for events with three jets in the upper
panels of Fig.~\ref{fig:allj}. For three jets in the final state, the
combination of detector effects and ISR has a visible effect on the
kinematics of the leading quark. This softening is correctly reversed
by the unfolding. For the sub-leading quark the problem and the
unfolding performance is similar to the exclusive 2-jet case.  The
situation becomes more interesting when we consider samples with two
to four jets all combined. Now the network has to flexibly resolve the
combinatorial problem to extract the two $W$-decay jets from a mixed
training sample.  In Fig.~\ref{fig:allj} we show a set of unfolded
distributions from a variable jet number. Without showing them, it is
clear that the $p_{T,j}$-threshold at the detector level is corrected,
and the cINN allow for both $p_{T,q}$ values to zero.  Next, we see
that the comparably flat azimuthal angle difference at the parton
level is reproduced to better than 10\% over the entire
range. Finally, the $m_{jj}$ distribution with a additional MMD loss
re-generates the $W$-mass peak at the parton level almost
perfectly. The precision of this unfolding is not any worse than it is
for the INN generator in Sec.~\ref{sec:gen_inn_events}.  This means that
conditional generative networks unfold detector effects and jet
radiation for LHC processes very well, even through their network
architectures are more complex than the classifiers used in
Sec.~\ref{sec:cond_omni}.

As discussed in 
Eq.\eqref{eq:unfold_model}, we need to remove any unwanted model 
dependence from the learned unfolding probability. 
For instance, the leading model
dependence of the forward simulations and the learned unfolding 
probability could be some kind of mass scale $\mu$,
\begin{align}
  \psim(\xp|\mu) 
  &= \int d\xr \; p(\xp|\xr,\mu) \psim(\xr|\mu) \; .
\end{align}
This mass scale can be an unphysical nuisance parameter or an 
observable.
If we just apply the trained unfolding generator to the real data, 
we find
\begin{align}
  \punf(\xp|\mu,\bar{\mu}) 
  = \int d\xr \; \pmd(\xp|\xr,\mu) \pd(\xr|\bar{\mu}) 
    \qquad \text{with} \qquad 
    \mu \ne \bar{\mu} \; .
\end{align}
In this dual dependence of the unfolded distribution, $\bar{\mu}$ 
comes from the actual data, and $\mu$ is the prior from the 
simulated training data. If we cannot avoid an influence by the 
prior, we need to define a re-training algorithm which ensures that 
we can unfold with a probability learned with the correct top mass
$\pmd(\xp|\xr,\bar{\mu})$. One method to achieve that 
is classically 
referred to Bayesian iterative unfolding, and
similar to the Omnifold reweighting strategy described in
Sec.~\ref{sec:cond_omni}.
The iterative IcINN algorithm is illustrated in Fig.~\ref{fig:it_cINN}:
\begin{enumerate}
\item We start with simulated event pairs before and after detector,
  train the cINN on the two $\psim(x|\mu)$ (step~1), and unfold the measured
  data $\pd(\xr|\bar{\mu})$ (step~2). This gives events following
  $\punf(\xp|\mu,\bar{\mu})$.
\item Next, we train a classifier to learn the ratio between the
  $\punf(\xp|\mu,\bar{\mu})$ and $\psim(\xp|\mu)$ and reweight the
  latter (step~3). From the combined argument of $\punf$ we see that this will
  replace $\mu \to \bar{\mu}$ for part of $\psim$, but
  not all, so we will be left with $\psim(\xp|\mu,\bar{\mu})$.
\item Because the training data is paired events, we can transfer
  these weights to $\psim(\xr|\mu,\bar{\mu})$ and use the weighted, paired events to
  re-train the unfolding cINN $\pmd(\xp|\xr,\mu,\bar{\mu})$. 
\item The
  training-unfolding-reweighting steps can be repeated until the
  algorithm converges, $\mu \to \bar{\mu}$, the classifier returns a global value of
  0.5, and our learned unfolding probability only depends on $\bar{\mu}$,
\begin{align}
  \punf(\xp|\bar{\mu}) 
  = \int d\xr \; \pmd(\xp|\xr,\bar{\mu}) \pd(\xr|\bar{\mu}) 
\end{align}
\end{enumerate}
As a side remark, this algorithm only works if at the end of step~1 $\bar{\mu}$
appears in $\punf(\xp|\mu,\bar{\mu})$ and if in step~2 the reweighting is
numerically stable.
Technically, it turns out to be more efficient if the cINN is not
trained from scratch each time. To re-train the cINN on the weighted
paired events the loss function in Eq.\eqref{eq:cinn_loss0} is
supplemented with learned weights
\begin{align}
  \loss_\text{cINN}
  = -  \XLangle w(\xr) \log \pmd(\theta |\xp,\xr)
  \XRangle_{p_\text{parton} \sim p_\text{reco}}
  \; .
\end{align}
This iterative approach ensures that the unfolding network is trained
on events similar to the data we actually unfold, so there is no bias
from a difference between training data and measured data.  This
removes a major systematic uncertainty\index{systematic uncertainty} from unfolded experimental
results and it makes it easier to generalize the unfolding network
from one hard process to another.

\subsubsection{Unfolding top decays}
\label{sec:cond_inn_}

An especially challenging unfolding task is for mass measurements,
where we have to unfolding of strongly peaked kinematic distributions.
An example which appears all through these lecture notes is boosted
hadronic top decays,
\begin{align}
  t \to b W^{+*} \to b u \bar{d} \; .
\end{align}
A decaying boosted top forms a jet, as discussed in
Sec.~\ref{sec:class_cnn_tag}, which to first approximation consists of
three constituent subjets. The fundamental physics parameter which
describes the kinematics and which we want to measure is the top mass
$m_t$. It is approximately given by the peak position of the measured
invariant mass of these three constituent jets
\begin{align}
  m_t \sim 
  M_{jjj}^2 = M_{12}^2 + M_{23}^2 + M_{13}^2 - m_1^2 - m_2^2 - m_3^2 \; .
    \label{eq:trijetmass}
\end{align}
As we will see, the main challenge of top decay unfolding is the model
dependence from the top mass value assumed for the simulated training
data.

The simulated top events are inspired by a CMS analysis using
classical unfolding methods and contain lepton-hadron top pairs
\begin{align}
    pp 
    \to t\bar{t} 
    \to (b q \bar{q}') \; (\bar{b} \ell^- \bar{\nu}) + \text{c.c.}
    \quad \text{with} \quad \ell = e,\mu \; .
\end{align}
Similar to Eq.\eqref{eq:boosted_ttbar} we reconstruct a top jet and
require for the fat jet and the three constituent (sub-)jets with
\begin{align}
 p_{T,J} > 400\;\text{GeV}
 \qqquad \text{and} \qqquad 
 p_{T,j_{1,2,3}} > 30\;\text{GeV} 
 \qquad 
 |\eta_{j_{1,2,3}}| < 2.5 \; ,
\end{align}
As discussed in Sec.~\ref{sec:cond_inn_unfold}, the training events
contain paired gen-level and reco-level information.

The unfolding task is described in Eq.\eqref{eq:unfold_pic_1}, where we
learn the conditional unfolding probability
\begin{align}
  \pmd(\xp|\xr) \approx p(\xp|\xr) 
\end{align}
from the paired training data. We refer to the generated
particle-level phase space $\xp$, even though it does not really have
to be at the parton level.

For the top decays, the unfolding scheme shown in
Eq.\eqref{eq:unfold_pic_1} is missing a critical complication.  Our
simulation assumes a top mass value $m_s$, which we can tune to match
the actual data.  The dependencies of the four phase space
distributions on $m_s$ and its `correct' value in the data, $m_d$,
imply
\begin{alignat}{9}
  & \psim(\xp|m_s)
  && \punf(\xp|m_s,m_d)
  \notag \\
  & \hspace*{-9mm} {\scriptstyle p(\xr|\xp)} \Bigg\downarrow 
  && \hspace*{+6mm} \Bigg\uparrow {\scriptstyle \pmd(\xp|\xr,m_s)}
  \notag \\
  & \psim(\xr|m_s) 
  \quad \xleftrightarrow{\text{\; correspondence \;}} \quad 
  && \pd(\xr|m_d)
  \label{eq:scheme2} 
\end{alignat}
In the forward direction, $p(\xr|\xp)$ does not have an explicit
$m_s$-dependence, but in the inverse direction the simulated datasets
and Bayes' theorem induce such a dependence.  By assumption, $m_s =
m_d$ ensures that the simulated and actual data agree at the
reco-level,
\begin{align}
  \psim(\xr|m_s=m_d) \really \pd(\xr|m_d) \; .
\end{align}
This is the relation we use to infer $m_d$ at the reco-level. 
Alternatively, 
we can do the same inference at the gen-level, requiring
\begin{align}
  \psim(\xp|m_s=m_d) \really \punf(\xp|m_s=m_d,m_d) \; .
\end{align}
The problem with this unfolded inference is the dual dependence of
$\punf(\xp|m_s,m_d)$ through the reco-level data and the learned
conditional probability.  In fact, looking at Eq.\eqref{eq:scheme2}
the situation for top decays is even worse, in that the unfolded
distribution only depends on the prior
\begin{align}
  \punf(\xp|m_s,m_d) \approx \punf(\xp|m_s,\cancel{m_d}) \; .
\end{align}

A mix of physical dependence and prior could be targeted by the
reweighting-based iterative method from Fig.~\ref{fig:it_cINN}.
However, a shift of a peaked distribution immediately leads to large
density ratios between the true and the shifted distributions.  The
iterative method implicitly assumes that $\punf(\xp|m_s,m_d)$ depends
mostly on $m_d$ and at a reduced level on $m_s$, which is not at all
the case for top mass unfolding.  This is why we develop an
alternative, one-step strategy to remove the dominating top mass bias:
\begin{enumerate}
\item First, we 
increase the sensitivity on $m_d$ by 
pre-processing the data and
adding an estimator of $m_d$ 
to the representation of $\xr$. 
We simply use the weighted median of the 3-jet masses in a batch of events at reco-level, 
$M_{jjj}^\text{batch}$,
\begin{align}
 M_{jjj}^\text{batch} \approx m_d \equiv m_t \Bigg|_\text{data} \; .
\end{align}
This batch-wise kinematic information can be extracted at the level of
the loss evaluation, and it goes beyond the usual single-event
information, similar to the MMD loss from
Sec.~\ref{sec:gen_gan_events}.

\item Second, we weaken the bias from the training data by combining
  training data with different top masses, but without an additional
  label,
\begin{align}
m_t = \left\{ 169.5, 172.5, 175.5 \right\}~\text{GeV}
\qqquad \text{(combined training).}
\label{eq:training_masses}
\end{align}
The range has to ensure that top mass in the data is within the
training range, to avoid network extrapolation.
\end{enumerate}

\begin{figure}[t]
    \centering
    \begin{tikzpicture}[node distance=2cm, scale=0.6, every node/.style={transform shape}]

\node (part1) [txt] {$x_{\text{reco},1}$};
\node (part2) [txt, right of=part1, xshift=-0.5cm] {$...$};
\node (part3) [txt, right of=part2, xshift=-0.5cm] {$x_{\text{reco},n}$};

\node (emb_part1) [embed, below of=part1, yshift=-0.3cm, rotate=90]{Emb};
\node (emb_part2) [txt, below of=part2, yshift=-0.3cm] {$...$};
\node (emb_part3) [embed, below of=part3,yshift=-0.3cm,  rotate=90]{Emb};

\node (TE) [transformer, below of=emb_part2, yshift=-0.7cm, text width=4cm,
text depth=1.5cm, align=center] {Transformer-Encoder};
\node (TE_att) [attention, below of=TE, yshift=1.6cm] {Self-Attention \\
Reco-level correlations};

\node (reco1) [txt, right of=part3, xshift=1.75cm] {$x_{\text{gen},1}(t)$};
\node (reco3) [txt, right of=reco1] {$...$};
\node (reco6) [txt, right of=reco3] {$x_{\text{gen},N}(t)$};

\node (t) [txt, right of=reco6, xshift=0.5cm] {$t$};

\node (emb_reco1) [embed, below of=reco1, yshift=-0.3cm, rotate=90]{Emb};
\node (emb_reco3) [txt, below of=reco3, yshift=-0.3cm] {$...$};
\node (emb_reco6) [embed, below of=reco6, yshift=-0.3cm, rotate=90]{Emb};

\node (TD) [transformer, right of=TE, xshift=5.3cm, yshift=-0.8cm, text width=4cm,
text depth=3.1cm, align=center, minimum height=4cm] {Transformer-Decoder};
\node (TD_att) [attention, right of=TE_att, xshift=5.3cm] {Self-Attention \\
Part-level correlations};
\node (TD_crossatt) [attention, below of=TD_att, yshift=0.4cm] {Cross-Attention \\
Combinatorics};

\node (inn1) [small_cinn, below of=reco1, yshift=-8.2cm, rotate=90]{Linear};
\node (inn3) [txt, below of=reco3, yshift=-8.2cm]{$...$};
\node (inn6) [small_cinn, below of=reco6, yshift=-8.2cm, rotate=90]{Linear};


\node (prob1) [txt, below of=inn1, yshift=-0.3cm]{$\Big( v_\theta(c_1, t ),$ };
\node (prob2) [txt, below of=inn3, yshift=-0.3cm]{$...$};
\node (prob6) [txt, below of=inn6, yshift=-0.3cm]{$, \; v_\theta(c_{N}, t ) \Big)$};
\node (prob) [txt, left of=prob1, xshift=-2 cm]{$v_\theta(x_\text{gen}(t), t , x_\text{reco}) = \;$};

\draw [arrow, color=black] (part1.south) -- (emb_part1.east);
\draw [arrow, color=black] (part3.south) -- (emb_part3.east);

\draw [arrow, color=black] (emb_part1.west) -- (TE.north -| emb_part1.west);
\draw [arrow, color=black] (emb_part3.west) -- (TE.north -| emb_part3.west);

\draw [arrow, color=black] (TE.south -| emb_part1.west) --  ([yshift=-1cm]TE.south -| emb_part1.west) -- ([yshift=-1cm]TE.south -| TD.west) ; 
\draw [arrow, color=black] (TE.south -| emb_part3.west) --  ([yshift=-0.7cm]TE.south -| emb_part3.west) -- ([yshift=-0.7cm]TE.south -| TD.west);

\draw [arrow, color=black] ([xshift=-0.2cm]reco1.south) -- ([xshift=-0.2cm]emb_reco1.east);
\draw [arrow, color=black] ([xshift=-0.2cm]reco6.south) -- ([xshift=-0.2cm]emb_reco6.east);

\draw [arrow, color=black] (emb_reco1.west) -- (TD.north -| emb_reco1.west);
\draw [arrow, color=black] (emb_reco6.west) -- (TD.north -| emb_reco6.west);

(A) (B);

\draw [arrow, color=black] ([xshift=-0.2cm]TD.south -| emb_reco1.west)  -- node [text width=1.5cm, pos=0.3, font=\Large, right] {$c_{1}$} ([xshift=-0.2cm]inn1.east -| emb_reco1.west);
\draw [arrow, color=black] ([xshift=-0.2cm]TD.south -| emb_reco6.west)  -- node [text width=1.5cm,pos=0.3, font=\Large, right] {$c_{N}$}  ([xshift=-0.2cm]inn6.east -| emb_reco6.west);

\draw [arrow, color=black] (inn1.west -| emb_reco1.west) -- (prob1.north -| emb_reco1.west);
\draw [arrow, color=black] (inn6.west -| emb_reco6.west) -- (prob6.north -| emb_reco6.west);

\draw [arrow, color=black] (t.south) --  ([yshift=-0.5cm]t.south) -- ([yshift=-0.5cm, xshift=0.2cm]t.south -| reco1.center) -- ([xshift=0.2cm]emb_reco1.east); 
\draw [arrow, color=black] ([yshift=-0.5cm, xshift=0.2cm]t.south -| reco6.center) -- ([xshift=0.2cm]emb_reco6.east); 
\draw [arrow, color=black] (t.south) --  ([yshift=-8.1cm]t.south) -- ([yshift=-8.1cm, xshift=0.2cm]t.south -| reco1.center) -- ([xshift=0.2cm]inn1.east); 
\draw [arrow, color=black] ([yshift=-8.1cm, xshift=0.2cm]t.south -| reco6.center) -- ([xshift=0.2cm]inn6.east); 

\end{tikzpicture}
    \caption{Schematic representation of a parallel Transfusion
      network. Figure from Ref.~\cite{Favaro:2025psi}.}
    \label{fig:transfusion}
\end{figure}

Any kind of density estimation over LHC phase space is local, provided
we apply the correct Minkowski metric, and for instance
forward-simulating or unfolding detector effects do not lead to
multi-mode distributions over the conditional probabities. This is why
normalizing flows, as introduced in Sec.~\ref{sec:gen_inn}, and their
more expressive CFM cousins introduced in Sec.~\ref{sec:gen_diff_cfm}
are so successful. However, when we move from individual 4-momenta to
jets or events, \underline{combinatorics} lead to non-local
effects. These can be described well using transformers, or more
specifically self attention, as introduced in
Sec.~\ref{sec:class_graph_trans}.

Unfolding a set of subjets is a problem of this kind, where we
estimate the joint phase space distribution of a set of 4-vectors, and
the accuracy of the density estimation is the main challenge. This
suggests to combine an INN or CFM generator with transformer modules
one of two ways:
\begin{itemize}
\item First, we can encode the velocity $v_\theta(x,t)$ in
  Eq.\eqref{eq:CFM_loss} in a network that includes attention
  layers. We call such a network \underline{TraCFM}, and it allows the network to
  learn the joint multi-constituent phase space density including a
  combinatorial assignment.
\item Alternatively, we can factorize the sampling into CFMs for
  individual constituents and link these CFMs with a transformer. We
  call this combination \underline{Transfusion}.
\end{itemize}
For the unfolding of detector effects in top decays we can assume that
the conditional probabilities of the individual constituents are
essentially uncorrelated with each other, while the information on the
top mass and the $W$-mass comes from the (correct) combination of the
subjets.

\begin{figure}[t]
    \includegraphics[width=0.495\textwidth, page =5] {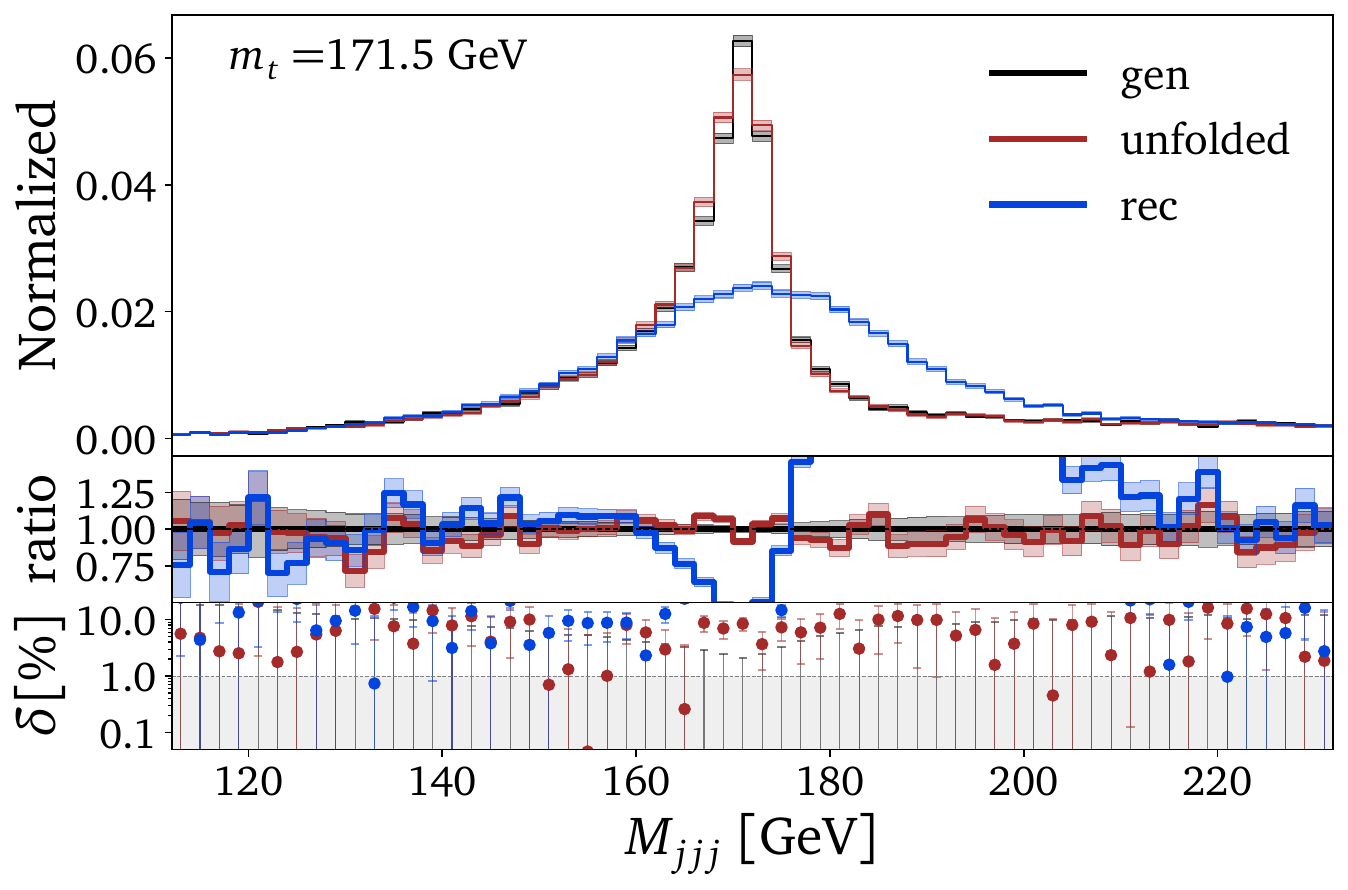}
    \includegraphics[width=0.495\textwidth, page =5] {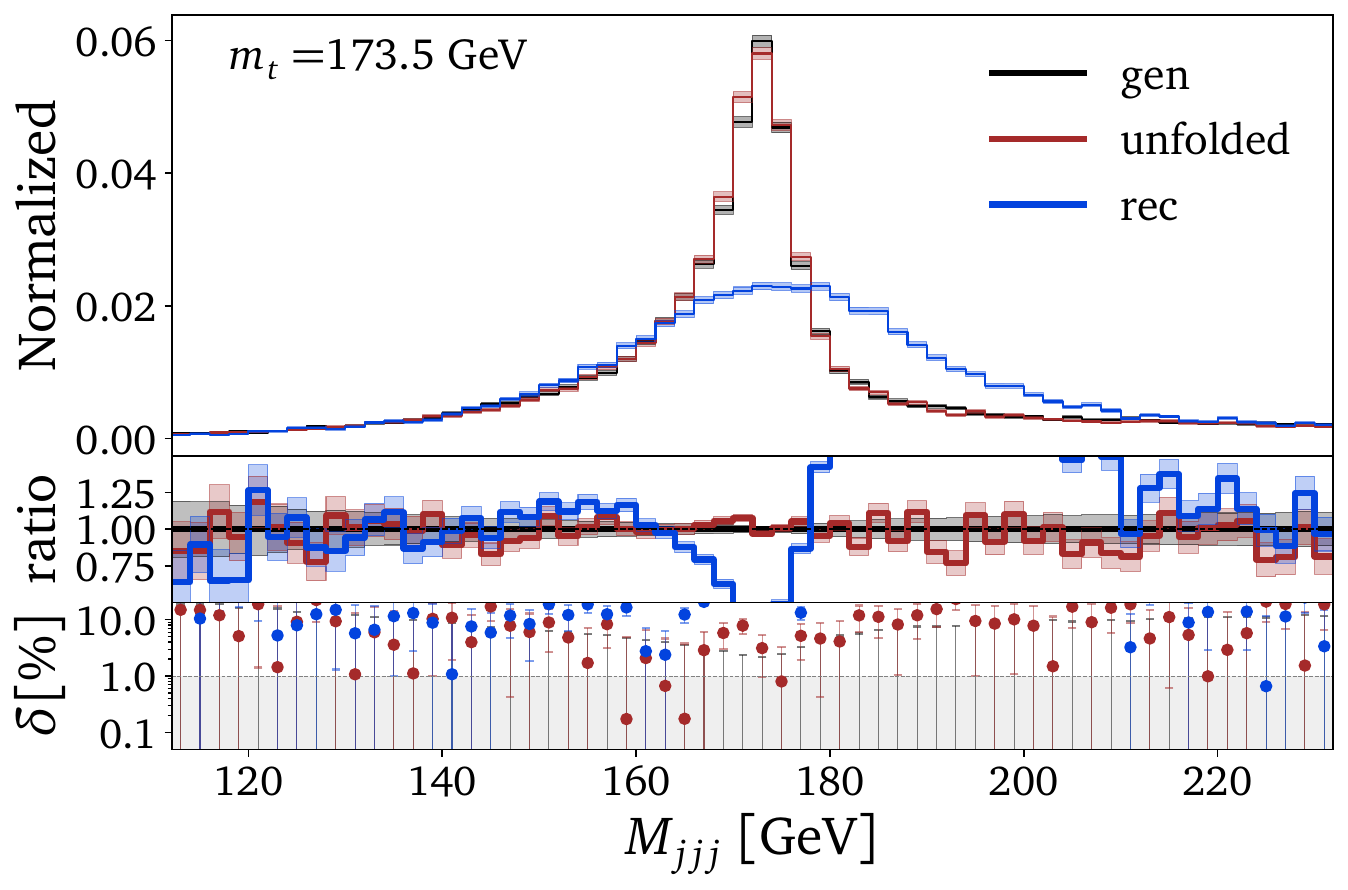}
    \caption{$M_{jjj}$-distribution from 4-dimensional unfolding of
      data with $m_t=171.5$~GeV (left) and $m_t=173.5$~GeV (right). We
      train the network combining samples with the three top masses
      given in Eq.\eqref{eq:training_masses}. Figure from
      Ref.~\cite{Favaro:2025psi}.}
    \label{fig:unbiased_results_4d}
\end{figure}

A generative transformer incorporating CFM blocks for individual
constituents is shown in Fig.~\ref{fig:transfusion}.  Each component
of the $n$-dimensional condition and of the $N$-dimen\-sional input
$x(t)$ is individually embedded by concatenating positional
information and zero padding.  The embedded conditions are passed
through the encoder part of a transformer, while the embedded input is
passed through the decoder.  In both transformer parts, we apply
self-attention and cross-attention to learn the correlations in the
condition and in the input.  The transformer output is a
high-dimensional embedding vector $c_i$, which is mapped back to a
1-dimensional component of the velocity field by a shared linear
layer.  This way we express the learned $N$-dimensional velocity field
as
\begin{align}
    v_\theta (x_\text{gen}(t),t,x_\text{reco} ) = \left(v_\theta (c_{1}, t), \dots, v_\theta (c_{N}, t) \right).
\end{align}
Using the Transfusion architecture we first unfold a 4-dimensional part of
the full phase space, defined by the di-jet and tri-jet invariant
masses,
\begin{align}
  x = \big( \; \{ M_{ik} \}, M_{jjj} \; \big)
    \qquad i,k=1,2,3 \; ,
\end{align}
This phase space includes enough information to measure the
top mass and calibrate the analysis using the $W$-mass peak.  The
results for two assumed top masses in $\pd$ are shown in
Fig.~\ref{fig:unbiased_results_4d}. Neither of these two mass values
appear in the training data.  In both panels we the top mass peak is
the main kinematic feature, and it is reproduced without a significant
shift in the relative deviation.  The fitted peak values are
$m_\text{peak} = (171 \pm 1)$~GeV and $m_\text{peak} = (173 \pm
1)$~GeV, respectively. While the bias will never vanish entirely, it
is well contained within the numerical uncertainties.

\begin{figure}[b!]
    \includegraphics[width=0.495\linewidth, page=13]{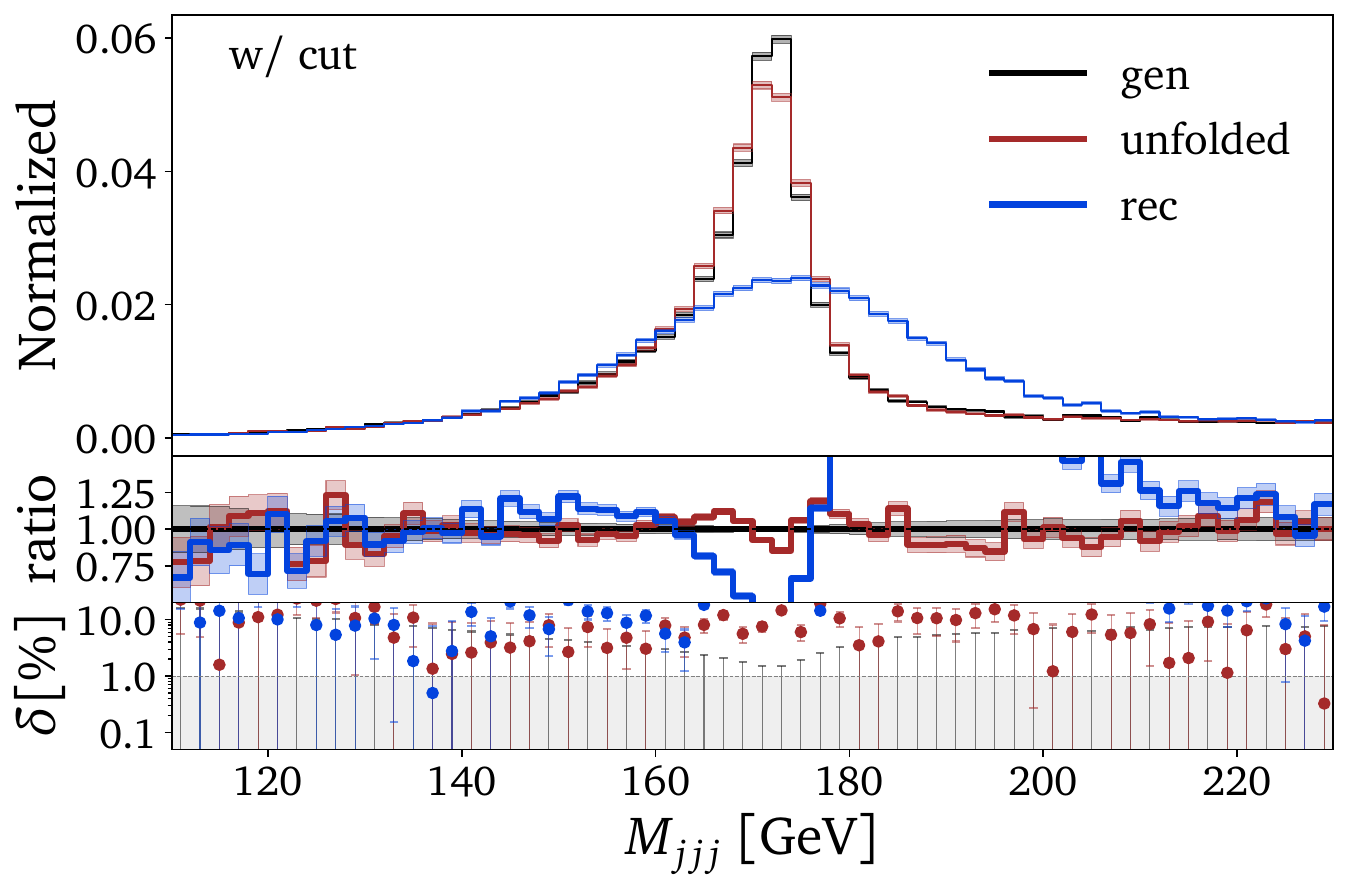} 
    \includegraphics[width=0.495\linewidth, page=2 ]{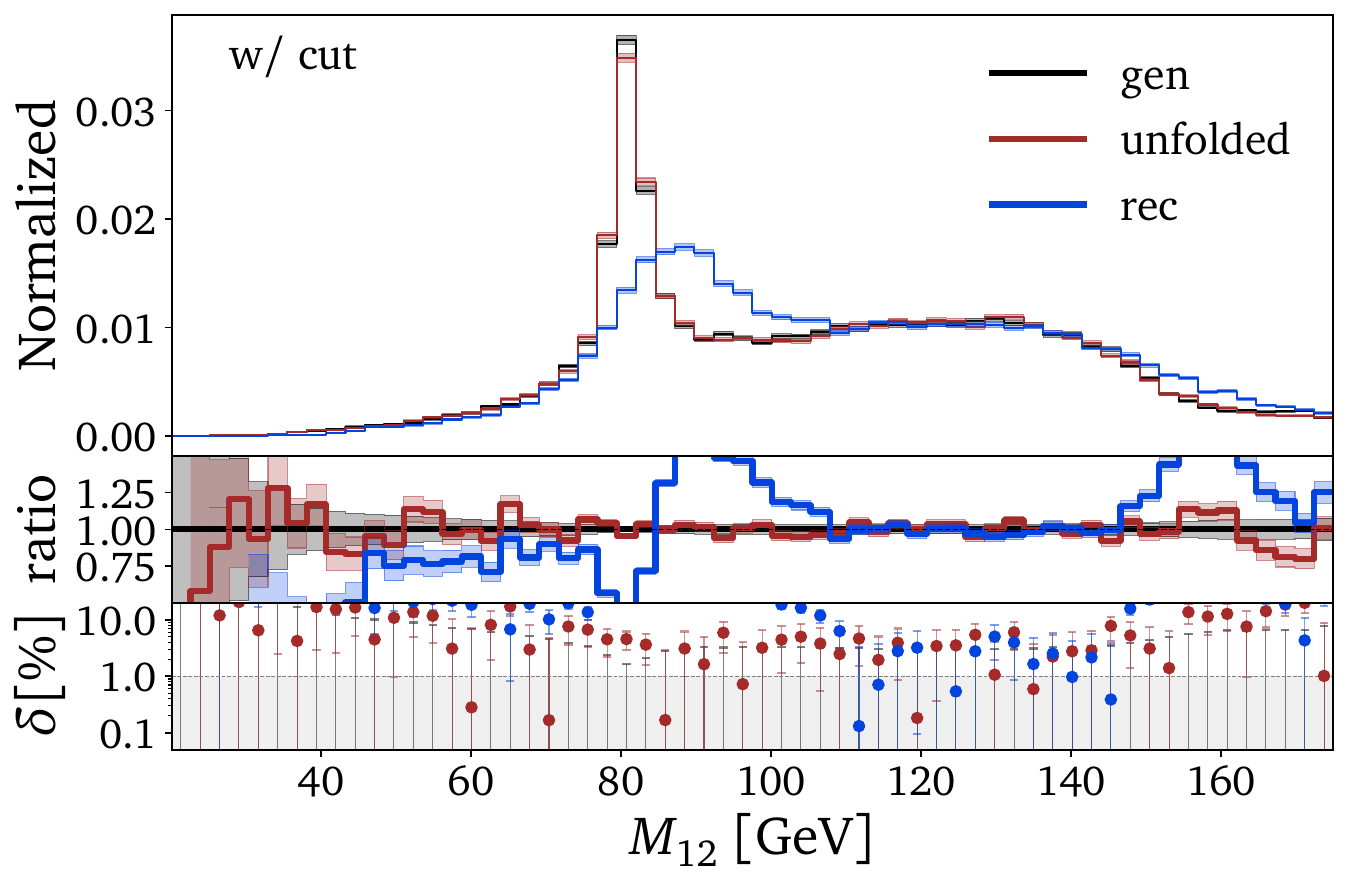} \\
    \includegraphics[width=0.495\linewidth, page=5 ]{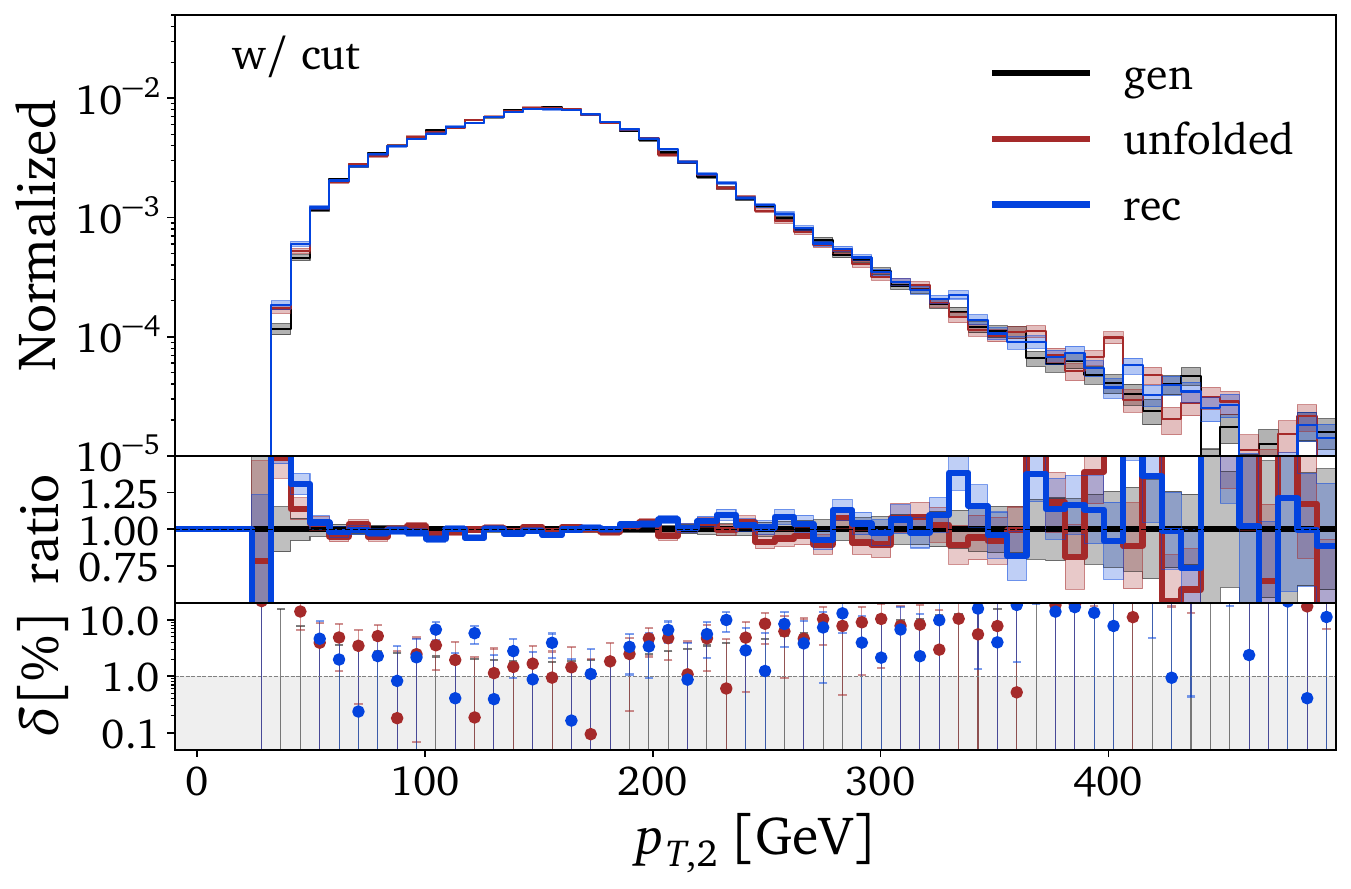} 
    \includegraphics[width=0.495\linewidth, page=38]{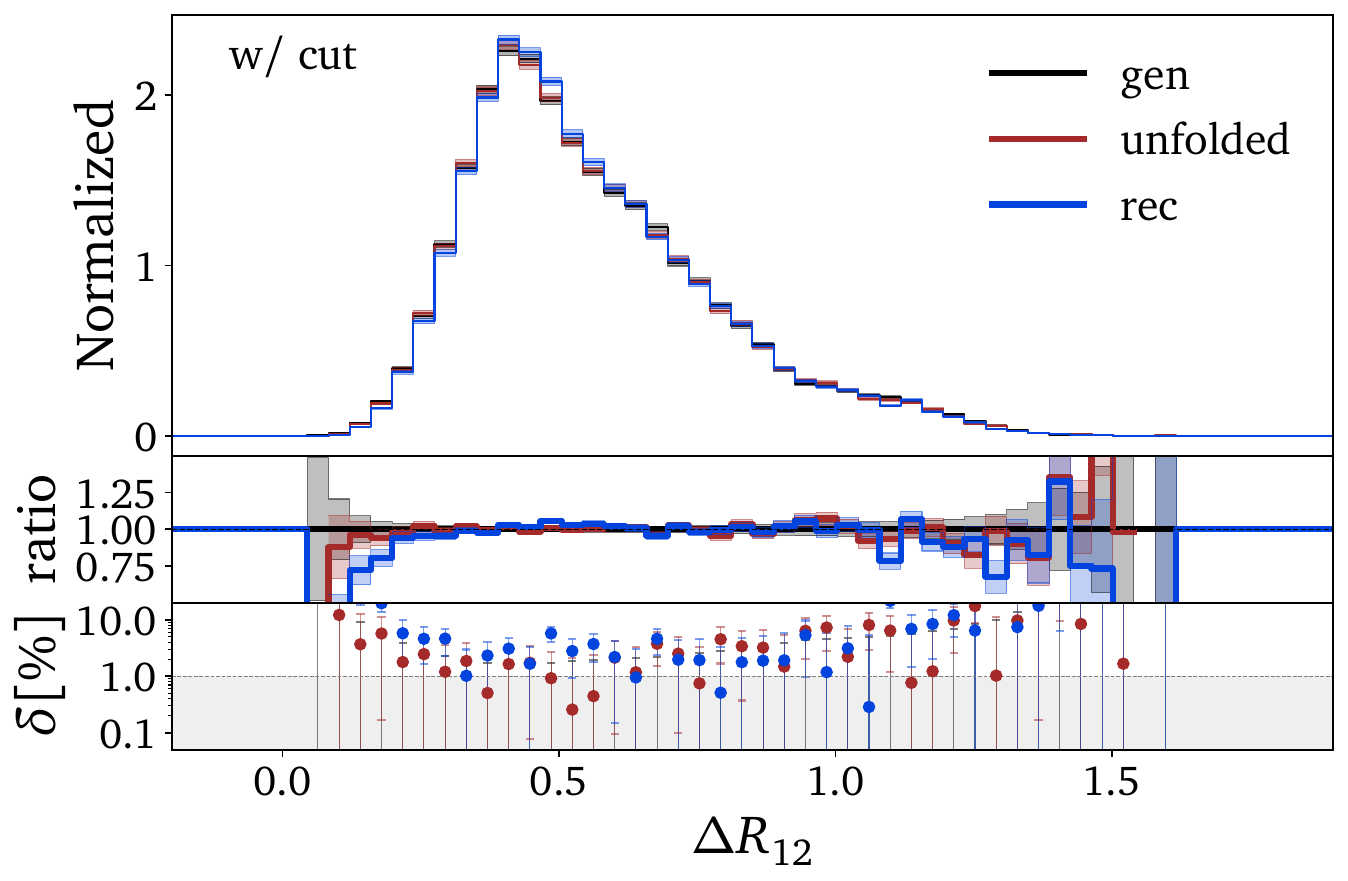} 
    \caption{Kinematic distributions after full, 12-dimensional
      unfolding. We show the target 3-jet distribution, one of the
      di-jet masses with the $W$-peak, and two subjet observables.
      Figure from Ref.~\cite{Favaro:2025psi}.}
    \label{fig:12d_deltaphi}
\end{figure}

If we want to not just measure the top mass, but search for any kind
of new physics effect in top decays, we need to unfold the full
12-dimensional phase space. If the top mass measurement is more
precise in the 4-dimensional subspace, we can assume that we know the
top mass in our dataset for the full unfolding. This factorization of
our two physics questions helps with the precision.  The full phase
space can be described as
\begin{align}
  x = \big( \; \{ m_i \}, \{ M_{ik} \}, \{p_{T,i}\}, \{\eta_i\} \; \big) 
  \qquad i,k=1,2,3 \; ,
\label{eq:cms_unfold_phase}
\end{align}
As a side remark --- an issue arises for the azimuthal angle between
the two leading jets. This angle is learned as a
correlation of seven phase space directions, and in terms of $\cos
\Delta \phi$.  Numerically, the network does not ensure the physical
range $\cos \Delta \phi = -1~...~1$.  We can enforce the physical
range by cutting the cosine for small angles, but this leads to a
slight mis-modelling of the small-$\Delta \phi$ regime.  Instead, we
accept that for unfolding the masses in the parametrization of
Eq.\eqref{eq:cms_unfold_phase} we might have to pay a prize in the
coverage of the angular correlations, and we apply an additional
acceptance cut $\Delta \phi_{ij} > 0.1$ to reco- and gen-levels
unfolded events.

An set of unfolded kinematic distribution from the 12-dimensional
phase space is shown in Fig~\ref{fig:12d_deltaphi}. Given
Eq.\eqref{eq:cms_unfold_phase}, most of the shown kinematic
observables are, in reality, complex correlations of our phase space
basis, so notion of learned observables and learned correlation does
not really hold.  Even for the full phase space, the ratio of
generated and unfolded kinematic observables is extremely close to
unity, the biggest deviation appearing in the triple jet mass.  This
include the three 2-jet masses, for which the condition $M_{ik}
\approx m_W$ leads to three distinct lines on top of the conbinatorial
continuum.

This study serves as a blueprint for an actual CMS analysis, both, for
a top mass measurement and for a wider use of the unfolded data, and
it has been confirmed that the results shown here are quantitatively
the same for the full CMS simulations. This means that, similar to the
situation in astrophysics, it will be possible to measure the top mass
in unfolded LHC data as an undergraduate project. Moreover,
phenomenologists can test their new physics models on unfolded LHC
data provided they can run a simple phase space generator like
Madgraph.

\subsubsection{Generative inference}
\label{sec:cond_inn_bayes}

If a cINN can invert a forward simulation by generating a probability
distribution in the target phase space from a Gaussian latent
distribution, we should be able to use the same network to generate
\underline{posterior probability distributions} in model space.  The
network would also be conditioned on observed events, but the INN
would link the Gaussian latent space to a multi-dimensional parameter
space.  We illustrate this inference task on the QCD splittings
building up the parton shower\index{QCD splittings}. These splitting kernels are given in
Eq.\eqref{eq:qcd_splittings}, and their common pre-factor has been
measured with LEP data. In terms of the \underline{color factors} these
classic measurements give
\begin{align}
  C_A \equiv N_c = 2.89 \pm 0.21
  \qquad \text{and} \qquad 
  C_F \equiv \frac{N_c^2-1}{2 N_c} = 1.30 \pm 0.09 \; .
\label{eq:lep}
\end{align}
For a more systematic approach to measuring the splitting kernels,
organized in terms of the relative transverse momentum of the daughter
particles in the splitting, we modify the strictly collinear splitting
kernels of Eq.\eqref{eq:qcd_splittings}. We keep the argument $z$
describing the momentum fraction of the leading daughter parton, and
parameterize the transverse momentum in the splitting using a new
observable $y$, defined by
\begin{align}
  p_T^2 = y \;  z (1-z) \; .
\end{align}
In terms of the soft and collinear variables the relevant splitting
kernels for massless QCD now read
\begin{align}
P_{g \leftarrow g}(z,y) &= C_A \bigg[ D_{gg} \left(\frac{z(1-y)}{1-z(1-y)} + \frac{(1-z)(1-y)}{1-(1-z)(1-y)}\right) 
          + F_{gg} z(1-z) + C_{gg} yz(1-z) \bigg] \notag \\
P_{q \leftarrow q}(z,y) &= C_F \left[ D_{qq} \frac{2z(1-y)}{1-z(1-y)} + F_{qq} (1-z) + C_{qq} yz(1-z) \right] \notag \\
P_{g \leftarrow q}(z,y) &= \frac{1}{2} \left[ F_{qq} \left(z^2 + (1-z)^2 \right) + C_{gq} yz(1-z) \right] \; .
\label{eq:qcd_kernels}
\end{align}
The collinear expressions from Eq.\eqref{eq:qcd_splittings} can be
recovered in the limit $p_T^2 \propto y \to 0$. We also include a set of
free parameters to allow for possible deviations from the QCD
results. To leading order in perturbative QCD they are
\begin{align}
  D_{qq,gg} = 1
  \qqquad
  F_{qq,gg} = 1
  \qqquad
  C_{qq,gg,gq} = 0 \; .
  \label{eq:def_params}
\end{align}
These parameters are a generalization of the $C_A$ vs $C_F$
measurements quoted in Eq.\eqref{eq:lep}.  The prefactors $D_{ij}$
would correspond to a universal correction like a two-loop anomalous
dimension and resums sub-leading logarithms arising from the collinear
splitting of soft gluons. Their measurements can largely be identified
as
\begin{align}
  D_{qq} \sim C_F
  \qquad \text{and} \qquad
  D_{gg} \sim C_A \; .
\end{align}
The $F_{ij}$ modify the leading terms in $p_T$, truncated in the
strong coupling. The rest terms $C_{ij}$ are defined through an
additional factor $p_T^2$. The modified splitting kernels of
Eq.\eqref{eq:qcd_kernels} can be included in a Monte Carlo generator,
for instance Sherpa.

\begin{figure}[t]
\centering
\usetikzlibrary{arrows.meta,shapes}

\definecolor{Rcolor}{HTML}{E99595}
\definecolor{Gcolor}{HTML}{C5E0B4}
\definecolor{Bcolor}{HTML}{9DC3E6}
\definecolor{Ycolor}{HTML}{FFE699}

\tikzstyle{network} = [thick, rectangle, minimum width=1.8cm, minimum height=1.5cm, text centered, align=center, draw]
\tikzstyle{data} = [thick, rectangle, rounded corners=0.4cm, minimum width=2.0cm, minimum height=1.5cm, text centered, align=center,draw ]
\tikzstyle{arrow} = [thick,-{Latex[scale=1.0]}, line width=0.5mm]

\begin{tikzpicture}[node distance=2cm, scale=0.8, every node/.style={transform shape}]
\node (cinn) [network, fill=Bcolor] {cINN};
\node (summary) [network, above of = cinn, yshift=0.5cm, fill=Bcolor] {Summary\\net};
\node (sherpa) [data, left of = summary, xshift = -1.2cm, fill=Ycolor] {Sherpa\\jets};
\node (model) [data, left of = cinn, xshift = -1.2cm, fill=Ycolor] {QCD\\model};
\node (random) [data, right of = cinn, xshift = 0.8cm, fill=Gcolor] {Gaussian};

\draw [arrow, color=black] ([yshift=0em]summary.south) -- ([yshift=0em]cinn.north) node[midway,right,xshift=0.2cm]{$h$};
\draw [arrow, color=black] ([yshift=0em]sherpa.east) -- ([yshift=0em]summary.west) node[midway,above,yshift=0.2cm]{$\{x_m\}$};
\draw [arrow, color=black] ([yshift=0em]model.north) -- ([yshift=0em]sherpa.south);
\draw [arrow, color=black] ([yshift=0em]model.east) -- ([yshift=0em]cinn.west) node[midway,above,yshift=0.2cm]{$m$};
\draw [arrow, color=black] ([yshift=0em]cinn.east) -- ([yshift=0em]random.west) node[midway,above,yshift=0.2cm]{$z$};

\node [below of = cinn, yshift=0.8cm] {$g(m;h)$};
\node [below of = random, yshift=0.8cm] {$P(z)$};

\node[above,font=\large\bfseries, yshift = 0.7cm] at (current bounding box.north) {Training};

\end{tikzpicture}\nobreak\hspace{1.5em}%
\begin{tikzpicture}[node distance=2cm, scale=0.8, every node/.style={transform shape}]
\node (cinn) [network, fill=Bcolor] {cINN};
\node (summary) [network, above of = cinn, yshift=0.5cm, fill=Bcolor] {Summary\\net};
\node (jets) [data, left of = summary, xshift = -1.2cm, fill=Ycolor] {LHC\\jets};
\node (measurement) [data, left of = cinn, xshift = -1.2cm, fill=Ycolor] {QCD\\measurement};
\node (sampling) [data, right of = cinn, xshift = 0.8cm, fill=Gcolor] {Gaussian \\ sampling};
\node[above,font=\large\bfseries, yshift = 0.7cm] at (current bounding box.north) {Inference};

\draw [arrow, color=black] ([yshift=0em]summary.south) -- ([yshift=0em]cinn.north) node[midway,right,xshift=0.2cm]{$h$};
\draw [arrow, color=black] ([yshift=0em]jets.east) -- ([yshift=0em]summary.west) node[midway,above,yshift=0.2cm]{$\{x\}$};
\draw [arrow, color=black] ([yshift=0em]cinn.west) -- ([yshift=0em]measurement.east) node[midway,above,yshift=0.2cm]{$m$};
\draw [arrow, color=black] ([yshift=0em]sampling.west) -- ([yshift=0em]cinn.east) node[midway,above,yshift=0.2cm]{$z$};

\node [below of = measurement, yshift=0.8cm] {$P(m|\{x\})$};
\node [below of = cinn, yshift=0.8cm] {$\bar g(z;h)$};
\node [below of = random, yshift=0.8cm] {$z \sim P(z)$};
\end{tikzpicture}
\caption{BayesFlow setup of the cINN for training and
  inference. Figure from Ref.\cite{Bieringer:2020tnw}.}
\label{fig:bayesflow}
\end{figure}
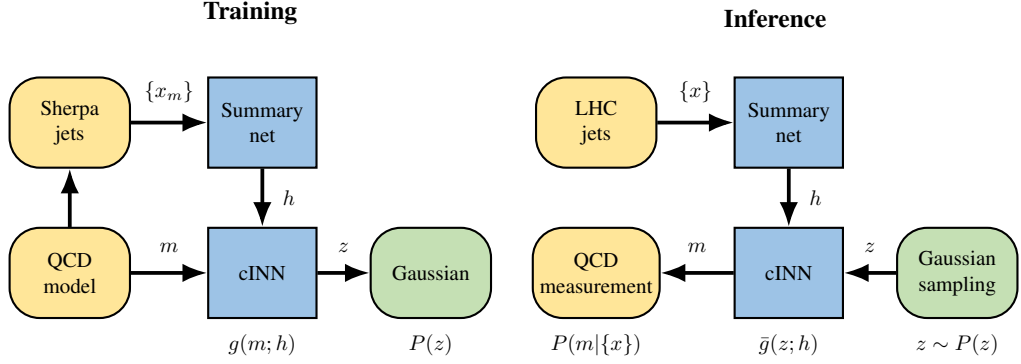

To extract the splitting parameters from measured jets $x$ we use the
cINN architecture described in Sec.~\ref{sec:cond_inn_unfold}.  The
cINN-inference framework, also referred to as \underline{BayesFlow}, is
illustrated in Fig.~\ref{fig:bayesflow}. In the training phase we scan
over model parameters $m$, in our case the modification factors with
the QCD values given in Eq.\eqref{eq:def_params}, and generate the
corresponding jets. The $m$-dependent simulated jet data is referred
to as $x_m$.  We then train a summary network combined with the cINN
to map the model parameters $m$ onto a Gaussian latent space. This
corresponds to the cINN unfolding in Fig.~\ref{fig:cinn}, with the
parton-level events replaced by the model parameters and the
conditional reconstruction-level events replaced by the simulated
jets. The technical challenge in this application is that the training
is fully amortized, which means we map the full set of training data
onto the Gaussian latent space, which turns into a memory limitation
to the amount of training data.

For the network evaluation or inference we sample from the Gaussian
latent space into the model parameter space $m$ to generate a
correlated posterior distribution of the allowed $D_{_ij}$, $F_{ij}$,
and $C_{ij}$. If we want to interpret this framework in a Bayesian
sense we can identify the starting distributions in model space, which
we use to generate the training jets, as the prior, and the final
inference outcome as the posterior.

Before applying BayesFlow to LHC jets including
\underline{hadronization} and \underline{detector simulation}, we test
the inference model on a simple toy shower.  We simulate jets using
the process
\begin{align}
  e^{+} e^{-} \to Z \to q \bar{q} \; ,
\end{align}
with massless quarks and a hard showering cutoff at 1~GeV.  Most jets
appear close to the phase space boundary $p_{T,j} < m_Z/2$. For each
event we apply the parton shower to one of the outgoing quarks, such
that the second quark acts as the spectator for the the first
splitting, and we only consider one jet.  The network then analyses
the set of outgoing momenta except for the initial spectator momentum.
For the training data we scan the parameters $\{ D_{ij},F_{ij},C_{ij}
\}$ in two or three dimensions.  The input to the summary network per
batch are a sets of constituent 4-vectors, and we typically train on
100k randomly distributed points in model space. This number is much
smaller than what we typically use for network training at the LHC.

\begin{figure}[t]
  \includegraphics[width=0.495\textwidth]{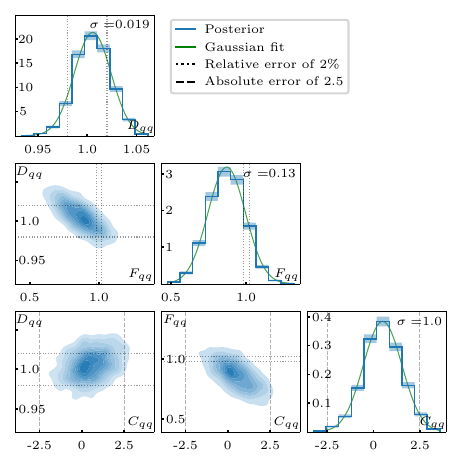}
  \includegraphics[width=0.495\textwidth]{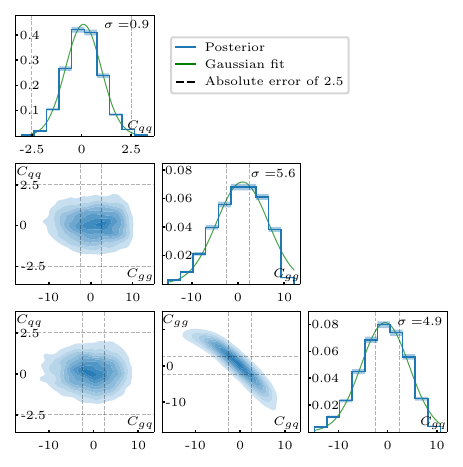}
  \caption{Posterior probabilities for the toy shower, gluon radiation
    only, $\{ D_{qq}, F_{qq},C_{qq} \}$ (left), and the
    $p_T$-suppressed rest terms for all QCD splittings, $\{ C_{qq},
    C_{gg}, C_{gq} \}$ (right).  We assume SM-like jets. Figure from
    Ref.~\cite{Bieringer:2020tnw}.}
  \label{fig:dqq_fqq_cqq}
\end{figure}

For our first test we restrict the shower to the $P_{qq}$ kernel, such
that a hard quark successively radiates collinear and soft
gluons. This way our 3-dimensional model space is given by
\begin{align}
  \{ D_{qq}, F_{qq},C_{qq} \} \; .
\end{align}
In the left panel of Fig.~\ref{fig:dqq_fqq_cqq} we show the posterior
probabilities for these model parameters assuming true SM-values. All
1-dimensional posteriors are approximately Gaussian.  The
best-measured parameter of the toy model is the regularized
divergence, followed by the finite terms, and then the rest term with
its assumed $p_T$-suppression.  This reflects the hierarchical
structure of the splitting kernel. The correlations between parameters
are small, but not negligible.

In a second step we include all QCD splitting kernels from
Eq.\eqref{eq:qcd_kernels} and target the unknown rest terms
\begin{align}
  \{ C_{qq},C_{qg},C_{gg} \} \; .
\end{align}
This means we assume our perturbative predictions for the leading
contributions to hold, so we need to estimate the uncertainty of the
perturbative description. For $C_{qq}$ the uncertainty band shrinks slightly,
in the absence of the dominant contributions to this kernel.  For the
other two rest terms, $C_{gg}$ and $C_{gq}$, we find significantly
larger uncertainty bands and a strong anti-correlation.

\begin{figure}[b!]
  \centering
  \includegraphics[width=0.35\textwidth]{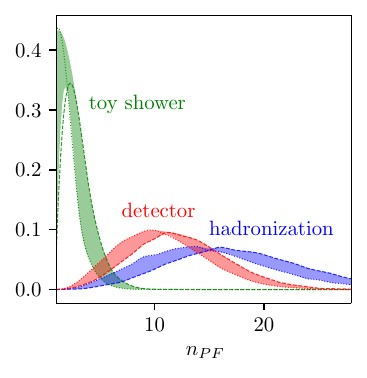}
  \hspace*{0.1\textwidth}
  \includegraphics[width=0.35\textwidth]{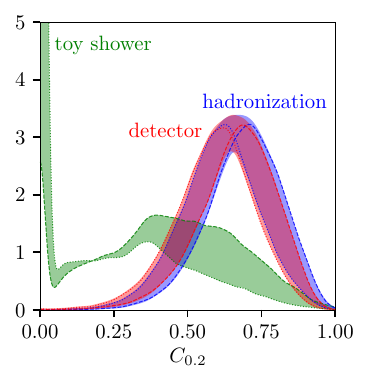} 
  \caption{High-level observables $n_\text{PF}$ and $C_{0.2}$ for 100k
    jets. We show the distributions for the toy shower, the Sherpa
    shower with hadronization, and including a detector simulation The
    bands indicate the variation of $D_{qq} = 0.5.~...~2$ (dotted and
    dashed).  Figure from Ref.~\cite{Bieringer:2020tnw}.}
  \label{fig:generator_comp}
\end{figure}

For a more realistic simulation we rely on a full Sherpa shower with
hadronization and a fast detector simulation. The dataset consists of
our usual particle flow objects forming the jet. Unlike for the
classification of boosted jets we study relatively soft jets, again
simulated from $Z$-decays with a spectrum
\begin{align}
  p_{T,j} = 20~\gev~...~\frac{m_Z}{2} \; .
\end{align}
To illustrate the physics limiting the measurement, we show two
high-level observables from Eq.\eqref{eq:qg_obs} for the toy shower,
after hadronization, and after detector effects in
Fig.~\ref{fig:generator_comp}. The shaded bands reflect a variation
$D_{qq} = 0.5.~...~2$.  The number of constituents $n_\text{PF}$
generally increases with $D_{qq}$. The toy shower with a high cutoff
and no hadronization does not generate a very large number of particle
flow objects. Hadronization increases the number of constituents
significantly, but is not related to QCD splittings. The finite
detector resolution and the detector thresholds decreases
$n_\text{PF}$ again.  The constituent-constituent correlation
$C_{0.2}$ loses all toy events at small values when we include
hadronization, while the broad feature around $C_{0.2} \sim 0.4$
narrows and moves to slightly larger values.
The main message from Fig.~\ref{fig:generator_comp} is that from a QCD
point of view the hadronization effects are qualitatively and
quantitatively far more important than the detector
effects.

\begin{figure}[t]
  \includegraphics[width=0.495\textwidth]{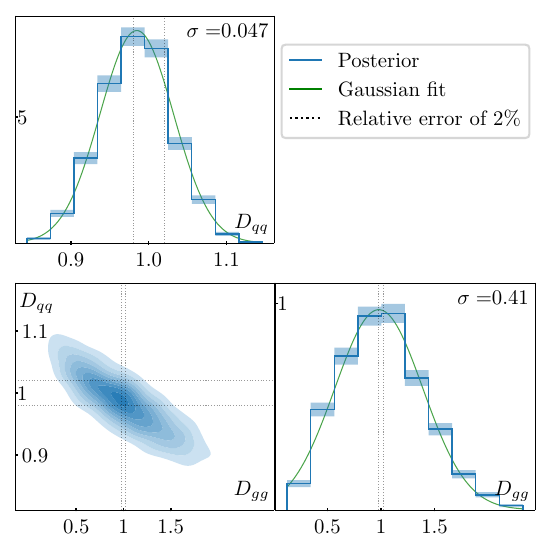}
  \includegraphics[width=0.495\textwidth]{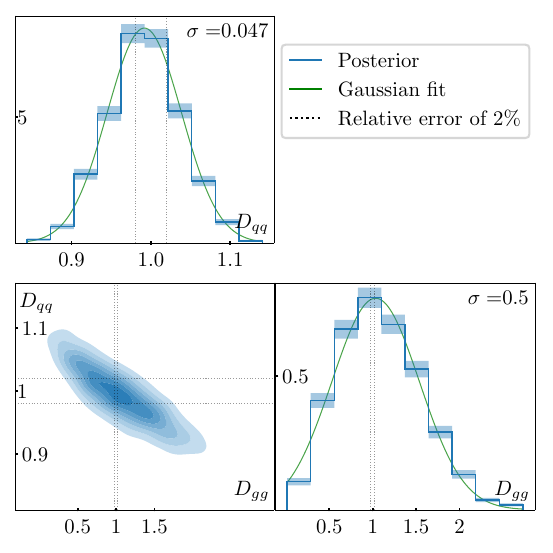}
  \caption{Posterior probabilities for the Sherpa shower,
    soft-collinear leading terms for all QCD splittings, $\{ D_{qq},
    D_{gg}\}$. We assume SM-like jets and show results without
    Delphes detector simulation (left) and including detector effects
    (right). Figure from Ref.~\cite{Bieringer:2020tnw}.}
  \label{fig:sherpa_dqq_dgg}
\end{figure}

To simplify the task in view of the hadronization effects we allow for
all QCD splittings, but only measure the leading soft-collinear
contributions, corresponding to $C_F$ and $C_A$,
\begin{align}
  \{ D_{qq}, D_{gg} \} \; .
\end{align}
The results are shown in Fig.~\ref{fig:sherpa_dqq_dgg}. First, we see
that the measurement after hadronization only and after hadronization
and detector effects is not very different.  In both cases we see a
significant degradation of the measurements, especially in $D_{gg}$,
and a strong correlation which reflects the fact that we are only
looking at quark-induced jets.  For an actual measurement this
correlation could be easily removed by combining quark-dominated and
gluon-dominated samples.

\subsection{Simulation-based inference}
\label{sec:cond_sim}

As simulation-based inference we consider a wide range of methods
which all share the basic idea that we want to use
\underline{likelihoods} to extract information from data, ideally
event by event, but the likelihood is not accessible
explicitly. Likelihoods have been a theme for all our ML-applications,
starting with the network training introduced in
Sec.~\eqref{sec:basics_deep_multi}. In this section we will be a
little more specific and introduce two ways we can use modern
ML-methods to infer fundamental (model) parameters from LHC data.

\subsubsection{Likelihood extraction}
\label{sec:cond_sim_like}

In spite of discussing likelihoods and likelihood ratios several
times, we have never written down the likelihood as we use it in an
analysis of datasets consisting of unweighted and uncorrelated
events. A likelihood for a counting experiment can be split into the
statistical probability of observing $n$ event with $b$ background
events expected, and a normalized probability of observing individual
events in a given phase space point. If we phrase the question in
terms of a hypothesis or set of model parameters $\theta_B$ we can
write
\begin{align}
  \boxed{
    p(x|\theta_B) = \text{Pois}(n|b) \; \prod_{i=1}^n f_B(x_i)
    } 
  \qquad \text{with} \qquad
  \text{Pois}(n|b) = \frac{b^n}{n!} \; e^{-b} \; .
\end{align}
The Neyman--Pearson lemma\index{Neyman-Pearson lemma} then tells us that if we want
to, for instance, decide between a background-only and
signal-plus-background hypothesis we will find optimal results
by using the likelihood ratio as the estimator
\begin{align}
  \frac{p(x|\theta_{S+B})}{p(x|\theta_B)}
  &= \frac{\text{Pois}(n|s+b) \; \prod_i f_{S+B}(x_i)}{\text{Pois}(n|b) \; \prod_i f_B(x_i)} \notag \\
  &= e^{-s} \left( \frac{s+b}{b} \right)^n \;
  \frac{\prod_i f_{S+B}(x_i)}{\prod_i f_B(x_i)} \notag \\
  &= e^{-s} \; 
  \frac{\prod_i (s+b) f_{S+B}(x_i)}{\prod_i b f_B(x_i)} \
  = e^{-s} \; 
  \frac{\prod_i [ s f_S(x_i) + b f_B(x_i)}{\prod_i b f_B(x_i)]} \; . 
\end{align}
We have used the assumption that $s+b$ signal plus background events
follow weighted signal and background distributions independently. We
can translate this into the log-likelihood ratio as a function of the
transition amplitudes $f(x)$ over phase space, noticing that the
log-likelihood ratio is additive for phase space points or events
$x_i$,
\begin{align}
  \log \frac{p(x|\theta_{S+B})}{p(x|\theta_B)}
  = -s
  + \sum_i \log \left[ 1 +  \frac{s f_S(x_i)}{b f_B(x_i)} \right] \; . 
\label{eq:llr}
\end{align}
If we work with simulation only, consider irreducible backgrounds at
the parton level, and limit ourselves to the same partons in the
initial state, we can compute the log-likelihood ratio as
\begin{align}
  \log \frac{p(x|\theta_{S+B})}{p(x|\theta_B)}
  = -s
  + \sum_i \log \left[ 1 +  \frac{|\mathcal{M}_S(x_i)|^2}{|\mathcal{M}_B(x_i)|^2} \right] \; . 
\end{align}
Prefactors relating the transition matrix element to the fully
exclusive cross sections, including parton densities, will drop out of
the ratio of signal and backgrounds. This means we can use
Eq.\eqref{eq:llr} to compute the maximum possible significance for
observing a signal over a background integrating over matched signal
and background event samples. Moving from two equal hypotheses, signal
vs background, to measuring a parameter $\theta$ around a reference
value $\theta_\text{ref}$ we can rewrite the corresponding likelihood
ratio
\begin{align}
  \log \frac{p(x|\theta)}{p_\text{ref}(x)}
  = -s
  + \sum_i \log \left[ 1 +  \frac{|\mathcal{M}(x_i|\theta)|^2}{|\mathcal{M}(x_i)|_\text{ref}^2} \right] 
  \qquad \text{with} \qquad
  p_\text{ref}(x) \equiv p(x|\theta_\text{ref}) \; .
\label{eq:llr_matrix}
\end{align}
If we are interested in parameter values $\theta$ close to a reference
point $\theta_\text{ref}$, we can simplify our problem and
\underline{taylor} the log likelihood around $\theta_\text{ref}$,
\begin{align}
  \log \frac{p(x|\theta)}{p(x|\theta_\text{ref})}
  =
  (\theta - \theta_\text{ref})
  \underbrace{\frac{\partial}{\partial \theta} \log p(x|\theta) \Bigg|_{\theta_\text{ref}}}_{t(x|\theta_\text{ref})}
  + \ord \left(  (\theta - \theta_\text{ref})^2 \right) \; .
\label{eq:def_score}
\end{align}
The leading term is called the \underline{score} in statistics or
\underline{optimal observable}\index{optimal observable} in particle physics. Neglecting the
higher-order terms we solve this equation and find
\begin{align}
  p(x|\theta) \approx e^{t(x|\theta_\text{ref}) (\theta - \theta_\text{ref})} \; p(x|\theta_\text{ref}) \,.
\end{align}
This relation to the likelihood function implies that the score
$t(x|\theta_\text{ref})$ are its sufficient statistics. If we measure
it we capture all information on $\theta$ included in the the full
event record $x$.  It is possible to show that the score is not only
linked to the Neyman-Pearson lemma for discrete hypotheses $H_{S+B}$
vs $H_B$, but also saturates the \underline{Cram\'er-Rao bound}\index{Cram\'er-Rao bound} for a
continuous parameter measurement $\theta \sim
\theta_\text{ref}$.

For the application of modern ML-methods to likelihood extraction we
follow the approach of the public analysis tool
MadMiner~\cite{Brehmer:2018eca,Brehmer:2019xox}.  In
Eq.\eqref{eq:gan_dopt} we have already seen the 
relation between classifiers and 
likelihood ratios.  The starting point is the
Neyman-Pearson lemma\index{Neyman-Pearson lemma}, telling us 
that an optimal
discriminator $D$ has to reproduce the likelihood ratio. To confirm this
we start from the optimally trained discriminator in
Eq.\eqref{eq:gan_dopt}, translate the output into our new conventions,
and solve for the likelihood ratio,
\begin{align}
  D_\text{opt}(x)
  &= \frac{p_\text{ref}(x)}{p_\text{ref}(x) + p(x|\theta)} 
  = \frac{1}{1 + \dfrac{p(x|\theta)}{p_\text{ref}(x)}}  \notag \\
  \Leftrightarrow \qquad 
  \frac{p(x|\theta)}{p_\text{ref}(x)}
  &= \frac{1 - D_\text{opt}(x)}{D_\text{opt}(x)} \; .
  \label{eq:lr_opt}
\end{align}
To construct this estimator for the likelihood ratio we start by
generating two training datasets, one following $p_\text{ref}(x)$ and
one following $p(x|\theta)$. In a slight variation of the
two-hypothesis classification we now train a $\theta$-dependent
classifier, where $\theta$ enters the training as a condition. After
training, we can use the output of this classifier for a given value
of $\theta$ as a numerical approximation to the likelihood ratio
$p(x|\theta)/p_\text{ref}(x)$. This approach has the advantage that we
only need to train a stable classification network. The advantage of
this method is that it is fast for a limited number of phase space
points, as long as the network has converged on a smooth classification
output. In the MadMiner implementation this approach is called
calibrated ratios of likelihoods (Carl).

For a second approach, we go beyond classification and train a
regression network to encode likelihoods or likelihood ratios over the
combined phase space and parameter space $(x,\theta)$.  For this
purpose we modify the relation between the likelihood ratio given in
Eq.\eqref{eq:llr_matrix} and the matrix elements using an approximate
factorization for LHC physics including a range of latent
variables. We start with the observation that we can write the
likelihood in $x$ as an integral over latent variables which describe
the hard process $z_p$, the step to the parton shower $z_s$, and to
the detector level $z_d$. We can also assume that our parameter of
interest only affects the hard scatting,
\begin{align}
  p(x|\theta)
  = \int dz \; p(x,z|\theta)
  \approx \int dz_d dz_s dz_p \; p(x|z_d) \; p(z_d|z_s) \; p(z_s|z_p) \; p(z_p|\theta) \; .
  \label{eq:lr_fact}
\end{align}
Using this factorization we see that the problem with the likelihood
ratio is that it is given by a ratio of two integrals, for which there
is no easy and efficient way to evaluate it, even using modern
ML-methods,
\begin{align}
  \frac{p(x|\theta)}{p_\text{ref}(x)}
  = \frac{\int dz \; p(x,z|\theta)}{ \int dz \; p_\text{ref}(x,z)} \; .
\end{align}
However, we can simplify the problem using the property of the
\underline{joint likelihood ratio} being an unbiased estimator of the
likelihood ratio. This means we can evaluate the ratio of the joint
likelihoods instead of the ratio of the likelihoods,
\begin{align}
  \frac{p(x,z|\theta)}{p_\text{ref}(x,z)}
  \approx \frac{p(z_p|\theta)}{p_\text{ref}(z_p)}
  =\frac{|\mathcal{M}(z_p|\theta)|^2}{|\mathcal{M}_\text{ref}(z_p)|^2}
  \; \frac{\sigma_\text{ref}}{\sigma(\theta)} \; .
  \label{eq:joint_lr}
\end{align}
In the last step we do not assume that all prefactors in the relation
between matrix element and rate cancel, so we ensure the correct
normalization explicitly. The advantage of using the joint likelihood
ratio as an estimator for the likelihood ratio is that we can encode
it in a network comparably easily, using the factorization properties
of LHC rates.

Inspired by the fact that the joint likelihood ratio can serve as an
estimator for the actual likelihood ratio we can try to numerically
extract the likelihood ratio from the joint likelihood ratio.  If we
assume that we have a dataset encoding the likelihood ratio over phase
space we still need to encode it in a network. We simplify our
notation from Eq.\eqref{eq:lr_fact} to
\begin{align}
  p(x,z|\theta) = p(x|z_p) \; p(z_p|\theta) 
  \qquad \text{or} \qquad 
  p(x|\theta)
  &= \int dz_p \; p(x,z_p|\theta) \notag \\
  &\approx \int dz_p \; p(x|z_p) \; p(z_p|\theta) \; .
  \label{eq:lr_simp}
\end{align}
Slightly generalizing the problem we can ask if it is possible to
construct proxies for $(x,z_p)$-dependent distributions as
$x$-dependent distributions, with a given test function. This is
similar to the variational approximation introduced in
Sec.~\ref{sec:basics_deep_bayes}. We will see that the test function
$p(x|z_p) p(z_p|\theta)$ combined with an L2 norm over the two
functions will turn out useful
\begin{align}
F(x) = \int dz_p \; \left[ f(x,z_p) - \hat{f}(x) \right]^2 \; p(x|z_p) \; p_\text{ref}(z_p) \; .
\label{eq:def_distri}
\end{align}
The variational condition defines the approximation $\hat{f}_*(x)$
through a minimization of $F(x)$,
\begin{align}
  0 = \frac{\delta F}{\delta \hat{f}} 
  &= \frac{\delta}{\delta \hat{f}} 
  \int dz_p \; \left[ f(x,z_p) - \hat{f}(x) \right]^2 \; p(x|z_p) \; p_\text{ref}(z_p) \notag \\
  &= 
  \int dz_p \; p(x|z_p) \; p_\text{ref}(z_p) \; \frac{\delta}{\delta \hat{f}} \left[ f(x,z_p) - \hat{f}(x) \right]^2  \notag \\
  &= 
  - 2 \int dz_p \; p(x|z_p) \; p_\text{ref}(z_p) \; \left[ f(x,z_p) - \hat{f}(x) \right]  \notag \\
  \Leftrightarrow \qquad
  \hat{f}_*(x)
  &= \frac{\int dz_p \; f(x,z_p) \; p(x|z_p) \; p_\text{ref}(z_p)}{\int dz_p \; p(x|z_p) \; p_\text{ref}(z_p)} 
   = \frac{\int dz_p \; f(x,z_p) \; p(x|z_p) \; p_\text{ref}(z_p)}{p_\text{ref}(x)} \; .
   \label{eq:l2_variational}
\end{align}
We can apply this method to the joint likelihood ratio from
Eq.\eqref{eq:joint_lr} and find
\begin{align}
f(x,z_p) 
&= \frac{p(z_p|\theta)}{p_\text{ref}(z_p)}
\approx \frac{p(x|z_p) \; p(z_p|\theta)}{p(x|z_p) \; p_\text{ref}(z_p)} \notag \\
\Rightarrow \qquad 
\hat{f}_*(x) 
&= \frac{\int dz_p \; f(x,z_p) \; p(x|z_p) \; p(z_p|\theta)}{p_\text{ref}(x)} 
= \frac{p(x|\theta)}{p_\text{ref}(x)} \; .
\end{align}
This means that by numerically minimising Eq.\eqref{eq:def_distri} as
a loss function we can train a regression network for reproduce the
likelihood ratio from Eq.\eqref{eq:llr_matrix}.

The third, and final approach to extract likelihoods for the LHC
starts with remembering that according to Eq.\eqref{eq:def_score} the
derivative of the log-likelihood ratio is a sufficient statistics for
a parameter of interest $\theta$ in the region around
$\theta_\text{ref}$. This leads us to computing the \underline{joint
  score}
\begin{align}
  t(x,z|\theta) &= \frac{\partial}{\partial \theta} \log  p(x,z|\theta) 
  = \frac{1}{p(x,z|\theta)} \; \frac{\partial}{\partial \theta} p(x,z|\theta) \notag \\
  &\approx \frac{p(x|z_d) \; p(z_d|z_s) \; p(z_s|z_p)}{p(x|z_d) \; p(z_d|z_s) \; p(z_s|z_p) \; p(z_p|\theta)} \; \frac{\partial}{\partial \theta} p(z_p|\theta) 
  = \frac{\partial p(z_p|\theta)/\partial \theta}{\p(z_p|\theta)} \notag \\
  &= \frac{\sigma(\theta)}{|\mathcal{M}(z_p|\theta)|^2}
  \; \frac{\partial}{\partial \theta} \frac{|\mathcal{M}(z_p|\theta)|^2}{\sigma(\theta)}
  = \frac{\partial|\mathcal{M}(z_p|\theta)|^2/\partial \theta}{|\mathcal{M}(z_p|\theta)|^2}
  - \frac{\partial \sigma(\theta)/\partial \theta}{\sigma(\theta)} \; ,
\end{align}
and limit our analysis to $\theta \sim \theta_\text{ref}$. Just as for
the joint likelihood ratio we use the variational approximation from
Eq.\eqref{eq:l2_variational} to train a network for the score. This
time we define, in the simplified notation of Eq.\eqref{eq:lr_simp},
\begin{align}
  F(x) &= \int dz_p \; \left[ f(x,z_p) - \hat{f}(x) \right]^2 \; p(x|z_p) \; p(z_p|\theta) \notag \\
  f(x,z_p) &= t(x,z_p|\theta)
  = \frac{p(x|z_p) \; \partial p(z_p|\theta)/\partial \theta}{p(x|z_p) \; p(z_p|\theta)} \notag \\
  \Rightarrow \qquad 
  \hat{f}_*(x) 
  &= \frac{\int dz_p \; f(x,z_p) \; p(x|z_p) \; p(z_p|\theta)}{\int dz_p \; p(x|z_p) \; p(z_p|\theta)} 
  = \frac{\int dz_p \; p(x|z_p) \; \partial p(z_p|\theta)/\partial \theta}{p(x|\theta)} 
= t(x|\theta) \; .
\end{align}
This means we can also encode the score as the local summary
statistics for a model parameter $\theta$ in a network using a
variational approximation. In MadMiner this method is called score
approximates likelihood locally or Sally.

To illustrate this score extraction we look at the way heavy new
physics affects the Higgs production channel
\begin{align}
  pp \to  WH \to \ell \bar{\nu} \; b \bar{b} \; .
  \label{eq:wh_proc}
\end{align}
If we assume that new particles are heavy and not produced on-shell,
the appropriate QFT description is \underline{higher-dimensional
  operators} and the corresponding Wilson coefficients. The effective
Lagrangian we use to describe LHC observation then becomes
\begin{align}
\lag = \lag_\text{SM}+ \sum_{d,k} \frac{ C_k^d}{\Lambda^{d-4}} \ope{k}^d
\; .
\label{eq:def_eft}
\end{align}
We can organize the effective Lagrangian by the power $d$ of the scale
of the unknown new physics effects, $\Lambda$. The coupling parameters
$C_k$ are called Wilson coefficients. The Lagrangian of the
renormalizable Standard Model has mass dimension four, there exists
exactly one operators at dimension five, related to the neutrino mass,
and 59 independent operators linking the Standard Model particles at
dimension six. Adding flavor indices increases this number very
significantly. The $WH$ production process mostly tests three of them,
\begin{align}
  \ope{HD}
&= (\phi^\dagger \phi)\square (\phi^\dagger\phi)
  - \frac{1}{4} (\phi^\dagger D^\mu\phi)^* (\phi^\dagger D_\mu \phi) \notag \\
\ope{HW} &= \phi^\dagger\phi W_{\mu\nu}^a W^{\mu\nu a} \notag \\
\ope{Hq}^{(3)}&= (\vpjt)({\overline Q}_L \sigma^a \gamma^\mu Q_L)\; ,
\label{eq:wh_ops}
\end{align}
where $\phi$ is the Higgs doublet, $D_\mu$ the full covariant
derivative, $W_{\mu\nu}^a$ the weak field strength, and
$\vpjt=i\phi^\dagger(\frac{\sigma^a}{2}D_{\mu} \phi)
-i(D_{\mu}\phi)^\dagger \frac{\sigma^a}{2}\phi$.  The three operators
affect the transition amplitude for $WH$ production in different
ways. In principle, dimension-6 operators can come with up to two
derivatives, which after a Fourier transformation turn into two powers
of the momentum transfer in the interaction, $p^2/\Lambda^2$. Of the
three operators in Eq.\eqref{eq:wh_ops} it turns out that $\ope{HD}$
induces a finite and universal contribution to the Higgs wave
function, turning into a rescaling to all single-Higgs interactions.
In contrast, $\ope{HW}$ changes the momentum structure of the $WWH$
vertex and leads to momentum-enhanced effects in the $WH$ final state.
The operator $C_{Hq}^{(3)}$ has the unusual feature that it induces a
$q {\overline{q}}' WH$ 4-point interaction which avoids the
$s$-channel suppression of the SM-process and also leads to a momentum
enhancement in the $WH$ kinematics.

\begin{figure}[t]
  \centering
  \includegraphics[width=.40\linewidth]{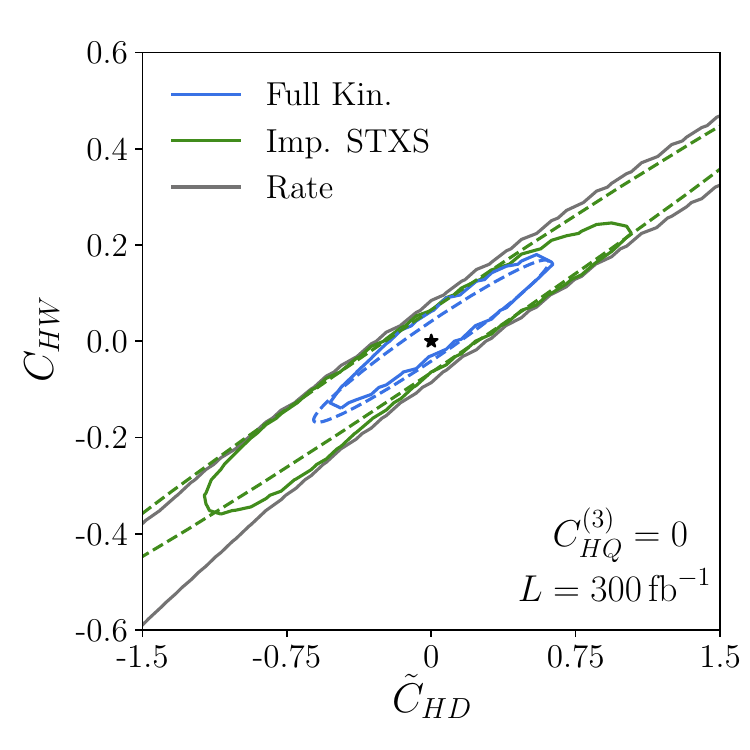}
  \hspace*{0.1\textwidth}
  \includegraphics[width=.40\linewidth]{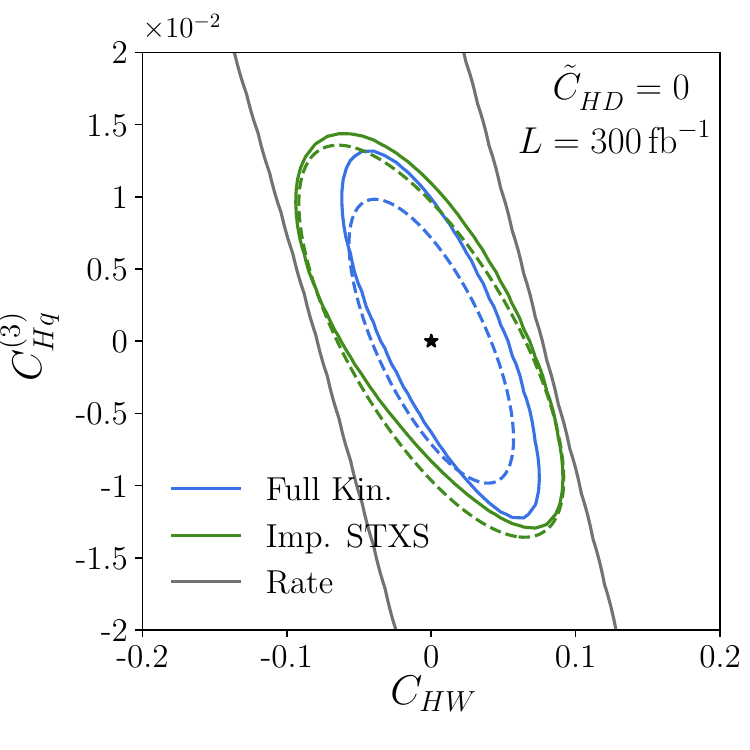}
  \caption{Expected exclusion limits from rates only (grey),
    simplified template cross sections (green), and using Sally to
    extract the score. Limits based on a linearized description in the
    Wilson coefficients are shown as dashed lines, while the solid
    lines also include squared contributions of the operators defined
    in Eq.\eqref{eq:wh_ops}. Figure from Ref.~\cite{Brehmer:2019gmn}.}
\label{fig:stxs}
\end{figure}

Based on what we know about the effects of the operators, we can study
their effects in the total production rate and a few specific
kinematic distributions,
\begin{align}
  \sigma_{WH}
  \qqquad
  p_{T,W} \approx p_{T,H} 
  \qqquad
  m_{T,\text{tot}} \sim m_{WH} \; .
  \label{eq:wh_obs}
\end{align}
Because of the neutrino in the final state it is hard to reconstruct
the invariant mass of the $WH$ final state, so we can use a transverse
mass construction as the usual proxy. These transverse mass
constructions typically replace the neutrino 3-momentum with the
measured missing transverse momentum and have a sharp upper cutoff at
the invariant mass, more details are given in
Ref.~\cite{Plehn:2009nd}.

For the simple $2 \to 2$ signal process given in Eq.\eqref{eq:wh_proc}
we would expect that the number of independent kinematic variables is
limited. At the parton level a $2 \to 2$ scattering process can be
described by the scattering angle, the azimuthal angle represents a
symmetry, and the embedding of the partonic process into the hadronic
process can be described by the energy of the partonic scattering as a
second variable. This means, measuring Wilson coefficients from
2-dimensional kinematic correlations should be sufficient. This leads
to a definition of a set of improved simplified template cross
sections (STXS) in terms of the two kinematic variables given in
Eq.\eqref{eq:wh_obs}. The extraction of the score using the Sally
method allows us to benchmark this 2-dimensional approach and to
quantify how much of the entire available information is captured by
them.

In Fig.~\ref{fig:stxs} we show two slices in the model space spanned
by the operators and Wilson coefficients given in
Eq.\eqref{eq:wh_ops}. In the left panel we first see that even for two
operators the rate measurement leaves us with an approximately flat
direction in model space. Adding 2-dimensional kinematic information
breaks this flat directions, but this breaking really requires us to
also include the squared contributions of the Wilson coefficients,
specifically $\ope{HW}$. Using the full phase space information leads
to significantly better results, even for the simple $2 \to 2$
process. The reason is that we are not really comparing the
SM-predictions and the dimension-6 signal hypothesis, but are also
including the continuum background to $\ell \bar{\nu} b \bar{b}$
production, and in the tails of the signal kinematics this continuum
background becomes our leading limitation. In the right panel of
Fig.~\ref{fig:stxs} we show another slice through the model parameter
space, now omitting the operator $\ope{HD}$ which is only constrained
by the total $WH$ rate. Again, we have a perfectly flat direction from
the rate measurement, but the two remaining operators can be
distinguished already in the linearized description from their effects
over phase space, and the 2-dimensional kinematic information captures
most of the effects encoded in the full likelihoods.  This example is
not an actual measurement, but only a study for the LHC Run~3, but it
shows how we can employ simulation-based inference in a more classical
way, with binned kinematic distributions, and using the full
likelihood information. It should be obvious which methods does
better.

\subsubsection{Flow-based anomaly detection}
\label{sec:gen_cathode}

Now that we can extract likelihoods from event samples, we need to ask
what kind of data we would train on. In the last section we used
supervised training on simulated event samples for this purpose. An
alternative way would be to use measured data to extract
likelihoods. Obviously, this has to be done with some level of
\underline{unsupervised} training.  In Sec.~\ref{sec:auto_dense} we
introduced the CWoLa method and its application to bump hunt searches
for BSM physics. Our goal is to use unsupervised likelihood extraction
to enhance such bump hunts\index{anomaly detection}.

\begin{figure}[t]
\centering
\includegraphics[width=0.60\textwidth]{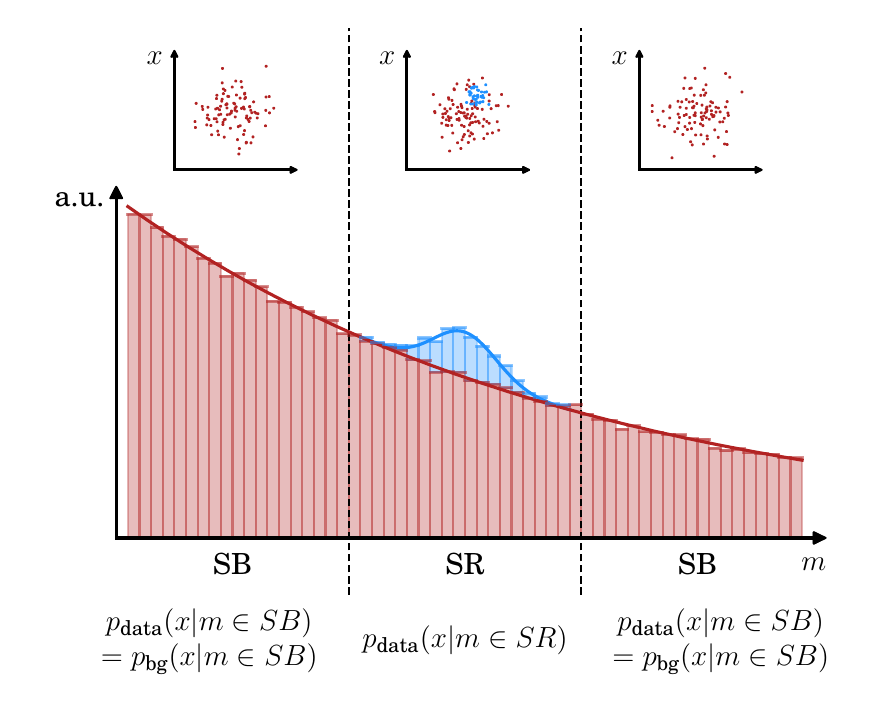}
\caption{Definition of signal region (SR) and side band (SB).  The red
  distribution is the assumed background, the blue is the
  signal. Features of the events other than $m$ are called $x$. Figure
  from from Ref.\cite{Hallin:2021wme}.}
\label{fig:CATHODE_SR}
\end{figure}

The general setup of such searches is illustrated in
Fig.~\ref{fig:CATHODE_SR}: We start with a smooth feature $m$, for
example the di-jet or di-lepton invariant mass, and look for signal
bumps rising above the falling background. In addition, we assume that
the signal corresponds a local overdensity in other event features,
collectively referred to as $x$. The problem we want to address is how
to estimate for the background in a data-driven way all over phase
space. Usually, one defines a signal region (SR) and a side band (SB),
and estimates the amount of background in the signal region based on
the side bands. One then scans over different signal region
hypotheses, covering the full feature space in $m$. A simple technical
method is to fit the $m$-distribution, remove individual bins, and
look for a change in $\chi^2$.  This data-driven approach has the
advantage that it does not suffer from imperfect background
simulations. In the Higgs discovery plots it looks like this is what
ATLAS and CMS did, but this is not quite true, because the information
on the additional features in $x$ was crucial to enhance the
statistical power of the side band analysis. Using the
machine-learning techniques we discussed so far, we can approach this
problem from several angles.

As always, the Neyman-Pearson lemma\index{Neyman-Pearson lemma} ensures that the best
discriminator between two hypotheses is the likelihood ratio, with the
specific relation given in Eqs.\eqref{eq:gan_dopt}
and~\eqref{eq:lr_opt}.  The key point of the \underline{CWoLa}\index{CWoLa} method
in Sec.~\ref{sec:auto_dense} is that a network classifier learns a
monotonic function of this ratio, if the two training dataset have
different compositions of signal and background. If we identify the
two training datasets with a data-driven modelling of the background
likelihood and a measured signal+background likelihood we can use
CWoLa to train a signal vs background classifier using the link
\begin{align}
\frac{x \sim \pd(x|m \in SR)}{x \sim \pd(x|m \in SB)}
  \; \stackrel{\text{class}}{\longrightarrow}
\frac{p_{S+B}(x)}{p_B(x)}
\; \to \; 
\frac{p_S(x)}{p_B(x)}  \; .
\label{eq:cathode_cwola}
\end{align}
In the last \underline{CWoLa} step we train a classifier to distinguish events
from the signal and background regions, only using the features $x$
and not $m$. The challenge with this method is the fine print in the
CWoLa method, which essentially requires $m$ and $x$ to not be
correlated, so we can ignore the fact that in
Eq.\eqref{eq:cathode_cwola} the numerator and denominator are
evaluated in different phase space regions.

If we could use some kind of network to access likelihoods directly,
this would allow us to construct the likelihood ratio simply as a
ratio of two independently learned likelihoods. This method is called
ANOmaly detection with Density Estimation or \underline{Anode}. Here,
one flow learns the assumed background density from the side bands,
$\pmd(x | m\in SB)$, where the label `model' indicates that we will
use this distribution to model the background in the signal
region. Since the label $m$ is a continuous variable, we can extract
the density of events in the signal region from $\pmd(x | m)$ simply
by passing events with $m\in SR$ through it and using the network to
interpolate in $m$.  For this to happen, the flows have to learn the
conditional density $p(x | m)$ instead of the joint density $p(x, m)$
because we need to be able to interpolate into the signal region to
compute $\pmd(x | m\in SR)$. If a flow has learned $p(x, m)$ and then
sees $m\notin SR$, it will return $p(x, m\in SR)=0$ instead of
interpolating from the side bands into the signal region. This means
Anode approximates the data distribution in the background regions and
then relies on an interpolation to reach the signal region,
\begin{align}
  \pd(x|m \in SB)
  \; \stackrel{\text{train}}{\longrightarrow}
  \pmd(x|m \in SB)
  \; \stackrel{\text{interpolate}}{\longrightarrow}
  \pmd(x|m \in SR) \; .
\end{align}
The interpolated $\pmd(x | m)$ for all $m$ is given by the red
distribution in Fig.~\ref{fig:CATHODE_SR}.  The second network
directly learns a hypothetical signal-plus-background density from the
signal region $\pd (x | m\in SR)$. The learned density in the signal
region includes the hypothetical signal, trained in an unsupervised
manner. We can now form the ratio of the two learned densities in the
signal region, each encoded in a network,
\begin{align}
  \frac{\pd(x | m\in SR)}{\pmd(x | m\in SR)} \; ,
  \label{eq:lr_cathode}
\end{align}
which approaches the desired log-likelihood ratio and can be used for
analysis.

Finally, we can go beyond classification and density estimation and
make use of the fact that normalizing flows are not only density
estimators, but generative networks. The corresponding method of
enhancing a bump hunt is to combine the ideas of CWoLa\index{CWoLa} and Anode to
Classifying Anomalies THrough Outer Density Estimation or
\underline{Cathode}. First, we learn the distribution of events in the
side band, $\pmd(x|m \in SB)$, just like in the Anode approach.
Second, using this density estimator, $\pmd(x|m)$, we sample
artificial events in the signal region
\begin{align}
  \pd(x|m \in SB)
  \; \stackrel{\text{train}}{\longrightarrow}
  \pmd(x|m \in SB)
  \; \stackrel{\text{sample}}{\longrightarrow}
  x \sim \pmd(x|m \in SR) \; .
\end{align}
To guarantee the correct distribution of the continuous condition $m$
we use a \underline{kernel density estimator}. This is a
parameter-free method for density estimation, where for instance a
Gaussian kernel is placed at each element of the dataset and the sum
of all Gaussians at a given point is an estimator for the points
density. Selecting a Gaussian at random and sampling from it draws
samples from this distribution. The method is less efficient than
normalizing flows, especially at high dimensions and for large
datasets, but it is well-suited to model the $m$-dependence in the
signal region from the side bands.

\begin{figure}[t]
\centering
\includegraphics[width=0.30\textwidth]{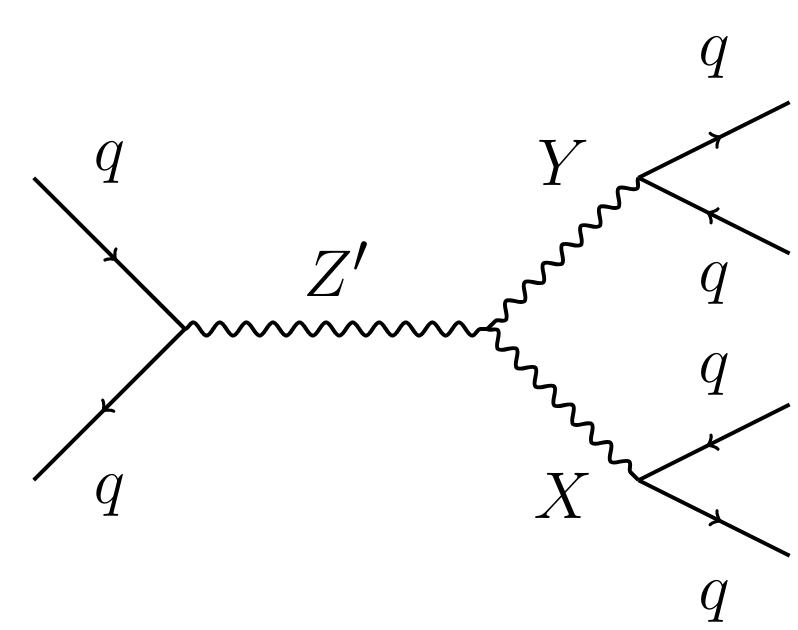}
\caption{Signal process in the LHC Olympics R\&D dataset, from
  Ref.~\cite{Kasieczka:2021xcg}.}
\label{fig:CATHODE_process}
\end{figure}

The sampled events will follow the distribution $\pmd(x|m \in SR)$,
expected for background-only events in the signal region and again
corresponding to the red distribution in Fig.~\ref{fig:CATHODE_SR}. If
there are actual signal events in the signal region we can use the two
datasets following $\pmd(x | m\in SR)$ and $\pd(x | m\in SR)$ to train
a CWoLa classifier, as illustrated in Eq.\eqref{eq:cathode_cwola}.
This means the third step of Cathode is to apply the CWoLa method: we
train a classifier to distinguish the generated events from the data
in the signal region. If there is a signal present, the classifier
learns to distinguish the two sets based on the log-likelihood ratio
argument of Eq.\eqref{eq:cwola}.

Cathode has several advantages over the other bump-hunt enhancing
algorithms.  First, unlike for CWoLa the data and the generated
samples have $m\in SR$, so even when the features $x$ are correlated
with $m$, the classifier will not learn to distinguish the sets by
deducing the value of $m$ from $x$.  Second, in contrast to CWoLa, the
amount of training data is not limited.  Cathode can oversample events
in the signal region. which improves the quality of the classifier the
same way we have seen with GANplification in
Sec.~\ref{sec:gen_gan_amp} and super-resolution in
Sec.~\ref{sec:gen_gan_super}. Finally, compared to Anode, the
likelihood-ratio in Eq.\eqref{eq:lr_cathode} is learned from a
classifier and not constructed as a ratio of two learned
log-likelihoods. This is easier and gives more stable results.

\begin{figure}[t]
\centering
\includegraphics[width=0.60\textwidth]{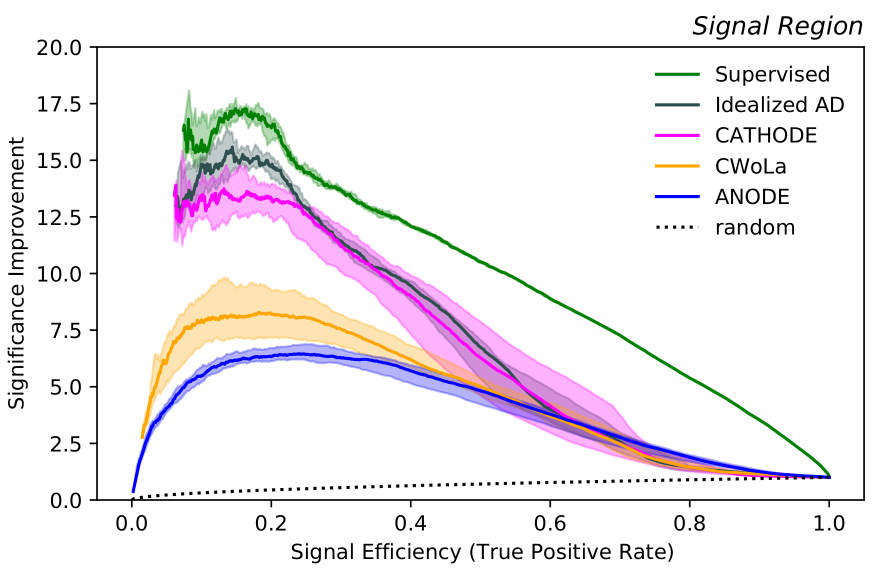}
\caption{Significance improvement characteristic (SIC) for different
  bump-hunting methods. Figure from Ref.~\cite{Hallin:2021wme}.}
\label{fig:CATHODE_SIC}
\end{figure}

As an application we compare CWola, Anode, and Cathode to a standard
new-physics searches dataset, namely the \underline{LHC Olympics} R\&D
dataset. This dataset consists of simulated di-jet events from Pythia
and with the fast detector simulation Delphes. It includes 1M QCD di-jet
events as background and 1k signal events describing the process
\begin{align}
  Z' \to X (\to qq) Y(\to qq)
  \qquad \text{with} \quad
  m_{Z'} = 3.5~\tev, m_X = 500~\gev, m_Y = 100~\gev \; .
\end{align}
All events are required to satisfy a single-jet trigger with
$p_T>1.2$~TeV.  We use the kinematic training features
\begin{align}
  \{ \; m_{j_1j_2}, m_{j_1}, m_{j_2}-m_{j_1},
  \tau_{21}^{j_1}, \tau_{21}^{j_2} \; \} \; .
\end{align}
Here $j_1$ and $j_2$ refer to the two highest-$p_T$ jets ordered by
jet mass ($m_{j_1}<m_{j_2}$), and $\tau_{ij} \equiv \tau_i/\tau_j$ are
their $n$-subjettiness ratios defined in Eq.\eqref{eq:tau_N}. The
di-jet invariant mass $m_{j_1j_2}$ defines the signal and side band
regions through
\begin{align}
  SR: \quad m_{jj} = 3.3~...~3.7~\tev \; .
\end{align}  
In Fig.~\ref{fig:CATHODE_SIC} we see how the three methods perform in
terms of the \underline{significance improvement characteristic}
(SIC), which gives the improvement factor of the statistical
significance $S/\sqrt{B}$. The supervised anomaly detector as a
reference is given by a classifier trained to distinguish perfectly
modeled signal from background. The idealized anomaly detector shows
the best possible performance of a data-driven anomaly detector. It is
trained to distinguish perfectly modeled background from the signal
plus background events in the signal region. Of the three methods
Cathode clearly outperforms Anode and CWoLa and approaches the
idealized anomaly detection over a wide range of signal efficiencies,
indicating that the distribution of artificial events in the signal
region follows the true background distribution closely.

\subsubsection{Symbolic regression of optimal observables}
\label{sec:cond_sim_sr}

After learning how to extract likelihood ratios for a given inference
task from simulations and from data, we want to return to particle
theory and ask what we can learn from these numerical methods for
particle theory. This is an example where we expect a numerical code
or a neural network to correspond, at least approximately, to a
formula. At least for physics, such an approximate formula would be
the perfect explanation or illustration of our neural
network. Formulas also have practical advantages over neural networks
--- we know that neural networks are great at interpolating, but
usually provide very poor extrapolations. If we combine networks with
formulas as models, these formulas will provide a much better
\underline{extrapolation}. This means is that in a field where all
improvements in precision involve a step from analytic to numerical
expressions, we need a way to at least approximately invert this
direction. This is a little different from typical applications of
symbolid regression, where we start from some kind of data, use a
neural network to extract the relevant information, and then turn this
relevant information into a formula. We really just want to see a
formula for a neural network in the sense of explainable AI.

In Eq.\eqref{eq:def_score} we have introduced the score or the optimal
observable\index{optimal observable} for the model parameter $\theta$ in a completely abstract
manner, and in Sec.~\ref{sec:gen_cathode} we have deliberately avoided
all ways to understand the likelihood ratio in terms of theory or
simulations.  However, in many LHC applications, for example in
measuring the SMEFT Wilson coefficients defined in
Eq.\eqref{eq:wh_ops}, we can at least approximately extract the score
or optimal observables. For those Wilson coefficients the natural
reference point is the Standard Model, $\theta_\text{ref} = 0$. At
parton level and assuming all particle properties can be observed, we
know from Eq.\eqref{eq:llr_matrix} that the likelihood is proportional
to the transition amplitude,
\begin{align}
p(x|\theta)
\propto |\mat(x|\theta)|^2
= |\mat(x)|^2_\text{ref} + \theta |\mat(x)|^2_\text{int} + \ord (\theta^2) \; ,
\label{eq:opt_obs1}
\end{align}
where $|\mat|^2_\text{int}$ denotes the contribution of the
interference term between Standard Model and new physics to the
complete matrix element. In that case the score becomes 
\begin{align}
  t(x|\theta_\text{ref}=0) 
= \frac{|\mat(x)|^2_\text{int}}{|\mat(x)|^2_\text{ref}} \; .
\label{eq:opt_obs2}
\end{align}
This formula illustrates that the score does not necessarily have to
be an abstract numerical expression, but that we should be able to
encode it in a simple formula, at least in a perturbative approach.

\begin{figure}[t]
  \centering
  \includegraphics[width=0.4\textwidth]{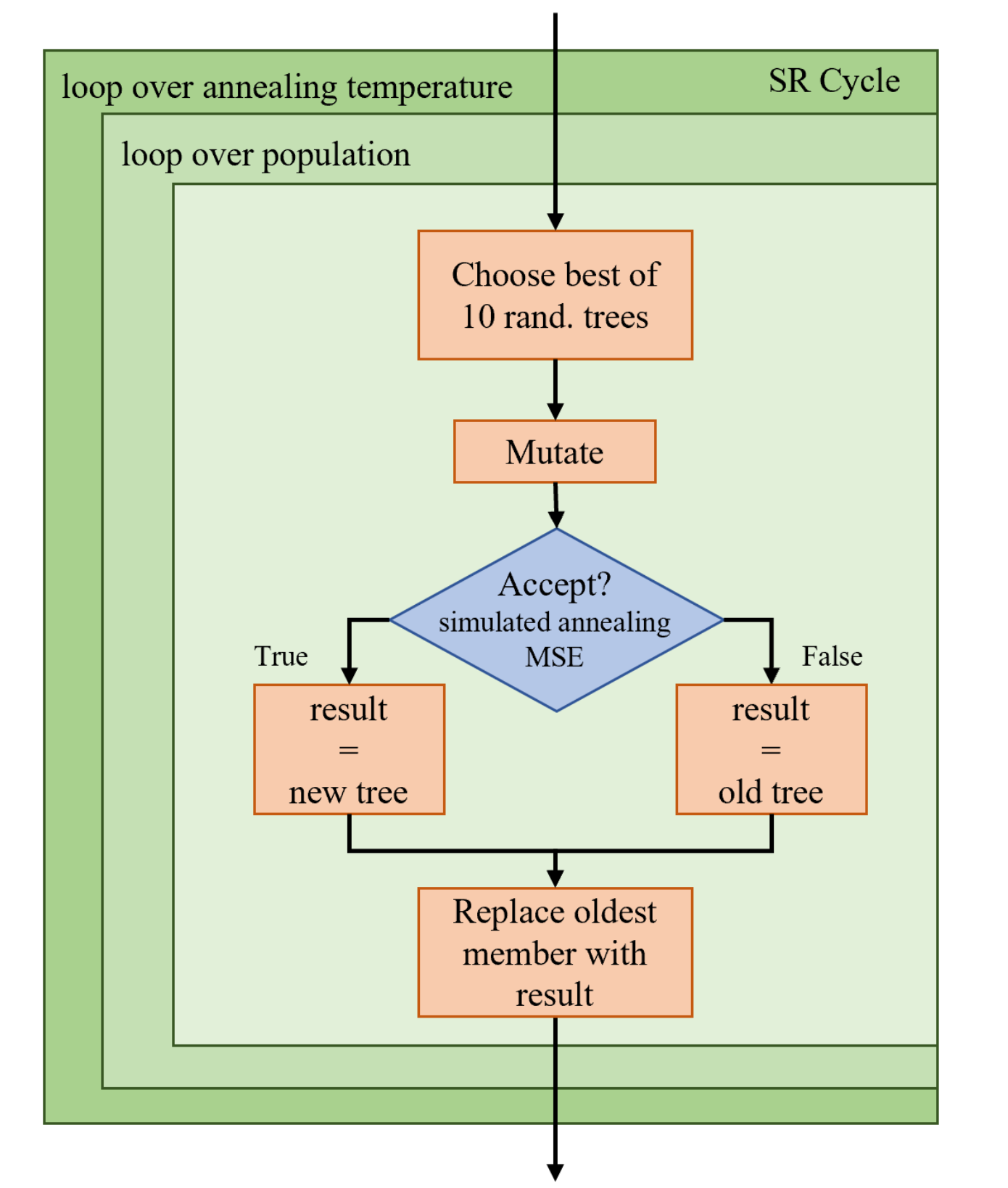}
  \caption{Illustration of the PySR algorithm.}
\label{fig:pysr}
\end{figure}

The approximate parton-level description of the score suggests that it
should be possible to derive a closed formula in terms of the usual
phase space observables. As the numerical starting point we use an
event generator\index{event generators} like MadGraph to generate a dataset of score values
over phase space.  Extracting formulas from numerical data is called
symbolic regression.  The standard application of symbolic regression
is in combination with a neural network, where a complex dataset is
first described by a neural network, extracting its main feature and
providing a fast and numerical powerful surrogate. This neural network
is then transformed into approximate formulas. In our case, we
directly approximate the numerically encoded score of
Eq.\eqref{eq:opt_obs2} with compact formulas.

The public tool \underline{PySR} uses a genetic algorithm to find a
symbolic expression for a numerically defined function in terms of
pre-defined variables.  Its population consists of symbolic
expressions, which are encoded as trees and consist of nodes which can
either be an operand or an operator function. The operators we need
are addition, subtraction, multiplication, squaring, cubing and
division, possibly sine and cosine. The algorithm is illustrated in
Fig.~\ref{fig:pysr}.  The tree population evolves when new trees are
created and old trees are discarded. Such new trees are created
through a range of mutation operators, for instance exchanging,
adding, or deleting nodes.  The figure of merit which we use to
evaluate the new trees is the MSE between the score values
$t(x_i|\theta)$ and the PySR approximation $g_i$, as defined in
Eq.\eqref{eq:mse_loss}
\begin{align}
  \text{MSE} \sim \sum_i \left[ g(x_i)-t(x_i|\theta) \right]^2 \; ,
\label{eq:def_mse}
\end{align}
Unlike for a network training we do not want an arbitrarily complex
and powerful formula, so we balance the MSE with the number of nodes a
complexity measure.  The PySR figure of merit is then defined with the
additional parsimony parameter through the regularized form
\begin{align}
  \boxed{
  \text{MSE}^*
  = \text{MSE}
  +\text{parsimony} \cdot \#_\text{nodes}
  } \; .
\label{eq:parsimony}
\end{align}
This combination will automatically find a good compromise between
accuracy and complexity.

Simulated annealing is a standard minimum finder algorithm. Its key
feature is a temperature $T$ which allows the algorithm to sample
widely in the beginning and then focus on a local minimum. It accepts
a mutation of a tree with the probability
\begin{align}
  p =
  \min \left[ \exp\left(-\frac{1}{T} \; \frac{\text{MSE}^*_\text{new}-\text{MSE}^*_\text{old}}{\text{MSE}^*_\text{old}}\right) , 1 \right] \; .
\label{eq:mutation1}
\end{align}  
If the new tree is better than the old tree, $\text{MSE}^*_\text{new}
< \text{MSE}^*_\text{old}$, the exponent is larger than zero, and the
probability to accept the new tree is one. If the new tree is slightly
worse than the old tree, the exponent is smaller than zero, and the
algorithm can keep the new tree as an intermediate step to improve the
population, but with a finite probability. This probability is rapidly
decreased when we dial down the temperature.  The output of the PySR
algorithm is a set of expressions given by the surviving populations
once the algorithm is done. This \underline{hall of fame} (HoF)
depends on the MSE balanced by the number of nodes.  If we are really
interested in the approximate function. we need to supplement PySR
with an optimization fit of all parameters in the HoF functions using
the whole dataset. Such a fit is too slow to be part of the actual
algorithm, but we need it for the final form of the analytic score and
for the uncertainties on this form.

\begin{figure}[t]
  \includegraphics[width=0.325\textwidth,page=1]{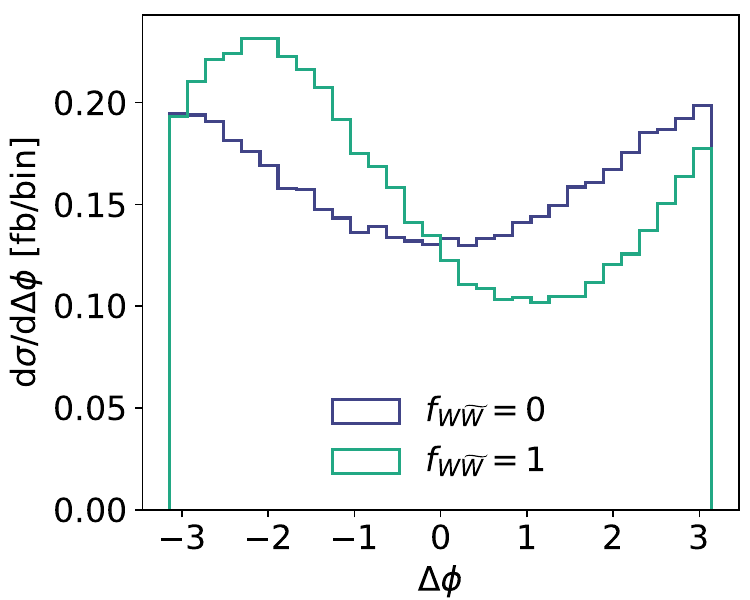}
  \includegraphics[width=0.325\textwidth,page=2]{WBF_histograms}
  \includegraphics[width=0.325\textwidth,page=3]{WBF_histograms}
  \caption{Kinematic distributions for the two tagging jets in WBF
    Higgs production at parton level for two Wilson coefficients
    $f_{W\widetilde{W}}$.}
  \label{fig:wbf_hist_obs}
\end{figure}

There are cases where optimal observables or scores are used for
instance by ATLAS to test fundamental properties of their datasets.
One such fundamental question is the CP-symmetry\index{symmetries} of the $VVH$ vertex,
which can for instance be tested in weak boson fusion Higgs
production. It turns out that for this application we know the
functional form of the score at leading order and at the parton level,
so we can compare it to the PySR result going beyond these simple
approximations. The signal process we need to simulate is 
\begin{align}
  pp \to Hjj  \; .
\end{align}
To define an optimal observable\index{optimal observable} we choose the specific CP-violating
operator at dimension six, in analogy to Eq.\eqref{eq:def_eft},
\begin{align}
  \lag = \lag_\text{SM} + \frac{f_{W\widetilde{W}}}{\Lambda^2} \ope{W \widetilde{W}}
  \qquad \text{with} \qquad 
  \ope{W \widetilde{W}} = - (\phi^{\dagger}\phi) \; \widetilde{W}_{\mu\nu}^kW^{\mu\nu k} \; .
\label{eq:def_fwwt}
\end{align}
We know that the signed azimuthal angle between the tagging jets
$\Delta \phi$ is the appropriate genuine CP-odd observable. This
observable has the great advantage that it is independent of the Higgs
decay channels and does not involve reconstructing the Higgs
kinematics.  In Fig.~\ref{fig:wbf_hist_obs} we show the effect of this
operator on the WBF kinematics. First, the asymmetric form of $\Delta
\phi$ can most easily be exploited through an asymmetry
measurement. Second, the additional momentum dependence of $\ope{W
  \widetilde{W}}$ leads to harder tagging jets This is unrelated to
CP-violation, and we have seen similar effects in
Sec.~\ref{sec:cond_sim_like}. Finally, there is no effect on the jet
rapidities.

\begin{figure}[b!]
  \includegraphics[width=0.325\textwidth,page=1]{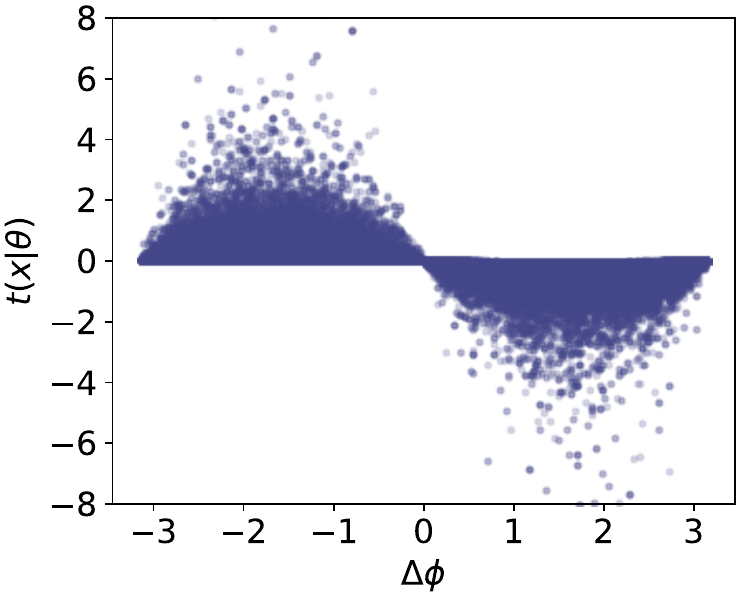}
  \includegraphics[width=0.325\textwidth,page=2]{WBF_score_fw0}
  \includegraphics[width=0.325\textwidth,page=3]{WBF_score_fw0}
  \caption{Score for simplified WBF Higgs production at parton level
    and with $f_{W\widetilde{W}} = 0$. The kinematic observables
    correspond to Fig.~\ref{fig:wbf_hist_obs}.}
  \label{fig:fww0_wbf}
\end{figure}

For the leading partonic contribution to $WW$-fusion, $u d \to H d u$
we can compute the score at the Standard Model point,
\begin{align}
t(x|f_{W\widetilde{W}} = 0 ) \approx -\frac{8 v^2}{m_W^2} \frac{(k_d k_u)+(p_u p_d)}{(p_d p_u)(k_u k_d)} \; \epsilon_{\mu\nu\rho\sigma} \; k_d^\mu k_u^\nu p_d^\rho p_u^\sigma \; ,
\end{align}
where $k_{u,d}$ are the incoming and $p_{u,d}$ the outgoing quark
momenta.  We can assign the incoming momenta to a positive and
negative hemisphere, $k_\pm =(E_\pm,0,0,\pm E_\pm)$, do the same for
the outgoing momenta $p_\pm$, and then find in terms of the signed
azimuthal separation
\begin{align}
t(x|f_{W\widetilde{W}} = 0) \approx -\frac{8 v^2}{m_W^2} \frac{2E_+E_-+(p_+ p_-)}{(p_+ p_-)}p_{T+}p_{T-} \; \sin\Delta\phi \; .
\label{eq:score_wbf}
\end{align}
The dependence $t \propto \sin \Delta\phi$ reflects the CP-sensitivity
while the prefactor $t \propto p_{T+}p_{T-}$ reflects the dimension-6
origin.

\begin{table}[t]
	\begin{minipage}{0.66\textwidth}
		\begin{small}
			\begin{tabular}{cc|lr}
			\toprule
      compl& dof & function& MSE \\
			\midrule
3 & 1 &$a \, \Delta\phi$ & $1.30\cdot 10^{-1}$ \\
4 & 1 &$\sin(a \Delta\phi)$ & $2.75\cdot 10^{-1}$ \\
5 & 1 &$a\Delta\phi x_{p,1}$& $9.93\cdot 10^{-2}$ \\
6 & 1 &$-x_{p,1}\sin(\Delta\phi +a)$& $1.90\cdot 10^{-1}$ \\
7 & 1 &$(-x_{p,1}-a)\sin(\sin(\Delta\phi))$& $5.63\cdot 10^{-2}$ \\
8 & 1 &$(a-x_{p,1})x_{p,2}\sin(\Delta\phi)$& $1.61\cdot 10^{-2}$ \\
14& 2 &$x_{p,1}(a\Delta\phi -\sin(\sin(\Delta\phi)))(x_{p,2}+b)$& $1.44\cdot 10^{-2}$ \\
15& 3 &$-(x_{p,2}(a\Delta\eta^2+x_{p,1})+b)\sin(\Delta\phi+c)$& $1.30\cdot 10^{-2}$ \\
16& 4 &$-x_{p,1}(a-b\Delta\eta)(x_{p,2}+c)\sin(\Delta\phi+d)$ & $8.50\cdot 10^{-3}$ \\
\multirow{2}{*}{28} &\multirow{2}{*}{7} & $(x_{p,2}+a)(bx_{p,1}(c-\Delta\phi)$ & \multirow{2}{*}{$8.18\cdot 10^{-3}$} \\
  & &$\; \; -x_{p,1}(d\Delta\eta+ex_{p,2}+f)\sin(\Delta\phi+g))$& \\
			\bottomrule
			\end{tabular}
		\end{small}
	\end{minipage}
	\begin{minipage}{0.34\textwidth}
		\centering
		\includegraphics[width=\textwidth,page=1]{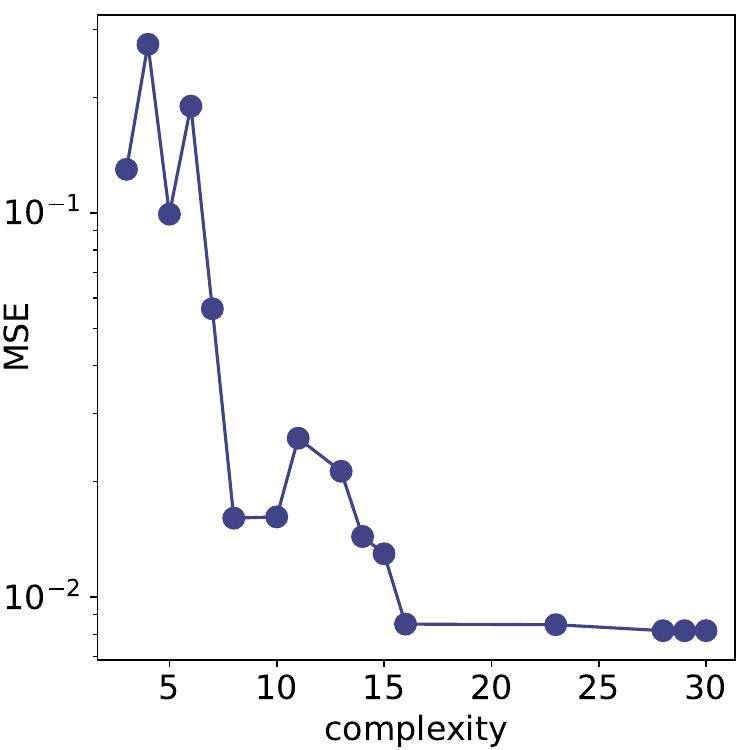}
	\end{minipage}
\caption{Score hall of fame for simplified WBF Higgs production with
  $f_{W\widetilde{W}} =0$, including a optimization fit.}
\label{tab:fww0_wbf}
\end{table}

For small deviations from the CP-conserving Standard Model we show the
dependence on the score in Fig.~\ref{fig:fww0_wbf}. The kinematic
observables are the same as in Fig.~\ref{fig:wbf_hist_obs}, and we see
how $f_{W\widetilde{W}} >0$ moves events from $\Delta \phi > 0$ to
$\Delta \phi < 0$. The dependence on $p_{T,1}$ indicates large
absolute values of the score for harder events, which will boost the
analysis when correlated with $\Delta \phi$. The dependence on $\Delta
\eta$ is comparably mild.  To encode this score dependence in a
formula we use PySR on the observables
\begin{align}
  \left\{ \; x_{p,1}, \; x_{p,2}, \; \Delta \phi, \; \Delta \eta \; 
  \right\}
  \qquad \text{with} \qquad x_{p,j} = \frac{p_{T,j}}{m_H}  \; .
\label{eq:def_four_obs}
\end{align}
In Tab.~\ref{tab:fww0_wbf} we show the results, alongside the
improvement in the MSE. Starting with the leading dependence on
$\Delta \phi$, PySR needs 8 nodes with one free parameter to derive $t
\approx p_{T,1} p_{T,2} \sin \Delta\phi$. Beyond this point, adding
$\Delta \eta$ to the functional form leads to a further improvement
with a 4-parameter description and 16 nodes. The corresponding formula
for the score is
\begin{align}
&  t(x_{p,1},x_{p,2},\Delta \phi,\Delta \eta|f_{W\widetilde{W}}=0) =
  - x_{p,1} \left( x_{p,2}+c \right) \left( a - b\Delta\eta \right) \sin(\Delta\phi+d) \notag \\
  &  \quad \text{with}
  \quad a=1.086(11)
  \quad b=0.10241(19)
  \quad c=0.24165(20)
  \quad d =0.00662(32) \; .
\label{eq:score_wbf_fww0}
\end{align}
The numbers in parentheses give the uncertainty from the optimization
fit. While $d$ comes out different from zero, it is sufficiently small
to confirm the scaling $t \propto \sin \Delta \phi$. Similarly, the
dependence on the rapidity difference $\Delta \eta$ is suppressed by
$b/a \sim 0.1$. This simple picture will change when we move away from
the Standard Model and evaluate the score are finite
$f_{W\widetilde{W}}$. However, for this case we neither have a
reference results nor an experimental hint to assume such CP-violation
in the $HVV$ interaction.

Until now, we have used PySR to extract the score at parton level. The
obvious question is what happens when we add a fast detector
simulation to the numerical description of the score.  To extract the
(joint) score from the MadGraph event samples we can use MadMiner.  In
general, detector effects will mostly add noise to the data, which
does affect the PySR convergence. We find the same formulas as without
detector effects, for instance the 4-parameter formula of
Eq.\eqref{eq:score_wbf_fww0}, but with different parameters,
\begin{align}
&  t(x_{p,1},x_{p,2},\Delta \phi,\Delta \eta|f_{W\widetilde{W}}=0) =
  - x_{p,1} \left( x_{p,2}+c \right) \left( a - b\Delta\eta \right) \sin(\Delta\phi+d) \notag \\
  &  \quad \text{with}
  \quad a=0.9264(20)
  \quad b=0.08387(35)
  \quad c=0.3542(20)
  \quad d =0.00911(67) \; .
\end{align}
The absolute differences are small, even though the pull for instance
of $c$ exceeds 100.  Still, $d \ll 1$ ensures $t \propto \sin \Delta
\phi$ also after detector effects, and $b/a \ll 1$ limits the impact
of the rapidity observable. Indeed, detector effects do not introduce
a significant bias to the extracted score function.

\begin{figure}[t]
  \centering
  \includegraphics[width=0.60\textwidth,page=1]{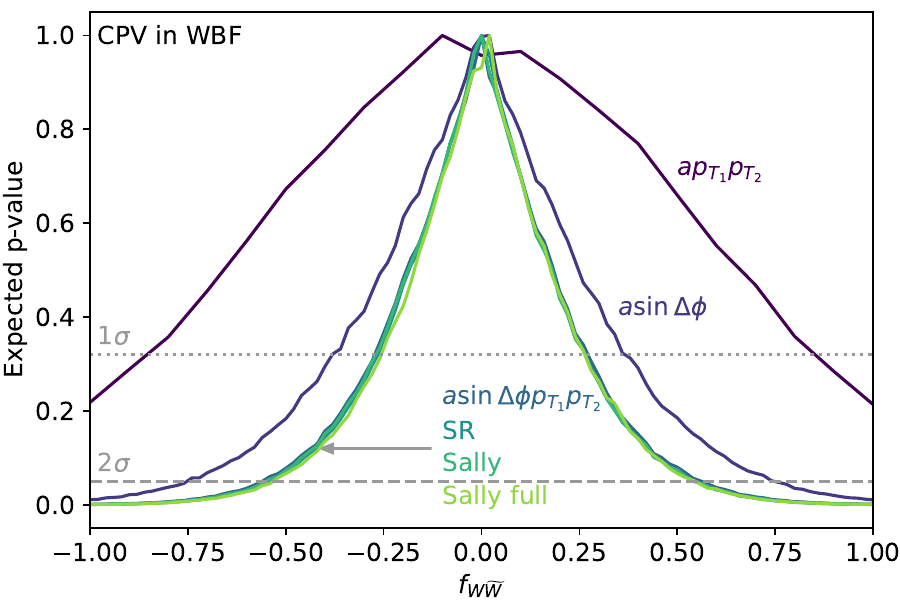}
  \caption{Projected exclusion limits assuming $f_{W\widetilde{W}}=0$ for different
    (optimal) observables. The Sally network uses $p_{T_1}$, $p_{T_2}$,
    $\Delta\phi$ and $\Delta\eta$, Sally full uses 18 kinematic
    variables.}
    \label{fig:limits}
\end{figure}

Finally, we can switch from the $\text{MSE}^*$ figure of merit
something more realistic, for instance the expected reach of the
different approximations in an actual analysis.  To benchmark the
reach we can compute the log-likelihood distributions and extract the
$p$-value for an assumed $f_{W\widetilde{W}} = 0$ including detector
effects and for an integrated Run~2 LHC luminosity of $139~\ifb$. The
analytic functions we use from the HoF in Fig.~\ref{tab:fww0_wbf} are
\begin{align}
  a_1p_{T,1}p_{T,2} \qqquad
  a_2\sin\Delta\phi \qqquad 
  a_3p_{T,1}p_{T,2}\sin\Delta\phi \; .
\end{align}
The first is just wrong and does not probe CP-violation at all. The
second taylors the proper results for small azimuthal angles, which
does appear justified looking at Fig.~\ref{fig:wbf_hist_obs}. The
third function is what we expect from the parton level.
We compare these three results to the Sally method using the four PySR
observables in Eq.\eqref{eq:def_four_obs}, and using the full set of
18 observables. All exclusion limits are shown in
Fig.~\ref{fig:limits}. Indeed, for all score approximations the
likelihood follows a Gaussian shape. Second, we find that beyond the
minimal reasonable form $ap_{T1}p_{T2}\sin\Delta\phi$ there is only
very little improvement in the expected LHC reach for the moderate
assumed luminosity.

This application of symbolic regression confirms our hope that we can
use modern numerical methods to reverse the general shift from
formulas to numerically encoded expressions. In the same spirit as
using perturbative QCD we find that symbolic regression provides us
with useful and correct approximate formulas, and formulas are the
language of physics.  The optimal observable for CP-violation is the
first LHC-physics formula re-derived using modern machine learning,
and at least for Run~2 statistics it can be used in experiment without
any loss in performance. However, when analyszing more data we have to
extract a more precise formula, again a situation which is standard in
physics, where most of our simple formulas rely on a taylor series or
a perturbative expansion.\bigskip

With this appliation we are at the end of our tour of modern machine
learning and its LHC applications. We have established neural networks
as extremely powerful numerical tools, which are not black boxes, but
offer a lot of control. First, an appropriate loss function tells us
exactly what the network training is trying to achieve. Second, neural
network output comes with an uncertainty band, in some cases like NNPDF even
with a comprehensive uncertainty treatment. And, finally, trained
neural networks can be transformed into formulas. Given what we can do
with modern machine learning at the LHC, there is no excuse to not
play with them as new and exciting tools. What we have not talked
about is the unifying power of data science and machine learning for
the diverging fields of particle theory and particle experiment --- to
experience this, you will have to come to a conference like ML4Jets or
a workshop like Hammers and Nails and watch for yourselves.

\clearpage
\bibliographystyle{tepml}
\bibliography{tilman,literature}
\end{fmffile}

\printindex

\end{document}